\documentclass[aps,prb,twocolumn]{revtex4}
\usepackage{graphicx} %Include figure files
\newcommand{\be}{\begin{equation}}
\newcommand{\ee}{\end{equation}}
\newcommand{\bea}{\begin{eqnarray}}
\newcommand{\eea}{\end{eqnarray}}

\begin{document}

\title{The Puzzle of High Temperature Superconductivity in \\
Layered Iron Pnictides and Chalcogenides}

\author{David C. Johnston}
\altaffiliation[email: ]{johnston@ameslab.gov}
\affiliation{
Ames Laboratory and Department of Physics and Astronomy, Iowa State University, Ames, Iowa 50011, USA\\
\\
{\bf  This is a preprint of an article accepted for publication in Advances in Physics}  
} 
\date{\today}

\begin{abstract}
The response of the worldwide scientific community to the discovery in 2008 of superconductivity at $T_{\rm c} = 26$~K in the Fe-based compound LaFeAsO$_{1-x}$F$_x$ has been very enthusiastic.  In short order, other Fe-based superconductors with the same or related crystal structures were discovered with $T_{\rm c}$ up to 56~K\@.  Many experiments were carried out and theories formulated to try to understand the basic properties of these new materials and the mechanism for $T_{\rm c}$.  In this selective critical review of the experimental literature, we distill some of this extensive body of work, and discuss relationships between different types of experiments on these materials with reference to theoretical concepts and models.  The experimental normal-state properties are emphasized, and within these the electronic and magnetic properties because of the likelihood of an electronic/magnetic mechanism for superconductivity in these materials.

\end{abstract}

%\pacs{PACS numbers go here}

\maketitle

\tableofcontents

\clearpage

\section{\label{SecIntro} Introduction and Overview}

\begin{figure}
\includegraphics [width=2.5in]{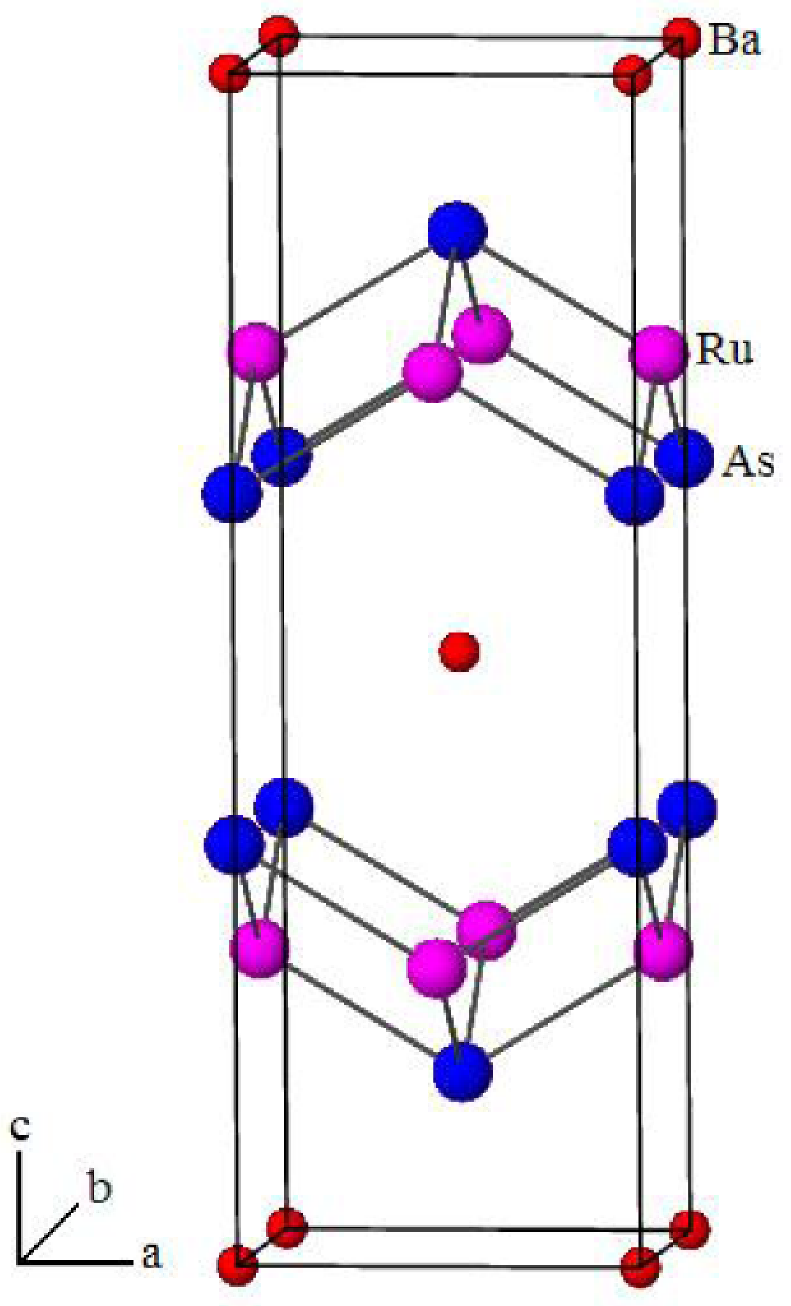}
\caption{(Color online) Crystal structure of BaRu$_2$As$_2$ with the body-centered-tetragonal ThCr$_2$Si$_2$-type structure (space group \emph{I}4/\emph{mmm}).\cite{Nath2009} The Ru atoms form a planar square lattice in the $a$-$b$ plane where each Ru atom is at the center of a distorted tetrahedron of As atoms, and the Ru$_2$As$_2$ layers alternate with Ba layers along the $c$-axis.  Reprinted with permission from Ref.~\onlinecite{Nath2009}.  Copyright (2009) by the American Physical Society.}
\label{Struct122} 
\end{figure}

\begin{figure*}
\includegraphics [width=2.1in,angle=-90]{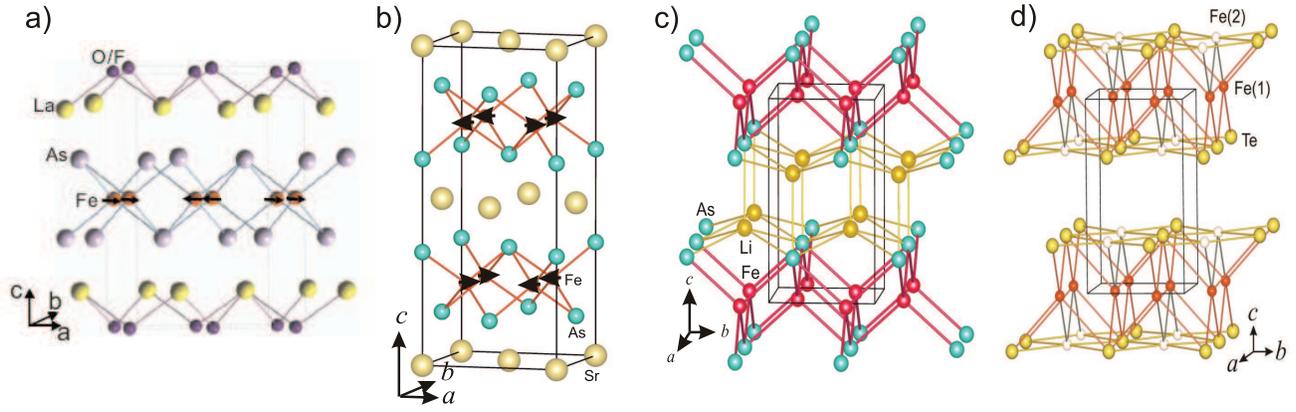}
\caption{(Color online) Comparison of the crystal structures of (a) LaFeAsO$_{1-x}$F$_x$, (b) ${\rm SrFe_2As_2}$, (c) LiFeAs, and (d) Fe$_{1+x}$Te.  Each of these structures contains a square lattice of Fe atoms at high temperatures that can distort at low temperatures. Each Fe atom is tetrahedrally coordinated by As (a,b,c) or Te (d).  In (b), the outline of the low-temperature orthorhombically distorted unit cell is shown, and ordered magnetic moments on the Fe atoms below the magnetic ordering temperature are shown by arrows.  In (d), the Fe(2) atoms are the extra $x$ atoms in Fe$_{1+x}$Te, with $x\sim 1$--10\%. Reproduced with permission from Ref.~\onlinecite{Lynn2009}, Copyright (2009), with permission from Elsevier. }
\label{Structures} 
\end{figure*}

\begin{figure}
\includegraphics [width=1.5in,viewport= 00 0 330 360,clip]{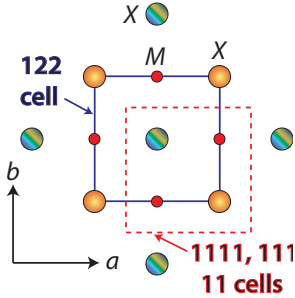}
\caption{(Color online) Projection of an $M$-$X$ ($M$ = metal; \emph{X = Pn, Ch}) layer onto the $ab$-plane of the 1111- (e.g., LaFeAsO), 111- (e.g., LiFeAs), 11- (e.g., FeSe) and 122-type (e.g., ${\rm BaFe_2As_2)}$ Fe-based compounds.  The structure of the $M$-$X$ layer is identical in all cases, with the composition $M_2X_2$ within the unit cell.  The basal plane unit cell of the 11-, 111- and 1111-type structures are simply shifted by $(\frac{a}{4},\frac{a}{4})$ with respect to the 122-type cell. The yellow $X$ atoms with a radial gradient are on one side of the $M$ layer and the $X$ atoms with diagonal stripes are on the other.  This puts each $M$ atom in a (generally distorted) tetrahedral coordination by $X$.}
\label{1111_122_layers} 
\end{figure}

%\squeezetable
\begin{table}
\caption{\label{Tcdata} Superconducting transition onset temperatures $T_{\rm c}$ for several FeAs-based and related compounds.}
\begin{ruledtabular}
\begin{tabular}{lcc}
 Compound   & $T_{\rm c}$ & Ref. \\
 & (K)  \\ \hline
LaFeAsO$_{0.89}$F$_{0.11}$ & 26 & \onlinecite{kamihara2008} \\
CeFeAsO$_{0.85}$ & 46.5 & \onlinecite{ren2008a}\\
NdFeAsO$_{1-y}$ & 54 & \onlinecite{Kito2008}\\
SmFeAsO$_{1-x}$F$_{x}$ & 55.0 & \onlinecite{ren2008}\\
Gd$_{0.8}$Th$_{0.2}$FeAsO & 56.3 & \onlinecite{Wang2008}\\
LaFePO$_{1-x}$F$_x$ & 5 & \onlinecite{Kamihara2006} \\
Ba$_{0.6}$K$_{0.4}$Fe$_2$As$_2$ & 38  & \onlinecite{rotter2008a}\\
Sr$_{0.6}$K$_{0.4}$Fe$_2$As$_2$ & 35.6  & \onlinecite{sasmal2008, chen2008d}\\
KFe$_2$As$_2$ & 3.8  & \onlinecite{sasmal2008}\\
LiFeAs & 18 & \onlinecite{Tapp2008R, Wang2008S, Pitcher2008C} \\
FeSe & 8 & \onlinecite{Hsu2008} \\
${\rm (Sr_4Sc_2O_6)Fe_2P_2}$ & 17  & \onlinecite{Ogino2009}\\
\end{tabular}
\end{ruledtabular}
\end{table}

The discovery\cite{kamihara2008} in 2008 of superconducting transition temperatures up to $T_{\rm c} =  26$~K in LaFeAsO$_{1-x}$F$_x$ with $x \sim 0.11$ with the primitive tetragonal ZrCuSiAs-type (1111-type) structure,\cite{Johnson1974} following the same group's earlier discovery in 2006 of superconductivity at $T_{\rm c}$ up to $\sim 5$~K in isostructural LaFePO$_{1-x}$F$_x$,\cite{Kamihara2006} captured the imaginations of physicists and chemists worldwide.  The crystal structure contains FeAs layers with Fe atoms in a square planar lattice arrangement, and these layers alternate with LaO layers along the $c$-axis.  Fe metal is a ferromagnet, and few would have anticipated that an Fe-containing material could show such an extraordinary $T_{\rm c}$.  Applying pressure increases the $T_{\rm c}$ even further to 43~K.\cite{Takahashi2008}  Remarkably, it was found that replacement of the nonmagnetic La by \emph{magnetic} rare earth elements substantially increased $T_{\rm c}$ (\emph{e.g.}, Refs.~\onlinecite{ren2008, ren2008a, Kito2008}) to its current record\cite{Wang2008} of 56.3~K for this structure class.  In contrast to the lower $T_{\rm c}$ materials, applying pressure to these higher $T_{\rm c}$ materials results in a \emph{decrease} in $T_{\rm c}$.\cite{Takeshita2008}  Subsequent work led to the identification of superconductors with similar FeAs layers in the body-centered-tetragonal ThCr$_2$Si$_2$-type (122-type) structure, shown in Fig.~\ref{Struct122} for the compound BaRu$_2$As$_2$.\cite{Nath2009}  The maximum $T_{\rm c}$ for this structure class so far is 38~K for Ba$_{0.6}$K$_{0.4}$Fe$_2$As$_2$.\cite{rotter2008a}  Other similar superconducting materials such as LiFeAs (111-type) containing FeAs layers have been reported.\cite{Tapp2008R, Wang2008S, Pitcher2008C}  More exotic superconducting compounds such as ${\rm (Sr_4Sc_2O_6)Fe_2P_2}$ contain thicker layers in between the Fe layers,\cite{Ogino2009} but the superconducting transitions of this class of materials are all very broad and these materials will therefore not be further discussed.  Even the binary compound $\alpha$-FeSe (11-type) with a layered structure becomes superconducting with $T_{\rm c} = 8$~K.\cite{Hsu2008} Attention has largely shifted from the 1111- to the 122- and 11-type compounds, even though the latter two classes of materials have lower $T_{\rm c}$, mainly because large single crystals of the latter compounds can be grown which allows more definitive characterizations of the properties, especially by neutron scattering, compared to $\lesssim 1$~mg size crystals of the 1111-type compounds.  Single crystals of the 1111-type compounds with mm size were recently grown by Yan et al.\ using a NaAs flux.\cite{Yan2009}   

The crystal structures of four of these classes of materials are compared in Fig.~\ref{Structures}.\cite{Lynn2009}  The basal planes of the 11-type, 111-type, 1111-type and 122-type compounds at room temperature are compared in more detail in Fig.~\ref{1111_122_layers}.  One sees that irrespective of the structure type, the layers are identical and the composition of a layer within the unit cell is $M_2X_2$, where $M$ is a metal atom and $X$ is either a pnictogen \emph{Pn} = P, As, Sb or chalcogen \emph{Ch} = S, Se, Te.  For this reason, and by comparison with Fig.~\ref{Structures}, one sees that there are two formula units (f.u.)\ per unit cell for all four types of compounds, but only one $M$-$X$ layer per unit cell in the 11-type and 1111-type compounds, because their formula unit contains only one $M$ atom, and two layers per unit cell in the 122-type compounds because this formula unit contains two $M$ atoms.  This property correlates with the $c$-axis parameters which increase in the order $\sim$~6~\AA, 8.5~\AA, and 12--13 \AA\ for the 11-type, 1111-type and 122-type compounds, respectively, whereas the $a$-axis is about the same ($\sim 4.0$~\AA)  (see the tables in the Appendix).  In all these compounds, the superconductivity is believed to be associated with the Fe square lattice layers.  A selection of $T_{\rm c}$ values obtained at ambient pressure is given in Table~\ref{Tcdata}.\cite{kamihara2008, Kamihara2006, ren2008a, Kito2008, ren2008, Wang2008, rotter2008a, Tapp2008R, Wang2008S, Pitcher2008C, Hsu2008, sasmal2008, chen2008d}

In this review we will emphasize the FeAs-based 11-, 122- and 1111-type compounds because they have been the most extensively studied among the Fe-based superconductors.  We note that the Fe(S,Se,Te) compounds tend to have significant nonstoichiometry, disorder and clustering problems that often result in smeared out properties such as wide superconducting transitions.  For large single crystals of Fe$_{1+y}$(Te$_{1-x}$Se$_{x}$), bulk superconductivity was observed for $x\approx 0.5,y\approx 0$, but the crystals were highly inhomogeneous.\cite{Sales2009a}  

A high resolution single crystal structure refinement of FeTe$_{0.56}$Se$_{0.44}$ indicated that the positions of the Se and Te atoms, which are crystallographically equivalent, are at different heights in the unit cell (see Table~\ref{data6} in the Appendix), arising from the quite different sizes of the Se and Te atoms.\cite{Tegel2010}  This Se/Te positional disorder results in disorder in the Fe atomic positions that is reflected in an elongation of the Fe thermal ellipsoid in the $c$-axis direction.\cite{Tegel2010}  Similar results were obtained by Louca and coworkers.\cite{Louca2010}

\begin{figure}
\includegraphics [width=3in]{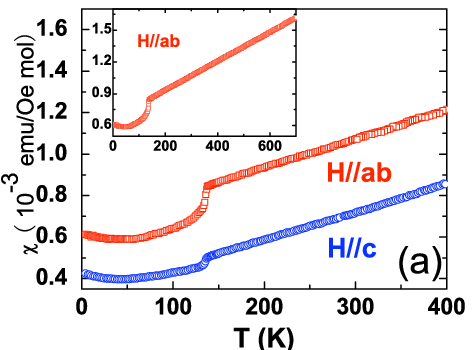}
\includegraphics [width=3.3in]{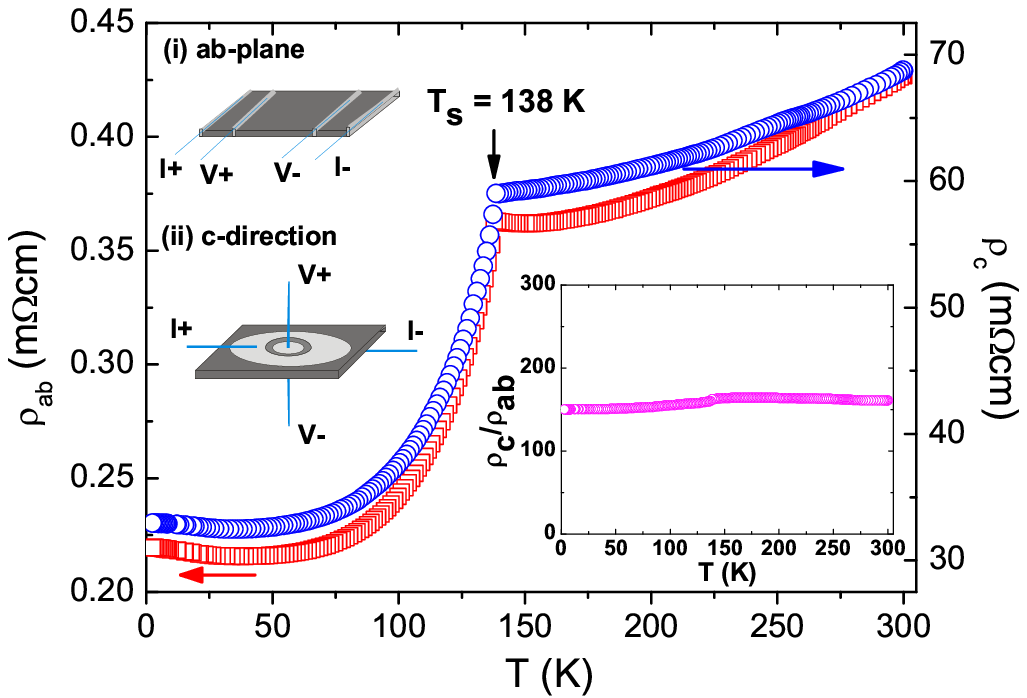}
\caption{(Color online) Anisotropic magnetic susceptibility $\chi$ and electrical resistivity $\rho$ versus temperature $T$ for single crystals of BaFe$_2$As$_2$.\cite{wang2008}  The coupled structural and magnetic transition at 138~K is apparent in both measurements.  Reprinted with permission from Ref.~\onlinecite{wang2008}.  Copyright (2009) by the American Physical Society.}
\label{FigChiRhoBaFe2As2} 
\end{figure}

\begin{figure*}
\includegraphics [width=3.5in]{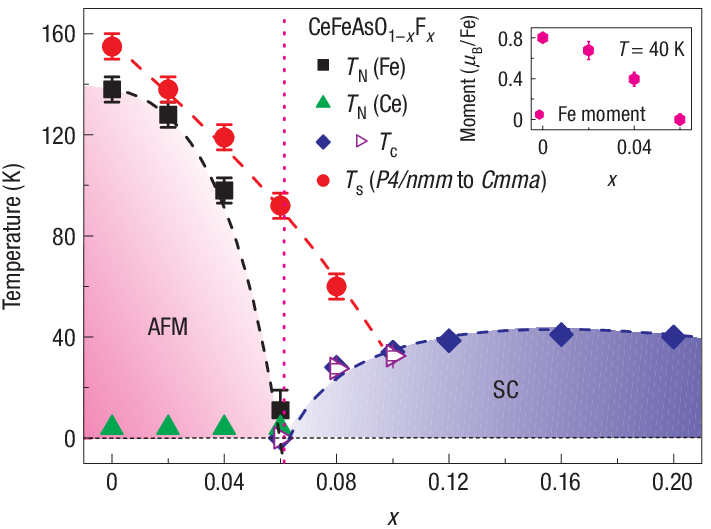}\hspace{0.1in}
\includegraphics [width=2.8in]{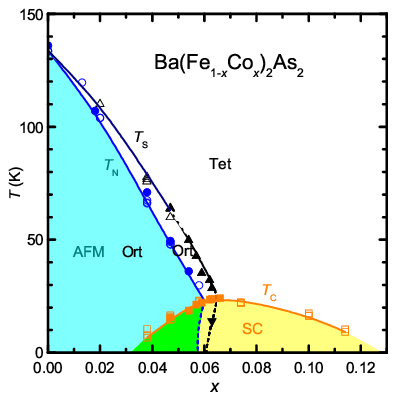}
\caption{(Color online) Temperature $T$ versus composition $x$ phase diagrams of illustrative 1111-type electron-doped polycrystalline CeFeAsO$_{1-x}$F$_x$ (Ref.~\onlinecite{Zhao2008}) and 122-type electron-doped single crystalline Ba(Fe$_{1-x}$Co$_x$)$_2$As$_2$.\cite{Nandi2010} The regions of the phase diagrams are: paramagnetic tetragonal (high-temperature regions) for $T > T_{\rm S},\ T_{\rm c}$; orthorhombic structural distortion between temperatures $T_{\rm S}$ and $T_{\rm N}$(Fe), Ort; orthorhombic distortion and long-range antiferromagnetic ordering occurring together (AFM, Ort); and superconductivity (SC).  The inset of the phase diagram for CeFeAsO$_{1-x}$F$_x$ shows the low-temperature ordered Fe moment $\mu$ versus $x$; the $\mu(x)$ data for Ba(Fe$_{1-x}$Co$_x$)$_2$As$_2$ are very similar.\cite{Lester2009}  In both systems, superconductivity can coexist with the orthorhombically distorted structure.  In CeFeAsO$_{1-x}$F$_x$, SC and AF do not coexist. In BaFe$_{2-x}$Co$_x$As$_2$, SC and AF coexist at low temperatures over the restricted composition range $0.035 \lesssim x \lesssim 0.06$; note the re-entrant behaviors on the right side of this region (see Fig.~\ref{Nandi_ortho_split} in Sec.~\ref{SecOrthoSC} and Fig.~\ref{FigBaFeCoAsMT} in Sec.~\ref{Sec_IntSCMag} below).  In both systems, the optimum superconducting transition temperatures are not reached until the long-range structural and magnetic transitions are both completely suppressed.  The highest $T_{\rm c}$ occurs at ``optimum'' doping $x$.  The lower $T_{\rm c}$ region at smaller $x$ is called the ``underdoped'' region and the lower $T_{\rm c}$ region at larger $x$ is called the ``overdoped'' region.  See also Refs.~\onlinecite{Lester2009}, \onlinecite{Chauviere2010} and~\onlinecite{Hardy2010b} for very similar Ba(Fe$_{1-x}$Co$_x$)$_2$As$_2$ phase diagrams derived from single crystal studies.  Reprinted with permission from Refs.~\onlinecite{Zhao2008} and~\onlinecite{Nandi2010}.  Figure from Ref.~\onlinecite{Zhao2008}: Reprinted by permission from Macmillan Publishers Ltd, Copyright (2008).  Figure from Ref.~\onlinecite{Nandi2010}: Copyright (2010) by the American Physical Society.}
\label{FigBaKFe2As2_phase_diag} 
\end{figure*}

\begin{figure}
\includegraphics [width=3.3in]{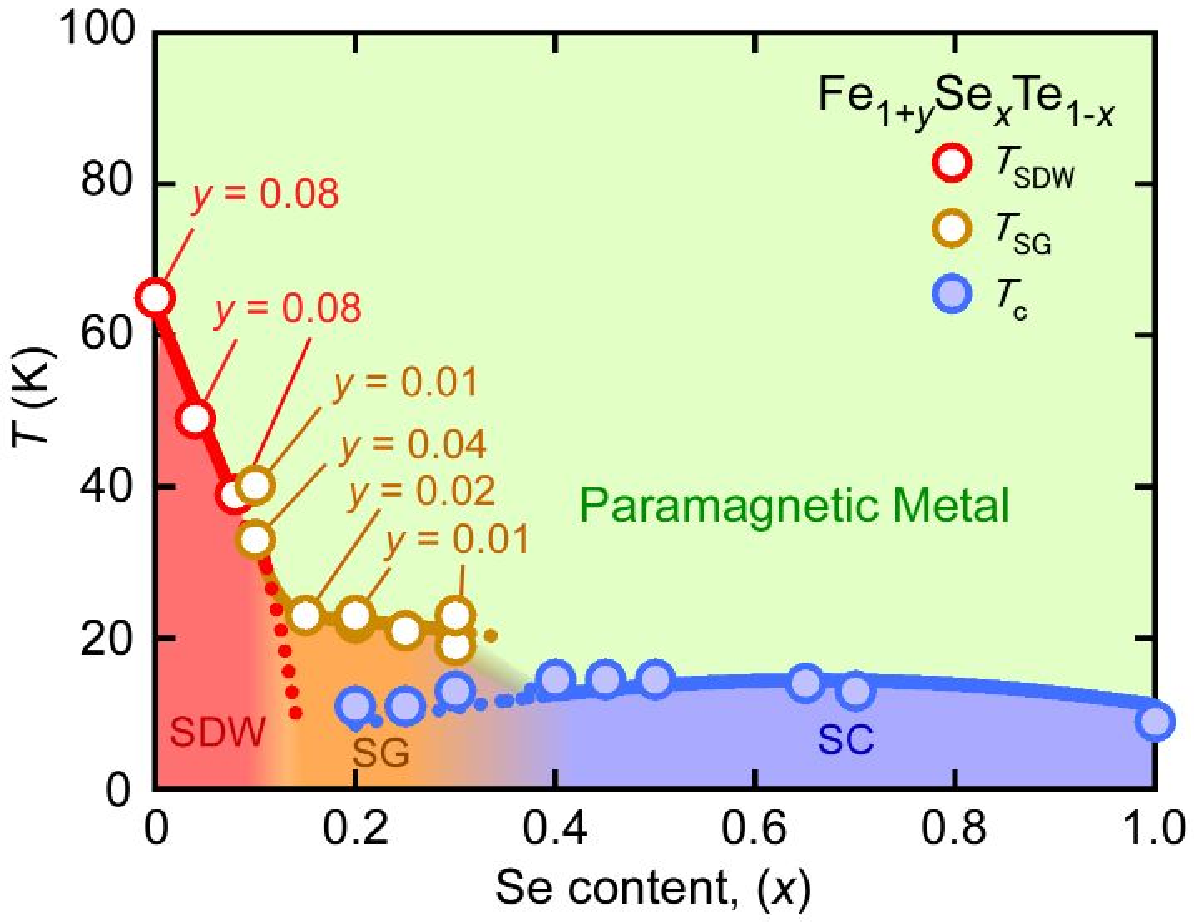}
\caption{(Color online) Temperature-composition phase diagrams for the Fe$_{1+y}$(Te$_{1-x}$Se$_x$) system determined from measurements on single crystals.\cite{Katayama2010}  The phase regions determined from magnetic susceptibility measurements are tetragonal paramagnetic metal at high temperatures for all compositions, spin density wave (SDW), superconductivity (SC), and short-range spin glass-like (SG) static magnetic ordering.  Nonzero excess iron concentrations $y$ are given in the figure.  The minimum value of $y$ decreases as $x$ increases.  The long-range SDW ordering occurs together with a monoclinic or orthorhombic lattice distortion, depending on $y$, whereas the SG ordering occurs in the tetragonal phase.  The term SDW implies itinerant magnetism.  As discussed in Sec.~\ref{SecChi}, the antiferromagnetism is most likely due to local magnetic moments, instead of itinerant magnetism as for the other Fe-based parent compounds.  The superconducting transition temperatures in the figure are onset temperatures.  Reprinted with permission from Ref.~\onlinecite{Katayama2010}.}
\label{KatayamaFig2} 
\end{figure}

The undoped nonsuperconducting ``parent'' compounds of the high $T_{\rm c}$ FeAs-based materials have high electrical resistivities $\rho$ at room temperature, as illustrated for BaFe$_2$As$_2$ in Fig.~\ref{FigChiRhoBaFe2As2}, where the $ab$-plane resistivity at 300~K is 430~$\mu\Omega$-cm and the resistivity anisotopy ratio $\rho_c/\rho_{ab}$ is about 150.\cite{wang2008}  For comparison, the resistivity of copper metal at room temperature is about 1.6~$\mu\Omega$~cm.\cite{Blatt1968}  Other measurements of $\rho(T)$ of single crystals by Tanatar et al.\ find a much smaller anisotopy $\rho_c/\rho_{ab} \sim 5$ and a $d\rho_c/dT < 0$ at 300~K for BaFe$_2$As$_2$,\cite{Tanatar2009} and $\rho_c/\rho_{ab} \sim 2$ to~5 for optimally doped superconducting ${\rm Ba(Fe_{0.926}Co_{0.0.074})_2As_2}$ between 25 and 300~K, depending on the temperature.\cite{Tanatar2009c}  Similarly, Moon et al.\ obtained $\rho_c/\rho_{ab} \sim 4$ for a single crystal of EuFe$_2$As$_2$.\cite{Moon2010a} For a single crystal of LiFeAs, Song et al.\ find $\rho_c/\rho_{ab} = 1.4$ at 300~K, increasing to 3.3 at 20~K\@.\cite{Song2010}  On the other hand, Kashiwaya et al.\ find a resistivity anisotropy $\rho_c/\rho_{ab} \sim 120$ at 50~K for a PrFeAsO$_{0.7}$ crystal.\cite{Kashiwaya2010}  

The parent 1111 and 122 compounds exhibit apparently coupled structural and commensurate antiferromagnetic (AF) transitions, also called ``spin density wave'' (SDW) transitions, at $\sim 100$--200~K\@.  These transitions result in distinct changes in the magnetic, thermal and electrical properties such as illustrated for BaFe$_2$As$_2$ in Fig.~\ref{FigChiRhoBaFe2As2} where these transitions are both at the same temperature of 138~K.\cite{wang2008}  An AF ground state of LaFeAsO was predicted theoretically early on from first principles electronic structure calculations.\cite{Ma2008a}

Superconductivity is induced upon ``doping'' the parent compounds by changing the composition, as in LaFeAsO$_{1-x}$F$_x$, LaFeAsO$_{1-y}$, or Ba$_{1-x}$K$_{x}$Fe$_2$As$_2$ as discussed above, or upon application of pressure as in SrFe$_2$As$_2$ or BaFe$_2$As$_2$ which reach $T_{\rm c}$'s up to 35~K,\cite{Alireza2009, Colombier2009, Kotegawa2009, Matsubayashi2009, Ishikawa2009, Yamazaki2010} with a concommitant suppression or elimination of the structural and magnetic ordering transitions.  The $T_{\rm c}$ of FeSe remarkably increases from 8~K at ambient pressure to 37~K at $\sim 9$~GPa.\cite{Margadonna2009a, Medvedev2009} The temperature-composition phase diagrams of the illustrative CeFeAsO$_{1-x}$F$_x$ and BaFe$_{2-x}$Co$_x$As$_2$ systems are shown in Fig.~\ref{FigBaKFe2As2_phase_diag}.\cite{Zhao2008, HChen2009, Ni2008g, Chu2009, Wang2008c, Lester2009, Nandi2010, Chauviere2010, Hardy2010b}  The phase diagram for LaFeAsO$_{1-x}$F$_x$ (Refs.~\onlinecite{Huang2008}, \onlinecite{Luetgens2009}, \onlinecite{Kamihara2010}) is similar to that in the left panel except that the magnetic and structural transition temperatures show a sharp drop to zero at the onset of superconductivity at $x\approx 0.04$--0.05.  The phase diagrams for SmFeAsO$_{1-x}$F$_x$,  \cite{Drew2009, Margadonna2009} the hole-doped 122-type system Ba$_{1-x}$K$_{x}$Fe$_2$As$_2$,\cite{HChen2009} and the 122-type electron-doped systems  BaFe$_{2-x}$Rh$_x$As$_2$ and BaFe$_{2-x}$Pd$_x$As$_2$,\cite{Ni2009} SrFe$_{2-x}$Ni$_x$As$_2$,\cite{Saha2009} SrFe$_{2-x}$Rh$_x$As$_2$, SrFe$_{2-x}$Ir$_x$As$_2$, and SrFe$_{2-x}$Pd$_x$As$_2$,\cite{Han2009} are qualitatively similar to that in the right panel.  Interestingly, polycrystalline samples in the nominally hole-doped system SrFe$_{2-x}$Mn$_x$As$_2$ exhibit no superconductivity for compositions $0 \leq x \leq 0.30$, and the crystallographic transition is only depressed from 205~K to about 130~K over this composition range.\cite{Kasinathan2009}  Similarly, single crystals of BaFe$_{2-x}$Cr$_x$As$_2$ exhibit no superconductivity for compositions $0 \leq x \leq 0.75$, and here again, the crystallographic transition is only slowly depressed from 140~K to about 60~K at $x = 0.36$.\cite{Sefat2009c}  Cheng et al.\ have studied the variation of $T_{\rm c}$ in polycrystalline samples of 25\% hole-doped ${\rm (Ba_{0.5}K_{0.5})Fe_2As_2}$ with $T_{\rm c} =34$~K upon partially replacing the Fe by Mn or Zn.\cite{Cheng2010}  They found that Mn subsitution rapidly reduces $T_{\rm c}$ to 12~K at 2.0\% doping and to $< 2$~K by 4\% doping.  In contrast, Zn replacement reduced $T_{\rm c}$ at a much smaller rate, with $T_{\rm c} = 33$~K at 10\% doping.\cite{Cheng2010}

The induction of superconductivity by doping Co or other transition metals into the Fe site indicates that atomic disorder in the superconducting Fe layer ostensibly does not suppress superconductivity, contrary to the behaviors of layered cuprate high $T_{\rm c}$ superconductors where doping onto the Cu sublattice is always detrimental to $T_{\rm c}$.  On the other hand, the maximum $T_{\rm c}$ in these systems (38~K for 122-type systems and 56~K for 1111-type systems) is not reached by substituting onto the Fe site, suggesting that there may indeed be a suppression of the maximum $T_{\rm c}$ due to Fe-sublattice disorder.

Remarkably, the phase diagram of the BaFe$_{2}$As$_{2-x}$P$_x$ system shows a similar phase diagram as in the right panel of Fig.~\ref{FigBaKFe2As2_phase_diag}, with $T_{\rm c}$ up to 30~K, even though there is ostensibly no charge doping involved in the substitution of As by P.\cite{Jiang2009a}  Similarly, isoelectronic substitution of Ru for Fe in SrFe$_{2}$As$_2$ (Ref.~\onlinecite{Schnelle2009}) or BaFe$_{2}$As$_2$ (Refs.~\onlinecite{paulraj2009, Rullier-Albenque2010, Thaler2010}) also induces superconductivity at temperatures up to 20~K at $\sim 30$--55\% replacement of Fe by Ru.  In contrast, replacing Cu by any other element at these high concentrations in the high-$T_{\rm c}$ cuprates destroys any possibility of superconductivity.  Thus it is evidently not necessary to dope additional carriers into the parent compounds to induce superconductivity, but only to suppress the crystallographic and long-range-antiferromagnetic transitions.  On the other hand, P substitution for As in the system CeFeAs$_{1-x}$P$_x$O does not induce superconductivity even though it does suppress the crystallographic and Fe antiferromagnetic transitions to zero temperature by $x = 0.4$.\cite{Luo2010, Cruz2010}

K\"ohler and Behr have compared the nominal and measured F contents in the LaFeAsO$_{1-x}$F$_x$ and SmFeAsO$_{1-x}$F$_x$ samples and concluded that the actual F contents can be significantly different from the nominal ones used in constructing the phase diagrams, and therefore that the previous phase diagrams for F-containing systems may need to be revised.\cite{Kohler2009}  

The phase diagram obtained for the Fe$_{1+y}$(Te$_{1-x}$Se$_x$) system\cite{Katayama2010} is somewhat different than discussed above, containing a spin glass phase between the antiferromagnetic and superconducting phases, as shown in Fig.~\ref{KatayamaFig2}.  A similar phase diagram was obtained in Ref.~\onlinecite{Liu2010}.  Long-range antiferromagnetic ordering ceases for $x \gtrsim 0.1$, and bulk superconductivity does not set in until $x$ exceeds $\sim 0.3$ to~0.4. Structural studies indicate that the low-temperature long-range structural distortion ceases with increasing $x$ at the same composition at which long-range antiferromagnetic ordering ceases, and further suggest that the spin glass phase at low temperatures is accompanied by the onset of lattice disorder of some kind.\cite{Katayama2010}  As discussed in Sec.~\ref{SecChi} below, it appears that the SDW/AFM region corresponds to local moment, rather than itinerant, antiferromagnetism.  Additional evidence for short-range antiferromagnetic ordering was observed in other investigations.\cite{Bao2009, Wen2009, Khasanov2009, Babkevich2010, Xu2010}  The phase diagram of Ref.~\onlinecite{Khasanov2009} shows the short-range ordering regime extending all the way to $x = 0.45$, overlapping the region of bulk superconductivity.  On the other hand, Ref.~\onlinecite{Xu2010} also finds short-range ordering up to $x = 0.45$, but with no bulk superconductivity.  At $x = 0.50$, bulk superconductivity is found, but with no static magnetic ordering.  The authors suggest that bulk static magnetic order and bulk superconductivity may be mutually exclusive in this system; when they do appear to occur simultaneously, they may occur in different spatial regions.\cite{Xu2010}

The emergence of high $T_{\rm c}$ upon destruction of long-range AF order as illustrated in Fig.~\ref{FigBaKFe2As2_phase_diag} is qualitatively similar to observations in the layered cuprate high $T_{\rm c}$ superconductors.\cite{Johnston1997}  The close association of AF ordering and superconductivity in both types of materials suggests that the superconductivity may have an electronic/magnetic mechanism.  However, the cuprate parent compounds are antiferromagnetic insulators rather than metals, which is an important distinction between these two classes of high $T_{\rm c}$ superconductors.

Many research papers have been written on the properties of the above Fe-based and related materials and their theoretical interpretations since the spring of 2008.  About 2\,000 experimental papers and 500 theoretical papers have been published in journals and/or posted on the arXiv\cite{arXiv} so far. Much of this research is driven by the following questions: What is the mechanism for $T_{\rm c}$?  Is new physics involved in the properties?  What is the upper limit of $T_{\rm c}$ for this class of materials?  What materials properties control $T_{\rm c}$?  Where should we look next to find new superconductors with high $T_{\rm c}$?  Should we consider the materials to be strongly correlated electron systems?  Where are these materials situated with respect to the two limits of strongly correlated localized magnetism\cite{Si2008} and weakly correlated itinerant magnetism?\cite{Yildirim2009}  What is the relationship between the properties of the Fe-based materials and the layered high $T_{\rm c}$ copper oxides\cite{Johnston1997} which contain a Cu square lattice of localized spins 1/2?  Is the superconducting mechanism the same or different compared to the cuprates?  Can these materials lead to technological breakthroughs in the widespread utilization of superconductors?  Also, in view of the ongoing lack of consensus about the superconducting mechanism in the cuprates, there is hope that study of the Fe-based materials might provide insights into the mechanism in the cuprates. 

Reviews of the Fe-based layered superconductor field have previously appeared.\cite{Norman2008, Sadovskii2008, Ivanovskii2008, Izyumov2008, Ishida2009, Paglione2010} Reviews specifically on ${\rm BaFe_2As_2}$ (Ref.~\onlinecite{Mandrus2010}) and on the Fe(S, Se, Te) superconductors\cite{Mizuguchi2010a} were written.  Many topical invited reviews, most of which are reviews of results from specific research groups, are collected in a special issue of Physica~C.\cite{PhysicaC}  Reviews of layered 1111-type and similar oxypnictides and oxychalcogenides have been published.\cite{Clarke2008, Ozawa2008, Pottgen2008, Volkova2008}  A listing and classification of 600 ThCr$_2$Si$_2$-type compounds has been given.\cite{Just1996}  A conference on Fe-based superconductors was held in December 2008 with published proceedings.\cite{RomeConf08}  A focus issue of New J.\ Phys.\ (Feb.\ 2009) was devoted to these materials.\cite{NewJPhysFocusIssue} A topical review on the magnetism in the Fe-based superconductors and parent compounds has appeared.\cite{Lumsden2010}  A survey of chalcogenide superconductors is available.\cite{Nagata1999}  A review of the effects of transition metal doping onto the Fe site in ${\rm BaFe_2As_2}$ is given in Ref.~\onlinecite{Canfield2010}.

Herein we provide a selective overview of the large and growing field of FeAs-based superconductors and related materials with different emphases than in  the above previous reviews.  In particular, we critically evaluate the experimental results from the literature for the various properties and the relationships of those properties to each other and to $T_{\rm c}$, with reference to theoretical concepts and models.  Although the superconducting properties are discussed, we emphasize the normal state properties, and within these the electronic and magnetic properties as determined from various kinds of studies because of their likely importance to an electronic/magnetic mechanism for superconductivity in these materials.  We do not even touch on applied aspects of superconductivity in these materials such as measurements of the critical currents or the construction of superconducting wires.  Many tutorials are given as introductions to various subjects.  We hope that these will be useful not only to students but also to researchers not specializing in the described subjects.  Many tables and figures of data have been included for various properties.  An overview of the present state of the field and a summary of the basic scientific issues remaining to be resolved is given at the end in Sec.~\ref{SecConclusions}. For a quick overview of the main accomplishments in this field, the reader can jump directly to Sec.~\ref{SecConclusions}.

A summary of the most common symbols used in this review is given in Table~\ref{Symbols} in the Appendix.  In addition to data tables in the text, we will utilize the data tables in the Appendix that give listings of various crystallographic and physical properties of the compounds along with the associated references.  Although no index is provided, one can use the table of contents and/or do a search of the pdf document for specific keywords.

\section{\label{SecStruct}Structural Properties}

\subsection{\label{SecStructOverview} Overview}

\begin{figure}
\includegraphics [width=2.in]{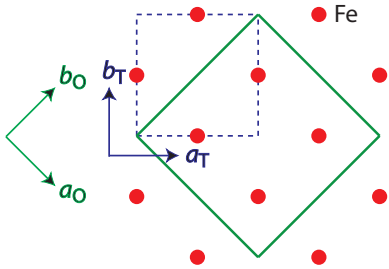}
\caption{(Color online) Relationships between the basal plane $a$ and~$b$ axes of the high-temperature tetragonal (T, dashed blue outline, see Fig.~\ref{Struct122}) and the low-temperature orthorhombic (O, solid green outline) structures of the 1111-type and 122-type FeAs-based compounds.  For clarity, only the Fe atoms in a single layer parallel to the $a$-$b$~plane of the structures are shown in the figure.  The basal-plane lattice parameters are related to each other by $a_{\rm T} = b_{\rm T}$, $a_{\rm O} \approx b_{\rm O} \approx \sqrt{2}\,a_{\rm T}$.  In this review, we consistently quote orthorhombic lattice parameters such that $c_{\rm O} > a_{\rm O} > b_{\rm O}$ for both the 1111-type and 122-type compounds.}
\label{FigTetrag_Ortho_struct} 
\end{figure}

All of the FeAs-based high $T_{\rm c}$ superconductors contain square lattice layers of Fe atoms where each Fe atom is at the center of a (usually) distorted As tetrahedron to form an equiatomic FeAs layer such as shown in Fig.~\ref{Struct122} for isoelectronic Ru in place of Fe.  These FeAs layers are separated by spacer/charge donation layers along the $c$-axis such as Ba layers in body-centered-tetragonal BaFe$_2$As$_2$ or LaO layers in primitive tetragonal LaFeAsO.  The same structures are sometimes formed when P replaces As, and/or when Co, Ni or other transition metals partially and/or completely replace the Fe. As seen in Fig.~\ref{Structures}, the 11-type Fe$_{1+y}$(Te$_{1-x}$Se$_x$) system has no charge reservoir layer. 

The high $T_{\rm c}$s in the Fe-based materials are observed for compounds that must be doped or put under pressure to reduce or eliminate the combined structural and magnetic transitions seen in the phase diagrams exemplified in Figs.~\ref{FigBaKFe2As2_phase_diag} and~\ref{KatayamaFig2}.  In all these systems of parent compounds, the high temperature structures are tetragonal and the low-temperature structures are distorted variants.  The high-temperature tetragonal lattice parameters for the four main classes of Fe-based materials are given in the Appendix.  

As noted above, partially replacing isoelectronic P for As in ${\rm BaFe_2As_2}$ suppresses the long-range structural and antiferromagnetic transitions and induces superconductivity.\cite{Jiang2009a}  Since P is smaller than As, this substitution results in a shrinking of the unit cell, corresponding to what is called ``chemical pressure.'' On the other hand, it is known that by substituting isoelectronic Sr for Ba, the unit cell shrinks to the same volume for which BaFe$_2$(As$_{1-x}$P$_x)_2$ becomes superconducting, but the Sr substitution does not suppress the crystallographic/antiferromagnetic transition temperature $T_0$ or induce superconductivity.\cite{Wang2009a}  In fact, Sr substitution monotonically enhances $T_0$ from 137~K for the pure Ba compound to 205~K for the pure Sr compound.  Therefore, the unit cell volume is not the only parameter determining whether the crystallographic and magnetic transitions are suppressed and superconductivity is induced.  

Rotter, Hieke and Johrendt have discovered a significant crystallographic difference between the (Ba$_{1-x}$Sr$_x$)Fe$_2$As$_{2}$ and BaFe$_2$(As$_{1-x}$P$_x)_2$ systems.\cite{Rotter2010}  They find that due to the large size mismatch between P and As, these two atoms are at different heights from the Fe layers, even though they nominally occupy the same crystallographic position (see Table~\ref{data5A} in the Appendix).  This is the same situation as noted above in the Fe$_{1+y}$(Te$_{1-x}$Se$_x$) system, in which the Se and Te atoms have different heights from the Fe layers.\cite{Tegel2010, Louca2010}  From the crystallographic data and band calculations, Rotter et al.\ infer that the different heights of the P and As layers in the BaFe$_2$(As$_{1-x}$P$_x)_2$ system have a dramatic influence on suppressing the magnetic and crystallographic transitions in favor of superconductivity, due to the giant magnetoelastic coupling described in Sec.~\ref{SecMagnetoElastic} below.  Quoting Ref.~\onlinecite{Rotter2010}, ``Phosphorus doping suppresses the SDW state by increasing the width of the $d$-bands, which in turn leads to shorter Fe-As bonds due to its strong coupling to the magnetic state.''

Low-temperature structure data are given for the 1111-type and 122-type FeAs-based materials in Appendix Tables~\ref{LoTStructData1111} and~\ref{LoTStructData122}, respectively.  The low temperature structures are distortions of the high temperature structures, rather than a complete rearrangement of the atoms.  Remarkably, even though second order transitions between the two structures are allowed by symmetry since the orthorhombic space groups are, respectively, subgroups of the tetragonal space groups, some of these transitions are reported to be first order such as in $\rm CaFe_2As_2$,\cite{Ni2008b, Goldman2008} $\rm SrFe_2As_2$,\cite{Loudon2010} and $\rm BaFe_2As_2$.\cite{Huang2008b}  On the other hand, Wilson et al.\ found that their magnetic and structural neutron diffraction data on a single crystal of $\rm BaFe_2As_2$ near the SDW transition temperature were consistent with a second order phase transition.\cite{Wilson2009} 

The relationship between the $a$-$b$ plane axes in the tetragonal and orthorhombic structures of the 122- and 1111-type compounds is shown in Fig.~\ref{FigTetrag_Ortho_struct}.  One would expect twinning to occur because the orthorhombic distortion is small ($\lesssim 1$\%, see the tables in the Appendix).  Twins have indeed been observed optically below the respective tetragonal-orthorhombic transition temperature in $A$${\rm Fe_2As_2}$ ($A$ = Ca, Sr, Ba)\cite{Tanatar2009b} and ${\rm Ba(Fe_{0.985}Co_{0.015})_2As_2}$.\cite{Chu2009b}  The twin boundaries run along the orthorhombic [110] and [1$\bar{1}$0] directions (tetragonal [100] and [010] directions) and form planes that traverse the materials parallel to the $c$-axis and are separated in the $a$-$b$ plane by $\sim 10$--50~$\mu$m. Transmission electron microscopy of the $A$${\rm Fe_2As_2}$ compounds gives similar results except that the twin boundaries are separated by only 0.1--0.4~$\mu$m.\cite{Ma2009}   In addition, a fine tweed pattern is found in Ca${\rm Fe_2As_2}$.\cite{Ma2009}  It is not clear why optical and electron microscopies give different results for the twin boundary spacing.  The reason is possibly associated with the sample preparation needed for the TEM measurements that require extremely thin samples.

Within an \emph{MX}$_4$ tetrahedron where $M$ is the transition metal atom and \emph{X} is a pnictogen (\emph{Pn} = P, As, Sb, Bi) or chalcogen (\emph{Ch} = S, Se, Te), there is a twofold \emph{X-M-X} bond angle where the two \emph{X} atoms are on the same side of the $M$ atom layer along the $c$ axis, and a fourfold \emph{X-M-X} bond angle where the two \emph{X} atoms are on opposite sides of the $M$ layer (see Fig.~\ref{Struct122}). ÊThe twofold and fourfold \emph{X-M-X} bond angles are given by\cite{Nath2009}
\bea
\theta_2 &=& {\rm arccos}\left[\frac{-\frac{a^2}{4} + (z - \alpha)^2 c^2}{r^2}\right] \ \ \ \ ({\rm twofold})\nonumber\\
\theta_4 &=& {\rm arccos}\left[\frac{-(z - \alpha)^2 c^2}{r^2}\right] \ \ \ \ \ \ \ \ \ \ \ ({\rm fourfold})\nonumber\\
{\rm where} \label{Eqtheta}\\
r^2 &=& \frac{a^2}{4} + \left(z - \alpha\right)^2 c^2\nonumber
\eea
and $\alpha = 0$, 1/4, 1/2 and~1 for the FeSe-type (11-type), BaFe$_{2}$As$_{2}$-type (122-type), LaFeAsO-type (1111-type) and LiFeAs (111-type) structures, respectively.  Here $a$ and $c$ are the tetragonal lattice parameters, $z$ is the $c$-axis position parameter of the \emph{X} atom in a unit cell of the respective structure (e.g., $z\approx 0.25$ in FeSe, $z\approx 0.35$ in BaFe$_{2}$As$_{2}$, $z\approx0.65$ in LaFeAsO, and $z\approx0.75$ in LiFeAs), and $r$ is the nearest-neighbor \emph{M-X} distance within an $M$-centered $MX_4$ tetrahedron (all four $M$-$X$ nearest-neighbor distances are the same in each of the  structures).  The average bond angle for all six \emph{X-M-X} bonds is close to the value of 109.47$^\circ$ for an undistorted tetrahedron.  Thus if $\theta_2 > 109.47^\circ$ then $\theta_4 < 109.47^\circ$, and \emph{vice versa}.   The $M$ atoms in each structure form a square lattice where the fourfold nearest-neighbor $M$-$M$ distance in all four structures is $d_{M-M} = a/\sqrt{2}$.  The distance (height $h$) between an $M$ layer and either adjacent \emph{X} layer is $h =|z - \alpha|c$.   

\subsection{Relationship between \emph{Pn-T-Pn} Bond Angle and $T_{\rm c}$}

Studies of structure-property relationships of the Fe pnictide-based superconductors have suggested correlations with the superconducting transition temperature $T_{\rm c}$.  Early on, Lee and coworkers\cite{Lee2008, Lee2008b} and Zhao and coworkers\cite{Zhao2008} reported a correlation for a wide range of parent compounds Ba(Fe,Ni)$_2$P$_2$, $R$(Fe,Ni)(P,As)O and TbFeAsO$_{0.9}$F$_{0.1}$ where $R$ is a rare earth element, that the highest $T_{\rm c}$ occurred for the doped materials in which the respective Fe\emph{Pn}$_4$ or Ni\emph{Pn}$_4$ tetrahedra were least distorted (\emph{Pn} = P, As).  The angle plotted in Refs.~\onlinecite{Lee2008} and~\onlinecite{Lee2008b} that is correlated with $T_{\rm c}$ is the twofold \emph{Pn-M-Pn} bond angle in Eqs.~(\ref{Eqtheta}).  More recently, from measurements of the structure versus applied pressure for BaFe$_2$As$_2$, Kimber et al.\ concluded that the structure is more important than doping in inducing superconductivity in this compound.\cite{Kimber2009}  Another correlation was found by Mikuda and coworkers among $R$FeAsO$_{1-\delta}$ compounds, where the $T_{\rm c}$ was found to increase monotonically with increasing $^{75}$As nuclear quadupole resonance frequency $^{75}\nu_{\rm Q}$, which in turn reflects the local coordination and bonding of the As atoms.\cite{Mukuda2008, Mukuda2009}  

\begin{figure}
\includegraphics [width=3.3in]{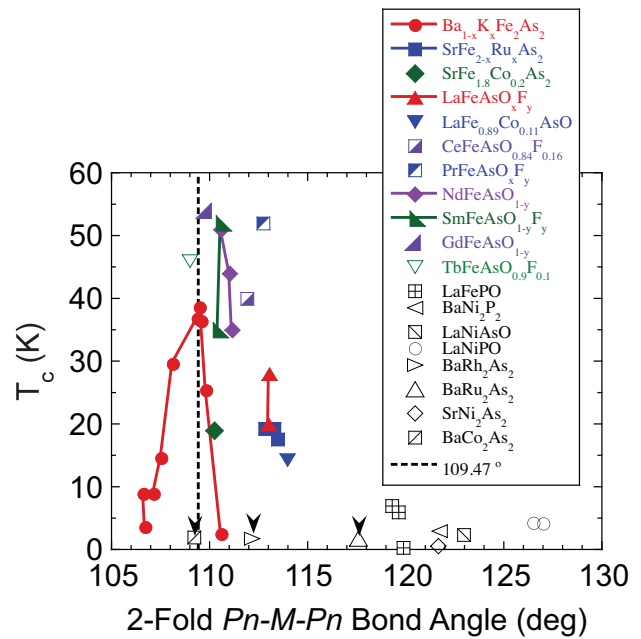}
\caption{\label{TcAngle} (Color online) Superconducting transition onset temperature $T_{\rm c}$ versus room-temperature two-fold $Pn$-$M$-$Pn$ angle for a variety of layered transition metal pnictides, where $Pn$ = P, As, and $M$ is the 3$d$ transition metal atom.  Each plotted point represents a sample with an individually measured $T_{\rm c}$ and crystal structure refinement.  The vertical arrows pointing downwards indicate that superconductivity is not observed above the indicated temperature for the respective compound.  The dashed vertical line indicates the equal twofold and fourfold $Pn$-$M$-$Pn$ bond angles of arccos($-1/3) \approx 109.47^\circ$ for an undistorted $M$-centered $M$\emph{Pn}$_4$ tetrahedron.}
\end{figure}

\begin{figure}
\includegraphics [width=3.3in]{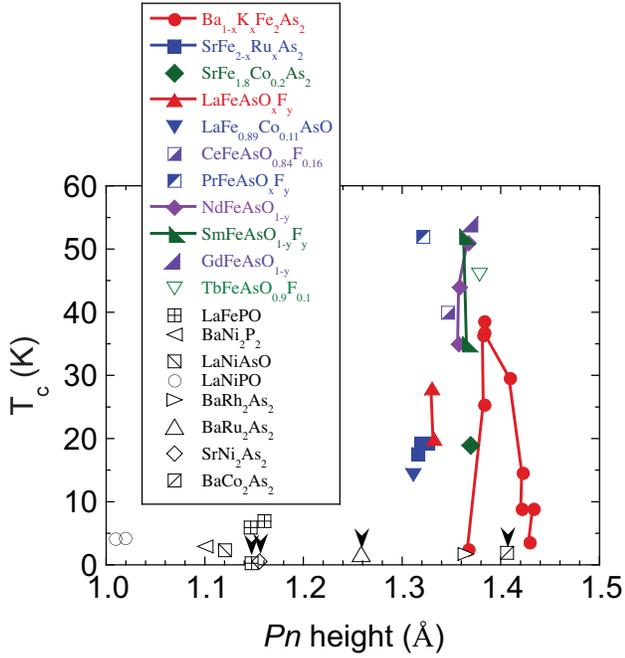}
\caption{\label{TcHeight} (Color online) Superconducting transition onset temperature $T_{\rm c}$ versus the distance (height) between the transition metal $M$ layer and either adjacent pnictogen $Pn$ layer, where $Pn$ = P or As.  Each plotted point represents a sample with an individually measured $T_{\rm c}$ and crystal structure refinement.  The vertical arrows pointing downwards indicate that superconductivity is not observed above the indicated temperature for the respective compound.}
\end{figure}

Shown in Fig.~\ref{TcAngle} is a plot of $T_{\rm c}$ versus the twofold \emph{Pn-M-Pn} angle for a variety of 122-type and 1111-type compounds, where each data point corresponds to a sample with both a measured $T_{\rm c}$ and a measured structure.  From the figure, a unique correlation between $T_{\rm c}$ and bond angle is not present.  On the other hand, the envelope containing all the data in the figure has an obvious peak near the angle of 109.47$^\circ$ corresponding to an undistorted pnictogen tetrahedron centered by the transition metal $M$.  This suggests that the \emph{potential} for high $T_{\rm c}$ is greatest for undistorted $MPn_4$ tetrahedra, although in that view other factors are clearly also affecting $T_{\rm c}$.  Horigane et al.\ have shown that FeSe$_{1-x}$Te$_x$ samples also follow the overall behavior in Fig.~\ref{TcAngle}, where the $T_{\rm c}$ is between 6 and 14~K and the twofold bond angle is between 96$^\circ$ and 104$^\circ$.\cite{Horigane2009}

However, it is not obvious that the non-FeAs-based compounds plotted have the same mechanism for $T_{\rm c}$ as the FeAs-based ones, and thus it is not clear that they should be considered together with the FeAs-based materials with respect to the structure-$T_{\rm c}$ relationship.  When viewed in this light, the only correlation remaining from the plot is that all of the FeAs-based compounds have about the same bond angle, irrespective of $T_{\rm c}$, which may simply be a reflection of the specific chemistry of the FeAs-based materials.  We note that within the Ba$_{1-x}$K$_x$Fe$_2$As$_2$ system plotted as the filled red circles, it appears that the highest $T_{\rm c}$ within this system corresponds to an angle equal to the angle of 109.47$^\circ$ for an undistorted FeAs$_4$ tetrahedron.  However, the two data points on the right-hand side of the vertical dashed line, which define the existence of the peak in $T_{\rm c}$, are for samples where, coincidentally(?), the magnetic/structural transitions have not yet been completely suppressed by K doping and which might therefore have suppressed $T_{\rm c}$s due to these transitions.

\subsection{Relationship between Pnictogen Height and $T_{\rm c}$}

Several authors have suggested a related alternative correlation between the $T_{\rm c}$ and the distance (height $h$) mentioned above between the Fe layer and either adjacent pnictogen layer for a specific selected subset of Fe-based layered superconductors.\cite{Mizuguchi2010, Okabe2010}  In particular, they considered only the maximum $T_{\rm c}$ that occurs within a given system, and omitted certain compounds such as $\rm{KFe_2As_2}$ in their plotted data.  This work was motivated by a theoretical study by Mizuguchi et al.\ discussed in the following paragraph.\cite{Kuroki2009}  In Fig.~\ref{TcHeight}, $T_{\rm c}$ is plotted versus $h$ for each of the samples plotted in Fig.~\ref{TcAngle}.  Here again, there is no unique correlation, although the envelope of the data has a peak around $h \sim 1.38$~\AA, as previously noted.\cite{Mizuguchi2010, Okabe2010}  Huang and coworkers have found a correlation between the Se height parameter $z$ and the $T_{\rm c}$ of ${\rm FeSe_{0.5}Te_{0.5}}$ epitaxial thin films.\cite{Huang2010}

On the theoretical side, Kuroki and coworkers have calculated that for spin-fluctuation-mediated superconductivity, the value of $h$ is ``a possible switch'' between high-$T_{\rm c}$ nodeless sign reversing $s^\pm$ pairing (large $h$) and low-$T_{\rm c}$ nodal (either $d$-wave or nodal $s$-wave) pairing (small $h$) in the iron-based superconductors.\cite{Kuroki2009}  Here, they compare high-$T_{\rm c}$ FeAs-based superconductors that appear to be nodeless with LaFePO with a low $T_{\rm c}$ that appears to have nodes in the superconducting wave function. In these calculations, since the spin fluctuation mechanism for superconductivity in both the low- and high-$T_{\rm c}$ Fe-based compounds is assumed to be the same, the envelope function in Fig.~\ref{TcHeight} in this case is physically meaningful over the whole range of $h$ encompassing all of the plotted \emph{Fe-based} compounds.  It is unclear whether or not the 1111-type and 122-type compounds in the figure that do not contain Fe would fit into this scenario.

\subsection{\label{SecOrthoSC} Relationship between Orthorhombic Distortion and Superconductivity}

\begin{figure}
\includegraphics [width=3.3in]{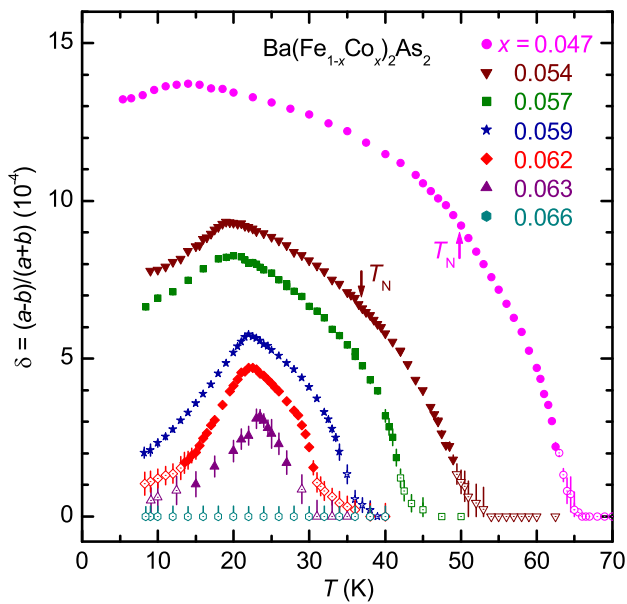}
\caption{(Color online) Orthorhombic distortion $\delta = (a-b)/(a+b)$ versus temperature $T$ for single crystals of Ba(Fe$_{1-x}$Co$_x$)$_2$As$_2$.\cite{Nandi2010}  As $x$ increases, the N\'eel temperature $T_{\rm N}$ and the tetragonal-to-orthorhombic structural transition temperature $T_{\rm S}$ decrease (see the phase diagram in Fig.~\ref{FigBaKFe2As2_phase_diag}).  These samples are in the region of the phase diagram where coexistence of superconductivity, long-range antiferromagnetism, and the orthorhombic crystallographic distortion occurs.  The sharp cusps in the temperature dependence occur at the respective superconducting transition temperature $T_{\rm c}$.  Reprinted with permission from Ref.~\onlinecite{Nandi2010}.  Copyright (2010) by the American Physical Society.}
\label{Nandi_ortho_split} 
\end{figure}

According to the phase diagram for Ba(Fe$_{1-x}$Co$_x$)$_2$As$_2$ in Fig.~\ref{FigBaKFe2As2_phase_diag}, there is a small composition ($x$) region where superconductivity coexists with the orthorhombic structural distortion at low temperatures.  Nandi and co-workers have investigated the temperature $T$ dependence of the orthorhombic distortion $\delta \equiv (c-a)/(c+a)$ in crystals of Ba(Fe$_{1-x}$Co$_x$)$_2$As$_2$ in this region of the phase diagram, as shown by the remarkable data in Fig.~\ref{Nandi_ortho_split}.\cite{Nandi2010}  Two significant features were found.  First, the long-range antiferromagnetic transition at $T_{\rm N}$ has no discernable influence on $\delta(T)$.  Second, $\delta$ is found to couple strongly to the superconductivity: $\delta$ is suppressed upon cooling through the superconducting transition temperature $T_{\rm c}$, with the suppression increasing with increasing $x$.  In fact, for $x = 0.063$, the distortion is completely suppressed by the superconductivity at the lowest temperatures, which is a so-called re-entrant behavior of $\delta$.  A similar suppression of $\delta$ below $T_{\rm c}$ was subsequently found in single crystals of Ba(Fe$_{0.961}$Rh$_{0.039}$)$_2$As$_2$.\cite{Kreyssig2010}

\subsection{\label{SecFeSeTe} The Fe$_{1+y}$(Te$_{1-x}$Se$_{x})$ System}

These materials crystallize in the anti-PbO-type primitive tetragonal structure shown in Fig.~\ref{Structures}.  The crystal data are given in the Appendix in Table~\ref{data6}.  Here we discuss the 11-type Fe$_{1+y}$(Te$_{1-x}$Se$_{x})$ system separately because of complications in its crystal structure.  The formula is written this way instead of Fe(Te$_{1-x}$Se$_{x})_{1-z}$ because structure refinements and mass density measurements for Fe$_{1+y}$Te show that (i) there are excess ($y$) Fe atoms present beyond those needed to fill the fully-occupied Fe positions in the Fe square lattice layers, and (ii) the excess Fe atoms go into interstitial positions within the Te layers as shown in Fig.~\ref{Structures}, distinct from the Fe positions in the Fe square lattice layers.\cite{Gronvold1954}  The minimum amount $y$ of excess Fe incorporated into the crystal structure decreases as $x$ increases.  Furthermore, we write the composition of this system as Fe$_{1+y}$(Te$_{1-x}$Se$_{x})$ instead of Fe$_{1+y}$(Se$_{1-x}$Te$_{x})$ because the undoped $x = 0$ composition, Fe$_{1+y}$Te, is an antiferromagnetic metal for which the antiferromagnetic transition is driven to zero temperature and superconductivity appears by partially substituting Se for Te.  Thus by writing Fe$_{1+y}$(Te$_{1-x}$Se$_{x})$, the sequence of phases with increasing $x$ is the same as in the other classes of Fe-based superconductors.

\begin{figure}
\includegraphics[width=3.in]{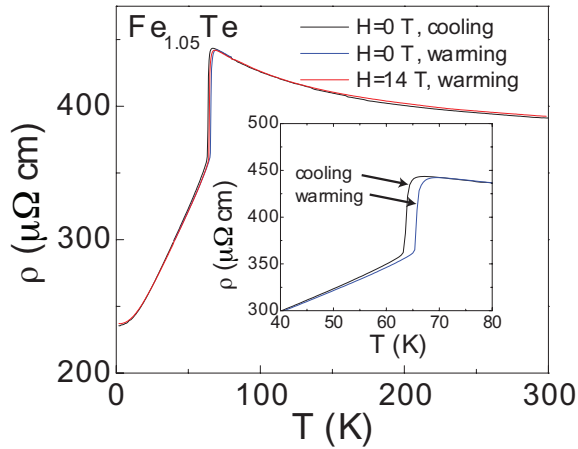}
\caption{(Color online) Electrical resistivity $\rho$ in the $ab$-plane versus temperature $T$ for a single crystal of Fe$_{1.05}$Te.\cite{Chen2009a}  The inset shows an expanded plot of the 2~K hysteresis observed upon heating and cooling through the combined crystallographic and antiferromagnetic first order transition at $T_{\rm N} \approx 65$~K\@.  Reprinted with permission from Ref.~\onlinecite{Chen2009a}.  Copyright (2009) by the American Physical Society.}
\label{Fe1.05Te_rho} 
\end{figure}

Statements are sometimes made in the literature that Fe$_{1+y}$Te$_{1-x}$Se$_x$ compounds with $x = 0$ are ``nonmetallic''\cite{Chen2009a} and others are metallic. Such statements may be based on the negative temperature coefficient of the in-plane resistivity $\rho$ for the former materials at low temperatures, such as shown for a single crystal of Fe$_{1.05}$Te in Fig.~\ref{Fe1.05Te_rho} for temperatures above the N\'eel temperature $T_{\rm N} = 65$~K\@.\cite{Chen2009a}   However, there is no evidence in the literature that any of the Fe$_{1+y}$Te compounds are insulating for $T \to 0$ as would be expected for a semiconductor (i.e., non-metal).  Indeed, the magnitude of the normal state in-plane resisitivity of Fe$_{1.05}$Te in Fig.~\ref{Fe1.05Te_rho} is similar to that of ${\rm BaFe_2As_2}$ in Fig.~\ref{FigChiRhoBaFe2As2} above.  Liu et al.\ have suggested that weak charge carrier localization occurs in samples with excess Fe ($y \sim 0.11)$.\cite{Liu2009a}  However, this suggestion was based on $\rho(T)$ data that increased by only $\sim 15$--50\% upon cooling over the discussed temperature ranges.  In addition, their heat capacity measurements below 12~K showed very similar large electronic specific heat coefficients $\gamma$ for both Fe$_{1.04}$Te and Fe$_{1.11}$Te single crystals (see the Appendix).\cite{Liu2009a}  These $\gamma$ values presumably both reflect the presence of a degenerate Fermi liquid.  Furthermore, it is well-known that crystalline disorder in metallic chalcogenides can induce a negative temperature coefficient of resistivity, such as was shown to occur in the gold-colored superconducting NaCl-structure defect-compound Zr$_{1-x}$S with $T_{\rm c} = 2.5$--4.5~K.\cite{Johnston1972, Moodenbaugh1978}  In Zr$_{1-x}$S, the lattice disorder arises from disordered vacancies on the Zr sites for large $x$ and on both the Zr and S sites for smaller $x$.\cite{Moodenbaugh1978}  Thus we suggest that the term ``nonmetallic'' should not be applied to the Fe$_{1+y}$Te$_{1-x}$Se$_x$ system until convincing evidence for $\rho\to\infty$ as $T \to 0$ is demonstrated.

According to Ref.~\onlinecite{Sales2009a}, bulk superconductivity only occurs in the PbO-type system Fe$_{1+y}$(Te$_{1-x}$Se$_{x})$ for $x \approx 0.5$ according to their heat capacity measurements.  However, heat capacity and low-field magnetic susceptibility measurements for polycrystalline Fe$_{1.01}$Se ($x = 1)$ by McQueen et al.,\cite{McQueen2009b} magnetic susceptibility measurements of FeSe$_{0.974}$ by Pomjakushina et al.,\cite{Pomjakushina2009} and heat capacity measurements of a ${\rm Fe_{1.03}Te_{0.63}Se_{0.37}}$ single crystal by Liu et al.\cite{Liu2009a} indicate bulk superconductivity for these compositions as well.

In Fe$_{1+y}$(Te$_{1-x}$Se$_{x})$ compounds with compositions $x \gtrsim 0.3$ for which bulk superconductivity is observed, Sales and coworkers noticed that the superconductivity can become non-bulk if the amount of excess Fe $y$ becomes large.\cite{Sales2009a}  Furthermore, Yang et al.\ found bulk superconductivity via magnetic susceptibility measurements for a Fe$_{1.04}$(Te$_{0.67}$Se$_{0.33})$ single crystal but not for a Fe$_{1.12}$(Te$_{0.70}$Se$_{0.30})$ single crystal\cite{Yang2009a} (see Table~\ref{FeTeSeData} in the Appendix).  Similar results were found by Liu et al.\ from heat capacity measurements\cite{Liu2009a} and by Viennois et al.\cite{Viennois2010}

\subsubsection*{Fe$_{1+y}$Se}

In this section the Fe$_{1+y}$Se system is considered in more detail because much information of various kinds is available for it. The phase diagram of the Fe-Se system was obtained by Okamoto in 1991.\cite{Okamoto1991}  The tetragonal anti-PbO-type Fe$_{1+y}$Se phase of interest in this review decomposes on heating to 457~$^\circ$C and has a small homogeneity range from 49.0 to 49.4~at\%~Se [misquoted as 45 to 49.4~at\%~Se in Ref.~\onlinecite{Hsu2008}] at 380~$^\circ$C,\cite{Okamoto1991} which corresponds to the formulas Fe$_{1.02}$Se to Fe$_{1.04}$Se, respectively.  Various studies refine the structure with either excess Fe or with Se vacancies, but irrespective of that, the results all indicate nearly stoichiometric (i.e., 1:1) FeSe.  From Ref.~\onlinecite{Okamoto1991}, Fe$_{1.04}$Se is in equilibrium with metallic Fe.  This means that if the samples are more Fe-rich than Fe$_{1.04}$Se, they will be contaminated by ferromagnetic Fe metal impurities.  Therefore it is puzzling that Refs.~\onlinecite{Hsu2008}, \onlinecite{Margadonna2008} and~\onlinecite{Fang2008} studied the superconducting and other properties of iron-rich polycrystalline samples with nominal compositions Fe$_{1.22}$Se, Fe$_{1.09}$Se and Fe$_{1.22}$(Te$_{1-x}$Se$_{x})$ ($0 \leq x \leq 1$), respectively.  In the first and third studies that report magnetic susceptibility data, the authors did not report $M(H)$ isotherm measurements (see Sec.~\ref{SecFMImpurities} below) to assess the contribution of ferromagnetic impurities to the measured magnetizations.  The less-studied phase diagram of the Fe-Te system also shows that metallic Fe is in equilibrium with the PbO-type Fe$_{1+y}$Te phase.\cite{Chiba1955}  Thus potential problems associated with contamination of the samples by ferromagnetic Fe metal impurities are evidently present throughout the Fe$_{1+y}$(Te$_{1-x}$Se$_{x})$ system of compounds.

McQueen and coworkers carried out a definitive study of the variation of the structural and physical properties of polycrystalline PbO-type Fe$_{1+y}$Se versus $y$ for $y \approx 0$,\cite{McQueen2009b}  although they did not report measurements of the normal state magnetic susceptibility.  They found that optimum bulk superconductivity with $T_{\rm c} = 9$~K was obtained for $1+y = 1.01$, and that neither this superconducting composition nor the nonsuperconducting composition $1+y = 1.03$ exhibit long-range magnetic ordering as deduced from $^{57}$Fe M\"ossbauer measurements at 295 and 5~K.\cite{McQueen2009b}  Their revised phase diagram for $y \approx 0$ indicates a homogeneity range of this phase of $1+y = 1.01$--1.03, slightly lower than the previous range of 1.02--1.04,\cite{Okamoto1991} with decomposition of the phase occurring below 300~$^\circ$C (a new result) and above 450~$^\circ$C.\cite{McQueen2009b}  They also found that unless care is exercised to prevent oxygen contamination, the samples contain ferromagnetic Fe$_3$O$_4$ impurities.  They suggested that this oxygen contamination is the reason that excess Fe (above $1+y = 0.03$) was included into previous preparations in order to obtain reasonably pure polycrystalline samples of PbO-type Fe$_{1+y}$Se, i.e., the extra Fe absorbs (``getters'') the oxygen impurities. Other refinements of the FeSe crystal structure also show nearly stoichiometric occupancies of the Fe and Se sites (see Table~\ref{data6} in the Appendix).

Pomjakushina et al.\ reported a complementary study of the phase stability of FeSe.\cite{Pomjakushina2009}  They found that irrespective of starting composition, this phase is nearly a line compound (a compound with only one composition) with composition ${\rm FeSe_{0.974\pm0.005}}$.  Note that this is written as Se-deficient rather than Fe-rich.  The authors attempted to refine their neutron powder diffraction data assuming an Fe-rich model in which the excess Fe atoms go into the Se layer as the Fe atoms go into the Te layer in Fig.~\ref{Structures} for Fe$_{1+x}$Te, but ruled that structural model out because of poor fits to their diffraction data.  Interestingly, the authors found that all of their samples were superconducting with about the same $T_{\rm c} \approx 8.3$~K, consistent with a narrow homogeneity range for the compound.  Similar results were obtained by Li et al.\cite{Li2010b}  These results conflict with those of McQueen et al.,\cite{McQueen2009b} who found instead that there existed at least one composition within the homogeneity range of FeSe that is not superconducting.

Fe$_{1+y}$Se can exhibit a lattice distortion from primitive tetragonal at room temperature (\emph{P}4/\emph{nmm}) to base-centered orthorhombic (\emph{Cmma}) upon cooling to low temperatures.\cite{Margadonna2008, McQueen2009c, Phelan2009}   McQueen et al.\ found that this transition occurs for $y = 0.01$ at 90~K, but not for $y = 0.03$.\cite{McQueen2009c}  Pomjakushina et al.\cite{Pomjakushina2009} and Khasanov et al.\cite{Khasanov2010} found this distortion to occur at about 100~K in each of the two samples they examined at low temperatures.  The lattice distortion is the same type as seen in the 1111-type FeAs-based compounds at low temperatures.   

\section{\label{SecNormalState}Normal State Properties}

The key to understanding the superconducting mechanism in the Fe-based superconductors is to understand their normal-state properties, since the superconducting state grows out of it and the microscopic interactions determining the former are likely to be responsible for the latter.  Therefore the most extensive part of this review is a description and discussion of the normal state properties.

\subsection{\label{Sec_BandStruct} Reciprocal Lattices, Brillouin Zones, Band Structures and Fermi Surfaces}

\subsubsection{\label{SecBanStrucIntro} Introduction}

Many band structure calculations have been carried out for the Fe-based high $T_{\rm c}$ superconductors (see the Appendix).  For a transition metal cation at the center of a tetrahedron of electronegative anions, due to crystalline electric field effects one nominally expects the five $d$-orbitals to split into a low energy set of two so-called $e_g$ orbitals and a set of three $t_{2g}$ orbitals at higher energy.  In oxides, this splitting is clearly seen.  However, in the FeAs-type materials all five $d$-orbitals are at about the same energy, and hence the bands formed from them overlap near the Fermi energy to a large extent. Furthermore, the electronic density of states at the Fermi energy is primarily derived from the Fe 3$d$ orbitals, and the As 4$p$ orbitals do not contribute much (see below).  This indicates that direct hopping of electrons from Fe atom to Fe atom is the main mechanism for metallic conduction, rather than hopping from Fe to As to Fe.  Indeed, the Fe-Fe distance in the $A$Fe$_2$As$_2$ compounds is only about 10\% larger than in pure elemental Fe metal.  Neutron diffraction measurements of the magnetic form factor of ${\rm SrFe_2As_2}$ showed that the form factor is nearly isotropic and similar to that of elemental iron.\cite{Ratcliff2010, Lee2010}  Evidence for Fe-As bonding was obtained from the spatial distribution of the magnetization density.\cite{Ratcliff2010}  

This situation is entirely different than in the high $T_{\rm c}$ cuprates, where an oxygen atom is situated midway between each pair of Cu atoms.  Thus, in order for conduction to occur on the Cu sublattice the conduction carriers have to hop from Cu to O to Cu, which results in a large contribution of the O $2p$ orbitals to the density of states at the Fermi energy and a very strong antiferromagnetic superexchange interaction between the Cu spins 1/2.  A lucid and more extended discussion of relationships between the FeAs-based and cuprate materials is given by Sawatzky \emph{et al}.\cite{Sawatzky2008}  The density of states at the Fermi energy obtained from band structure calculations on the Fe-based materials is an important quantity determining the electronic properties such as the magnetic susceptibility and is therefore discussed separately in the following Sec.~\ref{Sec_N(EF)}.

A critically important feature of the Fe-based superconductors as revealed by band structure calculations and angle-resolved photoemission spectroscopy (ARPES) measurements is that they are \emph{semimetals}.  This term has a very special and specific meaning in condensed matter physics.  It does not mean ``almost a metal'' or ``similar to a metal.''  A semimetal is a true metal, which is defined as a material having finite conductivity as the absolute temperature $T$ approaches zero Kelvins.  Furthermore, this term must be distinguished from an equally confusing term ``half-metal'', which is also a true metal but one in which the conduction band is completely spin-polarized (ferromagnetic).  We will not further discuss half-metals in this review.

\begin{figure}
\includegraphics [width=2.5in]{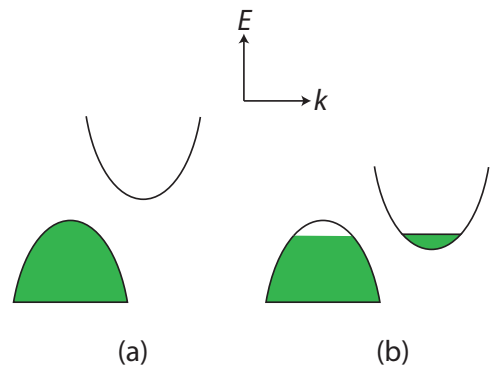}
\caption{(Color online) A sketch showing how a semimetal comes about in 2 dimensions.  The energy $E$ of an electron is plotted versus the $x$ component $k$ of the crystal momentum ($\sim $ mass times speed) of the electron.  The $y$ momentum component axis points into the page.  The pictures are vertical cuts through the centers of two-dimensional bands that are paraboloids of revolution about their vertical symmetry axes.  The upper band is the conduction band and the lower band is the valence band.  The parts of each band filled by electrons are shown as solid green.  (a) An indirect band gap semiconductor with a filled lower valence band and an empty upper conduction band at temperature $T = 0$.  Thus at $T = 0$ the material is an insulator.  (b) A semimetal.  Here, the valence and conduction bands overlap slightly in energy and some electrons from the valence band spill over to the conduction band until the energies of the highest occupied electron states in each band become the same (the Fermi energy), as shown.  This is a true metal with finite conductivity at $T = 0$.  The current carriers in the valence band are ``holes'' because they arise from holes in the electron population at the top of the band, and are ``electrons'' at the bottom of the conduction band, and the numbers of each are the same.  Thus one speaks of hole and electron pockets, respectively, as measured by ARPES.  If the material is ``doped'' with excess electrons or holes, then the electron and hole concentrations are no longer the same.  In this case, in the absence of electron correlation effects even the material in (a) is metallic at $T = 0$ because either there are electrons in the conduction band or holes in the valence band at $T = 0$, respectively.}
\label{semimetal_BS}
\end{figure}

\begin{figure}
\includegraphics [width=3.3in]{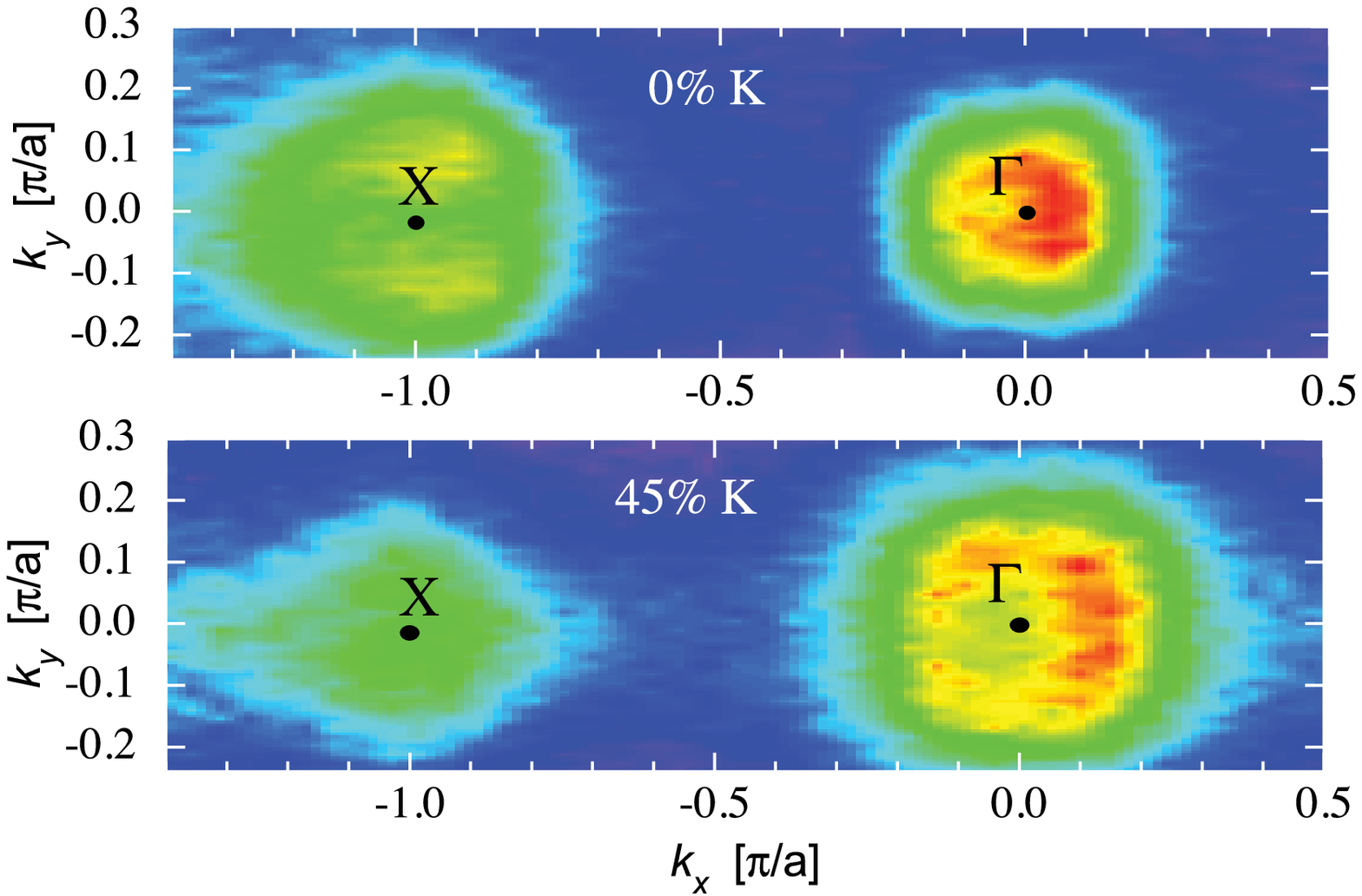}
\caption{(Color online) The Fermi surfaces of Ba$_{1-x}$K$_x$Fe$_2$As$_2$ for $x = 0$ (top panel) and $x = 0.45$ (bottom panel) as measured by ARPES, in the $k_x$-$k_y$ plane of the Fe layers, where $a$ is the basal plane tetragonal lattice constant of the unit cell.\cite{Liu2008}  The hole pocket(s) is at the $\Gamma$ point and the electron pocket(s) is at the X~point of the Brillouin zone.  Band structure calculations indicate that there are two concentric electron pockets about the $\Gamma$ point and two concentric hole pockets about the X point, but these early ARPES measurements do not resolve the members of either pair.  Higher resolution data are shown below in Fig.~\ref{FigNakayama_BaKFeAs_ARPES}.  Hole doping occurs by replacing some of the Ba by K as $x$ increases from~0 to~0.45, and the Fermi energy decreases so that the hole pockets at the $\Gamma$ point get larger and the electron pockets at the X point shrink, as expected from Fig.~\ref{semimetal_BS}(b).  Reprinted with permission from Ref.~\onlinecite{Liu2008}.  Copyright (2008) by the American Physical Society.}
\label{FigBaKFe2As2_FS} 
\end{figure}

To explain what a semimetal is, Fig.~\ref{semimetal_BS}(a) shows a sketch of the band structure of a two-dimensional semiconductor.  By definition, the band structure of a material is the set of dependencies of the energies of itinerant (movable) current carriers as a function of their momenta $k$ (mass times velocity).  In Fig.~\ref{semimetal_BS}(a), the conduction and valence bands are separated vertically by an energy gap and their centers of momenta are displaced horizontally from each other which results in an ``indirect band gap'' semiconductor that is an electrical insulator at temperature $T = 0$.  A semimetal is a metal that derives its metallic character from the same type of horizontally displaced valence and conduction bands but that overlap slightly in energy as illustrated in Fig.~\ref{semimetal_BS}(b) so there is no longer an energy gap between the two bands.  Therefore electrons ``spill over'' from the previously filled valence band into the conduction band until the energy of the highest-energy occupied state in each band (the Fermi energy) is the same.  This results in small electron and hole ``pockets'' of current carriers as described in the figure caption.  These pockets are directly observed in ARPES measurements of the Fermi surface as shown in Fig.~\ref{FigBaKFe2As2_FS}.\cite{Liu2008} The ``Fermi surface'' is the locus of all points in momentum space separating the occupied from the unoccupied electron states at $T = 0$ [see Fig.~\ref{semimetal_BS}(b)].  Thus the undoped FeAs-based parent compounds are ``fully compensated'' metals, where the electron and hole concentrations $n_{\rm e}$ and $n_{\rm h}$, respectively, are identical.  The existence of the electron and hole pockets allows a new type of superconducting electron pairing to occur, known as $s^\pm$ pairing.\cite{mazin2008}  This pairing is specific to the type of semimetallic electronic structure shown in Figs.~\ref{semimetal_BS}(b) and~\ref{FigBaKFe2As2_FS} where there is a pairing interaction between the electron and hole pockets in momentum space, and the superconducting order parameter has opposite signs on the electron and hole pockets, hence the designation $s^{\pm}$.

\subsubsection{\label{SecBZ} Reciprocal Lattices and Brillouin Zones}

\begin{figure}
\includegraphics [width=3.3in]{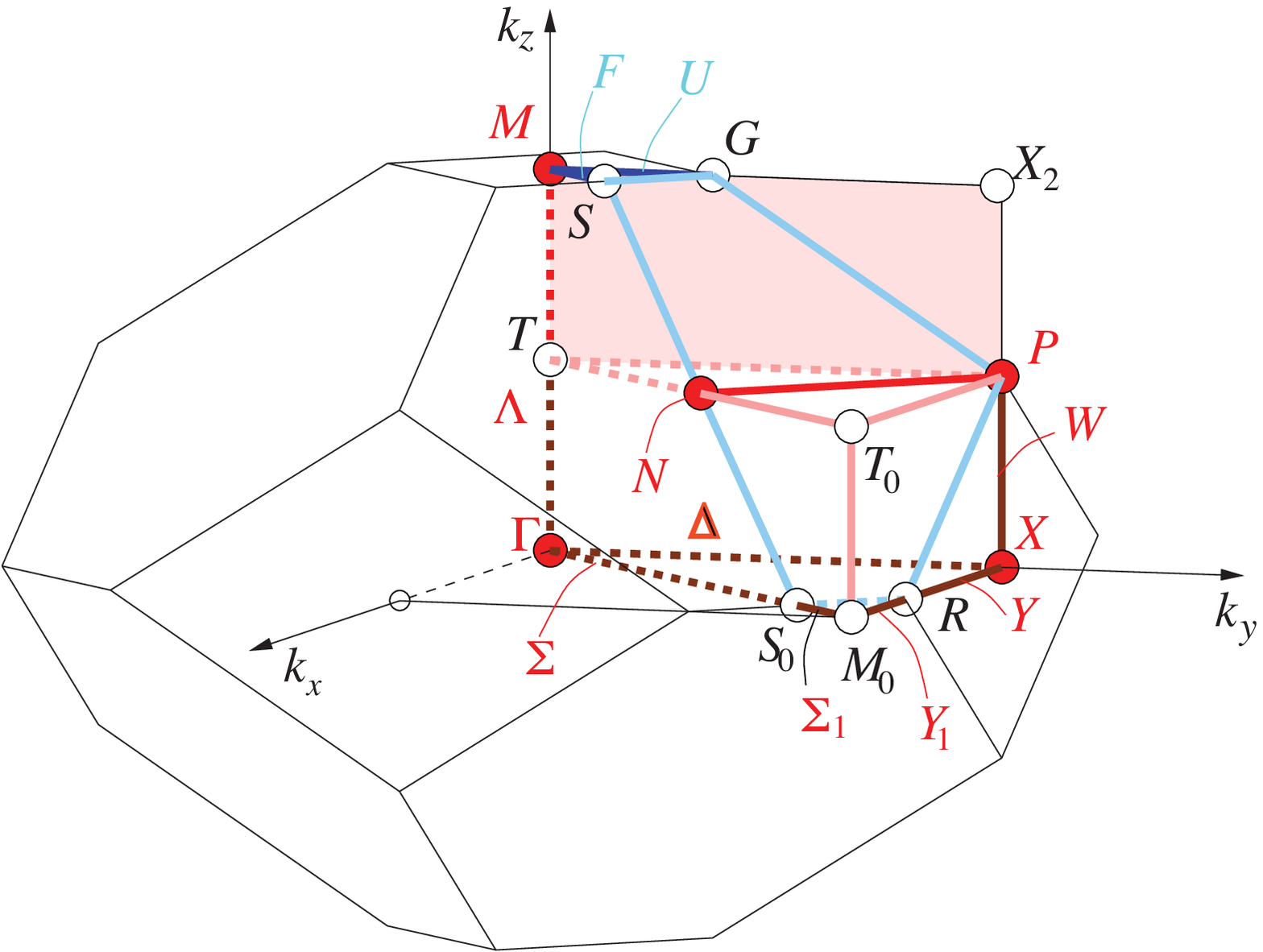}
\includegraphics [width=3.3in,viewport=00 0 360 300,clip]{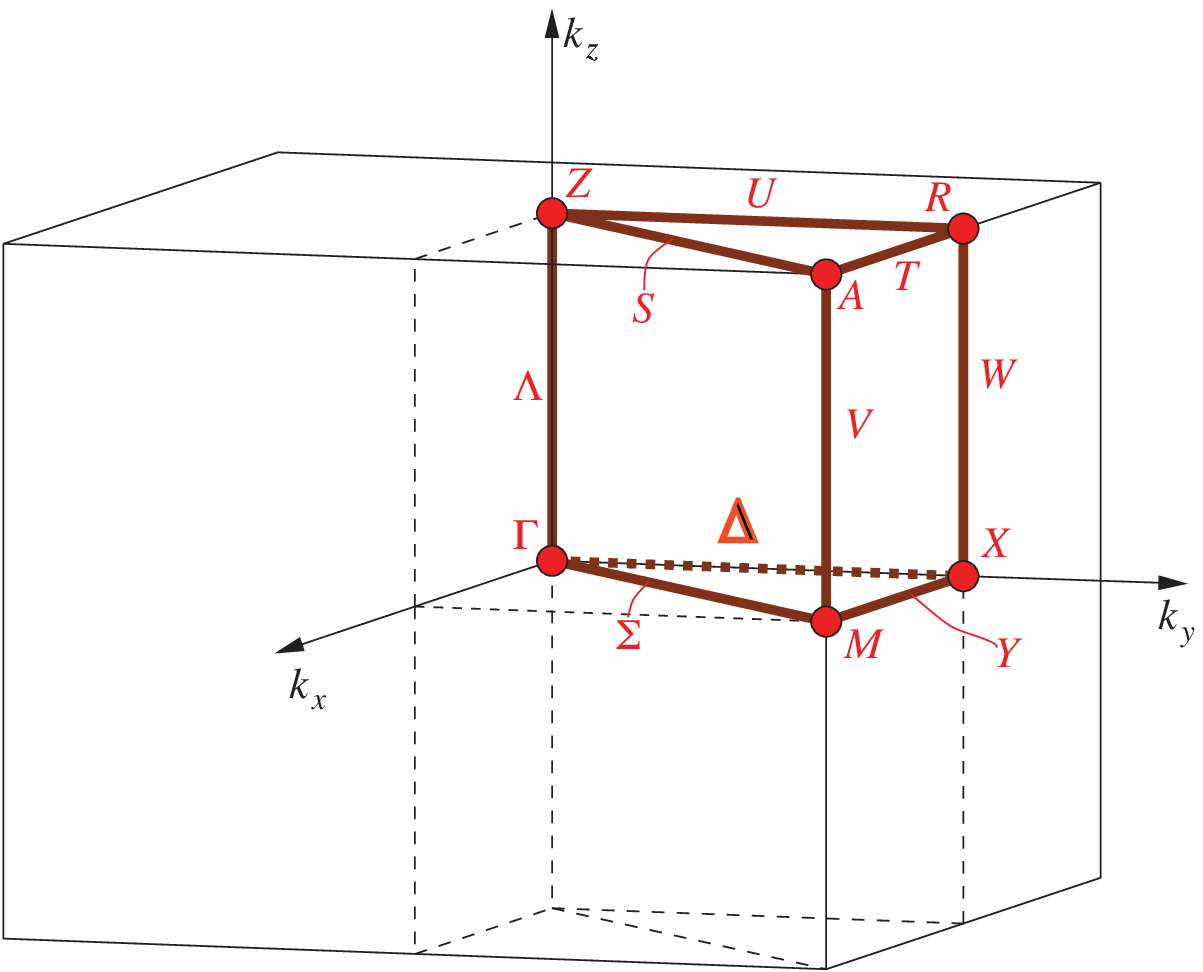}
\caption{(Color online) Topology of the first Brillouin zones (BZ) for the body-centered-tetragonal (bct) space group \emph{I}4/\emph{mmm} (No.~139, for $c/a \approx 2.2 > 1$, top, 122-type) and primitive tetragonal (pt) \emph{P}4/\emph{nmm} (No.~129, bottom, 11-, 111- and 1111-types) that Fe-based compounds crystallize in at room temperature, respectively.  The center of the BZ at {\bf k} = 0 in each structure is denoted as the $\Gamma$ point.  For the pt reciprocal lattice in the bottom figure, the reciprocal lattice translation vectors are ${\bf k} = \left(n_1\frac{2\pi}{a}, n_2\frac{2\pi}{a},n_3\frac{2\pi}{c}\right)$, where $n_1,n_2,n_3$ are integers. Ambiguities in the creation and labeling of the reciprocal lattice and BZ for the bct direct lattice in the top figure are discussed in the text. The most important of these is that, for pt and bct direct lattices with the same lattice parameters oriented in the same way in real space, the X~point in the top panel is at the same point in reciprocal space as the M~point in the bottom panel.  Reprinted with permission from Ref.~\onlinecite{BZFigs}. }
\label{BZs} 
\end{figure}

\begin{figure}
\includegraphics [width=1.5in]{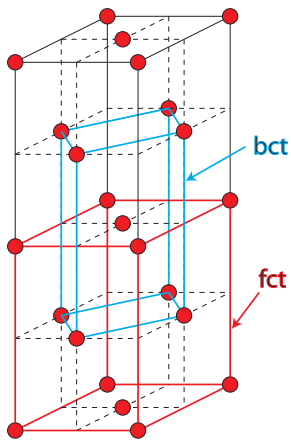}
\caption{(Color online) Reduction of a face-centered-tetragonal (fct) reciprocal lattice (red outlined unit cell) to a smaller body-centered-tetragonal (bct) reciprocal Bravais lattice (blue outlined unit cell).  The filled red circles are the reciprical lattice points.}
\label{BZ-fct_to_bct} 
\end{figure}

A Brillouin zone (BZ) is a Wigner-Seitz primitive cell of the reciprocal lattice of a crystal.\cite{kittel1966}  The (first) BZs of the room-temperature body-centered-tetragonal (bct) 122-type and primitive tetragonal (pt) 11-, 111- and 1111-type Fe-based tetragonal crystal structures are shown in Fig.~\ref{BZs}.\cite{BZFigs}  The top and bottom BZs look different from each other because the symmetries of the two reciprocal lattices  are different.  The reciprocal lattice of the pt real-space (``direct'') lattice is another pt lattice, and the reciprocal lattice of the bct direct lattice is another bct lattice (see below!).   Note that the M point in the bottom panel of Fig.~\ref{BZs} is at a qualitatively different position in reciprocal space from the M point in the top panel.   The notations for the symmetry lines and points in the figures are defined in Ref.~\onlinecite{Cracknell1979}, together with pictures of the first BZs of all fourteen\cite{kittel1966} three-dimensional direct Bravais lattices and the symmetry line and point designations for each.  

With regard to notation, the adjective ``primitive'' when applied to describing a unit cell in real space means that there is one lattice point per unit cell.  To each lattice point is attached a ``basis'' of atoms.\cite{kittel1966}  For example, for the primitive tetragonal LaFeAsO, there are two formula units per unit cell, which means that the basis consists of two formula units, together containing two Fe atoms.  Since each Fe layer contains two Fe atoms within the unit cell ($\equiv$~Fe$_2$), the 1111-type (and 11- and 111-type) Fe-based superconductors have one Fe layer per unit cell.  For the 122-type bct compound ${\rm BaFe_2As_2}$, on the other hand, the direct unit cell contains two lattice points, one at a corner and the other at the body center.  Since there are again two formula units per unit cell, the basis consists of one formula unit, again containing two Fe atoms.  Since there are four Fe atoms per unit cell, there are two Fe$_2$ layers per unit cell, which is also seen directly from Fig.~\ref{Struct122}.  This is an important distinction between the structures of the 122-type versus the 11-, 111- and 1111-type Fe-based materials.  Similarly, in reciprocal space, a primitive unit cell contains one lattice point per unit cell, and a bct unit cell again contains two lattice points per unit cell.  Both real space and reciprocal space lattices are described by the same set of fourteen three-dimensional Bravais lattices.

Figure~\ref{Struct122} shows that the bct unit cell of the 122-type compounds contains two Fe layers per unit cell, as just discussed.  The 214-type cuprates such as ${\rm La_2CuO_4}$ at high temperatures $T > 530$~K and the doped 214-type superconducting compositions at all temperatures also have a bct unit cell (space group $I$4/\emph{mmm}).\cite{Jorgensen1997} However, due to the very weak electronic and magnetic coupling between the CuO$_2$ layers, the magnetic and electronic transport properties are quasi-two-dimensional and one can treat the two layers per unit cell as equivalent and can thus ignore for most purposes the body-centering of the tetragonal unit cell.\cite{Johnston1997}  This is not the case with the 122-type FeAs compounds, because their interlayer coupling is much stronger.  Indeed, Yaresko et al.\ showed for the 122-type (Ba,K)Fe$_2$As$_2$ system that ``the bct symmetry mixes $k_z$ dispersion into the $(k_x, k_y)$ dispersions.''  Furthermore, Graser et al.\ showed theoretically that interlayer coupling has an influence on the spin fluctuations and superconducting pairing strength in BaFe$_2$As$_2$.\cite{Graser2010}  Many experiments indicate significant interlayer electronic coupling in the 122-type compounds such as in Fig.~\ref{FigBa(FeCo)2As2_ARPES} below.  Thus there is physics in the statement that there are two layers per unit cell for the 122-type Fe-based compounds. 

In the Fe-based superconductor field, the BZs in published calculated figures of the Fermi surfaces for the bct 122-type compounds sometimes do not look like the BZ in the top panel of Fig.~\ref{BZs}.  Instead, the BZ is plotted for convenience in these publications as a rectangular parallelopiped, such as in Figs.~\ref{LaFeAsOBS} and~\ref{FigFS} below, with the same volume as the actual Brillouin zone (I. I. Mazin, private communication).  

In the bottom panel of Fig.~\ref{BZs} showing the pt BZ of the \emph{primitive tetragonal} direct lattice, the ${\bf k}_x$, ${\bf k}_y$, and ${\bf k}_z$ translation vectors of the reciprocal lattice point along the conventional pt direct lattice axes, respectively, according to\cite{kittel1966}
\bea 
{\bf k}_x &=& \frac{2\pi}{a}\hat{\bf a},\nonumber\\
{\bf k}_y &=& \frac{2\pi}{a}\hat{\bf b}, \hspace{0.2in}{\rm (primitive\ tetragonal)}\label{ptkaxes}\\
{\bf k}_z &=& \frac{2\pi}{c}\hat{\bf c}.\nonumber
\eea

However, the labeled ${\bf k}_x$ and ${\bf k}_y$ axes of the conventional (as opposed to primitive) reciprocal lattice of the \emph{body-centered-tetragonal} direct lattice are at an angle of $45^\circ$ to the ${\bf a}$ and ${\bf b}$ axes of the direct lattice, respectively.  \emph{This important fact is not obvious from the top panel of Fig.~\ref{BZs}.}  The following explanation was provided by Andreas~Kreyssig (private communication).  The reciprocal lattice of a body-centered direct lattice is in general face-centered.  Thus, the reciprocal lattice of a bct direct lattice is a face-centered tetragonal (fct) lattice, but the fct lattice is not a conventional Bravais lattice.  However, as shown in Fig.~\ref{BZ-fct_to_bct}, a fct reciprocal lattice unit cell can be transformed into a smaller bct Bravais reciprocal lattice unit cell by translating the reciprocal lattice by half the fct reciprocal unit cell height along the ${\bf k}_z$ axis, rotating the lattice in the ${\bf k}_x$-${\bf k}_y$~plane by 45$^\circ$, and decreasing the ${\bf k}_x$ and ${\bf k}_y$ fct reciprocal lattice vector lengths by a factor of $\sqrt{2}$.  Thus, \emph{the reciprocal lattice of a bct direct lattice is another bct lattice} in which the ${\bf k}_x$ and ${\bf k}_y$ reciprocal lattice translation vectors and axes are at angles of 45$^\circ$ to the ${\bf a}$ and ${\bf b}$ conventional direct lattice vectors, respectively.  This is a very confusing aspect of the BZ of the bct direct lattice.  Fortunately, the \emph{reciprocal} lattices of only two of the fourteen three-dimensional direct Bravais lattices have the directions of their conventional lattice vectors in different directions than the directions of the conventional lattice vectors of the respective \emph{direct} lattices.  One of these exceptions is the bct direct lattice and the other is the rhombohedral direct lattice.\cite{Cracknell1979} 

\begin{figure}
\includegraphics [width=1.in]{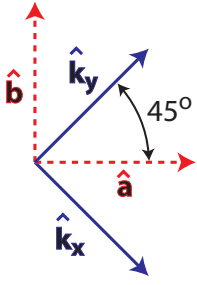}
\caption{(Color online) Relationship of the conventional direct lattice unit vectors $\hat{\bf a}$ and $\hat{\bf b}$ for a body-centered-tetragonal direct lattice with the conventional reciprocal lattice unit vectors $\hat{\bf k}_x$ and $\hat{\bf k}_y$ of its body-centered-tetragonal reciprocal lattice.}
\label{bct_BZ_axes} 
\end{figure}

The Wigner-Seitz primitive cell of the bct reciprocal lattice is the first BZ of the bct reciprocal lattice, which is shown in the top panel of Fig.~\ref{BZs} for a direct bct lattice with $c/a > 1$ as for the 122-type FeAs-based compounds (the topology of the BZ is different if $c/a < 1$).\cite{BZFigs}   The conventional reciprocal lattice translation vectors ${\bf k}_\alpha~(\alpha = x,y,z)$ of the conventional bct direct lattice are written in terms of the conventional lattice vectors {\bf a}, {\bf b} and {\bf c} of the bct direct lattice with magnitudes $a$, $a$, and $c$, respectively, as\cite{Cracknell1979}
\bea
{\bf k}_x &=& \frac{2\pi}{a}(\hat{\bf a} - \hat{\bf b}),\nonumber\\
{\bf k}_y &=& \frac{2\pi}{a}(\hat{\bf a} + \hat{\bf b}), \hspace{0.2in}{\rm (body\ centered\ tetragonal)}\label{Eqka}\\
{\bf k}_z &=& \frac{4\pi}{c}\hat{\bf c}\nonumber,
\eea    
with magnitudes
\bea
|{\bf k}_x| &=& \frac{2\sqrt{2}\pi}{a},\nonumber\\
|{\bf k}_y| &=& \frac{2\sqrt{2}\pi}{a},\label{Eqka2}\\
|{\bf k}_z| &=& \frac{4\pi}{c}\nonumber,
\eea    
respectively.  Note the differences between these magnitudes for the bct reciprocal lattice translation vectors and those for the pt reciprocal translation lattice vectors in Eqs.~(\ref{ptkaxes}).  The relationships between the in-plane bct direct lattice unit vectors and bct reciprocal lattice unit vectors are shown in Fig.~\ref{bct_BZ_axes}.  

In the bottom panel of Fig.~\ref{BZs}, the ${\bf k}_x$-axis of the reciprocal lattice for the pt 11-type, 111-type, and 1111-type FeAs compounds points in the $\hat{\bf a}$ direction and intersects a face of the BZ at the X point, and the M point of the BZ is in the corner with in-plane wave vector $\left(\frac{1}{2},\frac{1}{2}\right)$~r.l.u.\ [for the notation, see Eq.~(\ref{EqRLV2}) below].  On the other hand, the $\hat{\bf a}$-direction of the direct bct lattice is at a 45$^\circ$ angle to the bct reciprocal lattice ${\bf k}_x$ and ${\bf k}_y$ axes in the top panel of Fig.~\ref{BZs} as shown in Fig.~\ref{bct_BZ_axes} and intercepts the BZ boundary at the S$_0$ point.  A comparison of Figs.~\ref{BZs} and~\ref{bct_BZ_axes} reveals the important fact that, with respect to the conventional tetragonal direct lattices with the same lattice parameter $a$, \emph{the M point in the pt BZ is at the same position in reciprocal space as the X point in the bct BZ}\@.  This confusing notational difference appears in band structure calculations for the different Fe-based compounds in the next section.  For a pt and a bct direct lattice  with the same basal plane lattice parameter $a$, the magnitudes of the reciprocal lattice vectors from the $\Gamma$ point to the M or X point, respectively, have the same value $Q = \sqrt{2}\pi/a$.  This fact will be important when considering Fermi surface nesting, long-range magnetic ordering, and the neutron spin resonance in the superconducting state of the 122-type versus 11-, 111-, and 1111-type Fe-based materials.

Electron, nonmagnetic neutron, or x-ray diffraction (elastic scattering) patterns of crystals are Fourier transforms in wave vector space of the atomic positions in the material and are directly related to the reciprocal lattice.  A peak in the diffraction pattern (in wave vector space) occurs for an electron, neutron, or x-ray wave vector \emph{change} ${\bf Q}$ satisfying\cite{kittel1966}
\be
{\bf Q} = h{\bf k}_x + k{\bf k}_y + \ell {\bf k}_z,
\label{EqDiffMI}
\ee
where ${\bf k}_\alpha$ ($\alpha = x,y,z$) are the reciprocal lattice translation vectors of the crystal and $(hk\ell)$ are  integers called Miller indices that label the particular reflection.  One can easily show that in a Bragg diffraction geometry in which the angle of incidence equals the angle of reflection from planes of atoms separated by a distance $d$, the magnitude of ${\bf Q}$ is $Q = (4\pi/\lambda)\sin\theta$, where $\lambda$ is the electron, neutron or x-ray wavelength and $\theta$ is the Bragg angle.  Then using the additional relation $Q = 2\pi/d$ gives the Bragg law $\lambda = 2d\sin\theta$.  This is often written as $n\lambda = 2d\sin\theta$ with integer $n$, but when Miller indices are used to label the diffraction peaks, the $n$ is absorbed into the values of the Miller indices.

Finally, we mention that the BZs of the tetragonal Fe-based superconductors have been discussed in the literature in terms of both ``folded'' and ``unfolded'' BZs.  Irrespective of the type of Fe-based superconductor, whether it is 1111-, 111-, 11-, or 122-type, a layer has two Fe atoms (Fe$_2$) per unit cell.  The ``folded'' BZ of the Fe$_2$ direct lattice is the actual BZ utilizing the actual crystal structures with two Fe atoms and two chalcogen or pnictogen atoms per Fe layer per unit cell.  An ``unfolded'' BZ arises when the direct lattice is considered to be a pt Fe square lattice, with only one Fe atom in each Fe layer per unit cell and ignoring the presence of any other atoms in the unit cell such as the pnictogen or chalcogen atoms.  From Fig.~\ref{1111_122_layers}, the translation vectors of the Fe square lattice direct unit cell are rotated by $45^\circ$ with respect to those of the Fe$_2$ cell, with a unit cell edge that is a factor of $\sqrt{2}$ smaller.  The in-plane BZ of the Fe square lattice is also rotated by $45^\circ$ with respect to the BZ of the Fe$_2$ one and is a factor of $\sqrt{2}$ larger than the Fe$_2$ one.  This has been a source  of confusion in the literature, in addition to that discussed above deriving from the 45$^\circ$ rotation of the bct reciprocal lattice with respect to the bct direct lattice.

\begin{figure}
\includegraphics [width=3.3in]{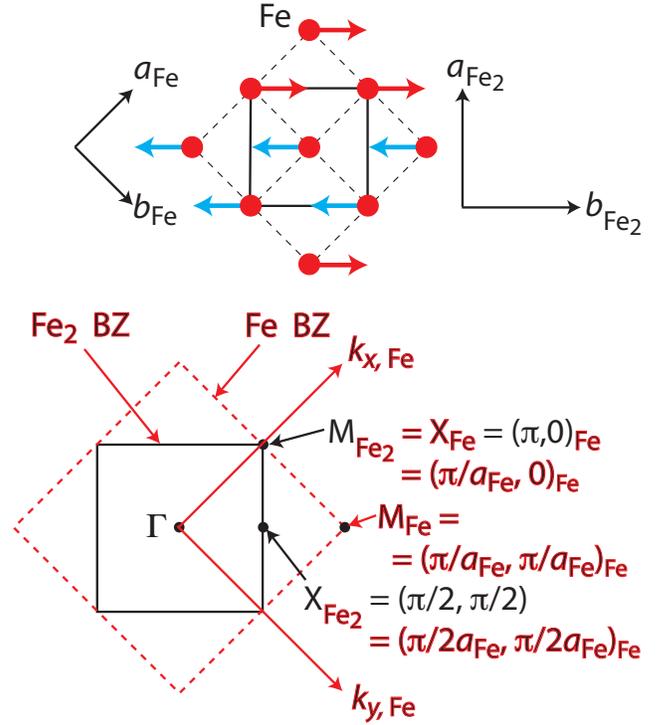}
\caption{(Color online) Relationships between a mixture of in-plane Brillouin zone notations used for FeTe in Ref.~\onlinecite{Xia2009}.  The top figure gives the magnetic structure in the ordered state.  The Fe square lattice unit cells in dashed lines in the top part was given in the paper as the direct lattice to be referred to.  The black small square Brillouin zone and black notations in the lower part of the figure is a mixture of notations for the  ``unfolded'' Fe BZ for the Fe square lattice containing one Fe atom per cell per layer (red dashed box), and the tetragonal ``folded'' Fe$_2$ unit cell containing two Fe atoms per unit cell per layer (black solid box).  The notations in red were added by us to solve the puzzle of the notations of these two Brillouin zones that are \emph{both used in the same figure} and elsewhere in Ref.~\onlinecite{Xia2009}, as described in the text.  Even though the black BZ given is the Fe$_2$ BZ, the $k_x$ and $k_y$ axes referred to in Fig.~3 of the paper are actually for the Fe square lattice BZ as shown by the red $k_{x\,{\rm Fe}}$ and $k_{y\,{\rm Fe}}$ BZ axes here.}
\label{TetragVsFeBZ} 
\end{figure}

It is especially challenging to follow the discourse of a paper when the reciprocal lattice notations of both the Fe$_2$ and Fe square direct lattices are used interchangably in the same figure and/or elsewhere in the paper as for example occurs in Ref.~\onlinecite{Xia2009}, in which ARPES data for a crystal of FeTe are presented.  We discuss the notations in this paper to illustrate how to convert between the various notations.  At room temperature, FeTe has a primitive tetragonal crystal structure with two Fe atoms per lattice point and in-plane lattice parameter $a_{{\rm Fe}_2} = b_{{\rm Fe}_2} = 3.822$~\AA\ (see Appendix).  Reference~\onlinecite{Xia2009} instead gives the direct lattice as the Fe square lattice with lattice parameters $a_{\rm Fe} = b_{\rm Fe} = a_{{\rm Fe}_2}/\sqrt{2}$ with one Fe atom per lattice point, which is reproduced at the top of Fig.~\ref{TetragVsFeBZ} as the dashed square lattice.  Then in another figure, the BZ is shown as the black square at the bottom of Fig.~\ref{TetragVsFeBZ} with the M and X points labeled in black as shown.  The authors also show the hole pockets to be at the $\Gamma$ point at the center (0,0) of the black BZ and the electron pockets at the corner M points.  Thus, the black BZ corresponds to the BZ of the Fe$_2$ direct lattice, because that is the lattice for which the electron pockets are at the M points of the BZ (see Fig.~\ref{LaFeAsOBS} below), even though the direct lattice is given in Ref.~\onlinecite{Xia2009} as the Fe square lattice.  The Fe square lattice BZ is shown as the red dashed square at 45$^\circ$ to the black one in the bottom panel of Fig.~\ref{TetragVsFeBZ}.  Then, even though the letter designations M and X correspond to points in the Fe$_2$ BZ, the numerical values given in Ref.~\onlinecite{Xia2009} for their coordinates are the coordinates with respect to the Fe square lattice BZ\@.  Thus instead of giving the M point coordinates in the  Fe$_2$ BZ as $(\pi/a_{{\rm Fe}_2},\pi/a_{{\rm Fe}_2})$, the coordinates are given as $(\pi/a_{\rm Fe},0)$ in the notation of the Fe square lattice BZ\@.  Similarly, the coordinates of the black X point\cite{Xia2009} given in Fig.~\ref{TetragVsFeBZ} for the Fe$_2$ BZ are the coordinates with respect to the Fe square lattice BZ, as shown in the figure.  Additionally, in Fig.~3 of Ref.~\onlinecite{Xia2009}, the labeled reciprocal space $k_x$ and $k_y$ translation vectors go from the $\Gamma$ point through two M points of the BZ\@.  As shown by the red axes in the bottom panel of Fig.~\ref{TetragVsFeBZ}, these reciprocal lattice vectors are those of the red dashed Fe square lattice BZ, not of the solid black square Fe$_2$ BZ, even though it is the latter BZ to which the designation M in Ref.~\onlinecite{Xia2009} refers.  

Reference~\onlinecite{Lee2009} presents ARPES data, neutron scattering data and band structure calculations on FeTe$_{0.5}$Se$_{0.5}$ and is similarly hard to follow, because the direct lattice is assumed to be the Fe square lattice, whereas the band calculations presumably also include the contributions of the Se/Te atoms, which requires the Fe$_2$ lattice, which in turn requires a folding of the Fe square lattice Brillouin zone.

As another example of the ambiguities that can occur due to various Brillouin zone notations, the authors of Ref.~\onlinecite{Lumsden2010a} present inelastic neutron scattering measurements of Fe$_{1+y}$Se$_{1-x}$Te$_x$ and show in their Fig.~1 the BZs of both the Fe square lattice and the conventional tetragonal (Fe$_2$) direct unit cell.  The labeling of the points in the two BZs is different.  For the Fe$_2$ BZ, the labeling is in reciprocal lattice units (r.l.u.), whereas for the Fe square lattice BZ it is in \AA$^{-1}$ with the basal plane lattice parameter $a$ set to unity [see Eqs.(\ref{EqRLV1}) and~(\ref{EqRLV2}) and Table~\ref{AFPropVectors} below].  Evidently the authors do this to help differentiate certain points in the two respective BZs.  Then in the captions to Figs.~2--4, it is not stated which of the two BZs the BZ notation refers to.  However, because the units of the reciprocal lattice in the latter three figures are r.l.u.'s instead of \AA$^{-1}$, one can infer from the notation in their Fig.~1 that the BZ being referred to in their Figs.~2--4 is the Fe$_2$ BZ\@.  

In the remainder of this review, to avoid the above multiplicity of ambiguities, reciprocal lattice vectors are usually written with respect to the conventional tetragonal unit cell axis unit vectors $\hat{\bf a},\hat{\bf b},\hat{\bf c}$ as 
\bea
{\bf Q}({\rm \AA}^{-1}) &=& \frac{H2\pi}{a}\hat{\bf a} + \frac{K2\pi}{a}\hat{\bf b} + \frac{L2\pi}{c}\hat{\bf c} \nonumber\\
&\equiv& \left(\frac{H2\pi}{a},\frac{K2\pi}{a},\frac{L2\pi}{c}\right),\label{EqRLV1}
\eea
where $H,K,L$ are real variables (not necessarily integers).  Note the difference in notation for the $\hat{\bf c}$ component compared with that for the bct BZ in Eq.~(\ref{Eqka}).  In the literature, $a$ and $c$ in Eq.~(\ref{EqRLV1}) are usually set to unity.  A {\bf Q} is often expressed in reciprocal lattice units (r.l.u.) implicitly defined by Eq.~(\ref{EqRLV1}) as 
\be
{\bf Q} = (H,K,L)~{\rm r.l.u.}
\label{EqRLV2}
\ee
For example, the X and M points in the bottom panel of Fig.~\ref{BZs} have in-plane coordinates $\left(\frac{1}{2},0\right)$ and $\left(\frac{1}{2},\frac{1}{2}\right)$~r.l.u., respectively.  As noted above, it is very common in the Fe-based superconductivity literature to use both notations in Eqs.~(\ref{EqRLV1}) and~(\ref{EqRLV2}) in the same paper, sometimes in the same sentence or in the same figure.  The difference in notations is that the former one always has ``$\pi$'' in it, whereas the r.l.u.\ notation never does.  For additional conversions between notations for particular reciprocal lattice points, see Fig.~\ref{Ordering_Wave_Vectors} below.

The exception to using tetragonal reciprocal lattice notation in this review is when we discuss magnetic ordering in the orthorhombically distorted low-temperature structures of the 122- and 1111-type compounds, where the magnetic structure is most naturally discussed in terms of the orthorhombic reciprocal lattice notation.  The low-temperature orthorhombic basal plane direct $a$ and $b$ axes are rotated by 45$^\circ$ and the lattice parameters are a factor of $\sqrt{2}$ longer compared with the high-temperature tetragonal $a$- and $b$-axes in both classes of compounds.

\subsubsection{Band Structures: 122- and 1111-Type Compounds}

\begin{figure}
\includegraphics[width=3.2in]{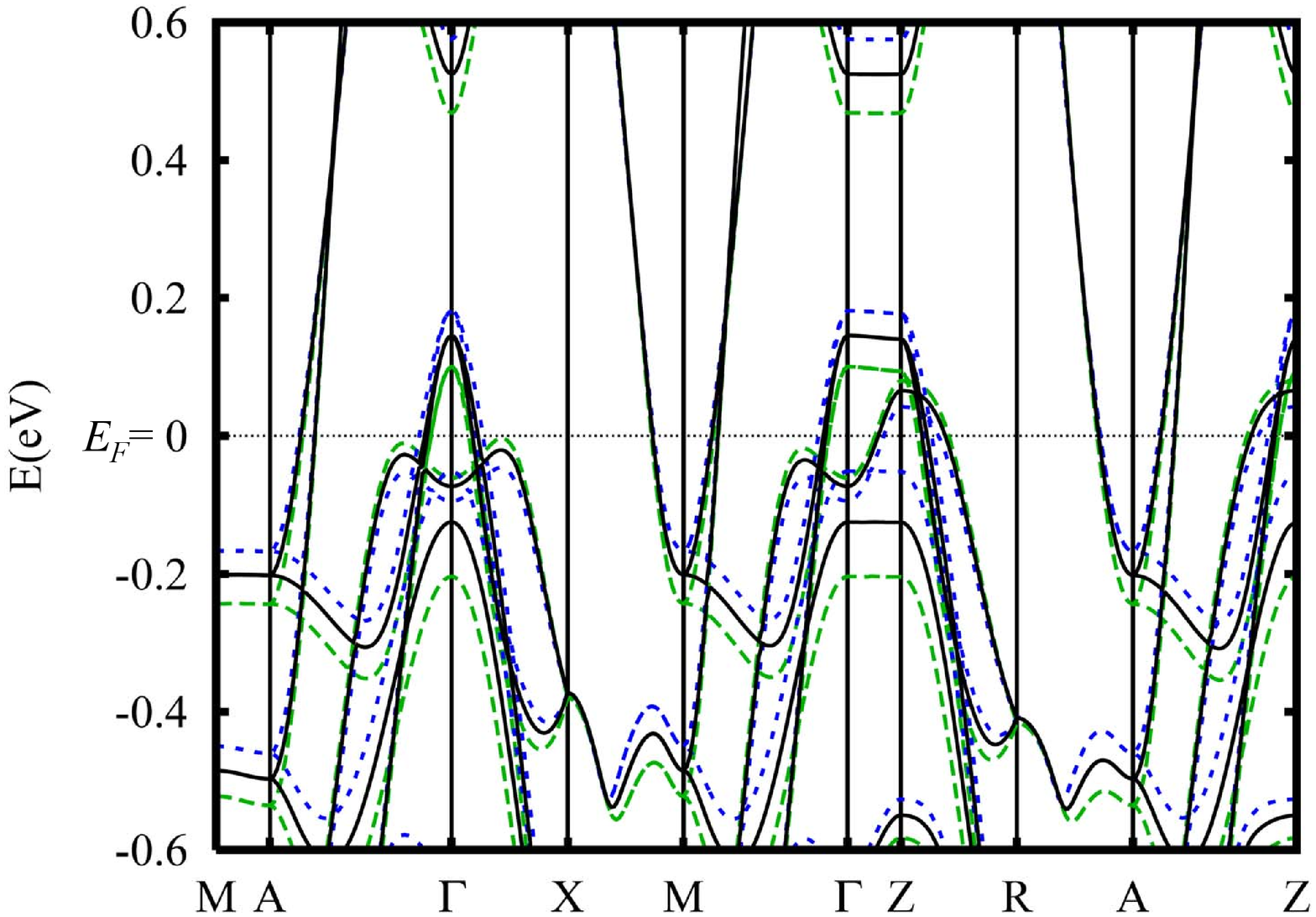}
\includegraphics[width=3in]{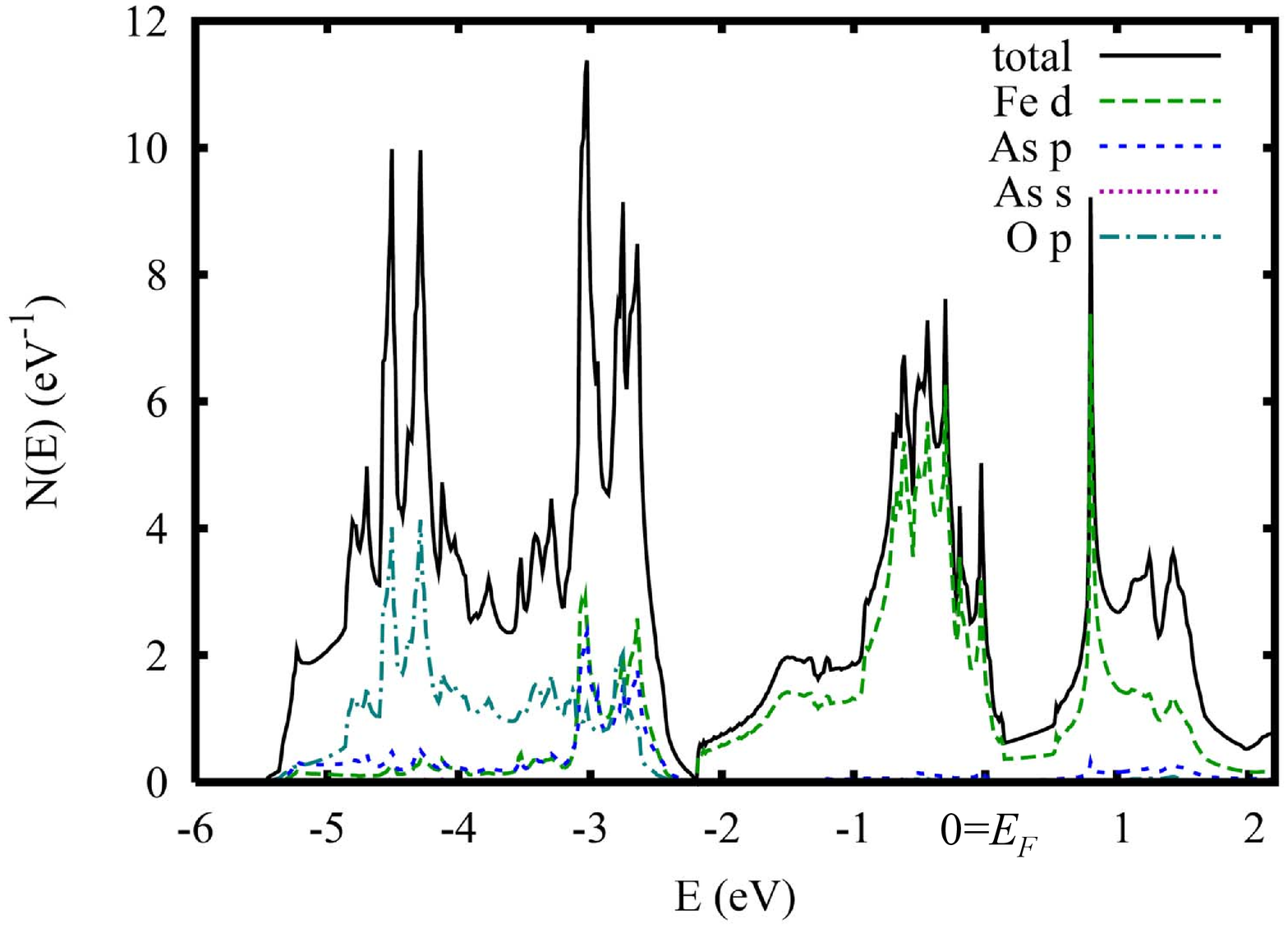}
\includegraphics[width=2.25in,viewport=-30 00 470 380,clip]{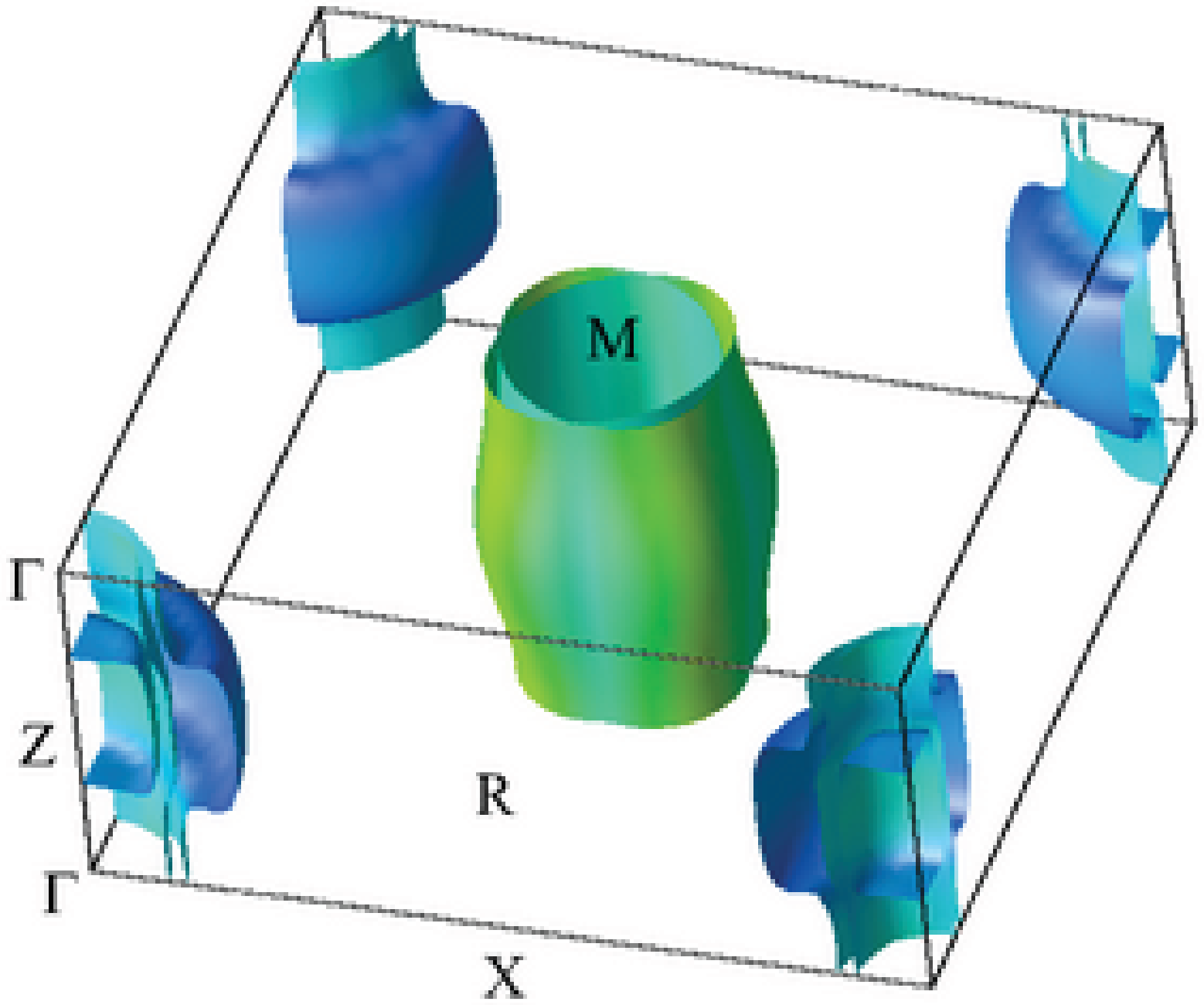}
\caption{(Color online)  Top: LDA band structure of LaFeAsO.\cite{djsingh2008b}  The symmetry designations are given in the bottom panel of Fig.~\ref{BZs}.  The solid black curves give the band structure for the observed crystal structure, whereas the blue dotted lines and green dashed lines are for the As planes shifted by 0.035~\AA\ towards and away from the Fe planes, respectively.  This shows the sensitivity of the band structure to the As positions.  From the figure, the hole bands are centered in the $k_x$-$k_y$ plane at $\Gamma$, and the electron bands at M\@. Middle: Total density of states $N$ versus energy $E$ (solid black curve) and orbital projected densities of states (dashed lines).  Most of the density of states at the Fermi energy, $N(E_{\rm F})$, arises from the Fe orbitals.  Bottom: Fermi surfaces, shaded by velocity [dark (blue) is low velocity]. Note that the center $\Gamma$ of the Brillouin zone is shifted to the corners of the figure.  The hole Fermi surfaces (pockets) are at the corners of the figure, through the $\Gamma$ point, and the electron pockets are in the middle, through the M point.  Reprinted with permission from Ref.~\onlinecite{djsingh2008b}.  Copyright (2008) by the American Physical Society.}
\label{LaFeAsOBS} 
\end{figure}

Calculations of the band structure using density functional theory in the local density approximation (LDA) for the room temperature primitive tetragonal structure of LaFeAsO are shown in Fig.~\ref{LaFeAsOBS} (top).\cite{djsingh2008b}  The calculations\cite{Ma2008a, djsingh2008b} clearly show the presence of hole bands near the Fermi energy $E_{\rm F}$ at the $\Gamma$ point of the Brillouin zone and electron bands at the M points, which make a 45$^\circ$ angle with respect to the $\hat{\bf a}$ and $\hat{\bf b}$ directions and are at the corners of the BZ\@.  Thus, the electron bands are located on the axes of the Fe square lattice with respect to the $\Gamma$ point (see Fig.~\ref{1111_122_layers}).  The calculations in the top panel of Fig.~\ref{LaFeAsOBS} also show that the energy bands are very sensitive to the distance of the As planes from the Fe planes.  The density of states versus energy $N(E)$ is shown in the middle panel of Fig.~\ref{LaFeAsOBS}.  The states at $E_{\rm F}$ are primarily of Fe character as noted above, indicating that the predominent electronic conduction mechanism is direct Fe-Fe hopping.  The main weight of the As 4$p$ bands is $\sim 3$~eV below $E_{\rm F}$, justifying the often quoted formal oxidation state of As in the undoped parent compounds as As$^{-3}$. The strong bonding between the Fe and As atoms is also reflected in the hybridized Fe and As bands $\sim 3$~eV below the Fermi level.  The bottom panel of Fig.~\ref{LaFeAsOBS}  shows the hole Fermi surfaces at $\Gamma$ and the electron Fermi surfaces at M\@.  A significant dispersion occurs along the $k_z$ ($\hat{\bf c}$) axis, which indicates significant electronic coupling between adjacent FeAs layers. 

\begin{figure}
\includegraphics[width=3.in,viewport= 00 40 550 420,clip]{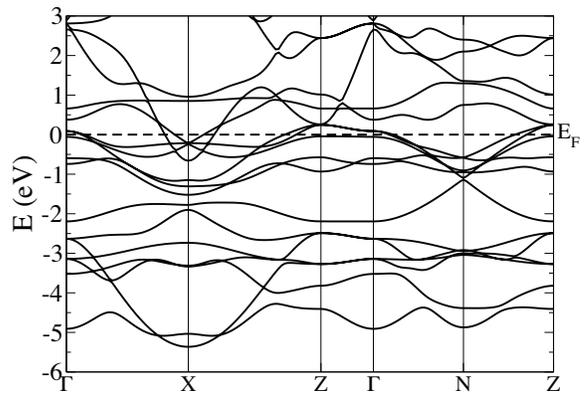}
\caption{LDA band structure around the Fermi energy $E_{\rm F}$ for body-centered-tetragonal (bct) ${\rm BaFe_2As_2}$.\cite{Zhang2009} One can clearly see hole bands at the $\Gamma$ point and electron bands at the X point(s) (see the Brillouin zone for the bct lattice in Fig.~\ref{BZs}). The X point in the bct Brillouin zone is at the same position in reciprocal space as the M point for the primitive tetragonal 11-, 111- and 1111-type compounds.  Reprinted with permission from Ref.~\onlinecite{Zhang2009}.  Copyright (2009) by the American Physical Society.}
\label{BaFe2As2BS} 
\end{figure}

A corresponding LDA band calculation for the 122-type parent compound ${\rm BaFe_2As_2}$ is shown in Fig.~\ref{BaFe2As2BS} for energies near $E_{\rm F}$.\cite{Zhang2009}  Here, hole bands occur at the $\Gamma$ point, as in LaFeAsO, but the electron bands are now at the X point instead of at the M ($\sim$ M$_0$) point (see~Fig.~\ref{BZs}).  That is, the electron bands appear to have rotated their positions in the BZ by 45$^\circ$ with respect to their positions in LaFeAsO\@.  Indeed, all published calculations for the 122-type FeAs-based compounds show that the electron pockets are at the X points of the BZ\@.  

However, as discussed above in Sec.~\ref{SecBZ}, the primitive tetragonal M point is equivalent to the body-centered-tetragonal X point due to a 45$^\circ$ rotation of the bct reciprocal lattice with respect to the direct lattice.  As also noted there, the point to remember is that the electron pockets in both types of compounds are at the same positions in reciprocal space with respect to the respective direct lattices.  This means that the nesting wave vector between the electron and hole pockets for all of the Fe-based superconductors and parent compounds is the same ${\bf Q}_{\rm nesting} = \left(\frac{1}{2},\frac{1}{2}\right)$~r.l.u. in tetragonal notation.  We will be returning to this fact in the discussion of the antiferromagnetic ordering of the parent compounds in Sec.~\ref{SecLRMO}, the spin fluctuations in both the parent compounds and superconductors in Sec.~\ref{SecNeutInel}, and the neutron spin resonance mode in the superconducting state in Sec.~\ref{ResonanceMode}.

\begin{figure}
\includegraphics[width=2.in,viewport=-30 00 560 460,clip]{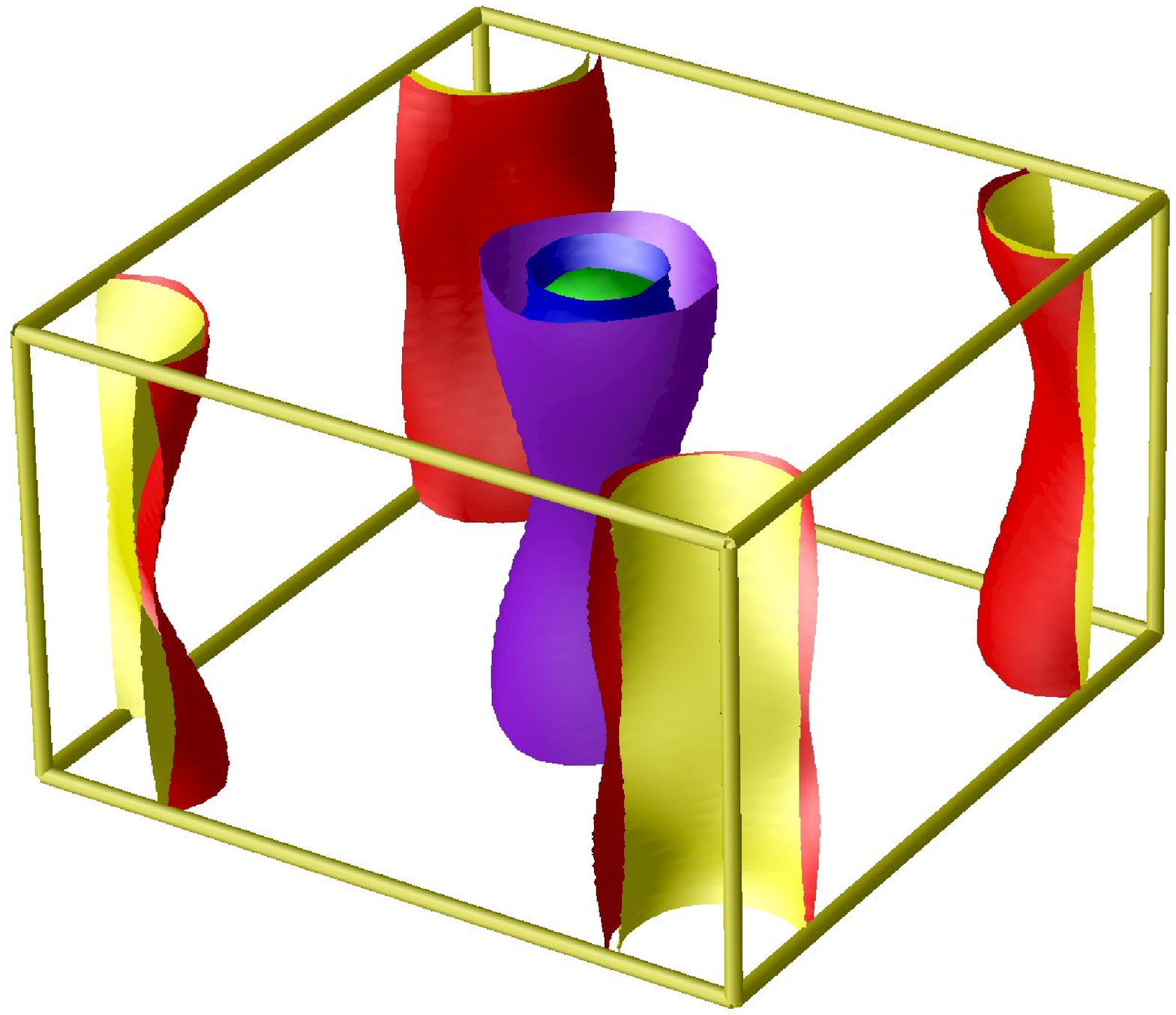}
\includegraphics[width=2.in,viewport=40 40 560 390,clip]{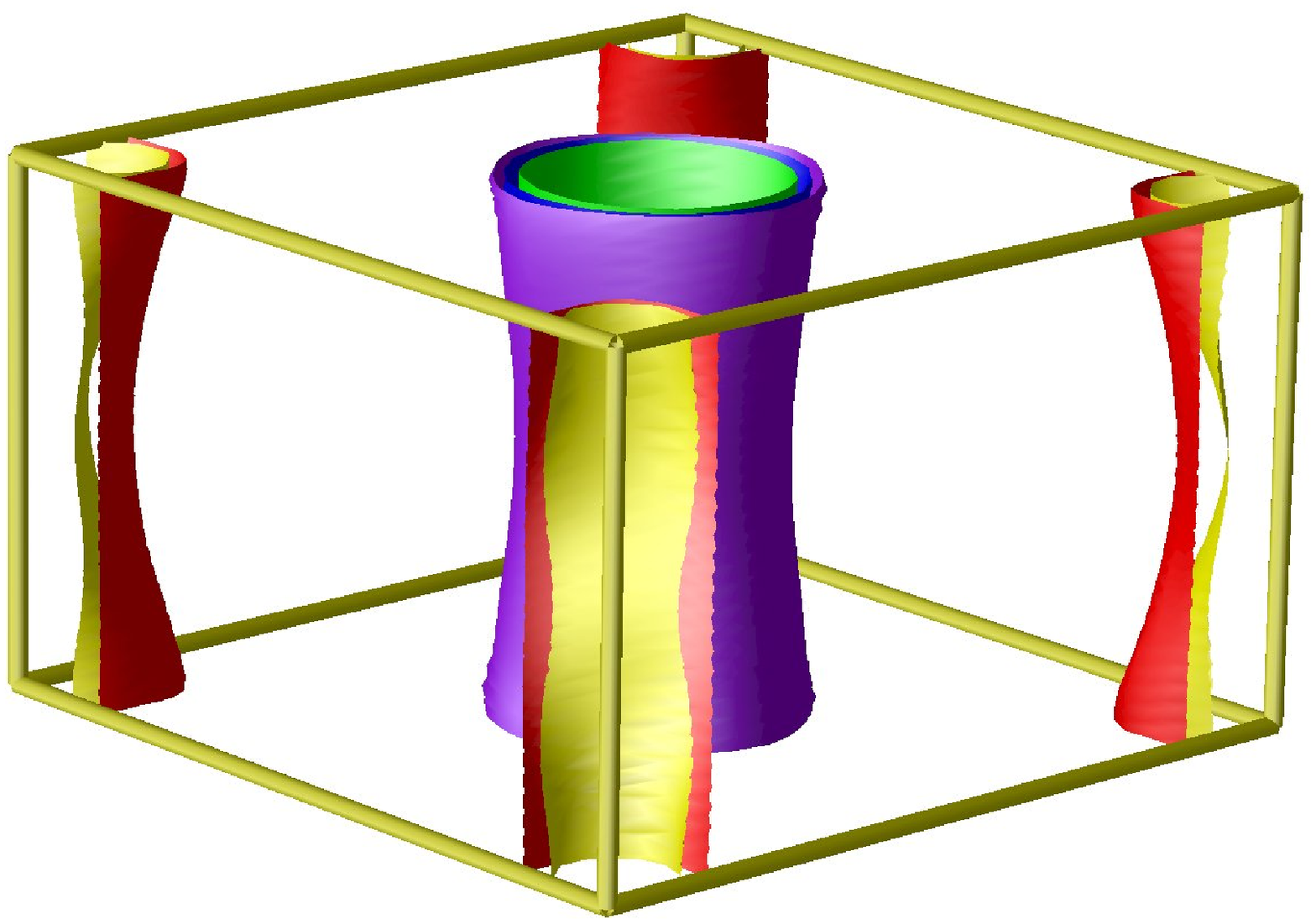}
\caption{(Color online) The Fermi surfaces of the tetragonal nonmagnetic 10\% electron-doped ${\rm Ba(Fe_{0.9}Co_{0.1})_2As_{2}}$ (top) and 10\% hole-doped ${\rm (Ba_{0.8}Cs_{0.2})Fe_2As_{2}}$ (bottom) obtained using LDA and the virtual crystal approximation.\cite{Mazin2009} The horizontal plane is in the $a$-$b$ tetragonal crystal plane, and the vertical axis is parallel to the crystallographic $c$-axis.  The hole Fermi surfaces are in the center and the electron Fermi surfaces are at the corners of the Brillouin zone.  The figure shows that the hole Fermi surfaces shrink and expand as electrons and holes are doped into ${\rm BaFe_2As_{2}}$, respectively.  Reprinted with permission from Ref.~\onlinecite{Mazin2009}, Copyright (2009), with permission from Elsevier.  }
\label{FigFS} 
\end{figure}

\begin{figure}
\includegraphics [width=2.5in,viewport= 0 00 560 710,clip]{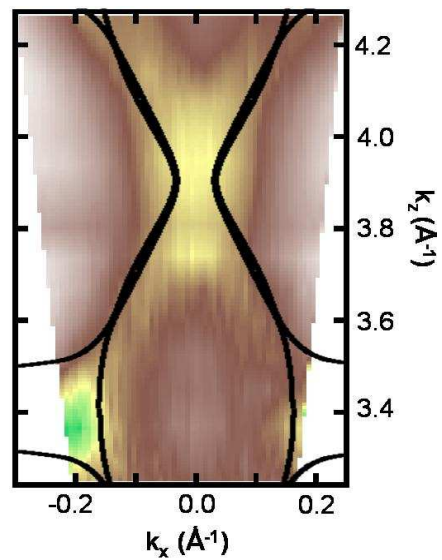}
\vskip -0.3in
\caption{(Color online) Fermi surface contours in the $k_x$-$k_z$ plane in ${\rm Ba(Fe_{0.9}Co_{0.1})_2As_2}$ as measured by ARPES, where $k_x$ is in the plane of the Fe atoms and $k_z$ is along the $c$-axis.\cite{Vilmercati2009} The positions of the $\Gamma$ and Z points in the Brillouin zone are at $k_x = 0$ and $k_z \approx 3.88$ and 3.39~\AA$^{-1}$, respectively.  The black curves are the hole band dispersions calculated using LDA in the virtual-crystal approximation and are seen to be in good agreement with the data.  These data show that the hole Fermi surface pockets strongly disperse in the $k_z$ direction and therefore that the compound has three-dimensional, rather than two-dimensional, hole Fermi surfaces.  Reprinted with permission from Ref.~\onlinecite{Vilmercati2009}.  Copyright (2009) by the American Physical Society.}
\label{FigBa(FeCo)2As2_ARPES} 
\end{figure}

LDA band calculations for the Fermi surfaces of 10\% electron and hole doped ${\rm BaFe_2As_2}$ are shown in Fig.~\ref{FigFS}.\cite{Mazin2009}  The band dispersion along the vertical $k_z$-axis is quite noticable for both cases, similar to the case of LaFeAsO in Fig.~\ref{LaFeAsOBS} above.  Shown in Fig.~\ref{FigBa(FeCo)2As2_ARPES} are experimental contour plots of the occupied states in the $k_x$-$k_z$ plane in 10\% electron-doped ${\rm Ba(Fe_{0.9}Co_{0.1})_2As_2}$ determined using ARPES.\cite{Vilmercati2009}  These data show a strong dispersion of the hole Fermi surface(s) along $k_z$, which means that the hole Fermi surfaces in this compound are three-dimensional rather than two-dimensional.  This dispersion was confirmed from ARPES measurements by other groups on crystals of the same system.\cite{Malaeb2009, Thirupathaiah2010}  Another important point regarding these measurements is that the observation of $k_z$-axis hole band dispersion proves that the ARPES measurements on the 122-type FeAs-based compounds, which probe only $\sim 3$--5~\AA\ into a material, are reflecting bulk properties of the material.  

In the bottom panel of Fig.~\ref{FigFS}, one sees that hole doping by partially replacing Ba$^{+2}$ by Cs$^{+1}$ results in an expansion of the hole FSs about the $\Gamma$ point in the middle of the BZ and a contraction of the electron FSs about the X point in the corners.  The compound ${\rm KFe_2As_2}$ corresponds to a large doping level of 0.5 hole/Fe.  From ARPES and quantum oscillation studies of the Fermi surface and band structure calculations,\cite{Sato2009, Terashima2010} one finds the two previous hole pockets and an additional hole pocket about $\Gamma$, and a small 3D pocket about $(00\pi)$.  The previous electron pockets centered at X completely disappear and are replaced by two small quasi-cylindrical hole pockets with elliptical cross section nearby that are not centered at X\@.\cite{Terashima2010}

Utfeld and coworkers reported Compton scattering measurements of the projection of the Fermi surfaces onto the $k_xk_y$-plane of an optimally doped crystal of ${\rm Ba(Fe_{0.93}Co_{0.07})_2As_2}$, although no other experimental characterizations of the crystal were given.\cite{Utfeld2010}  These bulk data indicate substantial dispersion of one or both hole Fermi surfaces along $k_z$ that are in agreement with LDA band calculations in the virtual crystal approximation with the optimized As position, with the best agreement obtained after rigidly shifting the electron and hole bands in energy in opposite directions.\cite{Utfeld2010}

Singh\cite{djsingh2008} and Shein and Ivanovskii\cite{Shein2010} have carried out electronic structure calculations for the 111-type compounds LiFeP and LiFeAs and found Fermi surfaces very similar to those described above for the 1111- and 122-type parent compounds.

\subsubsection{\label{Sec11BSARPES} Band Structures: 11-Type {\rm Fe}$_{1+y}${\rm Te}$_{1-x}${\rm Se}$_x$ Compounds}

The band structures and Fermi surfaces of these compounds calculated using LDA are similar to those of the 122- and 1111-type compounds.\cite{Subedi2008a}  

ARPES studies of a single crystal of Fe$_{1+y}$Te with $y < 0.05$ by Xia et al.\cite{Xia2009} at 10~K (below $T_{\rm N}$) showed metallic character at the Fermi energy, where circular hole pockets at the $\Gamma = (0,0)$ point of the tetragonal Brillouin zone and elliptical electron pockets at the M point at $\left(\frac{1}{2},\frac{1}{2}\right)$~r.l.u.\ were found, in agreement with the band calculations.\cite{Xia2009}  As expected for an undoped semimetal, the numbers of electron and hole carriers were found to be about the same; i.e., any doping by excess Fe atoms was too small to be resolved.  The bands found from ARPES are in fairly good agreement with LDA band calculations, but where the bands are compressed in energy (``renormalized'') by the usual factor of two, presumably reflecting the presence of many-body correlations of unstated origin that increase the effective mass by roughly a factor of two above the band mass.

Thus the nesting vector between the hole and electron pockets is $\left(\frac{1}{2},\frac{1}{2}\right)$~r.l.u.  On the other hand, the antiferromagnetic propagation vector below the N\'eel temperature $T_{\rm N}$ is ${\bf Q}_{\rm AF} =\left(\frac{1}{2},0\right)$~r.l.u.\ (see Table~\ref{AFPropVectors} and Fig.~\ref{FeTe_Mag_Struct} below), which is at a 45$^\circ$ angle to the nesting vector, suggesting that the AF order is not due to Fermi surface nesting of itinerant electrons/holes.  Furthermore, no gaps in the energy bands were found from the ARPES experiments\cite{Xia2009} that would be expected from band folding effects if the antiferromagnetic ordering were due to an itinerant spin density wave.  On the other hand, the ARPES data did show weak ``shadow Fermi surfaces'' at the X points [= ${\bf Q}_{\rm AF} =\left(\frac{1}{2},0\right)$~r.l.u.] that ``might arise from a band folding due to long-range magnetic order'' of local magnetic moments.  The authors also point out that the large linear electronic specific heat (Sommerfeld) coefficient of 34~mJ/mol~K$^2$ observed\cite{Chen2009a} for Fe$_{1.05}$Te argues against severe gapping of the Fermi surface due to an itinerant spin density wave.\cite{Xia2009}

Superconducting single crystals with composition FeTe$_{0.58}$Se$_{0.42}$ and $T_{\rm c} = 11.5$~K were also studied using  ARPES by Tamai et al.\cite{Tamai2010}  As in Fe$_{1+y}$Te above, the hole and electron concentrations are equal to within $\pm0.01$ electrons/unit cell, consistent with an undoped semimetal.  At low temperatures, the shapes of the bands were very similar to those computed using LDA, which allowed precise estimates of the renormalizations of the various bands.  The authors found remarkably large Fermi surface sheet-dependent mass enhancements of from six to twenty times the band mass, qualitatively consistent with the large electronic specific heat coefficient of 39~mJ/mol~K$^2$ measured by Sales et al.\cite{Sales2009a} for a crystal with a similar composition.  The large mass enhancement is also qualitatively consistent with the mass enhancement for a similar composition of about a factor of eight higher than the band mass at zero frequency from an extended-Drude analysis of the optical properties by Homes et al.\cite{Homes2010} as discussed below in Sec.~\ref{OpticsExp}.  The microscopic mechanism for the mass enhancement remains to be determined.

\subsubsection{Implications of the Semimetallic Behavior of the FeAs-type Parent Compounds}

The occurrence of high $T_{\rm c}$ superconductivity in semimetals suggests that a strategy for searching for new high $T_{\rm c}$ superconductors of the type seen in the Fe-based materials is to look for new semimetals with particular normal state properties such as the tendency towards antiferromagnetic SDW transitions and high magnetic susceptibility as discussed below in Sec.~\ref{Sec_Stoner}.  From the author's personal vantage point of having dealt extensively with the high $T_{\rm c}$ copper oxide superconductors, this is an unfamiliar approach.  The reason is that semimetallic parent compounds have the same numbers of electron and hole conduction carriers, which can only come about if the compound is a ``valence compound'' where one can formally assign (if somewhat arbitrarily) integer oxidation states to the constituent atoms.  For example, for the FeAs-type parent compound BaFe$_2$As$_2$, the oxidation states are often assigned to be Ba$^{+2}$, Fe$^{+2}$ and As$^{-3}$.  This situation of integral oxidation states for transition metal cations in oxides usually results in an insulating ground state due to electron correlation effects (``Mott-Hubbard physics'') instead of a (semi)metallic ground state.  Thus the existence of a metallic ground state for the FeAs-based superconductor parent compounds indicates that such electron correlation effects are not as strong as in the layered high $T_{\rm c}$ cuprates.  However, we note that the very similar isostructural compound BaMn$_2$As$_2$ is an antiferromagnetic insulator for $T\to 0$ instead of a metal,\cite{singh2009, an2009} and with a high ordered moment $\mu = 3.9\,\mu_{\rm B}$/Mn and N\'eel temperature $T_{\rm N} = 625$~K,\cite{singh2009, YSingh2009} which all indicate that electron correlation effects are significantly stronger than in the FeAs-based materials.  This compound will be discussed further near the end of this review in Sec.~\ref{TheSearch}.

\subsection{\label{Sec_CondCarrierOptics}Conduction Carrier Properties from Optics Measurements}

\subsubsection{\label{SecOpticsUnits} Introduction: Units}

%\squeezetable
\begin{table}
\caption{\label{OpticsUnits2} Useful unit conversion expressions associated with optics measurements.  Unless otherwise stated, the conventional optics units are Gaussian (cgs) units, where, e.g., $c$ is the speed of light in vacuum in units of cm/s.  The exception is conductivity $\sigma$ units where the two relevant expressions convert SI units to cgs wave number units.  The default conversion expressions for energy and temperature are for frequencies in cm$^{-1}$.  {\bf Caution}: to convert scattering rates $1/\tau^\prime$ in cm$^{-1}$ to energy or temperature, as shown, the default conversion factors in the table for frequencies are multiplied by $2\pi$ because of the extra factor of $2\pi$ on the right-hand sides of Eqs.~(\ref{Eqtaup}) and (\ref{Eqtaup2}).}
\begin{ruledtabular}
\begin{tabular}{l|cc}
Physical Quantity & Conversion Expression\\\hline
energy & 1~eV = 8\,065.6~cm$^{-1} = 11\,605$~K\\
 & 1 cm$^{-1}$ = 0.12398 meV\\
 & $1/\tau^\prime$: \hspace{0.1in} 1 cm$^{-1}$ = 0.77901 meV\\
temperature & 1~cm$^{-1} = 1.4388$~K\\
 & $1/\tau^\prime$: \hspace{0.1in} 1 cm$^{-1}$ = 9.0401 K\\ 
angular frequency & \hspace{-0.5in}${\rm 1~cm^{-1}= 2\pi c~{\rm rad/s}}$ \\
& \hspace{0.45in} $= 1.8836\times 10^{11}$ rad/s\\
conductivity $\sigma$& 1 $(\Omega~{\rm cm})^{-1}\ ({\rm SI})\footnotemark[1] = 9 \times 10^{11}$~s (cgs) \\
&\hspace{0.78in}= ${\rm 59.96~cm^{-1}}$\\
spectral weight of $\sigma$ & $1~(\Omega~{\rm cm})^{-1}~{\rm cm}^{-1} = 59.96~{\rm cm}^{-2}$
\end{tabular}
\end{ruledtabular}.
\footnotetext[1]{Here ``cm'' is a unit of convenience; the SI (mks) unit for length is m.}
\end{table}

Before discussing the formalism and the experimental results of optics measurements and their application to understand features of the current carriers in the Fe-based superconductors and parent compounds, it is necessary to first discuss the units and unit conversions used by practitioners in the condensed matter physics optics community.

Frequencies and relaxation rates from optics experiments are usually expressed in the same Gaussian (cgs) wave number (inverse wavelength) units of cm$^{-1}$.  Several unit conversion expressions are given in Table~\ref{OpticsUnits2}.  The basic relationship between wavelength $\lambda$ and frequency $f$ of light in vacuum is $\lambda f = c$, where $c$ is the speed of light in vacuum in cgs units of cm/s.  Thus the wave number of light is defined as $\frac{1}{\lambda}=\frac{f}{c}$ and has cgs units of cm$^{-1}$.  The frequency $f$ is expressed in wave numbers via 
\be
f^\prime[{\rm cm^{-1}}] \equiv \frac{f[{\rm s^{-1}}]}{c[{\rm cm/s}]} = \frac{\omega[{\rm rad/s}]}{2\pi c[{\rm cm/s}]}, 
\label{eqfprime}
\ee
where the angular frequency is $\omega = 2 \pi f$.  Thus we also have
\be
\omega = 2\pi c f^\prime .
\label{eqfprime2}
\ee
Alternatively, the frequency or wave number can be expressed in energy units of eV: $E = hf = \hbar\omega = hc\frac{1}{\lambda}$, where $h$ is Planck's constant and $\hbar = h/(2\pi)$.  For sinusoidal waves described by the dispersion relation $\omega = vk$ in some medium, $v$ is the phase velocity (speed) of the wave and $k = 2\pi/\lambda$ is the \emph{angular} wave number which is a factor of $2\pi$ larger than the above definition of wave number as $1/\lambda$.  Sometimes, in optics papers the symbol ``$\omega$'' is used to represent frequency in Hz or in wave numbers instead of the conventional meaning of the symbol as angular frequency in rad/s. In this review we adhere to the conventional definition of the symbol $\omega$ as angular frequency in rad/s.

An ambiguity in optics units is due to an extra factor of $2\pi$ in the definition of the conduction carrier relaxation rate that is consistently used by the optics community.  It arises in the following way (T. Timusk, private communication).  With the current carrier relaxation time $\tau^\prime$ and relaxation rate $f_\tau^\prime \equiv 1/\tau^\prime$ in wave number (${\rm cm^{-1}}$)  units, one has $1/\tau^\prime = f_\tau^\prime = f_\tau/c$ where $f_\tau$ is the relaxation rate in Hz.  Now if $\tau$ is the relaxation time in seconds, one might expect that $f_\tau = 1/\tau$ and therefore
\be
\frac{1}{\tau} = f_\tau = cf_\tau^\prime  = c\frac{1}{\tau^\prime}.\ \ \ ({\rm Not~used})
\label{EaXform1}
\ee
However, this conversion expression is not the one used by the optics community.  In Eq.~(\ref{EqDrudeSigma1}) below for the Drude contribution to the real part $\sigma_1$ of the optical conductivity from the conduction carriers, this contribution drops to one-half its zero-frequency value when $\omega = \omega_{1/2} = 1/\tau$.  Since the left side of this expression has units of rad/s, $1/\tau$ is considered by the optics community to be an angular frequency too (T. Timusk, private communication).  Therefore, one must include an extra factor of $2\pi$ into Eq.~(\ref{EaXform1}) to convert this angular frequency to frequency, yielding
\be
\frac{1}{\tau} = \omega_{1/2} = 2\pi f_{1/2} = 2\pi c f_{1/2}^\prime = 2\pi c\frac{1}{\tau^\prime},
\label{Eqtaup}
\ee
instead of Eq.~(\ref{EaXform1}).  
To confirm whether or not the factor of $2\pi$ on the right-hand side is included in a specific paper's definition of 1/$\tau^\prime$ in units of cm$^{-1}$, one can check whether or not the observed $\sigma_1(\omega \to 0)$ agrees with the value of the measured dc conductivity $\sigma_{\rm dc}$ and/or the value of $\sigma_0$ calculated using Eq.~(\ref{Eqsigma01}) below.  This is one reason that $\sigma_1(\omega \to 0)$, $\sigma_0$ from Eq.~(\ref{Eqsigma01}) and the measured dc $\sigma_{\rm dc}$ are all listed in Table~\ref{Opticsdata3} below, if available.  From Eq.~(\ref{Eqtaup}) we also have
\be
\tau = \frac{\tau^\prime}{2\pi c}.
\label{Eqtaup1}
\ee

An important feature of the definitions of $f^\prime$ [Eq.~(\ref{eqfprime2})] and $\tau^\prime$ [Eq.~(\ref{Eqtaup1})] is that 
\be
 \omega \tau = f^\prime \tau^\prime.
\label{Eqfpot}
\ee
At the half-height of the Drude peak in $\sigma_1(\omega)$ one has $\omega_{1/2}\tau = f^\prime_{1/2}\tau^\prime = 1$.  Therefore in a plot of $\sigma_1$ versus $f^\prime$, a plotting convention adherred to by the optics community, one can directly read off the scattering rate in cm$^{-1}$ from a plot of the Drude part of the conductivity versus $f^\prime$ as 
\be
\frac{1}{\tau^\prime} = f^\prime_{1/2}.
\label{Eqtphalf}
\ee

We make the following comments regarding Eq.~(\ref{Eqtaup}).  First, ``rad'' is not a conserved unit; it is a notation that says that a circle is broken up into $2\pi$ parts, as opposed to 360 parts for the notation ``degrees'' or 1 part for the notation ``revolutions''.  The definition of radian measure of an angle $\theta$ is $\theta = s/r$ where $s$[cm] is an arc length and $r$[cm] is the radius of the circle.  Thus $\theta$ is dimensionless.  One must put in the notation ``rad'' by hand to designate that a circle is divided into $2\pi$ pieces when radian measure of angle is used.  Second, ``rad'' is put in or taken out according to whether or not an angular quantity is being described.  The speed $v$[cm/s] of a particle going around a circle of radius $r$[cm] at angular speed $\omega$[rad/s] is $v = \omega r$.  Thus when computing $v$, one throws away ``rad'' from the right-hand side.  When computing $\omega = v/r$, one puts in ``rad'' by hand on the right-hand side, as in $\theta = s/r$ above.  Third, since angular notations like ``rad'' are not conserved, the fact that $\omega\tau$ is dimensionless, where the angular frequency $\omega$ has ``units'' of rad/s, does not require $1/\tau$ to be an angular frequency with ``units'' of rad/s.  In particular, $1/\tau$ can be in units of 1/s (Hz), and in that case, in the dimensionless product $\omega\tau$ one simply throws away the notation ``rad''.

Finally, in simple derivations of the dc conductivity expression in Eq.~(\ref{EqSigma3}) below,\cite{kittel1966} the current carrier mean collision or relaxation time $\tau$ (in s) does not have any angular aspect associated with it and therefore the relaxation or collision rate 1/$\tau$ is a frequency in Hz and not an angular frequency in rad/s.  In this case one would use the conversion expression in Eq.~(\ref{EaXform1}).  

In any event, Eq.~(\ref{Eqtaup}) is consistently used by the optics community. This definition leads to Eq.~(\ref{Eqfpot}) which is useful in practice as shown in Eq.~(\ref{Eqtphalf}).

A similar ambiguity occurs in the THz spectroscopy community.  In some papers, the numerical values stated to be the relaxation time ``$\tau$'' [in s] are actually $2\pi\tau$.  The extra factor of $2\pi$ is included without telling the reader.  To find out that the factor of $2\pi$ is present, one has to go through the calculation of $\tau$ starting from the original experimental conductivity data that are given.

The conversion expression between conductivity units of $(\Omega~$cm)$^{-1}$~(SI) and ${\rm cm^{-1}}$(cgs) that is widely used by the optics community and given above in Table~\ref{OpticsUnits2}, is implicit in and derived from Eq.~(\ref{Eqsigma01}) below.  In particular, the conversion factor in parentheses on the right-hand side of Eq.~(\ref{Eqsigma01}) is, by definition of ``conversion factor'', equal to unity, so we have
\[
\frac{1}{59.96~\Omega~{\rm cm}} = \frac{1}{{\rm cm}}
\]
or
\be
1~(\Omega~{\rm cm})^{-1} = 59.96~{\rm cm}^{-1}.
\label{EqConvocc}
\ee
This conversion expression assumes the presence of the extra factor of $2\pi$ in the relationship between $\tau$ and $\tau^\prime$ in Eqs.~(\ref{Eqtaup}) and~(\ref{Eqtaup1}).  

Usually when the conductivity $\sigma_1$ is plotted versus some kind of frequency in optics papers, the $\sigma_1$ has SI units of $(\Omega~{\rm cm})^{-1}$ and the frequency is $f^\prime$ in cm$^{-1}$.  Thus the spectral weight (area under the curve of $\sigma_1$ versus $f^\prime$) has units of $(\Omega~{\rm cm})^{-1}$cm$^{-1}$.  However, spectral weights are often quoted in optics papers in units of cm$^{-2}$ instead.  To get these spectral weight units, one must convert $(\Omega~{\rm cm})^{-1}$ to cm$^{-1}$ using Eq.~(\ref{EqConvocc}) to get
\be
1~(\Omega~{\rm cm})^{-1}{\rm cm}^{-1} = 59.96~{\rm cm}^{-2},
\ee
as listed in Table~\ref{OpticsUnits2}.

\subsubsection{\label{SecOpticsThy} Formalism: Drude Optical Conductivity, dc Conductivity, and Conduction Carrier Relaxation Rate}

The total carrier density (electrons plus holes) can be estimated experimentally from the real part of the optical conductivity versus light angular frequency $\sigma_1(\omega)$ measurements.  This is a distinct advantage over Hall effect measurements where the contributions from electrons and holes tend to cancel.  Because of the difficulty in reaching quantitative conclusions from Hall effect or thermoelectric power measurements on multiband conductors such as semimetals, we will not attempt discussions or analyses of these measurements in this review.

All optical conductivity measurements on the FeAs-based materials (but not on all the 11-type compounds, see Sec.~\ref{OpticsExp} below) have shown a Drude peak centered at zero frequency that is associated with an at least partially (see Sec.~\ref{SecOptCorr} below) coherent metallic state, given by\cite{Harrison1970} 
\begin{equation}
\sigma_1(\omega) = \frac{\sigma_0}{1 + \omega^2\tau^2},
\label{EqDrudeSigma1}
\end{equation}
where the (real) zero-frequency electrical conductivity $\sigma(0) \equiv \sigma_0$ can be written as\cite{kittel1966, Harrison1970} 
\begin{equation}
\sigma_0 = \frac{ne^2 \tau}{m^*},
\label{EqSigma3}
\end{equation}
and $\tau$ is the mean free scattering time for the current carriers, $n$ is the conduction carrier number density, $e$ is the fundamental charge, and $m^*$ is the effective (band) mass of the current carriers.  A plot of the Drude conductivity $\sigma_1$ versus $\omega$ in Eq.~(\ref{EqDrudeSigma1}) is given below in Fig.~\ref{SC_density}.  If multiple bands contribute to the conductivity, the total conductivity $\sigma_0$ is the sum of the conductivities of the individual bands: $\sigma_0 = \sum_i \sigma_{0i}$.  Thus when deriving optical parameters of the current carriers, these are each an appropriate combination over all contributing bands.  

Analysis of optical conductivity data can yield values of $\tau$ and of the plasma angular frequency $\omega_{\rm p}$ of the conduction carriers.\cite{Dressel2002}  The $\omega_{\rm p}$ is the angular frequency at which the entire electron gas would uniformly oscillate in response to an initial uniform displacement from its equilbrium position with respect to the positive charge distribution of the nuclei.  It is expressed in terms of $n$ in SI units as\cite{kittel1966}
\begin{equation}
\omega_{\rm p}^2 = \frac{n e^2}{\varepsilon_0 m^*},
\label{EqOmegaP}
\end{equation}
where $\varepsilon_0 = 8.8542 \times 10^{-12}~{\rm C^2/(J~m)} = 8.8542 \times 10^{-14}~{\rm s/(\Omega~cm)}$ is the dielectric permittivity of free space.  In general, if more than one electron band contributes to the conductivity, then a different $\omega_{\rm p}$ could be associated with each band, since $n/m^*$ is in general different for different bands.  Using Eq.~(\ref{EqOmegaP}), one can rewrite the dc conductivity in Eq.~(\ref{EqSigma3}), assuming an effective single-band model, as
\be
\sigma_0 = \varepsilon_0 \omega_{\rm p}^2 \tau = \varepsilon_0 \frac{\omega_{\rm p}^2}{\frac{1}{\tau}}.
\label{EqSigma0}
\ee
Note that this expression includes the influence of $m^*$ since it is contained within $\omega_{\rm p}$.  

As discussed above in Sec.~\ref{SecOpticsUnits}, the reported plasma frequencies from optics experiments are frequencies $f_{\rm p}^\prime$ expressed in wave number units of cm$^{-1}$, not angular frequencies, so the \emph{angular} plasma frequency in rad/s is given by
\be
\omega_{\rm p} = 2\pi c f_{\rm p}^\prime.
\label{Eqfpp}
\ee
From Eq.~(\ref{Eqtaup1}) we have
\be
\frac{1}{\tau} = 2\pi c\frac{1}{\tau^\prime}.
\label{Eqtaup2}
\ee
Substituting Eqs.~(\ref{Eqfpp}) and~ (\ref{Eqtaup2}) into Eq.~(\ref{EqSigma0}) gives
\be  
\sigma_0 =2\pi c\varepsilon_0 \frac{{f_{\rm p}^\prime}^2}{\frac{1}{\tau^\prime}} = \left(\frac{1~{\rm cm}}{59.96~\Omega~{\rm cm}}\right)\frac{{f_{\rm p}^\prime}^2}{\frac{1}{\tau^\prime}},
\label{Eqsigma01}
\ee
where as discussed above and in Sec.~\ref{SecOpticsUnits}, both $f_{\rm p}^\prime$ and $1/\tau^\prime$ are in optics units of cm$^{-1}$.  

Instead of fitting the observed low-frequency $\sigma_1(\omega)$ by the expected Drude behavior in Eq.~(\ref{EqDrudeSigma1}) (usually together with additional terms arising from interband transitions, which together form the so-called ``Drude-Lorentz model'', see Sec.~\ref{OpticsExp} below) to obtain $\omega_{\rm p}$ and $\tau$, the value of $\omega_{\rm p}$ is sometimes determined from the ``spectral weight'' of the Drude peak in $\sigma_1(\omega)$, which is the integral of $\sigma_1(\omega)$ from $\omega = 0$ up to some cutoff angular frequency $\omega_{\rm c}$ that is applied to avoid including higher frequency interband transitions in the integral.  The optical Drude spectral weight ${\rm SW_{Drude}}$ is then defined to be
${\rm SW_{Drude}} = \int_0^{\omega_{\rm c}}\sigma_1(\omega)d\omega$.

If only the Drude component to the optical conductivity were present, one could set $\omega_{\rm c} = \infty$ in ${\rm SW_{Drude}} = \int_0^{\omega_{\rm c}}\sigma_1(\omega)d\omega$.  If in addition $\tau$ were independent of frequency, one would obtain using Eqs.~(\ref{EqDrudeSigma1}) and~(\ref{EqSigma0}) the spectral weight
\be
\int_0^\infty \sigma_1(\omega)d\omega = \frac{\pi}{2}\varepsilon_0 \omega_{\rm p}^2,
\label{EqSW}
\ee
where $\tau$ drops out of the expression on the right-hand side.  This single-band expression remains valid even when $\tau$ and $m^*$ are frequency-dependent (T. Timusk, private communication).  In practice, experimentalists sometimes define an effective plasma angular frequency $\omega_{\rm p}$, via Eq.~(\ref{EqSW}), by
\be
\frac{\pi}{2}\varepsilon_0 \omega_{\rm p}^2 \equiv \int_0^{\omega_{\rm c}}\sigma_1(\omega)d\omega,\label{EqOmegaPeff}
\ee
where the cutoff angular frequency $\omega_{\rm c}$ is chosen to include as much of the Drude conductivity as possible, but at the same time to exclude as much as possible contributions from higher frequency interband transitions that usually overlap with the Drude conductivity to some extent [see Figs.~\ref{FigHu_optics}(c) and~\ref{FigHu_optics}(d) below].  The cutoff $\omega_{\rm c}$ in Eq.~(\ref{EqOmegaPeff}) is therefore somewhat arbitrary and the accuracy of the derived $\omega_{\rm p}$ is uncertain.  

Using the definition of $\omega_{\rm p}$ in Eq.~(\ref{EqOmegaP}), one can rewrite Eq.~(\ref{EqOmegaPeff}) to obtain the carrier density $n$ that contributes to the carrier conduction up to frequency $\omega$ by
\be
n(\omega) = \frac{\hbar}{e^2} \frac{2m^*}{\pi \hbar} \int_0^{\omega}\sigma_1(\omega)d\omega,
\label{eqnomega1}
\ee
where the first factor on the right-hand side is $\hbar/e^2 = 4108~\Omega$ in SI units.  The other quantities in the prefactor to the integral are evaluated in cgs units if $\sigma_1$ is in SI units of $(\Omega\,{\rm cm})^{-1}$.    Alternatively, defining the carrier density $n$ as the number of conduction carriers per Fe atom (or any other type of specified atom), $n = {\cal N}_{\rm Fe}/V_{\rm Fe}$, where $V_{\rm Fe}$ is the volume per Fe atom, using Eq.~(\ref{eqnomega1}) one can define the number of carriers per Fe atom contributing to  $\sigma_1$ up to angular frequency $\omega$ as
\be
{\cal N}_{\rm Fe}(\omega) = \frac{\hbar}{e^2} \frac{2m^*V_{\rm Fe}}{\pi \hbar} \int_0^{\omega}\sigma_1(\omega)d\omega.
\label{eqnomega2}
\ee

Sometimes a so-called ``extended-Drude'' analysis of the optical properties is carried out such as was done for ${\rm Ba_{0.55}K_{0.45}Fe_2As_2}$,\cite{JYang2009} EuFe$_2$As$_2$,\cite{Wu2009}  LaFePO,\cite{Qazilbash2008} ${\rm Fe_{1.06}Te_{0.88}S_{0.14}}$,\cite{Stojilovic2010} ${\rm FeTe_{0.55}Se_{0.45}}$,\cite{Homes2010} and ${\rm Ba(Fe_{0.92}Co_{0.08})_2As_2}$ and ${\rm Ba(Fe_{0.95}Ni_{0.05})_2As_2}$,\cite{Barasic2010} which allows a frequency-dependent relaxation rate $\tau^{-1}(\omega)$ and effective mass $m^*(\omega)$ of the current carriers to be derived.  In this analysis, writing the complex conductivity as $\sigma(\omega) = \sigma_1(\omega) + i \sigma_2(\omega)$, one defines\cite{Stojilovic2010, Puchkov1996} [see Eq.~(\ref{EqSigma0}); SI units]
\be
\frac{1}{\tau(\omega)}=\varepsilon_0 \omega_{\rm p}^2\Re\left[\frac{1}{\sigma(\omega)}\right],
\ee
and 
\be
\frac{m^*(\omega)}{m_{\rm b}}= \frac{\varepsilon_0 \omega_{\rm p}^2}{\omega}\Im\left[\frac{1}{\sigma(\omega)}\right],
\ee
where $\omega_{\rm p}$ can be obtained using Eq.~(\ref{EqOmegaPeff}) and $m_{\rm b}$ is the band mass.

In cgs units, $\varepsilon_0$ in the above formulas is replaced by 1/($4\pi$).

\subsubsection{\label{OpticsExp} Experimental Results}

For an early review of optical and Raman spectroscopy studies of the Fe-based superconductors, see Ref.~\onlinecite{Hu2009}.

\subsubsection*{a. 122- and 1111-Type Compounds}

\begin{figure*}
\includegraphics [width=6.8in]{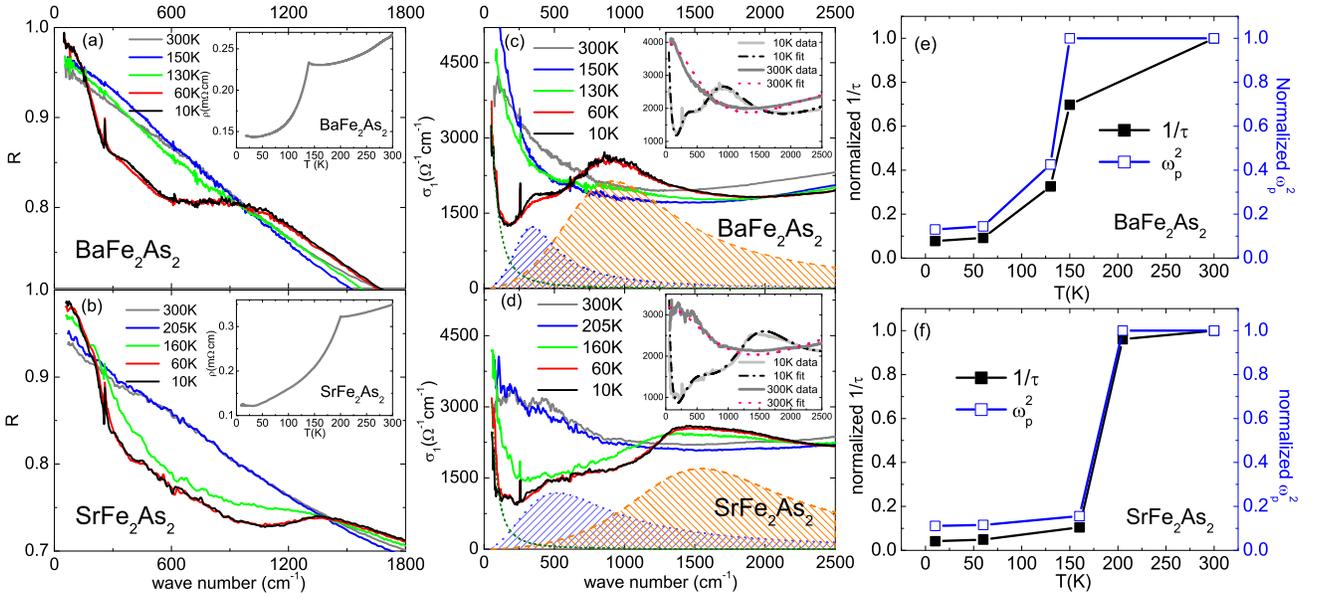}
\caption{(Color online)   Illustration of the results and analysis of optical experiments on layered FeAs-type materials.\cite{Hu2008}  Panels (a) and (b) show the raw experimental $ab$~plane optical reflectivity $R$ versus wave number $f^\prime$ (in our notation) data from the $ab$~planes of ${\rm BaFe_2As_2}$ and ${\rm SrFe_2As_2}$, respectively.  The insets show the respective behaviors of the dc electrical resistivity $\rho$ versus temperature $T$.  Panels (c) and (d) show the real part $\sigma_1$ of the optical conductivity obtained from Kramers-Kronig transformation of the data in parts (a) and (b), respectively.  The Drude term (short-dashed green curve) and the first two Lorentz contributions are also shown at the bottom for the fits at $T = 10$~K, respectively.  The insets show the respective $\sigma_1(f^\prime)$ data and total fits up to $f^\prime = 2500$~cm$^{-1}$ for $T = 10$ and~300~K\@.  Parts (e) and (f) show the relative changes versus $T$ of the conduction carrier relaxation rate $1/\tau$ and plasma angular frequency squared $\omega_{\rm p}^2 \sim n/m^*$ for the two compounds, obtained from the Drude parts of the fits in parts (b) and (c), respectively.  Reprinted with permission from Ref.~\onlinecite{Hu2008}.  Copyright (2008) by the American Physical Society.}
\label{FigHu_optics} 
\end{figure*}

%\squeezetable
\begin{table}
\caption{\label{OpticsdataTheory} Theoretical parameters for layered Fe-based materials.  Here $\omega_{\rm p}$ is the in-plane plasma angular frequency, $n$ is the carrier concentration in units of both carriers per cm$^3$ and in carriers per formula unit [(f.u.)$^{-1}$], $h$ is Planck's constant, and $\hbar = h/(2\pi)$.  If only $\omega_{\rm p}$ and not $n$ is given in the reference, $n$ is determined using Eq.~(\ref{EqOmegaP2}) by taking the ratio of the band mass to the free electron mass to be  $m^*/m_{\rm e} \equiv 1$. If both $\omega_{\rm p}$ and $n$ are given in the reference, then $m^*/m_{\rm e}$ is calculated using Eq.~(\ref{EqOmegaP2}).  The number of carriers per formula unit is the number of carriers per unit volume times the volume per formula unit determined using the lattice parameters in the tables in the Appendix, with two formula units per tetragonal unit cell for both the 1111- and 122-type compounds. The carrier density per unit volume of LaFeAsO is expected to be smaller those of the 122-type compounds because the former contains half the number of Fe atoms per unit cell than the latter does, whereas the unit cell volume is about 2/3 times that of the latter.}
\begin{ruledtabular}
\begin{tabular}{l|ccccc}
 Compound   & $\hbar\omega_{\rm p}$  &  $n$ & $n$ & $m^*/m_{\rm e}$ & Ref.\\
& (eV) & ($10^{21}$~cm$^{-3}$) & (f.u.)$^{-1}$ \\ \hline
LaFeAsO & & 3.7\footnotemark[1] &   0.26\footnotemark[1] & & \onlinecite{djsingh2008b} \\
LaFeAsO & & 3.9\footnotemark[1] &   0.28\footnotemark[1]  &  & \onlinecite{Ma2008a} \\
CaFe$_2$As$_2$ & 2.60 & 8.42 & 0.74 & 1.72 & \onlinecite{Ma2008}\\
 & 3.02  & 6.60 & 0.58 & $\equiv 1$ & \onlinecite{Tanatar2009} \\
SrFe$_2$As$_2$ & 3.40 & 6.66 & 0.63 & 0.80 & \onlinecite{Ma2008}\\
& 2.80 & 5.68 & 0.54 & $\equiv 1$ & \onlinecite{Tanatar2009}\\
BaFe$_2$As$_2$ & 3.15 & 5.08 & 0.52 & 0.71 & \onlinecite{Ma2008}\\
\end{tabular}
\end{ruledtabular}
\footnotetext[1]{Directly from electronic structure calculations.}
\end{table}

%\squeezetable
\begin{table*}
\caption{\label{Opticsdata2} Experimental in-plane optical parameters for layered Fe-based materials.  All the experimental results listed were obtained on single crystals.  Here $T$ is the temperature of the experiment, $f^\prime_{\rm c}$ is the cutoff frequency in Eq.~(\ref{EqOmegaPeff}) if applicable, $f^\prime_{\rm p}$ is the plasma frequency, $n$ is the total carrier concentration in two different units (see the caption to Table~\ref{OpticsdataTheory}), $1/\tau^\prime$ is the mean free scattering rate of the current carriers, $h$ is Planck's constant, and $\hbar = h/(2\pi)$.  The definitions of the symbols are $f^\prime \equiv f/c$ and $1/\tau^\prime \equiv 1/(2\pi c \tau)$, where $f$ is the frequency in Hz, $\tau$ is the relaxation time in s, and $c$ is the speed of light in vacuum in cm/s.  See Sec.~\ref{SecOpticsUnits} for an extended discussion of the units used by the optics community.  The conversion of $f^\prime_{\rm p}$ values to $n$ values is accomplished using Eq.~(\ref{Eqnfp}) assuming a band effective mass $m^*/m_{\rm e} = 1$.  The conversion factors for $1/\tau^\prime$ in cm$^{-1}$ to meV and K in the last two columns are given in Table~\ref{OpticsUnits2}.}
\begin{ruledtabular}
\begin{tabular}{l|ccccccccc}
 Compound   &  $T$ & $f_{\rm c}^\prime$ &$f_{\rm p}^\prime$ &  $n_{\rm}$ & $n_{\rm}$ & $1/\tau^\prime$ & Ref. & $1/\tau^\prime$ & $1/\tau^\prime$\\
 & (K) & (cm$^{-1}$) & (cm$^{-1}$) & ($10^{21}$~cm$^{-3}$) & (f.u.$^{-1}$) & (cm$^{-1}$) & & (meV) & (K)\\ \hline
LaFeAsO & 300 & 1\,400 & 10\,400 & 1.20 & 0.098 & 640 & \onlinecite{Chen2009} & 500 & 5790\\
SrFe$_2$As$_2$ & 300 &  & 13\,840 & 2.13 & 0.203 & 950 & \onlinecite{Hu2008} & 740 & 8590\\
 & 10 &  & 4\,750 & 0.251 & 0.024 & 40 & \onlinecite{Hu2008} & 31 & 360 \\
BaFe$_2$As$_2$ & 300 &  & 12\,900 & 1.85 & 0.189 & 700 & \onlinecite{Hu2008} & 550 & 6330\\
 & 10 &  & 4\,660 & 0.242 & 0.025 & 55 & \onlinecite{Hu2008} & 43 & 500\\
BaFe$_2$As$_2$\footnotemark[1] & & & 12\,700\footnotemark[1] & 1.81\footnotemark[1]  & 0.185\footnotemark[1] & & \onlinecite{Drechsler2009} \\
BaFe$_2$As$_2$ & 295 & & 8\,630 & 0.829  & 0.085 & 398 & \onlinecite{Akrap2009} & 310 & 3600\\
Ba$_{0.55}$K$_{0.45}$Fe$_2$As$_2$\footnotemark[1] & 295 &  &  12\,900\footnotemark[1] & 1.85\footnotemark[1] & 0.189\footnotemark[1] & 880\footnotemark[1] & \onlinecite{JYang2009} & 690 & 7960\\
 & 28 &  &   12\,900\footnotemark[1] & 1.85\footnotemark[1] & 0.189\footnotemark[1] & 280\footnotemark[1] & \onlinecite{JYang2009} & 230 & 2620 \\
${\rm Ba(Fe_{0.92}Co_{0.08})_2As_2}$ & 300 &  & 4\,500\footnotemark[5]  & 0.23\footnotemark[5] & 0.011\footnotemark[5] & 460(40)\footnotemark[5] & \onlinecite{Wu2010} & 360 & 4160\\
 & 100 &  & 4\,500\footnotemark[5]  & 0.23\footnotemark[5] & 0.011\footnotemark[5] & 115(15)\footnotemark[5] & \onlinecite{Wu2010} & 90 & 1040 \\
${\rm Ba(Fe_{0.92}Co_{0.08})_2As_2}$ & 200 & 2\,500 & 12\,000  & 1.6 & 0.082 & 880\footnotemark[2] & \onlinecite{Barasic2010} & 690 & 7960\\
 & 30 & 2\,500 & 12\,000  & 1.6 & 0.082 & 450\footnotemark[2] & \onlinecite{Barasic2010} & 350 & 4070 \\
${\rm Ba(Fe_{0.9}Co_{0.1})_2As_2}$\footnotemark[3] & 30 &  & 8\,500\footnotemark[4]  & 0.80\footnotemark[4] & 0.04\footnotemark[4] & 200\footnotemark[4] & \onlinecite{Fischer2010} & 160 & 1810 \\
${\rm Ba(Fe_{0.9}Co_{0.1})_2As_2}$\footnotemark[3] & 200 &  & 4\,800(1000)  & 0.26 & 0.013 & 115(40) & \onlinecite{Nakamura2009} & 90 & 1040\\
${\rm Ba(Fe_{0.95}Ni_{0.05})_2As_2}$ & 300 &  & 6\,600\footnotemark[5]  & 0.48\footnotemark[5] & 0.025\footnotemark[5] & 500\footnotemark[5] & \onlinecite{Wu2010} & 390 & 4520\\
 & 100 &  & 6\,600\footnotemark[5]  & 0.48\footnotemark[5] & 0.025\footnotemark[5] & 220(20)\footnotemark[5] & \onlinecite{Wu2010} & 170 & 1990\\
${\rm Ba(Fe_{0.95}Ni_{0.05})_2As_2}$ & 200 & 2\,500 & 15\,000  & 2.5 & 0.128 & 1\,000\footnotemark[2] & \onlinecite{Barasic2010} & 780 & 9040 \\
& 35 & 2\,500 & 15\,000  & 2.5 & 0.128 & 800\footnotemark[2] & \onlinecite{Barasic2010} & 620 & 7230 \\
EuFe$_2$As$_2$ & 300 & 2\,500 &  13\,800  & 2.12 & 0.196 & 960 & \onlinecite{Wu2009} & 750 & 8680 \\
LaFePO  & 298 & 3\,000 & 14\,900 & 1.34 & 0.090 & 400\footnotemark[2] & \onlinecite{Qazilbash2008} & 310 & 3620 \\
 & 10 & 3\,000 & 14\,900 & 1.34 & 0.090 & 150\footnotemark[2] & \onlinecite{Qazilbash2008} & 120 & 1360\\
FeTe$_{0.55}$Se$_{0.45}$ & 295 &  & 7200(360) & 0.58(6) & 0.025(3) & 414(21) & \onlinecite{Homes2010} & 322 & 3750 \\
& 200 &  & 7200(360) & 0.58(6) & 0.025(3) & 363(18) & \onlinecite{Homes2010} & 283 & 3280 \\
& 100 &  & 7200(360) & 0.58(6) & 0.025(3) & 317(16) & \onlinecite{Homes2010} & 247 & 2870\\
\end{tabular}
\end{ruledtabular}
\footnotetext[1]{Grown in Sn flux: contaminated by Sn impurities.}
\footnotetext[2]{From a generalized Drude analysis in the low frequency limit.}
\footnotetext[3]{THz conductivity measurement of thin film.}
\footnotetext[4]{Large uncertainty in this parameter.}
\footnotetext[5]{From a fit containing two Drude terms; this value is for the narrow Drude part.}
\end{table*}

%\squeezetable
\begin{table*}
\caption{\label{Opticsdata3} Conduction electron transport properties of Fe-based layered materials derived from optical measurements.  The experimental in-plane optical parameters $f_{\rm p}^\prime$ and $1/\tau^\prime$ for layered Fe-based single crystals are from Table~\ref{Opticsdata2}.  The dc conductivities $\sigma_0^{\rm calc}$ are calculated from these optical parameters using Eq.~(\ref{Eqsigma01}).  The measured zero-frequency limit of the conductivity $\sigma_1(\omega \to 0)$ from optical measurements is listed if available.  The dc conductivity obtained using conventional dc four-probe resistivity measurements of the in-plane resistivity of single crysals is given in the last column, where usually a different crystal was measured than in the optical conductivity measurements.}
\begin{ruledtabular}
\begin{tabular}{l|ccccccc}
 Compound   &  $T$ &$f_{\rm p}^\prime$ & $1/\tau^\prime$ & $\sigma_0^{\rm calc}$ & $\sigma_1(\omega \to 0)$  & Ref. & $\sigma_{\rm dc}$\\
 & (K) & (cm$^{-1}$) & (cm$^{-1}$) & $(\Omega~$cm)$^{-1}$ & $(\Omega~$cm)$^{-1}$ && $(\Omega~$cm)$^{-1}$\\ \hline
LaFeAsO & 300 & 10\,400 & 640 & 2\,800 &  2\,700&\onlinecite{Chen2009} & 2\,900 [\onlinecite{Chen2009}]\\
SrFe$_2$As$_2$ & 300 & 13\,840 & 950 & 3\,400 & 3\,200 &\onlinecite{Hu2008} & 2\,500 [\onlinecite{Tanatar2009}]\\
 & 10   & 4\,750   & 40 & 9\,400 &&\onlinecite{Hu2008} & 13\,000 [\onlinecite{Tanatar2009}]\\
BaFe$_2$As$_2$ & 300   & 12\,900 & 700 & 4\,000 & 4\,100 &\onlinecite{Hu2008} & 2\,300 [\onlinecite{wang2008}] \\
 & 10   & 4\,660   & 55 & 6\,600 &&\onlinecite{Hu2008} & 4\,500 [\onlinecite{wang2008}]\\
BaFe$_2$As$_2$ & 295  & 8\,630  & 398 & 3\,100 & 3\,400 &\onlinecite{Akrap2009} & 2\,300 [\onlinecite{wang2008}] \\
Ba$_{0.55}$K$_{0.45}$Fe$_2$As$_2$\footnotemark[1] & 295 & 12\,900\footnotemark[1]  & 880\footnotemark[1] & 3\,100\footnotemark[1] &  3\,200\footnotemark[1] & \onlinecite{JYang2009} & 1\,200\footnotemark[1] [\onlinecite{Ni2008h}]\\
 & 28   & 12\,900\footnotemark[1] & 280\footnotemark[1] & 9\,900\footnotemark[1] & 6\,000\footnotemark[1]  &\onlinecite{JYang2009} & 2\,500\footnotemark[1]  [\onlinecite{Ni2008h}] \\
${\rm Ba(Fe_{0.92}Co_{0.08})_2As_2}$ & 300 & 4\,500\footnotemark[5]  & 460(40)\footnotemark[5] & 730 & & \onlinecite{Wu2010} & 2\,500 [\onlinecite{Wu2010}] \\
 & 100 &   4\,500\footnotemark[5]  & 115(15)\footnotemark[5] & 2\,900\footnotemark[5] & 4050 & \onlinecite{Wu2010} & 4\,800 [\onlinecite{Wu2010}] \\
${\rm Ba(Fe_{0.92}Co_{0.08})_2As_2}$ & 200 & 12\,000  & 880\footnotemark[2] & 2\,700 & 2\,800 & \onlinecite{Barasic2010} & 3\,300 [\onlinecite{Barasic2010}]\\
 & 30 & 12\,000 & 450\footnotemark[2] & 5\,300 & 5\,200 &\onlinecite{Barasic2010} & 6\,700 [\onlinecite{Barasic2010}] \\
${\rm Ba(Fe_{0.9}Co_{0.1})_2As_2}$\footnotemark[3] & 30 &   8\,500\footnotemark[4]  & 200\footnotemark[4] & 6000 & 6\,000 & \onlinecite{Fischer2010} \\
${\rm Ba(Fe_{0.95}Ni_{0.05})_2As_2}$ & 300 &  6\,600\footnotemark[5]  & 500\footnotemark[5] &  &  & \onlinecite{Wu2010} & 2\,600 [\onlinecite{Wu2010}] \\
 & 100 &  6\,600\footnotemark[5]  & 220(20)\footnotemark[5] &  &  & \onlinecite{Wu2010} & 4\,600 [\onlinecite{Wu2010}] \\
${\rm Ba(Fe_{0.95}Ni_{0.05})_2As_2}$ & 200 & 15\,000  & 1\,000\footnotemark[2] & 3\,800 & 3\,800 &\onlinecite{Barasic2010} & 3\,300 [\onlinecite{Barasic2010}] \\
& 35 & 15\,000  & 800\footnotemark[2] & 4\,700 & 4\,950 &\onlinecite{Barasic2010} & 4\,800 [\onlinecite{Barasic2010}] \\
EuFe$_2$As$_2$ & 300   &  13\,800  & 960 & 3\,300 & 2\,600 &\onlinecite{Wu2009} & 2\,500 [\onlinecite{Wu2009}]\\
LaFePO  & 298  &  14\,900 & 400\footnotemark[2] & 9\,300 & 7\,000 &\onlinecite{Qazilbash2008} & 2\,200 [\onlinecite{Baumbach2008}]\\
 & 10 &  14\,900 & 150\footnotemark[2] & 25\,000 &&\onlinecite{Qazilbash2008} & 71\,000  [\onlinecite{Baumbach2008}]\\
FeTe$_{0.55}$Se$_{0.45}$ & 295 &   7200(360) & 414(21) & 2090 & 2100 &  \onlinecite{Homes2010} \\
& 200 &   7200(360) & 363(18) & 2380 & 2400 &  \onlinecite{Homes2010} \\
& 100 &   7200(360) & 317(16) & 2730 & 2800 &  \onlinecite{Homes2010} \\
& 18 &    & 60\footnotemark[2] &  &  &  \onlinecite{Homes2010} \\
\end{tabular}
\end{ruledtabular}
\footnotetext[1]{Grown in Sn flux: contaminated by Sn impurities.}
\footnotetext[2]{From a generalized Drude analysis in the low frequency limit.}
\footnotetext[3]{THz measurement of thin film.}
\footnotetext[4]{Large uncertainty in this parameter.}
\footnotetext[5]{From a fit containing two Drude terms; this value is for the narrow Drude part.}
\end{table*}

As expected for semimetals, the carrier concentrations $n$ of the undoped FeAs-based parent compounds and the lightly doped superconducting compositions are roughly an order of magnitude smaller than those of conventional metals like lithium, copper and aluminum where $n = 4.7$, 8.5 and $18.1 \times 10^{22}$~cm$^{-3}$, respectively.\cite{kittel1966}  For example, LDA band structure calculations for the total (electrons plus holes) carrier concentrations in undoped nonmagnetic LaFeAsO, CaFe$_2$As$_2$, SrFe$_2$As$_2$ and BaFe$_2$As$_2$ are listed in Table~\ref{OpticsdataTheory}.\cite{Tanatar2009, djsingh2008b, Ma2008} Perhaps surprisingly, the band masses are of order one free electron mass.  Note that when normalized per Fe atom, the theoretical carrier concentration $n \sim 0.3/$Fe is about the same in the 1111- and 122-type structure parent compounds.

High quality sets of experimental optical data and analyses for single crystals of the 122-type layered FeAs compounds ${\rm BaFe_2As_2}$ and ${\rm SrFe_2As_2}$ are shown in Fig.~\ref{FigHu_optics}.\cite{Hu2008}  This figure illustrates the steps to extract important information about the conduction carriers from optical measurements.  One starts with measured reflectivity versus frequency data from the $ab$~planes in panels (a) and (b), respectively.  Then by a Kramers-Kronig transformation, using appropriate extrapolations of the data to zero and infinite frequency, one obtains the optical conductivity, the real part $\sigma_1$ of which is plotted in panels (c) and (d), respectively.  Then to isolate the Drude contribution in Eq.~(\ref{EqDrudeSigma1}), one employs a so-called Drude-Lorentz fit where $\sigma_1(\omega)$ is the sum of the Drude term in Eq.~(\ref{EqDrudeSigma1}) plus several damped simple harmonic oscillator $\sigma_1$ terms corresponding to interband transitions, as exemplified in panels (c) and (d) for the data at temperature $T = 10$~K\@.  From the fits, one obtains the experimental Drude parameters which are the conduction carrier plasma frequency $f_{\rm p}^\prime$ and relaxation rate $1/\tau^\prime$, in addition to parameters describing the interband transitions.  Note that in Figs.~\ref{FigHu_optics}(c) and~(d), the Drude and lowest interband conductivities overlap in frequency to some extent, as noted above, which is why the value of the cutoff angular frequency $\omega_{\rm c}$ to be used in Eq.~(\ref{EqOmegaPeff}) is somewhat ambiguous.  The temperature dependences of the relative values of $f_{\rm p}^\prime$ and $1/\tau^\prime$ for the two compounds are shown in panels (e) and  (f), respectively.

From Eq.~(\ref{EqSigma3}), one can solve for the conduction carrier concentration $n$ in terms of $\omega_{\rm p}$ as
\bea
n &=& \frac{\varepsilon_0 m^*}{e^2} \omega_{\rm p}^2 = \frac{\varepsilon_0 m_{\rm e}}{\hbar^2 e^2}\left(\frac{m^*}{m_{\rm e}}\right)(\hbar\omega_{\rm p})^2\nonumber\\
&=& (0.724 \times 10^{21}~{\rm cm}^{-3})\left(\frac{m^*}{m_{\rm e}}\right)(\hbar\omega_{\rm p})^2,
\label{EqOmegaP2}
\eea
where $m_{\rm e}$ is the free electron mass and the second equality is for $\hbar\omega_{\rm p}$ in units of eV\@.  Instead, in Table~\ref{Opticsdata2} experimental values of $f_{\rm p}^\prime$ in units of cm$^{-1}$ are listed.\cite{Homes2010, Chen2009, Hu2008, Drechsler2009, Barasic2010, JYang2009, Wu2009, Akrap2009, Qazilbash2008, Fischer2010, Wu2010, Nakamura2009}  The carrier concentrations are then determined from these values using  an expression similar to Eq.~(\ref{EqOmegaP2}) given by 
\be
n = (0.724 \times 10^{21}~{\rm cm}^{-3})\left(\frac{m^*}{m_{\rm e}}\right)\left(\frac{f^\prime_{\rm p}}{8\,066}\right)^2,
\label{Eqnfp}
\ee
where we have assumed that $m^*/m_{\rm e} = 1$ and have used the conversion factor between wave numbers and eV in Table~\ref{OpticsUnits2}.  The theoretical values for $n$ in Table~\ref{OpticsdataTheory} are factors of 2--3 larger than the experimental values for the tetragonal (high-temperature) phases in Table~\ref{Opticsdata2}, indicating band enhancements of the current carrier effective mass above the free electron mass by a number of this order.  However, these enhancements are larger than those in the fourth column of Table~\ref{OpticsdataTheory}, suggesting the occurrence of many-body mass enhancements over the band theory values.  Note the widely different values of $n$ in Table~\ref{Opticsdata2} for ${\rm Ba(Fe_{0.92}Co_{0.08})_2As_2}$ and ${\rm Ba(Fe_{0.95}Ni_{0.05})_2As_2}$ depending on the method of deriving the plasma frequency.

When a semimetal is doped, the total carrier concentration of electrons plus holes is expected to remain approximately unchanged according to Figs.~\ref{semimetal_BS} and~\ref{FigBaKFe2As2_FS}, because hole doping reduces the concentration of electron carriers and vice versa.  This expectation is supported in Table~\ref{Opticsdata2} where the plasma frequencies of BaFe$_2$As$_2$ (Ref.~\onlinecite{Drechsler2009}) and Ba$_{0.55}$K$_{0.45}$Fe$_2$As$_2$ (Ref.~\onlinecite{JYang2009}) are seen to be nearly the same at room temperature (above the crystallographic/SDW transition temperature of $\approx 140$~K for BaFe$_2$As$_2$).  On the other hand, one also notices variations in the carrier density of the same compound from different groups such as for BaFe$_2$As$_2$.\cite{Drechsler2009, Akrap2009}

From the ratios between the high and low temperature $n$ data for SrFe$_2$As$_2$ and BaFe$_2$As$_2$ in Table~\ref{Opticsdata2}, one sees that the SDW transitions result in decreases in the carrier concentrations by factors of seven to eight, consistent with expectation due to Fermi surface gapping and reconstruction for a Fermi-surface-nesting-driven SDW as further discussed below in Sec.~\ref{SecItinLocMag}.
 
Values $\sigma_0^{\rm calc}$ of the dc conductivity calculated from the optical constants $f_{\rm p}^\prime$ and $1/\tau^\prime$ via Eq.~(\ref{Eqsigma01})  are given in Table~\ref{Opticsdata3},\cite{Homes2010, Wu2009, Qazilbash2008, Hu2008, Barasic2010, Chen2009, JYang2009, Akrap2009, Fischer2010, Wu2010} along with the measured $\sigma_1 (\omega \to 0)$ obtained directly from the optics experiments (if available) and with  experimental values $\sigma_{\rm dc}$ obtained from conventional four-probe resistivity measurements.\cite{wang2008, Tanatar2009,Wu2009, Barasic2010, Chen2009, Ni2008h, Baumbach2008, Wu2010}  One sees that the reason that the resistivity decreases as the temperature drops below the SDW transition temperatures of SrFe$_2$As$_2$ and BaFe$_2$As$_2$ (see bottom panel of Fig.~\ref{FigChiRhoBaFe2As2}), even though the carrier concentration drastically decreases, is that the scattering time $\tau$ increases even more than the carrier concentration decreases.

Several studies have analyzed the $ab$-plane optical conductivity of Ba(Fe$_{1-x}$Co$_x)_2$As$_2$ and other 122-type crystals in terms of a Drude-Lorentz model with \emph{two} Drude contributions.\cite{Moon2010, Barasic2010, Nakajima2010, Wu2010, Lucarelli2010, Chen2009b}  Both Drude contributions have the same functional form as in Eq.~(\ref{EqDrudeSigma1}), but the first ``narrow'' one has a relatively small relaxation rate $1/\tau$ and the second phenomenological ``broad'' one has a large $1/\tau$.  For ${\rm BaNi_2As_2}$ both terms are apparently considered to arise from coherent carriers.\cite{Chen2009b} For ${\rm EuFe_2As_2}$ the narrow Drude term is taken to arise from coherent electron carriers and the second broad one from coherent hole carriers.\cite{Moon2010}  

In other studies analyzing the data using two Drude terms, the first Drude term is assumed to arise from coherent conduction carriers and the second from ``incoherent'' carriers.  By ``incoherent'' we understand to mean that the wave vectors of those current carriers are not good quantum numbers, and the mean-free-path and mean-free-time $\tau$ are not concepts that can be applied to them.  Thus the Drude form of the conductivity applied to the second class of incoherent carriers is a fitting function with no clear physical meaning.  Both classes of carriers have nonzero densities of states at the Fermi energy $E_{\rm F}$, and hence the conduction states at $E_{\rm F}$ consist of both coherent and incoherent parts.  It is not stated in any of the relevant papers what causes one part of the conduction carriers to be coherent and the other part to be incoherent, or what the origins of the two classes of carriers are.  The narrow Drude part is found to give rise to a $1/T^2$ Fermi liquid-like temperature contribution to the conductivity.  One of the papers claims that both the coherent and incoherent current carriers contribute to the superconducting condensate, each with its own superconducting energy gap.\cite{Wu2010}  Strong-coupling theories predict the coexistence of coherent and incoherent current carriers, but the coherent carriers are low-energy carriers with energies near the Fermi energy, whereas the incoherent carriers are high-energy carriers.\cite{Dai2009, Si2009}  Thus in this case the incoherent carriers evidently could not be represented by a Drude term in the optical conductivity.

It seems that the two-Drude conductivity component description of the current carriers may be oversimplified when considering that all these materials are multiband systems where at least four Fermi surface sheets arising from the Fe atoms are likely to be important to the conduction properties.

The $c$-axis charge dynamics of ${\rm BaFe_2As_2}$ crystals was studied by Chen et al.\cite{Chen2010}  They found a small anisotropy ratio of the conductivity at 300~K given by $\sigma_{ab}/\sigma_c \approx 2.8$, similar to the anisotropy ratio of $3\pm1$ found by Tanatar et al.\ from dc resisitivity measurements.\cite{Tanatar2009}  The authors studied the changes in the charge dynamics upon cooling below $T_{\rm N} = 137$~K and found significant differences with the $ab$-plane charge dynamics, and suggested the origin of this difference in terms of Fermi surface reconstruction.

The conduction carrier optical properties of the 122-type FeAs-based superconductors are further utilized below  in Sec.~\ref{SecOptCorr} to obtain quantitative estimates of their degrees of conduction carrier correlation, where these values are compared with corresponding values for a wide variety of other materials.

\subsubsection*{b. 11-type {\rm Fe}$_{1+y}${\rm Te}$_{1-x}${\rm Se}$_{x}$ Crystals}

\begin{figure}
\includegraphics [width=3.4in]{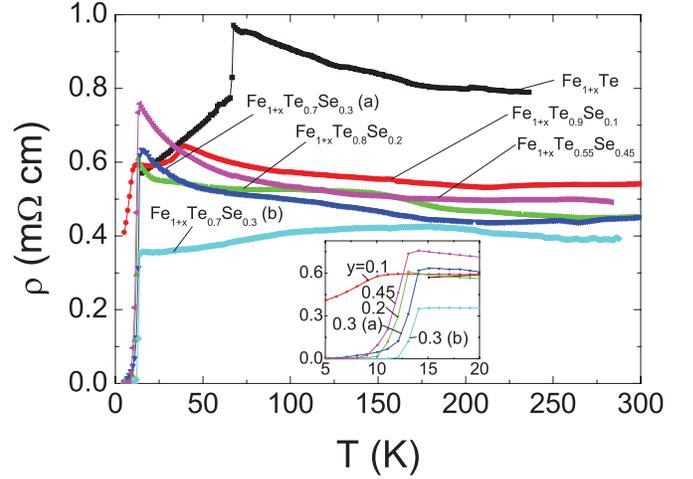}
\caption{(Color online)  In-plane electrical resistivity $\rho$ versus temperature $T$ for a series of Fe$_{1+x}$(Te$_{1-y}$Se$_{y})$ crystals (note the interchange of $x$ and $y$ compared with the text).  The temperature dependence of the data for $y=0$ is similar to that in Fig.~\ref{Fe1.05Te_rho} for a crystal from another group, although the magnitude is a factor of two larger than in Fig.~\ref{Fe1.05Te_rho}.  This illustrates the lack of reproducibility between measurements on different crystals.  Inset: Expanded plots of the data at low temperatures, showing more clearly the superconducting transitions.  Reprinted with permission from Ref.~\onlinecite{Pallecchi2009}.  Copyright (2009) by the American Physical Society.}
\label{Pallecchi_Fe(TeSe)_rho} 
\end{figure}

\begin{figure}
% For arXiv, uncomment the first one and comment out the 2nd.
\includegraphics [width=\columnwidth]{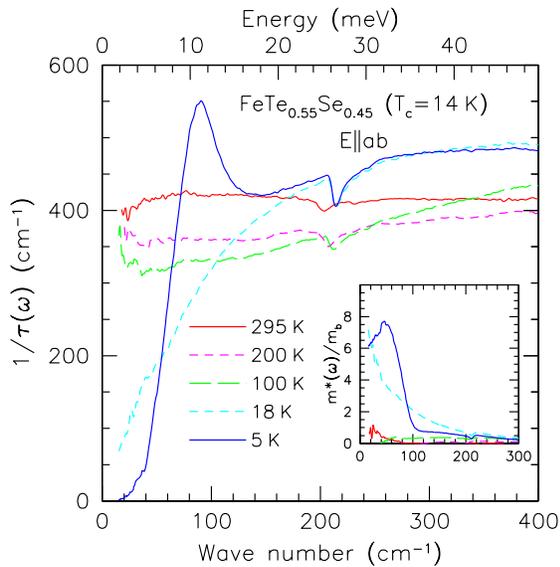}
\caption{(Color online)  In-plane free carrier relaxation rate $1/\tau$ obtained from an extended-Drude model  versus frequency $f^\prime$ in wave numbers for a single crystal of FeTe$_{0.55}$Se$_{0.45}$ with $T_{\rm c} = 14$~K.\cite{Homes2010}  In the analysis, the values $f_{\rm p}^\prime = 7200~{\rm cm^{-1}}$ and $\varepsilon_\infty = 4$ were used.  Data are plotted as solid and dashed curves for the temperatures indicated.  The dips at 204~cm$^{-1}$ are due to an infrared-active phonon mode.  Inset: Effective mass $m^*/m_{\rm b}$ versus frequency.  From the figure, one sees that the Drude-Lorentz analysis is not valid at the low temperature of 18~K~$> T_{\rm c}$ because $1/\tau$ becomes strongly frequency dependent, which is indicative of strong electronic correlations.  Reprinted with permission from Ref.~\onlinecite{Homes2010}.    Main figure (not inset): Copyright (2010) by the American Physical Society.}
\label{HomesFig3} 
\end{figure}

The in-plane optical properties of an Fe$_{1.05}$Te single crystal were studied by Chen et al.\cite{Chen2009a}  Above $T_{\rm N} = 65$~K, $\sigma_1$ was found to be rather flat, with no obvious signature of a Drude peak at zero frequency.  Thus they inferred that the charge transport was incoherent in this $T$ range.  However, below $T_{\rm N}$ a clear Drude peak developed, reflecting a significant decrease in the conduction carrier scattering rate.  They did not report values of $\omega_{\rm p}$ or $1/\tau$ in this low-$T$ range.

Homes and coworkers measured the in-plane optical properties versus temperature of a superconducting single crystal of FeTe$_{0.55}$Se$_{0.45}$ with $T_{\rm c} = 14$~K.\cite{Homes2010}  They fitted their $\sigma_1(\omega)$ data by a Drude-Lorentz model and obtained the plasma frequency and scattering rates of the Drude term that are listed in Tables~\ref{Opticsdata2} and~\ref{Opticsdata3} for $T = 295$, 200 and 100~K\@.  From Table~\ref{Opticsdata2}, the carrier concentration calculated from the plasma frequency assuming $m^*/m_{\rm e} = 1$ is only 0.025 carriers/f.u.  The literature data on the dc resistivity of {\rm Fe}$_{1+y}${\rm Te}$_{1-x}${\rm Se}$_{x}$ crystals, with which to compare the optical conductivity data, show variations in magnitude and temperature dependence even for crystals with about the same composition (see, e.g., Refs.~\onlinecite{Sales2009a} and \onlinecite{Pallecchi2009}).  The data by Pallecchi et al.\cite{Pallecchi2009} for a series of crystals is shown in Fig.~\ref{Pallecchi_Fe(TeSe)_rho}.  Irrespective of the temperature dependence of $\rho$ for the various compositions, the magnitude of $\rho$ is of order 0.5~m$\Omega$~cm below 300 K, so the conductivity is $\sigma_0 \sim 2000~(\Omega~{\rm cm})^{-1}$. This value is similar to the values calculated from optics data at 100--295~K for FeTe$_{0.55}$Se$_{0.45}$ in Table~\ref{Opticsdata3}.

At a lower temperature of 18~K~$> T_{\rm c}$, the peak in $\sigma_1(\omega)$ at $\omega = 0$ cannot be fitted any longer by the Drude model, and the extended-Drude model described above at the end of Sec.~\ref{SecOpticsThy} was used by Homes et al.\cite{Homes2010}  From that analysis they obtained the frequency dependence of the scattering rate $1/\tau^\prime$ shown for five different temperatures in Fig.~\ref{HomesFig3}.\cite{Homes2010}  At 100~K and above, $1/\tau^\prime$ is frequency independent up to a frequency of 300~${\rm cm^{-1}}$, and self-consistently agrees with the relaxation rate obtained at the respective temperature from the above Debye-Lorentz model.  From the strong frequency dependence of the data at 18~K~$> T_{\rm c}$, the authors inferred that strong electronic correlations have set in upon cooling from 100~K to 18~K\@.  This is an interesting result that the degree or strength of electronic correlations can be strongly temperature dependent in the Fe-based materials.  The strong electron correlations are reflected by a strong enhancement of the effective mass with decreasing temperature as shown in the inset of Fig.~\ref{HomesFig3}, which reaches a value of about 7--8 band masses at low frequency at 18~K\@.  This temperature dependent mass enhancement would presumably be reflected in other properties as well.  The $1/\tau^\prime$ data in the superconducting state at 6~K in Fig.~\ref{HomesFig3} were analyzed and indicated the presence of two superconducting gaps, presumably arising from different Fermi surface sheets, and that only about one-fourth of the current carriers participate in the superconducting condensate.\cite{Homes2010} 

A Drude peak was also observed in $\sigma_1(\omega)$ for a single crystal of ${\rm Fe_{1.06}(Te_{0.88}S_{0.14})}$ at low temperatures (note that this is sulphur-doped, not selenium-doped).\cite{Stojilovic2010}  From an extended-Drude analysis, the effective mass $m^*$ of the current carriers became negative below 200~K, which is to be contrasted with the positive values of $m^*$ in the inset of Fig.~\ref{HomesFig3} above.  The authors concluded that ${\rm Fe_{1.06}(Te_{0.88}S_{0.14})}$ contains an incoherent non-Fermi-liquid with a pseudogap below 200~K,\cite{Stojilovic2010} in agreement with theoretical predictions for FeTe$_{1-x}$Se$_x$.\cite{Crako2009, Craco2009a}  An interesting question is why this compound differs from other 11-type compositions that appear to contain an at least partially coherent Fermi liquid as judged from optical experiments such as those cited above for FeTe$_{0.55}$Se$_{0.45}$ and also differs from the 122-type compounds described in Sec.~\ref{SecOptCorr} below.

\subsection{\label{SecMagnetoElastic}  Giant Magnetostructural Coupling}

\begin{figure}
\includegraphics [width=3in]{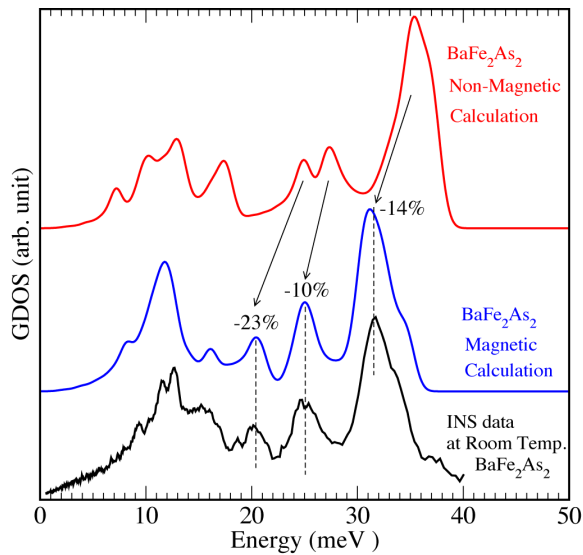}
\caption{(Color online) Generalized phonon density of states GDOS versus energy for BaFe$_2$As$_2$.  The top red curve shows a calculation with no spin on the Fe atoms, whereas the middle blue curve is a calculation including a spin on the Fe atoms (but no magnetic order).  The bottom black curve is a plot of experimental data from inelastic neutron scattering measurements at 300~K.\cite{Zbiri2009}  A comparison of the calculations with the data indicates that the Fe magnetism has to be included to explain the lattice vibration properties.  Reprinted with permission from Ref.~\onlinecite{Yildirim2009}, Copyright (2009), with permission from Elsevier.}
\label{FigPhononDOS} 
\end{figure}

A strong coupling between the magnetic state of the Fe and the lattice was inferred from early conventional electronic structure calculations.\cite{DJSingh2009a}  This sensitivity was identified as originating from the degree of covalent Fe-As bonding.\cite{Belashchenko2008}  When the experimental atomic positions in the crystal structures of the FeAs-based parent compounds are used in the calculations, the ordered Fe magnetic moments at $T = 0$ are calculated to be far larger than the experimental values.  However, for the experimental structure the total energy of the compound is calculated not to be at a minimum with respect to the adjustable height of the As atoms above and below the Fe planes, which depends on the As ``$z$-coordinate'' in its general position within the structure, and therefore does not agree with experiment.  If the total energy of a compound is minimized, one gets a somewhat different $z$ value from the experimental value but remarkably, a value of the ordered Fe moment is obtained that is much closer to the experimental value.  These calculated As positions are called ``relaxed'' or ``optimized'' positions in the literature.  The calculations described above indicated that the Fe magnetism and As atomic positions depend very sensitively on each other (see also Fig.~\ref{LaFeAsOBS} above).  

There are other indications of an anomalously strong coupling between the spin state of the Fe atom and the lattice structure in the FeAs-based materials.  For example, calculations for CaFe$_2$As$_2$ predict a $c$-axis parameter that is 10\% (1.1~\AA!) smaller if the Fe atoms are nonmagnetic than if they have a magnetic moment.\cite{Yildirim2009}  The phonon density of states at 300~K (above the SDW transition temperature of 140~K) as measured by inelastic neutron scattering\cite{Zbiri2009} can be very well explained if a magnetic state of the Fe atoms is included in the calculations, as shown in Fig.~\ref{FigPhononDOS}.\cite{Yildirim2009}  From calculations such as this, Yildirim concluded that there is a magnetic moment on the Fe atoms at all temperatures and in the superconducting as well as nonsuperconducting compositions, and that ``the iron magnetism could be in the form of fluctuating SDW type small magnetic domains or it could be at the atomic limit of paramagnetic Fe ions (\emph{i.e.}\ para-magnon).''\cite{Yildirim2009}  Similarly, Hahn and co-workers found that the phonon spectra of CaFe$_2$As$_2$ as measured at room temperature via inelastic x-ray scattering could be explained best if the calculations were carried out for the magnetically ordered state, consistent with the observation of short-range SDW order up to room temperature.\cite{Hahn2009}  In the BaFe$_{2-x}$Co$_x$As$_2$ system where the structural transition temperature is greater than the SDW transition temperature for $x > 0$, Lester \emph {et al.} found that the SDW transition affects the temperature dependence of the structural Bragg peak intensities from both x-ray and neutron diffraction measurements.\cite{Lester2009}

A dramatic example of the dependence of the spin state of the Fe on the crystal structure was evident from diffraction measurements of CaFe$_2$As$_2$ under applied pressure.  Neutron and x-ray diffraction measurements of the crystalline and magnetic structure showed that at the relatively low pressure of 0.3~GPa (3 kbar), the lattice volume at 50~K collapses by an astonishing 4.5\% with a concomitant loss of the SDW and Fe magnetic moment.\cite{Kreyssig2008, Goldman2009}  On the other hand, LaFeAsO$_{0.9}$F$_{0.1}$ with $T_{\rm c} = 28$~K and SmFeAsO$_{0.93}$F$_{0.07}$ with $T_{\rm c} = 37$~K show no lattice transformations up to  pressures of at least 32~GPa and 9~GPa, respectively.\cite{Garbarino2008, Martinelli2009b}

From ARPES measurements, Liu \emph{et al.}\ found that 3D dispersion along the $k_z$ axis of the hole pocket in undoped ${\rm CaFe_2As_2}$ occurs below, but not above, the SDW/structural transition temperature $T_{\rm N}$ where the dispersion is 2D, and that the electron pocket does not disperse in either temperature regime.\cite{Liu2009}  From polarization-dependent ARPES measurements on ${\rm BaFe_2As_2}$, Shimojima \emph{et al.} also found a drastic change in the Fermi surfaces below $T_{\rm N}$ to a 3D orbitally ordered state.\cite{Shimojima2009}  From comparison with band structure calculations, they deduced that the Fermi surface transformation is mainly caused by the antiferromagnetic ordering rather than the structural transition.  Quantum oscillation experiments on ${\rm SrFe_2As_2}$ (Ref.~\onlinecite{Sebastian2008}) and ${\rm BaFe_2As_2}$ (Ref.~\onlinecite{Analytis2009}) also indicate drastic reconstructions of the Fermi surfaces from 2D to 3D below the respective $T_{\rm N}$, with reductions in carrier concentrations.  These results suggest that spin correlations are important because in a SDW material like Cr the Fermi surface opens gaps below $T_{\rm N}$ but is not totally reconstructed.  These results also indicate that the competition between the SDW and superconductivity apparent from the phase diagrams in Fig.~\ref{FigBaKFe2As2_phase_diag} arises because they compete for the same conduction electrons.  Furthermore, due to the Fermi surface reconstruction from the SDW, the nesting between hole and electron Fermi surface pockets required for the $s^\pm$ inter-orbital pairing mechanism is largely reduced by the SDW.  On the other hand, the giant magnetoelastic coupling may provide an avenue for enhancing the electron-phonon interaction mechanism for superconductivity (see Sec.~\ref{Sec_SC_Mech} below).

The strong magnetostructural coupling in the FeAs-based materials has been concluded to occur also in the FeSe$_{1-x}$Te$_x$ system.\cite{Moon2010}  Here, the height of the Se/Te layers is concluded to control whether the antiferromagnetic ordering has the above stripe structure, or whether a different observed double-stripe structure is stable.

\subsection{\label{Sec_N(EF)}Density of States at the Fermi Energy}

\begin{figure}
\includegraphics [width=3.3in]{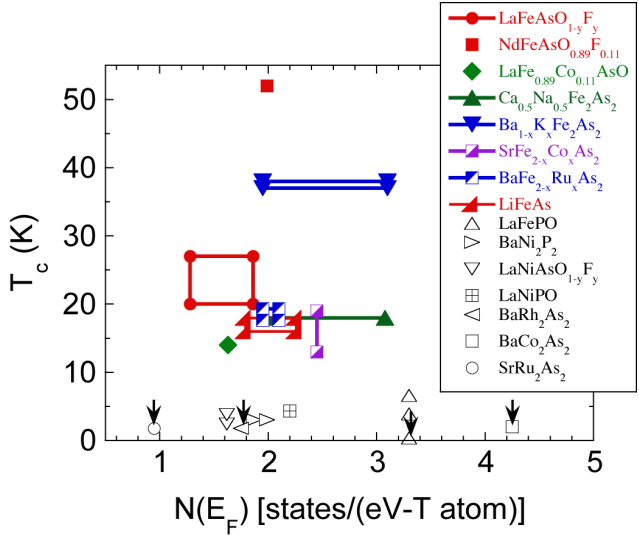}
\caption{\label{TcN(EF)} (Color online) Superconducting transition temperature $T_{\rm c}$ versus the bare nonmagnetic band structure density of conduction electron states at the Fermi energy for both spin directions $N(E_{\rm F})$ for a variety of 112, 1111, and 111 transition metal $T$ pnictides.  The vertical arrows pointing downwards indicate that superconductivity is not observed above the indicated temperature.  Multiple data points for the same compound (error boxes) indicate the range(s) of $T_{\rm c}$ and/or $N(E_{\rm F})$ reported for the compound.  The colored symbols are for compounds containing FeAs layers.  See the Appendix for the references and Table~\ref{FeTeSeData} for data for the Fe$_{1+y}$Te$_{1-x}$Se$_x$ system.}
\end{figure}

In the BCS theory of superconductivity,  the transition temperature arising from the electron-phonon interaction is $T_{\rm c} = 1.14\Theta_{\rm D} \exp[-1/N(E_{\rm F})V]$, where $\Theta_{\rm D}$ is the Debye temperature for the lattice, $N(E_{\rm F})$ is the bare band structure density of conduction electron states at the Fermi energy $E_{\rm F}$, and $V$ is an electron-phonon coupling constant.  By the latter we mean the strength of the coupling between the conduction electrons and lattice vibrations.  From this expression, one expects a strong positive correlation between $T_{\rm c}$ and $N(E_{\rm F})$.  Shown in Fig.~\ref{TcN(EF)} is a plot of $T_{\rm c}$ versus $N(E_{\rm F})$ for many of the Fe-based and related compounds.  Surprisingly, one sees that there is no correlation between $T_{\rm c}$ and $N(E_{\rm F})$, even within the subset of FeAs-based materials shown with colored symbols.  This is not to say that $N(E_{\rm F})$ is irrelevant to the value of $T_{\rm c}$ in these materials, but rather that other properties that are changing between the different compounds are at least as important if not more so.  Furthermore, the $N(E_{\rm F})$ values from band structure calculations do not include the influence of electron correlation effects such as spin fluctuations.  These fluctuations may strongly change between compounds with different $T_{\rm c}$'s.  Evidence that the bare $N(E_{\rm F})$ values can be too low by a factor of order five is suggested from the dependence of the magnetic susceptibility on $N(E_{\rm F})$ in Fig.~\ref{ChiN(EF)} below and by consideration of the change in the heat capacity at $T_{\rm c}$ for Ba$_{1-x}$K$_x$Fe$_2$As$_2$ crystals discussed below in Sec.~\ref{Sec_SC_Cp}.

\subsection{\label{Sec_Stoner}Magnetic Susceptibility}

\subsubsection{\label{SecChiUnits} Introduction: Units and Other Things}

Before discussing the results of magnetic susceptibility measurements on the FeAs-type materials, it is useful to discuss the units of such measurements.  Gaussian (cgs) units are almost universally used in the presentation of magnetization and magnetic susceptibility data in the condensed matter physics literature because of their simplicity.  The term ``magnetization'' is usually reserved for normalized magnetic moment, e.g., magnetic moment per mole, per gram, or per unit volume.  The Quantum Design, Inc., SQUID magnetometer is currently used worldwide to measure the magnetic moment of samples.  The software of this instrument outputs the magnetic moment of a sample in ``electromagnetic units'', or ``emu'', which is a Gaussian unit: 
\be
{\rm 1~emu = 1~G~cm^3 = 1~erg/G.}
\label{Convertemu}
\ee
The Gaussian units of magnetic induction $B$ and magnetic field $H$ have the same magnitude,
\be
{\rm 1~G = 1~Oe}, 
\ee
which is one reason this system of units is so transparent and widely used.  A unit of convenience in the Gaussian system of units is the Tesla (T): 
\be
{\rm 1~T = 10^4~G = 10^4~Oe}.  
\ee
When normalized to a mole of formula units of a material, the molar magnetization $M$ is in units of emu/mol = G\,cm$^3$/mol, and the molar magnetic susceptibility $\chi = M/H$ is in units of emu/(Oe~mol) = cm$^3$/mol.  Different authors sometimes quote magnetic moment and magnetic susceptibility of a sample, which are not the same quantities, in the same units of ``emu'' [compare, for example, the units for magnetic susceptibility in Fig.~\ref{FigChiRhoBaFe2As2}(a) above with those in Fig.~\ref{FigChi(T)} below that are both reproduced from the literature].  Because of this ambiguity and for ease of unit conversions, in this review (and in all of my other publications), ``emu'' units are avoided except in some figures reproduced from the literature.  The magnetization $M$ expressed as magnetic moment per unit volume  has units of emu/cm$^3$ = G\,cm$^3$/cm$^3$ = G\@.  The magnetic susceptibility $\chi = M/H$ per unit volume is then in units of G/Oe = 1, which is dimensionless.  

To convert from one normalization of $M$ or $\chi$ to another is straightforward if the proper Gaussian units are used.  For example, to convert from molar susceptiblity to (dimensionless) volume susceptibility one easily sees that one should divide the former in cm$^3$/mol by the molar volume in cm$^3$/mol.  In the expression relating the magnetic induction to the magnetic field in Gaussian units in the absence of demagnetization effects, $B = H + 4\pi M$, $M$ is the magnetic moment per unit volume in units of G\@.  A perfect diamagnet has $B = 0$ inside it and therefore has a dimensionless volume susceptibility $\chi = M/H = -1/(4\pi)$.  For this reason, the fraction of full diamagnetic susceptibility of a superconducting sample is often expressed for low fields as $-4\pi \chi$ with $\chi$ being the dimensionless Gaussian volume susceptibility. 

Ideally, if a material is cooled in a small magnetic field, the magnetic field is expelled from the material when it becomes superconducting.  This flux expulsion upon cooling through $T_{\rm c}$ is called the ``Meissner effect.''  If a metal were cooled in the same field and somehow became a perfect conductor with zero resisitivy, but not a superconductor, the magnetic flux would be trapped and not expelled.  Thus a superconductor is different than a perfect conductor.  On the other hand, there are very few materials that exhibit a 100\% Meissner effect, because the flux gets ``pinned'' and cannot escape when cooling below $T_{\rm c}$.  The Fe-based superconductors generally show small Meissner effects.  An upper limit to the superconducting volume fraction can be obtained by cooling the sample in zero field (``zero-field-cooling'') and then applying a small ($\sim 10$~Oe) field at a low temperature $T \ll T_{\rm c}$ and measuring the diamagnetic shielding volume susceptibility and comparing that to the ideal value of $-1/(4\pi)$.  This diamagnetic shielding measurement is sometimes improperly called a measurement of the Meissner effect and/or the shielding fraction as ``the Meissner fraction'', which instead must be measured on cooling in a fixed field.  With a demagnetization factor $N$, the ideal value of the volume susceptibility becomes $-1/[4\pi(1-N)]$, which is more negative than the previous value.  The reason that diamagnetic shielding only measures an upper limit to the superconducting volume fraction is that it only takes a small amount of superconducting material to provide 100\% diamagnetic shielding.  Such a result would occur, for example, for a nonsuperconducting copper ball coated with superconducting Pb ($T_{\rm c} = 7.2$~K).  An easy and reliable way to check for bulk superconductivity in a sample is by measuring the heat capacity jump at $T_{\rm c}$ (see Sec.~\ref{Sec_SC_Cp}).

In local moment systems in which, e.g., nearest-neighbor spins are coupled by an exchange interaction (energy) $J$, theorists often present the calculated magnetic susceptibility simply as ``$\chi$ versus $T$'', where both the magnetic susceptibility $\chi$ and the temperature $T$ are dimensionless.  What is actually plotted are the dimensionless quantities $\chi J/(Ng^2\mu_{\rm B}^2)$ versus $k_{\rm B}T/J$, where $N$ is the number of spins, $g$ is the spectroscopic splitting factor ($g$-factor) of the magnetic moment, $\mu_{\rm B}$ is the Bohr magneton, and $k_{\rm B}$ is Boltzmann's constant.  Thus, the theorist's ``$\chi$'' is the magnetic susceptibility per spin, in units of $1/J$, where $g\mu_{\rm B}$ is set equal to unity.  The theorist's ``$T$'' is the thermal energy $k_{\rm B}T$ normalized by the energy scale $J$.

The adjective ``diamagnetic'' or noun ``diamagnetism'' refer to a negative value of $\chi$ and ``paramagnetic'' or ``paramagnetism'' refer to a positive value of $\chi$.  The adjective ``paramagnetic'' is also sometimes used to refer to a state of a material that is not long-range magnetically ordered, irrespective of the sign of $\chi$, to distinguish this state from a magnetically-ordered state, perhaps of the same material.  Alternatively, a state of a material with no long-range order of any kind at a particular temperature or in a particular temperature range, especially if that material does exhibit superconductivity or long-range magnetic order at a lower temperature, is often called the ``normal state'' of the material.

\subsubsection{\label{SecFMImpurities} Influence of Ferromagnetic and Paramagnetic Impurities on Magnetic Susceptibility Measurements}

In the Fe-based superconductor field, it is very common when carrying out magnetic susceptibility measurements versus temperature $T$ to simply measure the magnetization versus temperature $M(T)$ at fixed applied magnetic field $H$ with $H \sim$ 1--5~T, and then report the susceptibility as $\chi(T) \equiv M(T)/H$.  A major problem with this procedure, especially with the Fe-containing materials of interest in this review, is that the samples often contain ferromagnetic impurities and their contribution to $M$ is only rarely separately measured and accounted for.   

To illustrate the high sensitivity of magnetization measurements to the presence of ferromagnetic impurities in a sample, suppose a measurement at room temperature in $H = 1$~T is being carried out of the magnetic moment of a typical 0.10~g ($3.5 \times 10^{-4}$~mol) sample of polycrystalline LaFeAsO containing only 0.10~mol\% (20~$\mu$g) of ferromagnetic Fe metal impurity.  The ferromagnetic ordering (Curie) temperature of Fe metal is $T_{\rm C} = 1043$~K and its ferromagnetic saturation moment is $\mu_{\rm sat} = 2.2~\mu_{\rm B}$/Fe~atom = $1.2 \times 10^4$~G~cm$^3$/mol~Fe.  Here we have used the very useful conversion expression
\be
N_{\rm A}\mu_{\rm B} = 5585~{\rm \frac{G~cm^3}{mol}}.
\label{mBGcm3Convert}
\ee
That is, one mole of Bohr magnetons has a magnetic moment of 5585~G~cm$^3$.  Using the intrinsic susceptibility of polycrystalline LaFeAsO at 300~K of $3.3\times 10^{-4}$~cm$^3$/mol (see Table~\ref{LaFeAsOFdata}),\cite{Klingeler2008} the intrinsic magnetic moment of the 0.1~g LaFeAsO sample in $H = 1$~T is 0.0012~G~cm$^3$, whereas that of the saturated Fe impurity is 0.0042~G~cm$^3$ which is a factor of 3.5 times larger than the intrinsic value.  Thus the measured $M/H$ for this hypothetical sample of LaFeAsO is a factor of 4.5 times larger than the intrinsic $\chi$ value.  Therefore, unless it is explicitly stated in a paper reporting magnetic susceptibility data that the influence of the contribution of ferromagnetic impurities to $M$ have been checked for \emph{and} taken into account,\cite{wang2008, Klingeler2008} one should consider reported $\chi(T)$ data such as those in the tables in the Appendix to be upper limits only.  This is true even for measurements of single crystals, because crystals can and often do contain inclusions and/or intergrowths of ferromagnetic impurities.  

\begin{figure}
\includegraphics [width=2.5in]{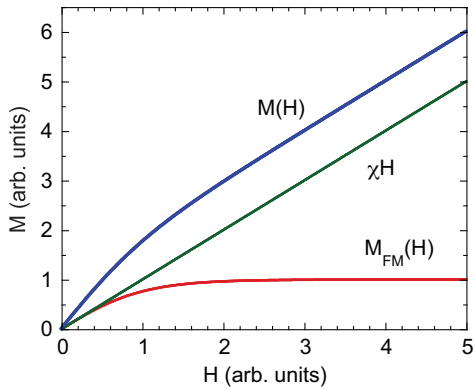}
\caption{(Color online) Sketch of the observed magnetization $M$ versus applied magnetic field $H$ at fixed temperature $T$ for a hypothetical sample containing ferromagnetic impurities.  The ferromagnetic impurities have magnetization $M_{\rm FM}(H)$, the intrinsic magnetization of the sample is $\chi H$, and the observed magnetization is $M(H) = M_{\rm FM}(H) + \chi H$.  The magnetization of the ferromagnetic impurities saturates to a constant value $M_{\rm S}$ above a saturation field $H_{\rm S}$ which in this example is $H_{\rm S} \sim 2$~units.  For $H > H_{\rm S}$, one has the linear relation $M(H) = M_{\rm S} + \chi H$.  The intrinsic susceptibility $\chi$ is thus the high-field slope of $M(H)$ and the $y$-intercept of the high-field linear behavior is $M_{\rm S}$.}
\label{FigFMimp}
\end{figure}

The (zero-field) magnetic susceptibility of a sample not exhibiting long-range magnetic ordering is formally defined to be $\chi = \lim_{H\to 0} M(H)/H$.  Of course, this definition assumes the absence of ferromagnetic impurities in the sample.  For materials with an intrinsic magnetization $M_{\rm intrinsic}$ that is proportional to $H$ via $M_{\rm intrinsic} = \chi H$ where $\chi$ is the intrinsic susceptibility, one can determine the temperature-dependent ferromagnetic impurity contribution to the measured $M$ by carrying out and analyzing $M(H)$ isotherm measurements at various temperatures as sketched in Fig.~\ref{FigFMimp}.  In this case one has $M(H)|_T =\chi(T) H + M_{\rm FM}(H)|_T$, where $M_{\rm FM}(H)|_T$ is the ferromagnetic contribution to $M(H)|_T$.  The $M_{\rm FM}(H)|_T$ is nonlinear and typically saturates to a constant value $M_{\rm S}(T)$ at ``high'' fields $H_{\rm S} \gtrsim 1$~T (in Fig.~\ref{FigFMimp}, $H_{\rm S} \sim 2$~units).  Then $\chi(T)$ is the slope of the high-field linear region of the measured $M(H)|_T$.  Alternatively, it is usually much easier to determine $\chi(T)$ from $M(T)|_H$ temperature sweeps at $H > H_{\rm S}$ as follows.  If $M_{\rm S}(T)$ has been determined separately from $M(H)|_T$ isotherms (typically five such isotherm measurements are sufficient), one can interpolate the $M_{\rm S}(T)$ data, usually using only a simple low-order polynomial in $T$, and then use
\be
\chi(T) = \frac{M(T)|_H - M_{\rm S}(T)}{H}.\hspace{0.2in}(H > H_{\rm S})
\label{EqChiInt}
\ee
However, even if one determines $\chi(T)$ using Eq.~(\ref{EqChiInt}) or more directly from $M(H)|_T$ isotherms, if the ferromagnetic impurity contribution to $M(T)|_H$ at the measurement field $H$ or at the highest $M(H)$ measurement field, respectively, is $\gtrsim 10$\%, then the value of the extracted ``intrinsic'' susceptiblity $\chi$ can still be significantly in error because of imperfect modeling of $M_{\rm FM}(H)$ [i.e., $M_{\rm FM}(H)$ does not become strictly constant even up to the highest field of the $M(H)$ measurement].  

Sometimes $M(T)|_H$ measurements of FeAs-type and related compounds are carried out at low fields $H \lesssim 0.1$~T $\ll H_{\rm S}$; this drastically accentuates the influence of the ferromagnetic impurities on the value of the reported $\chi \equiv M(T)/H$ (see Fig.~\ref{FigFMimp}).  Another  disadvantage to carrying out $M(T)|_H$ measurements at $H \ll H_{\rm S}$ is that one cannot use Eq.~(\ref{EqChiInt}) to extract $\chi(T)$.  In particular, in this field region the precise form of $M_{\rm FM}(H)|_T$ of the ferromagnetic impurities is generally temperature- and magnetic field-history dependent and therefore cannot be uniquely determined and corrected for.

For an explicit example where the above procedures were essential to derive the intrinsic anisotropic $\chi_{\alpha\alpha}(T)$ from $M_{\alpha\alpha}(H,T)$ ($\alpha = x, y, z$) data, see Ref.~\onlinecite{singh2009} describing magnetization measurements on ${\rm BaMn_2As_2}$ single crystals.  In this study, $\sim 0.1$\% of ferromagnetic MnAs impurities with Curie temperature $T_{\rm C} \approx 320$~K significantly affected the measurements at lower temperatures and their temperature-dependent magnetization contribution had to be corrected for.  

The ``intrinsic'' $\chi$ as defined above that is derived from $M(H)|_T$ isotherms or from Eq.~(\ref{EqChiInt}) may still not be intrinsic, since it can contain the contribution of paramagnetic local moment impurities that are not magnetically ordered.  These typically give a Curie-Weiss contribution $C/(T - \theta)$ to $\chi$ at $T \gtrsim \theta$ at which temperatures their magnetization is proportional to $H$.  The reported $\chi(T)$ data for Fe-based superconductor materials often exhibit such Curie-Weiss like contributions that are evident from low-temperature upturns in the reported $\chi(T)$, but these are rarely analyzed and corrected for.  Most of these reported upturns are likely not intrinsic.  The Curie constant $C$ could be used to obtain a rough estimate of the paramagnetic impurity concentration.  In addition, if the Weiss temperature $\theta$ is sufficiently small, the nonlinear $M(H)|_T$ behaviors of the paramagnetic impurities at low temperatures $T \lesssim \theta$ could be analyzed to yield accurate values for both the concentration and the spin of the paramagnetic impurities.  For an example of a high quality analysis that considers both ferromagnetic and paramagnetic impurities, see the discussion of the magnetization data for a Fe$_{1.12}$Te single crystal in the next section.

\subsubsection{\label{SecChi} Temperature Dependence of the Magnetic Susceptibility at High Temperatures}

\begin{figure}
\includegraphics [width=3in]{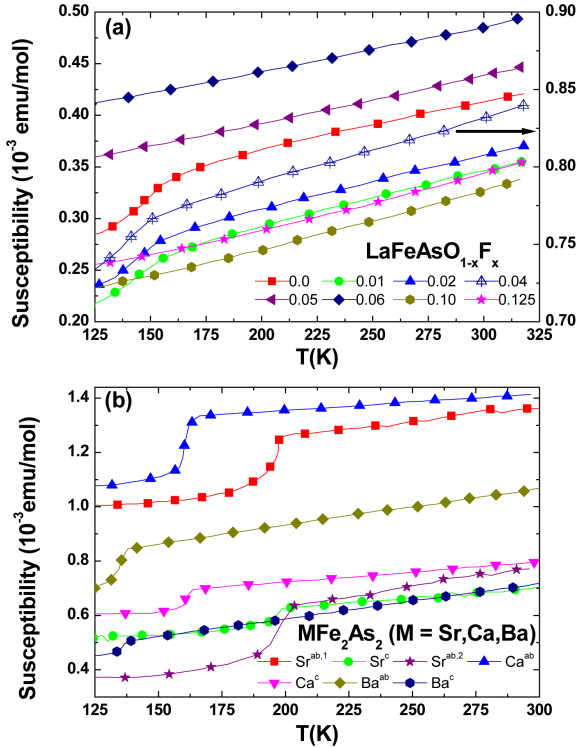}
\caption{(Color online) Magnetic susceptibility versus temperature $T$ for (a) LaFeAsO$_{1-x}$F$_x$ superconducting and nonsuperconducting compounds and (b) $M$Fe$_2$As$_2$ nonsuperconducting parent compounds as compiled from the literature.\cite{GMZhang2008}  The literature references are given in Ref.~\onlinecite{GMZhang2008}.  Reprinted with permission from Ref.~\onlinecite{GMZhang2008}.  Copyright (2009) by the European Physical Society.}
\label{FigChi(T)} 
\end{figure}

\subsubsection*{a. 122- and 1111-type FeAs-based Compounds }

In this section we consider the temperature dependence of the magnetic susceptibility $\chi$ of the FeAs-type parent compounds and doped superconductors.  At low temperatures, $\chi$ often shows an upturn, presumably due to paramagnetic impurities as discussed in the preceding section.  As shown above in Fig.~\ref{FigChiRhoBaFe2As2} for the BaFe$_2$As$_2$ parent compound and also for Ca(Fe$_{1-x}$Co$_{x})_2$As$_2$ ($0 \leq x \leq 0.125$, Ref.~\onlinecite{Klingeler2010}) and BaFe$_{1.83}$Co$_{0.18}$As$_2$,\cite{Wang2008c} at high temperatures $\chi$ increases with temperature $T$ above $\sim 200$~K up to the highest temperature measured of 700~K\@.  This temperature dependence has been observed for all of the 1111 and 122 compounds in the paramagnetic state at high temperatures, as described in the review by Zhang and coworkers and illustrated in Fig.~\ref{FigChi(T)}.\cite{GMZhang2008}  A similar increase in $\chi$ with $T$ was previously observed in the layered cuprate parent compounds in which a local moment description for the Cu$^{+2}$ spins 1/2 was conclusively proved.\cite{Johnston1997}  This positive temperature coefficient of $\chi$ up to the maximum measurement temperature of $\sim 1000$~K is due to the fact that the measurement temperature range was on the low-$T$ side of the broad maximum that hypothetically occurs at $\sim 1500$~K due to short-range two-dimensional antiferromagnetic ordering of the Cu spins.  

The observed increase of $\chi$ with $T$ in the FeAs-based compounds has been attributed to itinerant electron antiferromagnetic spin fluctuations,\cite{GMZhang2008} or alternatively to AF correlations between local moments in a strong coupling description.\cite{Laad2009}  In general, one would expect AF correlations of any sort to depress $\chi$, because those correlations resist the alignment of the electron spins with the applied field.  As temperature increases, the AF correlations are expected to decrease, which can cause the observed $\chi$ to increase.  The question then is the detailed origin of the apparent \emph{linear} temperature dependence, which in the inset of the top panel of  Fig.~\ref{FigChiRhoBaFe2As2} appears for BaFe$_2$As$_2$ from $T_{\rm s,N} = 138$~K all the way up to the maximum measurement temperature of 700~K\@.  

An increasing and approximately linear dependence of the conduction electron Pauli spin susceptibility $\chi^{\rm Pauli}$ on $T$ was predicted for the FeAs compounds by Korshunov et al.\ in a two-dimensional Fermi liquid picture.\cite{Korshunov2009}  The slope is proportional to the square of the SDW amplitude connecting the nested hole and electron pockets.  The authors claimed quantitative agreement with experiments.\cite{Korshunov2009}

\begin{figure}
\includegraphics [width=2.5in]{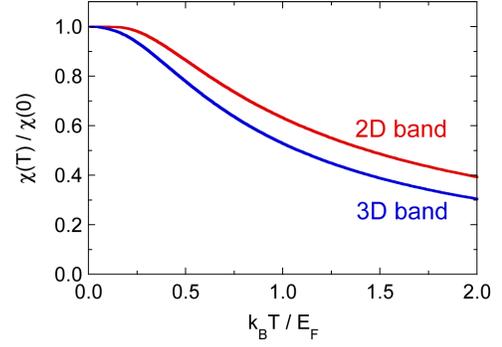}
\caption{(Color online) Magnetic spin susceptibility $\chi$ versus temperature $T$ for quasi-free-electron Fermi gases occupying a single parabolic two-dimensional (2D) or 3D electron band.  The Fermi energy $E_{\rm F}$ is the chemical potential at $T = 0$ and $k_{\rm B}$ is Boltzmann's constant.  At zero temperature, $\chi(0) = \mu_{\rm B}^2 N(E_{\rm F})$ in each case, where $N(E_{\rm F})$ is the respective density of electron states at $E_{\rm F}$ for both spin directions and $\mu_{\rm B}$ is the Bohr magneton.  The $\chi$ monotonically decreases with increasing temperature in both cases.  This behavior is qualitatively different from that observed for the FeAs-based layered compounds as illustrated in Fig.~\ref{FigChi(T)}, where $\chi$ \emph{increases} instead of decreases with increasing $T$ in the paramagnetic state.}
\label{Figchi_2D_3D} 
\end{figure}

An alternate proposal for the increase in $\chi$ with increasing $T$ was given by Rullier-Albenque et al.\cite{Rullier-Albenque2009}  They suggested that this results simply from strong thermal excitations of the electrons in the electron bands at the X~point because the Fermi energy is extremely small there ($\sim 25$--50~meV) as deduced from ARPES measurments.\cite{Yi2009}  For a two-dimensional (2D) quasi-free-electron gas in a parabolic band with energy $E$ versus wavevector $k$ given by $E = \hbar^2 k^2/(2 m^*)$, one straightforwardly obtain the following exact expression valid for all temperatures\cite{Johnston2010}
\be
\chi(T) = \mu_{\rm B}^2 N(E_{\rm F})\left[1 - e^{-E_{\rm F}/(k_{\rm B}T)}\right],
\label{Eqchi2D}
\ee
where the Fermi energy $E_{\rm F}$ is the chemical potential at $T = 0$, $\mu_{\rm B}$ is the Bohr magneton, and $N(E_{\rm F})$ is the density of electronic states at $E_{\rm F}$ for both spin directions.  A plot of Eq.~(\ref{Eqchi2D}) in Fig.~\ref{Figchi_2D_3D} shows that the susceptibility of a 2D quasi-free-electron gas \emph{decreases monotonically} with increasing $T$.  The same qualitative behavior as in Eq.~(\ref{Eqchi2D}) is found for a 3D parabolic band, as also shown in Fig.~\ref{Figchi_2D_3D}, where in this case we have calculated $\chi(T)$ numerically.  These behaviors of $\chi(T)$ for 2D and 3D quasi-free-electron gases both disagree with the observed behaviors in Fig.~\ref{FigChi(T)} where the susceptibilities for all compounds \emph{increase} with increasing temperature in the high-temperature paramagnetic (magnetically disordered) state instead of decrease.

On the other hand, Sales and coworkers calculated that for a three-dimensional (3D) two-band semimetallic band structure like that shown in Fig.~\ref{semimetal_BS}, one could obtain a $\chi(T)$ that increased with increasing temperature assuming parabolic bands.\cite{Sales2009}  The parameters of the model were also used to estimate the temperature dependent resistivity and Seeback coefficient, in reasonable agreement with the experimental data for the BaFe$_{2-x}$Co$_x$As$_2$ system.  For the susceptibility fit, the fitted overlap of the valence and conduction bands was 250~K, and the effective masses of the electron and hole bands were 17 and 30 times the free electron mass, respectively.  These latter values are much larger than the values of order 2--4 electron masses inferred from ARPES and quantum oscillation experiments and from band structure calculations.  They also found that the increase in the susceptibility with increasing temperature does not occur for a 2D electron gas semimetal with parabolic bands, so it is not clear that their calculations apply to the Fe-based compounds.

We found that if the conventional LDA band structure of ${\rm BaFe_2As_2}$ is populated by electrons versus $T$ according to the Fermi-Dirac distribution function, the $\chi$ slightly decreases by about 1.3\% with increasing $T$ between 200 and 400~K,\cite{Johnston2010} instead of strongly increasing as observed in Fig.~\ref{FigChiRhoBaFe2As2}.

Using LDA + DMFT (local density approximation combined with dynamical mean field theory) calculations, Skornyakov, Katanin and Anisimov predicted that the $\chi$ of LaFeAsO increases approximately linearly from 400 to 1000~K before bending over and becoming Curie-Weiss-like.\cite{Skornyakov2010a}  This is in qualitative agreement with the experimental data from 150--500~K,\cite{Klingeler2010} but the predicted slope is about a factor of four too small compared with the experimental data.  Interestingly, the predicted temperature dependence arises mainly because the single-particle band structure itself is temperature-dependent, rather than arising from antiferromagnetic correlations.\cite{Skornyakov2010a}

\subsubsection*{b. 11-type Fe$_{1+y}$Te$_{1-x}$Se$_x$ Compounds}

%\squeezetable
\begin{table*}
\caption{\label{Fe1+xTeCurieWeissDat} Curie constants $C_{\rm Curie}$ and Weiss temperatures $\theta$ from Curie-Weiss law fits, Eq.~(\ref{EqCurieWeissLaw}), to magnetic susceptibility data for Fe$_{1+y}$Te$_{1-x}$Se$_x$ single crystals.\cite{Yang2009a}  The first set of $C_{\rm Curie}$ and $\theta$ parameters are for all Fe atoms in the samples.  The second set of parameters is presumed to arise from the excess $y$ Fe(II) atoms.  There was no low-$T$ upturn in the data for $x = 0.45$ and 0.48 [see Fig.~\ref{YangFig2bc}(b)], so no fitting parameters are included here for those compositions.  The fitting ranges to determine the listed $\chi_0$ values were 45--300~K ($x = 0.33$) to 100--300~K ($x = 0$).  The $\chi_0$ values and respective fitting temperature ranges were provided by J. H. Yang (private communication). The data from Ref.~\onlinecite{Yang2009a} (all columns except the last one) are reproduced by permission of Ref.~\onlinecite{Yang2009a}.  }
\begin{ruledtabular}
\begin{tabular}{ccc|ccccccc}
Fe & Te & Se  & $T$ range & $\chi_0 $& $C_{\rm Curie}$ & $\theta$ & $C_{\rm Curie(II)}$ & $\theta_{\rm II}$ & $C_{\rm Curie(II)}/y$\\
 &&& (K) & (10$^{-3}$~cm$^{3}$/mol) & (cm$^{3}$\,K/mol) & (K)  & (cm$^{3}$\,K/mol) & (K) & (cm$^{3}$\,K/mol)\\ \hline
1.12 & 1 & 0 & 100--300 & $-0.97$ & 2.24 & 319 &  --- & --- & --- \\
1.00 & 0.95 & 0.05 & 100--300 & $-0.50$ & 1.6 & 260 & --- & --- & ---\\
1.01 & 0.88 & 0.12  & 20--50 & 1.05 & --- & --- & 0.10 & 52 & 10\\
1.07 & 0.80 & 0.20 & 20--50 & 1.29 & --- & --- & 0.02 & 5 & 0.3\\
1.12 & 0.70 & 0.30 & 20--50 & 1.15 & --- & --- & 0.12 & 24 & 1.0\\ 
1.04 & 0.67 & 0.33 & 20--50 & 1.05 & --- & --- & 0.05 & 23 & 1.3\\ 
\end{tabular}
\end{ruledtabular}
\end{table*}

\begin{figure}
\includegraphics [width=3in]{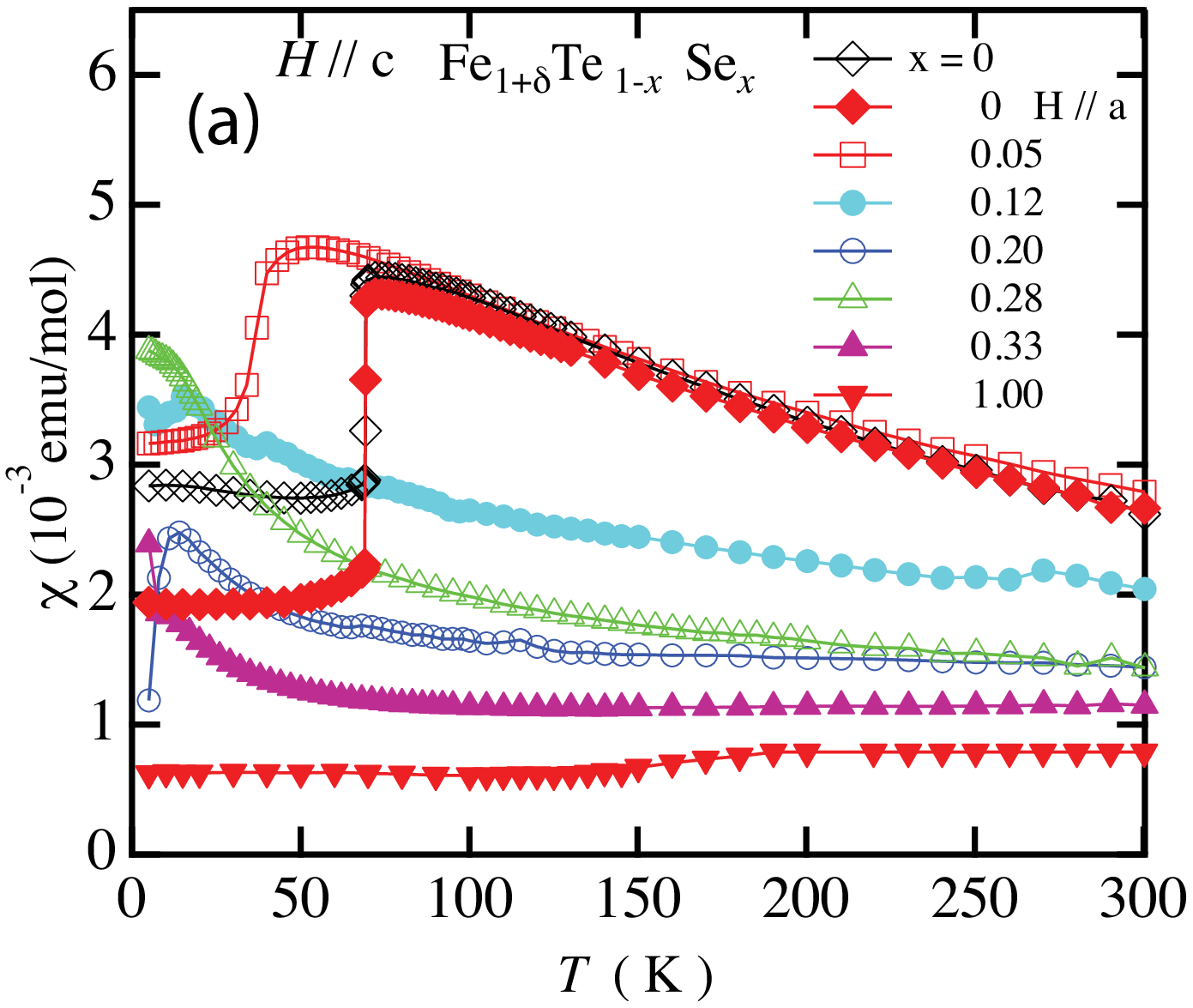}
\includegraphics [width=3.2in]{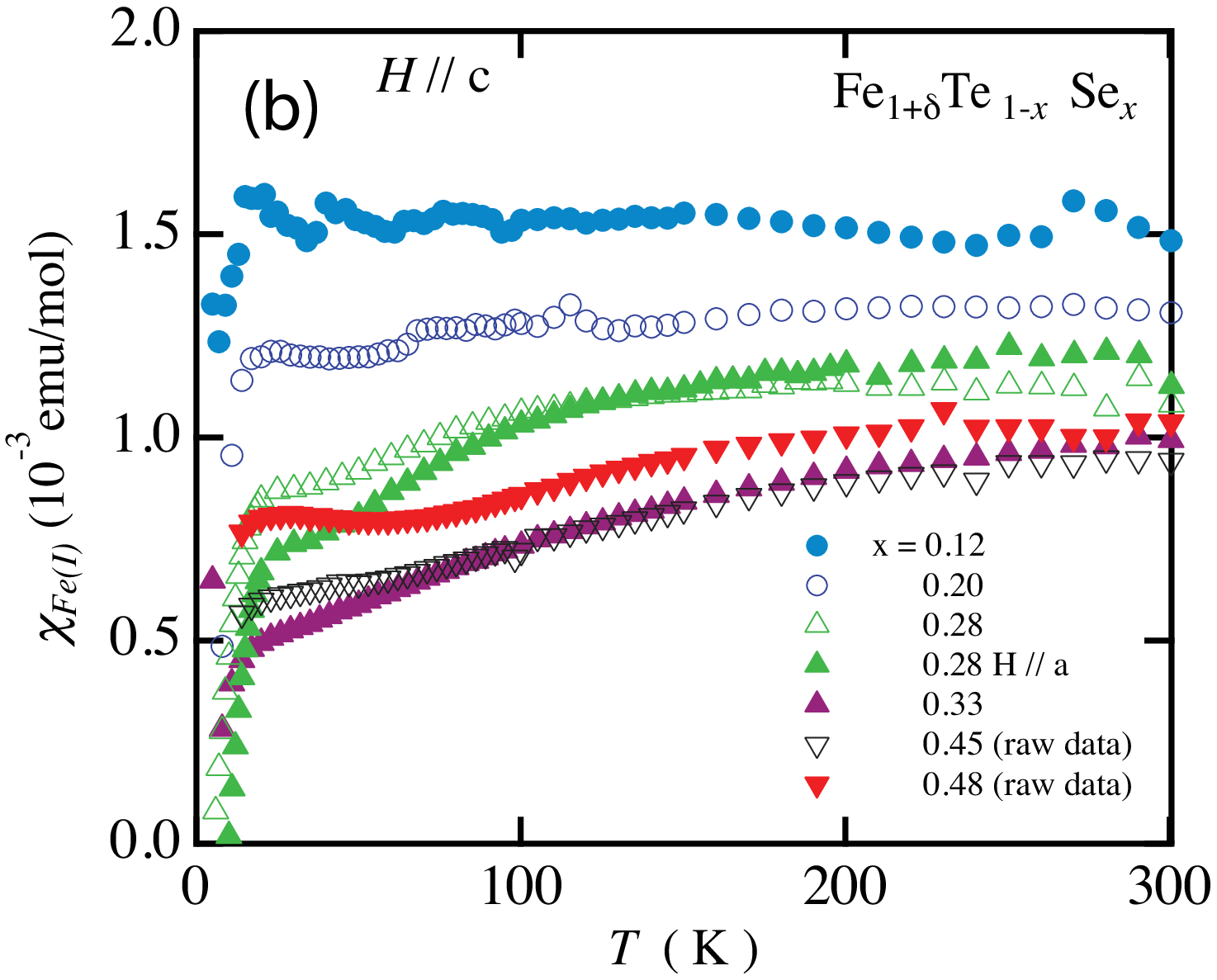}
\caption{(Color online) (a) Magnetic susceptibility $\chi$ (after correction for ferromagnetic impurities) versus temperature $T$ for single crystals of Fe$_{1+y}$Te$_{1-x}$Se$_x$.  Most data are for $H\parallel c$ but data for $H\parallel a$ are shown for $x = 0$.  (b) Contribution $\chi_{\rm Fe(I)}$ of the Fe atoms in the Fe square lattice layers, obtained by correcting the data in (a) for the contribution of the excess Fe(II) atoms in the Te/Se layers.  Most data are for $H\parallel c$ but data for $H\parallel a$ are shown for $x = 0.28$.  Reprinted with permission from Ref.~\onlinecite{Yang2009a}. }
\label{YangFig2bc} 
\end{figure}

The reported magnetic susceptibilities $\chi$ of Fe$_{1+y}$Te$_{1-x}$Se$_x$-type compounds generally show either an increase or decrease with decreasing temperature above $T_{\rm N}$, depending on the composition.  The absolute values tend to be very large compared to the data in Fig.~\ref{FigChi(T)} for the 122- and 1111-type compounds.  A major problem is that the samples often contain relatively large amounts of ferromagnetic impurities such as Fe metal as discussed above in Sec.~\ref{SecFeSeTe}.  As with most magnetic susceptibility studies of the Fe-based superconductors, very few studies of the Fe$_{1+y}$Te$_{1-x}$Se$_x$-type compounds report that they check for, and correct for if appropriate, the presence of ferromagnetic impurities.  The study below by Yang et al.\ is a notable exception.\cite{Yang2009a}

Yang and coworkers reported a definitive $\chi(T)$ study of the Fe$_{1+y}$Te$_{1-x}$Se$_x$ system.\cite{Yang2009a}  First, they carried out the measurements on single crystals with $0 \leq x \leq 1$.  Second, they determined the compositions of their crystals from energy-dispersive x-ray analysis as shown in Table~\ref{Fe1+xTeCurieWeissDat}, rather than reporting nominal compositions.  Third, they measured the susceptibility anisotropy for two representative compositions $x = 0$ and $x = 0.28$.  Fourth, they carried out $M(H)|_T$ isotherm measurements to obtain the contributions of ferromagnetic impurities to the measured magnetizations, and then corrected the observed $M(T)|_H$ data for these ferromagnetic impurity contributions to obtain $\chi(T)$ as described above in Sec.~\ref{SecFMImpurities}.  For example, the saturation magnetization $M_{\rm S}$ of the ferromagnetic impurites was found to be nearly independent of $T$ up to 200~K for Fe$_{1.12}$Te, with the value $M_{\rm S}\approx 15~{\rm G~cm^3/mol}$, which according to the conversion expression in Eq.~(\ref{mBGcm3Convert}) is equivalent to the contribution of 0.12~mol\% of ferromagnetic Fe metal impurities with an ordered moment of 2.2~$\mu_{\rm B}$/Fe atom.  They derived the $\chi(T)$ of Fe$_{1.12}$Te from the $M(T)|_H$ data at high fields of $H = 3$--5~T, for which $M_{\rm S}\lesssim 0.1 M$, thus allowing a reliable correction for $M_{\rm S}$.

The $\chi(T)$ data for the Fe$_{1+y}$Te$_{1-x}$Se$_x$ system obtained by Yang et al., after correction for ferromagnetic impurities, are shown in Fig.~\ref{YangFig2bc}(a).\cite{Yang2009a}  Most of the data are for $H\parallel c$ but data for $H\parallel a$ are also shown for $x = 0$.  For Fe$_{1.12}$Te ($x = 0$) a sharp first-order magnetic/crystallographic transition is seen at $T_{\rm N} = 69$~K\@.  This transition also appears as a sharp spike in the heat capacity of a Fe$_{1.05}$Te crystal at $T_{\rm N} = 65$~K,\cite{Chen2009a} and as a discontinuous and hysteretic resistivity change in Fig.~\ref{Fe1.05Te_rho} above.  Above $T_{\rm N}$, the $\chi(T)$ is nearly isotropic, as was also observed by Chen et al.\ for Fe$_{1.05}$Te.\cite{Chen2009a}  Below $T_{\rm N}$, the $\chi(T)$ becomes nearly independent of $T$, exhibiting an anisotropy with $\chi_a < \chi_c$,  again as also observed by Chen et al.\ for Fe$_{1.05}$Te,\cite{Chen2009a} suggesting that the major component of the ordered moment lies in the $ab$-plane, consistent with magnetic neutron diffraction measurements of the antiferromagnetic structures discussed below in Sec.~\ref{Sec11-type}.  For $x = 0.05$, the transition is broader, suggesting a possible second-order transition, and decreases in temperature to $T_{\rm N} \approx 40$~K\@.  The magnitude of $\chi(T)$ of the Fe$_{1.05}$Te crystal reported by Chen et al.\cite{Chen2009a} is roughly a factor of two larger than for any of the samples in Fig.~\ref{YangFig2bc}(a), evidently reflecting the presence of ferromagnetic impurities in the crystal that were not accounted and corrected for (N. L. Wang, private communication).

In Fig.~\ref{YangFig2bc}(a), the normal state $\chi(T)$ data for $x = 0$--0.12 have the general appearance of a single Curie-Weiss term.  Therefore Yang et al.\ fitted the data for $x = 0$ and~0.05 over the temperature range 100--300~K by a constant plus the Curie-Weiss law
\be
\chi = \chi_0 + \frac{C_{\rm Curie}}{T + \theta}
\label{EqCurieWeissLaw}
\ee
where $\chi_0$ is a temperature independent term arising from orbital (diamagnetic core and paramagnetic Van Vleck) contributions, plus perhaps $T$-independent diamagnetic Landau orbital and paramagnetic Pauli spin susceptibilities from conduction electrons, the Curie constant is
\be
C_{\rm Curie} = \frac{N_{\rm A}g^2\mu_{\rm B}^2S(S+1)}{3k_{\rm B}},
\ee
$N_{\rm A}$ is Avogadro's number, $g$ is the spectroscopic splitting factor ($g$-factor), $\mu_{\rm B}$ is the Bohr magneton, and $k_{\rm B}$ is Boltzmann's constant.  The fitted $\chi_0$ values, Curie constants and Weiss temperatures $\theta$ are listed in Table~\ref{Fe1+xTeCurieWeissDat}.\cite{Yang2009a}  These Curie-Weiss fits for these two crystals represent  the average paramagnetic behavior of all Fe atoms in the crystals.  

For $0.20 \leq x \leq 0.33$, it appears that there is a low-$T$ upturn riding on a more or less constant or weakly sloping background in Fig.~\ref{YangFig2bc}(a).  The authors attributed this upturn to the $y$ excess Fe(II) atoms per formula unit in the structure, fitted the data for $x = 0.12$ to~0.33 by Eq.~(\ref{EqCurieWeissLaw}) over the low-temperature range 20--50~K, and obtained the $\chi_0$, $C_{\rm Curie(II)}$ and $\theta_{\rm II}$ fitting parameters given in Table~\ref{Fe1+xTeCurieWeissDat}, where the Curie-Weiss term reflects the magnetic behavior of only the \emph{excess} Fe(II) atoms.  This analysis is supported by NMR shift measurements in Sec.~\ref{SecChiSpinNMR} below for the same system, which do not show a Curie-Weiss upturn, indicating that the spins giving rise to it in the susceptibility data are indeed dilute.  For $x = 0.45$ and~0.48, there was no sign of a Curie-Weiss type behavior over any $T$ range [see Fig.~\ref{YangFig2bc}(b)] so no Curie-Weiss-type fit to the data for these two crystals was done.  As discussed in Sec.~\ref{SecFeSeTe} above, the value of the excess Fe concentration $y$ of the PbO-type FeAs phase becomes essentially zero for $x = 1$, suggesting that the lack of a Curie-Weiss upturn at low temperatures for the $x = 0.45$ and~0.48 crystals is due to a similar negligible Fe excess.\cite{Yang2009a}

For $x = 0$ and~0.05, the Weiss temperatures $\theta$ in Table~\ref{Fe1+xTeCurieWeissDat} are large, with $\theta/T_{\rm N} \approx 4.6$ and~6.5 for $x = 0$ and~0.05, respectively, which in mean field theory for a local moment system would instead be unity.  These large ratios suggest that the excess Fe atoms and/or fluctuation effects associated with the low dimensionality of the Fe square lattice may be depressing $T_{\rm N}$.

Another important result from the fit for $x=0$ is the large (average) value of the Curie constant $C_{\rm Curie} = 2.24~{\rm cm^3~K/mol}$ for \emph{all} Fe atoms in the crystal, which is between the values $C_{\rm Curie} = 1.88$ and $3.00~{\rm cm^3~K/mol}$ expected for localized iron spins $S = 3/2$ and $S = 2$ with $g = 2$, respectively.  This correspondence suggests that the bulk Fe spins carry a \emph{local magnetic moment}, contrary to the itinerant electron picture most appropriate for the 122- and 1111-type FeAs compounds.  The large values of the ordered moment for Fe$_{1+y}$Te in Table~\ref{FeOrdering} below are also suggestive of a local moment system.  For $S = 2$ with $g = 2$ one expects an ordered moment of $4~\mu_{\rm B}$/Fe which is larger than the largest observed value $\approx 2.5~\mu_{\rm B}$/Fe for Fe$_{1+y}$Te in Table~\ref{FeOrdering}.  This reduction may be due to interactions with the excess statistically distributed Fe(II) atoms, covalency effects, the itinerant nature of at least some of the otherwise localized Fe valence electrons, and/or low-dimensionality of the Fe(I) square spin lattice.

The value of $C_{\rm Curie}$ for $x = 0$ is much too large to arise from only the $y = 0.12$ excess Fe(II) atoms per formula unit. In fact, from the data one cannot resolve whether the excess Fe atoms carry a magnetic moment or not because the temperature dependence is dominated by the \emph{bulk} Fe spins. There is only one temperature dependence, instead of the sum of two clear dependences which would differentiate between the square lattice Fe(I) spins and the excess Fe(II) spins.  In particular, there is no Curie-Weiss upturn in $\chi$ below $T_{\rm N}$ in Fig.~\ref{YangFig2bc}, which suggests that if the excess Fe atoms do have local magnetic moments, they must participate in the long-range antiferromagnetic ordering of the bulk Fe spins in the Fe square lattice layers.  

The last column of Table~\ref{Fe1+xTeCurieWeissDat} gives the Curie constant per mole of \emph{excess} Fe atoms, $C_{\rm Curie(II)}/y$, which ranges from 0.3 to 10~${\rm cm^3~K/mol}$, with an average value of 3.2~${\rm cm^3~K/mol}$, close to the value of 3.00~${\rm cm^3~K/mol}$ expected for a mole of spins $S = 2$ with $g = 2$.  A theoretical estimate of the magnetic moment of the excess Fe is $2.4~\mu_{\rm B}$/Fe(II)~atom,\cite{Zhang2009a} corresponding to an average spin $\langle S\rangle = 1.2$, for which one expects $C_{\rm Curie(II)}/y = 1.3~{\rm cm^3~K/[mol}$~Fe(II)], which in turn is close to two of the values in the last column of Table~\ref{Fe1+xTeCurieWeissDat}.

Shown in Fig.~\ref{YangFig2bc}(b) is $\chi_{\rm Fe(I)}(T)$ for the Fe(I) atoms in the Fe square lattice layers, obtained by subtracting the inferred Curie-Weiss contribution of the Fe(II) atoms, using the parameters in Table~\ref{Fe1+xTeCurieWeissDat}, from the observed data in Fig.~\ref{YangFig2bc}(a).\cite{Yang2009a}  For $x = 0.45$ and~0.48 in Fig.~\ref{YangFig2bc}(b), no corrections were made to the raw data (except corrections for ferromagnetic impurities) because as noted above they showed no Curie-Weiss behaviors to begin with.   

As $x$ increases into the superconducting composition range $x \gtrsim 0.3$ (see Fig.~\ref{KatayamaFig2}), the $\chi_{\rm Fe(I)}(T)$ data are similar in magnitude and temperature dependence to those for the 122- and 1111-type compounds in Fig.~\ref{FigChi(T)} above that are itinerant antiferromagnetic materials.  \emph{This suggests a crossover from local moment to itinerant magnetism with increasing} $x$.  Indeed, the phase diagram for the Fe$_{1+y}$Te$_{1-x}$Se$_x$ system in Fig.~\ref{KatayamaFig2} shows a static short-range ordering ($\sim$ spin glass) region for $0.1 \lesssim x \lesssim 0.4$, which is characteristic of local moment systems.

\subsubsection{Relationship between Magnetic Susceptibility and $T_{\rm c}$}

Until the advent of the cuprate high $T_{\rm c}$ superconductors in 1986, superconductivity was never observed on the same sublattice of a structure on which a dense array of transition metal magnetic moments resided.  Our experience with the cuprates taught us that magnetism (in particular, local magnetic moments with \emph{antiferromagnetic} correlations between them) and superconductivity are not necessarily mutually exclusive.  Indeed, a leading contender for the superconducting mechanism in the high $T_{\rm c}$ cuprates is currently an electronic/magnetic mechanism.  Even so, prior to the discovery in 2008 of high $T_{\rm c}$ superconductivity in the FeAs-based materials, \emph{ferromagnetic} correlations within a material were not viewed as portending the occurrence of superconductivity.  Since iron metal is a ferromagnet and compounds showing strong paramagnetic enhancement of the static magnetic susceptibility $\chi$ tend not to become superconducting, the discovery of high $T_{\rm c}$ superconductivity in FeAs-based materials came to many as quite a surprise.  

\begin{figure}
\includegraphics [width=3.3in]{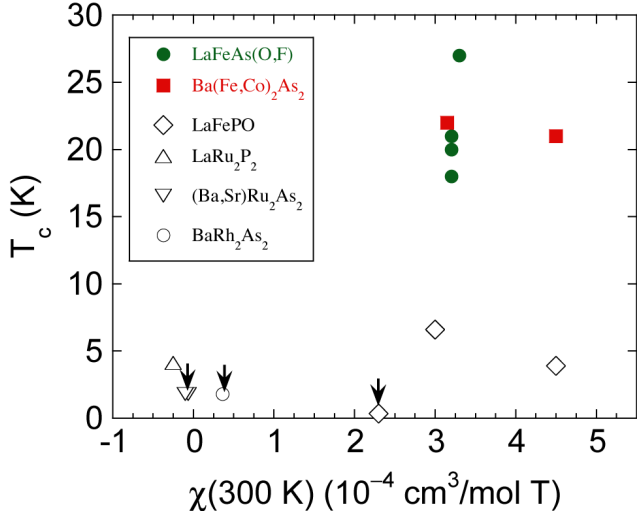}
\caption{\label{TcChi} (Color online) Superconducting transition temperature $T_{\rm c}$ versus powder-averaged magnetic susceptibility $\chi$ at 300~K for a variety of 122, 1111, and 111 pnictides that exhibit no structural or magnetic transitions below 300~K\@.  The vertical arrows pointing downwards indicate that superconductivity is not observed above the indicated temperature for the respective compound.  The colored filled symbols are for compounds containing FeAs layers.  Multiple data points for the same compound indicate different values of $T_{\rm c}$ and/or $\chi$(300~K) reported for the compound.  The $\chi$(300~K) data are normalized per mole of transition metal $T$ atoms.  See the Appendix for the references  and Table~\ref{FeTeSeData} for data for the Fe$_{1+y}$Te$_{1-x}$Se$_x$ system.  The data for this system are off-scale on the right-hand side.}
\end{figure}

Following up on this idea, shown in Fig.~\ref{TcChi} is a plot of $T_{\rm c}$ versus the powder-averaged susceptibility $\chi(300$~K) of various Fe-based compounds.  From Fig.~\ref{FigChiRhoBaFe2As2}, the susceptibility of single crystals is usually anisotropic with $\chi$ measured with the magnetic field $H$ along the $c$-axis smaller than when it is in the $a$-$b$ plane.  This anisotropy is likely mostly due to anisotropy in the orbital susceptibility rather than in the spin susceptibility.  We chose 300~K for a comparison temperature for $\chi$ in order to reduce impurity effects on $\chi$ that tend to become more pronounced with decreasing temperature.  The data points plotted are for samples that show no structural or magnetic transitions below room temperature.  Figure~\ref{TcChi} shows that all of the FeAs-based materials plotted have large susceptibilities, irrespective of their $T_{\rm c}$'s.  We therefore hypothesize that a large $\chi$ is a necessary but not sufficient condition to obtain high $T_{\rm c}$ in the Fe-based and related superconductors.

To investigate this correlation further, the magnetic susceptibility $\chi$ of a metal not containing local magnetic moments can be written
\begin{equation}
\chi = \chi_{\rm core} + \chi_{\rm VV} + \chi_{\rm L} + \chi_{\rm P},
\label{EqChiSum}
\end{equation}
where the first three terms on the right-hand-side are orbital susceptiblities and the last term is the Pauli spin susceptibility of the conduction electrons that can be enhanced from many-body effects.  The first three terms are the diamagnetic atomic core susceptibility, the paramagnetic Van Vleck susceptibility of the electrons, both of which are normally independent of temperature, and the Landau diamagnetic susceptibility of the conduction electrons.  The core and Van Vleck terms tend to compensate each other since they have opposite signs.  The third term is small compared to the fourth in transition metal compounds and is usually ignored.  These contributions have not yet been quantitatively estimated for the Fe-based superconductor materials.  However, as an example, BaRh$_2$As$_2$ with the ThCr$_2$Si$_2$ structure has values estimated to be $\chi_{\rm core} = -0.303 \times 10^{-4}$~cm$^3$/mol and $\chi_{\rm VV} = 1.43 \times 10^{-4}$~cm$^3$/mol, for a net value  $\chi_{\rm core} + \chi_{\rm VV}  = 1.13 \times 10^{-4}$~cm$^3$/mol.\cite{singh2008}

\begin{figure}
\includegraphics [width=3.3in]{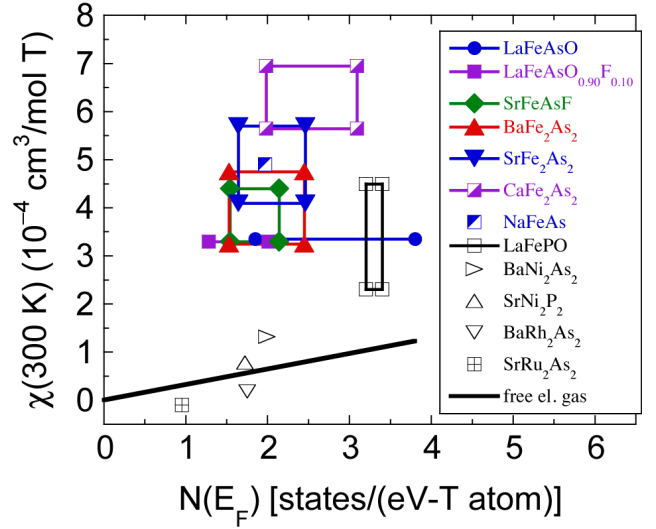}
\caption{(Color online) Magnetic susceptibility at 300~K $\chi(300~$K) versus the bare nonmagnetic band structure density of states for both spin directions $N(E_{\rm F})$ for a variety of 122, 1111, and 111 transition metal $T$ pnictides.  Multiple data points for the same compound indicate the range(s) of $\chi(300~$K) and/or $N(E_{\rm F})$ reported for the compound.  The colored symbols are for compounds containing FeAs layers.  The data for all the FeAs-based compounds are significantly above the sloping straight line shown that is expected from the degenerate quasi-free-electron gas model.  See the Appendix for the references  and Table~\ref{FeTeSeData} for data for the Fe$_{1+y}$Te$_{1-x}$Se$_x$ system.  The data for this system are vertically off-scale.}
\label{ChiN(EF)} 
\end{figure}

\subsubsection{\label{SecChiSpin} Pauli Spin Susceptibility}

The Pauli spin susceptibility $\chi_{\rm P}$ in Eq.~(\ref{EqChiSum}) can be enhanced from the ``bare'' value by many-body effects.  The bare value is defined to be the value calculated from the bare band structure density of states $N(E_{\rm F})$ according to\cite{kittel1966}
\begin{equation}
\chi_{\rm P} = \mu_{\rm B}^2 N(E_{F}) = \left(3.233 \times 10^{-5}\,\frac{\rm cm^3}{\rm mol}\right) N(E_{F}),
\label{EqChiP}
\end{equation}
where $\mu_{\rm B}$ is the Bohr magneton and the density of states at the Fermi energy $N(E_{F})$ on the far right-hand-side is expressed in states/(eV f.u.) for both spin directions.  As noted above, a distinguishing feature of the high $T_{\rm c}$ FeAs-based materials is their large $\chi$ values that evidently reflect signficant enhancement of the conduction electron spin susceptibility.\cite{djsingh2008, Kohama2008, Haule2009}  To broadly examine this feature, we show in Fig.~\ref{ChiN(EF)} a plot of $\chi(300$~K) versus $N(E_{\rm F}$) for a variety of FeAs-based and related materials.  It is important for the data for the FeAs-based parent compounds in the figure that the $\chi$ is quoted at 300~K, above the structural/antiferromagnetic transition temperatures at 200~K and below, because all of the $N(E_{\rm F})$ values plotted are calculated for the nonmagnetic (i.e., not long-range magnetically ordered) tetragonal states of the parent compounds.  The sloping straight line is a plot of Eq.~(\ref{EqChiP}) for the degenerate nearly-free-electron gas.  A striking result obvious from Fig.~\ref{ChiN(EF)} is that \emph{all} of the FeAs-based compounds have $\chi(300$~K) values significantly above the sloping line, suggesting enhancements of the spin susceptibility by factors of order five.  Such large enhancements often indicate nearness to itinerant ferromagnetism.  Indeed, in an early paper, Singh and Du noticed the strong enhancement of the susceptibility for LaFeAsO, theoretically found a marginally stable itinerant ferromagnetism ground state, and further stated, ``Here we show that LaOFeAs is in fact close to magnetism, with competing ferromagnetic and antiferromagetic fluctuations, with the balance controlled by doping.''\cite{djsingh2008b}  For precise quantitative estimates of the enhancement factor, one would need to take into account the orbital susceptibilities in Eq.~(\ref{EqChiSum}) and the temperature dependence of $\chi_{\rm P}$.  

Independent information about enhancement of $\chi^{\rm Pauli}$ was found from inelastic neutron scattering studies of ${\rm Ba(Fe_{0.92}Co_{0.08})_2As_2}$ crystals by Parshall and coworkers.\cite{Parshall2009}  They found that the SDW fluctuation column at orthorhombic wave vector (1,0,$L$)~r.l.u.\ extended up to high temperatures, with a wave vector width that was far smaller than expected from band structure calculations.  They rationalized this result in terms of exchange enhancement of the spin susceptibility at that wave vector.  One could speculate that this enhancement might also be present to some degree with respect to the the zero frequency, zero wave vector Pauli spin susceptibility.  The authors also suggested from their results that the magnetic properties are closer to the itinerant electron limit than the local moment limit.\cite{Parshall2009}

The observations suggest that both high $T_{\rm c}$ and enhancement of the Pauli spin susceptibility occur, or don't occur, together.  One can therefore speculate that the electronic interactions that cause the enhancement of the susceptibility and the interactions that are responsible for the high $T_{\rm c}$ of the FeAs-based materials are at the least closely related if not the same.  Thus a significant enhancement of the spin susceptibility may be a divining rod for an electronic mechanism for superconductivity in this class of materials.  

It appears that the normal state electronic heat capacity coefficient $\gamma_{\rm n}$ of the 35~K superconductor Ba$_{0.6}$K$_{0.4}$Fe$_{2}$As$_{2}$ is enhanced by roughly the same factor of 5--6 above band structure calculations (see Sec.~\ref{Sec_SC_Cp} below), suggesting that the enhancement of both $\chi^{\rm Pauli}$ and $\gamma_{\rm n}$ arise from the same many-body interactions, and that those are critically important to high $T_{\rm c}$.  The precise nature of these many-body interactions has not yet been identified, but we speculate that they are somehow associated with antiferomagnetic spin fluctuations as suggested from inelastic neutron scattering and NMR results below.  Antiferromagnetic spin fluctuations peak in intensity at a nonzero wave vector, but the enhancement might have a ``tail'' that increases the static uniform spin susceptibility at zero wave vector.  Recent theory that predicts a $T_{\rm c}$ of the right magnitude in the Fe-based materials explicitly includes antiferromagnetic spin fluctuations in an electronic $s^\pm$ pairing  model,\cite{JZhang2009} as will be discussed in more detail in Sec.~\ref{Sec_SC_Mech}.  However, as suggested in the following section, this tail is too small to account for the large measured susceptibilty values.  Alternatively, or in addition, Hund's rule on-site ferromagnetic coupling (Stoner enhancement) could play a role in enhancing the static uniform spin susceptibility.   

\subsubsection{\label{SecChiSpinNeuts} Magnetic Spin Susceptibility from Inelastic Magnetic Neutron Scattering Measurements}

It is of interest to see what the static susceptibility at the antiferromagnetic wave vector ${\bf Q}_{\rm AF}$ determined from neutron scattering measurements predicts for the uniform spin susceptibility $\chi^{\rm Pauli}$.  In a fit using theory for a nearly antiferromagnetic Fermi liquid, we reproduce here Eq.~(\ref{EqChippNAFFL}) from later in the review,
\[
\chi^\prime({\bf Q},0) = \frac{C}{T + \theta + \xi_0^2 |{\bf Q} - {\bf Q}_{\rm AF}|^2}.\nonumber
\]
Thus the static uniform susceptibility is
\be
\chi^{\rm Pauli} = \chi^\prime(0,0) = \frac{C}{T + \theta + \xi_0^2 |{\bf Q}_{\rm AF}|^2}.\nonumber
\ee
The parameters obtained from the fits to the neutron scattering data for superconducting Ba(Fe$_{0.925}$Co$_{0.075}$)$_{2}$As$_{2}$ were $C = 1.2~{\rm cm^3~K/mol}$, $\theta = 30$~K, and $\xi_0 = 160$~\AA~K$^{1/2}$.\cite{Inosov2010}  Here $Q_{\rm AF} = 2\pi/a_{\rm O} = 1.13$~\AA$^{-1}$, where the basal plane lattice parameter in orthorhombic notation is $a_{\rm O} = \sqrt{2}\,a_{\rm T}$ and the tetragonal lattice parameter from Table~\ref{data5} in the Appendix is $a_{\rm T} = 3.928$~\AA.  These values give an essentially temperature independent value for the susceptibility of
\be
\chi^{\rm Pauli} = \frac{1.2~{\rm cm^3~K/mol}}{33\,000~{\rm K}} = 3.6\times 10^{-5}~{\rm cm^3/mol},
\ee
which is about a factor of four too small (see Fig.~\ref{NingFig4}), but is of the right order of magnitude.  The above extrapolation from ${\bf Q}_{\rm AF}$ to {\bf Q} = 0 is not expected to be very accurate.  The lack of a significant predicted temperature dependence may arise from the assumption that the antiferromagnetic correlation length follows the mean field behavior $\xi \sim (T + \theta)^{-1/2}$ (see Sec.~\ref{SecChippNAFL} below).

An estimate of the static uniform spin susceptibility from neutron scattering data was made for ${\rm CaFe_2As_2}$ in the paramagnetic state by Diallo et al.\cite{Diallo2010}  From their Eq.~(\ref{EqChiFromNeuts}) below and the parameter values obtained from their fits to their data, they obtained the nearly temperature-independent value 
\[
\chi_{\rm spin} = 2.39(2) \times 10^{-4}~{\rm cm^3/mol},
\]
which is in reasonable agreement, but perhaps somewhat too low, compared with the measured values for this compound in Fig.~\ref{FigChi(T)} and in the Appendix.

\subsubsection{\label{SecChiSpinNMR} NMR Knight Shift and Conduction Electron Spin Susceptibility}

\subsubsection*{a. Introduction}

A metal has a paramagnetic Pauli spin susceptibility $\chi_{\rm spin}$ of the conduction electrons as discussed above in Sec.~\ref{SecChiSpin}, so when a metal is placed in a magnetic field $H$ it develops a positive spin magnetization $M_{\rm spin} = \chi_{\rm spin} H$ in the direction of the field.  This $M_{\rm spin}$ is coupled to the nuclei via the contact hyperfine interaction, which results in an effective positive change $\Delta H$ of the magnetic field seen by a nucleus compared to the applied field $H$.   The normalized effective magnetic field shift is called the Knight shift or spin shift $K_{\rm spin}$, which follows\cite{Slichter1963, Abragam1961}
\be
K_{\rm spin} = \frac{\Delta H}{H} = \frac{8\pi}{3}\langle|u_{\bf k}(0)|^2\rangle_{E_{\rm F}}~\chi_{\rm spin}^{\rm total} \equiv A_{\rm hf}\chi_{\rm spin},
\label{KnightShift}
\ee
where $\langle|u_{\bf k}(0)|^2\rangle_{E_{\rm F}}$ is the expectation value around the Fermi surface of the periodic part $u_{\bf k}({\bf r = 0})$ of the electronic wave function at a single nuclear site at ${\bf r} = 0$ (each nucleus interacts with every conduction electron in the sample due to the delocalized nature of the conduction electrons), and $\chi_{\rm spin}^{\rm total}$ is the total spin susceptibility of the sample.  The quantity $A_{\rm hf}$ is called the hyperfine coupling constant of the nuclear spins to the conduction electron spins, and $\chi_{\rm spin}$ is a normalized spin susceptibility (e.g., per mole of formula units) that is further discussed below.  Due to the periodic boundary conditions on the conduction electron wave functions, one has the normalizaton $\langle|u_{\bf k}(0)|^2\rangle_{E_{\rm F}} \sim 1/V$ where $V$ is the volume of the sample, and one also has $\chi_{\rm spin}^{\rm total} \sim V$, so Eq.~(\ref{KnightShift}) is independent of $V$.  Note also that the units of $\langle|u_{\bf k}(0)|^2\rangle_{E_{\rm F}}$ are ${\rm 1/cm^3}$ due to the normalization of $u_{\bf k}({\bf r})$, and the units of $\chi_{\rm spin}^{\rm total}$ are ${\rm cm^3}$ (see Sec.~\ref{SecChiUnits} above), so the right-hand side of Eq.~(\ref{KnightShift}) is dimensionless, as required by the left-hand side.  

The total Pauli spin susceptibility of a metallic sample can be written
\be 
\chi_{\rm spin}^{\rm total} = \frac{g^2 \mu_{\rm B}^2}{4} N(E_{\rm F})_{\rm total} = \frac{(\gamma_{\rm e}\hbar)^2}{4} N(E_{\rm F})_{\rm total},
\ee
where $g$ is the $g$-factor of the conduction electrons, $\mu_{\rm B}$ is the Bohr magneton, $N(E_{\rm F})_{\rm total}$ is the density of states at the Fermi energy $E_{\rm F}$ for both spin directions for the entire sample (the Zeeman degeneracy is included) and $\gamma_{\rm e} = g\mu_{\rm B}/\hbar$ is the gyromagnetic ratio of the electron.  Then Eq.~(\ref{KnightShift}) can be rewritten as
\be
K_{\rm spin} = \frac{2\pi}{3}(\gamma_{\rm e}\hbar)^2\langle|u_{\bf k}(0)|^2\rangle_{E_{\rm F}}~N(E_{\rm F})_{\rm total} \equiv A_{\rm hf}\chi_{\rm spin}.
\label{KnightShift2}
\ee

Ambiguity currently exists about how to write the relationship between the actual observed $K_{\rm spin}$ and the actual observed $\chi_{\rm spin}$ of a material using Eq.~(\ref{KnightShift}).  It would be highly desirable to quote both $A_{\rm hf}$ and $\chi_{\rm spin}$ in dimensionless units, which would allow a unique value of the hyperfine coupling constant $A_{\rm hf}$ to be derived from a measured Knight shift and spin susceptibility.  In the case of $\chi_{\rm spin}$, this would correspond to inserting $\chi_{\rm spin}$ as the dimensionless volume susceptibility (discussed above in Sec.~\ref{SecChiUnits}) into Eq.~(\ref{KnightShift}).  However, the NMR community writes Eq.~(\ref{KnightShift}) in terms of a hyperfine coupling constant $A_{\rm hf}$ that is (almost) always quoted in units of Oe$/\mu_{\rm B}$.  This means that $\chi_{\rm spin}$ has to be expressed in units of $\mu_{\rm B}/$Oe, but it is not always clear whether this is per atom, or per a certain type of atom, or per a certain group of atoms, the choice being called a formula unit (f.u.) here.  Thus the value of $A_{\rm hf}$ depends on the definition of ``f.u.''

If $A_{\rm hf}$ in Eq.~(\ref{KnightShift}) is expressed in units of Oe$/\mu_{\rm B}$, the question then is how to express $\chi_{\rm spin}$ in units of $\mu_{\rm B}$/Oe if the measured susceptibility is in units of cm$^3$/(mol~f.u.).  We note that when $\chi$ is written in units of cm$^3$/mol, it is implicitly understood that ``mol'' refers to a mole of f.u.  Here we have to include f.u.\ explicitly.  In the present context the units we get for $\chi$ are $\mu_{\rm B}$/(Oe~f.u.), and we have to include ``f.u.''\ in order that the units of $A_{\rm hf}$ can be properly defined.  Using the conversion expression in Eq.~(\ref{mBGcm3Convert}) one obtains
\bea
\chi_{\rm spin}\left[{\rm \frac{\mu_{\rm B}}{Oe~f.u.}}\right]  
&=&\chi_{\rm spin}\left[{\rm \frac{G~cm^3}{Oe~mol~f.u.}}\right]\times\frac{N_{\rm A}\mu_{\rm B}}{\rm 5585\frac{G~cm^3}{mol}}\nonumber\\
&=& \chi_{\rm spin}\left[{\rm \frac{cm^3}{mol~f.u.}}\right]\times\frac{N_{\rm A}\mu_{\rm B}}{\rm 5585\frac{G~cm^3}{mol~}}\label{EqConvert},
\eea
where we have used 1~G = 1~Oe and where Avogadro's number (a conversion factor) is $N_{\rm A} = 6.022\times 10^{23}/$mol.  This shows that the commonly used units for $A_{\rm hf}$ are not expressed correctly.  If one is going to use units of Oe/$\mu_{\rm B}$ for $A_{\rm hf}$, one has to include ``f.u.''\ in the numerator.  Then we get from Eqs.~(\ref{KnightShift}) and~(\ref{EqConvert})
\be
K_{\rm spin} = A_{\rm hf}\left[{\rm \frac{Oe~f.u.}{\mu_{\rm B}}}\right] \chi_{\rm spin}\left[{\rm \frac{cm^3}{mol~f.u.}}\right] \times \frac{N_{\rm A}\mu_{\rm B}}{\rm 5585\frac{G~cm^3}{mol~}}.
\label{EqKChiCorrect}
\ee
Thus, the value of $A_{\rm hf}$ depends on the definition of ``f.u.''  

The NMR community usually writes Eq.~(\ref{EqKChiCorrect}) instead as
\begin{equation}
K_{\rm spin} = \frac{A_{\rm hf}}{N_{\rm A}\mu_{\rm B}}\chi_{\rm spin}.
\label{EqKChi}
\end{equation}
Clearly, the units are not consistent between the left- and right-hand sides of this equation.  If $A_{\rm hf}$ has units of Oe/$\mu_{\rm B}$ = Oe/(G~cm$^3$) = 1/cm$^3$, $N_{\rm A}$ has units of mol$^{-1}$, and $\chi_{\rm spin}$ has units of cm$^3$/(mol~f.u.), then the right-hand side of Eq.~(\ref{EqKChi}) has units of 1/(G~cm$^3$~f.u.), whereas the left-hand side is dimensionless.  Evidently, Eq.~(\ref{EqKChi}) is written by the NMR community as a shorthand for Eq.~(\ref{EqKChiCorrect}).  But in any case one should not  forget about the ``f.u.'' issue.  To avoid all these complications,  Eq.~(\ref{KnightShift}) could be routinely written and used in analyzing data as $K_{\rm spin} = A_{\rm hf}\chi_{\rm spin}$, where both $A_{\rm hf}$ and $\chi_{\rm spin}$ are dimensionless.

\subsubsection*{b. {\rm Ba(Fe}$_{1-x}${\rm Co}$_x)_2${\rm As}$_2$}

\begin{figure}
\includegraphics[width=3.3in,viewport=35 40 730 570,clip]{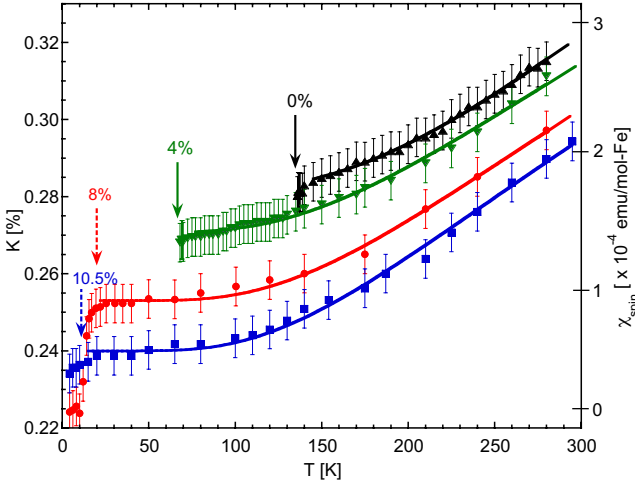}
\caption{(Color online) Left vertical axis: Temperature $T$ dependence of the $^{75}$As NMR Knight shift $K$ measured along the $c$-axis for four single crystals of Ba(Fe$_{1-x}$Co$_{x}$)$_{2}$As$_{2}$, where the percent doping listed is the value of $x$.\cite{Ning2009}  Data are only shown in the paramagnetic phase, above the tetragonal-to-orthorhombic transition temperature that decreases with increasing $x$.  The downturns at low $T$ arise from the onset of superconductivity.  The vertical solid and dashed arrows indicate spin density wave ($T_{\rm SDW}$) and superconducting ($T_{\rm c}$) transition temperatures, respectively.  The solid curves are fits by the expression (\ref{EqPG}) with pseudogap values $\Delta_{PG}/k_{B} = 711$~K ($x=0$), 570~K ($x=0.04$), 520~K ($x=0.08$), and 490~K ($x=0.105$), respectively.  Right vertical axis: Conduction electron Pauli spin susceptibility $\chi_{\rm spin}$ obtained from $K$ using Eq.~(\ref{EqKChi}).  Reprinted with permission from Ref.~\onlinecite{Ning2009}.  Copyright (2009) by the Physical Society of Japan.}
\label{NingFig4}
\end{figure}

Imai and coworkers\cite{Imai2008, Ning2009} have estimated the Pauli spin susceptibility of four single crystals of Ba(Fe$_{1-x}$Co$_{x}$)$_{2}$As$_{2}$ from a so-called $K$-$\chi$ analysis of the NMR shift $K$.  The spin part of the shift $K_{\rm spin}$ and the static uniform spin susceptibility $\chi_{\rm spin}$ [$\chi_{\rm P}$ in Eq.~(\ref{EqChiSum})] are related according to Eq.~(\ref{EqKChiCorrect}).  The total shift of the resonance is $K = K_{\rm spin} + K_{\rm chem}$, where $K_{\rm chem}$ is the ``chemical'' shift arising from the first three (orbital) susceptibility terms in Eq.~(\ref{EqChiSum}).  For their analysis of their $c$-axis $^{75}$As NMR measurements on BaFe$_{2-x}$Co$_x$As$_2$ crystals, they used the value of $^{75}A_{\rm hf}^c$ measured by Kitagawa \emph{et al.}\cite{Kitagawa2008} on a single crystal of BaFe$_{2}$As$_2$ above $T_{\rm N} = 135$~K\@.  In order to determine $K_{\rm chem}$, they assumed that $K_{\rm spin}$ goes to zero for $T \to 0$ in the superconducting state of a sample with $x = 0.08$, as in Fig.~\ref{FigLaFeAsO_NMR_Fe} below.  Thus they obtained a $\chi_{\rm spin}$ that increased monotonically with increasing $T$ in Fig.~\ref{NingFig4}, similar to the data in the top panel of Fig.~\ref{FigChiRhoBaFe2As2} and in Fig.~\ref{FigChi(T)}, with a value at 300~K of $2.9 \times 10^{-4}$~cm$^3$/mol Fe for $x = 0$, decreasing to  $2.1 \times 10^{-4}$~cm$^3$/mol Fe for $x = 0.11$.\cite{Ning2009}  The value for $x = 0$ is close to the error box for the total susceptibility for BaFe$_{2}$As$_2$ in Fig.~\ref{ChiN(EF)}, suggesting that the net orbital susceptibility from the first three terms of Eq.~(\ref{EqChiSum}) is relatively small.  However, it is not obvious that the value of $A_{\rm hf}$ for $x = 0$ is the same as for the doped compounds with $x > 0$.  A similar analysis of $ab$-plane As Knight shift data for LaFeAsO$_{0.9}$F$_{0.1}$ gave $\chi_{\rm spin} = 1.8 \times 10^{-4}$~cm$^3$/mol Fe at 300~K,\cite{Imai2008} significantly smaller than the total susceptibility for this compound in Fig.~\ref{ChiN(EF)}, suggesting the presence of a significant positive net orbital susceptibility.  

The authors of Ref.~\onlinecite{Ning2009} suggested that the increase in $\chi_{\rm spin}$ with $T$ in Fig.~\ref{NingFig4} is due to the presence of a spin pseudogap $\Delta_{\rm PG}$.  Thus they fitted their Knight shift data by 
\be
K = A + B\exp\left[-\frac{\Delta_{\rm PG}}{k_{\rm B}T}\right],
\label{EqPG}
\ee
where $A$ and $B$ are constants.  The fits are shown as solid curves in Fig.~\ref{NingFig4} and the respective $\Delta_{\rm PG}/k_{\rm B}$ values are given in the figure caption.  However, the large $\Delta_{\rm PG} \gtrsim 40$~meV was not observed above $T_{\rm c}$ in magnetic inelastic neutron scattering measurements (see, e.g., Fig.~\ref{FigInosov_3D_60K} below).\cite{Inosov2010}

\subsubsection*{c. {\rm Fe}$_{1+y}${\rm (Te}$_{1-x}${\rm Se}$_x${\rm )}}

\begin{figure}
\includegraphics[width=3.in]{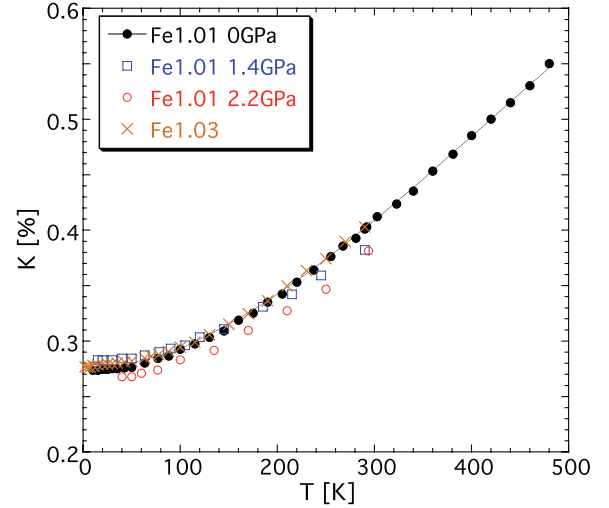}
\caption{(Color online)  Shift $K = K_{\rm chem} + K_{\rm spin}$ of the $^{77}$Se NMR resonance versus temperature $T$ in polycrystalline Fe$_{1+y}$Se samples.\cite{Imai2009}  Data for the superconducting sample with $y = 1.01$ and $T_{\rm c} = 9$~K were obtained up to a pressure of 2.2~GPa as indicated, whereas the data for nonsuperconducting sample with $y = 1.03$ are at zero pressure.  Reprinted with permission from Ref.~\onlinecite{Imai2009}.  Copyright (2009) by the American Physical Society. }
\label{FeSe_Imai_Fig3}
\end{figure}

\begin{figure}
\includegraphics[width=3in]{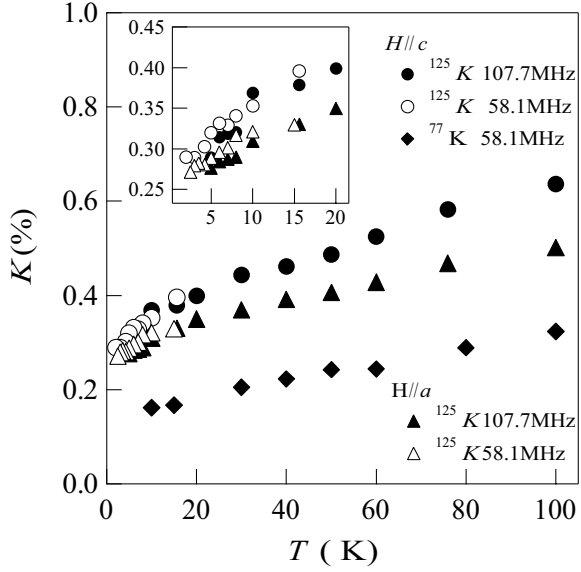}
\caption{(Color online)  Shift $K = K_{\rm chem} + K_{\rm spin}$ of the $^{77}$Se NMR resonance and $^{125}$Te NMR resonance versus temperature $T$ in a single crystal of Fe$_{1.04}$Te$_{0.67}$Se$_{0.33}$.\cite{Michioka2009}  Data for both $H\parallel c$ and $H\parallel a$ are shown.  Inset: Expanded plots of the data at low temperatures.  Reprinted with permission from Ref.~\onlinecite{Michioka2009}.  }
\label{MikiokaFig2}
\end{figure}

Imai and coworkers measured the $^{77}$Se NMR shift $K = K_{\rm chem} + K_{\rm spin}$ versus temperature and pressure of two polycrystalline samples of Fe$_{1+y}$Se with $y = 0.01$ and~0.03 as shown in Fig.~\ref{FeSe_Imai_Fig3}.\cite{Imai2009}  These are the same two samples of McQueen et al.\cite{McQueen2009b} that were discussed above in Sec.~\ref{SecFeSeTe}.  The $K$ is linearly related to $\chi_{\rm spin}$ but the hyperfine coupling constant between $K_{\rm spin}$ and $\chi_{\rm spin}$ was not determined, apparently because the normal state $\chi(T)$ was not measured for these two samples.\cite{McQueen2009b}  The $T$ dependence of $K$ is proportional to the $T$ dependence of $\chi_{\rm spin}$ and has a form very similar to that for the Ba(Fe$_{1-x}$Co$_{x}$)$_{2}$As$_{2}$ crystals in Fig.~\ref{NingFig4}.  Interestingly, whether a Fe$_{1+y}$Se sample is superconducting or not is not reflected in $K(T)$.  The magnitude and temperature dependence of $K$ for the nonsuperconducting sample with $y = 0.03$ are nearly identical to those of the superconducting sample with $y = 0.01$ and $T_{\rm c} = 9$~K\@.

Qualitatively similar behaviors of $K(T)$ were observed in the normal state of a single crystal of Fe$_{1.04}$Te$_{0.67}$Se$_{0.33}$ with $T_{\rm c} = 15$~K by Mishioka et al.\ from $^{77}$Se and $^{125}$Te NMR resonance shift measurements, as shown in Fig.~\ref{MikiokaFig2}.\cite{Michioka2009}  The $K(T)$ data look similar to the normal state $\chi(T)$ data for the 122- and 1111-type superconductors in Fig.~\ref{FigChi(T)} above.  Interestingly, these $K(T)$ data show no evidence for a Curie-Weiss-like upturn at low $T$ like that seen in the bulk $\chi(T)$ for a single crystal of about the same composition in Fig.~\ref{YangFig2bc}(a) above.  This indicates that the upturn in those $\chi(T)$ data is due to the low concentration of excess Fe atoms, thus confirming this assumption used in the analysis of those data by Yang et al.\cite{Yang2009a}  Another interesting aspect of the data in Fig.~\ref{MikiokaFig2} is that there is a difference in slope of $^{125}K_c(T)$ and $^{125}K_a(T)$ at 107.7~MHz, which indicates an anisotropy in the bulk spin susceptibility $\chi_{\rm spin}(T)$.  From a $K$-$\chi$ analysis using their $K(T)$ data and their $\chi(T)$ data (not shown in their paper) for their crystal, the authors obtained values for the $^{77}$Se and $^{125}$Te hyperfine coupling constants and for $K_{\rm chem}$.\cite{Michioka2009}  

From the decreases in the $K(T)$ data below $T_{\rm c}$ shown in the inset of Fig.~\ref{MikiokaFig2}, the authors inferred that the Cooper pairs are spin singlets,\cite{Michioka2009} consistent with the same deductions for other Fe-based superconductors (see Sec.~\ref{SecCooperPairSpin} below).

\subsection{\label{SecLRMO} Long-Range Magnetic Ordering: Energetics and Phase Diagrams}

\subsubsection{Classical Energies of the Ferromagnetic Structure and the Antiferromagnetic Stripe, Double Stripe, Diagonal Double Stripe and N\'eel (Checkerboard) Structures}

\begin{figure}
\includegraphics[width=3.3in]{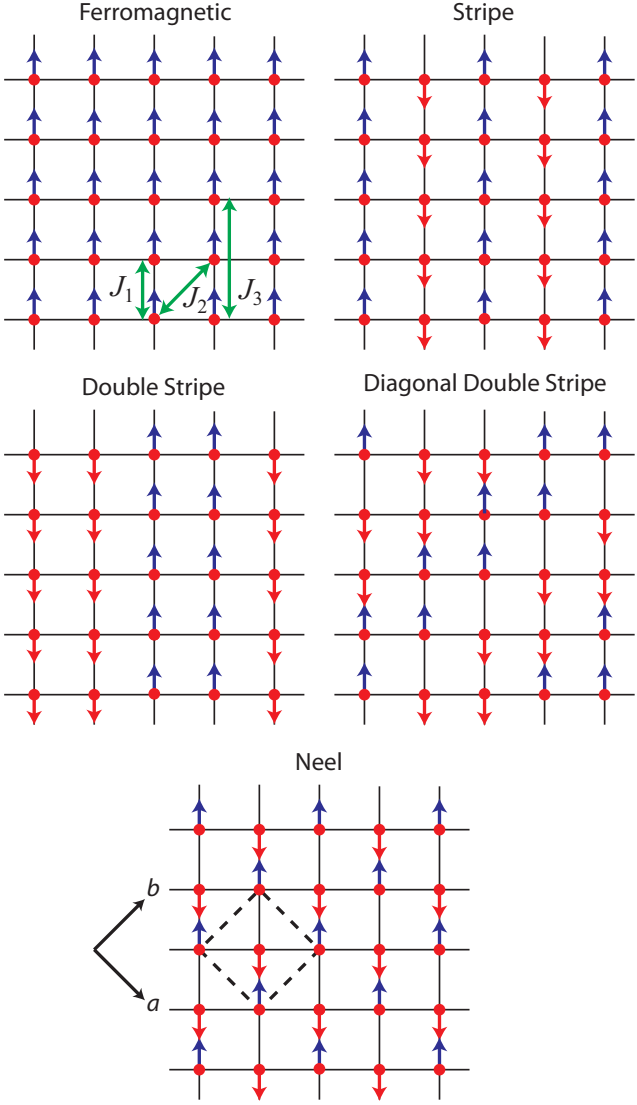}
\caption{(Color online) Five possible collinear commensurate in-plane magnetic orderings.  These include the ferromagnetic ordering in the upper left, and antiferromagnetic stripe, double stripe, diagonal double stripe, and N\'eel-type structures, as indicated.  Note that the N\'eel structure could also be called a diagonal stripe structure.  The nearest- ($J_1$), second-nearest-($J_2$) and third-nearest-neighbor ($J_3$) couplings between spins are shown in the top left panel.  The dashed box in the bottom panel is the basal plane of the tetragonal unit cell of all of the Fe-based superconductors from Fig.~\ref{1111_122_layers}.  This unit cell is tilted by 45$^\circ$ with respect to the Fe square lattice.}
\label{Possible_Mag_Structures_All_Reduce}
\end{figure}

\begin{table}
\caption{\label{AFPropVectors} Propagation vectors ${\bf Q}_{\rm AF}$ of the five magnetic structures in Fig.~\ref{Possible_Mag_Structures_All_Reduce} in the notation of the tetragonal high-temperature crystal structures that each of the Fe-based superconductors and parent compounds have.  Here $\lambda$ is the wavelength of the antiferromagnetic spin structure modulation, $a$ is the tetragonal lattice constant, and $\hat{\bf Q}_{\rm AF}$ is a unit vector in the direction of the antiferromagnetic propagation vector.  The values of ${\bf Q}_{\rm AF} = (2\pi/\lambda)\hat{\bf Q}_{\rm AF}$ are given both in absolute units of \AA$^{-1}$ as ${\bf Q}_{\rm AF} = (H2\pi/a)\hat{\bf a} + (K2\pi/a)\hat{\bf b} $ according to Eq.~(\ref{EqRLV1}) and in reciprocal lattice units ${\bf Q}_{\rm AF} = (H,K)$~r.l.u.\ according to Eq.~(\ref{EqRLV2}).  Most authors set $a = 1$ when they quote values of ${\bf Q}_{\rm AF}$ in absolute units.}
\begin{ruledtabular}
\begin{tabular}{lcccc}
Stable State &  $\lambda$ & $\hat{\bf Q}_{\rm AF}$ & ${\bf Q}_{\rm AF}$ & ${\bf Q}_{\rm AF}$ \\
&&  &  (\AA$^{-1}$) & (r.l.u.) \\
 \hline
Ferromagnet & $\infty$ &  --- & (0,0) & (0,0)  \\
Stripe & $\sqrt{2}\,a$ & $(\hat{\bf a} + \hat{\bf b})/\sqrt{2}$ & $(\frac{\pi}{a},\frac{\pi}{a})$ & $(\frac{1}{2},\frac{1}{2})$ \\
Double stripe & $2\sqrt{2}\,a$ & $(\hat{\bf a} + \hat{\bf b})/\sqrt{2}$ & $(\frac{\pi}{2a},\frac{\pi}{2a})$ & $(\frac{1}{4},\frac{1}{4})$  \\
Diagonal double stripe & $2a$ & $\hat{\bf a}$ & $(\frac{\pi}{a},0)$ & $(\frac{1}{2},0)$ \\
N\'eel (checkerboard) & $a$ & $\hat{\bf a}$ & $(\frac{2\pi}{a},0)$ & (1,0) \\
\end{tabular}
\end{ruledtabular}
\end{table}

\begin{figure}
\includegraphics[width=3.3in]{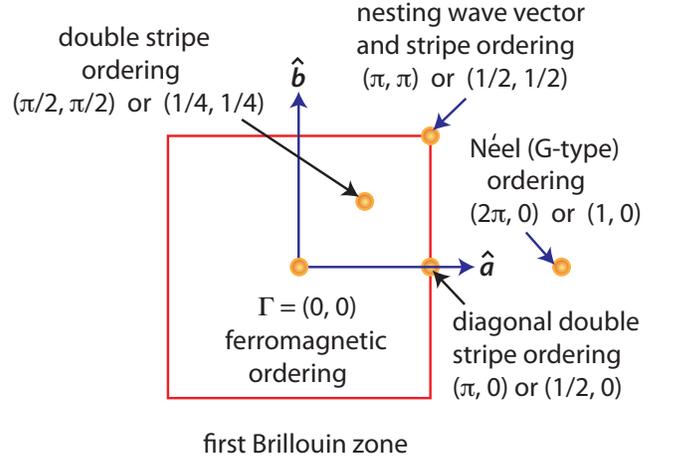}
\caption{(Color online) Ordering wave vectors as listed in Table~\ref{AFPropVectors} for the magnetic structures shown in Fig.~\ref{Possible_Mag_Structures_All_Reduce}.  The $\hat{\bf a}$ and $\hat{\bf b}$ are unit vectors in the directions of the axes of the conventional direct tetragonal unit cell containing two Fe atoms per layer per unit cell as in the bottom panel of Fig.~\ref{Possible_Mag_Structures_All_Reduce}.  The ``nesting wave vector'' is the wave vector connecting the electron and hole Fermi surfaces as in, e.g., Fig.~\ref{FigFS} above.  The first notation for a wave vector is in absolute units according to Eq.~(\ref{EqRLV1}) with the lattice parameter $a$ set to unity, and the second one is in reciprocal lattice units (r.l.u.) according to Eq.~(\ref{EqRLV2}).  The Brillouin zone has a side length of $2\pi/a$, which is the magnitude of the basal plane reciprocal lattice translation vectors along the $\hat{\bf a}$ and $\hat{\bf b}$ directions.}
\label{Ordering_Wave_Vectors}
\end{figure}

We first examine the classical energetics of the five types of collinear commensurate in-plane magnetic structures shown in Fig.~\ref{Possible_Mag_Structures_All_Reduce}, where the three exchange interactions considered are also defined in the top left panel.  Because of possible confusion between different low-temperature unit cells and their Brillouin zones, we will write down the antiferromagnetic propagation vectors {\bf Q}$_{\rm AF}$ in terms of the conventional in-plane tetragonal Brillouin zone.  The orientation of the high-temperature tetragonal basal plane for all of the Fe-based superconductors and parent compounds is shown as the dashed box in the bottom panel of Fig.~\ref{Possible_Mag_Structures_All_Reduce}, from Fig.~\ref{1111_122_layers}, where $a=b$.  It is tilted by 45$^\circ$ with respect to the Fe square lattice.  The antiferromagnetic propagation wave vector of the various magnetic structures shown has a magnitude $Q_{\rm AF} = 2\pi/\lambda$ where $\lambda$ is the wavelength of the antiferromagnetic modulation.  The direction of ${\bf Q}_{\rm AF}$ is perpendicular to lines or planes of spins with the same direction.  With these rules, the ${\bf Q}_{\rm AF}$ of the five different magnetic structures in Fig.~\ref{Possible_Mag_Structures_All_Reduce} are listed in Table~\ref{AFPropVectors}.  The ordering wave vectors are shown with respect to the Brillouin zone in conventional tetragonal notation in Fig.~\ref{Ordering_Wave_Vectors}.

The relative stabilities of the different magnetic structures in Fig.~\ref{Possible_Mag_Structures_All_Reduce} depend on the signs and relative strengths of the nearest- ($J_1$), second-nearest- ($J_2$) and third-nearest-neighbor ($J_3$) couplings.  We note that if, instead of classifying the exchange constants according to the physical distances between a central spin and its neighbors, one were instead counting the number of bonds from the central spin to its neighboring spins, both $J_2$ and $J_3$ connect spins separated by two bonds from the central spin, so both would be considered to be second-nearest-neighbor couplings.  The classical Hamiltonian in zero applied magnetic field is
\be
{\cal H} = J_{1} S^2 \sum_{\langle ij \rangle}\hat{{\bf S}}_i \cdot \hat{{\bf S}}_j + J_{2} S^2 \sum_{\langle ik \rangle}\hat{{\bf S}}_i \cdot \hat{{\bf S}}_k 
+ J_3 S^2 \sum_{\langle i\ell \rangle}\hat{{\bf S}}_i \cdot \hat{{\bf S}}_\ell ,
\label{EqJ1J2J3}
\ee
where the sums are over distinct pairs of spins, $S$ is the magnitude of the spin and $\hat{{\bf S}}$ is a spin unit vector.  Referring to Fig.~\ref{Possible_Mag_Structures_All_Reduce}, there are four of each type of neighbor to a given spin.   In Eq.~(\ref{EqJ1J2J3}) and elsewhere in this review, a positive $J$ means antiferromagnetic coupling and a negative $J$ means ferromagnetic coupling.  

The classical energies of the five different collinear commensurate magnetic structures in Fig.~\ref{Possible_Mag_Structures_All_Reduce} are
\bea
E_{\rm ferromagnetic} &=& 2NS^2 (J_1 + J_2 + J_3),\nonumber\\
E_{\rm stripe} &=& 2NS^2 (-J_2 + J_3),\nonumber\\
E_{\rm double\ stripe} &=& 2NS^2 \frac{J_1}{2},\label{OrdEnergies}\\
E_{\rm diagonal\ double\ stripe} &=& 2NS^2 (-J_3),\nonumber\\
E_{\rm Neel} &=& 2NS^2 (-J_1 + J_2 + J_3),\nonumber
\eea
where $N$ is the number of spins in the spin lattice, and a factor of 1/2 was included on the right-hand sides to avoid double-counting pairs of spins.  From Eqs.~(\ref{OrdEnergies}), the stripe structure is completely frustrated with respect to $J_1$, the double stripe structure is completely frustrated with respect to both $J_2$ and $J_3$, and the diagonal double stripe structure is completely frustrated with respect to both $J_1$ and $J_2$.  

Interesting differences in the ordered energies of several of the magnetic structures are
\be
E_{\rm ferromagnetic} - E_{\rm Neel} = 2NS^2 (2J_1),
\label{FerroNeel}
\ee
which only depends on $J_1$,
\be
E_{\rm Neel} - E_{\rm stripe} = 2NS^2 (2J_2 - J_1),
\label{NeelStripe}
\ee
which is independent of $J_3$, and 
\be
E_{\rm stripe} - E_{\rm diagonal\ double\ stripe} = 2NS^2 (2J_3 - J_2),
\label{StripeDouble}
\ee
which is independent of $J_1$.

\subsubsection{\label{SecJ1J2Model} The $J_1$-$J_2$ Model and Phase Diagram}

\begin{table}
\caption{\label{StabilityCondxJ1J2} Classical stability conditions for three types of commensurate collinear in-plane magnetic orderings in Fig.~\ref{Possible_Mag_Structures_All_Reduce} with respect to each other in the  $J_1$-$J_2$ model with $J_3 = 0$.  An antiferromagnetic $J$ is positive and a ferromagnetic $J$ is negative.}
\begin{ruledtabular}
\begin{tabular}{llcc}
&Stable State &  Condition on $J_1$& Condition on $J_2$\\
 \hline
1&Ferromagnet & $J_1 < 0$ & $J_2 < -\frac{J_1}{2}$  \\
%\hline
2&Stripe &  no restriction & $J_2 > \frac{|J_1|}{2}$  \\
%\hline
3&N\'eel & $J_1 > 0$ & $J_2 < \frac{J_1}{2}$ \\
\end{tabular}
\end{ruledtabular}
\end{table}

\begin{figure}
\includegraphics[width=3.3in,viewport= 0 40 400 340,clip]{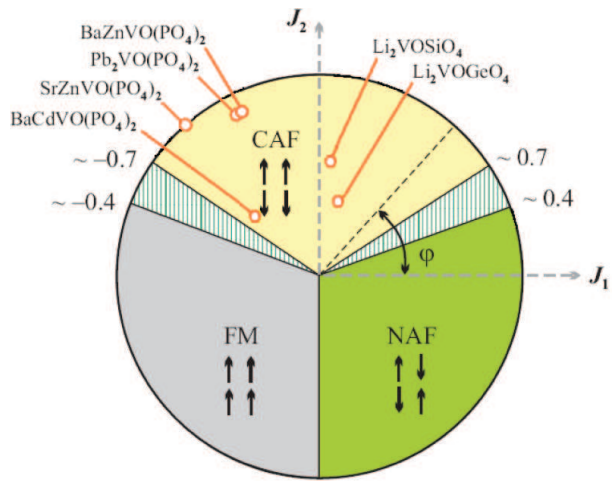}
\caption{(Color online) The $J_{1}$-$J_{2}$ phase diagram of the spin-1/2 Heisenberg square spin lattice showing different ordered phases and the quantum spin liquid (QSL) regimes.\cite{Nath2009a} The fractional numbers on the periphery of the circle are values of $J_2/J_1$.  The two QSL regimes are the two vertically hatched areas. The boundaries of the QSL regimes are not precisely known. The other ordered regions of the phase diagram contain the ferromagnetic (FM) phase, the columnar antiferromagnetic (CAF) phase which is termed the stripe phase in this review, and the N\'{e}el antiferromagnetic (NAF) phase. The locations of various investigated compounds in the CAF part of the phase diagram are shown. The radius of the circle is $J/k_{\rm B} = 12.2$~K\@.  Reprinted with permission from Ref.~\onlinecite{Nath2009a}.  Copyright (2009) by the American Physical Society.}
\label{Nathfig1}
\end{figure}

Setting $J_3 = 0$, one obtains the $J_1$-$J_2$ model where the collinear commensurate ferromagnetic, stripe antiferromagnetic and N\'eel antiferromagnetic phases compete.   The conditions on the two exchange constants in the classical $J_1$-$J_2$ model for a given structure to have the lowest energy with respect to the other two structures are given in Table~\ref{StabilityCondxJ1J2}.  According to the criteria in Table~\ref{StabilityCondxJ1J2}, the stripe state is stable for
\be
J_2 > \frac{|J_1|}{2},
\label{EqClassJ2J1}
\ee
which means that $J_2$ has to be positive (antiferromagnetic), whereas $J_1$ can be either ferromagnetic or antiferromagnetic.  

The $J_1$-$J_2$ model on a square lattice has a long history.  The phase diagram for a spin $S = 1/2$ square lattice is shown in Fig.~\ref{Nathfig1}.\cite{Nath2009a}  One sees that the classical criterion for the stripe phase [or the columnar antiferromagnetic (CAF) phase as designated in Fig.~\ref{Nathfig1}], in Eq.~(\ref{EqClassJ2J1}) is not quite correct in this case.  According to Fig.~\ref{Nathfig1}, the classical stability ratio $J_2/|J_1| = 0.5$ in Eq.~(\ref{EqClassJ2J1}) instead results in a quantum-disordered phase, with a somewhat larger value $J_2/|J_1| \gtrsim 0.7$ required to stabilize the spin stripe phase.  This model has been applied to the Fe-based materials.\cite{Yildirim2009, Si2008}

\subsubsection{The $J_1$-$J_2$-$J_3$ Model and Phase Diagram}

%\squeezetable
\begin{table*}
\caption{\label{StabilityCondxJ1J2J3} Classical stability conditions for the five types of in-plane magnetic orderings in Fig.~\ref{Possible_Mag_Structures_All_Reduce} when they all compete with each other, where the three exchange constants are $J_1$, $J_2$ and $J_3$.  An antiferromagnetic $J$ is positive and a ferromagnetic $J$ is negative.  From the table, if $J_1 = 0$, only the stripe and diagonal double stripe states occur.}
\begin{ruledtabular}
\begin{tabular}{rlccc}
&Stable State &  Condition on $J_1$ & Condition on  $J_2$ & Condition on $J_3$ \\
 \hline
1 & Ferromagnet & $J_1 < 0$ & $J_2 \leq 0$ & $J_3 < \frac{1}{2} (-J_1 - J_2)$ \\
2 &  & & $0 < J_2 < -\frac{J_1}{2}$ & $J_3 < \frac{1}{2} (-J_1 - 2 J_2)$ \\
3 & Double stripe & $J_1 < 0$ & $0 < J_2 \leq -\frac{J_1}{2}$ & $\frac{1}{2} (-J_1 - 2 J_2) < J_3 <  -\frac{J_1}{2}$\\
4 &  &  & $-\frac{J_1}{2} < J_2 < -J_1$ & $\frac{1}{2} (J_1 + 2 J_2) < J_3 <  -\frac{J_1}{2}$\\
5 & Diagonal double stripe & $J_1 \leq 0$ & $J_2 \leq 0$ & $J_3 > \frac{1}{2} (-J_1 - J_2)$\\
6 &  &  & $0 < J_2 \leq -J_1$ & $J_3 >  -\frac{J_1}{2}$\\
7 &  &  & $J_2 > -J_1$ & $J_3 >  \frac{J_2}{2}$\\
8 &  & $J_1 > 0$ & $ J_2 \leq \frac{J_1}{2}$ & $J_3 >  \frac{1}{2} (J_1 - J_2)$\\
9 &  &  & $ J_2 > \frac{J_1}{2}$ & $J_3 >  \frac{J_2}{2}$\\
10 & Stripe &  $J_1 \leq 0$ & $-\frac{J_1}{2} < J_2 \leq -J_1$ &  $J_3 < \frac{1}{2} (J_1 + 2 J_2)$ \\
11 &  &  & $J_2 > -J_1$ &  $J_3 < \frac{J_2}{2} $ \\
12 & & $J_1 > 0$ & $J_2 > \frac{J_1}{2}$ & $J_3 < \frac{J_2}{2} $ \\
13 & N\'eel & $J_1 > 0$ & $J_2 < \frac{J_1}{2}$ & $J_3 < \frac{1}{2} (J_1 - J_2)$ \\
\end{tabular}
\end{ruledtabular}
\end{table*}

\begin{figure}
\includegraphics[width=3.3in]{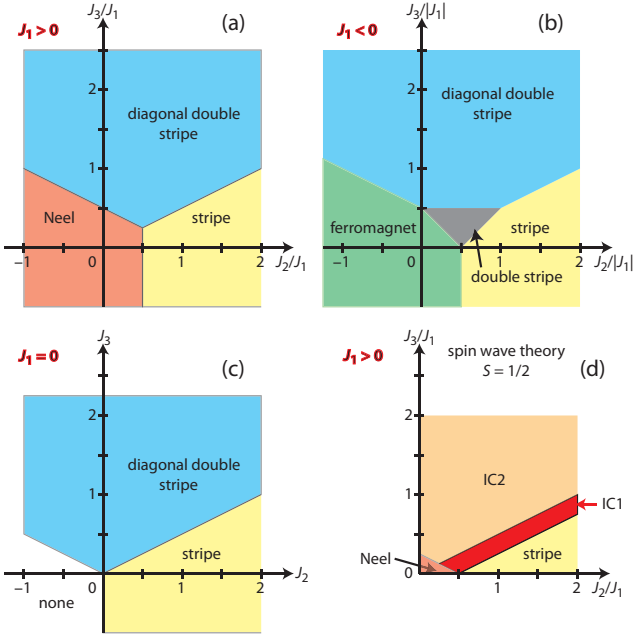}
\caption{(Color online) The zero-temperature $J_{1}$-$J_{2}$-$J_{3}$ phase diagrams of the Heisenberg square spin lattice showing different ordered phases. Panels (a), (b) and (c) are the classical predictions based on the classical stability conditions for the five competing commensurate collinear phases in Table~\ref{StabilityCondxJ1J2J3}.  For $J_1 = 0$ in panel~(c), the only stable phases are the stripe and double stripe phases.  Panel (d) gives the phase diagram determined by Moreo et al.\ from spin wave theory for spin $S = 1/2$.\cite{Moreo1990}  Similar to (a) for $J_1 > 0$, the phase diagram in (d) contains the N\'eel and stripe phases over similar parts of the phase diagram, but in (d) two different incommensurate phases IC1 and IC2 occur than are not in (a) by construction, and the diagonal double stripe structure is missing in (d) compared to~(a).}
\label{J1J2J3_Phase_diagramAllReduced}
\end{figure}

A refinement or extension of the $J_1$-$J_2$ model is the $J_1$-$J_2$-$J_3$ model.  The conditions on the three exchange constants in the $J_1$-$J_2$-$J_3$ Heisenberg model for the classical ground state structure within the set of five competing commensurate collinear magnetic structures in Fig.~\ref{Possible_Mag_Structures_All_Reduce} are given in Table~\ref{StabilityCondxJ1J2J3}.  The classical zero-temperature phase diagrams constructed from these stability conditions are shown in Figs.~\ref{J1J2J3_Phase_diagramAllReduced}(a)--\ref{J1J2J3_Phase_diagramAllReduced}(c).  In addition, the $J_1$-$J_2$-$J_3$ phase diagram determined for spin $S = 1/2$ and $J_1 > 0$  from spin wave theory by Moreo et al.\ is shown in Fig.~\ref{J1J2J3_Phase_diagramAllReduced}(d), which includes incommensurate phases that we have not considered.\cite{Moreo1990}  From the phase diagrams in Figs.~\ref{J1J2J3_Phase_diagramAllReduced}(a)--\ref{J1J2J3_Phase_diagramAllReduced}(c), one sees that a positive (antiferromagnetic) $J_3$ is needed to stabilize either the double stripe or the diagonal double stripe structure.  Interesting features of Figs.~\ref{J1J2J3_Phase_diagramAllReduced}(a)--\ref{J1J2J3_Phase_diagramAllReduced}(c) are that when $J_1$ changes sign from positive to negative, the N\'eel structure is replaced by the ferromagnetic structure over a similar part of the phase diagram, and the double stripe structure becomes stable over a limited part of the phase diagram.  At the same time, the diagonal double stripe structure remains stable over roughly the same region of the phase diagram.  The diagonal double stripe structure has been previously studied in the context of the manganites.\cite{Hotta2003, Dong2008}  The lack of this structure in the phase diagram for the $S = 1/2$ system in Fig.~\ref{J1J2J3_Phase_diagramAllReduced}(d) evidently occurs because this commensurate structure has a higher energy than the two incommensurate structures shown.  However, it might occur in an Ising model for $S = 1/2$ that would enforce collinear antiferromagnetic ordering (E. Dagotto, private communication).

\subsection{Long-Range Magnetic Ordering: Experimental Magnetic Structures of the Fe-Based Compounds}

\subsubsection{Introduction}

We will consider the antiferromagnetic ordering in the  1111-, 111-, and 122-type FeAs-based parent compounds separately from those of the 11-type Fe$_{1+y}$Te$_{1-x}$Se$_x$ compounds because these two classes of materials have quite different low temperature long-range ordered magnetic structures.  Mainly we will be considering the magnetic structure refinements from neutron diffraction measurements.  In any of these materials, at and below the long-range magnetic ordering temperature $T_{\rm N}$, every compound is distorted from the respective high-temperature tetragonal structure to either an orthorhombic structure or a monoclinic structure.  This distortion causes crystallographic and magnetic twins to form as discussed above in Sec.~\ref{SecStructOverview}, which can potentially change the magnetic neutron diffraction intensities and hence significantly affect the inferred ordered Fe moments if the twinning is not properly taken into account.  The occurrence and impact of twinning is rarely\cite{Goldman2008} explicitly discussed in reports of the magnetic structures of the Fe-based compounds obtained from magnetic neutron diffraction measurements.

\subsubsection{\label{1111-111-122} 1111-, 111-, and 122-type FeAs-Based Compounds}

\subsubsection*{In-plane Component of the Magnetic Structures}

\squeezetable
\begin{table}
\caption{\label{FeOrdering} Iron N\'eel temperatures $T_{\rm N}$ for long-range antiferromagnetic ordering and low-temperature ordered Fe moments $\mu$ for 1111-, 122-, 111-, and 11-type Fe-based compounds.}
\begin{ruledtabular}
\begin{tabular}{l|ccc}
 Compound   &  $T_{\rm N,~Fe}$ &$\mu$ & Ref.\\
 & (K) & ($\mu_{\rm B}/$Fe) \\ \hline
LaFeAsO & 137 & 0.36(5) & \onlinecite{Huang2008}, \onlinecite{Cruz2008} \\
 &  & 0.63(1) & \onlinecite{Qureshi2010} \\
 & & $\approx 0.35$ & \onlinecite{mcguire2008}\\
 & & 0.78(8) & \onlinecite{Li2010c}\\
CeFeAsO & 138(4) & 0.94(3) & \onlinecite{Zhao2008} \\
 & 140 & 0.8$_5$ & \onlinecite{Cruz2010} \\
PrFeAsO & 127 & 0.48(9) & \onlinecite{Zhao2008b} \\
 & $\sim 85$ & 0.35(5) & \onlinecite{Kimber2008} \\
NdFeAsO & 141(6) & 0.25(7) & \onlinecite{Chen2008}\\
 &150 & 0.9(1) & \onlinecite{Qiu2008} \\
 &137 & 0.54(1) & \onlinecite{Tian2010} \\
\hline
CaFeAsF & 114(3) & 0.49(5) & \onlinecite{Xiao2009c} \\
SrFeAsF & 133(3) & 0.58(6) & \onlinecite{Xiao2009b} \\
\hline
${\rm CaFe_2As_2}$ & 173 & 0.80(5) & \onlinecite{Goldman2008}\\
${\rm SrFe_2As_2}$ & 220 & 0.94(4) & \onlinecite{Zhao2008c}\\
 & 205 & 1.01(3) & \onlinecite{Kaneko2008}\\
 &  & 1.04(1) & \onlinecite{Lee2010}\\
${\rm BaFe_2As_2}$ & 142 & 0.87(3) & \onlinecite{Huang2008b}\\
 & 136 & 0.93(6) & \onlinecite{Wilson2009}\\
 & 136(1) & 0.91(21) & \onlinecite{Matan2009}\\
${\rm EuFe_2As_2}$ & 190 & 0.98(8) & \onlinecite{Xiao2009}\\
\hline
${\rm NaFeAs}$ & 37 & 0.09(4) & \onlinecite{Li2009b} \\
\hline
Fe$_{1.05}$Te & 72.5(10) & 2.54(2) & \onlinecite{Martinelli2010} \\
Fe$_{1.068}$Te & 67\footnotemark[3] & 2.25(8)\footnotemark[1] & \onlinecite{Li2009c} \\
Fe$_{1.09}$Te & $\sim 70$ & 1.86(2)\footnotemark[2] & \onlinecite{Iikubo2009} \\
Fe$_{1.14}$Te &  & 0.76(2)\footnotemark[2] & \onlinecite{Bao2009} \\
Fe$_{1.02}$Te$_{0.96}$Se$_{0.04}$ & 49 & 2.09(3) & \onlinecite{Liu2010} \\
Fe$_{1.02}$Te$_{0.95}$Se$_{0.05}$ & 47 & 1.68(6) & \onlinecite{Liu2010} \\
Fe$_{1.02}$Te$_{0.92}$Se$_{0.08}$ & 22 & 0.33(2) & \onlinecite{Liu2010} \\
Fe$_{1.09}$Te$_{0.75}$Se$_{0.25}$ & none &  & \onlinecite{Iikubo2009} \\
Fe$_{1.08}$Te$_{0.67}$Se$_{0.33}$ & none & & \onlinecite{Bao2009}\\ 
FeTe$_{0.584}$Se$_{0.416}$ & none & & \onlinecite{Li2009c}\\ 
FeTe$_{0.507}$Se$_{0.493}$& none & & \onlinecite{Li2009c}\\
\end{tabular}
\end{ruledtabular}
\footnotetext[1]{The cited value is the magnitude of the ordered moment.  The components of the ordered moment along the monoclinic $a$- $b$- and $c$-axes are $-0.7(2)$, 2.0(7), and $0.7(1)~\mu_{\rm B}$/Fe, respectively.  Thus the in-plane ordered moment is oriented along the $b$-axis.  The authors speculate that the $c$-axis component is due to the presence of the interstitial Fe(2) spins.}
\footnotetext[2]{Incommensurate antiferromagnetic structure.}
\footnotetext[3]{The simultaneous structural/magnetic transition is reported to be first order.}
\end{table}

As noted with respect to Fig.~\ref{FigChiRhoBaFe2As2} above, most of the undoped 1111-, 111- and 122-type FeAs-based parent compounds exhibit structural transitions at $T_{\rm S}$ and commensurate antiferromagnetic (AF or spin density wave SDW) ordering transitions at N\'eel temperatures $T_{\rm N} \sim 100$--200~K\@.  Magnetic neutron diffraction measurements have determined the magnetic structures and ordered magnetic moments per Fe ($\mu$).\cite{Lynn2009}  Low-temperature crystal and magnetic structure data for the 1111-type and 122-type FeAs-based compounds are listed in Tables~\ref{LoTStructData1111} and~\ref{LoTStructData122} in the Appendix, respectively.  A summary of the Fe antiferromagnetic ordering temperatures and ordered Fe magnetic moments is given in Table~\ref{FeOrdering}.\cite{Zhao2008, Cruz2010, Goldman2008, Huang2008b, Wilson2009, Cruz2008, Qureshi2010, mcguire2008, Li2010c, Zhao2008b, Kimber2008, Chen2008, Qiu2008, Tian2010, Xiao2009c, Xiao2009b, Zhao2008c, Kaneko2008, Matan2009, Xiao2009, Li2009b, Martinelli2010, Li2009c, Iikubo2009, Bao2009}  For the $R$LaFeAsO compounds, whether the rare earth element $R$ is magnetic or not has little influence on the Fe ordering temperature.  Interestingly, the $T_{\rm N}$ for the $A{\rm Fe_2As_2}$ compounds shows a nonmonotonic variation with alkaline earth $A$ radius.  

One sees from Table~\ref{FeOrdering} that the Fe ordered moment is small and variable for the FeAs-based compounds.  This is often an indicator for itinerant electron magnetism instead of local moment magnetism.  Indeed, the ordered moments (0.1--1~$\mu_{\rm B}$/Fe atom) in Table~\ref{FeOrdering} are anomalously small compared to values found for Fe$^{+2}$ ions in ionic insulators which are nominally $d^6$ ions with high spin $S = 2$, with a corresponding ordered moment $\mu = gS\mu_{\rm B} = 4~\mu_{\rm B}$/Fe atom, respectively, where $g \approx 2$ is the spectroscopic splitting factor ($g$-factor) of the Fe ions and $\mu_{\rm B}$ is the Bohr magneton.  For example, the ordered moment of Fe in FeO is $3.32~\mu_{\rm B}$/Fe atom.\cite{Roth1958}  The observed ordered moments in the FeAs-type materials would imply a spin $S \lesssim 1/2$ per Fe atom, which is hard to understand in a local moment model.  Alternative views have been put forward that the low ordered moments result from magnetic frustration\cite{Si2008, Rodriguez2009, Manousakis2010} and/or fluctuation\cite{Dai2009, Mazin2009a, Hansmann2010} effects in a large-local-moment system.  However, the small effective moment of Fe observed by inelastic neutron scattering experiments \emph{in the paramagnetic state} of ${\rm CaFe_2As_2}$\cite{Diallo2010} argues against this possibility and instead indicates that the magnetic ordering transitions are spin density wave (SDW) transitions associated with itinerant electron antiferromagnetism.  Schmidt et al.\ have studied frustrated local-moment models for the magnetism of the Fe-based superconductors and parent compounds, and concluded that ``the anomalously low moment in the pnictides is not explained by quantum fluctuations in effective localized moment models but needs a more microscopic viewpoint including the itinerant multiorbital nature of the magnetic state. ... This does not invalidate, however, the exceptinal usefulness of the simple $J_{1a,b}$-$J_2$ model to describe the low-energy spin excitations.''\cite{Schmidt2010}

For the Fe$_{1+y}$Te compounds, the ordered moments for the smaller $y$ values are signficantly larger than for the FeAs-based compounds, suggesting an important difference with the latter materials.  Indeed, analysis of the magnetic susceptibility data in Sec.~\ref{SecChi} above suggests that a local moment picture for the magnetism is more appropriate for these materials.

The structural and magnetic transitions are apparently coupled, since the transition temperatures $T_{\rm S}$ and the $T_{\rm N}$ values are either the same as observed in the 11-, 111- and 122-type Fe-based parent compounds, or with $T_{\rm S} \gtrsim T_{\rm N}$ as found for the 1111-type and some doped 122-type compounds. 
 
\begin{figure}
% For arXiv, uncomment the first one and comment out the 2nd.
\includegraphics [width=0.7\columnwidth]{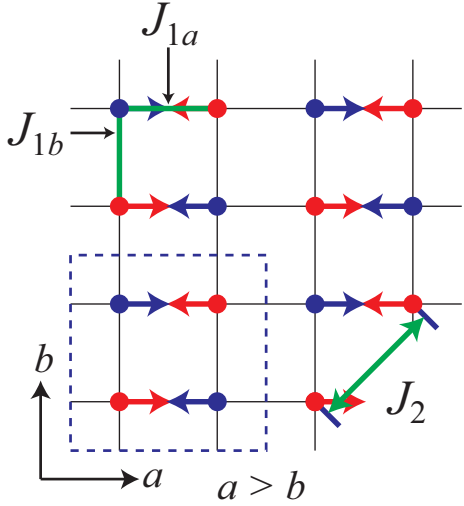}
\caption{(Color online) In-plane commensurate antiferromagnetic structure of the 111-, 1111- and 122-type FeAs-based parent compounds.  The Fe atoms (filled circles) are on the corners of a square lattice at temperatures $T > T_{\rm s} \geq T_{\rm N}$ and of a slightly ($\sim 1$\%) orthorhombically distorted ($a > b$) square lattice for $T < T_{\rm s}$, where $T_{\rm s}$ is the tetragonal-to-orthorhombic structural transition temperature and $T_{\rm N}$ is the antiferromagnetic, or spin-density-wave, transition temperature.  The basal plane orthorhombic unit cell is shown as the dashed box.  The room temperature tetragonal basal plane has edges that are smaller by a factor of $\sqrt{2}$ and are rotated by 45$^\circ$ with respect to the orthorhombic axes.  The collinear ordered moment direction is in the $a$-$b$ plane of the orthorhombic structure and is directed along the longer $a$-axis.  The red and blue arrows represent spins on the red and blue sublattices of Fe atoms, that respectively consist of next-nearest-neighbors.  Each sublattice is individually antiferromagnetically ordered in a commensurate collinear N\'eel (Ising-like) configuration.  When both spin lattices are considered together, this intralattice ordering causes spin stripes to form along the $b$-axis, which also causes magnetic frustration between the two sublattices irrespective of whether the intersublattice coupling is ferromagnetic or antiferromagnetic.  The nearest-neighbor exchange coupling constants $J_{1a}$ and $J_{1b}$ (between sublattices) and the next-nearest-neighbor coupling constant $J_2$ (within each sublattice) in a local moment description of the orthorhombic phase are shown. An anisotropy between $J_{1a}$ and $J_{1b}$ is needed for the system to choose whether the stripe orientation is vertical (as observed) or horizontal.}
\label{Stripe_Mag_Struct}
\end{figure}

Within the Fe planes, the magnetic structure in the orthorhombic AF-ordered state of the undoped  111- (NaFeAs), 1111- and 122-type parent compounds has been found to be the same  interesting collinear ``stripe'' configuration from both magnetic neutron diffraction measurements (see the references in Tables~\ref{LoTStructData1111} and~\ref{LoTStructData122} in the Appendix) and $^{75}$As NMR measurements,\cite{Kitagawa2008, Kitagawa2009} as depicted in Fig.~\ref{Stripe_Mag_Struct}.  In particular, this magnetic structure is not a collinear checkerboard N\'eel state where the nearest-neighbor spins of each Fe spin are antiparallel to that of the given spin, but is the stripe state in Fig.~\ref{Possible_Mag_Structures_All_Reduce}.  As shown in Tables~\ref{LoTStructData1111} and~\ref{LoTStructData122} and Fig.~\ref{Stripe_Mag_Struct}, the ordered Fe moment in the FeAs-based materials is always directed along the longer orthorhombic $a$-axis in the $a$-$b$~plane, as also shown from $^{75}$As NMR measurements in the magnetically ordered states of ${\rm BaFe_2As_2}$ (Ref.~\onlinecite{Kitagawa2008}) and ${\rm SrFe_2As_2}$.\cite{Kitagawa2009}  

\begin{figure}
% For arXiv, uncomment the first one and comment out the 2nd.
\includegraphics [width=0.70\columnwidth]{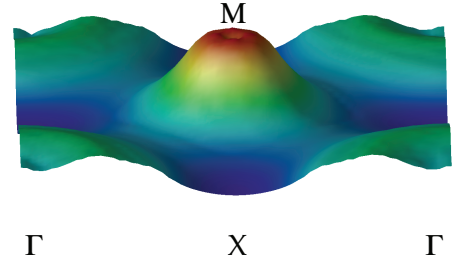}
\caption{(Color online) Real part of the in-plane wave vector-dependent noninteracting static susceptibility $\chi({\bf Q}_{ab})$ for primitive tetragonal LaFeAsO calculated at $Q_z = \pi/c$ using density functional theory.\cite{mazin2008}  The Brillouin zone letter symmetry point notations are given in the lower panel of Fig.~\ref{BZs}.  The $\chi({\bf Q}_{ab})$ peaks at the same wave vector ${\bf Q}_{\rm nesting} = \left(\frac{1}{2},\frac{1}{2}\right)$~r.l.u.\ (tetragonal notation) at which Fermi surface nesting occurs in Figs.~\ref{LaFeAsOBS} and~\ref{FigFS}, which is also the in-plane propagation vector for the observed antiferromagnetic stripe magnetic structure.  Reprinted with permission from Ref.~\onlinecite{mazin2008}.  Copyright (2008) by the American Physical Society.}
\label{Mazin_chi}
\end{figure}

The nesting wave vector for the hole and electron Fermi surface pockets in Figs.~\ref{LaFeAsOBS} and~\ref{FigFS} is the same as ${\bf Q}_{\rm AF} = \left(\frac{1}{2},\frac{1}{2}\right)$~r.l.u.\ in tetragonal notation for the observed stripe antiferromagnetic propagation vector in Table~\ref{AFPropVectors}, but does not match that of any of the other magnetic structures.  This suggests that the antiferromagnetic ordering is a spin density wave (SDW)  due to Fermi surface nesting of itinerant electrons rather than to local magnetic moments.  This interpretation is confirmed by a calculation of the wave vector dependent static susceptibility $\chi({\bf Q})$ for LaFeAsO by Mazin et al.\ in Fig.~\ref{Mazin_chi},\cite{mazin2008} which shows a peak at ${\bf Q}_{\rm AF}$, which is at the corner M point of the primitive tetragonal Brillouin zone in the bottom panel of Fig.~\ref{BZs} above.  A similar enhancement of $\chi({\bf Q})$ at the M point was found for LaFeAsO in more refined calculations by Monni et al.\cite{Monni2010}  Utfeld et al.\ have calculated that the noninteracting $\chi({\bf Q})$ for optimally doped ${\rm Ba(Fe_{0.93}Co_{0.07})_2As_2}$ still has a pronounced peak near the nesting wave vector, in spite of significant 3D dispersion of one of the hole Fermi pockets,\cite{Utfeld2010} and Yaresko et al.\ also found peaks in the noninteracting $\chi({\bf Q})$ for both doped and undoped LaFeAsO$_{1-x}$F$_x$ and (Ba$_{1-x}$K$_x$)${\rm Fe_2As_2}$,\cite{Yaresko2009} thus supporting the $s^\pm$ superconducting pairing model (see below).  An early unbiased numerical investigation of a two orbital model gave robust evidence for the stability of the stripe state in the undoped iron arsenides.\cite{Daghofer2008}  The in-plane component of the nesting and AF wave vector in tetragonal notation can be written in absolute units (as opposed to r.l.u.\ above) as ${\bf Q}_{\rm AF}({\rm \AA}^{-1}) = \left(\frac{\pi}{a},\frac{\pi}{a}\right)$, which is usually  abbreviated as ${\bf Q}_{\rm AF} = (\pi,\pi)$ by setting $a=1$.

An alternate interpretation is that the antiferromagnetic ordering can be understood using a local moment picture.\cite{Si2008}  The interpenetrating blue and red sublattices in Fig.~\ref{Stripe_Mag_Struct} each separately exhibit collinear N\'eel antiferromagnetic order.  Considering both sublattices together, ferromagnetically aligned ``stripes'' of spins occur that are oriented vertically in the figure, along the shorter $b$ axis of the orthorhombic basal plane.  However, in a local moment picture this does not necessarily mean that the interactions between spins along the vertical stripes on the two different sublattices are ferromagnetic.  On the contrary, this spin structure can be due to a dominant antiferromagnetic (AF) next-nearest-neighbor exhange interaction $J_2 \gtrsim |J_1|/2$ \emph{within each of the two sublattices} as shown above in Fig.~\ref{Nathfig1}.\cite{Si2008, Ma2008b}  For equal antiferromagnetic or ferromagnetic nearest-neighbor AF interactions $J_{1a} = J_{1b}  \equiv J_1$ \emph{between opposite sublattices}, the interactions are fully frustrated: the spins on either sublattice could collectively rotate by any angle with respect to the spins in the other sublattice without changing the energy of the system.  However, Yaresko et al.\ found from local spin density approximation calculations on LaFeAsO and ${\rm BaFe_2As_2}$ that this local moment description fails.\cite{Yaresko2009}  In particular, they found that the energy of the system strongly depends on the angle between the two spin sublattices.  They further state, ``The dependence of the energy on the angle between the Fe moments in the two sublattices cannot be described by the simple $J_1$-$J_2$ Heisenberg model, but may be reproduced by a biquadratic term proportional to $({\bf S}_i\cdot {\bf S}_j)^2$, which favors collinear stripe AFM order.''\cite{Yaresko2009}  To our knowledge, no experimental results have been modeled using the biquadratic term.

Continuing with the local moment description, each Fe spin in an approximately square lattice layer has four nearest neighbor spins and four next-nearest-neighbor spins (see Fig.~\ref{Stripe_Mag_Struct}), and two nearest-neighbor spins along the $c$-axis.  One can extend the Heisenberg spin Hamiltonian to three dimensions via a $J_1$-$J_2$-$J_c$ model as
\be
{\cal H} = J_1 \sum_{\langle ij \rangle_{a,b}}{\bf S}_i \cdot {\bf S}_j + J_2 \sum_{\langle ik \rangle_{ab}}{\bf S}_i \cdot {\bf S}_k + J_{1c} \sum_{\langle i\ell \rangle_{c}}{\bf S}_i \cdot {\bf S}_\ell 
\label{EqJ1J2}
\ee
with periodic boundary conditions, where $J > 0$ corresponds to AF interactions and $J < 0$ to ferromagnetic (FM) interactions as usual, the first sum is over nearest-neighbor spin pairs in the $a$-$b$~plane, the second sum is over next-nearest-neighbor spin pairs in the $a$-$b$~plane, the third sum is over nearest-neighbor spin pairs along the $c$-axis, and each spin pair is only counted once.  

To lift the frustration between the two sublattices within the stripe state and obtain the particular orientation of the stripes in Fig.~\ref{Stripe_Mag_Struct} that occurs for all of the 111-, 1111- and 122-type FeAs-based parent compounds requires $J_{1a} > J_{1b}$, which is an extension of the $J_1$-$J_2$ model discussed above in Sec.~\ref{SecJ1J2Model}.  This in turn suggests that the observed tetragonal to orthorhombic structural distortions in the FeAs-based parent compounds may be driven by magnetic interactions.\cite{Yildirim2009}  Electronic structure calculations alternatively indicate that the observed stripe state is in itinerant spin density wave (SDW) state that arises from Fermi surface nesting of the itinerant electron and hole Fermi surfaces.\cite{mazin2008, dong2008b}

If $J_{1a}$ and $J_{1b}$ are different in the low-temperature orthorhombic structure, the spin Hamiltonian~(\ref{EqJ1J2}) becomes
\bea
{\cal H} &=& J_{1a} \sum_{\langle ij \rangle_{a}}{\bf S}_i \cdot {\bf S}_j + J_{1b} \sum_{\langle ij \rangle_{b}}{\bf S}_i \cdot {\bf S}_j \nonumber\\
&+& J_2 \sum_{\langle ik \rangle_{ab}}{\bf S}_i \cdot {\bf S}_k + J_{1c} \sum_{\langle i\ell \rangle_{c}}{\bf S}_i \cdot {\bf S}_\ell.
\label{EqJ1aJ1bJ2Jc}
\eea
The stripe state in Fig.~\ref{Stripe_Mag_Struct}, now termed here the ``stripe-$b$'' state with the stripes running along the $b$-axis as observed, competes not only with the N\'eel state, but also with the ``stripe-$a$'' state with the stripes running along the $a$-axis.  The classical energies of the three ordered states are
\bea
E_{\rm stripe\ b} &=& NS^2( -2J_{1a} + 2J_{1b} -4 J_2)/2,\nonumber\\
E_{\rm stripe\ a} &=& NS^2( 2J_{1a} - 2J_{1b} -4 J_2)/2,\label{Eqsstripes}\\
E_{\rm Neel} &=& NS^2(-2J_{1a} -2J_{1b} + 4J_2)/2,\nonumber
\label{EqENeel}
\eea
where $J_2 > 0$.  From Eqs.~(\ref{Eqsstripes}) one obtains two classical stability conditions for the stripe-$b$ state in Fig.~\ref{Stripe_Mag_Struct} as
\be
J_{1a} > J_{1b}\ \ \ \ {\rm and}\ \ \ J_2 > J_{1b}/2.
\label{EqStripeStab}
\ee
This solution allows $J_{1b}$ to be negative (ferromagnetic).  As seen from Fig.~\ref{Stripe_Mag_Struct}, a situation with antiferromagnetic $J_{1a},\ J_2 > 0$ and ferromagnetic $J_{1b} < 0$ would result in a nonfrustrated stripe-$b$ magnetic structure.

\subsubsection*{The $c$-axis Component of the Magnetic Structures}

Regarding the $c$-axis ordering, the ordering between nearest-neighbor Fe atoms along the $c$-axis in adjacent planes is generally found to be antiferromagnetic (AF, $J_c > 0$), as shown by the experimental data in Tables~\ref{LoTStructData1111} and~\ref{LoTStructData122} in the Appendix.  For ferromagnetic alignment along the $c$-axis, the ordering wave vector in the 122-, 111-, and 1111-type compounds is (100)~r.l.u.\ in orthorhombic notation and in the 11-type compounds (see also Sec.~\ref{Sec11-type} below) is (100)~r.l.u.\ in tetragonal notation.  

There is some confusion in the literature about how to write the three-dimensional (3D) AF propagation vector for the Fe-based compounds if the Fe spin alignment along $c$ is AF\@.  Let the Fe$_2$ interplane distance be $d$.  In all the Fe-based compounds, when the interplane Fe spin alignment is AF the wavelength of the AF propagation vector component in the $c$-axis direction is $2d$.  As discussed above in Sec.~\ref{SecBZ}, the 122-type compounds are body-centered-tetragonal with two Fe$_2$ layers per unit cell, so that $2d = c$.  Therefore for the 122-type compounds, the 3D AF stripe-$b$ propagation vector is (1,0,1)~r.l.u.\ in orthorhombic notation.  On the other hand, the 11-, 111-, and 1111-type compounds are primitive tetragonal containing one Fe$_2$ layer per unit cell.  Therefore with respect to the $c$-axis component of antiferromagnetic propagation vector, the wavelength is $2d= 2c$.  The corresponding 3D propagation vectors for AF interplane spin alignment are then $\left(1,0,\frac{1}{2}\right)$~r.l.u.\ in orthorhombic notation for stripe-$b$ in-plane ordering in the 111- and 1111-type compounds and the same wave vector in tetragonal notation for diagonal double stripe in-plane ordering in the 11-type compounds (see also Sec.~\ref{Sec11-type} below).  

In some papers on the 1111-type compounds, the AF propagation vector for AF interplane spin alignment is instead given as (1,0,1)~r.l.u.,\cite{Xiao2009c} which is implicitly with respect to the reciprocal lattice of the \emph{magnetic} real space lattice instead of the conventional notation as the reciprocal lattice of the \emph{crystallographic} lattice.  These two reciprocal lattices are different when $2d = 2c$ for AF Fe interlayer spin alignment.  In this review and in the tables in the Appendix, the AF ordering wave vector is always given in terms of the reciprocal lattice of the crystallographic lattice, in agreement with convention.

Interestingly, below 15~K the $c$-axis Fe spin alignment in NdFeAsO ($T_{\rm N~Fe} = 137$~K) switches from antiferromagnetic to ferromagnetic, even though the Nd moments order at a lower temperature of 6~K.\cite{Tian2010}

\subsubsection{${\rm EuFe_2As_2}$}

\begin{figure}
\includegraphics[width=3.3in]{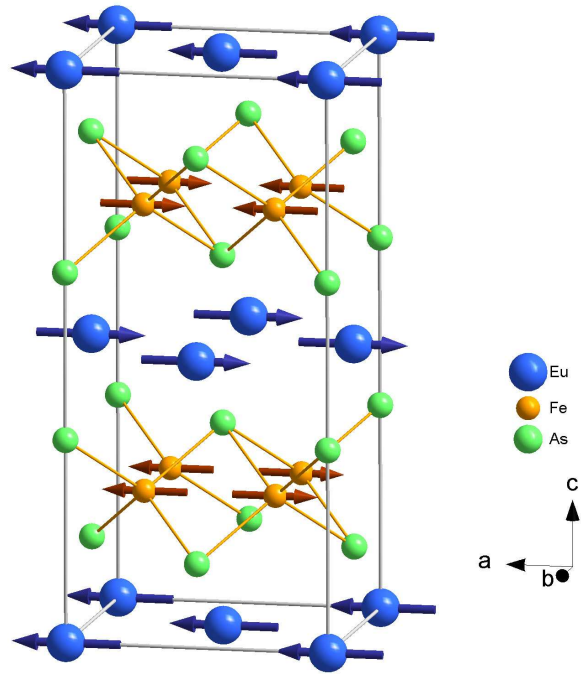}
\caption{(Color online) Magnetic structure of ${\rm EuFe_2As_2}$ at 2.5~K determined from neutron diffraction measurements.\cite{Xiao2009}  The unit cell shown is the orthorhombic chemical and magnetic unit cell.  All spins point along the $a$-axis.  The Fe spins order in an antiferromagnetic stripe structure below 190~K with orthorhombic ordering wave vector (101), with the ferromagnetically aligned stripes running along the $b$-axis. The Eu spins $S = 7/2$ order in an A-type antiferromagnetic structure below 19~K with wave vector (001), where the Eu spins are ferromagnetically aligned within a plane parallel to the $a$-$b$ plane, and the spins in adjacent planes are antiferromagnetically aligned.  Reprinted with permission from Ref.~\onlinecite{Xiao2009}.  Copyright (2009) by the American Physical Society.}
\label{EuFe2As2_mag_struct}
\end{figure}

The magnetic ordering in the 122-type compound ${\rm EuFe_2As_2}$ is interesting because it contains two different magnetic species.  In addition to the Fe sublattice, the Eu sublattice consists of spin-only Eu$^{+2}$ ions with spin $S = 7/2$.  The Fe sublattice exhibits long-range SDW ordering below the tetragonal-orthorhombic transition temperature of 190~K, whereas the Eu lattice exhibits an independent antiferromagnetic ordering transition at $T_{\rm N} = 19$~K.\cite{Xiao2009, Herrero-Martin2009}  The Fe sublattice in Fig.~\ref{EuFe2As2_mag_struct} orders in a stripe structure with the ordered moment along the orthorhombic $a$-axis as in the other $A{\rm Fe_2As_2}$ ($A =$ Ca, Sr, Ba) compounds.  However, the Eu sublattice orders in a so-called A-type antiferromagnetic arrangement, where the Eu spins within a layer are all ferromagnetically aligned, but the alignment between layers is antiferromagnetic.  The ordered Eu moment is along the long orthorhombic $a$~axis as is the ordered Fe moment.  Interestingly, the neutron diffraction data indicated weak or negligible coupling between the Fe and Eu spin lattices.\cite{Xiao2009}

\subsubsection{\label{Sec11-type} 11-type {\rm Fe}$_{1+y}${\rm (Te}$_{1-x}${\rm Se}$_x${\rm )} Compounds}

\begin{figure}
\includegraphics[width=2.5in]{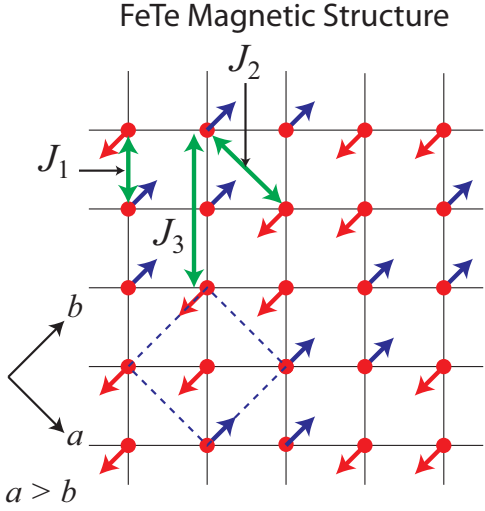}
\caption{(Color online) Commensurate collinear in-plane antiferromagnetic diagonal double stripe structure of Fe$_{1+y}$Te for small values of the excess Fe concentration $y$.  The in-plane ordering wave vector is ${\bf Q}_{\rm AF} = \left(\frac{1}{2},0\right)$~r.l.u.\ in tetragonal notation (see Table~\ref{AFPropVectors}).  The tetragonal/monoclinic basal plane $a$- and~$b$-axes and basal plane crystallographic unit cell outline (dashed lines) are indicated.  }
\label{FeTe_Mag_Struct}
\end{figure}

\begin{table}
\caption{\label{FeTeJsEs} Values of the exchange constants in the $J_1$-$J_2$-$J_3$ Heisenberg model for FeTe and FeSe derived from spin-polarized band theory.\cite{Ma2009a}  Also included are the energies of the four magnetic structures in Fig.~\ref{Possible_Mag_Structures_All_Reduce} calculated from spin-polarized band calculations (bottom),\cite{Ma2009a} and our calculations of the classical energies from Eqs.~(\ref{OrdEnergies}) using the listed exchange constants.  Here $S$ is the spin of the Fe atom and $N$ is the number of Fe atoms.  One sees that the antiferromagnetic diagonal double stripe phase is predicted to be the most stable structure for FeTe and and the single stripe structure for FeSe in the local moment picture.  The former prediction is in agreement with experiment, as shown in Fig.~\ref{FeTe_Mag_Struct}.  The itinerant polarized band calculations also correctly predict the ground state magnetic structure for FeTe, where the listed energy values are with respect to the energy of the nonmagnetic state.}
\begin{ruledtabular}
\begin{tabular}{l|cc}
Quantity&  FeTe & FeSe \\
\hline
Heisenberg $J_1$-$J_2$-$J_3$ model\\
$J_1S^2$ (meV) &  2.1 & 71 \\
$J_2S^2$ (meV) &  15.8 & 48 \\
$J_3S^2$ (meV) &  10.1 & 8.5 \\
\hline
$E_{\rm ferromagnetic}/N$ (meV) & 28.0 & 128 \\
$E_{\rm double\ stripe}/N$ (meV) & 1.1 & 35.5 \\
$E_{\rm doub\ diag\ stripe}/N$ (meV) & $-10.1$ & $-8.5$ \\
$E_{\rm stripe}/N$ (meV) & $-5.7$ & $-39.5$ \\
$E_{\rm Neel}/N$ (meV) & 23.8 & $-14.5$ \\
\hline
Spin-polarized band theory\\
$E_{\rm ferromagnetic}/N$ (meV) & $-90$ & 183 \\
$E_{\rm doub\ diag\ stripe}/N$ (meV) & $-166$ & $-89$ \\
$E_{\rm stripe}/N$ (meV) & $-156$ & $-152$ \\
$E_{\rm Neel}/N$ (meV) & $-98$ & $-101$ \\
\end{tabular}
\end{ruledtabular}
\end{table}

The magnetism of Fe$_{1+y}$Te$_{1-x}$Se$_x$ is complicated because of the $y$ excess Fe atoms in interstitial Fe(2) positions in the Te/Se layers (see Fig.~\ref{Structures}), which appear to carry local moments that affect both the paramagnetic and magnetically ordered states, and because both incommensurate and commensurate antiferromagnetic structures occur for $x = 0$ depending on the excess iron concentration $y$.\cite{Lynn2009}  The experimental magnetic susceptibiities for Fe$_{1+y}$Te compounds exhibit a Curie-Weiss component that is evidently associated with magnetic moments on the excess Fe atoms as discussed above in Sec.~\ref{SecChi}, and a theoretical density functional study of this system also strongly indicated that the excess Fe atoms are magnetic.\cite{Zhang2009a}  

\subsubsection*{a.  {\rm Fe}$_{1+y}${\rm Te}}

Bao and coworkers have shown that in Fe$_{1+y}$Te crystals, incommensurate antiferromagnetic (AF) ordering at large $y= 0.165$ and~0.141 gives way to commensurate AF ordering at smaller $y = 0.076$.\cite{Bao2009}  The commensurate antiferromagnetic in-plane component of the AF structure of Fe$_{1.05}$Te (Ref.~\onlinecite{Martinelli2010}), Fe$_{1.068}$Te (Ref.~\onlinecite{Li2009c}) and Fe$_{1.076}$Te (Ref.~\onlinecite{Bao2009}) is shown in Fig.~\ref{FeTe_Mag_Struct}.  Along the $c$-axis, the spins are antiferromagnetically aligned with each other.  The in-plane structure is an interesting antiferromagnetic diagonal double stripe structure as also shown above in Fig.~\ref{Possible_Mag_Structures_All_Reduce}.  Referring to Table~\ref{AFPropVectors}, the three-dimensional antiferromagnetic propagation vector is $\left(\frac{1}{2},0,\frac{1}{2}\right)$~r.l.u.  This magnetic structure is in strong contrast with that of the 11-, 122- and 1111-type compounds in Fig.~\ref{Stripe_Mag_Struct}, because even though the latter three classes of compounds have an in-plane stripe structure, those stripes are single stripes and they run along the Fe square lattice axis directions with in-plane propagation vector $\left(\frac{1}{2},\frac{1}{2}\right)$~r.l.u., whereas in FeTe, there are double stripes running along the Fe square lattice diagonal directions.  Furthermore, the ordered moment direction is along the shorter monoclinic $b$-axis in FeTe, but is along the longer orthorhombic $a$-axis in the other compounds.  On the other hand, an important similarity is that the stripe axis, which is the $b$-axis in both the low-temperature monoclinic structure here and in the orthorhombic structure for the other materials, is the shorter of the two basal plane axes, suggesting that a similar type of magnetostriction effect occurs in the antiferromagnetically ordered phase of both classes of materials.

A similar Fe antiferromagnetic diagonal double stripe structure was found from neutron diffraction measurements on a polycrystalline sample of the compound ${\rm La_2O_2Fe_2OSe_2}$, even though there was no obvious evidence of a temperature-induced lattice distortion.\cite{Free2010}  In this compound the Fe atoms form square lattice layers as in the Fe-based superconductors, and O atoms are incorporated into the Fe$_2$O layers via an anti-CuO$_2$ layer structure. Thus, O atoms are at the centers of one-half of the Fe$_4$ squares, and are arranged in an ordered way in the Fe layer.  The N\'eel temperature is $T_{\rm N} \approx 90$~K and the ordered moment is 2.83(3)~$\mu_{\rm B}$ per Fe atom, far larger than in the FeAs-based superconductor materials but similar to values for the Fe$_{1+y}$Te compounds, as listed in Table~\ref{FeOrdering}.  The ordered moments are in the $ab$-plane but the relative orientation of the ordered moments in the two Fe spin sublattices was not determined.  A resistivity measurement on a polycrystalline sample by Zhu et al.\ indicated semiconducting behavior with an activation energy of 0.19~eV, and their $\chi(T)$ data indicated $T_{\rm N} = 93$~K, consistent with the neutron diffraction measurements.\cite{Zhu2010}

From Table~\ref{AFPropVectors} the in-plane antiferromagnetic propagation vector for the diagonal double stripe structure ${\bf Q}_{\rm AF} = \left(\frac{1}{2}, 0\right)$~r.l.u.\ is at a 45$^\circ$ angle to the Fermi surface nesting vector $\left(\frac{1}{2}, \frac{1}{2}\right)$~r.l.u.\ between the electron and hole pockets from Figs.~\ref{LaFeAsOBS} and~\ref{FigFS}.  From this one might infer that the magnetic ordering in FeTe has to arise from interactions between local moments.  However, from spin-polarized electronic structure calculations, Ma et al.\ correctly predicted the observed antiferromagnetic diagonal double stripe structure in FeTe as shown at the bottom of Table~\ref{FeTeJsEs}.\cite{Ma2009a}  In addition, their band calculations predict an ordered moment of 2.2--2.6~$\mu_{\rm B}$/Fe,\cite{Ma2009a} which is in agreement with the experimental data for FeTe in Table~\ref{FeOrdering} above.  

To further address this issue, Ma et al.\ also calculated from their spin polarized band theory results the effective exchange constants between the Fe magnetic moments in FeTe and FeSe in the local moment $J_1$-$J_2$-$J_3$ Heisenberg model, as listed in Table~\ref{FeTeJsEs}.\cite{Ma2009a} We calculated the classical energies of the different ordered states that are listed in Table~\ref{FeTeJsEs}, using their exchange constants and our energy expressions in Eqs.~(\ref{OrdEnergies}).  Our results reproduce the results given by Ma et al.  In particular, one sees that the antiferromagnetic  diagonal double stripe structure has the lowest energy for FeTe, and the (single) stripe structure as in the 1111-, 111- and 122-type FeAs-based compounds has the lowest energy for FeSe.  Due to the lack of Fermi surface nesting in the ${\bf Q}_{\rm AF}$ direction and the success of the local moment model in explaining the observed magnetic structure of FeTe, the authors concluded that the local moment model is the most viable model to explain the observed antiferromagnetic structure of FeTe.\cite{Ma2009a}  In addition to the results described in their paper, Ma et al.\ also calculated the wave vector dependent susceptibility $\chi({\bf Q})$ of the itinerant electrons and found that it peaks at the nesting wave vector(s) ${\bf Q} = (\frac{1}{2},\frac{1}{2})$~r.l.u.\ in tetragonal reciprocal lattice units, as in Fig.~\ref{Mazin_chi} above for LaFeAsO, but not at the observed diagonal double stripe wave vector ${\bf Q} = (\frac{1}{2},0)$~r.l.u., thus confirming their local moment magnetism interpretation (Z.-Y.~Lu, private communication).  As noted above in Sec.~\ref{Sec11BSARPES}, ARPES data for FeTe are consistent with local moment magnetism rather than itinerant magnetism.

Han and Savrasov have suggested that all eight of the 4$s$ and 3$d$ valence electrons of the $y$ excess Fe atoms in Fe$_{1+y}$Te are donated to the FeTe layers, leaving those Fe atoms in a nonmagnetic [Ar]$^{+8}$ electronic configuration.\cite{Han2010}  This changes the Fermi surface nesting from the above ${\bf Q}_{\rm nesting} = \left(\frac{1}{2},\frac{1}{2}\right)$ to the observed magnetic ordering wave vector $\left(\frac{1}{2},0\right)$~r.l.u.\ in tetragonal notation, in which case the magnetic ordering results from itinerant magnetism.\cite{Han2010}  This scenario conflicts with both experiment (see Sec.~\ref{SecChi} above) and theory which indicate that the excess Fe atoms have a formal oxidation state Fe$^{+1}$ and carry a large magnetic moment of 2.4~$\mu_{\rm B}$.\cite{Zhang2009a}  Furthermore, as discussed above in Sec.~\ref{Sec11BSARPES}, ARPES experiments on a single crystal of Fe$_{1+y}$Te found no evidence for doping by excess Fe atoms to within the resolution, no evidence for Fermi surface gapping expected from an itinerant SDW, and no evidence for Fermi surface nesting at the AF ordering wave vector.  Furthermore, the magnetic susceptibility data for Fe$_{1+y}$Te discussed above in Sec.~\ref{SecChi} indicate that a local moment model for the Fe magnetism is appropriate.

Thus the preponderance of evidence indicates, amazingly, that in contrast to the itinerant antiferromagnetic spin density wave ordering in the 1111- 111- and 122-type compounds, the antiferromagnetic ordering in Fe$_{1+y}$Te arises from ordering of local magnetic moments.

In spite of the success of Ma et al.'s predictions for the ordered magnetic structure of FeTe,\cite{Ma2009a} their prediction of antiferromagnetic stripe ordering in FeSe (see Table~\ref{FeTeJsEs}), and the same earlier prediction for FeSe by Subedi et al.\ from Fermi surface nesting,\cite{Subedi2008a} has not held up to scrutiny.   All experimental studies to date on pure FeSe and solid solutions of FeSe with FeTe indicate that long-range antiferromagnetic ordering only occurs near the Te-rich end of the phase diagram in Fig.~\ref{KatayamaFig2}, and in particular, not for FeSe.  Upon replacing Te by Se,  experimentally it is found that there is no long-range magnetic order in the compositions FeTe$_{0.584}$Se$_{0.416}$ and FeTe$_{0.507}$Se$_{0.493}$.\cite{Li2009c}  However, evidence was found for short-range magnetic order in these compositions.  

In a theoretical study, Fang et al.\ explained the transition from commensurate to incommensurate ordering with increasing $y$ in Fe$_{1+y}$Te using a local moment $J_{1a}$-$J_{1b}$-$J_{2a}$-$J_{2b}$-$J_{3}$-$J_{c}$ model.\cite{Fang2009a}  

\subsubsection*{b.  {\rm Fe}$_{1+y}${\rm (Te}$_{1-x}${\rm Se}$_x${\rm )}, $x > 0$}

From Fig.~\ref{KatayamaFig2} and Table~\ref{FeOrdering}, both the N\'eel temperature and the ordered moment decrease with increasing $x$ from $x = 0$ to $x \approx 0.1$.  For $0.1 \lesssim x \lesssim 0.45$, this long-range diagonal double stripe order at wave vector $\left(\frac{1}{2},0,\frac{1}{2}\right)$~r.l.u.\ is replaced by static short-range ``spin-glass'' order\cite{Bao2009, Katayama2010, Liu2010, Wen2009, Khasanov2009, Babkevich2010, Xu2010} with a well-defined but slightly incommensurate wave vector $\left(\frac{1}{2}-\delta,0,\frac{1}{2}\right)$~r.l.u.\ with $\delta \sim 0.05$, that is thus similar to the long-range diagonal double stripe ordering wave vector that occurs for $x \lesssim 0.1$ and small $y$.   Xu et al.\ report that the short-range ordering becomes more two-dimensional with increasing $x$.\cite{Xu2010}  The in-plane component of the observed static short-range ordering wave vector is at an approximately 45$^\circ$ angle to the in-plane nesting wave vector $\left(\frac{1}{2},\frac{1}{2}\right)$~r.l.u.\  between the hole and electron Fermi surfaces (see Fig.~\ref{Ordering_Wave_Vectors}).  

\subsection{Electron Correlation Strength}

\subsubsection{Introduction}

In this section we discuss where the FeAs-based materials lie with respect to the limits of strongly correlated systems like the layered cuprate high-$T_{\rm c}$ superconductors and weakly correlated metals such as Cr.  In strongly-correlated magnetic systems, one expects to see well-defined local magnetic moments that exhibit magnetic ordering due to interactions between them.  For weakly correlated systems, the magnetic ordering is driven by Fermi surface features such as nesting between different Fermi surface sheets, which means that the Lindhard susceptibility of the conduction electrons $\chi({\bf Q})$ peaks at the ordering wavevector $Q_{\rm AF}$.  One estimate of the degree of electron correlation can be defined by the ratio of the Hubbard on-site Coulomb repulsion $U$ to the conduction electron bandwidth $W$, where the dividing line between weakly and strongly correlated systems is at $U/W \sim 1$.  For FeAs-based superconductors, the width of the entire $d$-band manifold near $E_{\rm F}$ is $\sim 4$~eV.\cite{djsingh2008, djsingh2008b}  Theoretical arguments have been made that these materials should be viewed as weakly correlated,\cite{Yildirim2009} strongly correlated where the parent compounds are ``bad metals'' on the verge of a metal-insulator transition,\cite{Si2008, Haule2009, Dai2009, Haule2008, Craco2008, Si2009} or somewhere in between.\cite{Johannes2009}  Several authors have argued theoretically that the on-site Hund's coupling $J_{\rm H}$ rather than $U$ is primarily responsible for the electron correlations.\cite{Haule2009, Wang2009, Johannes2009, Mazin2009}  An interesting and informative early overview of these issues was given by Tesanovic.\cite{Tesanovic2009}

Anisimov, Skornyakov and coworkers have extensively evaluated theoretically the degree of electron correlations in the Fe-based materials.\cite{Anisimov2009, Skornyakov2009}  They carried out combined LDA + DMFT (local density approximation combined with dynamical mean field theory) calculations of the electronic structure and electronic correlations in LaFeAsO, from which they obtained an average Coulomb repulsion $\bar{U} = 3$--4~eV and $J_{\rm H} = 0.8$~eV.\cite{Anisimov2009}  However, they noted that the $\bar{U}$ is strongly screened due to strong covalent bonding between the Fe and As atoms.  These authors compared their theoretical predictions with the observed x-ray and photoemission spectroscopy results and concluded that the electron correlations are moderate, with a relatively small effective mass enhancement $m^*/m_{\rm b} \approx 2$, where $m_{\rm b}$ is the band mass, consistent with the renormalization of the LDA band structure needed to fit ARPES data.  Furthermore, they observed that the spectra for the FeAs-based materials are very different than for the Mott insulator FeO, and showed no evidence for the existence of a lower Hubbard band that would be expected in a strongly correlated electron system.  They carried out similar calculations for LaFePO, with similar results.\cite{Skornyakov2010}  Aichhorn et al.\ have carried out similar LDA + DMFT calculations on the same compound LaFeAsO and also concluded that this compound has intermediate strength correlations with many-body mass enhancements of order two.\cite{Aichhorn2009}  They concluded that LaFeAsO is not close to a Mott-Hubbard metal-insulator transition.

\subsubsection{\label{BadMetals?} Electrical Conductivity: ``Bad Metals''?}

The FeAs-based materials have sometimes been characterized as ``bad metals'' simply because the electrical conductivity at room temperature is low compared with metals like Cu (see Table~\ref{Opticsdata3} above).  One can write the conductivity as $\sigma = ne\mu$, where $n$ is the carrier concentration, $e$ is the electron charge, and $\mu$ is the mobility.  Therefore, low conductivities can come simply from low carrier concentations $n$ which is the case for the Fe-based semimetals.  

In condensed matter physics, the term ``bad metal'', originally coined by Emery and Kivelson,\cite{Emery1995} has a specific meaning whereby the calculated mean free path $\ell$ for conduction electron scattering is of order or less than an interatomic distance: $\ell/d \lesssim 1$.\cite{Emery1995, Allen2002}  In this case, the concept of mean free path loses its meaning, the wavevector is no longer a good quantum number (the electron excitations are ``incoherent'') and the so-called quasiparticle weight at the Fermi energy falls to a small value or zero.  A complementary criterion for a bad metal is $k_{\rm F}\ell \lesssim 1$, where $k_{\rm F}$ is the Fermi wavevector.  However, $k_{\rm F}$ is ill-defined if there is more than one conduction band and/or if the Fermi surface(s) is(are) not spherical (3D bands) or cylindrical (2D bands).  Bad metal behavior is often considered to be indicative of strong electron correlations.

According to Emery and Kivelson,\cite{Emery1995} ``bad metals'' do not exhibit saturation of the resistivity with increasing temperature.  Therefore, one can pass from a ``good metal'' regime where $k_{\rm F}\ell \gg 1$ at low temperatures to the bad metal regime $k_{\rm F}\ell \lesssim 1$ at high temperatures.  However, quoting Emery and Kivelson: ``The failure of bad metals to exhibit resistivity saturation strongly suggests that any theory based on conventional quasiparticles with more or less well-defined crystal momenta suffering occasional scattering events does not apply. \emph{Since there is no crossover in the temperature dependence of the resistivity as the temperature is lowered, this conclusion applies by continuity even at lower temperatures where the putative mean free path deduced from the measured values of the resistivity would not, of itself, rule out the possibility of quasiparticle transport.}  In other words, a bad metal behaves as if it is a quasiparticle insulator which is rendered metallic by collective fluctuations.''\cite{Emery1995} (Their emphasis).  However, Haule and Kotliar have found that the FeAs-based materials exhibit a coherent to incoherent conduction carrier crossover on increasing the temperature above a coherence scale $T^*$ (see also Sec.~\ref{SecItinLocMag} below).\cite{Haule2009}  In the context of this review of the puzzle of superconductivity in the Fe-based materials, it seems most logical and relevant to classify specific Fe-based materials as bad metals, or not, depending on their \emph{low-temperature} ($T \gtrsim T_{\rm c}$) normal state properties, because the low-temperature properties are most relevant to the occurrence and mechanism of superconductivity in these materials.  It remains to be seen if this is a valid approach or not in the context of ``bad metals.''  Allen emphasized in 2002 that there was  no accepted theory for bad metals at that time.\cite{Allen2002}

The usual expression for the electrical conductivity $\sigma$ of a one-band conductor was given above in Eq.~(\ref{EqSigma3}).  To make contact with $\ell$, one makes the replacement $\tau = \ell/v_{\rm F}$, where $v_{\rm F} = \hbar k_{\rm F}/m^*$ is the Fermi velocity and $\hbar$ is Planck's constant $h$ divided by $2\pi$.  Then Eq.~(\ref{EqSigma3}) becomes
\begin{equation}
\sigma = \frac{n e^2\ell}{m^*v_{\rm F}} = \frac{n e^2\ell}{\hbar k_{\rm F}}.
\label{EqSigma2}
\end{equation}
For $n_{\rm band}$ equally conducting bands with the same $k_{\rm F}$, the two expressions on the right would each be multiplied by $n_{\rm band}$. 

Here we will make an estimate of the product $k_{\rm F}\ell$ for a two-dimensional (2D) band with a cylindrical Fermi surface, because the expression obtained is simple with only two clearly defined parameters contained in it, as opposed to the 3D case.  In this 2D case one obtains $n = k_{\rm F}^2/(2\pi \Delta c$), where $\Delta c$ is the distance between conducting layers, yielding\cite{Qazilbash2008} from Eq.~(\ref{EqSigma2})
\begin{equation}
k_{\rm F}\ell = \frac{h \Delta c}{\rho_{ab}e^2} = 0.258 \frac{\Delta c}{\rho_{ab}},\ \ \ ({\rm 2D\ conduction})
\label{EqkFl}
\end{equation}
where the second equality on the right is for $\Delta c$ in \AA\ and the electrical resistivity $\rho_{ab}$ ($= 1/\sigma_{ab}$) in m$\Omega$~cm.  Remarkably, this expression only contains two easily measured and unambiguous quantities $\rho_{ab}$ and $\Delta c$.  For $n_{\rm band}$ equally conducting bands with the same $k_{\rm F}$, the value of $k_{\rm F}\ell$ from Eq.~(\ref{EqkFl}) would be divided by $n_{\rm band}$.

\begin{figure}
\includegraphics[width=3.3in]{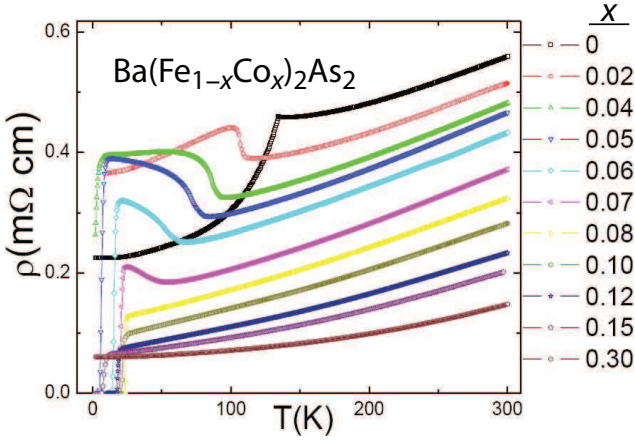}
\caption{(Color online) In-plane resistivity $\rho$ versus temperature $T$ of Ba(Fe$_{1-x}$Co$_x$)$_2$As$_2$ crystals for various values of $x$ as indicated on the right edge of the figure.  Reprinted with permission from Ref.~\onlinecite{Fang2009}.  Copyright (2009) by the American Physical Society.}
\label{Ba(FeCo)2As2_rho}
\end{figure}

A typical range of in-plane resistivity values for the FeAs-based systems is shown for the Ba(Fe$_{1-x}$Co$_x$)$_2$As$_2$ system in Fig.~\ref{Ba(FeCo)2As2_rho} for $x$ values from the undoped value $x = 0$ to the optimum doping $x \sim 0.08$ to heavily overdoped $x = 0.3$.\cite{Fang2009}  Note that the $ab$-plane resistivity at 300~K for $x = 0$ is about 30\% larger than in the different crystal in the bottom panel of Fig.~\ref{FigChiRhoBaFe2As2} above, indicating the variability between different crystals and measurements of nominally the same material.  For the optimum superconducting composition with $x = 0.08$, the normal state resistivity at low temperatures is seen to be $\rho_{ab} \approx 0.12~$m$\Omega$~cm.  Then utilizing Eq.~(\ref{EqkFl}) with $\Delta c = 6.5$~\AA\ gives $k_{\rm F}\ell \approx 14$.  According to the Hall coefficient data in Refs.~\onlinecite{Rullier-Albenque2009} and \onlinecite{Fang2009}, in a two-band model with one band an electron band and the other a hole band, the electron band contributes most strongly to the conductivity in this system.  The same conclusion was reached from resistivity and Hall coefficient measurements on single crystals of BaFe$_2$(As$_{1-x}$P$_x$)$_2$.\cite{Kasahara2009}  Therefore the one-band estimate $k_{\rm F}\ell \approx 14 \gg 1$ appears to be reasonable and indicates that the Ba(Fe$_{1-x}$Co$_x$)$_2$As$_2$ system is a coherent metal, i.e., not a ``bad metal.''  

Equation~(\ref{EqkFl}) is identical to the equation Si and Abrahams used early on to prove that the iron arsenides are bad metals, using LaFeAsO as an example.\cite{Si2008}  They used $\Delta c = 8.7$~\AA\ and $\rho(300~{\rm K}) = 5$~m$\Omega$~cm to obtain $k_{\rm F}\ell \approx 0.5$ from Eq.~(\ref{EqkFl}), and thus claimed that this compound is a bad metal.  However, those resistivity measurements were for a polycrystalline sample, and it is now clear that their $\rho(300~{\rm K})$ value for the in-plane resistivity that they used for the calculation of $k_{\rm F}\ell$ was at least an order of magnitude too large, and that the actual value is $k_{\rm F}\ell \gtrsim 5$ at room temperature.  The value of $k_{\rm F}\ell$ would further increase on cooling because the resistivity decreases.  The other criterion used in Ref.~\onlinecite{Si2008} to claim that LaFeAsO is a bad metal was that there was no Drude peak in the in-plane optical conductivity of LaFeAsO, which we now know is not correct from more recent optical measurements on single crystals.\cite{Chen2009}  

\subsubsection{\label{SecQuantumOsc}  Quantum Oscillation Experiments}

Quantum oscillations in the magnetization (de Haas van Alphen effect) and/or in the resistivity (Shubnikov-de Haas effect) versus applied magnetic field have been observed for single crystals of superconducting LaFePO with $T_{\rm c} = 6$~K (Ref.~\onlinecite{Coldea2008}) and nonsuperconducting SrFe$_2$As$_2$,\cite{Sebastian2008} BaFe$_2$As$_2$,\cite{Analytis2009} CaFe$_2$P$_2$,\cite{Coldea2009} and SrFe$_2$P$_2$.\cite{Analytis2009a}  \emph{The quantum oscillations cannot be observed unless the conduction electron states are coherent}.  In the cases of SrFe$_2$As$_2$ and BaFe$_2$As$_2$, the low-temperature Fermi surfaces (below the SDW transition temperatures) are in general agreement with LDA band structure calculations of the reconstructed Fermi surfaces arising from a nested-Fermi-surface driven SDW\@.  The mean-free-paths for three bands observed in CaFe$_2$P$_2$ were found to be 1900, 710, and 860~\AA, respectively, much larger than a lattice parameter.\cite{Coldea2009}  These quantum oscillation measurements and large mean-free paths for these five compounds indicate that these compounds are coherent metals.  The many-body conduction carrier mass enhancements found in the measurements are rather small, of order 1~to 2~times the LDA band structure values.

de Haas-van Alphen (dHvA) magnetization oscillation measurements versus applied magnetic field were also carried out on superconducting ${\rm KFe_2As_2}$ crystals with $T_{\rm c} = 3$~K.\cite{Terashima2010}  These measurements are important because, as for LaFePO above, there is no intrinsic crystallographic disorder in this superconducting compound.  The measurements indicated very strong enhancements of the current carrier masses of from 2.7 to 24 times the respective band mass, depending on the band.\cite{Terashima2010}  These large enhancements are qualitatively consistent with the large enhancement of the normal state electronic linear heat capacity coefficient $\gamma = 69$~mJ/mol\,K$^2$ (Ref.~\onlinecite{Fukazawa2009}) by a factor of almost seven above the LDA band structure value\cite{Terashima2010} of 10.1~mJ/mol\,K$^2$.  The authors stated, ``we do not ascribe the main origin of the mass enhancements specifically to low-energy spin fluctuations, but we ascribe it to band narrowing due to more general electronic correlations arising from the local Coulomb interaction on the Fe $3d$ shell.''\cite{Terashima2010}  It is interesting and important that despite these sometimes very large carrier mass enhancements, the corresponding conduction carrier conduction must still be coherent since that is required for observation of the dHvA effect.

\subsubsection{\label{SecItinLocMag} Itinerant versus Local Moment Magnetism}

The degree of electron correlation in a material is often associated with whether the material exhibits itinerant (weak electron correlation) or local moment (strong electron correlation) magnetism.  In real materials, however, because the possible degree of electron correlation forms a continuum, the possible magnetic behaviors also form a continuum between these two extreme limits.

The ordered Fe magnetic moments range from 0.25(7)~$\mu_{\rm B}$ for NdFeAsO (Ref.~\onlinecite{Ychen2008}) [$T_{\rm S} = 150$~K (Ref.~\onlinecite{Qiu2008}), $T_{\rm N} = 141(6)$~K (Ref.~\onlinecite{Ychen2008})]  to $0.94(4)~\mu_{\rm B}$ in SrFe$_2$As$_2$ [$T_{\rm S} = T_{\rm N} = 220(1)$~K].\cite{Zhao2008c}  In  these parent compounds, one expects a formal oxidation state Fe$^{+2}$, which would formally give a $d^6$ electronic configuration.  If the $d$-electrons were localized in a high-spin state in the five $d$-orbitals, one expects a spin $S = 2$ and an ordered moment $\mu = gS\mu_{\rm B} = 4~\mu_{\rm B}$ assuming a $g$-factor of $g = 2$, which is a far larger ordered moment than the observed ones.  

The relatively small and variable $\mu$ values observed for the parent compounds instead suggest that the AF order is a spin-density-wave (SDW) arising from itinerant electrons, although alternative views based on fluctuation and/or frustration effects in a large-local-moment system have been advanced as outlined above in Sec.~\ref{1111-111-122}.  This SDW has been argued to arise from Fermi surface nesting of the hole and electron Fermi surface sheets, and/or from one-electron band effects.\cite{Johannes2009}  Furthermore, below $T_{\rm N}$, optical measurements on EuFe$_2$As$_2$, SrFe$_2$As$_2$ and BaFe$_2$As$_2$ show a strong reduction in the carrier density.\cite{Moon2010, Hu2008}  A reconstruction of the Fermi surface as expected for a conventional itinerant SDW is observed in BaFe$_2$As$_2$ by ARPES\cite{Shimojima2009, Kondo2009} that is consistent with the results of quantum oscillation experiments on SrFe$_2$As$_2$ (Ref.~\onlinecite{Sebastian2008}) and BaFe$_2$As$_2$.\cite{Analytis2009}  These studies thus indicate that the magnetic transition is due to itinerant electrons rather than to exchange interactions between local magnetic moments with fixed magnitude.  

A theoretical study by Ferndandes and Schmalian of the optical properties of both doped and undoped FeAs-based materials, and in particular, the interaction between superconductivity and long-range antiferromagnetic ordering, strongly indicated that the magnetic ordering is itinerant.\cite{Fernandes2010b}  Additional qualitative theoretical support for this conclusion is that the calculated superexchange interactions in the local moment picture (unphysically) depend on the antiferromagnetic structure.\cite{Yildirim2009, Belashchenko2008, Yaresko2009, Yin2008}  Three- and five-band itinerant magnetism models\cite{Knolle2010,Kaneshita2010} can quantitatively reproduce the results of magnetic inelastic neutron scattering measurements such as those shown in Fig.~\ref{CaFe2As2SpinWaves} below. 

Haule and Kotliar have argued that the FeAs-based materials exhibit a coherent itinerant magnetism to incoherent local moment magnetism crossover on increasing the temperature above a coherence scale $T^*$.\cite{Haule2009}  Thus the magnetic susceptibility is predicted to monotonically decrease with increasing temperature, from temperature-independent Pauli paramagnetism to Curie-Weiss-like local moment magnetism.  However, the observed susceptibility behavior for both the undoped and doped materials is qualitatively different from this prediction, increasing with increasing temperature as illustrated above in Fig.~\ref{FigChi(T)}.

From inelastic magnetic neutron scattering measurements, direct estimates of lower limits of the instantaneous effective moments in the paramagnetic state of ${\rm CaFe_2As_2}$ ($\mu_{\rm eff} = 0.47~\mu_{\rm B}$/Fe at a maximum integrated energy of 100~meV) and ${\rm Ba(Fe_{1.935}Co_{0.065})_2As_2}$ ($\mu_{\rm eff} = 0.31~\mu_{\rm B}$/Fe at a maximum integrated energy of 80~meV) have been obtained by Diallo et al.\cite{Diallo2010} and by us from the data in Fig.~5 of Lester et al.,\cite{Lester2010} respectively (see Sec.~\ref{Sec122typeneuts} below).  These values are much smaller than the value $\mu_{\rm eff} = 4.90~\mu_{\rm B}$/Fe expected for a localized spin $S = 2$ with $g$-factor $g = 2$.

When the parent 1111 and 122 FeAs-based compounds are doped, both the structural and magnetic transitions are completely suppressed at the doping level at which optimum $T_{\rm c}$ is attained as illustrated above in Fig.~\ref{FigBaKFe2As2_phase_diag}.  However, all experiments on the FeAs-based materials so far indicate that the magnetic structure is still, perhaps surprisingly, commensurate with the lattice upon doping towards the SDW-superconductivity phase boundary.  If the magnetic ordering is a SDW due to Fermi surface nesting, one might expect that the ordering wavevector should depend on the doping level.  However, electronic structure calculations indicate that the commensurability of the SDW ordering can indeed be retained on doping.  Yaresko et al.\ found that commensurate ordering is the most stable SDW state for up to 10\% electron doping of LaFeAsO$_{1-x}$F$_x$ (i.e., up to $x = 0.1$) and up to 25\% hole doping in (Ba$_{1-y}$K$_y$)Fe$_2$As$_2$ (i.e., up to $y = 0.5$).\cite{Yaresko2009}  Singh et al.\ have given a qualitative description in terms of the doping dependence of the Lindhard susceptibility $\chi(Q)$ of why commensurate ordering is stable up to a finite critical doping level when the nesting wave vector between the centers of the electron and hole Fermi surface pockets does not depend on doping as in the present systems at low doping.\cite{DJSingh2009b}

We note that from magnetic and electrical resistivity measurements, the related compounds LaCoAsO and LaCoPO with neighboring Co replacing Fe were found to be \emph{itinerant ferromagnets} with Curie temperatures of 66~K and 43~K and low ordered moments of 0.39 and 0.33~$\mu_{\rm B}$/Fe, respectively.\cite{Yanagi2008}  Above the respective Curie temperatures, the magnetic susceptibilities of these compounds were found to strongly decrease with increasing temperature, in contrast to the increase with increasing temperature seen for the Fe-based compounds in Figs.~\ref{FigChi(T)} and~\ref{YangFig2bc} above.

\subsubsection{\label{SecOptCorr} Optical Spectroscopy}

Haule and coworkers have predicted that the FeAs-based materials are strongly correlated and are near a correlation-induced metal-insulator transition as noted above.\cite{Haule2009, Haule2008}  Using dynamical mean field theory, they calculated that the metallic state has no Drude peak at zero frequency in the frequency-dependent optical conductivity.\cite{Haule2008}  This is in disagreement with the experimental observations for the 122- and 1111-type compounds which clearly show a Drude peak for different Fe-based compounds as discussed above in Sec.~\ref{Sec_BandStruct}.  However, no clear Drude peak was observed for Fe$_{1.05}$Te above $T_{\rm N}$.  They also predicted a much larger carrier scattering rate than observed experimentally from optical studies and the existence of a Hubbard band 0.4~eV below $E_{\rm F}$ which has also not been observed in the 122- and 1111-type compounds but may have been seen in Fe$_{1.05}$Te.

More recently, Qazilbash and coworkers have obtained a quantitative and intuitive systematic classification of the degree of conduction electron correlation in a wide variety of materials ranging from Mott insulators to free electron metals, including the Fe-based pnictides LaFePO and ${\rm BaFe_2As_2}$, from the ratio of the experimentally determined optical spectral weight of the conduction carrier conductivity (the low-frequency Drude contribution $K_{\rm exp}$) to that ($K_{\rm band}$) predicted by band theory.\cite{Qazilbash2008}  For example, the strongly correlated integer-valent layered cuprate parent compound La$_2$CuO$_4$ is predicted by conventional band theory to be a good metal, but instead it is an electrical insulator due to Coulomb (Mott-Hubbard) electron correlations and therefore does not have a Drude peak in the real part of the optical conductivity, thus giving $K_{\rm exp}/K_{\rm band} = 0$.  Therefore one would expect $K_{\rm exp}/K_{\rm band}$ to decrease as the degree of electron correlation increases, with $0 \leq K_{\rm exp}/K_{\rm band} \leq 1$.

The optical spectral weight is defined as the area under a plot of the real part of the optical conductivity $\sigma_1$ versus angular frequency $\omega$ up to a cutoff angular frequency $\omega_{\rm c}$ as discussed above in Sec.~\ref{Sec_CondCarrierOptics}.  Following Millis \emph{et al.},\cite{Millis2005} Qazilbash et al.\ wrote the Drude spectral weight in energy units which is therefore termed an optical ``kinetic energy'' $K$ of the conduction carriers.\cite{Qazilbash2008}  This $K$ is not in general the same as the total kinetic energy of the conduction carriers, and the terminology is therefore unfortunate but has been propagated in the literature.  For example, the total kinetic energy of a quasi-free electron gas of density $n$ is proportional to $n^{5/2}$, whereas the optical ``kinetic energy'' $K$ is directly proportional to $n$ [see Eq.~(\ref{EqKoptical2}) below].  The explicit definition of $K$ is\cite{Qazilbash2008, Millis2005}
\be
K = \frac{\hbar}{e^2}\frac{2}{\pi}\hbar\, \Delta c \int_0^{\omega_{\rm c}} \sigma_1(\omega)\,d\omega.
\label{EqKoptical}
\ee
In SI units, the first factor is an electrical resistance $\hbar/e^2 = 4108~\Omega$.  The second factor 2/$\pi$ cancels a factor $\pi/2$ that arises when the integrand of Eq.~(\ref{EqKoptical}) is the Drude form of $\sigma_1(\omega)$ in Eq.~(\ref{EqDrudeSigma1}) with a frequency-independent relaxation time $\tau$ with $\omega_{\rm c} = \infty$, as shown in Eq.~(\ref{EqSW}).  Indeed, using Eqs.~(\ref{EqOmegaP}) and~(\ref{EqOmegaPeff}) one can rewrite  $K$ in Eq.~(\ref{EqKoptical}) in terms of the (effective)  plasma angular frequency $\omega_{\rm p}$ or conduction carrier density $n$ and band mass $m^*$ as
\be
K = \frac{\hbar}{e^2}\hbar\,\Delta c\,  \varepsilon_0 \omega_{\rm p}^2 = \frac{\hbar^2 n\, \Delta c }{m^*}.
\label{EqKoptical2}
\ee
For a two-dimensional system like the layered iron pnictides or the layered cuprates, the distance $\Delta c$ is the distance between adjacent conducting layers, which is not the same as the $c$-axis lattice parameter if there is more than one conducting layer per unit cell.  For example, the 1111- and 122-type compounds each contain two formula units per unit cell, and each layer contains two Fe atoms per unit cell, so the 1111-type compounds have one layer per unit cell and hence $\Delta c = c$, whereas the 122-type compounds contain two layers per unit cell and therefore $\Delta c = c/2$, where $c$ is the respective $c$-axis lattice parameter.

\begin{figure}
\includegraphics[width=3.in]{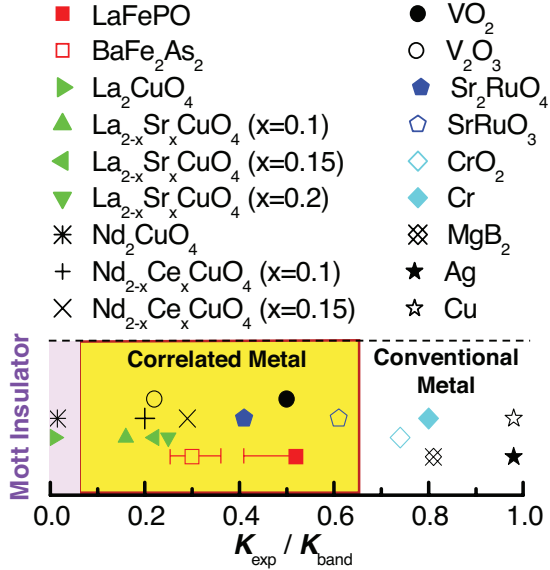}
\caption{(Color online) Ratio of the experimental conduction carrier optical spectral weight (expressed as a ``kinetic energy'') $K_{\rm exp}$ to that ($K_{\rm band}$) calculated using band theory for a variety of materials.  This ratio is the square of the ratio of the experimental to the band theoretical plasma frequency [see Eq.~(\ref{EqKoptical2})].  The degree of conduction electron correlation in a material is the degree of suppression of this ratio from the value of unity.  Mott-Hubbard insulators are on the far left and have ratios close to zero because the measured conduction carrier density and therefore plasma frequency are close to zero, whereas band theory predicts that they are metals.  At the opposite extreme, conventional free electron metals like Cu and Ag are on the far right and have ratios close to unity.  According to this scheme, the Fe-based layered pnictides LaFePO (filled red square) and ${\rm BaFe_2As_2}$ (open red square) are classified as moderately correlated metals, with the former less so than the latter.  Reproduced by permission from Ref.~\onlinecite{Qazilbash2008} and from Macmillan Publishers Ltd: Ref.~\onlinecite{Qazilbash2008}, Copyright (2009).}
\label{FigOptics_Correlations}
\end{figure}

Qazilbash et al.\ found $K_{\rm exp}/K_{\rm band} \approx 0.5$ and~0.3 for LaFePO and BaFe$_2$As$_2$, respectively, and compared these values with corresponding data for a wide variety of other compounds with varying degrees of electron correlation as shown in Fig.~\ref{FigOptics_Correlations}.\cite{Qazilbash2008}  The FeAs-based materials are found from this analysis to have an intermediate level of electron correlation.  A similar ratio $K_{\rm exp}/K_{\rm band} \sim 0.30$--0.38 was subsequently found from the $ab$~plane Drude conductivity for a single crystal of LaFeAsO by Chen et al.\cite{Chen2009}  

The reduction in the Drude weight ($K_{\rm exp}/K_{\rm band} < 1$) in a correlated electron system does not necessarily mean that the missing fraction of $K_{\rm exp}/K_{\rm band}$ corresponds to the fraction of incoherent current carriers.  To quote a sentence from the end of Sec.~\ref{SecQuantumOsc} above, ``It is interesting and important that despite these sometimes very large carrier mass enhancements, the corresponding conduction carrier conduction must still be coherent since that is required for observation of the dHvA effect.''  Instead, Eq.~(\ref{EqKoptical2}) shows that a reduction of $K_{\rm exp}/K_{\rm band}$ from unity can occur if there is a many-body enhancement of the coherent carrier band mass $m^*$ that is not accounted for by the LDA band calculations.  For LaFePO and ${\rm BaFe_2As_2}$, many-body mass enhancements of two to three are sufficient to give the data in Fig.~\ref{FigOptics_Correlations}.  As noted above, from LDA + DMFT calculations for LaFePO, Skornyakov et al.\ obtained mass enhancements of this order that agree with the results from these optics experiments as well as from a variety of other measurements on this compound.\cite{Skornyakov2010}

Lucarelli and coworkers have carried out an analysis of the in-plane optical conductivity of Ba(Fe$_{1-x}$Co$_x)_2$As$_2$ crystals with a Drude term from coherent current carriers, plus a phenomenological Drude term from incoherent current carriers,  plus a mid-infrared band, and have estimated the degree of correlation as the ratio of the spectral weights of the two Drude terms to the total spectral weight of the three terms.\cite{Lucarelli2010}  However, the source of the incoherent Drude-like conduction carriers is unclear.

\subsubsection{Other Spectroscopies}

An extensive experimental and theoretical study of x-ray absorption spectroscopy and resonant inelastic x-ray scattering measurements was carried out for nonsuperconducting BaFe$_2$As$_2$ and LaFe$_2$P$_2$ and superconducting SmFeAsO$_{0.85}$ by Yang et al.\cite{Yang2009}    The authors deduced that $U \lesssim 2$~eV,  and the Hund's coupling $J_{\rm H} \approx 0.8$~eV.  The data showed no features characteristic of Fe-based strongly-correlated electronic insulators such as $\alpha$-Fe$_2$O$_3$.  Thus they categorized the FeAs-based materials as weakly correlated, especially as compared with the layered high-$T_{\rm c}$ cuprates.\cite{Yang2009}  

The ARPES study of the reconstructed Fermi surfaces below $T_{\rm N}$ of CaFe$_2$As$_2$ and BaFe$_2$As$_2$ in Ref.~\onlinecite{Kondo2009} cited above were in agreement with calculations assuming interband coupling between the high-temperature hole and electron Fermi surface sheets.  Furthermore, these authors found a three-dimensional (3D) dispersion of the reconstructed Fermi surface along $k_z$ from the ARPES data, and that the reconstruction does not occur for $k_z$ values where interband coupling between the hole and electron Fermi surfaces is not efficient, thus demonstrating that interband magnetic scattering has an important role in the magnetic transition.  Despite the 3D dispersion, there are long cylindrical segments along $k_z$ of the reconstructed Fermi surfaces that evidently make this role possible.  Furthermore, the authors discovered \emph{incommensurate} nesting vectors in the reconstructed Fermi surface not explained by their model calculations, which they speculate may be associated with a ``failed density wave order such as the (charge density wave) predicted by renormalization group studies.''  

A combined STM and \emph{polarization-dependent} ARPES study of CaFe$_2$As$_2$ concluded that only two electron bands participate in the SDW, and that the SDW is driven by Fermi surface nesting and is not due to ordering of local magnetic moments.\cite{Hsieh2008}

\subsubsection{Summary}

The available evidence indicates that most Fe-based superconductors are moderately correlated, where the conduction electron correlations arising from Coulomb repulsion and/or Hund's coupling are significant but not as strong as in ionic materials like the layered cuprate high $T_{\rm c}$ superconductor family.  Various experiments consistently demonstrate that most of the Fe-based superconductors are at least partially coherent metals.  On the other hand, Te-rich compounds in the Fe$_{1+y}$(Te$_{1-x}$Se$_x$) system appear to be local moment antiferromagnets with stronger electron correlations than in other Fe-based materials.

\subsection{Spin Dynamics from Inelastic Magnetic Neutron Scattering Measurements: Formalism}

\subsubsection{\label{SpinIntro}Introduction}

\begin{figure}
\includegraphics[width=2.5in]{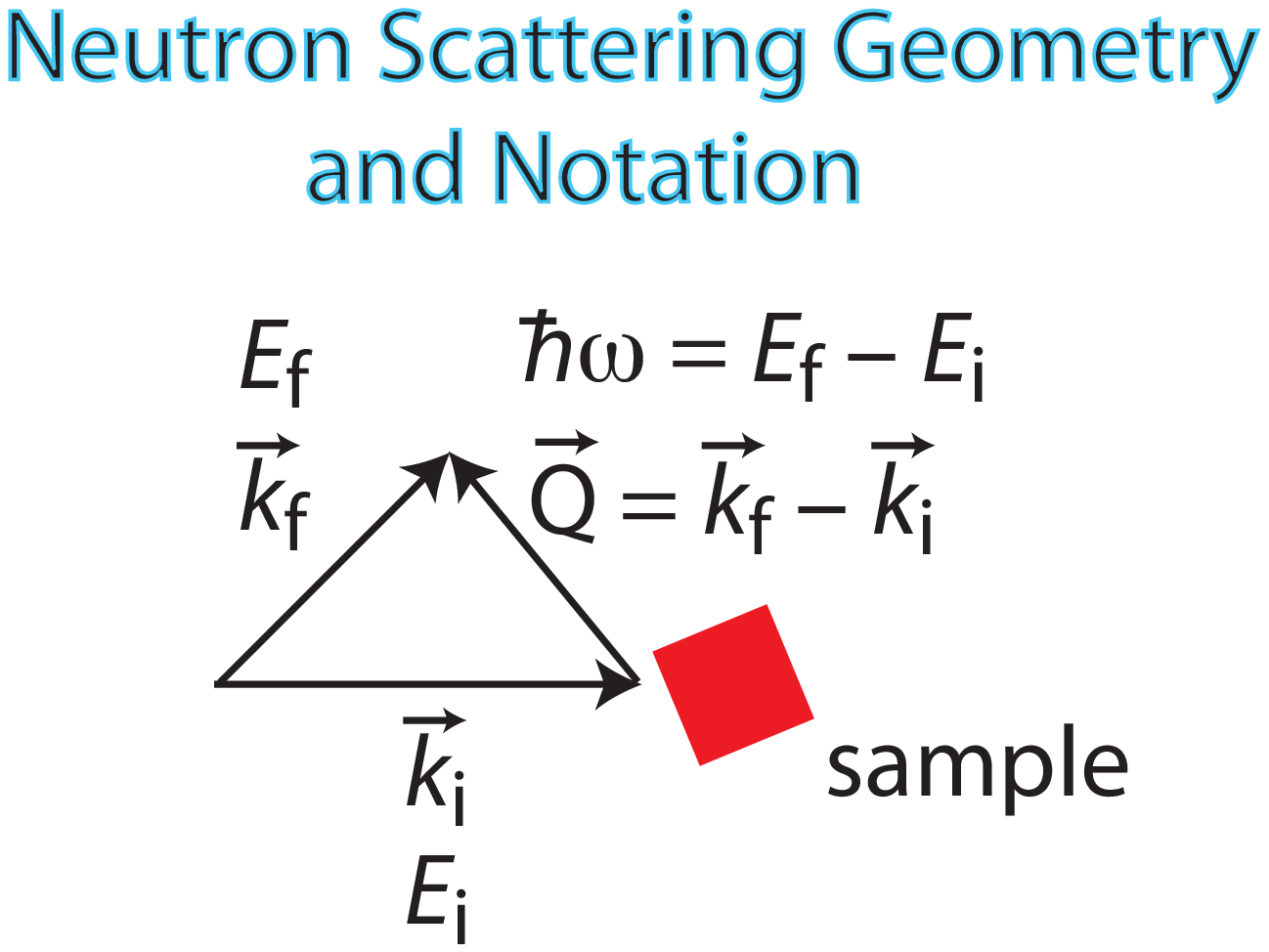}
\caption{(Color online) Configuration and notation used for neutron scattering from a sample.  Here $E_{\rm i}$ and $E_{\rm f}$ are the initial and final neutron kinetic energies, and ${\bf k}_{\rm i}$ and ${\bf k}_{\rm f}$ are the initial and final neutron wave vectors.  The magnitude of a neutron wave vector is $k \equiv 2\pi/\lambda$ where $\lambda$ is the de Broglie wavelength of the neutron.  The kinetic energy of a neutron is $E = p^2/(2m) = \hbar^2k^2/(2m)$ where $m$ is the neutron mass and $\hbar$ is Planck's constant divided by $2\pi$.  The energy transfer to the sample by each neutron is $\hbar\omega = E_{\rm f} - E_{\rm i}$, where $\omega$ is an angular frequency, and the momentum transfer to the sample per neutron is $\hbar Q = \hbar{\bf k}_{\rm f} - \hbar{\bf k}_{\rm i}$.  If $\omega = 0$, then the scattering is elastic and the experiment is usually called ``neutron diffraction'' instead of neutron scattering.  Usually the term ``scattering'' refers to inelastic collisions of the neutrons with the sample where $\omega \neq 0$.  For inelastic scattering, in addition to the expected positive energy transfer from the neutrons to the sample, energy can also be transferred \emph{to} the neutron beam \emph{from} a sample at a finite temperature, which would give a negative value of $\omega$.}
\label{NeutScattConfig}
\end{figure}

Many authors have emphasized the potential importance of magnetism to the superconducting properties of the Fe-based compounds.  The neutron has a magnetic moment and can therefore interact with electronic magnetic moments in a sample.  The technique of magnetic neutron scattering is a quintessential tool to quantitatively measure both static and dynamic magnetic properties of the electrons in a system and has been extensively used to measure the magnetic properties of the Fe-based superconductors and parent compounds.  Therefore we will discuss these measurements and their interpretations in detail.  

Many inelastic neutron scattering studies of the doped superconducting compositions have concentrated so far on the so-called ``spin resonance mode'' that occurs below the superconducting transition temperature and which will be discussed below in Sec.~\ref{ResonanceMode}.  The data discussed there indicate that the resonance mode with an energy gap evolves out of an antiferromagnetic spin fluctuation background upon cooling below $T_{\rm c}$.  The present section is concerned with the spin dynamics of mainly, but not exclusively, the nonsuperconducting Fe-based parent compounds.  Before reviewing  experimental data and their interpretations, an introduction to the formalism is given.\cite{Lovesey1984, Collins1989, Johnston1997, Ewings2008, Lester2010}  

A sketch of the experimental geometry with associated notation and nomenclature is shown in Fig.~\ref{NeutScattConfig} and described in the figure caption.  In general, the intensity $I$ of the neutrons scattered by a sample, which is the number of neutrons per second scattered into solid angle increment $d\Omega$ and into a final neutron energy increment $dE_{\rm f}$, is given by
\be
I = I_0 \frac{d^2\sigma_{\rm s}}{d\Omega\,dE_{\rm f}}\,d\Omega\,dE_{\rm f},
\label{EqNeutI}
\ee
where $I_0$ is the incident neutron intensity, $\sigma_{\rm s}$ is the scattering cross section and $d^2\sigma_{\rm s}/(d\Omega\,dE_{\rm f})$ is called the partial differential cross section.  For spin-only magnetic scattering by $N$ identical and spatially distinct atoms, one has
\be
\frac{d^2\sigma_{\rm s}}{d\Omega\,dE_{\rm f}} = A({\bf Q})\frac{k_{\rm f}}{k_{\rm i}}
\sum_{\alpha,\beta}\left(\Delta_{\alpha\beta} - \hat{\bf Q}_\alpha\hat{\bf Q}_\beta\right) S^{\alpha\beta}({\bf Q},\omega),
\label{NeutCrossSect}
\ee
where 
\be
A({\bf Q}) = N\hbar(\gamma_{\rm n} r_0)^2 \left(\frac{g}{2}\right)^2|f({\bf Q})|^2e^{-2W({\bf Q})},
\label{EqA}
\ee
$\hat{\bf Q}_\alpha$ is the $\alpha$ component of the unit vector $\hat{\bf Q}$ in the direction of ${\bf Q}$, $\Delta_{\alpha\beta}$ is the Kroneker delta function, $\gamma_{\rm n} r_0 = 5.391$~fm, $g\approx 2$ is the $g$-factor of the electronic magnetic moments the neutrons are  scattering from, $f({\bf Q})$ is the magnetic form factor of the magnetic atom which is the Fourier transform of the real-space spin density, $e^{-2W({\bf Q})}$ is the Debye-Waller factor that results in a reduction in the scattered intensity due to temperature-induced atomic motion but which is approximately unity at low temperatures, and $\alpha,~\beta = x,~y,~z$.  

The function $S^{\alpha\beta}({\bf Q},\omega)$ in Eq.~(\ref{NeutCrossSect}) contains the information on the magnetic properties of the sample that are of interest and can be directly extracted from neutron scattering measurements.  This quantity is variously called the magnetic scattering or response function, or the dynamic magnetic structure factor.  It is the space-time Fourier transform of the two-spin correlation function $\langle S_\alpha({\bf 0},0)S_\beta({\bf R},t)\rangle$ which is often diagonal ($\beta = \alpha$ only), where ${\bf R}$ denotes the magnetic atom position and $\langle\cdots\rangle$ denotes a thermal average, according to
\be
S^{\alpha\beta}({\bf Q},\omega) = \frac{1}{2\pi}\sum_{{\bf R}}e^{i{\bf Q}\cdot{\bf R}}\int_{-\infty}^\infty e^{i\omega t}\langle S_\alpha({\bf 0},0)S_\beta({\bf R},t)\rangle dt.
\label{Eq2spinCorrFcn}
\ee
An important point about Eqs.~(\ref{NeutCrossSect}) and~(\ref{Eq2spinCorrFcn}) is that if $S^{\alpha\beta}$ is diagonal, magnetic neutron scattering only occurs for the component of the scattering vector ${\bf Q}$ that is perpendicular to the local spin direction, sometimes referred to as transverse scattering.  Another important point is that all neutron scattering measurements have energy and wave vector resolution functions that are always convoluted with the theoretical $S^{\alpha\beta}({\bf Q},\omega)$ when fitting experimental data.

\subsubsection{\label{SecChiAOmega} Generalized Dynamical Magnetic Susceptibility $\chi({\bf Q},\omega)$}

It is often more convenient to express the magnetic neutron scattering response in terms of a generalized dynamical magnetic susceptibility $\chi({\bf Q},\omega)$ than in terms of a two-spin correlation function as in Eq.~(\ref{Eq2spinCorrFcn}).  The complex $\chi({\bf Q},\omega) = \chi^\prime({\bf Q},\omega) + i\chi^{\prime\prime}({\bf Q},\omega)$ describes the linear response of the electronic magnetization $M({\bf Q},\omega)$ to a magnetic field $H$ with wave vector ${\bf Q}$ and angular frequency $\omega$, via
\be
M_\alpha({\bf Q},\omega) = \chi_{\alpha\beta}({\bf Q},\omega) H_\beta({\bf Q},\omega).
\ee
Henceforth considering only diagonal forms of $S^{\alpha\beta}$ ($\beta = \alpha$) that are most often encountered in the present context, the fluctuation-dissipation theorem allows one to write $S^{\alpha\alpha}({\bf Q},\omega)$ in Eqs.~(\ref{NeutCrossSect}) and~(\ref{Eq2spinCorrFcn}) in terms of the imaginary (dissipative) part of $\chi({\bf Q},\omega)$ as\cite{White1970}
\bea
S^{\alpha\alpha}({\bf Q},\omega) &=&\frac{\hbar}{N\pi g_\alpha^2 \mu_{\rm B}^2} [1+n(\omega)]\chi^{\prime\prime}_{\alpha\alpha}({\bf Q},\omega)\nonumber \\
&=& \frac{\hbar}{N\pi g_\alpha^2 \mu_{\rm B}^2} \ \frac{\chi^{\prime\prime}_{\alpha\alpha}({\bf Q},\omega)}{1-\exp[-\hbar\omega/(k_{\rm B}T)]},
\label{EqSChi}
\eea 
so that one can fit the neutron scattering intensity data by varying the parameters in the chosen expression for $\chi^{\prime\prime}({\bf Q},\omega)$.  Here $g_\alpha$ is the spectroscopic splitting factor, or $g$-factor (usually $g_\alpha \approx 2$), of the electronic spins with the magnetic field in the $\alpha$ direction and $n(\omega) +1$ is the so-called detailed balance factor where 
\be
n(\omega) = \frac{1}{\exp[\hbar\omega/(k_{\rm B}T)] - 1}
\label{EqBosonNumber}
\ee
is the boson (magnon) occupation number for a magnetic mode at energy $\hbar\omega$.  If the $\omega$ values of interest satisfy $\hbar\omega \ll k_{\rm B}T$, then in this ``high-temperature approximation'' one can Taylor expand the exponential in Eq.~(\ref{EqSChi}) and retain only the first two terms, giving
\be
S^{\alpha\alpha}({\bf Q},\omega) = \frac{k_{\rm B}T}{N\pi g_\alpha^2 \mu_{\rm B}^2} \ \frac{\chi^{\prime\prime}_{\alpha\alpha}({\bf Q},\omega)}{\omega}.\ \ \  (\hbar\omega \ll k_{\rm B}T)
\label{EqSChi2}
\ee 

Using the Kramers-Kronig relations that embody causality, one can obtain the real zero-frequency ${\bf Q}$-dependent susceptibility $\chi^\prime({\bf Q}, 0)$ from $\chi^{\prime\prime}({\bf Q},\omega)$ as
\be
\chi^\prime({\bf Q}, 0) = \frac{1}{\pi} {\cal P}\int_{-\infty}^\infty \frac{\chi^{\prime\prime}({\bf Q},\omega)}{\omega} d\omega.
\label{EqKramersKronig}
\ee
The conventional uniform (${\bf Q} = 0$) static ($\omega = 0$) magnetic spin susceptibility is $\chi \equiv \chi^\prime(0,0)$.  Alternatively, using $\lim_{\omega\to 0}\chi^{\prime\prime}({\bf Q},\omega) = 0$ [the dissipation must go to zero at zero frequency since $\chi^{\prime\prime}({\bf Q},\omega)$ is an odd function of $\omega$], one can write $\chi$ in terms of the full complex dynamical susceptibility as
\be
\chi = \lim_{\omega\to 0}\chi({\bf Q} = 0,\omega).
\label{Eqchi00}
\ee
However, if the expression for $\chi({\bf Q},\omega)$ is used to describe antiferromagnetic (AF) fluctuations around a nonzero ${\bf Q}_{\rm AF}$, then Eq.~(\ref{Eqchi00}) is not expected to be accurate because it is then an extrapolation from ${\bf Q} = {\bf Q}_{\rm AF}$ to ${\bf Q} = 0$.

\subsubsection{Instantaneous Fluctuating Magnetic Moment Magnitude}

It is of particular importance for the Fe-based superconductors and parent compounds to be able to measure the instantaneous value of the fluctuating magnetic moment of the Fe atoms in the paramagnetic state because of the ongoing debate about whether a strong correlation large local moment picture or a weak correlation itinerant small magnetic moment picture is more appropriate for these materials.  Inelastic magnetic neutron scattering measurements offer a way to directly measure a lower limit on this quantity, as follows.

One considers only the diagonal $\alpha\alpha$ components of the dynamical structure factor in Eq.~(\ref{Eq2spinCorrFcn}).  Then one sums over all {\bf Q} within the first Brillouin zone of the Fe square lattice, multiplies both sides by $\exp(-i\omega t^\prime)$ and integrates over $\omega$ to obtain
\be
\int_{-\infty}^{\infty}d\omega\ \sum_{\bf Q}S^{\alpha\beta}({\bf Q},\omega) = N\langle S_\alpha^2 \rangle,
\label{EqFluctmoment1}
\ee
where $N$ is the number of spins, $\langle S_\alpha^2 \rangle$ is the expectation value of the square of the magnitude of the $\alpha$ component of the Fe spin, we have used $\sum_{{\bf Q}}e^{i{\bf Q}\cdot{\bf R}} = N\Delta_{\bf R,0}$, $\int_{-\infty}^{\infty}\exp(i\omega x)d\omega = 2\pi \delta(x)$, and have denoted $\langle S_\alpha^2({\bf 0},0)\rangle$ by $\langle S_\alpha^2\rangle$.  Usually one assumes Heisenberg spins, for which $\langle S_\alpha^2\rangle = S(S+1)/3$, so that Eq.~(\ref{EqFluctmoment1}) becomes 
\be
\int_{-\infty}^{\infty}d\omega\ \sum_{\bf Q}S^{\alpha\alpha}({\bf Q},\omega) = N\frac{S(S+1)}{3}.
\label{EqFluctmoment2}
\ee
Equivalently, one can replace the dynamic structure factor by the imaginary part of the generalized susceptibility according to Eq.~(\ref{EqSChi}) to obtain
\be
\int_{-\infty}^{\infty}d\omega\ \sum_{\bf Q} \ \frac{\chi^{\prime\prime}_{\alpha\alpha}({\bf Q},\omega)}{1-\exp[-\hbar\omega/(k_{\rm B}T)]} = \frac{N^2\pi g^2 \mu_{\rm B}^2}{\hbar}\frac{S(S+1)}{3}.
\label{EqFluctmoment3}
\ee
Using the definition of the effective moment of a spin given by $\mu_{\rm eff}^2 = g^2S(S+1)\mu_{\rm B}^2$ as it occurs in the Curie Weiss law for the magnetic susceptibility $\chi = N\mu_{\rm eff}^2/[3k_{\rm B}(T - \theta)]$, Eq.~(\ref{EqFluctmoment3}) becomes
\be
\int_{-\infty}^{\infty}d\omega\ \sum_{\bf Q} \ \frac{\chi^{\prime\prime}_{\alpha\alpha}({\bf Q},\omega)}{1-\exp[-\hbar\omega/(k_{\rm B}T)]} = \frac{N^2\pi}{\hbar}\frac{\mu_{\rm eff}^2}{3}.
\label{EqFluctmoment4}
\ee
One can define the imaginary part of the \emph{local} frequency-dependent susceptibility $\chi^{\prime\prime}(\omega)$ by
\be
\chi^{\prime\prime}(\omega) = \sum_{\bf Q}\chi^{\prime\prime}_{\alpha\alpha}({\bf Q},\omega).
\label{EqDefineLocalchi}
\ee
Then Eq.~(\ref{EqFluctmoment4}) can be more simply written as
\be
\int_{-\infty}^{\infty}d\omega\ \frac{\chi^{\prime\prime}_{\alpha\alpha}(\omega)}{1-\exp[-\hbar\omega/(k_{\rm B}T)]} = \frac{N^2\pi}{\hbar}\frac{\mu_{\rm eff}^2}{3}.
\label{EqFluctmoment5}
\ee

Thus one can determine the magnitude of the instantaneous effective moment of an Fe atom in the paramagnetic state by integrating the inelastic neutron scattering spectra over both wave vector and energy.  Note that the effective moment of a localized spin $S$ is larger than the expectation value $gS\mu_{\rm B}$ of the magnitude of the moment itself.  This method was used by Diallo et al.\ to determine a lower limit on the instantaneous magnitude of the Fe effective moment in ${\rm CaFe_2As_2}$ crystals by integrating up to the limit $\hbar \omega = 200$~meV,\cite{Diallo2010} and we used it to calculate a lower limit on $\mu_{\rm eff}$/Fe from the $\chi^{\prime\prime}(\omega)$ up to 80~meV by Lester et al.\ for ${\rm Ba(Fe_{1.935}Co_{0.065})_2As_2}$ crystals,\cite{Lester2010} as discussed in Sec.~\ref{Sec122typeneuts} below.  These experimental values are lower limits because of experimental limits on the ranges of the energy integrations.

\subsubsection{Dynamical Magnetic Susceptibility of Spin Waves}
Various forms for $\chi({\bf Q},\omega)$ have been used in the literature to fit experimental magnetic neutron scattering data, depending on the application. The low-energy elementary excitations of a magnetically ordered material are spin waves.  A particular quantized energy $\hbar\omega$ is associated with a particular wave vector ${\bf Q}$ of the spin wave, where $\omega$ is the angular frequency of the spin wave. To describe damped spin waves, the response function of a harmonically driven, damped simple harmonic oscillator, adapted to the spin wave problem, is often used\cite{Lester2010}
\be
\chi({\bf Q},\omega) = \chi^\prime({\bf Q},0)~\frac{\omega_0^2}{\omega_0^2 - \omega^2  - i\gamma\omega},
\label{EqDynamicChiSHO}
\ee
yielding
\be
\chi^{\prime\prime}({\bf Q},\omega) =\chi^\prime({\bf Q},0)~f({\bf Q},\omega),
\label{EqChppQO}
\ee
where 
\be
f({\bf Q},\omega) = \frac{\omega_0^2 \gamma \omega}{(\omega_0^2 - \omega^2)^2 + \gamma^2\omega^2}
\label{EqfomQ}
\ee
is a dimensionless scalar function of ${\bf Q}$ and $\omega$, $\omega_0 \equiv \omega_0({\bf Q})$ is the spin wave dispersion relation and $\gamma \equiv \gamma({\bf Q})$ characterizes the damping of the spin waves at wave vector ${\bf Q}$.  The $\chi^{\prime\prime}({\bf Q},\omega)$ in Eq.~(\ref{EqChppQO}) identically satisfies the Kramers-Kronig relation~(\ref{EqKramersKronig}).

Consider the function $f(\omega)$ at fixed ${\bf Q}$, $\gamma$ and $\omega_0$ in Eq.~(\ref{EqfomQ}).  For weak damping $\gamma\omega_0 \ll 1$, the full width at half maximum (FWHM) of the peak in $f(\omega)$ at $\omega \approx \omega_0$ is $\gamma$ and the peak height is $\approx \omega_0/\gamma$.  One therefore expects that $\int_0^\infty f(\omega)d\omega \sim \omega_0$ and therefore that $\int_0^\infty\chi^{\prime\prime}({\bf Q},\omega)\,d\omega$ for a given ${\bf Q}$ is independent of $\gamma$ for small $\gamma$.  Indeed, we find that the integral is given by
\be
\lim_{\gamma \to 0}\int_0^\infty f(\omega)d\omega = \frac{\pi}{2}\,\omega_0 \approx 1.571\, \omega_0.
\ee
This result is approximately correct even when $\gamma$ becomes of the order of $\omega_0$.  For example, $\int_0^\infty f(\omega)d\omega =1.566\,\omega_0$, $1.523\,\omega_0$, $1.361\,\omega_0$ and~$1.209\,\omega_0$ for $\gamma/\omega_0 = 0.01$, 0.1, 0.5 and~1, respectively. 

When  probing energy transfers $\hbar\omega \approx \hbar\omega_0$, one can expand the first term in the denominator of Eq.~(\ref{EqfomQ}) as $(\omega_0^2 - \omega^2)^2 = (\omega_0 + \omega)^2 (\omega_0 - \omega)^2 \approx 4\omega^2(\omega_0 - \omega)^2$, yielding
\be
f({\bf Q},\omega) \approx \frac{\omega_0^2\gamma/(4\omega)}{(\omega_0 - \omega)^2 + \gamma^2/4} = \frac{\omega_0^2\Gamma/(2\omega)}{(\omega_0 - \omega)^2 + \Gamma^2},
\label{EqfomQ2}
\ee
where $\Gamma \equiv \gamma/2$ is the half-width at half maximum (HWHM) of the peak in $f(\omega)$ when $\gamma\omega_0 \ll 1$.  This form for $\chi^{\prime\prime}({\bf Q},\omega)$ was used to analyze inelastic neutron scattering measurements of the spin wave excitations with energies up to 25~meV in co-aligned single crystals of ${\rm CaFe_2As_2}$ in Ref.~\onlinecite{McQueeney2008} and at 10--12~meV  of ${\rm BaFe_2As_2}$ crystals in Ref.~\onlinecite{Matan2009}.  However, note that the form~(\ref{EqfomQ2}) is not valid at low frequencies, since one requires that $f({\bf Q},\omega) \to 0$ as $\omega \to 0$.

In addition to its use in analyzing magnetic inelastic neutron scattering data, the imaginary part $\chi^{\prime\prime}({\bf Q},\omega)$ of the generalized susceptibility is important in the analysis of the electronic spin dynamics as reflected by the nuclear spin-lattice relaxation rates $1/T_1$ in NMR experiments, as discussed in Sec.~\ref{SecNMRDynamics} below.

\subsubsection{Local Moment Model for the Spin Wave Dispersion in the Orthorhombic Phase of the 122-Type FeAs-Based Compounds}

Irrespective of whether the layered FeAs-based parent compounds are itinerant or local moment antiferromagnets, the low-energy magnetic excitations are spin waves that can always be parametrized using a local moment spin Hamiltonian.  Most neutron scattering studies of these materials to date have used such a parametrization at low energies (perhaps in addition to other approaches), and we therefore consider such a local moment model in detail in this section.

The antiferromagnetic/SDW ordering in the FeAs-based materials always occurs in the low-temperature orthorhombic phase with lattice parameters $a,b,c$ with the convention $c > a > b$.  As discussed previously, the Fe layers are in the basal $a$-$b$ plane and are stacked along the $c$-axis.  The neutron scattering wave vector ${\bf Q}$ is formally defined in terms of the orthorhombic reciprocal lattice vectors (${\bf a}^*,{\bf b}^*,{\bf c}^*$).  However, in orthorhombic symmetry, these are parallel to the respective real-space axis unit vectors ($\hat{\bf a},\hat{\bf b},\hat{\bf c}$).  Therefore, one can write
\be
{\bf Q}\left({\rm \AA}^{-1}\right) = H\frac{2\pi}{a}\hat{\bf a} + K\frac{2\pi}{b}\hat{\bf b} + L\frac{2\pi}{c}\hat{\bf c},
\label{Eqrlu}
\label{EqQortho}
\ee
 which can be expressed in the orthorhombic reciprocal lattice units (r.l.u.) implicitly defined in Eq.~(\ref{Eqrlu}) as 
\be
{\bf Q} = (H,K,L)~{\rm r.l.u.}
\ee
Thus using, e.g., ${\bf a} = a\hat{\bf a}$ and Eq.~(\ref{EqQortho}), one obtains the relations
\be
{\bf Q}\cdot {\bf a} = 2\pi H,\ \ \ {\bf Q}\cdot {\bf b}= 2\pi K,\ \ \ {\bf Q}\cdot {\bf c} = 2\pi L
\label{EqQdot}
\ee
that we will need to use shortly.  The lower-case notation $(h,k,\ell)$ in reciprocal lattice units will be used to denote \emph{deviations} from a given $(H,K,L)$ point in the first magnetic Brillouin zone.

The Heisenberg spin Hamiltonian in Eq.~(\ref{EqJ1aJ1bJ2Jc}) has been extended by Ewings et al.\cite{Ewings2008} to specifically apply to the orthorhombic phase of the 122-type FeAs-based materials as
\bea
{\cal H} &=& J_{1a} \sum_{\langle ij \rangle_a}{\bf S}_i \cdot {\bf S}_j + J_{1b} \sum_{\langle ij \rangle_b}{\bf S}_i \cdot {\bf S}_j + J_2 \sum_{\langle ik \rangle_{ab}}{\bf S}_i \cdot {\bf S}_k \nonumber\\
&+& J_{1c} \sum_{\langle il \rangle_c}{\bf S}_i \cdot {\bf S}_l  + \sum_i[D(S_z^2)_i  + E(S_x^2 - S_y^2)_i]
\label{EqJ1aJ1bJ2}
\eea
where the $x$, $y$ and $z$ axes are along the $a$, $b$ and $c$~axes, respectively, and each pair of spins in the first four sums is only counted once.  The new term is the last sum that includes both axial ($D$) and orthorhombic ($E$) terms in the single-ion anisotropy energy where one expects $|D| > |E|$.  The authors diagonalized Hamiltonian~(\ref{EqJ1aJ1bJ2}) and obtained dispersion relations for two branches of spin waves that are nondegenerate due to the single-ion anisotropy terms.  The resulting spin wave dispersion relations, with slight changes in notation from Ref.~\onlinecite{Ewings2008} and using Eq.~(\ref{EqQdot}), are\cite{Ewings2008}
\be
\hbar \omega_\pm({\bf Q}) = \sqrt{A_{\bf{Q}}^2 - (C \pm B_{\bf{Q}})^2},
\label{EqDispRln}
\ee
where
\bea
A_{\bf{Q}} = 2S\{J_{1b}[\cos(\pi K) &-&1] + J_{1a}+2J_2+J_{1c} \nonumber\\
&+& (D - 3E)/2\}\nonumber,\\
\  \label{EqAB}\\
B_{\bf{Q}} = 2S\{J_{1a}\cos(\pi H) &+& 2J_2 \cos(\pi H)\cos(\pi K) \nonumber\\
&+& J_{1c} \cos(\pi L)\},\nonumber\\
C = -S(D + E), \nonumber
\eea
and $S$ is the ordered spin which is related to the ordered magnetic moment $\mu$ by $S = \mu/(g\mu_{\rm B})$ where $g$ is the spectoscopic splitting factor ($g$-factor) and $\mu_{\rm B}$ is the Bohr magneton. The ordered moment can be obtained from solving the ordered magnetic structure at low temperatures. As noted in the following Sec.~\ref{SecSpinGap}, the solution $\omega_+({\bf Q})$ in Eq.~(\ref{EqDispRln}) is likely to be unphysical for realistic exchange and anisotropy parameters.

A slightly different Hamiltonian was utilized by Diallo et al.\cite{Diallo2009} in which the single-ion anisotropy term was written as $\sum_iD(S_x)_i^2$ (R. J. McQueeney, private communication), where the uniaxial single-ion anisotropy axis is along the $a$ ($x$) direction in the plane, instead of along the $c$~axis, and is thus along the moment ordering direction.  The resulting dispersion relation was therefore slightly different from Eq.~(\ref{EqDispRln}), given by
\be
\hbar \omega_0({\bf Q}) = \sqrt{A_{\bf{Q}}^2 - B_{\bf{Q}}^2},
\label{McQueeneyDispRln}
\ee
where 
\be
A_{\bf{Q}} = 2S\{J_{1b}[\cos(\pi K) -1] + J_{1a}+2J_2+J_{1c} + D\}
\ee
and $B_{\bf{Q}}$ is the same as given above in Eqs.~(\ref{EqAB}).  Thus Diallo et al.\ predict a single dispersion relation instead of two potentially nondegenerate ones as in Eq.~(\ref{EqDispRln}).  In the absence ($D = E = 0$) of single-ion anisotropy that can lead to energy gaps in the spin wave excitation spectrum, the dispersion relations (\ref{EqDispRln}) and~(\ref{McQueeneyDispRln}) derived in the two studies\cite{Ewings2008, Diallo2009} are identical.  In their analysis of the high-energy spin wave dispersion in ${\rm CaFe_2As_2}$ crystals measured with inelastic neutron scattering (see Sec.~\ref{Sec122typeneuts} below), Zhao et al.\cite{JZhao2009} used the dispersion relation in Eq.~(\ref{McQueeneyDispRln}) after setting $D = 0$.

Ewings et al.\cite{Ewings2008} also calculated the magnetic response functions $S^{\alpha\beta}({\bf Q},\omega)$ per formula unit of 122-type FeAs-based materials for magnon creation in Eq.~(\ref{NeutCrossSect}) and found that only the diagonal correlations transverse ($yy$ and $zz$) to the ordered magnetic moment $\vec{\mu}||x$ contribute to the scattering function, as expected from the discussion at the end of Sec.~\ref{SpinIntro} above.  Their expressions for these are\cite{Ewings2008}
\bea
S^{yy}({\bf Q},\omega) &=& S_{\rm eff}\,\frac{A_{\bf Q}-B_{\bf Q}-C}{\hbar \omega_+({\bf Q})}[n(\omega)+1]\delta[\omega-\omega_+({\bf Q})]\nonumber\\
{\rm and}\label{EqSaa}\\
S^{zz}({\bf Q},\omega) &=& S_{\rm eff}\,\frac{A_{\bf Q}-B_{\bf Q}+C}{\hbar \omega_-({\bf Q})}[n(\omega)+1]\delta[\omega-\omega_-({\bf Q})],\nonumber
\eea
where $y||b$, $z||c$, $S_{\rm eff}$ is the ``effective spin'' which is expected to be the same as $S$ in the linear spin wave approximation, $n(\omega)$ is the boson occupation number in Eq.~(\ref{EqBosonNumber}) and $\delta(x)$ is the Dirac delta function.  These expressions~(\ref{EqSaa}) assume that the spin waves are not damped, which is a good approximation if the damping energy $\hbar\gamma$ in Eq.~(\ref{EqDynamicChiSHO}) is much less than the neutron spectrometer energy resolution.

In their analysis of the low-energy spin wave spectra for ${\rm Ba(Fe_{0.96}Co_{0.04})_2As_2}$, Christianson~et~al.\ used the dispersion relation~(\ref{McQueeneyDispRln}) and explicitly included an energy width via the expression for the damped simple harmonic oscillator form~(\ref{EqfomQ}) for $\chi^{\prime\prime}({\bf Q},\omega)$ in their expression for $S({\bf Q},\omega)$, according to\cite{Christianson2009}
\be
S({\bf Q},\omega) \propto \frac{A_{\bf Q} - B_{\bf Q}}{\omega_0} \frac{4}{\pi}  \frac{\Gamma\omega_0\omega}{(\omega_0^2 -\omega^2)^2 + 4(\Gamma\omega)^2},
\label{EqChris}
\ee
where $\Gamma \equiv \gamma/2$.

Applegate et al.\ have also made a study of spin waves in the orthorhombic square lattice Heisenberg $J_{1a}$-$J_{1b}$-$J_{2}$ model.\cite{Applegate2010}

\subsubsection{\label{SecSpinGap} Spin Wave Gap: 122-Type Compounds}

The energy gaps, or ``spin gaps'', in the spin wave spectrum at the minima of the dispersion relation at different points in the Brillouin zone can be obtained by inserting the appropriate ${\bf Q}= (H,K,L)$~r.l.u.\ orthorhombic reciprocal lattice units into the dispersion relation.  For the 122-type FeAs-based compounds, all studies so far show the orthorhombic antiferromagnetic propagation vector to be ${\bf Q}_{\rm AF} = (1,0,1)$~r.l.u.\ (see Table~\ref{LoTStructData122} in the Appendix).  The dispersion relations~(\ref{EqDispRln}) by Ewings et al.\cite{Ewings2008} give the spin wave gaps at ${\bf Q}_{\rm AF} = (1,0,1)$~r.l.u.\ as
\bea
\Delta_+(101) &=& 2^{3/2}S \sqrt{E[-(D-E)-2(2J_2 + J_{1a} + J_{1c})]}\nonumber \\
\label{EqDminus}\\
\Delta_-(101) &=& 2^{3/2}S \sqrt{(D-E)(2J_2 + J_{1a} + J_{1c}-E)}.\nonumber
\eea
These spin wave gaps do not depend on $J_{1b}$ because the coefficient of $J_{1b}$ in the expression for $A_{\bf Q}$ in Eqs.~(\ref{EqAB}) is zero for $K = 0$.  The solution $\Delta_+(101)$ is unphysical (imaginary) for likely values of the parameters inside the square root (see below), in which case the dispersion relation $\omega_+({\bf Q})$ itself in Eq.~(\ref{EqDispRln}) is also unphysical.

The spin wave gap at ${\bf Q}_{\rm AF} = (1,0,1)$~r.l.u.\ obtained from the  $\omega({\bf Q})$ of Diallo et al.\cite{Diallo2009} in Eq.~(\ref{McQueeneyDispRln}) is 
\be
\Delta(101) = 2S \sqrt{D[D + 2(2J_2 + J_{1a} + J_{1c})]}.
\label{EqDeltaDiallo}
\ee
This expression is different from either of the above two expressions by Ewings et al.\ in Eqs.~(\ref{EqDminus}) with $E = 0$.  However, Eq.~(\ref{EqDeltaDiallo}) reduces to $\Delta_-(101)$ in Eqs.~(\ref{EqDminus}) with $E = 0$ if $D \ll 2(2J_2 + J_{1a} + J_{1c})$.

On the scale of the full dispersion throughout the Brillouin zone, the spin gap is experimentally found to be relatively small (see, Table~\ref{NeutDispersionData2} and Fig.~\ref{CaFe2As2SpinWaves} below).  Also, the anisotropy parameter $D$ is found to be very small compared with the exchange constants $J$, and can be safely set to zero except when considering the spin gap that arises from the anisotropy and/or the neutron scattering intensity in Eq.~(\ref{EqChris}).  For example, in Eq.~(\ref{EqDeltaDiallo}), the second occurrence of $D$ can be set to zero, but not the first.  In that case, one can solve for $SD$ in terms of the observed $\Delta(101)$ as
\be
SD = \frac{[\Delta(101)]^2}{8S(J_{1a} + J_{1c} + 2 J_2)}.
\label{EqSD}
\ee

\subsubsection{\label{SecSpinGap2} Zone-Boundary Spin Wave Energies, Dispersion Relations, and Neutron Scattering Intensity: 122-Type Compounds}

\begin{figure*}
\includegraphics[width=3.4in]{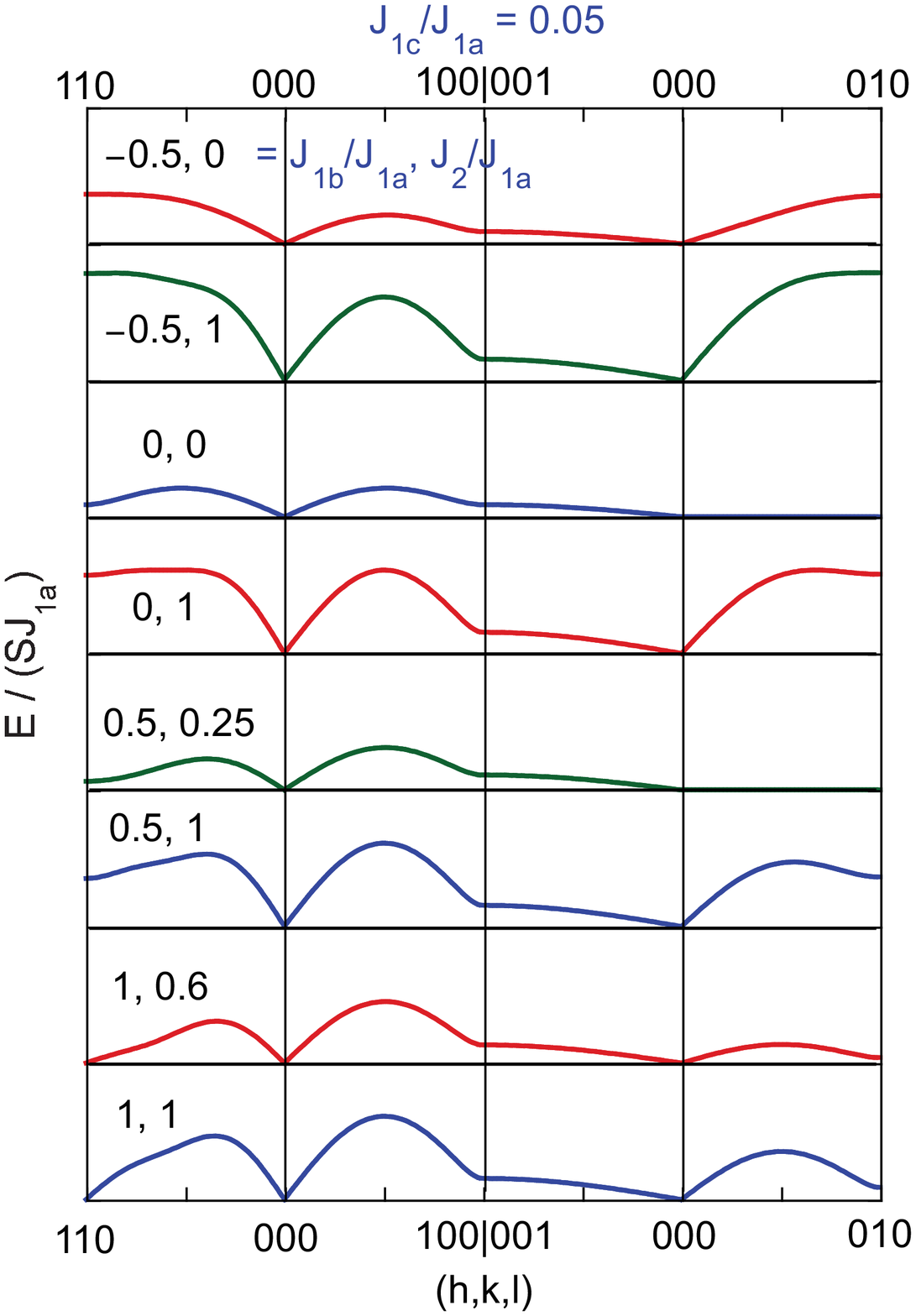}\hspace{0.2in}
\includegraphics[width=3.4in]{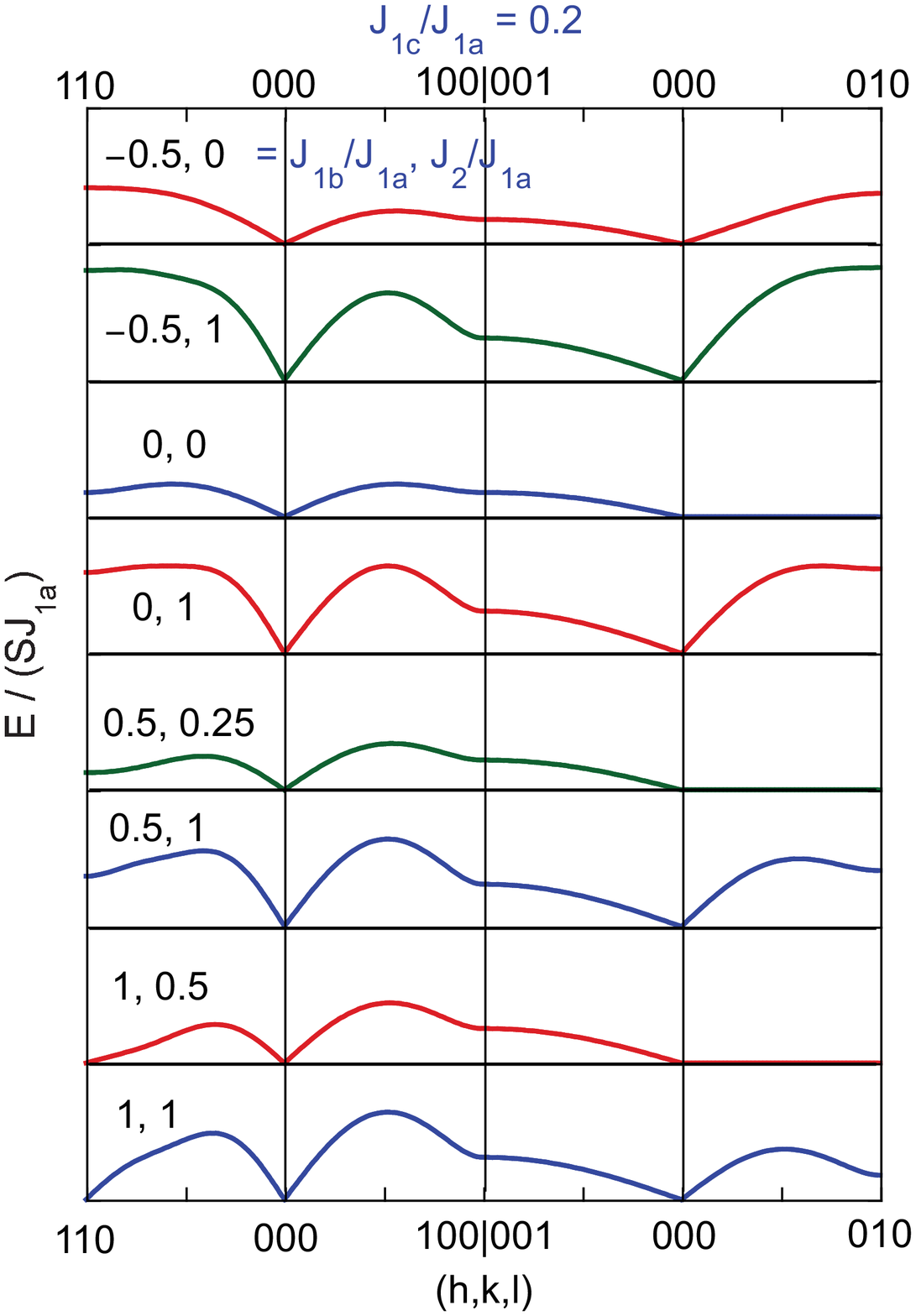}
\caption{(Color online) Theoretical spin wave dispersion relations $E$ versus relative wave vector ${\bf q} = (h,k,\ell)$~r.l.u.\ for different pairs of parameters $J_{1a},\ J_{1b},\ J_{1c}$ and~$J_2$ in Eq.~(\ref{McQueeneyDispRln}) for the Heisenberg model.  Here, $E$ is the energy of the spin waves in units of $SJ_{1a}$ and the single-ion anisotropy energies $D$ and $E$ are set to zero since their influence is small relative to the vertical energy scales of the figures.  The horizontal axes for the different parameter sets are each at $E = 0$ and are separated vertically by 10~units of $E/(SJ_{1a})$.  The orthorhombic wave vectors are measured relative to the antiferromagnetic ordering wave vector (1,0,1), i.e., ${\bf q} = (h,k,\ell)~{\rm r.l.u.} = [{\bf  Q} - (1,0,1)]$~r.l.u.  The left and right panels are for $J_{1c}/J_{1a} = 0.05$ and~0.2, respectively.  Each plot is for a specific set of $J_{1b}$ and~$J_2$ values, as shown.  By comparing the two figures, one sees that increasing $J_c$ by a factor of four has a significant influence on the dispersion relations.}
\label{SW_Disp_Reln}
\end{figure*}

\begin{figure*}
\includegraphics[width=3.4in]{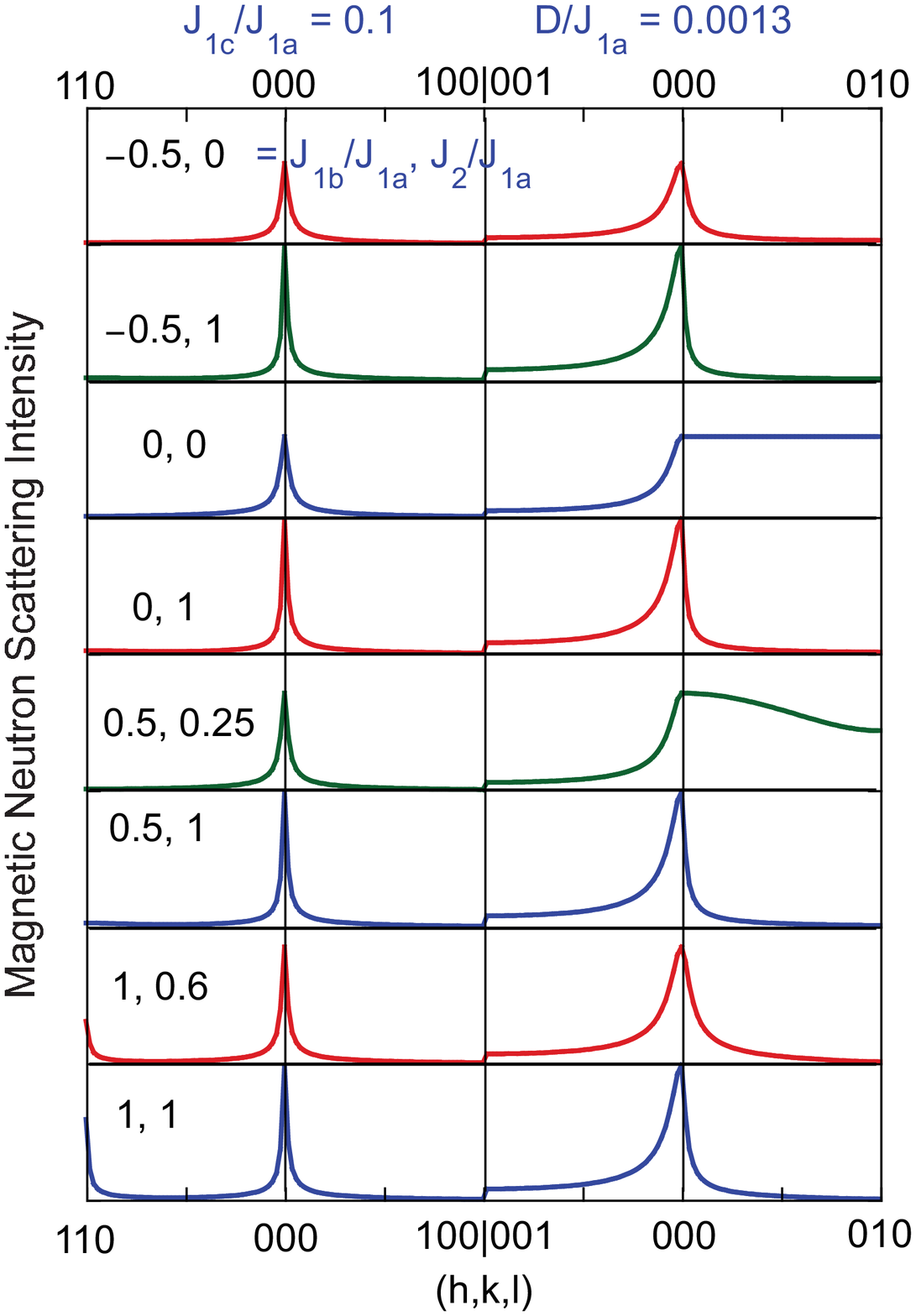}\hspace{0.2in}
\includegraphics[width=3.4in]{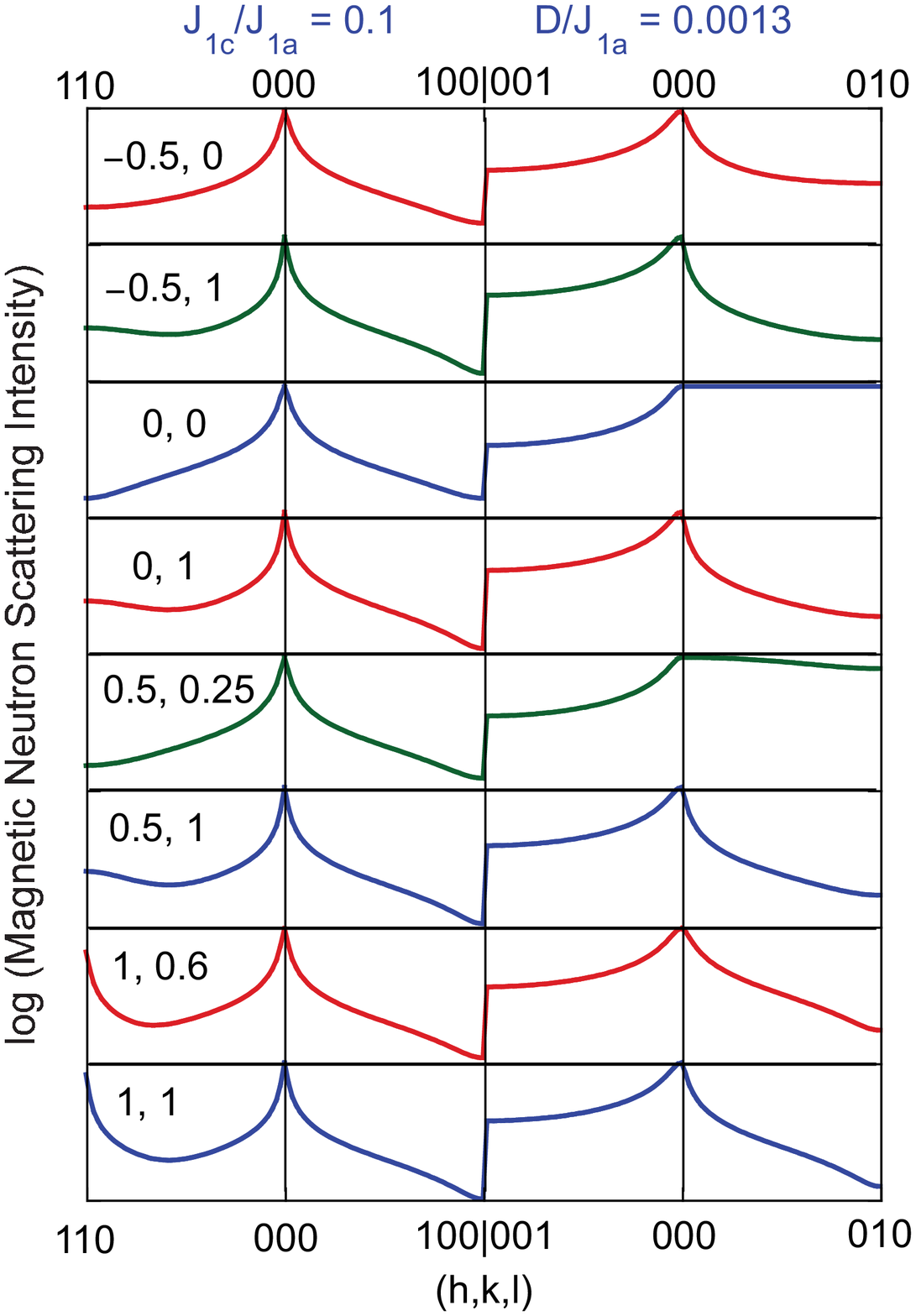}
\caption{(Color online) Linear (left) and logarithmic (right) plots of the theoretical magnetic neutron scattering intensity at $\omega = \omega_0({\bf Q})$ by spin waves of frequency $\omega_0({\bf Q})$ calculated using $S({\bf Q},\omega) \propto (A_{\bf Q} - B_{\bf Q})/\omega_0$ from Eqs.~(\ref{McQueeneyDispRln}) and~(\ref{EqChris}) for different pairs of ratios of $J_{1b}/J_{1a}$ and~$J_2/J_{1a}$, as shown, with fixed ratios $J_{1c}/J_{1a} = 0.1$ and $D/J_{1a} = 0.0013$.  Here ${\bf q} = (h,k,\ell)~{\rm r.l.u.} = [{\bf  Q} - (1,0,1)]$~r.l.u., and the variable parameter regimes are the same as in Fig.~\ref{SW_Disp_Reln}.  The vertical intensity scales are the same within each set of eight plots.  In the linear plots on the left-hand-side, the horizontal axes are at zero intensity. In the logarithmic plots on the right-hand-side, the intensity ratio in the vertical direction between adjacent horizontal axes is a factor of 380.}
\label{SW_Intensity}
\end{figure*}

The values of the spin wave energies at the magnetic zone boundaries ${\bf Q}_{\rm ZB}$ for AF ordering with orthorhombic ${\bf Q}_{\rm AF} = (1,0,1)$~r.l.u.\ can give additional information on the values of the parameters in the spin Hamiltonian.  By inserting values of ${\bf Q} = {\bf Q}_{\rm ZB}$ into the dispersion relation~(\ref{McQueeneyDispRln}) of Diallo et al., one obtains
\bea
\hbar\omega(100) &=&\hbar\omega(001) =\nonumber\\ 
 &\ & 2S \sqrt{(D + 2 J_{1c}) (D + 2 J_{1a} + 4 J_2)} \nonumber\\
\hbar\omega(000)&=& \Delta(101) = \label{EqEZB}\\
&\ & 2S \sqrt{D [D + 2 (J_{1a} + J_{1c} + 2 J_2)]} \nonumber \\
\hbar\omega(111) &=& \hbar\omega(010) = \nonumber\\
&\ & 2S \sqrt{[D + 2 (J_{1a} - J_{1b} + J_{1c})] }\nonumber\\
&\ &\ \ \times\sqrt{(D + 2( 2J_2 - J_{1b}) )}\nonumber\\
\hbar\omega(011) &=& 2S \sqrt{[D + 2 (J_{1a} - J_{1b})] }\nonumber\\
&\ &\ \ \times\sqrt{[D  +  2(2 J_2 - J_{1b} + J_{1c})]}.\nonumber
\eea
With regard to $\hbar\omega(111)$ and $\hbar\omega(011)$, recall that the stability conditions for the stripe-$b$ state in Eqs.~(\ref{EqStripeStab}) are $J_{1a} > J_{1b}$ and $2J_2 > J_{1b}$.  If one sets $J_{1a} = J_{1b} \equiv J_1$, one obtains the $J_1$-$J_2$ model.  

The band widths are not usually equal to the differences between any of these values in Eq.~(\ref{EqEZB}) and the gap $\Delta(101)$, because the dispersion relation can vary nonmonotonically with {\bf Q}.  The band width of a particular band for a particular combination of $J$s and $D$ has to be determined in general by evaluation of the full dispersion relation, as shown in Fig.~\ref{SW_Disp_Reln}.  In this figure, plots are made of the dispersion relation for ratios of the exchange constants that are relevant to the FeAs-based materials (see Table~\ref{NeutDispersionData2} below).  Here one must have $J_{1b}/J_{1a} \leq 1$ and $J_2/J_{1b} \geq 1/2$ for classical stability of the stripe-$b$ phase as shown above in Eq.~(\ref{EqStripeStab}).  We also include plots for the ratio $J_{1b}/J_{1a} = 1$ relevant to the $J_1$-$J_2$ model.  Note that in this case one must have $J_2/J_1 \geq 1/2$ in Eq.~(\ref{EqClassJ2J1}) for the stripe state to be stable according the classical model.  

The magnetic neutron scattering intensity as a function of the wave vector of the spin waves is shown in Fig.~\ref{SW_Intensity}.  One sees that in general, the scattered intensity dies off rather rapidly as the wave vector of the spin waves moves away from the magnetic Brillouin zone center at $(h,k,\ell) = (0,0,0)$ (compare Fig.~\ref{SW_Intensity} with Figs.~\ref{SW_Disp_Reln}).  Exceptions are in the third and fifth panels from the top of Fig.~\ref{SW_Intensity} for the dispersion along the $b$-axis (the stripe axis) for $(h,k,\ell)$ between (0,0,0) and (0,1,0).  The first case is for $J_2 = 0$ and the second is for $J_2$ at its stability edge $J_2/J_{1b} = 1/2$, where in general one requires $J_2/J_{1b} > 1/2$ for classical stability of the stripe-$b$ spin state.  Referring to Fig.~\ref{SW_Disp_Reln}, these two anomalous dependences of intensity along (0,1,0) correspond to the cases where there is no significant dispersion along this axis, i.e.\ $\omega_0({\bf Q}) \approx 0$.

\subsubsection{\label{SecSWCalcfromJ} Spin Wave Velocity at Low Energy: 122-Type Compounds}

The following generic expression describes the spin wave dispersion about a magnetic zone center $\bf{Q}_{\rm AF}$ at low energy
\begin{equation}
E_{\bf q} = \hbar \omega({\bf q}) = \sqrt{\Delta^2 + \hbar^2[v_{a}^2q_a^2 + v_{a}^2q_b^2 + v_c^2 q_c^2]},
\label{Eqdispersion}
\end{equation}
where ${\bf q} \equiv {\bf Q} - {\bf Q}_{\rm AF} = (h,k,\ell)$~r.l.u.\ is the ``reduced'' spin wave wave vector, $\hbar$ is Planck's constant divided by 2$\pi$, $\Delta$ is the spin wave energy gap described in the previous section,  and $v_{a}$, $v_{b}$ and $v_{c}$ are the spin wave velocities in the respective directions.  The spin wave dispersion at small energies about the observed orthorhombic ordering wave vector ${\bf Q}_{\rm AF}=(1,0,1)$ is derived for the 122-type FeAs-based compounds by expanding $\omega({\bf Q})$ in Eq.~(\ref{McQueeneyDispRln}) as in Eq.~(\ref{Eqdispersion}) using $(H,K,L) = (1+h,0+k,1+\ell)$ and $q_a a = 2\pi h$, etc., to obtain
\bea
\hbar v_a &=& aS\sqrt{(2J_2 + J_{1a})(2J_2 + J_{1a}+J_{1c})},\nonumber\\
\ \label{EqSWvel}\\
\hbar v_b &=& bS\sqrt{(2J_2- J_{1b})(2J_2 + J_{1a}+ J_{1c}) - DJ_{1b}},\nonumber\\
\ \nonumber\\ 
\hbar v_c &=& cS\sqrt{J_{1c}(2J_2 + J_{1a} + J_{1c})}.\nonumber
\eea
where $a,\ b$ and~$c$ are the orthorhombic lattice parameters.  

In some cases, only $\hbar v_a$ and $\hbar v_c$ are measured.  In that case, one can determine $SJ_{1c}$ and the combination $S(J_{1a}+ 2J_2)$ from Eqs.~(\ref{EqSWvel}) as
\bea
S(J_{1a}+2J_2) &=& \frac{c (\hbar v_a)^2}{a \sqrt{c^2 (\hbar v_a)^2 + a^2 (\hbar v_c)^2}},\nonumber\\
\label{EqGetJ1aJ2} \\
SJ_{1c} &=& \frac{a (\hbar v_c)^2}{c \sqrt{c^2 (\hbar v_a)^2 + a^2 (\hbar v_c)^2}}.\nonumber
\eea

Another possibility is that only $\hbar v_a$ and $\hbar v_b$ are measured to have different nonzero values, but $v_c$ (and therefore $J_{1c}$) is measured to be close to zero.  In that case, one can determine $SJ_{1c}$ and the combination $S(2J_2 + J_{1a})$ and $S(2J_2 - J_{1b})$ from Eqs.~(\ref{EqSWvel}) as
\bea
S(2J_2 + J_{1a}) &=& \frac{\hbar v_a}{a},\nonumber\\
\label{EqGetJ1aJ1b} \\
S(2J_2 - J_{1b}) &=& \frac{a (\hbar v_b)^2}{b^2 \hbar v_a}.\nonumber
\eea

Equations~(\ref{EqSWvel}) place restrictions on the ranges of the respective exchange constants that allow spin waves to propage in the different directions (the arguments of the square roots must be positive).  The restriction $J_2 > J_{1b}/2$  that is required for $D = 0$ to obtain a real $v_b > 0$ in Eqs.~(\ref{EqSWvel}) is the same as one of the two classical stability conditions in Eq.~(\ref{EqStripeStab}) for the observed magnetic stripe-$b$ phase.  The stripe magnetic structure becomes unstable with respect to the N\'eel state when $v_b \to 0$, and $v_b$ is especially sensitive to the value of $J_{1b}$.  From Eqs.~(\ref{EqSWvel}), when $J_{1a} = J_{1b}$ one obtains the perhaps counterintuitive result that $v_a \neq v_b$ even when $D = 0$.  This difference between $v_a$ and $v_b$ occurs because the two axes $a$ and~$b$ are not magnetically equivalent in the stripe-$b$ phase, as is clear from Fig.~\ref{Stripe_Mag_Struct}.

From the above equations, only the products of the ordered spin $S$ with the exchange constants and anisotropy parameters can be determined from fitting the  experimental dispersion relations by the theoretical ones.  However, $S$ ($S_{\rm eff}$) can be obtained by fitting the absolute intensity of the scattered neutrons using Eqs.~(\ref{EqSaa}).  Alternatively, as noted above, $S$ is determined independently from the fitted ordered moment when solving the low-temperature ordered magnetic structure if one assumes a value for the $g$-factor.

\subsubsection{\label{SecChippNAFL} Dynamical Susceptibility of the Nearly Antiferromagnetic Fermi Liquid}

Several groups\cite{Inosov2010, Diallo2010} have analyzed their magnetic inelastic neutron scattering results for $\chi^{\prime\prime}({\bf Q},\omega)$ of Fe-based superconductors and parent compounds in terms of the quasielastic response for a nearly antiferromagnetic Fermi liquid.\cite{Nakayama1987}  Here, the term ``quasielastic'' means that the response is diffusive and incoherent, with a peak in $\chi^{\prime\prime}({\bf Q},\omega)/\omega$ at $\omega = 0$, so there is no restoring force for spin waves ($\omega_0 = 0$) and hence no collective spin wave excitations.  However, the form for $\chi^{\prime\prime}({\bf Q},\omega)$ for a nearly antiferromagnetic Fermi liquid cannot be obtained by a simple modification of the expression (\ref{EqfomQ}) for a damped simple harmonic oscillator.  Instead, the general form of $\chi({\bf Q},\omega)$ is derived in independent calculations to be
\be
\chi({\bf Q},\omega) = \chi^\prime({\bf Q},0)\,\frac{G}{G - i(\hbar\omega)},
\label{EqGenChi}
\ee
where $\chi^\prime({\bf Q},0)$ is again the static susceptibility at wave vector {\bf Q}, $G\equiv G({\bf Q})$, and both $\chi^\prime$ and $G$ depend on temperature $T$.  Then $\chi^{\prime\prime}({\bf Q},\omega)$ is
\be
\chi^{\prime\prime}({\bf Q},\omega) = \chi^\prime({\bf Q},0)\,\frac{G\hbar\omega}{(\hbar\omega)^2 + G^2},
\label{EqChippNAFL}
\ee
To fit their inelastic neutron scattering data on the tetragonal optimally-doped superconductor ${\rm Ba(Fe_{0.925}Co_{0.075})_2As_2}$, Inosov et al.\ successfully used the expression\cite{Inosov2010}
\bea
\chi^{\prime\prime}({\bf q},\omega)_T &=& \frac{\chi_T \Gamma_T \, \hbar\omega}{(\hbar\omega)^2 + \Gamma_T^2(1 + \xi_T^2q^2)^2}\label{EqChippInosov}\\
{\rm with}\ \ \ \chi_T &=& \frac{C}{T + \theta},\nonumber\\
\Gamma_T &=& \Gamma_0(T + \theta),\nonumber\\
\xi_T &=& \frac{\xi_0}{\sqrt{T + \theta}},\nonumber\\
{\rm and}\ \ \ \ \ {\bf q} &\equiv& {\bf Q} - {\bf Q}_{\rm AF},\nonumber
\eea
where $\chi_T$ is a mean-field measure of the strength of the antiferromagnetic correlations in the normal state (i.e., the staggered susceptibility), $C$ is the Curie constant and $\theta({\bf q})$ is the Weiss temperature in the Curie-Weiss law for $\chi_T$, $\Gamma_T$ characterizes the magnetic excitation damping strength, $\xi_T$ is the mean-field magnetic correlation length, and ${\bf Q}_{\rm AF}$ is the antiferromagnetic ordering wave vector.

One can cast the expression~(\ref{EqChippInosov}) for $\chi^{\prime\prime}({\bf Q},\omega)$ into the general form~(\ref{EqChippNAFL}).  The static susceptibility is  
\bea
\chi^\prime(q,0) &=& \frac{\chi_T}{1 + \xi_T^2 q^2}\label{EqChippNAFFL}\\
&=& \frac{C}{T + \theta + \xi_0^2 q^2}.\nonumber
\eea
Thus the static susceptibility at a given {\bf Q} is just the mean-field Curie-Weiss law, but where the Weiss temperature $\theta + \xi_0^2 q^2$ depends on {\bf Q}.  At a given $T$, the $\chi^\prime({\bf Q},0)$ is largest for ${\bf q} = 0$, i.e., for ${\bf Q} = {\bf Q}_{\rm AF}$, as expected.  The relaxation rate $G$ in Eq.~(\ref{EqChippNAFL}) is obtained directly from Eqs.~(\ref{EqChippInosov}) as
\bea
G &=& \Gamma_T(1 + \xi_T^2 q^2) \\
&=& \Gamma_0(T + \theta + \xi_0^2 q^2).\nonumber
\eea
Note that $\Gamma_0$ has units of eV/K and $\xi_0^2$ has units of \AA$^2$\,K\@.  The relaxation rate at fixed ${\bf q}$ decreases with decreasing $T$ as expected.  From the staggered susceptibility for the ordering wavevector ${\bf q} = 0$ in Eq.~(\ref{EqChippNAFFL}), the expression~(\ref{EqChippInosov}) for $\chi^{\prime\prime}({\bf q},\omega)$ allows for long-range antiferromagnetic order to occur if $\theta({\bf q} = 0)$ is negative (this $\theta$ is for the staggered {\bf q} = 0, not the uniform {\bf Q} = 0, susceptibility).

Diallo et al.\ used a somewhat different and more complicated expression for $G({\bf Q})$ in Eq.~(\ref{EqGenChi}) to analyze their inelastic neutron scattering data on ${\rm CaFe_2As_2}$ single crystals in the paramagnetic state with $T > T_{\rm N}$.\cite{Diallo2010}  The reasons for the different treatment are (1) anisotropy between the orthorhombic $a$ and $b$ axes continues into the tetragonal phase above $T_{\rm N}$ (!)\ and this in-plane (2D) anisotropy has to be included in $\chi({\bf Q},\omega)$, and (2) weak dispersion is observed in the magnetic excitations along the $c$-axis that also has to be included.  With respect to the 2D part, the authors used the  generalized susceptibility in \emph{tetragonal} reciprocal lattice notation as
\be  
\chi_{\rm 2D}({\bf Q}^{\rm 2D}_{\rm AF} + {\bf q}_{\rm 2D},\omega) = \frac{\chi_0}{(q_{\rm 2D}^2 + \eta q_xq_y)a_{\rm T}^2 +\left(\frac{\xi_T}{a_{\rm T}}\right)^{-2} - i\frac{\hbar\omega}{\gamma}},
\label{EqChiGenFern}
\ee 
where ${\bf q}_{\rm 2D} = (q_x,q_y) = \frac{2\pi}{a_{\rm T}}(h,k)$ is the deviation of the in-plane component of the wave vector from the in-plane component of the antiferromagnetic wave vector ${\bf Q}^{\rm 2D}_{\rm AF} = \frac{2\pi}{a_{\rm T}}\left(\frac{1}{2},\frac{1}{2}\right)$, $\eta$ is the in-plane anisotropy parameter mentioned above, $\gamma$ is the Landau damping parameter, $\chi_0$ is the staggered susceptibility, $a_{\rm T}$ is the $a$-$b$~plane tetragonal lattice parameter, and $\xi_T$ is a temperature-dependent correlation length.  

The effect of the weak interlayer coupling $J_c$ on the dynamical susceptibility is taken into account via 
\be
\frac{1}{\chi({\bf Q}_{\rm AF} + {\bf q},\omega)} = \frac{1}{\chi_{\rm 2D}({\bf Q}^{\rm 2D}_{\rm AF} + {\bf q}_{\rm 2D},\omega)} + 2J_c \sin^2\left(\frac{q_zc}{4}\right)
\label{EqDialloChi3D}
\ee
where $\chi_{\rm 2D}({\bf Q}^{\rm 2D}_{\rm AF} + {\bf q}_{\rm 2D},\omega)$ is given in Eq.~(\ref{EqChiGenFern}) and the scattering vector is ${\bf Q} = {\bf Q}_{\rm AF} + {\bf q}_{\rm 2D} + q_z\hat{\bf c}$.  The imaginary part of the dynamical susceptibility used to fit the neutron scattering data is obtained by taking the imaginary part of Eq.~(\ref{EqDialloChi3D}), which we will not write down.

Setting ${\bf Q} = 0$ and $\omega = 0$ in Eq.~(\ref{EqDialloChi3D}) gives a prediction for the static 
uniform spin susceptibility\cite{Diallo2010}
\be
\chi_{\rm spin} = \chi(0,0) = \frac{\chi_0}{2\pi^2\left(1 + \frac{\eta}{2} \right) +\left(\frac{\xi_T}{a_{\rm T}}\right)^{-2} + 2J_c\chi_0}.
\label{EqChiFromNeuts}
\ee

\subsection{\label{SecNeutInel} Spin Dynamics from Inelastic Magnetic Neutron Scattering Measurements: Experimental Results }

\subsubsection{Introduction}

\begin{figure}
\includegraphics[width=3.3in]{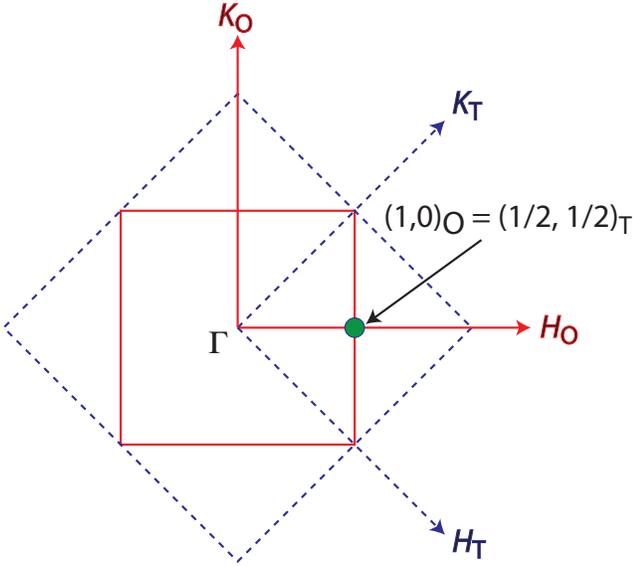}
\caption{(Color online) A sketch of the Brillouin zones (BZ) in the $Q_x$-$Q_y$~plane in orthorhombic (O, solid red axes and BZ) and tetragonal (T, dashed blue axes and BZ) notation.  The center of the BZ is at (0,0) in either notation and is designated by the symbol $\Gamma$.  The first BZ boundaries are at ${\bf Q}_{\rm BZ}({\rm \AA}^{-1}) = (\pm\frac{2\pi}{a}\hat{{\bf a}},0)$ and $(0,\pm\frac{2\pi}{b}\hat{{\bf b}})$, where $a$ and $b$ are the respective $a$-axis and $b$-axis lattice parameters in tetragonal or orthorhombic notation.  In reciprocal lattice units (r.l.u.), these BZ boundaries are designated as $(H,K) = (\pm 1,0)$ and $(0, \pm 1)$, respectively. The first BZ boundaries along the $\hat{{\bf c}}~{\rm or}~L$ direction (not shown) are at $Q_{{\rm BZ},c}({\rm \AA}^{-1}) = \pm\frac{2\pi}{c}$ or $L = \pm 1$~r.l.u.\ in either tetragonal or orthorhombic notation.  The in-plane component of the antiferromagnetic ordering vector for the FeAs-based layered compounds is shown as a filled green circle, which in orthorhombic notation is $(H,K) = (1,0)$ and in tetragonal notation is $(H,K) = \left(\frac{1}{2},\frac{1}{2}\right)$.}
\label{FigBrillouin_Zones}
\end{figure}

Before discussing the experiments, it is important to point out a difference in notation used in different neutron scattering papers to describe spin waves and/or spin fluctuations in the 122-type and 1111-type FeAs-based layered compounds.  As shown above in Fig.~\ref{FigTetrag_Ortho_struct}, the $a$-$b$~plane of the low-temperature orthorhombic structure is rotated by 45$^\circ$ with respect to the high-temperature tetragonal structure, and the orthorhombic $a$ and $b$~lattice parameters are about a factor of $\sqrt{2}$ larger than the tetragonal lattice parameter $a$.  The $c$-axis parameter is about the same in the two structures.  Long-range antiferromagnetic ordering is only observed in the low-temperature orthorhombic structure.  What this means is that the Brillouin zone in the orthorhombic structure is rotated by 45$^\circ$ in the $a$-$b$~plane with respect to that in tetragonal notation, and the basal plane Brillouin zone edges are a factor of $\sqrt{2}$ smaller in orthorhombic notation as shown in Fig.~\ref{FigBrillouin_Zones}.  Thus, the AF ordering wave vector in orthorhombic notation is (1,0,1)~r.l.u., whereas in tetragonal notation it is $(\frac{1}{2},\frac{1}{2},1)$~r.l.u.  Furthermore, a longitudinal scan parallel to the basal plane through the antiferromagnetic orthorhombic (1,0,1) peak along the $a$~direction would be a ($H$,0,1) scan, whereas in tetragonal notation it would be a $(\frac{H}{2},\frac{H}{2},1)$ scan according to Fig.~\ref{FigBrillouin_Zones}.  Similarly, a transverse scan in the basal plane through the orthorhombic (1,0,1)~r.l.u.\ peak along the $b$~direction would be a (1,$K$,1)~r.l.u.\ scan, whereas in tetragonal notation it would be a $(-\frac{K}{2},\frac{K}{2},1)$~r.l.u.\ scan.  

With respect to the FeAs-based compounds, it makes sense to use orthorhombic r.l.u.\ notation in the orthorhombic phase because of its simplicity.  In particular, as seen in Fig.~\ref{Stripe_Mag_Struct} the ordered moment and the antiferromagnetic propagation vector component in the basal plane are both along the orthorhombic $a$-axis, and the spin stripes are oriented along the orthorhombic $b$-axis.  Furthermore, the spin wave dispersion relations~(\ref{EqDispRln}) and~(\ref{McQueeneyDispRln}) are expressed in orthorhombic  notation.  Therefore we have used orthorhombic notation in the above sections and will continue to do so unless tetragonal notation is required to discuss particular literature data.  Some groups prefer tetragonal notation and other groups prefer orthorhombic notation for the orthorhombic phase. 

\subsubsection{\label{Sec122typeneuts} 122-Type FeAs-based Compounds}

\subsubsection*{a. Low-Energy Spin Waves at Temperatures $T < T_{\rm N}$}

\begin{figure}
\includegraphics[width=3.in]{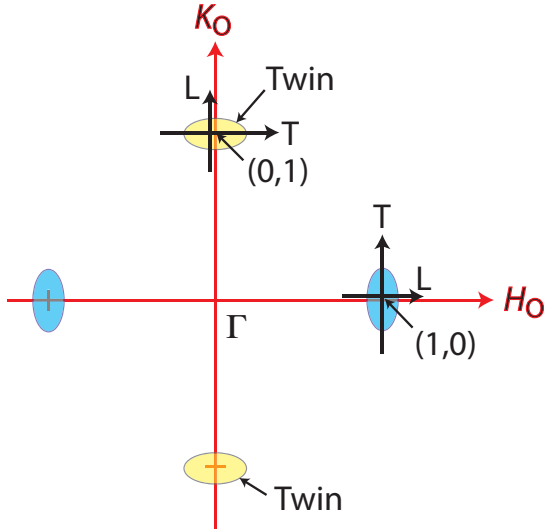}
\caption{(Color online) Sketch showing the influence of twinning on the in-plane magnetic scattering peak at fixed energy transfer in the first Brilloin zone of the orthorhombic (O) 122-type FeAs compounds.  The blue ellipses at $(H,K) = (\pm 1,0)$~r.l.u.\ are the spin wave scattering contours for an untwinned crystal at the \emph{magnetic} zone centers.  Here ``L'' refers to a longitudinal (in-plane radial) scan at fixed $L$ and ``T'' to a transverse (in-plane tangential) scan at fixed $L$ through the ellipse.  The ellipse is elongated in the direction of the smaller spin wave velocity component.  Here, since it is known that $v_a > v_b$ in the 122-type $A{\rm Fe_2As_2}$ compounds, the minor axis of the ellipse is in the radial ($\hat{\bf a}$) direction and the major axis is in the transverse ($\hat{\bf b}$) direction. The yellow ellipses at $(H,K) = (0,\pm 1)$~r.l.u.\ arise from twins that occur upon cooling through the tetragonal-orthorhombic transition temperature, in which the $\hat{\bf a}$ and $\hat{\bf b}$ axis unit vectors are switched.  In the figure, from the relative intensities of the blue and yellow ovals, the populations of the two twins are about the same in this example.  The anistropy in the response at (1,0)~r.l.u.\ should be the same as at (0,1)~r.l.u.\ for equal twin populations.  However, it is possible that some scattering from the twin appears at zone center of the untwinned domain, thus each distorting the spin response near each magnetic zone center of the other.  On the other hand, except when $J_{1a} \approx J_{1b}$, the plots of the intensity versus spin wave wave vector in Fig.~\ref{SW_Intensity} indicate that there is likely little interference between the magnetic scattering from the twinned and untwinned domains if the domains are macroscopic.}
\label{FigTwins}
\end{figure}

Studies of the spin dynamics in the FeAs-based materials have been carried out using inelastic neutron scattering techniques.\cite{Lynn2009, Osborn2009}  As seen above in Eqs.~(\ref{EqSWvel}), the spin wave velocities along the orthorhombic $a$- and $b$-axes are not expected to be the same.  As shown in Fig.~\ref{FigTwins}, it is clear from the type of neutron scan made (longitudinal or transverse) which spin wave velocity $v_a$ or $v_b$ is being measured.  Due to time and resource constraints, usually only $v_a$ is measured [e.g., longitudinal $(H,0,1)$ scans through the AF wave vector (1,0,1)] in single crystal experiments.  To measure $v_b$, one would need to carry out transverse scans through the AF wave vector such as $(1,K,1)$ scans.

Since the in-plane component of the antiferromagnetic (AF) propagation vector in the 122- and 1111-type FeAs-based materials is $Q_{{\rm AF},{ab}} = (\pm 1,0)$, for a single domain orthorhombic crystal at low temperatures there should be no AF peaks at $Q_{ab} = (0,\pm 1)$ in Fig.~\ref{FigTwins} as required by the twofold rotation axes in orthorhombic symmetry.  However, most crystals are twinned below the tetragonal-orthorhombic structural transition temperature, which switches the $\hat{\bf a}$ and $\hat{\bf b}$ lattice directions, resulting in a false apparent fourfold rotational symmetry if the populations of the two twins are the same.  It is possible that there can be scattering at the untwinned (1,0) magnetic zone center from the scattering from twin Brillouin zone centers at (0,1), and \emph{vice versa}.  As seen below in Table~\ref{NeutDispersionData2}, the exchange constant ranges inferred in different experiments are approximately
\bea
\frac{J_{1c}}{J_{1a}} &\sim& 0.01\ {\rm to\ 2},\nonumber\\
\frac{J_{1b}}{J_{1a}} &\sim& -0.2\ {\rm to\ 0.5},\label{JParPranges}\\
\frac{J_{2}}{J_{1a}} &\sim& 0.5\ {\rm to\ 1}.\nonumber
\eea
By reference to the spin wave dispersion relation data at {\bf q} = (110)~r.l.u.\ in Fig.~\ref{SW_Disp_Reln}, one sees that it is possible that there could be interference from the untwinned domains at the magnetic Brillouin zone centers of the twins, and vice versa.  However, the plots of the intensity versus spin wave wave vector in Fig.~\ref{SW_Intensity} show that there is likely little interference if the twins are macroscopic.  There has been little explicit discussion in the literature of the influence of twinning on the observed inelastic neutron scattering spectra.

As discussed above in Sec.~\ref{SecStructOverview}, twins have been observed optically below the respective tetragonal-orthorhombic transition temperature in $A$${\rm Fe_2As_2}$ ($A$ = Ca, Sr, Ba)\cite{Tanatar2009b} and ${\rm Ba(Fe_{0.985}Co_{0.015})_2As_2}$,\cite{Chu2009b}  with in-plane separations $\sim 10$--50~$\mu$m. Transmission electron microscopy of the $A$${\rm Fe_2As_2}$ compounds gives similar results except that the twin boundaries are separated by only 0.1--0.4~$\mu$m.\cite{Ma2009}   In addition, a tweed pattern is found in Ca${\rm Fe_2As_2}$.\cite{Ma2009}   

%\clearpage
%\squeezetable
\begin{table*}
\caption{\label{NeutDispersionData} Parameters characterizing the dispersion of spin waves in single crystals of FeAs-based compounds according to Eq.~(\ref{Eqdispersion}).  The quantities listed are the superconducting and/or crystallographic/magnetic transition temperature $T_{\rm c}$, $T_{\rm N}$ and/or $T_{\rm S} \equiv T_0$, the temperature $T$ at which the measurements were taken, the energy transfer $\hbar\omega$ range of the measurements, the spin-wave energy gap $\Delta$, and the $ab$-plane and $c$-axis spin wave velocities multiplied by $\hbar$, $\hbar v_{ab}$ or $\hbar v_a$, $\hbar v_b$ and $\hbar v_c$.  The $v_a$ is parallel to the AF propagation vector and perpendicular to the spin stripes, whereas $v_b$ is parallel to the spin stripes.  The superconducting compound ${\rm Ba(Fe_{0.935}Co_{0.065})_2As_2}$ is tetragonal at all temperatures and exhibits no long-range magnetic ordering above 2~K\@. To obtain the spin wave velocities $v_{a,b}$ and $v_c$ in conventional units, the conversion is 1~eV~\AA$/\hbar = 1.52 \times 10^7$~cm/s.  An error in parentheses is the error in the last digit of the preceding value.  }
\begin{ruledtabular}
\begin{tabular}{l|cccccccc}
Compound   &  $T_0$ & $T$ & $\hbar\omega$ range & $\Delta$ & $\hbar v_{a}$ &$\hbar v_{b}$  & $\hbar v_{c}$  & Ref. \\
& (K) & (K) & (meV) & (meV) & (eV~\AA) &  (eV~\AA) &(eV~\AA) & \\ \hline
${\rm CaFe_2As_2}$ & 172 & 14 & 10--25& 6.9(2) & 0.42(7) &  & 0.27(10) &  \onlinecite{McQueeney2008} \\
 & --- & --- & --- & 2.1\footnotemark[2] & 0.47\footnotemark[2] & 0.29\footnotemark[2] & 0.19\footnotemark[2] &  \onlinecite{McQueeney2008} \\
${\rm CaFe_2As_2}$ &  & 10 & 25--210 & $\equiv 0$ & 0.50(9)\footnotemark[4] & 0.35(4)\footnotemark[4] & 0.26(6)\footnotemark[4] &  \onlinecite{JZhao2009} \\
${\rm SrFe_2As_2}$ & 200--220 & 160 & 1--16 & 3.5 & 0.57(11)\footnotemark[4] &  & 0.28(6)\footnotemark[4] &  \onlinecite{JZhao2008} \\
 &  & 7 & 1--16 & 6.5 &  &  &  &  \onlinecite{JZhao2008} \\
${\rm BaFe_2As_2}$ &  & 5 & 10--12 & 9.8(4) & 0.28(15)   & & 0.057(7) & \onlinecite{Matan2009} \\
${\rm Ba(Fe_{0.96}Co_{0.04})_2As_2}$ & 11, 58.0(6) & 1.6, 20 & 3--11 & 8.1(2)\footnotemark[4] & 0.181(1)\footnotemark[4] &  & 0.043(1)\footnotemark[4]  & \onlinecite{Christianson2009} \\
${\rm Ba(Fe_{0.953}Co_{0.047})_2As_2}$ & 17,47,60 & 25 & 5--10 & 8(1) & $\geq 0.123$ &  &  0.043(9) & \onlinecite{Pratt2010} \\
${\rm Ba(Fe_{0.935}Co_{0.065})_2As_2}$ & 23,--- & 7 & 9--75 & 10.0(5)\footnotemark[1] & 0.58(6) & 0.23(3) &  \large\footnotemark[3] & \onlinecite{Lester2010} \\
\end{tabular}
\end{ruledtabular}
\footnotetext[1]{This spin gap is due to the ``spin resonance'' in the superconducting state (see Sec.~\ref{ResonanceMode}), rather than due to single-ion spin anisotropy and long-range antiferromagnetic order.}
\footnotetext[2]{Calculated using an itinerant model.}
\footnotetext[3]{Too small to measure.}
\footnotetext[4]{Calculated here from the $SJ$ and $SD$ values in Table~\ref{NeutDispersionData2} using Eqs.~(\ref{EqDeltaDiallo}) and~(\ref{EqSWvel}).}
\end{table*}

%\clearpage
%\squeezetable
\begin{table*}
\caption{\label{NeutDispersionData2} Exchange and anisotropy parameters characterizing the dispersion of spin waves in single crystals of FeAs-based compounds in a local moment model.  Here $S$ is the spin per Fe atom.  For the compounds that exhibit antiferromagntic ordering, $S = \mu/(g\mu_{\rm B})$ is the ordered Fe spin where $\mu$ is the ordered moment per Fe atom and assuming the $g$-factor $g = 2$.  Also, $J_1$, $J_2$ and $J_{1c}$ are the exchange constants in a $J_1$-$J_2$-$J_c$ model, and $J_{1a}$, $J_{1b}$, $J_2$ and $J_{1c}$ are the exchange constants in a $J_{1a}$-$J_{1b}$-$J_2$-$J_{1c}$ model.  The parameter $D$ is the single-ion axial anisotropy coefficient.  The superconducting compound ${\rm Ba(Fe_{1.935}Co_{0.065})_2As_2}$ is tetragonal at all temperatures and exhibits no long-range magnetic ordering above 2~K\@. An error in parentheses is the error in the last digit of the preceding value.}
\begin{ruledtabular}
\begin{tabular}{l|cccccccccc}
Compound      & $T$ & $\hbar\omega$ range & $S(2J_2-J_{1b})$ & $S(2J_2+J_{1a})$ & $SJ_1$ or $SJ_{1a}$ & $SJ_{1b}$ & $SJ_2$ & $SJ_c$ & $SD$  & Ref. \\
  & (K) & (meV) & (meV) & (meV) & (meV)& (meV) &  (meV) & (meV) & (meV) & \\ \hline
${\rm CaFe_2As_2}$   & 14 & 10--25&  & 76(14)\footnotemark[5] &  &  &  & 6.7(30)\footnotemark[5] & 0.07(2)\footnotemark[4] &\onlinecite{McQueeney2008} \\
  &  & &  &  & 41\footnotemark[2]$^,$\footnotemark[7] & 10\footnotemark[7] & 21\footnotemark[7] & 3\footnotemark[7] &  &\onlinecite{McQueeney2008} \\
${\rm CaFe_2As_2}$   &  & & 31.4\footnotemark[7] & 56.9\footnotemark[7] & 27.6\footnotemark[2]$^,$\footnotemark[7] & $-2.1$\footnotemark[7] & 14.6\footnotemark[7] &  &  &\onlinecite{Han2008a} \\
${\rm CaFe_2As_2}$   & 10 & 24--147&  & & 31(7)\footnotemark[2]$^,$\footnotemark[8] & 13(8)\footnotemark[8] & 31(3)\footnotemark[8] & 4.5(1) & 0.063 &\onlinecite{Diallo2009} \\
${\rm CaFe_2As_2}$   & 10 & 25--210&  & & 49.9(99)\footnotemark[2] & $-5.7(45)$ & 18.9(34) & 5.3(13) & $\equiv 0$ &\onlinecite{JZhao2009} \\
${\rm SrFe_2As_2}$   & 160 & 1--16 & & 100(20) &    &    & & 5(1)  & 0.018(4)\footnotemark[4] &\onlinecite{JZhao2008} \\
${\rm SrFe_2As_2}$   &  & & 24.8\footnotemark[7] & 62.5\footnotemark[7] & 35.5\footnotemark[2]$^,$\footnotemark[7] & $2.2$\footnotemark[7] & 13.5\footnotemark[7] &  &  &\onlinecite{Han2008a} \\
${\rm BaFe_2As_2}$   & 5 & 10--12 & & 71(37)\footnotemark[5] &    &    & & 0.27(5)\footnotemark[5]  & &\onlinecite{Matan2009} \\
${\rm BaFe_2As_2}$   &  & & 26.6\footnotemark[7] & 60.1\footnotemark[7] & 36.1\footnotemark[2]$^,$\footnotemark[7] & $-2.6$\footnotemark[7] & 12.0\footnotemark[7] &  &  &\onlinecite{Han2008a} \\
${\rm Ba(Fe_{0.96}Co_{0.04})_2As_2}$ & 1.6, 20 & 3--11 &  & 32.0(1) &  & && 0.34(1) & 0.25(1) & \onlinecite{Christianson2009} \\
${\rm Ba(Fe_{0.935}Co_{0.065})_2As_2}$ &   7 & 9--75 & 3(4)\footnotemark[6] & 104(10)\footnotemark[6] & 43(7)\footnotemark[1]$^,$\footnotemark[3] & --- & 30(3)\footnotemark[3] & $\approx 0$ & &\onlinecite{Lester2010} \\
\end{tabular}
\end{ruledtabular}
\footnotetext[1]{This value is $J_1\equiv J_{1a}=J_{1b}$.}
\footnotetext[2]{This value is $J_{1a}$.}
\footnotetext[3]{Uses Eqs.~(\ref{EqJ1J2vavb}).}
\footnotetext[4]{Calculated here using Eq.~(\ref{EqSD}), the $J$ values in this table, and the $\Delta$ value in Table~\ref{NeutDispersionData}.}
\footnotetext[5]{Calculated here using Eqs.~(\ref{EqGetJ1aJ2}) and the $v_a$ and $v_c$ data in Table~\ref{NeutDispersionData}.}
\footnotetext[6]{Calculated here using Eqs.~(\ref{EqGetJ1aJ1b}) and the $v_a$ and $v_b$ data in Table~\ref{NeutDispersionData}.}
\footnotetext[7]{Calculated using an itinerant model.}
\footnotetext[8]{Calculated assuming that $\hbar\omega[{\bf q} = (010)]$ is between 50 and 150~meV\@.}
\end{table*}

Experimental spin wave velocities determined from inelastic neutron scattering measurements using the low-energy dispersion relation~(\ref{Eqdispersion}) are summarized in Table~\ref{NeutDispersionData}.\cite{Matan2009, Lester2010, McQueeney2008, JZhao2009, Christianson2009, JZhao2008, Pratt2010}  Rather than report the spin wave velocities themselves in conventional units of cm/s, neutron scatterers always report spin wave velocities multiplied by $\hbar$ in units of eV~\AA\ [see Eq.~(\ref{Eqdispersion})].  The conversion factor is given in the caption to Table~\ref{NeutDispersionData}.  The measurements all show a steep initial dispersion with large energy gaps $\Delta \sim 6.5$--9.8~meV arising presumably from single-ion spin anisotropy.  The spin wave velocities for the $A{\rm Fe_2As_2}$ parent compounds are in the range $\hbar v_{ab} \sim 280$--560~meV\,\AA\ with $v_c/v_{ab} = 0.2$--0.5.  Thus the spin wave dispersion is three-dimensional but anisotropic, which contrasts with the nearly two-dimensional dispersion in the layered cuprates.\cite{Johnston1997}  The in-plane spin wave velocities for the 122-type FeAs-based materials are about a factor of two smaller than the value $\hbar v_{ab} = 0.85$~eV\,\AA\ observed for $T\to 0$ for the layered cuprate parent compound La$_2$CuO$_4$ below its N\'eel temperature of 325~K (Ref.~\onlinecite{Johnston1997}) and the coincidentlly nearly identical value of 0.85(10)~eV~\AA\ in the classic itinerant antiferromagnet Cr metal.\cite{Fawcett1988}  Exchange constants derived for various Fe-based compounds from neturon scattering data are given in Table~\ref{NeutDispersionData2},\cite{Matan2009, Lester2010, McQueeney2008, Diallo2009, JZhao2009, Christianson2009, JZhao2008} from which the spin wave velocities can be calculated using the expressions in Sec.~\ref{SecSWCalcfromJ}, which are then given in Table~\ref{NeutDispersionData}.  Values of the exchange constants calculated by Han et al.\ using itinerant models are also listed in Table.~\ref{NeutDispersionData}.\cite{Han2008a}  In the following we briefly discuss several low-energy spin wave scattering experiments on the undoped $A{\rm Fe_2As_2}$ parent compounds.  

\begin{figure}
\includegraphics[width=3.3in,viewport=19 10 130 145,clip]{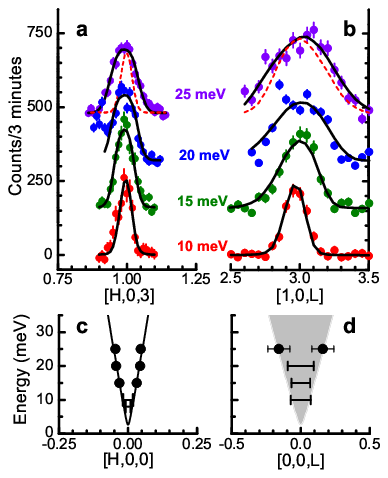}
\caption{(Color online) (a) and (b): Magnetic neutron scattering intensity versus wave vector at energy transfers from 10 to 25~meV.\cite{McQueeney2008}   Scans at different energies are vertically offset by 160 counts.  The solid black curves are fits to the data using Eq.~(\ref{EqfomQ2}) and the model dispersion relation in Eq.~(\ref{Eqdispersion}), convoluted with the instrumental resolution function.  The red dashed lines for $\hbar\omega = 25$~meV in (a) and~(b) are estimates of the resolution lineshapes, respectively. (c), (d): Dispersion relations derived from the fits to the data in (a) and~(b).  The notation $H$ and $L$ is different than in (a) and (b); here $H$ and $L$ refer to the \emph{deviation} in r.l.u.\ from the antiferromagnetic wave vector ${\bf Q}_{\rm AF} = (1,0,3)$ (R.~J.\ McQueeney, private communication).  In (d), the model dispersion along the [0,0,$L$] direction is plotted as a shaded region, indicating that the data for energies 10--20~meV only give a lower bound on the spin wave velocity along $L$.  Reprinted with permission from Ref.~\onlinecite{McQueeney2008}.  Copyright (2008) by the American Physical Society.}
\label{McQueeneyfig2}
\end{figure}

Low-energy spin-wave dispersion data obtained by McQueeney et al.\ for co-aligned single crystals of ${\rm CaFe_2As_2}$ are shown in Fig.~\ref{McQueeneyfig2}.\cite{McQueeney2008}  In Figs.~\ref{McQueeneyfig2}(a) and~(b), as the energy increases one does not see two peaks emerging from the data corresponding to counter-propagating spin wave branches, presumably because the dispersion is so steep that the observations are resolution-limited in ${\bf Q}$, but one can see that the widths of the peaks in ${\bf Q}$ increase with increasing energy.  The two counterpropagating spin wave peaks can be resolved in principle by increasing the energy transfer $\hbar\omega$.

The dispersion relation data in Figs.~\ref{McQueeneyfig2}(c) and~(d) were determined by fitting two peaks to the data in (a) and~(b) using Eqs.~(\ref{EqfomQ2}) and~(\ref{Eqdispersion}), as shown by the solid black curves.  The resulting spin wave velocities are listed in Table~\ref{NeutDispersionData}.  An important result emerging from this study\cite{McQueeney2008} is that the in-plane spin wave velocities are large, comparable to but roughly a factor of two smaller than in the layered cuprates as mentioned above,\cite{Johnston1997} and another is that the $c$-axis spin wave velocity is a large fraction of the in-plane spin wave velocity.  Thus the FeAs-type parent compounds should be considered to be anisotropic three dimensional antiferromagnets rather than quasi-two-dimensional antiferromagnets like the layered cuprates.  Furthermore, the authors found that the spin wave velocities could be quantitatively explained using band theory, as shown in Table~\ref{NeutDispersionData}.  They were also able to explain their subsequent high-energy spin wave dispersion and relaxation data on this compound\cite{Diallo2009} using an itinerant magnetism model (see the following section).  

Spin wave dispersion data for single crystal ${\rm BaFe_2As_2}$ (Ref.~\onlinecite{Matan2009}) are discussed below with reference to Fig.~\ref{Matanfig2}.

Spin wave excitations have also been observed by Christianson et~al.\ in single crystal ${\rm Ba(Fe_{0.96}Co_{0.04})_2As_2}$, which according to the phase diagram in Fig.~\ref{FigBaKFe2As2_phase_diag} is a composition where superconductivity and antiferromagnetism coexist below $T_{\rm c}$, in this case with $T_{\rm N} = 58.0(6)$~K and $T_{\rm c} = 11$~K.\cite{Christianson2009}  They carried out constant energy $(H,H,\bar{1})$ and $(\frac{1}{2},\frac{1}{2},L)$~r.l.u.\ scans in tetragonal notation at 3--11~meV, and analyzed their data using Eqs.~(\ref{McQueeneyDispRln}) and~(\ref{EqChris}).  Their spin wave velocities, and $SJ$ and~$SD$ values, are listed in Tables~\ref{NeutDispersionData} and~\ref{NeutDispersionData2}, respectively.  An interesting result from their measurements is that the spin wave dispersion is three-dimensional, as in undoped ${\rm BaFe_2As_2}$, but in contrast to optimally Co-doped Ba(Fe$_{1-x}$Co$_x)_2$As$_2$ where the dispersion is two-dimensional.  In their inelastic neutron scattering study of single crystal ${\rm Ba(Fe_{0.98}Ni_{0.02})_2As_2}$, Harriger et al.\ suggested that the observed decrease in the spin gap to 2~meV upon Ni doping may be due to a decrease in the $c$-axis coupling.\cite{Harriger2009}  

\subsubsection*{b. High-Energy Magnetic Excitations at Temperatures $T < T_{\rm N}$}

\begin{figure*}
\includegraphics [width=4.1in,viewport= 0 -30 700 300,clip]{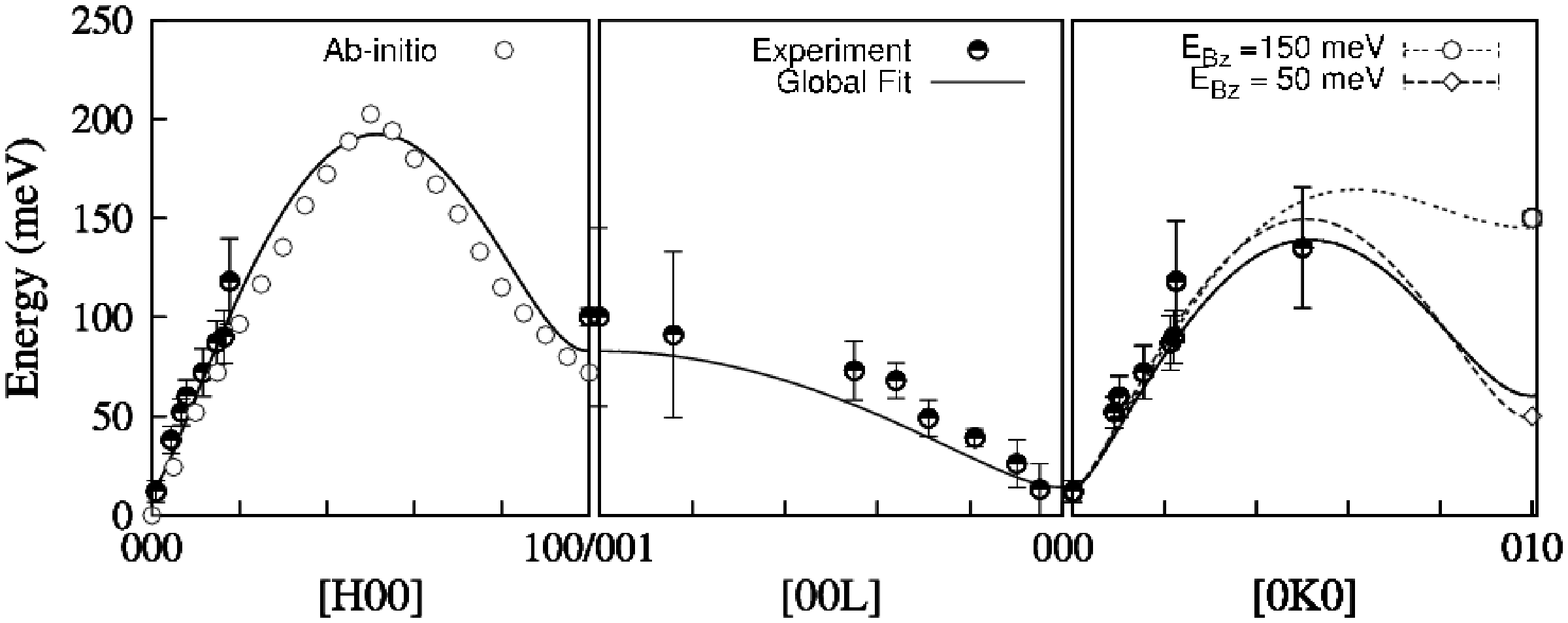}
\includegraphics [width=2.5in]{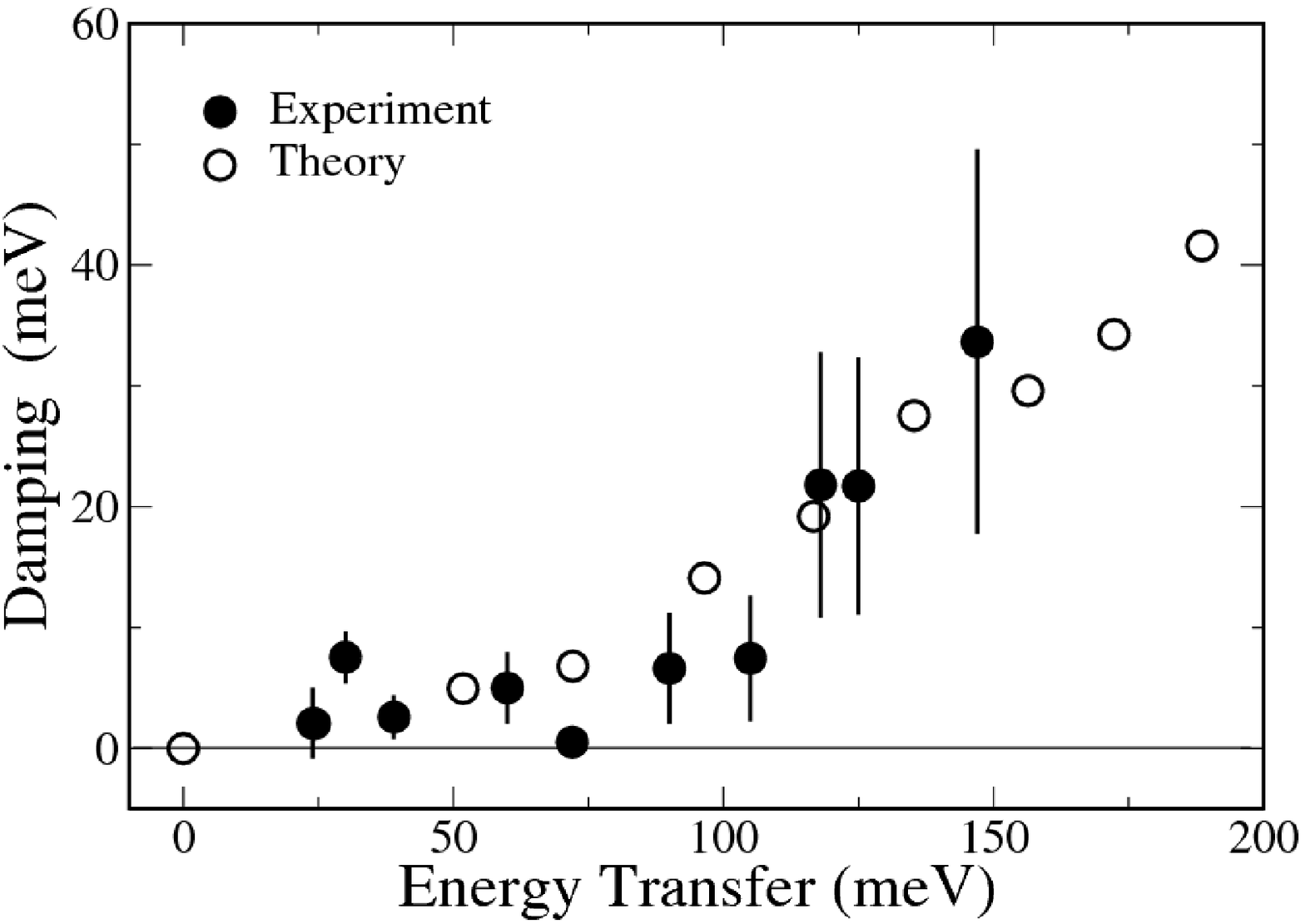}
\caption{Left panels:  Spin wave energy $\hbar\omega$ at 10~K in high symmetry directions for CaFe$_2$As$_2$ crystals.  The orthorhombic reciprocal lattice units on the horizontal axes are the \emph{deviations} ${\bf q}$ from the antiferromagnetic wave vector (1,0,1).\cite{Diallo2009}  Half-filled circles with error bars: experimental data.  Open circles: theoretical maxima in $\chi^{\prime\prime}({\bf q},\omega)$ from \emph{ab initio} band structure calculations.  Solid curves: global fit of spin wave theory [Eq.~(\ref{McQueeneyDispRln})] to the data.  Dashed curves in far right top subpanel: local moment dispersion relations for Brillouin zone energies $\hbar \omega (010) = 50$ and 150~meV\@.  Right panel:  Spin wave relaxation rate $\Gamma$ versus energy transfer $\hbar \omega$.\cite{Diallo2009}  The filled circles with error bars are experimental data and the open circles are Landau damping rates from density functional theory.  Reprinted with permission from Ref.~\onlinecite{Diallo2009}.  Copyright (2009) by the American Physical Society.}
\label{CaFe2As2SpinWaves2} 
\end{figure*}

For magnetic excitations at higher energies up to 200~meV, there are conflicting interpretations of inelastic neutron scattering data at low temperatures for single crystals of the same parent compound CaFe$_2$As$_2$.  The report of Diallo et al.\ favors an itinerant electron interpretation,\cite{Diallo2009} whereas that of Zhao \emph{et al.}\ favors a local moment Heisenberg model description.\cite{JZhao2009}  As emphasized by Diallo et al., one can always fit low-temperature spin wave dispersion relations for either a local or itinerant antiferromagnet state \emph{at low energies} by a local moment model.

The spin wave dispersion relations and damping parameter versus energy at 10~K in CaFe$_2$As$_2$ crystals determined by Diallo et al.\ from inelastic neutron scattering measurements are shown in Fig.~\ref{CaFe2As2SpinWaves2}.\cite{Diallo2009}  The authors fitted their data by the local moment spin wave dispersion relation~(\ref{McQueeneyDispRln}) and found good agreement as shown in the figure with exchange constants listed in Table~\ref{NeutDispersionData2}.  By fitting the absolute neutron scattering intensity, the authors obtained the ordered spin $S = \mu/(g\mu_{\rm B}) = 0.40(5)$, consistent with the ordered moment measured directly by solving the low temperature magnetic structure (see Table~\ref{LoTStructData122} in the Appendix).  It is hard to see how a spin $S = 1/2$ could arise in a local moment description of the magnetism, as also subsequently noted by Zhao et al.\cite{JZhao2009}  Indeed, Diallo et al.\ obtained an excellent quantitative description and understanding of all of their data using band theory as shown in Fig.~\ref{CaFe2As2SpinWaves2}, including the spin wave damping rates in the right-hand panel. Using this itinerant magnetism description, they were thus not confronted with explaining why the $J$ values take the values that they do or how a local spin~1/2 can come about in a local moment description.  They concluded, ``the present results favor an itinerant model for magnetic excitations in the parent ${\rm CaFe_2As_2}$ compound, as opposed to a local moment Heisenberg model. ... ${\rm CaFe_2As_2}$ is thus best described as an itinerant three-dimensional antiferromagnet.''  In subsequent publications, this research group continued to favor and extend the itinerant magnetism model for the FeAs-based superconductors and parent compounds.

\begin{figure}
\includegraphics [width=3.in]{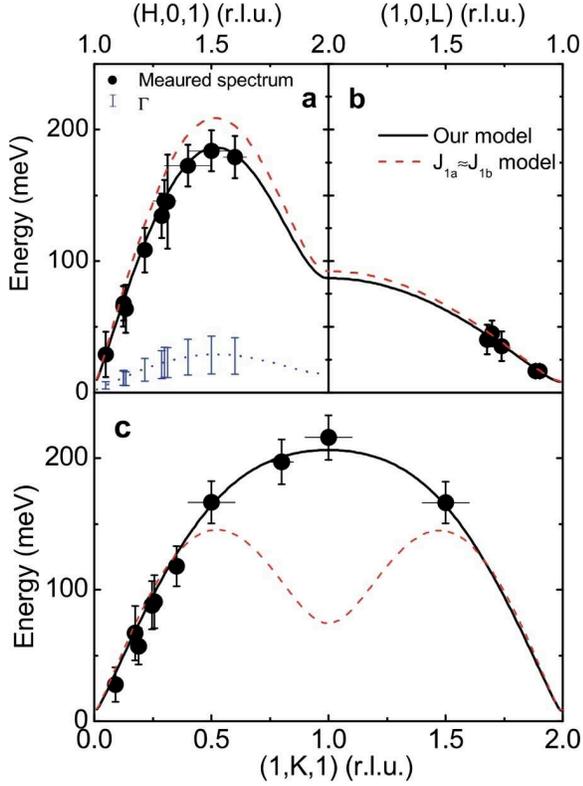}
\caption{(Color online) Spin wave energy $\hbar\omega$ in high symmetry directions for CaFe$_2$As$_2$ at 10~K in orthorhombic reciprocal lattice notation.\cite{JZhao2009}  The filled circles are data from inelastic neutron scattering measurements and the solid black curves are fits by a local moment Heisenberg model.  The dashed red curves are fits by an alternative $J_1$-$J_2$ Heisenberg model where $J_1 \equiv J_{1a}\approx J_{1b}$ and the blue vertical bars at the bottom of (a) are the experimental spin wave damping values $\Gamma \approx 0.15 \hbar\omega$.  Reproduced by permission from Ref.~\onlinecite{JZhao2009} and from Macmillan Publishers Ltd: Ref.~\onlinecite{JZhao2009}, copyright (2009).}
\label{CaFe2As2SpinWaves} 
\end{figure}

The  dispersions at 10~K in CaFe$_2$As$_2$ crystals measured by Zhao et al.\ along different high symmetry directions at 10~K are shown in Fig.~\ref{CaFe2As2SpinWaves}.\cite{JZhao2009} The total bandwidth of the spin wave excitations is about 200~meV\@.  The data were fitted  (solid curves) by a local moment Heisenberg model with anisotropy constant $D = 0$ and with exchange constants $J_{1a},\ J_{1b}$, $J_2$ and $J_{1c}$ as in Fig.~\ref{Stripe_Mag_Struct} using the dispersion relation~(\ref{McQueeneyDispRln}).\cite{JZhao2009}  The fitted $J$ values are listed in Table~\ref{NeutDispersionData2} and the low-energy spin wave velocities calculated from these $J$ values are given in Table~\ref{NeutDispersionData}.  The dashed red curves in Fig.~\ref{CaFe2As2SpinWaves} are fits by an alternative $J_1$-$J_2$ Heisenberg model to the low-energy data where $J_1 \equiv J_{1a}\approx J_{1b}$.  The large deviation of this fit from the high-energy data in Fig.~\ref{CaFe2As2SpinWaves}(c) shows that  setting $J_{1a}\approx J_{1b}$ is a poor approximation.  However, the large anisotropy between positive (antiferromagnetic) $SJ_{1a}$ and negative (ferromagnetic) $SJ_{1b}$ in Table~\ref{NeutDispersionData2} is very hard to understand in a local moment picture in view of the small orthorhombic distortion ($\sim 1$\%, see Table~\ref{LoTStructData122} in the Appendix) of this compound below $T_{\rm N} = T_{\rm S} = 170$~K\@.  Zhao et al.\ stated that this conundrum indicates that ``magnetism in the parent compounds of iron arsenide superconductors is neither purely local nor purely itinerant, rather it is a complicated mix of the two.''\cite{JZhao2009}  The energy-dependent damping of the spin waves is shown at the bottom of Fig.~\ref{CaFe2As2SpinWaves}(a).  The authors did not fit these data by theory, but they suggested that these data are not consistent with an itinerant magnetism description.

Inelastic magnetic neutron scattering measurements of single-phase LaFeAsO$_{1-x}$F$_x$ samples with $x = 0.057$ ($T_{\rm c} = 25$~K), $x = 0.082$ ($T_{\rm c} = 29$~K), and a strongly overdoped sample with $x = 0.158$ ($T_{\rm c} = 7$~K, with 10\% superconducting volume fraction) was carried out at a temperature of 4~K and at energy transfers up to 15~meV\@.\cite{Wakimoto2009}  Magnetic scattering was clearly present in the two superconducting samples with $x = 0.057$ and $x = 0.082$ at the antiferromagnetic wavevector and at an energy of $\sim 11$~meV, but was absent in the poorly superconducting sample with $x = 0.158$.  From these results the authors concluded that spin fluctuations arising from Fermi surface nesting are required for superconductivity to occur in the FeAs-based class of superconductors.

\subsubsection*{c. Magnetic Excitations in the Paramagnetic State above $T_{\rm N}$}

\begin{figure}
\includegraphics[width=3.3in]{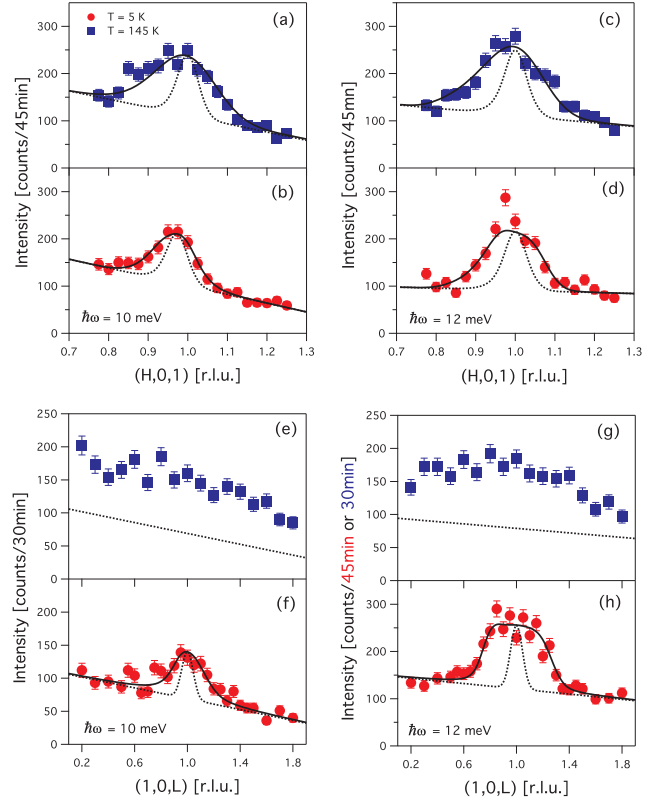}
\caption{(Color online) Constant energy scans at 10~meV (left panels) and 12~meV (right panels) through the orthorhombic magnetic Bragg wave vector (1,0,1)~r.l.u.\ at temperatures of 5~K and 145~K for a single crystal of ${\rm BaFe_2As_2}$ with $T_{\rm N} =  136(1)$~K.\cite{Matan2009}  The scans at 5~K in (b), (d), (f), and (h) were fitted by the low-energy spin wave dispersion relation~(\ref{Eqdispersion}) with parameters given previously in Table~\ref{NeutDispersionData} (solid curves).  The solid curve fits to the data at 145~K in (a) and~(b) are Gaussians convoluted with the instrumental resolution. The data show that the magnetic excitations at ${\rm 145~K } > T_{\rm N}$ are uncorrelated between layers.  However, the magnetic excitations at $T_{\rm N}$ and at a low energy of 0.7~meV do show dispersion along the $c$-axis (not shown). The dotted curves in (a)--(d) and (f) and (h) show the spectrometer resolution and the dotted lines in (e) and~(g) show the estimated background.   Reprinted with permission from Ref.~\onlinecite{Matan2009}.  Copyright (2009) by the American Physical Society.}
\label{Matanfig2}
\end{figure}

\begin{figure}
\includegraphics[width=3.3in]{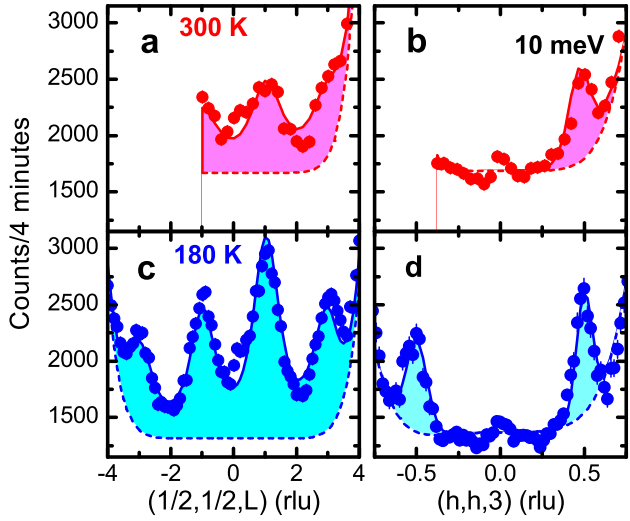}
\caption{(Color online) Wave vector scans $\left(\frac{1}{2},\frac{1}{2},L\right)$~r.l.u.\ (a,c) and $(H,H,3)$~r.l.u.\ (b,d) in tetragonal notation at fixed energy transfer $\hbar\omega = 10$~meV for single crystals of ${\rm CaFe_2As_2}$ at 180~K (c,d) and 300~K (a,b).\cite{Diallo2010}  The solid curves in (a) and (c) are fits to the data by the imaginary part of Eq.~(\ref{EqDialloChi3D}).  The solid curves in (b) and (d) are guides to the eye.  The dashed lines are estimates of the background scattering.  The shaded areas are the estimated magnetic scattering contributions.   Reprinted with permission from Ref.~\onlinecite{Diallo2010}.  Copyright (2010) by the American Physical Society.}
\label{DialloCaFig3}
\end{figure}

\begin{figure}
\includegraphics[width=3.3in]{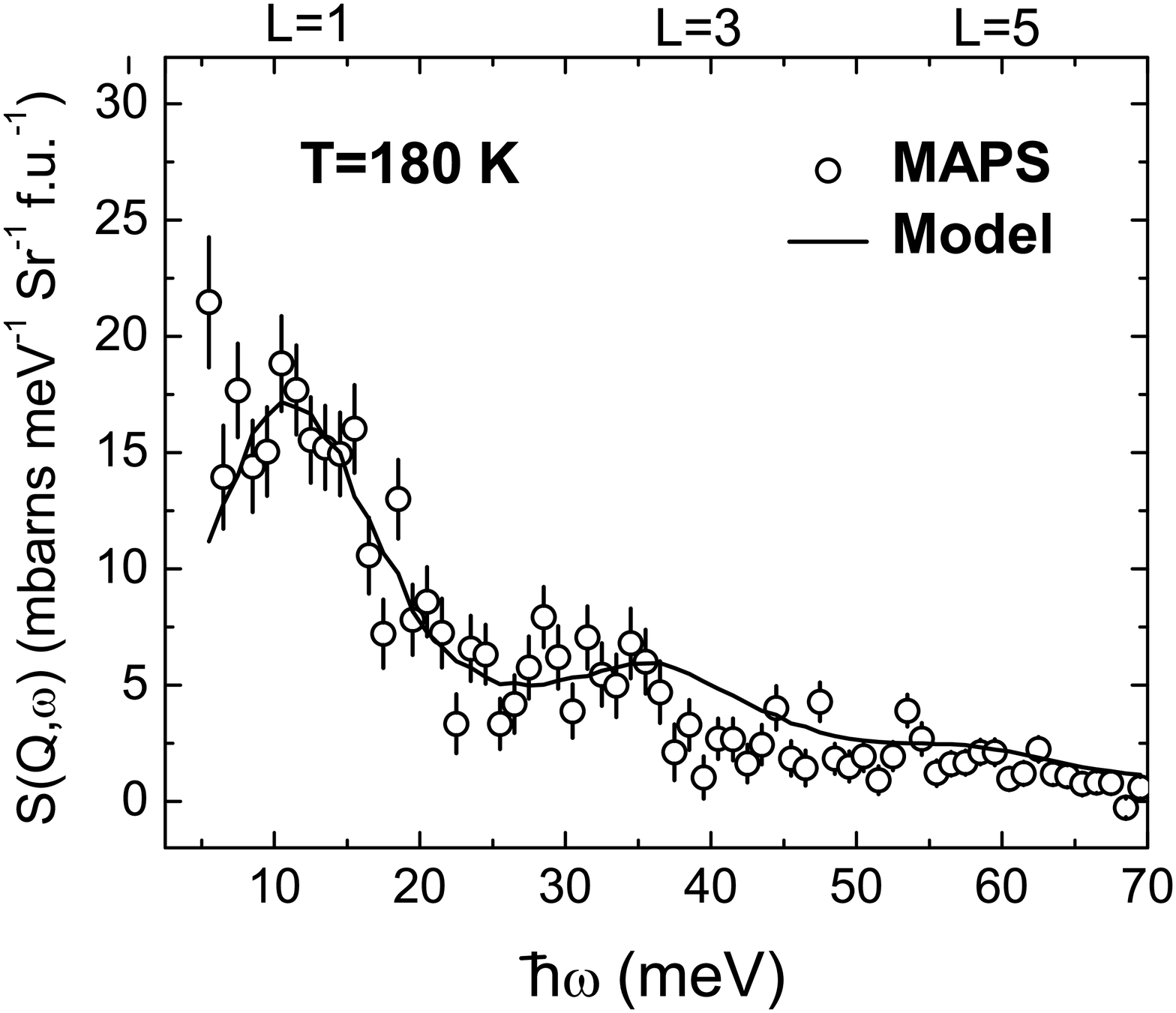}
\caption{Magnetic structure factor $S(Q,\omega)$ versus energy $\hbar\omega$ at tetragonal {\bf Q} = $\left(\frac{1}{2},\frac{1}{2},L\right)$~r.l.u.  An estimate of the nonmagnetic background has been subtracted.  The oscillations in the scattered intensity versus $L$ as seen in Fig.~\ref{DialloCaFig3} are present here but not as clearly defined.  The solid curve is a fit of the data by the imaginary part of Eq.~(\ref{EqDialloChi3D}) using the parameters in Eq.~(\ref{2DPars}). Reprinted with permission from Ref.~\onlinecite{Diallo2010}.   Copyright (2010) by the American Physical Society.}
\label{DialloCaFig8}
\end{figure}

Neutron scattering measurements at fixed energy transfer carried out on BaFe$_2$As$_2$ single crystals at temperature $T = 5$~K by Matan et al.\ showed spin wave dispersion both in the $a$-$b$ plane and along the $c$-axis as shown in Fig.~\ref{Matanfig2},\cite{Matan2009} as expected for anisotropic three dimensional propagation of spin waves.  However, the data in Fig.~\ref{Matanfig2} at $T = 145$~K $> T_{\rm N} = 136(1)$~K showed only propagation of magnetic excitations within the $a$-$b$ plane, indicating a transition from three-dimensional to two-dimensional spin correlations at 10--12~meV energy transfer on heating above $T_{\rm N}$.  Interestingly, these excitations were at the same wave vector as the spin wave excitations at 5~K\@. Additional measurements at $T_{\rm N} =136$~K showed a peak at (1,0,1) that is not present in the data in Fig.~\ref{Matanfig2} at 145~K that indicated intralayer magnetic correlation lengths of 15~\AA\ at energy 0.7~meV and 18~\AA\ at 12~meV, and critical scattering at and above $T_{\rm N}$ at 0.7~meV,\cite{Matan2009} in spite of several reports of a first order transition at $T_{\rm N}$ in this compound.

A first order phase transition has also been reported at the structural and magnetic transition in ${\rm CaFe_2As_2}$ at $T_{\rm N} = 172$~K\@.  However, magnetic correlations were observed by Diallo~et~al.\ from $T_{\rm N}$ up to their maximum measurement temperature of 300~K ($1.8\,T_{\rm N}$).\cite{Diallo2010}  They find that at $T_{\rm N}$, the spin gap that arises below  $T_{\rm N}$ in the spin wave excitation spectrum closes, and the spin wave excitations at tetragonal ${\bf Q} = \left(\frac{1}{2},\frac{1}{2},L\right)$~r.l.u.\ ($L =$~odd) are replaced by diffusive (quasielastic) excitations \emph{at the same wave vectors} that extend up to high energies of at least 60~meV.\cite{Diallo2010}  Contrary to the above results for BaFe$_2$As$_2$ crystals, modulations of the scattered intensity along $L$ occur all the way to 300~K, as shown in Fig.~\ref{DialloCaFig3}.\cite{Diallo2010}  The authors' analysis is consistent with itinerant spin fluctuations overdamped by particle-hole excitations.  They used a theory for a nearly antiferromagnetic Fermi liquid as outlined above in Sec.~\ref{SecChippNAFL} to fit their data.  From a fit to their high quality data at 180~K, they obtained the parameters in the expression~(\ref{EqDialloChi3D}) for the generalized 3D susceptibility as
\bea
\gamma &=& 43(5)~{\rm meV},\nonumber\\
\eta &=& 0.55(36),\label{2DPars}\\
\chi_0 &=& 0.20(5)~\frac{\mu_{\rm B}^2}{{\rm meV~f.u.}} = 0.007(2)~{\rm \frac{cm^3}{mol}},\nonumber\\
\xi_T &=& 7.90(10)~{\rm \AA},\nonumber\\
\eta_c &\equiv& J_c\chi_0 = 0.20(2),\nonumber
\eea
where the substitution $q_z = 2\pi(L-1)/c$ was made.  A fit to their data at 180~K by this theory is shown in Fig.~\ref{DialloCaFig8}.  Their theory for {\bf Q} near ${\bf Q}_{\rm AF}$ can be extrapolated to $\omega = 0$ and ${\bf Q} = 0$ to obtain a prediction for the static uniform susceptibility as discussed above in Secs.~\ref{SecChiSpin} and~\ref{SecChippNAFL}.  

From their fit to the data, the authors also estimated the instantaneous effective Fe moment $\mu_{\rm eff}$ according to Eq.~(\ref{EqFluctmoment4}).\cite{Diallo2010}  The fit model is not necessarily expected to be applicable in the high energy range $> 200$~meV\@.  With this caveat, the value obtained for $\mu_{\rm eff}$ depends on the cutoff energy on the fit function as follows (R. J. McQueeney, private communication).  The $\mu_{\rm eff}$ values obtained for the indicated cutoffs in parentheses are 0.47 (100), 0.66 (200),\cite{Diallo2010}  0.90 (400), 1.16 (800) and 1.41~$\mu_{\rm B}$/Fe~atom (1600~meV).  These values are all significantly smaller than the effective moment of 1.73~$\mu_{\rm B}$ for the smallest possible localized quantum spin ($S = 1/2$) with $g$-factor $g = 2$.  This result is thus inconsistent with proposed scenarios where the small ordered moments observed in the antiferromagnetically ordered states of the FeAs-based materials are suppressed due to frustration and/or fluctuation effects in a large-local-moment (e.g., $S = 2$, $\mu_{\rm eff} = 4.9~\mu_{\rm B}$/Fe) system.  Inosov et al.\ gave the value of the wave vector- and energy-integrated $\chi^{\prime\prime}$ up to 35~meV for optimally doped ${\rm Ba(Fe_{1.925}Co_{0.075})_2As_2}$ (see also below),\cite{Inosov2010} which together with Eq.~(\ref{EqFluctmoment4}) gives $\mu_{\rm eff} = 0.28~\mu_{\rm B}$/Fe, consistent with the above $\mu_{\rm eff}$ versus energy-cutoff behavior for ${\rm CaFe_2As_2}$.

For a polycrystalline sample of the 1111-type compound LaFeAsO with $T_{\rm N} \sim 140$~K, two-dimensional magnetic scattering was observed from $T_{\rm N}$ all the way up to 300~K.\cite{Ishikado2009}

An inelastic neutron scattering study of ${\rm Ba(Fe_{0.92}Co_{0.08})_2As_2}$ single crystals detected magnetic excitations above $T_{\rm c}$ at the SDW wavevector $(1,0, L)$~r.l.u.\ that were independent of $L$ for $-2.5 \lesssim L \lesssim 0.5$, again indicating that the spin fluctuations were quasi-two-dimensional.\cite{Parshall2009}  Furthermore, these authors found that the spin fluctuation intensity was present all the way up to 200~K, where the intensity was still $\sim 1/4$ of that at $T_{\rm c}$.\cite{Parshall2009}

\subsubsection*{d. Spin Excitations in Paramagnetic Tetragonal Doped Compounds}

\begin{figure}
% For arXiv, uncomment the first one and comment out the 2nd.
\includegraphics [width=\columnwidth]{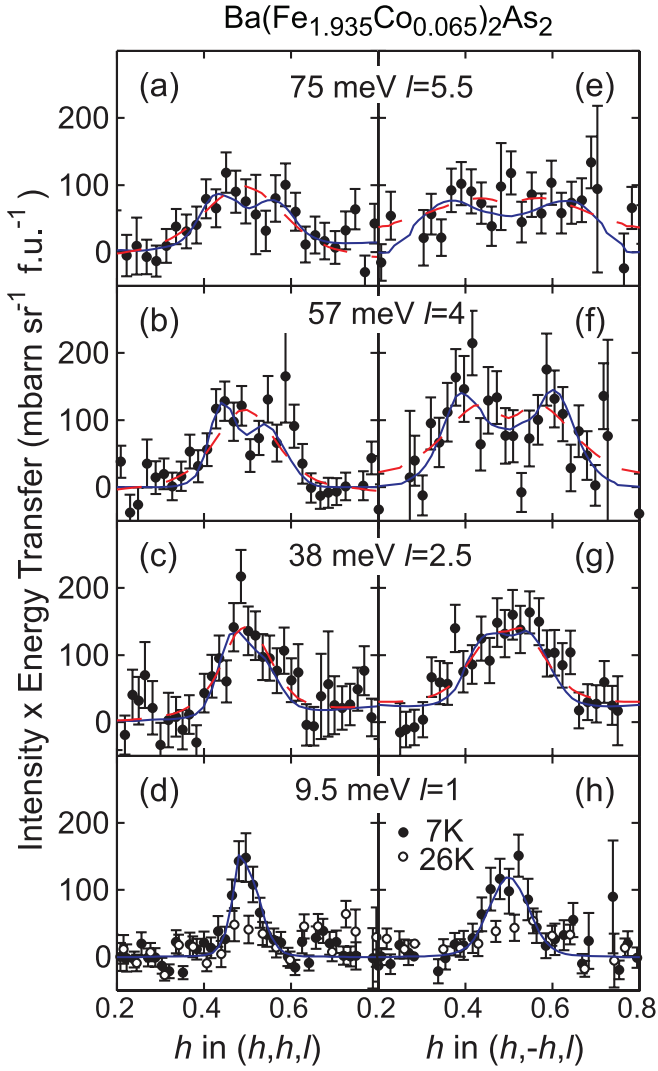}
\caption{(Color online)  Energy transfer $\hbar \omega$ versus tetragonal wave vector for magnetic excitations in co-aligned single crystals of tetragonal ${\rm Ba(Fe_{1.935}Co_{0.065})_2As_s}$ at a temperature of $T = 7$~K, which is below the superconducting transition temperature $T_{\rm c} = 23$~K\@.\cite{Lester2010}  Panels (a)--(d) show data for constant energy scans along tetragonal $(H,H,L)$ directions, which in orthorhombic notation are longitudinal $(H,0,L)$ scans through the in-plane wave vector (1,0,L).  Panels (e)--(h) show data for constant energy scans along tetragonal $(H,-H,L)$ directions, which in orthorhombic notation are transverse $(0,K,L)$ scans through the in-plane wave vector (1,0,L) (see Fig.~\ref{FigTwins}).  In panels (d) and (h) for $\hbar\omega = 9.5$~meV, data are also shown for $T = 26~K > T_{\rm c}$ to illustrate the development of the spin gap as part of the ``spin resonance'' below $T_{\rm c}$.  Reprinted with permission from Ref.~\onlinecite{Lester2010}.  Copyright (2010) by the American Physical Society.}
\label{FigLesterCuts}
\end{figure}

\begin{figure}
% For arXiv, uncomment the first one and comment out the 2nd.
\includegraphics [width=0.70\columnwidth]{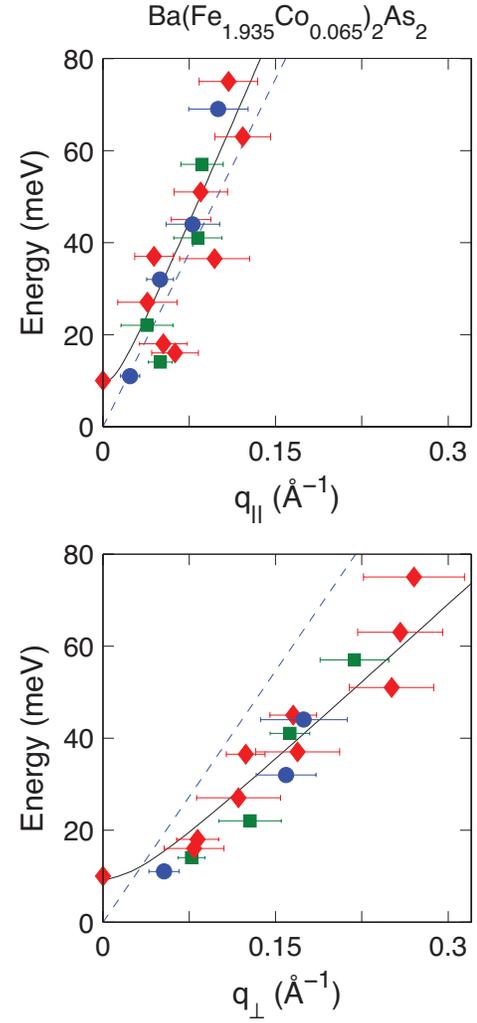}
\caption{(Color online) The dispersion of the magnetic excitations in aligned single crystals of tetragonal Ba(Fe$_{0.935}$Co$_{0.065}$)$_2$As$_2$ at temperature $T=7$~K ($T < T_{\rm c} = 23$~K) along (a) $q_{\parallel}$ and (b) $q_{\perp}$ as determined from fitting the data in Fig.~\ref{FigLesterCuts} by Eq.~(\ref{EqChppQO}).\cite{Lester2010}  Here, the longitudinal $q_{\parallel} = q_a$ is along the orthorhombic $a$-axis direction and the transverse $q_{\perp} = q_b$ is along the orthorhombic $b$-axis direction.  These $(q_a,q_b,L)$ values are the deviations from the orthorhombic $(1,0,0)$ point in the orthorhombic Brillouin zone.  There is no measurable dispersion along the $c$-axis in this study.  The solid curves are the fits of the data by Eq.~(\ref{Eqdispersion}) with $v_c = 0$.  Even without the fits, it is obvious from the figures that the longitudinal spin wave velocity is significantly larger than the transverse one.  This is an unexpected result in a tetragonal structure where $a_{\rm T} = b_{\rm T}$.  The symbols denote $L$ values:\ filled squares ($L$ even); filled circles ($L$ odd); filled diamonds ($L$ non-integer).   The spin gap at ${\bf q} = 0$ [${\bf Q} = (\frac{1}{2},\frac{1}{2},L)$~r.l.u.\ in tetragonal notation] is due to the resonance mode that develops below $T_{\rm c}$ (see Sec.~\ref{ResonanceMode}).  The dashed blue lines are the corresponding dispersions for CaFe$_2$As$_2$ by Zhao et al.\cite{JZhao2009} as quoted by Ref.~\onlinecite{Lester2010}.  Reprinted with permission from Ref.~\onlinecite{Lester2010}.  Copyright (2010) by the American Physical Society.}
\label{FigLesterDisp}
\end{figure}

\subsubsection*{\hspace{0.15in}d1. ${\rm Ba(Fe_{1.935}Co_{0.065})_2As_2}$}

Lester and coworkers observed spin wave-like excitations in the \emph{paramagnetic tetragonal} structure of superconducting ${\rm Ba(Fe_{1.935}Co_{0.065})_2As_2}$ single crystals as shown in Fig.~\ref{FigLesterCuts} (see Fig.~\ref{FigTwins}).\cite{Lester2010}  They did not observe any evidence of long-range magnetic order down to a temperature of 2~K\@.  By extending the energy range to 9--75~meV, they were able to clearly resolve two counterpropagating spin excitation branches in the $a$-$b$~plane at the higher energies as seen in the figure.  No dispersion along the $c$-axis was detected, which means that the spin excitations are purely two-dimensional and in the $a$-$b$~plane, consistent with an inelastic neutron scattering study of single crystal ${\rm Ba(Fe_{1.92}Co_{0.08})_2As_2}$,\cite{Lumsden2009} and in contrast to the strong $c$-axis dispersion of spin waves in the ${\rm BaFe_2As_2}$ parent compound as described above.  Remarkably, the excitations have the same in-plane wave vector $\left(\frac{1}{2},\frac{1}{2}\right)_{\rm T} = (1,0)_{\rm O}$ and similar energy versus wave vector dispersion as those in the antiferromagnetically ordered orthorhombic structure of the undoped 122-type parent compounds.  In particular, this wave vector suggests that spin waves are propagating in a short-range ordered antiferromagnetic structure that is locally the same as the long-range stripe-ordered antiferromagnetic structure in the orthorhombic parent compounds in Fig.~\ref{Stripe_Mag_Struct}.  From Fig.~\ref{FigTetrag_Ortho_struct}, if the spin correlations had instead been N\'eel-type in the $a$-$b$~plane, the wave vector of the fluctuations would have been $(1,1)$ in orthorhombic notation or $(1,0)$ and/or $(0,1)$ in tetragonal notation.  

Furthermore, as shown in Fig.~\ref{FigLesterCuts}, Lester et al.\ observed an anisotropy in the spin wave dispersion between the orthorhombic $(H,0,0)$ and~$(0,K,0)$ directions, with the same sign of the anisotropy as predicted by Eqs.~(\ref{EqSWvel}) for spin waves in the long-range ordered orthorhombic antiferromagnetic stripe-$b$ phase.\cite{Lester2010}  Because these spin waves occur in the tetragonal paramagnetic state in which there is no long-range magnetic ordering, these similarities suggest that there is a so-called nematic degree of freedom in the doped materials, where there is a distinct directional nature to the spin excitations that ``remember'' the specific nature of the stripe-$b$-type long-range antiferromagnetic ordering in the orthorhombic parent compounds, or \emph{vice versa}.\cite{Lester2010}  The authors suggested that this might arise from an ``order-from-disorder'' effect in a local moment picture but they point out that it can also arise in an itinerant description.

One might ask whether the in-plane anisotropy in the inelastic neutron scattering results for the nominally tetragonal phase might arise from inhomogeneous crystals containing macroscopic regions of orthorhombic, long-range antiferromagnetically ordered, material.  However, Lester et al.\ examined this issue and stated, ``Elastic neutron scattering revealed no evidence of magnetic order at this doping level at temperatures down to 2 K.''\cite{Lester2010}  There was also no evidence for an orthorhombic crystallographic distortion (S. Hayden, private communication).  These results indicate that the volume fraction of orthorhombic material, which would show long-range antiferromagnetic ordering, is negligible and that the deduction of a nematic degree of freedom in the tetragonal phase is correct.

Lester et al.\ analyzed their magnetic inelastic neutron scattering data for single crystal ${\rm Ba(Fe_{1.935}Co_{0.065})_2As_2}$ at 7~K $< T_{\rm c}$ in Fig.~\ref{FigLesterCuts} using the damped harmonic oscillator response function~(\ref{EqChppQO}) to determine the  dispersion of the magnetic excitations.  These data are plotted in Fig.~\ref{FigLesterDisp} for the two scan directions.  In Fig.~\ref{FigLesterDisp}, $q_\parallel$ is the same as $q_a$ in orthorhombic notation and $q_\perp$ is the same as $q_b$ in orthorhombic notation.  These data were fitted using the low-energy spin wave dispersion relation~(\ref{Eqdispersion}), where $v_c$ in Eq.~(\ref{Eqdispersion}) was taken to be zero, and the fits are shown as the solid curves in Figs.~\ref{FigLesterDisp}(a) and~(b).  To within the errors, there are no deviations of their data from Eq.~(\ref{Eqdispersion}).  The spin gap $\Delta$ and the anisotropic spin wave velocities $v_a$ and~$v_b$ obtained from the fits are listed in Table~\ref{NeutDispersionData}, where they are seen to be similar to the corresponding values for the undoped parent compounds.  The spin gap at ${\bf Q} = (\frac{1}{2},\frac{1}{2},L)$~r.l.u.\ (tetragonal notation) is due to the resonance mode that develops below $T_{\rm c}$ (see Sec.~\ref{ResonanceMode}).

One can attempt to interpret the spin wave velocities in terms of a local moment model.  However, with two spin wave velocities and six parameters in Eqs.~(\ref{EqSWvel}) one must make some assumptions and/or approximations in order to arrive at quantitative interpretations.   The approach of Lester et al.\cite{Lester2010} was to set $J_c = 0$ because they detected no spin wave dispersion along $c$, and set $D = 0$ which is a very good approximation.  They also set $J_{1a}=J_{1b}\equiv J_1$, which is an approximation with unknown error, forming the $J_1$-$J_2$ model discussed above in Sec.~\ref{SecLRMO}. Then Eqs.~(\ref{EqSWvel}) become\cite{Lester2010}
\bea
\hbar v_a &=& a_{\rm O}S(2J_2 + J_1)\nonumber\\
\ \\
\hbar v_b &=& a_{\rm O}S \sqrt{(2J_2 + J_1)(2J_2 - J_1)},\nonumber
\eea
where $a_{\rm O} = \sqrt{2}\,a_{\rm T}$ is the basal plane lattice parameter in orthorhombic notation, yielding
\bea
SJ_1 &=& \frac{\hbar v_a}{2a_{\rm O}} \left[1 - \left(\frac{v_b}{v_a}\right)^2\right],\nonumber\\
\label{EqJ1J2vavb} \\
SJ_2 &=& \frac{\hbar v_a}{4a_{\rm O}} \left[1 + \left(\frac{v_b}{v_a}\right)^2\right].\nonumber
\eea
Their values of $SJ_1$ and $SJ_2$ are listed in Table~\ref{NeutDispersionData2}.  

Alternatively, we have analyzed the spin wave velocities $v_a$ and $v_b$ using Eqs.~(\ref{EqGetJ1aJ1b}) to obtain $S(2J_2 + J_{1a})$ and $S(2J_2 - J_{1b})$, as listed in Table~\ref{NeutDispersionData2}.  Interestingly, the latter analysis indicates that $J_2$ is  close to $J_{1b}/2$, which according to Eq.~(\ref{EqStripeStab}) is the classical stability condition for stripe-$b$ type of long-range antiferromagnetic ordering.  However, it is not clear that spin wave theory should be applicable to metallic compounds not exhibiting long-range antiferromagnetic order.

Lester et al.\ integrated t eir $\chi^{\prime\prime}({\bf Q},\omega)$ data over {\bf Q} to obtain the local susceptibility $\chi^{\prime\prime}(\omega)$ according to Eq.~(\ref{EqDefineLocalchi}) and plotted their result in their Fig.~5.\cite{Lester2010}  We fitted their $\chi^{\prime\prime}(\omega)$ data for 26~K $> T_{\rm c}$ and for $\hbar\omega < 80$~meV by an odd-function polynomial in $\omega$, $\omega^3$ and $\omega^5$, and then calculated $\mu_{\rm eff}$ represented by this energy interval from $-80$ to~80~meV according to Eq.~(\ref{EqFluctmoment5}), yielding $\mu_{\rm eff} = 0.31~\mu_{\rm B}$/Fe atom.  This lower-limit value is a factor of 16 smaller than the value $\mu_{\rm eff} = 2\sqrt{6}~\mu_{\rm B} \approx 4.90~\mu_{\rm B}$Fe~atom expected for an Fe localized spin $S = 2$ with $g = 2$.

Li and coworkers have carried out detailed inelastic neutron scattering measurements on optimally-doped ${\rm Ba(Fe_{1.926}Co_{0.074})_2As_2}$ single crystals.\cite{Li2010}  They confirmed the in-plane anisotropy in the spin excitations found by Lester et al.\ and in addition found evidence for a crossover from diffusive to collective ``quasi-propagating'' spin wave excitations above an energy of $\sim 100$~meV\@.  They concluded that these spin waves disperse only in the transverse $(0K0)$~r.l.u.\ direction in orthorhombic notation, which is in the direction of the ferromagnetically aligned spin stripes in the magnetically ordered state of the undoped FeAs-based compounds.  The derived high-energy transverse spin wave velocity is $v_b \sim 0.40$~eV~\AA, which is similar to other values for the 122-type FeAs-based class of compounds in Table~\ref{NeutDispersionData} that were obtained from lower-energy measurements.

\subsubsection*{\hspace{0.15in}d2. ${\rm Ba(Fe_{1.925}Co_{0.075})_2As_2}$}

\begin{figure}
\includegraphics[width=3.3in]{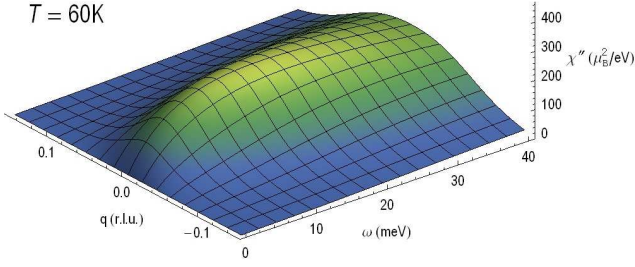}
\caption{Imaginary part $\chi^{\prime\prime}$ of the dynamical susceptibility of a ${\rm Ba(Fe_{0.925}Co_{0.075})_2As_2}$ single crystal at temperature $T = 60$~K versus 2D in-plane wave vector $q = |{\bf Q} - {\bf Q}_{\rm AF}|$ and magnetic excitation energy $\hbar \omega$, where ${\bf Q}_{\rm AF} = (\frac{1}{2}, \frac{1}{2}, L)$~r.l.u.\ in tetragonal notation and $\hbar \equiv 1$  in the figure.\cite{Inosov2010}  The units of $\chi^{\prime\prime}$ in the figure are $\mu_{\rm B}^2$~eV$^{-1}$~f.u.$^{-1}$, where f.u.\ means formula unit. Such 3D isothermal plots for temperatures from 5~to 300~K are available.\cite{Inosov2010}  The  wave vector scans from which the figure was constructed were longitudinal $(\frac{1}{2} + h, \frac{1}{2} + h, L)$~r.l.u.\ scans ($L = 1,3$), i.e.\ ${\bf q} = (h,h,0)$~r.l.u., and the anisotropy of the magnetic excitations in the ${\bf Q}_x$-${\bf Q}_y$ plane was not examined (D. S. Inosov, private communication).  Reprinted with permission from Ref.~\onlinecite{Inosov2010}.}
\label{FigInosov_3D_60K}
\end{figure}

\begin{figure}
\includegraphics[width=3.3in]{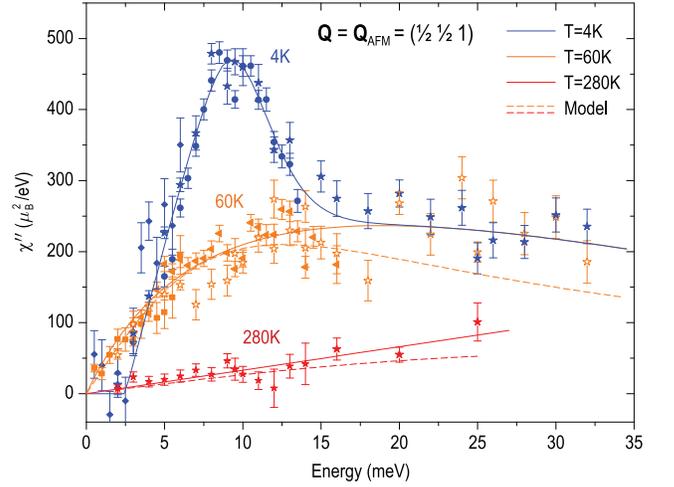}
\caption{(Color online) Imaginary part $\chi^{\prime\prime}$ of the dynamical susceptibility of a ${\rm Ba(Fe_{0.925}Co_{0.075})_2As_2}$ single crystal at temperatures $T = 4$, 60, and 280~K at in-plane wave vector $q = |{\bf Q} - {\bf Q}_{\rm AF}| = 0$ versus magnetic excitation energy $\hbar \omega$, where ${\bf Q}_{\rm AF} = (\frac{1}{2}, \frac{1}{2}, L)$~r.l.u.\ in tetragonal notation.\cite{Inosov2010}  The units of $\chi^{\prime\prime}$ in the figure are $\mu_{\rm B}^2$~eV$^{-1}$~f.u.$^{-1}$, where f.u.\ means formula unit.  The data at 60~K are from a vertical cut at $q=0$ of the data in Fig.~\ref{FigInosov_3D_60K}.  The solid curves are guides to the eye.  The dashed curves are cuts in a global fit to the normal state data by Eq.~(\ref{EqChippInosov}) based on the itinerant nearly antiferromagnetic Fermi liquid model, with parameter values given in Eq.~(\ref{EqFitPars}).  Reproduced by permission from Ref.~\onlinecite{Inosov2010} and from Macmillan Publishers Ltd: Ref.~\onlinecite{Inosov2010}, Copyright (2010).}
\label{FigChipp3Ts}
\end{figure}

Inosov and coworkers have carried out inelastic magnetic neutron scattering measurements on one large single crystal of ${\rm Ba(Fe_{1.925}Co_{0.075})_2As_2}$ over a range of temperature $T$, and energy $\hbar\omega$ and wave vector {\bf Q} transfer.\cite{Inosov2010}  An example of their normal state data for $\chi^{\prime\prime}(q,\omega)$ at $T = 60$~K is shown in the 3D plot in Fig.~\ref{FigInosov_3D_60K}.  The magnetic excitations disperse in $q$ with increasing energy but the energy range is evidently not sufficient to resolve the branches seen in Fig.~\ref{FigLesterCuts} at $\hbar\omega \geq 57$~meV for crystals with nearly the same composition.  A vertical cut through the data at $q = 0$ in Fig.~\ref{FigInosov_3D_60K} is shown in Fig.~\ref{FigChipp3Ts}, together with additional data at $q = 0$ for $T = 4$ and~280~K\@.  The data at 4~K, below $T_{\rm c} = 25$~K, reflect the presence of a ``spin resonance mode'' at 9.5~meV in the superconducting state that will be discussed later in Sec.~\ref{ResonanceMode}.

The authors analyzed the normal state ($T > T_{\rm c}$) data using the itinerant nearly antiferromagnetic Fermi liquid model according to the expression~(\ref{EqChippInosov}) for $\chi^{\prime\prime}({\bf Q},\omega)$ and obtained the values of the fitting parameters in Eq.~(\ref{EqChippInosov}) as
\bea
C &=& 3.8(10) \times 10^4~\frac{\mu_{\rm B}^2~{\rm K}}{{\rm eV~f.u.}} = 1.2(3)~{\rm \frac{cm^3~K}{mol~f.u.}},\nonumber\\
\theta &=& +30(10)~{\rm K},\nonumber\\
\xi_0 &=& 163(20)~{\rm \AA~K^{1/2}},\label{EqFitPars}\\
{\rm and}\ \ \Gamma_0 &=& 0.14(4)~{\rm \frac{meV}{K}}.\nonumber
\eea
The global fit is shown as the dashed lines in Fig.~\ref{FigChipp3Ts}.  Overall, the global fit to the data is very good.  The authors attributed the discrepancy in the fit to the data at the higher energies at 60~K in Fig.~\ref{FigChipp3Ts} possibly to ``the presence of several bands in the system, which shifts the maximum of $\chi_{\rm 60\,K}^{\prime\prime}({\bf Q},\omega)$  to a higher value of $\sim 20$~meV.''

We extrapolated the fitted normal state $\chi({\bf Q}\approx {\bf Q}_{\rm AF},\omega)$ to $({\bf Q} = 0,\omega = 0)$ to obtain a predicted estimated Pauli susceptibility in Sec.~\ref{SecChiSpin} above, and the prediction has the right order of magnitude but with no significant temperature dependence, contrary to observation.  However, the extrapolation is not expected to be accurate.

This study has an important bearing on the debate about local moment versus itinerant pictures for the magnetism in the FeAs-based compounds, since it showed that an itinerant model is viable over the temperature, wave vector and energy ranges examined.

\subsubsection{\label{Sec11neuts} 11-type {\rm Fe}$_{1+y}${\rm (Te}$_{1-x}${\rm Se}$_x${\rm )} Compounds}

\begin{figure*}
\includegraphics[width=7in]{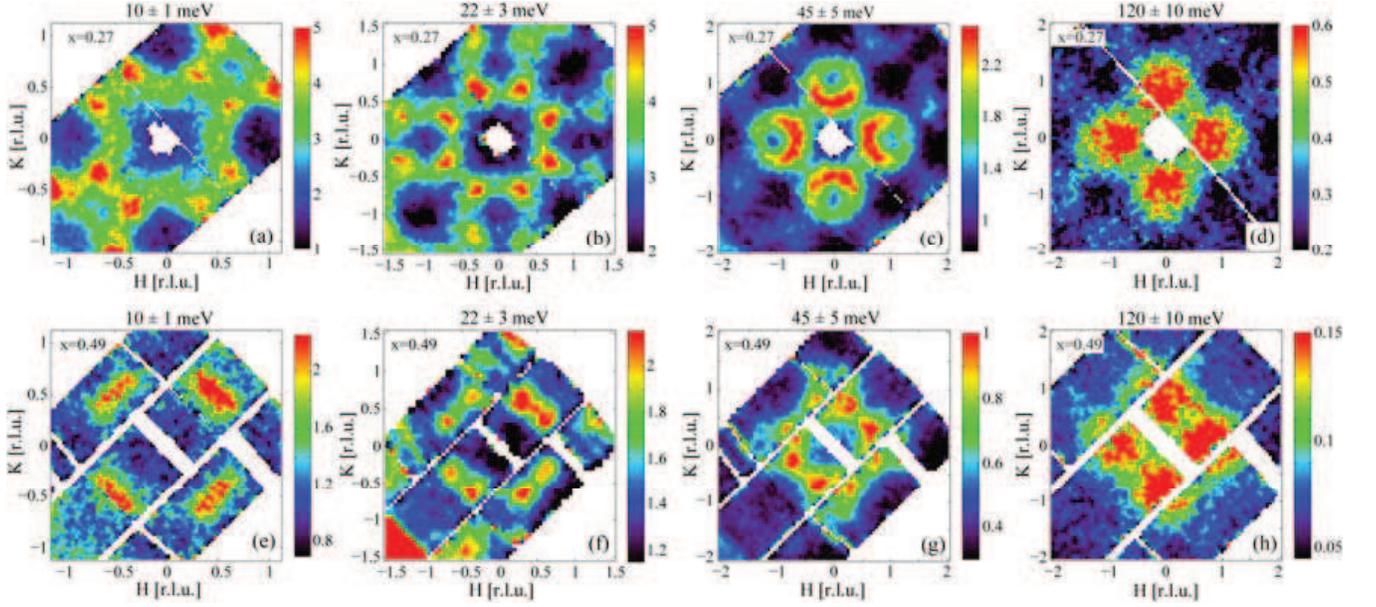}
\caption{(Color online)  Evolution of the inelastic neutron scattering plots projected onto the $H$-$K$ plane with increasing constant energy transfer (left to right) of single crystals of a non-bulk superconductor Fe$_{1.04}$Te$_{0.73}$Se$_{0.27}$ at 5~K (top four panels) and a bulk superconductor ($T_{\rm c} = 14$~K) FeTe$_{0.51}$Se$_{0.49}$ at 3.5~K (bottom four panels).\cite{Lumsden2010a}  Note the large differences in the calibrations of the color scales to the right of the respective figures.  Reproduced by permission from Ref.~\onlinecite{Lumsden2010a} and from Macmillan Publishers Ltd: Ref.~\onlinecite{Lumsden2010a}, Copyright (2010).}
\label{LumsdenFig2_0907.2417}
\end{figure*}

Inelastic neutron scattering studies on the Fe$_{1+y}$Te end member compound have not been reported.  Such measurements can give information on the spin wave dispersion relations of the magnetic excitations associated with the diagonal double stripe static ordering at wave vector $\left(\frac{1}{2},0,\frac{1}{2}\right)$~r.l.u.  Indeed, Babkevich et al.\ detected incommensurate magnetic excitations at wave vector $\left(\frac{1}{2}-\delta,0,\frac{1}{2}\right)$ ($\delta\approx 0.03$)~r.l.u.\ for a crystal of non-superconducting Fe$_{1.10}$Te$_{0.75}$Se$_{0.25}$ at relatively low energy transfers of 2--6~meV.\cite{Babkevich2010}  The data indicated ``a quasi-two-dimensional system with weak interactions along $c$.''  The measurements also demonstrated that the magnetic excitations persisted up to temperatures of at least 150~K\@.\cite{Babkevich2010}  At present there are no dispersion relations reported in either the long-range ordered or spin glass region for the spin wave excitations originating from the statically ordered diagonal double stripe structure with tetragonal in-plane wave vector $\left(\frac{1}{2},0\right)$~r.l.u.\

In addition to the above magnetic excitations at $\left(\frac{1}{2}-\delta,0,\frac{1}{2}\right)$~r.l.u.\ associated with spin wave excitations of the diagonal double stipe magnetic structure, inelastic neutron scattering studies have found excitations in Fe$_{1+y}$Te$_{1-x}$Se$_{x}$ crystals with $0.05 \lesssim x \lesssim 0.5$ at wave vectors with in-plane components $\sim \left(\frac{1}{2},\frac{1}{2}\right)$~r.l.u.\ that are similar to the nesting/stripe-ordering wave vector found in all the other Fe-based superconductors and parent compounds.\cite{Lumsden2010, Lee2009, Xu2010, Babkevich2010, Argyriou2009, Mook2009, Mook2009a, Li2010a}  Even in the long-range antiferromagnetically ordered regime of the phase diagram in Fig.~\ref{KatayamaFig2}, a crystal of Fe$_{1.02}$Te$_{0.95}$Se$_{0.05}$ with an ordered moment of $1.68(6)~\mu_{\rm B}$ showed incommensurate magnetic excitations at $\left(\frac{1}{2} + \varepsilon,\frac{1}{2}- \varepsilon\right)$~r.l.u.\ with $\varepsilon = 0.10(1)$.\cite{Liu2010}  Xu et al.\ found that bulk superconductivity only occurs if inelastic neutron scattering shows excitations at the nesting wave vector $\sim \left(\frac{1}{2},\frac{1}{2}\right)$~r.l.u.\cite{Xu2010} For superconducting compositions with $x \sim 0.3$--0.5, a main interest has been in the neutron spin resonance mode that develops in the superconducting state below $T_{\rm c}$ at or near the in-plane wave vector $\left(\frac{1}{2},\frac{1}{2}\right)$~r.l.u.\ as discussed below in Sec.~\ref{ResonanceMode}.  Here we further discuss the normal state aspects of the spin excitations.

A study of the magnetic excitations near $\left(\frac{1}{2},\frac{1}{2}\right)$~r.l.u.\ and their evolution with increasing energy was reported by Lumsden et al.\ as shown in Fig.~\ref{LumsdenFig2_0907.2417} for a non-bulk superconductor Fe$_{1.04}$Te$_{0.73}$Se$_{0.27}$ at a temperature of 5~K (top four panels) and a bulk superconductor ($T_{\rm c} = 14$~K) FeTe$_{0.51}$Se$_{0.49}$ at 3.5~K (bottom four panels).\cite{Lumsden2010a}  In addition to these data, the authors found that the excitations are quasi-two-dimensional, with little dispersion along the $c$-axis.  At low energy the excitations in Fig.~\ref{LumsdenFig2_0907.2417} are incommensurate (upper and lower left-hand panels), but with increasing energy the excitations disperse towards the (1,0)~r.l.u.\ points (upper and lower right-hand panels).  The authors emphasize the four-fold symmetry of the excitations around the (1,0)~r.l.u.\ point(s), which according to Fig.~\ref{Ordering_Wave_Vectors} corresponds to the wave vector for G-type (N\'eel or checkerboard) static antiferromagnetic ordering.  The data were successfully fitted by the so-called Sato-Maki function for $\chi^{\prime\prime}({\bf Q},\omega)$ that was previously applied to the cuprates, because it incorporates such a four-fold symmetry, which the authors suggest indicates a similarity between the superconducting mechanisms in the two types of compounds.  However, making this identification between the natures of the Fe-based and cuprate superconductors is puzzling.  The undoped and low-doped cuprates order antiferromagnetically at or near the (1,0)~r.l.u.\  wave vector and the spin waves for these and more highly doped samples disperse away from this wave vector at high energies.\cite{Johnston1997, Vignolle2007}  On the other hand, none of the Fe-based superconductors or parent compounds show any inclination to order antiferromagnetically  at or near (1,0)~r.l.u., and the magnetic excitations in Fig.~\ref{LumsdenFig2_0907.2417} disperse towards, not away from, the (1,0)~r.l.u.\ and equivalent points with increasing energy.

Other studies have established the occurrence of incommensurate magnetic excitations in the normal state of Fe$_{1+y}$Te$_{1-x}$Se$_{x}$ single crystals using inelastic neutron scattering measurements.  Argyriou et al.\ reported magnetic excitations in FeTe$_{0.6}$Se$_{0.4}$ ($T_{\rm c} = 14$~K) in the normal state at wave vectors $\left(\frac{1}{2}\pm \varepsilon, \frac{1}{2}\mp \varepsilon\right)$~r.l.u.\ with $\varepsilon = 0.09(1)$ at energies up to 80~meV.\cite{Argyriou2009}  These wave vectors are isolated wave vectors at a given energy and are not part of a spin-wave cone, suggesting  itinerant magnetism.\cite{Argyriou2009}  The authors qualitatively described these data using a four-band itinerant model.  Lee et al.\ found a similar discrete incommensurability of the magnetic excitations at wave vectors transverse to $\left(\frac{1}{2}, \frac{1}{2}\right)$~r.l.u.\ in FeTe$_{0.5}$Se$_{0.5}$ and suggested that this arises from coupled spin-orbital excitations.\cite{Lee2009}  They further suggested that ``If correct, it follows that these nematic fluctuations are involved in the (neutron spin) resonance and could be relevant to the pairing mechanism.''\cite{Lee2009}

Li and coworkers reported an hour-glass shaped dispersion of the spin excitations in the normal state of FeTe$_{0.6}$Se$_{0.4}$ at low energies of 1--10~meV.\cite{Li2010a}  The authors suggested that ``spin excitations and superconductivity in FeTe$_{0.6}$Se$_{0.4}$ are connected with that of the copper oxide superconductors in an unexpected way,'' and that ``among the different Fe-based superconductors, the FeSe$_x$Te$_{1-x}$ family is the closest material to the copper oxides.''\cite{Li2010a}  This latter assessment is consistent with the discussion in Sec.~\ref{SecChi} above for FeTe that suggested that this compound is a local moment antiferromagnet.  On the other hand, the susceptibility data for Se-doped crystals in Sec.~\ref{SecChi} suggested a crossover to itinerant magnetism, which is dissimilar to the local moment magnetism of the cuprates.  Further study of Se-doped crystals is needed to more definitively establish the nature of the magnetism in that region of the phase diagram.

\subsection{\label{SecNMRDynamics} Spin Dynamics from Nuclear Magnetic Resonance (NMR) Measurements}

\subsubsection{Introduction}

%\squeezetable
\begin{table}
\caption{\label{Nuclei} Some useful nuclei for NMR measurements on Fe-based superconductors and parent compounds.  The properties listed are  the natural abundance, the nuclear spin $I$, the gyromagnetic ratio $\gamma_{\rm n}/(2\pi)$, and the nuclear quadrupole moment in units of barns ($10^{-24}$~cm$^2$).\cite{HbPC1972}  Nuclei with spins~1/2 do not have a nuclear quadrupole moment.  The Korringa constant is $S_{\rm n} = (\gamma_{\rm e}/\gamma_{\rm n})^2[\hbar/(4\pi k_{\rm B})]$.  For a noninteracting electron gas, the Korringa ratio $K^2T_1T/S_{\rm n} = 1$, where $K$ is the Knight shift, $T_1$ is the nuclear spin-lattice relaxation rate and $T$ is the absolute temperature.  Also included is $\gamma_{\rm e}$ for the electron (e$^-$).}
\begin{ruledtabular}
\begin{tabular}{l|ccccc}
 nucleus   & abundance & $I$ & $\gamma/(2\pi)$ & $Q$ & $S_{\rm n}$ \\
 & (\%) & & (MHz/T) & (b) & (s K)\\ \hline
$^{19}$F & 100 & 1/2 & 40.054 & --- & $2.9756 \times 10^{-7}$ \\
$^{31}$P & 100 & 1/2 & 17.235 & --- & $1.6071 \times 10^{-6}$ \\
$^{57}$Fe & 2.19 & 1/2 & 1.3758 & --- & $2.5221 \times 10^{-4}$\\
$^{75}$As & 100 & 3/2 & 7.2919 & 0.3 & $8.9783 \times 10^{-6}$\\
$^{77}$Se & 7.58 & 1/2 & 8.118 & --- & $7.244 \times 10^{-6}$\\
$^{125}$Te & 6.99 & 1/2 & 13.45 & --- & $2.639 \times 10^{-6}$\\
e$^-$ & --- & --- & 28\,025.0 & --- & ---
\end{tabular}
\end{ruledtabular}
\end{table}

Here we first provide a brief introduction to the notation and implementation of nuclear magnetic resonance (NMR) measurements to probe electronic spin dynamics.\cite{Slichter1963, Abragam1961}  In a conventional pulsed NMR experiment, the nuclear magnetization ${\bf M}_{\rm n}$ is initially at equilibrium at a specified temperature, pointing in the direction of an applied dc magnetic field ${\bf H}$ which is normally taken to be the $z$ direction.  The equilibrium $z$~component of ${\bf M}_{\rm n}$ is given by the nuclear Curie law $M_{\rm n}^z(\infty) = C_{\rm n}/T$, where the notation $M_{\rm n}^z(\infty)$ is used for reasons discussed below and $C_{\rm n}$ is the nuclear Curie constant associated with the spin $I$ and gyromagnetic factor $\gamma_{\rm n}$ of the particular nucleus.  Then in a typical experiment one or more radio frequency (rf) pulses are applied that tip the nuclear magnetization from its equilibrium direction by 90$^\circ$, in which case the pulses are called ``$\pi/2$'' pulses.  The ${\bf M}_{\rm n}$ precesses around ${\bf H}$ at the nuclear Larmor angular frequency 
\be
\omega_{\rm n} = \gamma_{\rm n}H, 
\label{EqLarmor}
\ee
the $z$~component of which is monitored by the amplitude of the ac voltage at angular frequency $\omega_{\rm n}$ induced in a coil wound around the sample due to Faraday's law of induction.  The nuclear gyromagnetic  ratio $\gamma_{\rm n}$ is different for different nuclei, as shown in Table~\ref{Nuclei}.\cite{HbPC1972}  For a free electron, the Zeeman levels are split by energy $\hbar\omega = g_{\rm e}\mu_{\rm B}H$, where $g_{\rm e} = 2.002\,319$ is the spectroscopic splitting factor, or ``$g$-factor'', of the free electron and $\mu_{\rm B}$ is the Bohr magneton, so the gyromagnetic ratio of the electron is
\be
\gamma_{\rm e} = \frac{\omega_{\rm e}}{H} = \frac{g_{\rm e}\mu_{\rm B}}{\hbar},
\ee
with the numerical value given in Table~\ref{Nuclei}.

The energy of the nuclear spin system in the applied field is given by the conventional expression $E = -{\bf M}_{\rm n}\cdot {\bf H} = -{M}_{\rm n}^z H$ where ${\bf H} = H \hat{\bf z}$.  Thus the lowest energy state is the equilibrium state with the nuclear magnetization pointing in the direction of the applied field.  The process by which ${M}_{\rm n}^z$ relaxes to its equilibrium value $M_{\rm n}^z(\infty)$ is called ``longitudinal'', or ``nuclear spin-lattice'', relaxation and thus requires the net transfer of energy from the nuclear spin system to its environment (heat bath), which is called the ``lattice''.  Here, the term ``longitudinal'' means that the direction of relaxation of the $z$-axis component of the nuclear spins is in the direction of the applied magnetic field.  In the absence of nuclear electric quadrupole effects, this relaxation occurs exponentially with a relaxation time and rate denoted by $T_1$ and $1/T_1$, respectively.  For example, when the nuclear magnetization is initially tipped by $\pi/2$~rad using $\pi/2$ pulses, the component ${M}_{\rm n}^z(0)$ is initially zero (``saturated'') just after the application of the pulses.  Then as time $t$ passes, the magnetization relaxes towards its equilibrium value according to
\be
M_{\rm n}^z(t) = M_{\rm n}^z(\infty)\left(1 - e^{-t/T_1}\right).
\label{EqT1Relax}
\ee
In the literature, typically the superscript $z$ and subscript ${\rm n}$ are left off, with the understanding  that $M(t)$ then refers to the $z$~component of the nuclear magnetization.  One typically tests whether the nuclear spin-lattice relaxation is described by a single exponential by rewriting Eq.~(\ref{EqT1Relax}) as
\be
1 - \frac{M_{\rm n}^z(t)}{M_{\rm n}^z(\infty)} = e^{-t/T_1},
\label{EqT1Relax0}
\ee
or
\be
\ln\left[1 - \frac{{M}_{\rm n}^z(t)}{{M}_{\rm n}^z(\infty)}\right] = -\frac{t}{T_1}.
\label{EqT1Relax2}
\ee
Thus if a plot of the experimentally determined left-hand side of Eq.~(\ref{EqT1Relax2}) versus $t$ is linear, then one has single-exponential relaxation where a least square fit to the data gives the slope which is $-1/T_1$.  This is often how $1/T_1$ at a particular temperature is determined experimentally for single-exponential decays.  Sometimes a prefactor $A\ (\approx 1)$ is included before the exponential in Eq.~(\ref{EqT1Relax0}) to take into account possible incomplete saturation at time $t = 0$.

If different nuclei in a sample see different local magnetic fields due to, e.g., different magnetic environments in the sample, then their individual Larmor frequencies will be different, resulting in ``inhomogenous'' broadening of the NMR absorption line.  Therefore, after a saturation pulse, the different nuclear spins in the sample will ``fan out'' in the $xy$ plane and get out of phase with each other in their precession around the applied field.  This is an energy-conserving process since the $z$~component of the nuclear magnetization is not affected.  The relaxation rate of the $xy$~plane magnetization in a frame of reference rotating about the $z$ axis at the average Larmor frequency is called the ``transverse relaxation rate'' and is denoted by $1/T_2$, where the term ``transverse'' refers to relaxation of the spins in a plane perpendicular to the applied magnetic field. 

When the nuclear spin $I = 1/2$, the nucleus does not have an electric quadrupole moment.  In this case there are no crystalline electric field (CEF) effects on the nuclear Zeeman levels and the nuclear spin-lattice relaxation is a single exponential.  Similarly, for any $I$, if a nucleus with an electric quadrupole moment is in a site of cubic symmetry, there is no gradient in the CEF at the nuclear site, and therefore there is again no electric quadrupole effect on the nuclear Zeeman levels, and the longitudinal relaxation is again single-exponential.  However, when $I \geq 1$ and the nuclear site symmetry is lower than cubic, then the nuclear Zeeman energies at zero field are changed by the gradient in the CEF at the nuclear site and the decay of the nuclear magnetization following saturation is no longer necessarily a single exponential, but is usually a well-defined sum of several exponentials, each containing $1/T_1$ as a parameter.  Thus even in these cases a well defined value of $1/T_1$ can be obtained from fitting the nuclear magnetization relaxation versus time data.  For example, many NMR studies of the FeAs-type materials have been carried out using the $^{75}$As nucleus that has $I = 3/2$.  If only the central $I_z = 1/2 \leftrightarrow -1/2$ line is irradiated, then in place of Eq.~(\ref{EqT1Relax0}) the magnetization decay follows
\be
1 - \frac{M_{\rm n}^z(t)}{M_{\rm n}^z(\infty)} = 0.1~e^{-t/T_1} + 0.9~e^{-6t/T_1} .
\label{EqT1Relax3}
\ee
However, if the entire spectrum could be irradiated including the side bands $I_z =\pm 3/2 \leftrightarrow \pm 1/2$, the magnetization decay would again be a single exponential as in Eq.~(\ref{EqT1Relax0}).

\subsubsection{\label{SecNMRDynamics}  Electron Spin Dynamics from NMR $1/T_1$}

\begin{figure}
\includegraphics [width=2.in]{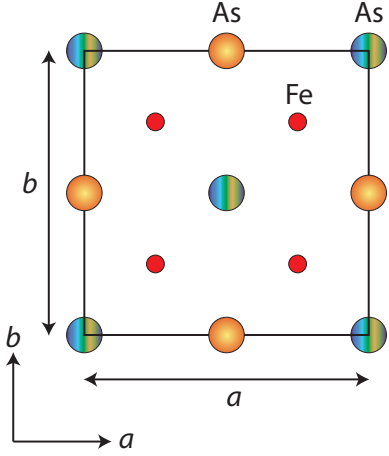}
\caption{(Color online) Projection down the $c$ axis of an FeAs plane in orthorhombic unit cell notation for the 1111-type and 122-type FeAs-based compounds.  The small red spheres are Fe atoms forming a square lattice in the $a$-$b$ plane.  The large circles are As atoms above (orange with radial gradient) and below (vertical stripes) the plane of the Fe atoms, placing the Fe atoms in distorted tetrahedral coordination by As.  The As atoms above and below the Fe plane are in planes that are, respectively, equal distances $(z_{\rm As}-\alpha)c$ along the $c$-axis from the Fe plane [see Eqs.~(\ref{Eqtheta}) and the tables in the Appendix], thus placing each Fe atom in distorted tetrahedral coordination by As (see also Fig.~\ref{Struct122}).}
\label{NMR_Hyperfine_A} 
\end{figure}

The electronic spin dynamics in a material are reflected by the longitudinal nuclear spin-lattice relaxation rate $1/T_1$ of embedded nuclei, because fluctuations in the electronic spin magnetization couple to the nuclear spins and induce transitions between the nuclear Zeeman levels, resulting in longitudinal relaxation.  The properties of several useful nuclei for NMR in the FeAs-based materials are listed in Table~\ref{Nuclei}.\cite{HbPC1972}  In general, $1/(T_{1}T)$ is expressed in terms of the component perpendicular to the applied field of the imaginary part $\chi_{M\perp}^{\prime\prime}({\bf Q}, \omega_{\rm n})$ of the dynamic susceptibility per mole of electronic spins at the nuclear Larmor angular frequency $\omega_{\rm n}$, as\cite{moriya1963,foot3,mahajan1998}
\begin{equation}
\frac{1}{T_{1}T} = \frac{2\gamma_{\rm n}^{2}k_{\rm B}}{N_{\rm A}^{2}}
\sum\limits_{{\bf Q}}\mid A({\bf Q})\mid
^{2}\frac{\chi^{''}_{M\perp}({\bf Q},\omega_{\rm n})}{\omega_{\rm n}},
\label{t1form}
\end{equation}
where $N_{\rm A}$ is Avogadro's number, the sum is over wave vectors ${\bf Q}$ within the first Brillouin zone, $A({\bf Q})$ is the form factor of the hyperfine interaction between the electronic and nuclear spins as a function of ${\bf Q}$ in units of Oe/$\mu_{\rm B}$, $\chi^{''}_{M\perp}({\bf q},\omega_{\rm n})$ is the imaginary part of the dynamical susceptiblity \emph{per mole of spins} (per mole of Fe in our case), and $\chi^{''}_{M\perp}({\bf q},\omega_{\rm n})/N_{\rm A}$ is expressed in (unconventional) units of $\mu_{\rm B}$/Oe.  An important aspect of Eq.~(\ref{t1form}) is that the relaxation rate depends on the integral of $\chi_{M\perp}^{\prime\prime}({\bf Q}, \omega_{\rm n})$ over ${\bf Q}$, including, in particular, a contribution for antiferromagnetic systems at a nonzero wave vector that can dominate the temperature dependence of $1/T_1T$.  Thus the temperature dependence of $1/T_1T$ can be quite different from that of the uniform static susceptibility $\chi \equiv \chi^\prime(0,0)$.  

For a given type of nucleus, the diagonal components (assumed isotropic) of the hyperfine coupling tensor of the nuclear spin to the nearest-neighbor Fe spins are
\be 
A({\bf Q}) = \sum_i B_i \exp({\bf Q}\cdot{\bf R}_i), 
\ee
where $B_i$ is the short-range transferred hyperfine coupling constant for coupling the nuclear spin to Fe spin $i$ and ${\bf R}_i$ is the position vector of Fe$_i$ with respect to the nucleus.  Thus, referring to Fig.~\ref{NMR_Hyperfine_A} in orthorhombic unit cell notation, for $^{57}$Fe and $^{75}$As nuclei, respectively, one has
\bea
\left|^{57}A({\bf Q})\right| &=& 2\ ^{57}B[\cos(Q_aa/2) + \cos(Q_bb/2)]\nonumber\\
\label{EqA1} \\
\left|^{75}A({\bf Q})\right| &=& 4\ ^{75}B\cos(Q_aa/4) \cos(Q_bb/4),\nonumber
\eea
where as noted above we have assumed that the coupling $B \equiv B_i$ is isotropic and independent of $i$ for each type of nucleus.  One can express the in-plane part of ${\bf Q}$ in terms of orthorhombic notation as ${\bf Q} = (H2\pi/a, K2\pi/b)$ and then Eqs.~(\ref{EqA1}) become
\bea
\left|^{57}A({\bf Q})\right| &=& 2\ ^{57}B[\cos(H\pi) + \cos(K\pi)]\nonumber\\
\label{EqA2} \\
\left|^{75}A({\bf Q})\right| &=& 4\ ^{75}B\cos(H\pi/2) \cos(K\pi/2).\nonumber
\eea

\begin{figure}
\includegraphics [width=3.3in,viewport=2 00 210 150,clip]{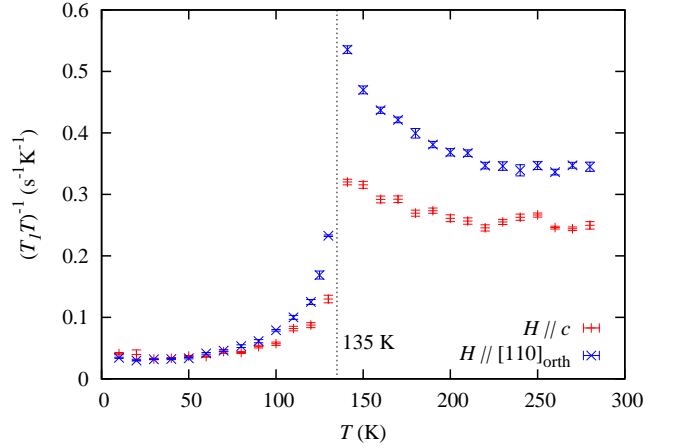}
\caption{(Color online) $^{75}$As nuclear spin-lattice relaxation rate divided by temperature $1/(T_1T)$ versus $T$ for single crystals of ${\rm BaFe_2As_2}$ grown from FeAs flux for the field oriented in the $a$-$b$~plane ($H\parallel[110]_{\rm orth}$) and parallel to the $c$-axis $(H\parallel c)$.\cite{Kitagawa2008}  The N\'eel temperature is $T_{\rm N} = 135$~K\@.  At low temperatures the relaxation is the Korringa relaxation of a Fermi liquid [$1/(T_1T) = $~constant] because the anisotropy gap of the spin waves freezes out relaxation from spin waves.  At high temperatures $T > T_{\rm N}$, the $H\parallel[110]_{\rm orth}$ data indicate the progressive development of stripe-type antiferromagnetic correlations in this plane on cooling towards $T_{\rm N}$.  Reproduced with permission from Ref.~\onlinecite{Kitagawa2008}.  Copyright (2008) by the Physical Society of Japan.}
\label{Kitagawa_t1t} 
\end{figure}

\begin{figure}
\includegraphics [width=3.3in]{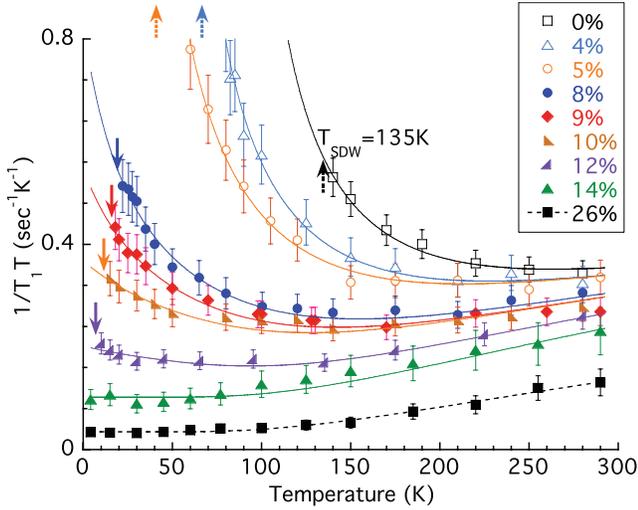}
\caption{(Color online) $^{75}$As nuclear spin-lattice relaxation rate divided by temperature $1/(T_1T)$ versus $T$ for a single crystal of Ba(Fe$_{1-x}$Co$_x)_2$As$_2$ grown from FeAs flux for $H\parallel ab$.\cite{Ning2010}  The compositions $x$ of the respective samples are given in the box.  The optimum doping concentration for superconductivity is about 7\% (see Fig.~\ref{FigBaKFe2As2_phase_diag}), so the data cover compositions from the highly underdoped to highly overdoped regimes.  The solid curves are fits by a Korringa term plus a Curie-Weiss term arising from the antiferromagnetic spin stripe fluctuations (see text).  Reprinted with permission from Ref.~\onlinecite{Ning2010}.  Copyright (2010) by the American Physical Society. }
\label{NingPRLFig4} 
\end{figure}

One sees from Eqs.~(\ref{EqA2}) that neither stripe-type [$(H,K) = (1,0)$] nor N\'eel-type [$(H,K) = (1,1)$] Fe spin correlations cause As nuclear relaxation to occur in the undistorted tetragonal phase from diagonal coupling of the As nuclear spins to the Fe electronic spins.  The reason is associated with the fact that an As atom is symmetrically located with respect to its four Fe nearest neighbors in the tetragonal phase.  However, Kitagawa and coworkers found that there exist off-diagonal components to the As hyperfine coupling tensor ${\bf B}$ defined by the internal field at the As site ${\bf H}_{\rm int} = {\bf B}\cdot {\bf M}^\dagger$ where ${\bf M}^\dagger$ is the Fe staggered magnetization.  The stripe-$b$ configuration of Fe spin moments oriented along the $a$-axis results in a net $c$-axis effective local internal magnetic field at the As site, which can be understood by replacing the Fe atoms by magnetic dipoles aligned along the $a$-axis.\cite{Kitagawa2008}  The off-diagonal As hyperfine coupling constant for stripe ordering in ${\rm BaFe_2As_2}$ at ${\bf Q} = (100)$ [or~(101)] is found to be $B_{ac} = 0.43$~T/$\mu_{\rm B}$ for coupling to each Fe spin.\cite{Kitagawa2008}  This analysis is consistent with, and indeed was motivated by, their observation that a stronger temperature variation of the As $1/(T_1T)$ occurred in the tetragonal phase for the applied field in the $ab$-plane than when it was parallel to the $c$-axis, as shown in Fig.~\ref{Kitagawa_t1t}.\cite{Kitagawa2008}  They also found the diagonal components in the paramagnetic state from a $K$-$\chi$ analysis ({\bf Q} = 0) to be $B_{aa} = B_{bb} = 0.66(2)$~T/$\mu_{\rm B}$ and $B_{cc} = 0.47(2)$~T/$\mu_{\rm B}$ for coupling to each Fe spin.  The relationship of these $B_{\alpha\beta}$ parameters with $|A({\bf Q})|$ in Eq.~(\ref{t1form}) is $|A_{\alpha\beta}({\bf Q})| = 4 B_{\alpha\beta}({\bf Q})$.  The authors thus concluded that their observed anisotropy in $1/(T_1T)$ in the tetragonal phase above $T_{\rm N}$ was due to spin-space anisotropy of the stripe-type (nematic) spin fluctuations, and suggested that such anisotropy might also be observed in the tetragonal doped superconducting compositions and indeed might have an important role in their superconductivity.  This anisotropy in the tetragonal structure of both the undoped and doped compounds was indeed found in inelastic magnetic neutron scattering measurements already discussed above in Sec.~\ref{SecNeutInel}.

Ning and coworkers carried out detailed $^{75}$As NMR $1/T_1$ measurements for $H\parallel ab$ (cf.~above discussion and Fig.~\ref{Kitagawa_t1t}) on single crystals of the electron-doped system Ba(Fe$_{1-x}$Co$_x)_2$As$_2$ covering the entire composition range from undoped ($x = 0$) to highly overdoped ($x = 0.26$) as shown in Fig.~\ref{NingPRLFig4} where $1/(T_1T)$ is plotted versus $T$.\cite{Ning2010}  The data are for so-called As(0) atoms that have four Fe nearest neighbors and no Co nearest neighbors.  A smooth evolution versus composition was found.  The data for $x = 0$ are in agreement with the corresponding data for $H\parallel ab$ in Fig.~\ref{Kitagawa_t1t}.  At the other extreme, the temperature dependence of the data for $x = 0.26$ is similar to the static uniform spin susceptibility $\chi_{\rm spin}(T)$ data for various compositions in this system in Fig.~\ref{NingFig4} and in Ref.~\ref{NingPRLFig4}, indicating that $(T_1T)^{-1}(T)$ for this $x$ range is due to Korringa relaxation by contact interaction of the nuclear spins with conduction electron spins giving $1/(T_1T) \sim \chi^2_{\rm spin}(T)$, and therefore that antiferromagnetic spin stripe correlations are not significant in highly overdoped samples.  The authors rationalized this result by considering that in the highly electron overdoped samples, the hole pockets have been filled up by electrons, so that there are no longer hole Fermi surface pockets at the $\Gamma$ point of the Brillouin zone.  Since the antiferromagnetic correlations in the itinerant magnetism scenario arise from Fermi surface nesting between the electron and hole pockets, the elimination of the hole pockets upon electron doping eliminates the spin stripe correlations.\cite{Ning2010}  The solid curves in Fig.~\ref{NingPRLFig4} are fits by a sum of a $T$-dependent Korringa term and a Curie-Weiss-like interpocket antiferromagnetic spin fluctuation term $C_{{\bf Q}_{\rm AF}}/(T + \theta)$ at the antiferromagnetic wave vector $Q_{\rm AF}$.\cite{Ning2010}  The $\theta$ value is found to become negative with decreasing $x$ at $x\approx 0.06$, which is the value at which long-range antiferromagnetic ordering begins with decreasing $x$, consistent with the assumptions of the model.

These results constitute strong evidence that antiferromagnetism in this system is itinerant rather than arising from local magnetic moments, and furthermore that spin stripe fluctuations are important to the superconducting mechanism in this system.\cite{Ning2010}  In this scenario, superconductivity is not suppressed by these spin fluctuations, but rather, enhanced.  On the other hand, the long-range spin density wave transition at $T_{\rm N}$ in the underdoped samples competes with superconductivity for the same conduction electrons, and hence optimum superconductivity is not observed until the long-range antiferromagnetic ordering is completely suppressed, in agreement with the experimental phase diagrams (see Fig.~\ref{FigBaKFe2As2_phase_diag}).  These two requirements for optimum $T_{\rm c}$ are therefore not mutually exclusive as might have otherwise been anticipated.

As noted in the Introduction, the compound FeSe has a $T_{\rm c}$ that increases from 8~K to 37~K under a pressure of 9~GPa.\cite{Margadonna2009a, Medvedev2009}  Imai and coworkers carried out a $^{77}$Se NMR study of FeSe under pressure to study how the spin dynamics evolve as $T_{\rm c}$ increases.\cite{Imai2009}  Similar to the data in Fig.~\ref{NingPRLFig4} for Ba(Fe$_{1-x}$Co$_x)_2$As$_2$, the authors found that the spin fluctuations are enhanced as the $T_{\rm c}$ increases, consistent with a spin-fluctuation mediated mechanism for superconductivity.

\subsubsection{\label{SecNeutsNMR} Relationships between Electron Spin Dynamics from NMR $1/T_1$ and from Inelastic Neutron Scattering: Ba(Fe$_{1-x}$Co$_x)_2$As$_2$}

It is useful to quantitatively relate the extensive results of inelastic magnetic neutron scattering measurements in Sec.~\ref{SecNeutInel} to expectation for NMR $1/T_1$ measurements.  First we note that the energy of a photon at the experimental nuclear Larmor frequency $\hbar\omega_{\rm n} \sim h(10$--100)~MHz~$\lesssim 0.4~\mu$eV~$\approx 5$~mK ($h$ is Planck's constant) is very small compared to the above inelastic neutron scattering energies, so Eq.~(\ref{t1form}) corresponds to a low-frequency limit in this regard.  Thus even a small gap $\Delta$ in the spin excitation spectrum can prevent NMR from probing the spin fluctuations if $k_{\rm B}T \ll \Delta$. To probe the bulk antiferromagnetic spin fluctuations one should therefore carry out measurements in the paramagnetic state.

For the doped Ba(Fe$_{1-x}$Co$_x)_2$As$_2$ system, the mean-field nearly antiferromagnetic Fermi liquid theory was successful in fitting most of the inelastic magnetic neutron scattering data for optimally doped ${\rm Ba(Fe_{1.925}Co_{0.075})_2As_2}$ as described above.\cite{Inosov2010}  Taking the limit $\omega \to 0$ of Eq.~(\ref{EqChippInosov}), as is appropriate in the paramagnetic state for the relatively low NMR frequencies as discussed above, gives
\bea
\frac{\chi^{\prime\prime}({\bf q},\omega)_T}{\omega} &=& \frac{\hbar \chi_T }{\Gamma_T(1 + \xi_T^2q^2)^2}\label{EqChippNMR}\\
{\rm with}\ \ \ \chi_T &=& \frac{C}{T + \theta},\nonumber\\
\Gamma_T &=& \Gamma_0(T + \theta),\nonumber\\
\xi_T &=& \frac{\xi_0}{\sqrt{T + \theta}},\nonumber\\
{\rm and}\ \ \ \ \ {\bf q} &\equiv& {\bf Q} - {\bf Q}_{\rm AF},\nonumber
\eea

We calculate the part of $1/T_1$ coming from antiferromagnetic fluctuations that are sharply peaked in wave vector at ${\bf Q}_{\rm AF} =(1,0)$~r.l.u.\ in orthorhombic notation, and we assume that only these contribute to the sum in Eq.~(\ref{t1form}), so we set the upper limit of the sum to $q = \infty$ giving
\bea
&\sum\limits_{{\bf Q}}|A({\bf Q})|
^{2}&\frac{\chi^{''}_{M\perp}({\bf Q},\omega_{\rm n})}{\omega_{\rm n}} \to\nonumber\\
&&|A_{ac}({\bf Q}_{\rm AF})|
^{2}\left(\frac{1}{\frac{2\pi}{L}}\right)^D\int_0^\infty \frac{\chi^{\prime\prime}({\bf q},\omega)_T}{\omega}d^Dq\nonumber \\
 &=& |A_{ac}({\bf Q}_{\rm AF})|
^{2}\frac{\hbar \chi_T }{\Gamma_T}\frac{L^2}{4\pi\xi_T^2}\ \ \ (D= 2)\nonumber\\
\label{EqInt}\\
&=& |A_{ac}({\bf Q}_{\rm AF})|
^{2}\frac{\hbar \chi_T }{\Gamma_T}\frac{V}{8\pi\xi_T^3},\ \ \ (D= 3)\nonumber
\eea
where $D$ is the space dimensionality of the spin lattice giving rise to the spin fluctuations, $L$ is the linear size, $L^2$ is the area and $V = L^3$ is the volume of the system. 

From Eqs.~(\ref{t1form}), (\ref{EqChippNMR}) and~(\ref{EqInt}), this mean-field theory predicts
\bea
\frac{1}{T_1T} &\sim& \frac{1}{T + \theta}\ \ \ \ \ \ \ \ \ (D=2)\label{EqNMRT12D3D} \\
\frac{1}{T_1T} &\sim& \frac{1}{(T + \theta)^{1/2}} \ \ \ \nonumber (D = 3).
\eea
The antiferromagnetic fluctuation parts of the fits to the $(T_1T)^{-1}(T)$ data in Fig.~\ref{NingPRLFig4} by Ning et al.\cite{Ning2010} are in agreement with the 2D result in Eq.~(\ref{EqNMRT12D3D}) with the Curie-Weiss form, and therefore also in agreement with the neutron scattering results which showed that the antiferromagnetic correlations in the doped systems are two-dimensional with weak or negligible spin coupling between Fe layers.  Thus the fits of the antiferromagnetic fluctuation part of $(T_1T)^{-1}(T)$ data by a Curie-Weiss like behavior\cite{Ning2010} is justified in mean field theory for two-dimensional fluctuations.

A more rigorous and demanding test of the relationship between the NMR and neutron scattering results is to predict the \emph{magnitude} of the temperature-dependent antiferromagnetic fluctuation part of the $^{75}$As $1/T_1$ from the inelastic neutron scattering fitting parameters\cite{Inosov2010} for optimally doped ${\rm Ba(Fe_{1.925}Co_{0.075})_2As_2}$.  From Eqs.~(\ref{t1form}), (\ref{EqChippNMR}) and the 2D result in Eq.~(\ref{EqInt}) we have
\be
\frac{1}{T_{1}T} = \frac{2\gamma_{\rm n}^{2}k_{\rm B}}{N_{\rm A}^{2}}
|A_{ac}({\bf Q}_{\rm AF})|^{2}\frac{\hbar \chi_T }{\Gamma_T}\frac{L^2}{4\pi\xi_T^2}= \frac{C_{{\bf Q}_{\rm AF}}}{T+\theta}.
\label{t12Dtot}
\ee
The NMR $1/T_1T$ Curie constant for ${\bf Q} = {\bf Q}_{\rm AF}$ is, using $L^2 = V/c$ where $c$ is the $c$-axis lattice parameter, 
\be
C_{{\bf Q}_{\rm AF}} = \frac{\gamma_{\rm n}^{2}\hbar k_{\rm B} V |A_{ac}({\bf Q}_{\rm AF})|^{2}C}{2\pi c N_{\rm A}^{2}\Gamma_0\xi_0^2}
\label{EqCAFNMR}
\ee
and $C$ is the Curie constant in Eq.~(\ref{EqChippNMR}) per mole of Fe atoms.  The values of the various parameters are obtained from the lattice parameter data in the Appendix, from the neutron scattering fitting parameters in Eqs.~(\ref{EqFitPars}), and from Table~\ref{Nuclei} as
\bea
\gamma_{\rm n} &=& 2\pi(7.292~{\rm MHz/T}) \ \ \ {\rm for\ ^{75}As}\nonumber\\
\frac{V}{N_{\rm A}c} &=& \frac{a^2c}{4c} = 15.7\ {\rm \AA}^2\ \ ({\rm area/Fe~atom})\nonumber\\
|A_{ac}({\bf Q}_{\rm AF})| &=& 1.72~{\rm T}/\mu_{\rm B}\nonumber\\
\frac{C}{N_{\rm A}} &=& 1.9 \times 10^4~\frac{\mu_{\rm B}^2~{\rm K}}{{\rm eV~Fe~atom}}\label{EqFitPars2}\\
\xi_0 &=& 160~{\rm \AA~K^{1/2}},\nonumber\\
{\rm and}\ \ \Gamma_0 &=& 0.14 ~{\rm \frac{meV}{K}}.\nonumber
\eea

Inserting all these values into Eq.~(\ref{EqCAFNMR}) gives
\be
C_{{\bf Q}_{\rm AF}} = 4.7~{\rm s}^{-1},
\ee
\emph{with no adjustable parameters}.  This value is of the same order as the value $C_{{\bf Q}_{\rm AF}} = 24(4)~{\rm s}^{-1}$ obtained by Ning et al.\ from their fits of their $(T_1T)^{-1}(T)$ data for Ba(Fe$_{1-x}$Co$_x)_2$As$_2$ for $x = 0$ to $x = 1$ in Fig.~\ref{NingPRLFig4} by a Curie-Weiss type behavior for the antiferromagnetic fluctuation part.\cite{Ning2010}  This is reasonable agreement considering the simplicity and assumptions of our model, the many fixed parameters used in the calculation, and the extrapolation of the inelastic neutron scattering fit at high energies to $\omega \approx 0$.  Indeed, this semiquantitative agreement  supports the applicability of the itinerant nearly antiferromagnetic Fermi liquid model that was used by Inosov et al.\ to analyze their neutron scattering data for ${\rm Ba(Fe_{1.925}Co_{0.075})_2As_2}$.\cite{Inosov2010}

\subsubsection{NMR Korringa Ratio $K^2T_1T/S_{\rm n}$ for Conduction Electrons}

\subsubsection*{a. Introduction}

The longitudinal nuclear magnetization can be relaxed by interactions with conduction electron spins in a metal, a process called Korringa relaxation.  The Korringa relaxation rate $1/T_1$ is given by\cite{Slichter1963}
\be
\frac{1}{T_1T} = \frac{16\pi^3 \hbar^3 k_{\rm B} \gamma_{\rm e}^2\gamma_{\rm n}^2}{9} \langle|u_{\bf k}(0)|^2\rangle_{E_{\rm F}} N^2(E_{\rm F}),
\label{EqKorringa}
\ee
where $\langle|u_{\bf k}(0)|^2\rangle_{E_{\rm F}}$ is the average over the Fermi surface of the square of the periodic part of the conduction electron wave function at the position of a nucleus as in Eq.~(\ref{KnightShift}) for the spin shift, and $N(E_{\rm F})$ is the conduction electron density of states at the Fermi energy for both spin directions.  Important features of Eq.~(\ref{EqKorringa}) are that $1/T_1$ is simply proportional to $T$ with a slope that goes as the square of $N(E_{\rm F})$.  

The factor $\langle|u_{\bf k}(0)|^2\rangle_{E_{\rm F}}$ in Eq.~(\ref{EqKorringa}) also appears in the expression~(\ref{KnightShift2}) for the Knight shift, and therefore can be eliminated from both Eqs.~(\ref{KnightShift2}) and~(\ref{EqKorringa}) yielding 
\be
\frac{K_{\rm spin}^2}{\frac{1}{T_1T}} = K_{\rm spin}^2T_1T = \frac{\hbar}{4\pi k_{\rm B}}\left(\frac{\gamma_{\rm e}}{\gamma_{\rm n}}\right)^2.
\label{EqKorr2}
\ee
The constant on the right-hand side is called the Korringa constant $S_{\rm n}$.  It is specific to the particular nucleus n with
\be
S_{\rm n} = \frac{\hbar}{4\pi k_{\rm B}}\left(\frac{\gamma_{\rm e}}{\gamma_{\rm n}}\right)^2.
\label{EqKorr3}
\ee
Values of $S_{\rm n}$ for several relevant nuclei are given above in Table~\ref{Nuclei}. The Korringa ratio is defined as
\be
R_{\rm K} \equiv \frac{K_{\rm spin}^2}{\frac{S_{\rm n}}{T_1T}} = \frac{K_{\rm spin}^2T_1T}{S_{\rm n}} \sim 1.
\label{EqKorr4}
\ee
A very appealing feature of Eq.~(\ref{EqKorr4}) is that the hyperfine coupling constant $A_{\rm hf} \sim\langle|u_{\bf k}(0)|^2\rangle_{E_{\rm F}}$ between the nuclei and conduction electrons was eliminated from the expression.  

The above expressions for $K_{\rm spin}$ and $1/(T_1T)$, and therefore for $R_{\rm K}$ in Eq.~(\ref{EqKorr4}), were derived assuming a noninteracting electron gas or a nearly-free-electron Fermi liquid.  Therefore deviations of $R_{\rm K}$ from unity can give information about many-body interactions affecting the conduction electrons.  A value $R_{\rm K} \sim 2$ has been observed in many heavy fermion compounds.  A value $R_{\rm K} < 1$ can result from antiferromagnetic conduction electron correlations because the dynamical conduction electron susceptibility can peak at finite wave vector, increasing $1/T_1$, but which can have little influence on $K_{\rm spin}$, which is proportional to the zero-frequency, zero-wave vector susceptibility.  The value of $R_{\rm K}$ can also serve as a diagnostic for situations where the generalized spin susceptibility can be deconvoluted into contributions from different ranges of {\bf Q} and $\omega$, as was done in the previous section to analyze $1/T_1$ versus composition in the Ba(Fe$_{1-x}$Co$_x)_2$As$_2$ system.\cite{Ning2010}  In the following we look again at this system, but this time with respect to the Korringa ratio.

\subsubsection*{b. {\rm Ba(Fe$_{1-x}$Co$_x)_2$As$_2$}}

\begin{figure}
\includegraphics [width=3.3in]{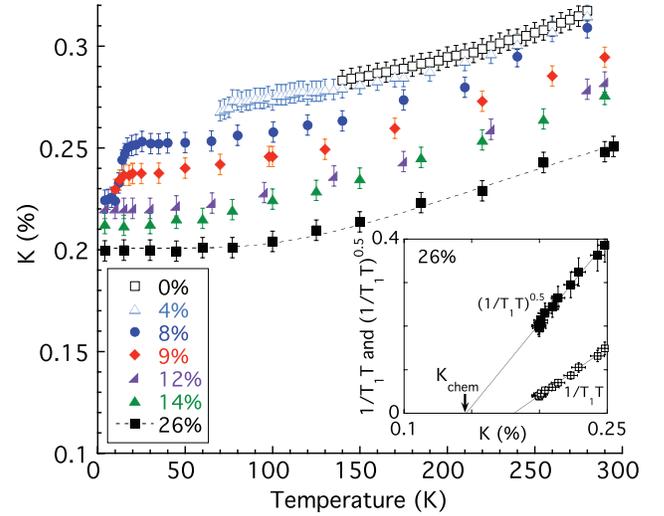}
\caption{(Color online)  Resonance shift $K$ versus temperature for $^{75}$As nuclei in single crystals of Ba(Fe$_{1-x}$Co$_x)_2$As$_2$ as in Fig.~\ref{NingPRLFig4},\cite{Ning2010} but with $H\parallel c$ (F. L. Ning, private communication).  The inset shows a plot of  $1/(T_1T)^{1/2}$ versus $K$ that was done in order to separate the spin and chemical (i.e., orbital) parts of $K$ [see Eq.~(\ref{Eqchemspin})], where the $K$ data are for $H\parallel c$ (unpublished; F. L. Ning, private communication). Also shown in the inset is a plot of $1/(T_1T)$ versus $K$.  Reprinted with permission from Ref.~\onlinecite{Ning2010}.  Copyright (2010) by the American Physical Society.}
\label{NingPRLFig3} 
\end{figure}

The $^{75}$As NMR shift versus temperature behaviors for single crystals of Ba(Fe$_{1-x}$Co$_x)_2$As$_2$ are shown in Fig.~\ref{NingPRLFig3}.\cite{Ning2010}  In general, like the magnetic susceptibility, the NMR shift $K$ has orbital and spin contributions
\be
K(T) = K_{\rm chem} + K_{\rm spin}(T),
\ee
where the subscript ``chem'' refers to the orbital part that comes from coupling of the nuclei to the orbital part of the magnetic susceptibility that is nominally independent of $T$.  For the highly overdoped sample with $x = 0.26$, the $K$ in Fig.~\ref{NingPRLFig3} increases with $T$, as does $1/T_1T$ for this crystal in Fig.~\ref{NingPRLFig4}, suggesting that the Korringa ratio may be independent of $T$ for this composition.  Presuming that these samples do not contain local moments, using Eq.~(\ref{EqKorr4}) gives
\be
\sqrt{\frac{1}{T_1T}} = K_{\rm chem} + \alpha K_{\rm spin},
\label{Eqchemspin}
\ee
with $T$ as an implicit parameter, where
\be
\alpha = \frac{1}{\sqrt{R_{\rm K}S_{\rm As}}}.
\label{EqAlpha}
\ee
Indeed, in the inset of Fig.~\ref{NingPRLFig3}, the authors plotted $1/\sqrt{T_1T}$ versus $K$ and obtained a linear behavior\cite{Ning2010} described by a horizontal intercept $K_{\rm chem}$ and slope $\alpha$, indicating a temperature independent Korringa ratio $R_{\rm K}$.  Using the observed $\alpha$, the value of the Korringa constant $S_{\rm As}$ from Table~\ref{Nuclei}, and  Eq.~(\ref{EqAlpha}), one obtains the Korringa ratio (F. L. Ning, private communication)
\be
R_{\rm K} = 0.85(10)\ \ \ {\rm for}\ x = 0.26.
\ee
The slight suppression of $R_{\rm K}$ from unity suggests that antiferromagnetic correlations are still not completely suppressed even at $x = 0.26$.  

Comparison of the data in Figs.~\ref{NingPRLFig4} and~\ref{NingPRLFig3} shows that as $x$ decreases from 0.26, $R_{\rm K}$ becomes $T$ dependent and decreases strongly with decreasing $T$, consistent with the increase with decreasing temperature of the generalized susceptibility at wave vector {\bf Q} = ${\bf Q}_{\rm AF}$ found from the above inelastic neutron scattering experiments on this system.

\subsection{\label{ElecCp} Electronic Heat Capacity}

\begin{figure}
\includegraphics [width=3.3in]{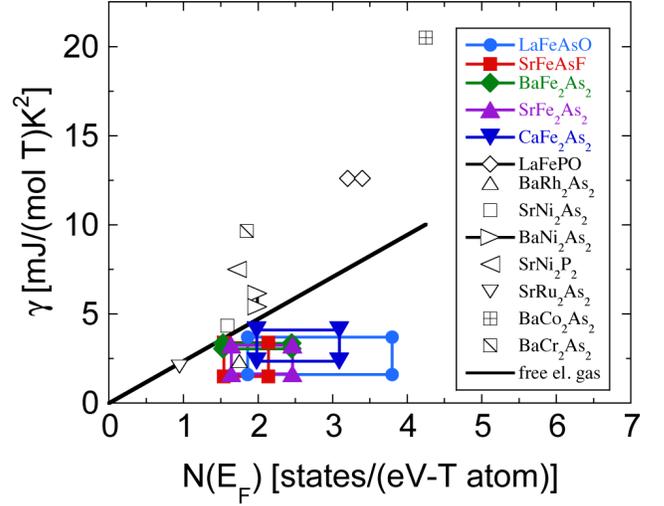}
\caption{(Color online) Experimental Sommerfeld electronic linear heat capacity coefficient $\gamma$ versus the calculated bare nonmagnetic band structure density of states for both spin directions $N(E_{\rm F})$ for undoped 1111 and 122 transition metal $T$ pnictides.  Multiple data points for the same compound indicate the range(s) of $\gamma$ and/or $N(E_{\rm F})$ reported for the compound.  The colored filled symbols are for compounds containing FeAs layers.  The data for the FeAs-based compounds are below the sloping straight line shown that is expected for a nearly-free-electron gas, indicating reductions in $N(E_{\rm F})$ due to SDW transitions in these undoped compounds.  See the Appendix for the references.}
\label{gammaN(EF)} 
\end{figure}

The ``bare'' normal-state electronic heat capacity of a degenerate nearly-free-electron gas is proportional to absolute temperature, $C_{\rm e} = \gamma_0 T$, with a slope $\gamma_0$ that is proportional to the bare $N(E_{\rm F})$ according to\cite{kittel1966}
\begin{equation}
\gamma_0 = \frac{\pi^2k_{\rm B}^2}{3}N(E_{\rm F}) = (2.359)N(E_{\rm F}),
\label{EqGamma0}
\end{equation}
where the right-hand equality is for $\gamma_0$ in mJ/mol~K$^2$ and $N(E_{\rm F})$ in states/(eV f.u.) for both spin directions.  In the presence of electron-phonon coupling, the measured $\gamma$ becomes
\begin{equation}
\gamma = \gamma_0 (1 + \lambda_{\rm e-ph}),
\label{EqGamma}
\end{equation}
where $\lambda_{\rm e-ph} \geq 0$ is the electron-phonon coupling parameter.  Thus one expects the observed $\gamma$ to be greater than or equal to $\gamma_0$.

We have plotted the experimentally measured $\gamma$ values versus the bare band structure $N(E_{\rm F})$ for a variety of FeAs-based and other related materials in Fig.~\ref{gammaN(EF)}.  These $\gamma$ values are typically obtained at temperatures between roughly 2 and 10~K\@.  We note that with the single exception of BaRh$_2$As$_2$, where the experimental $\gamma$ value may be  in error, the data points for the non-FeAs compounds in Fig.~\ref{gammaN(EF)} (black open symbols) are all on or above the sloping line, in agreement with expectation.  All of the data for the FeAs-based materials (colored filled symbols) are for the nonsuperconducting undoped parent compounds because the superconducting state interferes with the determination of the normal-state $\gamma$ (but see Sec.~\ref{Sec_SC_Cp} below).  A striking feature from Fig.~\ref{gammaN(EF)} is that \emph{all} of the FeAs-based parent compounds have $\gamma$ values \emph{below} the minimum sloping line predicted for the quasi-free-electron gas in Eqs.~(\ref{EqGamma0}) and (\ref{EqGamma}).  This situation is in stark contrast to the \emph{enhanced} spin susceptibilities at 300~K discussed above.  

Optical spectroscopy measurements versus temperature indicated that a large fraction of the conduction carrier density is removed by the SDW transitions below 140--200~K in the parent compounds SrFe$_2$As$_2$ and BaFe$_2$As$_2$ as discussed above in Sec.~\ref{Sec_BandStruct}.  Similarly, quantum oscillation experiments on ${\rm SrFe_2As_2}$ (Ref.~\onlinecite{Sebastian2008}) and ${\rm BaFe_2As_2}$ (Ref.~\onlinecite{Analytis2009}) indicate  reconstructions of the Fermi surfaces from 2D to 3D below the respective $T_{\rm N}$, with reductions in carrier concentrations.  Therefore one also expects a significant reduction in $N(E_{\rm F})$ at low temperatures compared with the bare nonmagnetic values, consistent with the data in Fig.~\ref{gammaN(EF)} for the nonsuperconducting FeAs-based parent compounds.  Indeed, electronic structure calculations by Ma, Lu and Xiang\cite{Ma2008} give $N(E_{\rm F})$ values in the nonmagnetic (stripe-ordered AF) state of 3.95 (1.22), 3.27 (1.92), and 3.93 (2.41) states/(eV~f.u.) for both spin directions for CaFe$_2$As$_2$, SrFe$_2$As$_2$, and BaFe$_2$As$_2$, respectively.  On the other hand, the same calculations yield ordered moments in the AF phase to be 2.0, 2.2, and 2.3~$\mu_{\rm B}$/Fe, respectively, whereas the observed values are all $\lesssim 1$~$\mu_{\rm B}$/Fe.

\subsection{\label{SecNematic} Nematic Correlations}

In several places in this review, the concepts of nematic correlations, or nematic degrees of freedom, have appeared, or will appear, which will be summarized here.  Additional evidence for nematicity will also be discussed.  Kimber et al.\cite{Kimber2010} provided a definition of what the term nematic means.  We quote, ``Nematic phases are frequently found in organic matter.  The defining characteristic of these phases is orientational order in the absence of long range positional order, resulting in distinctive uniaxial physical properties.  It has also been proposed that nematic order exists in some electronic systems, and may even play a role in mediating high temperature superconductivity. ... \emph{This breaking of the electronic symmetry compared to that of the underlying lattice is a hallmark of electronic nematic phases.}''\cite{Kimber2010}  (My emphasis).  In the present context, a nematic behavior is one in which there is anisotropy in a property that locally breaks the symmetry of the crystal and/or reciprocal lattice structure but globally preserves it, which will be specifically illustrated below.

As noted above in Sec.~\ref{Sec122typeneuts}, from inelastic neutron scattering measurements Lester et al.\ observed an anisotropy in the magnetic excitation dispersion relation in optimally-doped \emph{tetragonal} ${\rm Ba(Fe_{1.935}Co_{0.065})_2As_2}$ between the orthorhombic $(H,0,0)$ and~$(0,K,0)$ directions, with the same sign of the anisotropy as predicted by Eqs.~(\ref{EqSWvel}) for spin waves in the long-range ordered orthorhombic antiferromagnetic stripe-$b$ phase.\cite{Lester2010}  This anisotropy corresponds to \emph{magnetic} nematic correlations.  In particular, Fig.~\ref{FigLesterDisp} shows that there is an anisotropy in the magnetic dispersion relations between in-plane longitudinal and transverse scans through the $\left(\frac{1}{2},\frac{1}{2},L\right)$~r.l.u.\ position with fixed $L$ in tetragonal notation, which are the $(1,0,L)$~r.l.u.\ and equivalent positions in orthorhombic notation as noted above.  This anisotropy is the same as illustrated in Fig.~\ref{FigTwins}, where the blue and yellow ellipse neutron scattering intensity contours for the orthorhombic twins are equivalent to each other in the tetragonal structure.  Perhaps counterintuitively, according to Fig.~\ref{FigTwins} this orthorhombic-like local symmetry in reciprocal space still preserves the global fourfold rotational symmetry about the $c$-axis of the tetragonal reciprocal lattice.  In real space, these magnetic correlations correspond to short-range stripe antiferromagnetic order as found in the orthorhombic long-range ordered state of the undoped parent compounds.  Thus, the microscopic interactions giving rise to the stripe antiferromagnetic structure in the low-temperature orthorhombic crystal structure of the undoped parent compound ${\rm BaFe_{2}As_2}$ are still present and are causing short-range antiferromagnetic stripe order in the optimally-doped tetragonal ${\rm Ba(Fe_{1.935}Co_{0.065})_2As_2}$ compound.

Additional evidences of magnetic nematic behaviors were found in other studies.  From Sec.~\ref{Sec11neuts}, Lee et al.\ found discrete incommensurability of the magnetic excitations at in-plane wave vectors transverse to $\left(\frac{1}{2}, \frac{1}{2}\right)$~r.l.u.\ in tetragonal FeTe$_{0.5}$Se$_{0.5}$ and suggested that this arises from coupled spin-orbital excitations.\cite{Lee2009}  They further suggested that ``If correct, it follows that these nematic fluctuations are involved in the (neutron spin) resonance (see Sec.~\ref{ResonanceMode} below) and could be relevant to the pairing mechanism.''\cite{Lee2009}  Li et al.\ showed from inelastic neutron scattering measurements of ${\rm Ba(Fe_{1.926}Co_{0.074})_2As_2}$ single crystals that the in-plane anisotropy of the spin fluctuations in the normal state, reflecting a magnetic nematic degree of freedom as discussed above, is very similar in the neutron spin resonance mode.\cite{Li2010}

As discussed above in Sec.~\ref{SecNMRDynamics}, Kitagawa and coworkers concluded that their observed anisotropy in the NMR $1/(T_1T)$ in the tetragonal phase of ${\rm BaFe_2As_2}$ above $T_{\rm N}$ was due to spin-space anisotropy of the stripe-type (nematic) spin fluctuations, and suggested in advance of the above neutron scattering measurements that such anisotropy might also be observed in the tetragonal doped superconducting compositions and indeed might have an important role in their superconductivity.\cite{Kitagawa2008}

As discussed below in Sec.~\ref{SecRSInhomo}, Chuang et al.\ detected one-dimensional inhomogeneities on the surface of an underdoped ${\rm Ca(Fe_{0.97}Co_{0.03})_2As_2}$ crystal using tunneling microscopy measurements at $T = 4.3$~K.\cite{Chuang2010}  They interpreted their data in terms of nematic electronic correlations.  However, according to the above definition of Kimber et al.,\cite{Kimber2010} the correlations were not nematic because their symmetry was consistent with the twofold rotational symmetry about the $c$-axis of the low-temperature orthorhombic crystal structure at the temperature (4.3~K) of the measurements.

Decisive evidence for electronic nematic behavior was found by Chu et al.\ via anisotropic electrical resistivity measurements on single crystals of Ba(Fe$_{1-x}$Co$_x)_2$As$_2$.\cite{Chu2010}  First, using a uniaxial pressure clamp device, they were able to prevent orthorhombic twins from forming on cooling below the structural transition temperature $T_{\rm S}$ (see the phase diagram in Fig.~\ref{FigBaKFe2As2_phase_diag}).  Because the $b$-axis lattice parameter of the orthorhombic structure is smaller than the $a$-axis lattice parameter, the compressive stress favors $b$-axis alignment along the direction of application of the stress.  Second, for the composition range that exhibits a low-temperature structural and magnetic transition ($x = 0$--0.051), they found a large (up to a factor of 2!) anisotropy between the in-plane resistivity along the short $b$-axis (along the ferromagnetically aligned Fe spin stripes, see Fig.~\ref{Stripe_Mag_Struct}) and along the long $a$-axis of the orthorhombic structure (along the direction of antiferromagnetically aligned nearest-neighbor spins), with $\rho_b > \rho_a$.  This large in-plane resistivity anisotropy is, at first sight, very surprising given the small ($\leq 0.4\%$, depending on $x$)\cite{Prozorov2009b} in-plane distortion between the $a$- and $b$-axis lattice parameters in the orthorhombic phase.  The large anisotropy evidently results from the different nearest-neighbor spin alignments along the $a$- and $b$-axes as described above.  This resistivity anisotropy described so far is not a result of a nematic degree of freedom, because the low temperature crystal structure has the same symmetry as the resistivity anisotropy.  

The electronic nematic degree of freedom becomes apparent from the in-plane resistivity measurements by Chu et al.\ on the above Ba(Fe$_{1-x}$Co$_x)_2$As$_2$ crystal with $x = 0$ in the tetragonal phase at $T > T_{\rm S}$.\cite{Chu2010}  Here, the same sign of the anisotropy in $\rho_b / \rho_a$ was observed under applied uniaxial pressure as for $T < T_{\rm S}$, albeit with a smaller magnitude that diminished with increasing $T$ for $T > T_{\rm S}$.  The same nematic electronic behavior was revealed under uniaxial pressure in a tetragonal crystal with $x = 0.07$ that did not show a structural transition.  The authors concluded, ``These results indicate that the structural transition in this family of compounds is fundamentally electronic in origin.''\cite{Chu2010}  That is, the structural transition is driven by the coupled electronic charge and magnetic nematic degrees of freedom.  

Untwinned crystals of CaFe$_{2}$As$_2$ and BaFe$_{2}$As$_2$ were produced by Tanatar et al.\ by applying a uniaxial tensile stress, which favors $a$-axis alignment (instead of the above $b$-axis alignment from compressive stress) along the axis of the applied stress.\cite{Tanatar2010b}  The authors found $\rho_b / \rho_a = 1.2$ and~1.5 just below $T_{\rm S}$ for the Ca and Ba compounds, respectively, which is the same sign of anisotropy as found by Chu et al.\ above.  They observed  nematic electronic anisotropy above $T_{\rm S}$ for BaFe$_{2}$As$_2$ but not for CaFe$_{2}$As$_2$, which reflects the fact that the phase transition in the Ca compound is strongly first order whereas that in the Ba compound is second order or weakly first order.\cite{Tanatar2010b}

The anisotropic $T$-dependent optical properties of untwinned Ba(Fe$_{1-x}$Co$_x)_2$As$_2$ crystals with $x = 0$ and~0.025 under uniaxial pressure were studied.\cite{Dusza2010}  As with the anisotropic resistivity measurements above, anisotropy in the optical properties along the orthorhombic $a$- and $b$-axes was observed \emph{in the high-temperature tetragonal phase}, confirming the above electronic nematic degree of freedom.  Interestingly, the authors found that the antiferromagnetic transition partially gaps the optical conductivity $\sigma_1$ versus angular frequency $\omega$ for the $b$-axis while at the same time enhancing the optical conductivity along the $a$-axis.  From the observed anisotropy in $\sigma_1(\omega)$ extending to high frequencies $\hbar\omega \gg k_{\rm B}T_{\rm S}$, the authors further stated, ``These results are consistent with a scenario in which orbital order plays a significant role in the tetragonal-to-orthorhombic structural transition,'' thus adding yet another degree of freedom to the possible origins of the structural transition.

A pronounced softening of the shear modulus on cooling towards $T_{\rm S}$ (for $x=0$) or towards $T_{\rm c}$ (for $x=0.08$) was observed by Fernandes and coworkers from resonant ultrasound spectroscopy measurements on single crystals of Ba(Fe$_{1-x}$Co$_x)_2$As$_2$.\cite{Fernandes2010d} The authors developed a theory for the softening, consistent with the data, which indicated that the orthorhombic structural transition is driven by magnetic fluctuations and that the unconventional superconductivity in the iron pnictides is associated with the magnetic and electronic nematic degrees of freedom.\cite{Fernandes2010d}

\section{\label{SecSCProps} Superconducting Properties}

We have already considered the relationships between (i) $T_{\rm c}$ and doping in Sec.~\ref{SecIntro} (e.g., Figs.~\ref{FigBaKFe2As2_phase_diag} and~\ref{KatayamaFig2}); (ii) $T_{\rm c}$ and structural characteristics in Figs.~\ref{TcAngle}, \ref{TcHeight} and~\ref{Nandi_ortho_split}; (iii) $T_{\rm c}$ and the density of states at the Fermi energy in Fig.~\ref{TcN(EF)}; and (iv) $T_{\rm c}$ and the magnetic susceptibility in Fig.~\ref{TcChi}.  Here we discuss additional aspects of the superconducting properties of the Fe-based materials.  We do not discuss superconducting doped crystals of ${\rm BaFe_2As_2}$ grown in Sn flux because such crystals are contaminated by Sn impurities that drastically affect the properties of the crystals.  Applied aspects such as critical currents and the synthesis of superconducting wires are also not addressed.

\subsection{\label{SecCooperPairSpin} Spin of a Cooper Pair}

In the Bardeen-Cooper-Schrieffer (BCS) theory of superconductivity\cite{BCS57} that was originally developed for conventional superconductors with the electron-phonon mechanism for superconductivity, the superconducting state consists of bound electron pairs (Cooper pairs)\cite{Cooper56} in which both the intrinsic angular momenta (spins) and wavevectors ($\sim$ velocities) of the two electrons are in opposite directions to each other, respectively.  Each electron by itself is a fermion with intrinsic spin 1/2 but the Cooper pair has a net spin of zero.  Most known superconductors are of this ``conventional'' type, called spin singlet superconductors.  Quantum mechanics  tells us that the wave function of a Cooper pair has to be antisymmetric with respect to interchange of the two electrons in the pair, which means that the integer $n$ in the expression for the orbital angular momentum $L = n\hbar$ of a Cooper pair must be an even positive integer or zero, where $\hbar$ is Planck's constant divided by 2$\pi$.  If $n = 0$ or~2, one has an $s$-wave or $d$-wave pairing, respectively, where the notation is taken from atomic physics.  All conventional superconductors like Sn or Pb are spin singlet $s$-wave superconductors, but the layered cuprate high-$T_{\rm c}$ superconductors are spin singlet $d$-wave superconductors.\cite{Van_Harlingen1995}  On the other hand, if the two spins of the electrons in a Cooper pair line up in the same direction, then the superconductor is called a triplet superconductor and the integer $n$ has to be an odd positive  integer giving, \emph{e.g.},  a ``$p$-wave superconductor'' if $n = 1$ or ``$f$-wave superconductor'' if $n = 3$.  The compound Sr$_2$RuO$_4$ with $T_{\rm c} \approx 1.5$~K is thought to be a spin-triplet $p$-wave superconductor.\cite{Maeno2001}  All known superconductors contain Cooper pairs of one type or the other, and the type of Cooper pair in the new FeAs-based superconductors is important to determine.  

\begin{figure}
\includegraphics[width=3.3in]{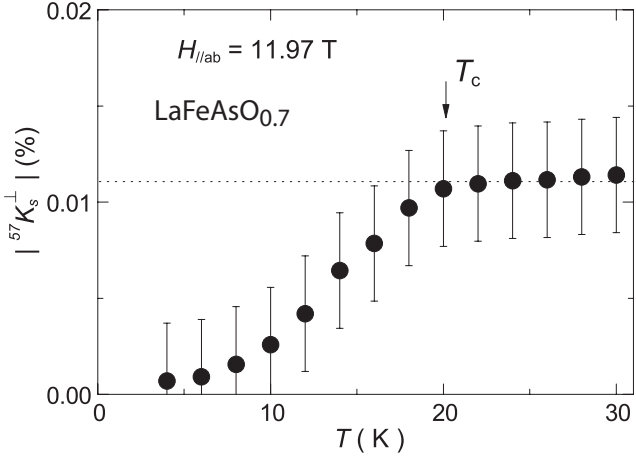}
\caption{Knight shift of $^{57}$Fe in LaFeAsO$_{0.7}$ versus temperature $T$.  The observation that  the Knight shift goes to zero for $(T \to 0)$ provides evidence that the spin state of a Cooper pair is a nonmagnetic singlet.  Reproduced with slight modification by permission of Ref.~\protect\onlinecite{Terasaki2009}.  Copyright (2009) by the Physical Society of Japan.}
\label{FigLaFeAsO_NMR_Fe}
\end{figure}

The most direct indication of the net spin of a Cooper pair is a measurement by NMR of the Knight shift of a  nucleus in a superconducting material.  As discussed previously in Sec.~\ref{SecChiSpinNMR}, the Knight shift is the fractional shift of the NMR resonance frequency of an embedded nucleus arising from the polarization of the conduction electron spin magnetization induced by the applied magnetic field.  In a spin singlet superconductor, the Knight shift $K$ goes to zero at $T = 0$ because the relatively small NMR applied magnetic field cannot polarize the nonmagnetic Cooper pairs, whereas in a spin triplet superconductor it does not.\cite{Maeno2001}  $^{57}$Fe NMR Knight shift measurements in LaFeAsO$_{0.7}$ ($T_{\rm c} = 28$~K) showed that $^{57}K$ decreased strongly below $T_{\rm c}$ and that $^{57}K(T\to 0) \approx 0$ (Fig.~\ref{FigLaFeAsO_NMR_Fe}), indicating that the net spin of a Cooper pair is zero.\cite{Terasaki2009}  The same conclusion was reached from $^{75}$As Knight shift measurements for $H\parallel ab$ and $H\parallel c$ on a single crystal of ${\rm Ba(Fe_{0.9}Co_{0.1})_2As_2}$ ($T_{\rm c} = 22$~K),\cite{Ning2008} on PrFeAsO$_{0.89}$F$_{0.11}$ (Ref.~\onlinecite{Matano2008}) and LaFeAsO$_{0.89}$F$_{0.11}$ ($T_{\rm c} = 28$~K),\cite{Kawabata2008} and from $^{57}$Fe Knight shift measurements on Ba$_{0.6}$K$_{0.4}$Fe$_2$As$_2$ ($T_{\rm c} = 38$~K).\cite{Yashima2009}

\begin{figure}
\includegraphics [width=3.3in]{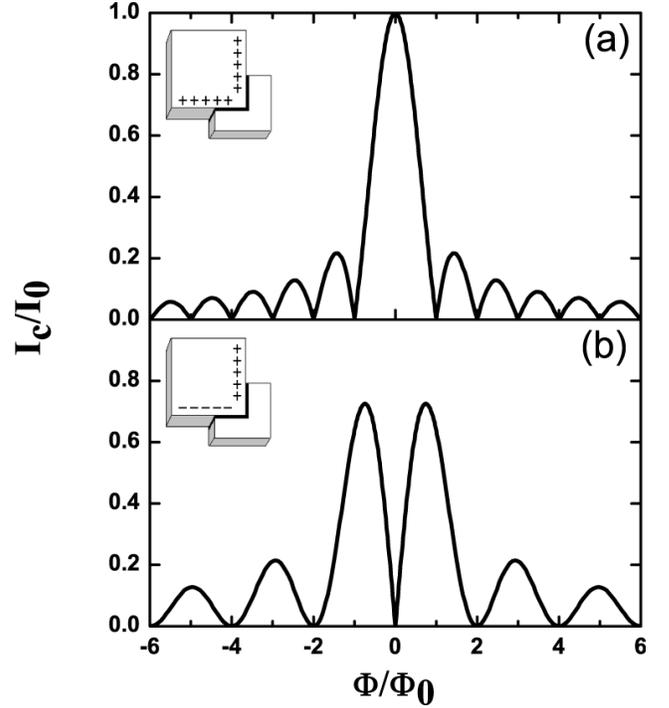}
\caption{Josephson critical current $I_{\rm c}$ versus magnetic flux threading the junction $\Phi$ for (a) an $s$-wave superconductor with zero phase shift between the two edges as shown and (b) a superconductor with a $\pi$~rad phase shift between the two edges.\cite{Zhou2008}  Reproduced by permission from Ref.~\onlinecite{Zhou2008}.}
\label{Ba(FeCo)2As2_corner_jcnA}
\end{figure}

\begin{figure}
\includegraphics [width=3.3in]{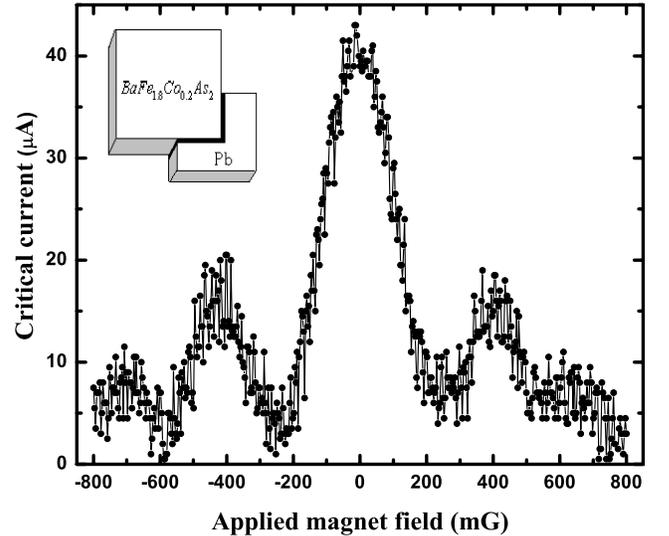}
\caption{Josephson critical current versus applied magnetic field threading a corner SNS tunneling junction $s$-wave superconductor Pb ($T_{\rm c} = 7$~K)/Au/BaFe$_{1.8}$Co$_{0.2}$As$_2$ ($T_{\rm c} = 22$~K) at a temperature of 1.2~K\@.\cite{Zhou2008}  Reproduced by permission from Ref.~\onlinecite{Zhou2008}.}
\label{Ba(FeCo)2As2_corner_jcnB}
\end{figure}

\subsection{\label{SecGapsetc} Superconducting Energy Gaps $2\Delta$ and Symmetry of the Superconducting Order Parameter in Momentum Space}

\subsubsection{Phase-Sensitive and Tunneling Measurements}

Since the net Cooper pair spin in the FeAs-based superconductors appears to be zero, the Cooper pair orbital angular momentum should correspond to $s$-wave or $d$-wave pairing as discussed above.  To determine which case is correct, experiments measuring the relative phase of the superconducting order parameter on adjacent edges of a crystal via quantum mechanical tunneling of Cooper pairs (``Josephson tunneling'') between an $s$-wave superconductor and the superconductor under study are directly relevant.  Figure~\ref{Ba(FeCo)2As2_corner_jcnA} shows the expected results for the critical current ``diffraction pattern'' for a corner junction between an $s$-wave superconductor and (a) another $s$-wave superconductor and (b) a $d_{x^2-y^2}$-wave superconductor.\cite{Zhou2008}  For the layered cuprate high $T_{\rm c}$ superconductors, such measurements demonstrated that the superconducting state is a $d_{x^2 - y^2}$-wave state.\cite{Van_Harlingen1995}  Such a ``corner junction'' tunneling measurement at 1.2~K between a 300 nm thick layer of superconducting Pb ($T_{\rm c} = 7$~K) and the $a$-$b$ plane of a superconducting single crystal of BaFe$_{1.8}$Co$_{0.2}$As$_2$ ($T_{\rm c} = 22$~K) is shown in Fig.~\ref{Ba(FeCo)2As2_corner_jcnB}.\cite{Zhou2008}  The Pb and sample were separated by 40 nm of Au, forming a superconductor-normal metal-superconductor (SNS) junction.  The result is consistent with an $s$-wave or $d_{xy}$ symmetry of the order parameter in this compound.  

Josephson tunneling experiments along the $c$-axis of Ba$_{1-x}$K$_x$Fe$_2$As$_2$ crystals ($T_{\rm c} = 26$--29~K) from Pb point contacts and Pb:In planar contacts ruled out pure $p$-wave and $d$-wave superconducting pairing symmetries, but were consistent with $s$-wave symmetry (including the currently favored $s^\pm$ type\cite{mazin2008}).  These measurements also confirmed that the tunneling entities were Cooper pairs with a net charge magnitude of $2e$, where $e$ is the magnitude of the charge on an electron.  $^{75}$As Knight shift measurements in BaFe$_{1.8}$Co$_{0.2}$Fe$_2$As$_2$ crystals below $T_{\rm c} = 22$~K were consistent with an $s$-wave pairing state but inconsistent with $p$-wave pairing.\cite{Ning2008}  

Observation of a persistent current in a superconducting loop consisting of a wire of the $s$-wave superconductor Nb ($T_{\rm c} = 9$~K) in series with a polycrystalline sample of NdFeAsO$_{0.88}$F$_{0.12}$ ($T_{\rm c} = 43$~K) indicated that NdFeAsO$_{0.88}$F$_{0.12}$ is a spin-singlet superconductor.\cite{CTChen2009}  Furthermore, these authors also observed jumps in the magnetic flux threading the superconducting loop by \emph{half-integral} multiples of the flux quantum $\Phi = hc/2e$, demonstrating the existence of a sign change in the superconducting order parameter between the surfaces of adjoining grains in  the NdFeAsO$_{0.88}$F$_{0.12}$ sample, and thus providing strong evidence for the $s^\pm$-wave pairing scenario.\cite{CTChen2009}

\subsubsection{\label{ARPESSC} ARPES Measurements}

\begin{figure}
\includegraphics [width=3.3in]{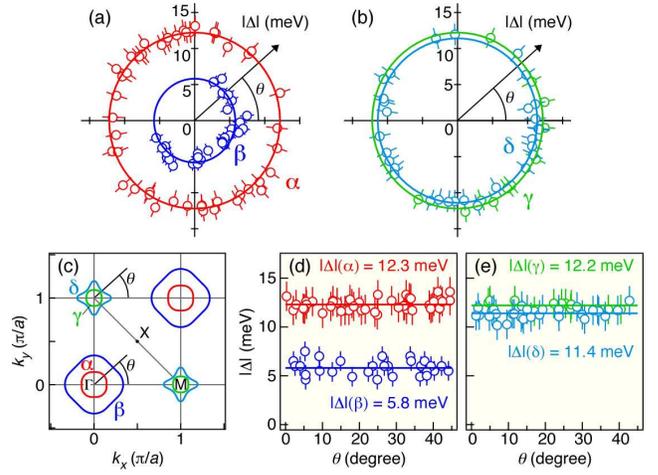}
\caption{(Color online) Superconducting gap magnitude $|\Delta|$ versus angle $\theta$ in the $k_x$-$k_y$ plane around (a) the $\alpha$ and $\beta$ hole pockets at the $\Gamma$ point of the Brillouin zone and (b) the $\delta$ and $\gamma$ electron pockets at the M point of the Brillouin zone that is shown in (c) (see also Figs.~\ref{semimetal_BS} and~\ref{FigBaKFe2As2_FS}) of single crystal ${\rm Ba_{0.6}K_{0.4}Fe_2As_2}$.\cite{Nakayama2009}  Panels (d) and (e) explicitly show $|\Delta|$ versus $\theta$ for the hole and electron pockets, respectively, illustrating that $|\Delta|$ is independent of $\theta$ for each Fermi surface sheet, and that $|\Delta|$ on the $\beta$ hole pocket is significantly smaller than on the other three Fermi surface sheets.  Reproduced with permission from Ref.~\onlinecite{Nakayama2009}.  Copyright (2009) by the European Physical Society.}
\label{FigNakayama_BaKFeAs_ARPES}
\end{figure}

%\clearpage
%\squeezetable
\begin{table}
\caption{\label{SCGapValues} Minimum ($\Delta_{\rm min}$) and maximum ($\Delta_{\rm max}$) superconducting gap values for $T \to 0$ determined from different measurements on single crystals at temperatures $T < T_{\rm c}$ for several compounds from representative measurements.  Note that the energy gap for the breakup of a Cooper pair, producing two quasiparticles, is $2\Delta$.  The measurement methods are angle resolved photoemission spectroscopy (ARPES), scanning tunneling spectroscopy (STS), optical reflectivity (optics), temperature $T$-dependent heat capacity $C_{\rm p}(T)$, and $T$-dependent magnetic penetration depth $\lambda(T)$.}
\begin{ruledtabular}
\begin{tabular}{l|ccccc}
Compound & $T_{\rm c}$ & $\Delta_{\rm min}$ & $\Delta_{\rm max}$  & method  & Ref. \\
  & (K) & (meV) & (meV) \\ \hline
${\rm SmFeAsO_{0.8}F_{0.2}}$  & 44 & 4.2(4) & 6.5(8) & $\lambda(T)$ & \onlinecite{Malone2009} \\
${\rm NdFeAsO_{0.9}F_{0.1}}$  & 53 & 15(2) & 15(2) & ARPES & \onlinecite{Kondo2008} \\
${\rm Ba_{0.68}K_{0.32}Fe_2As_2}$  & 38.5 & 3.5 & 12 & $C_{\rm p}(T)$  & \onlinecite{Popovich2010} \\
${\rm Ba_{0.6}K_{0.4}Fe_2As_2}$  & 37 & 5.8 & 12.3 & ARPES  & \onlinecite{Nakayama2009} \\
${\rm Ba_{0.6}K_{0.4}Fe_2As_2}$  & 37 & 6 & 12 & ARPES  & \onlinecite{Ding2008} \\
${\rm Ba_{0.6}K_{0.4}Fe_2As_2}$  & 37 & 5 & 13 & ARPES  & \onlinecite{Xu2010a} \\
${\rm Ba_{0.6}K_{0.4}Fe_2As_2}$  & 38 & 4 & 12 & ARPES  & \onlinecite{Zhang2010} \\
${\rm Ba(Fe_{0.95}Co_{0.05})_2As_2}$  & 18.5 & 1.1(2) & 4.0(5) & $\lambda(T)$  & \onlinecite{Luan2010} \\
${\rm Ba(Fe_{0.942}Co_{0.058})_2As_2}$  & 24 & 2.1 & 5.1 & $C_{\rm p}(T)$  & \onlinecite{Hardy2010b} \\
${\rm Ba(Fe_{0.935}Co_{0.065})_2As_2}$  & 24.5 & 3.3 & 9.7 & optics  & \onlinecite{Kim2010a} \\
${\rm Ba(Fe_{0.93}Co_{0.07})_2As_2}$  & 21 & 1.65 & 3.75 & $C_{\rm p}(T)$  & \onlinecite{Gofryk2010a} \\
${\rm Ba(Fe_{0.926}Co_{0.074})_2As_2}$  & 22.2 & 1.50 & 3.60 & $\lambda(T)$  & \onlinecite{Williams2009} \\
${\rm Ba(Fe_{0.925}Co_{0.075})_2As_2}$  & 21.4 & 1.8 & 4.1 & $C_{\rm p}(T)$  & \onlinecite{Hardy2010} \\
${\rm Ba(Fe_{0.92}Co_{0.08})_2As_2}$  & 22.5 & 1.92(6) & 6.8(9) & optics  & \onlinecite{Perucchi2010} \\
LiFeAs  & 17 & 3.1(3) & 3.1(3) & ARPES  & \onlinecite{Inosov2010b} \\
${\rm Fe_{1.05}Te_{0.85}Se_{0.15}}$  & 14 & 2.3 & 2.3 & STS  & \onlinecite{Kato2009} \\
${\rm Fe_{1.03}Te_{0.7}Se_{0.3}}$  & 13 & 4 & 4 & ARPES  & \onlinecite{Nakayama2009a} \\
${\rm FeTe_{0.55}Se_{0.45}}$  & 14 & 2.5 & 5.1 & optics  & \onlinecite{Homes2010} \\
${\rm FeTe_{0.5}Se_{0.5}}$  & 14.6 & 0.51(3) & 2.61(9) & $\lambda(T)$  & \onlinecite{Bendele2010} \\
\end{tabular}
\end{ruledtabular}
\end{table}

The temperature-dependent superconducting order parameter below $T_{\rm c}$ can usually be identified with an energy gap $2\Delta$ at the Fermi surface for single electron or hole (``quasiparticle'') excitations.  Breaking up a Cooper pair produces two quasiparticles, which costs an energy $2\Delta$.  An $s$-wave superconductor has no nodes (zeros) in $\Delta$ around the Fermi surface.  In the BCS theory for $s$-wave superconductivity, the relationship $2\Delta/k_{\rm B}T_{\rm c} = 3.53$ is obtained.  In the above $s^\pm$-type model, uniform $s$-wave order parameters develop on each of the electron and hole Fermi surface pockets (or sheets), but where they have opposite signs.\cite{mazin2008}   Doping changes the sizes of the Fermi surface sheets and in any case they need not have the same $N(E_{\rm F})$, so in general one does not expect the two gaps to have the same magnitude.  Measurements of various types have indeed broadly indicated the existence of two distinct and nearly uniform energy gap magnitudes on the Fermi surface sheets in the FeAs-type superconductors.  
Many angle-resolved photoelectron spectroscopy (ARPES) measurements of the \emph{magnitude} of the superconducting energy gap (not including the sign of the order parameter corresponding to a phase of $\pm \pi$ rad) have been carried out and are unanimous in their conclusion that there are no nodes in the energy gap in the observed Fermi surface sheets within the basal $k_x$-$k_y$ plane in the Brillouin zone, strongly supporting $s$-wave-like pairing.  A high-resolution ARPES study of ${\rm Ba_{0.6}K_{0.4}Fe_2As_2}$ ($T_{\rm c} = 37$~K) is shown in Fig.~\ref{FigNakayama_BaKFeAs_ARPES},\cite{Nakayama2009} where the superconducting gaps are found to be independent of angle around each of the four Fermi surface sheets (two electron and two hole pockets) within the plane of the Fe layers, with values of 5.8 and 12.3~meV on the hole pockets and 11.4 and 12.2 meV on the electron pockets of the Brillouin zone.  These gap values correspond to the ratios $2|\Delta|/k_{\rm B}T_{\rm c} = 3.6$, 7.7, 7.1, and 7.6.  Thus this compound is a ``multi-gap'' superconductor as previously observed in, \emph{e.g.} NbSe$_2$,\cite{Yakoya2001} MgB$_2$,\cite{Tsuda2003} and V$_3$Si,\cite{Kogan2009} which significantly influences the temperature dependences of a variety of other measurements in the superconducting state.  Evtushinsky \emph{et al.}\ compiled the values of $2|\Delta|/k_{\rm B}T_{\rm c}$ inferred from 26 different measurements on FeAs-based and FeSe$_{1-x}$ materials and found that they could be grouped into two ranges $2|\Delta|/k_{\rm B}T_{\rm c} = 2.5 \pm 1.5$ and $7 \pm 2$,\cite{Evtushinsky2009} which straddle the BCS $s$-wave value of 3.53.  Superconducting gap values at $T \to 0$ determined using ARPES and other measurements are listed in Table~\ref{SCGapValues}.\cite{Homes2010, Malone2009, Kondo2008, Popovich2010, Nakayama2009, Ding2008, Xu2010a, Zhang2010, Luan2010, Kim2010a, Gofryk2010a, Williams2009, Hardy2010, Perucchi2010, Inosov2010b, Kato2009, Nakayama2009a, Bendele2010}  From the table, the maximum gaps determined by optics measurements are generally larger than from other measurements on the same material.

From Fig.~\ref{FigNakayama_BaKFeAs_ARPES}(c), the inner hole $\alpha$ Fermi surface pocket in hole-doped ${\rm Ba_{0.6}K_{0.4}Fe_2As_2}$ clearly has better nesting with the $\gamma$ and $\delta$ electron Fermi surface sheets than does the outer hole $\beta$ Fermi surface pocket.  The degree of nesting is correlated with the superconducting gap $2\Delta$, since the $\alpha$ pocket has a larger gap than the $\beta$ pocket.  On the other hand, in electron-doped ${\rm BaFe_{1.85}Co_{0.15}As_2}$ the nesting is better between the $\beta$ hole pocket and the two electron pockets and also the superconducting gap is larger on the $\beta$ hole pocket than on the $\alpha$ hole pocket.\cite{Terashima2009b}  Together, these results strongly support the proposal that the high temperature superconductivity in these systems is due to an inter-orbital pairing mechanism between the hole and electron pockets arising from Fermi surface nesting between these pockets.

Additional ARPES measurements have indicated that the band structure near the Fermi energy of ${\rm BaFe_{1.8}Co_{0.2}As_2}$ is three-dimensional (3D) as shown previously in Fig.~\ref{FigBa(FeCo)2As2_ARPES}, where a strong dispersion of the hole pockets along $\Gamma$-Z is found.\cite{Vilmercati2009}  In turn, the degree of nesting between the hole and electron pockets thought necessary for superconducting pairing in the $s^\pm$ model would depend on $k_z$, which in turn suggests that the magnitude of the superconducting gap could strongly depend on $k_z$.  The dispersion of the superconducting  gap along $k_z$ has not yet been directly measured for this system.  However, the magnetic neutron scattering resonance in BaFe$_{1.9}$Ni$_{0.1}$As$_2$ single crystals ($T_{\rm c} = 20$~K) was found to disperse along $k_z$ (Ref.~\onlinecite{Chi2009}) as discussed in Sec.~\ref{ResonanceMode} below, which may be an indication that the superconducting gap also disperses along $k_z$ in this compound.\cite{Chi2009}  The ARPES results in Fig.~\ref{FigBa(FeCo)2As2_ARPES} indicate that strongly reduced dimensionality as in the layered cuprate superconductors is not required for high $T_{\rm c}$ superconductivity, at least for $T_{\rm c} \lesssim 56$~K\@.

Evidence for dispersion of the superconducting energy gap along $k_z$ has been obtained from two separate ARPES measurements of high quality crystals of ${\rm Ba_{0.6}K_{0.4}Fe_2As_2}$ with $T_{\rm c} = 37$--38~K.\cite{Xu2010a, Zhang2010}  Both studies resolved a third hole sheet at the center of the Brillouin zone that had not been previously observed.  Both studies are in general agreement with each other regarding the dispersion of the superconducting gap.  The gap on the electron pocket studied does does not disperse with $k_z$, whereas the gap on one of the hole pockets shows rather strong dispersion of order 50\%.\cite{Xu2010a, Zhang2010} 

Graser et al.\ have reported a theoretical study of the influence of three-dimensionality in the electronic structure on the superconducting pairing interaction in undoped and doped 122-type ${\rm BaFe_2As_2}$ and have found a significant influence.\cite{Graser2010}

\subsubsection{\label{SecThermCond} Thermal Conductivity Measurements}

\begin{figure}
\includegraphics [width=3.3in]{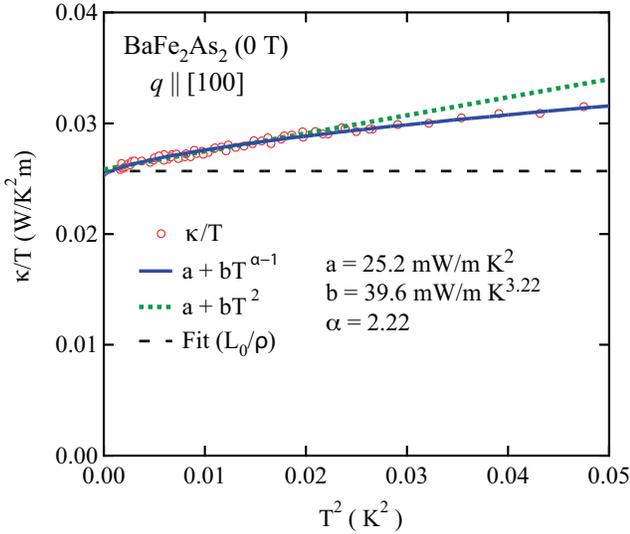}
\caption{(Color online)   Thermal conductivity $\kappa$ divided by temperature $T$ versus $T^2$ for an undoped crystal of ${\rm BaFe_2As_2}$ in zero applied magnetic field, with the heat flow in the $a$-$b$~plane.\cite{Kurita2009a}  The blue solid curve is a fit to the data (open red circles) by the power law shown over the whole temperature range with the listed values of fitting parameters.  The dotted green line assumes an exponent $\alpha = 3$ and the fit is for $T < 0.1$~K\@.  The horizontal dashed line is the value calculated from the Wiedemann-Franz law using the measured zero temperature electrical resistivity, which shows that this law holds in the present case at $T = 0$ where the phonon thermal conductivity is zero.  Reprinted with permission from Ref.~\onlinecite{Kurita2009a}.  Copyright (2009) by the American Physical Society.}
\label{KuritaFig2}
\end{figure}

\begin{figure}
\includegraphics [width=3.in,viewport=-30 00 300 600,clip]{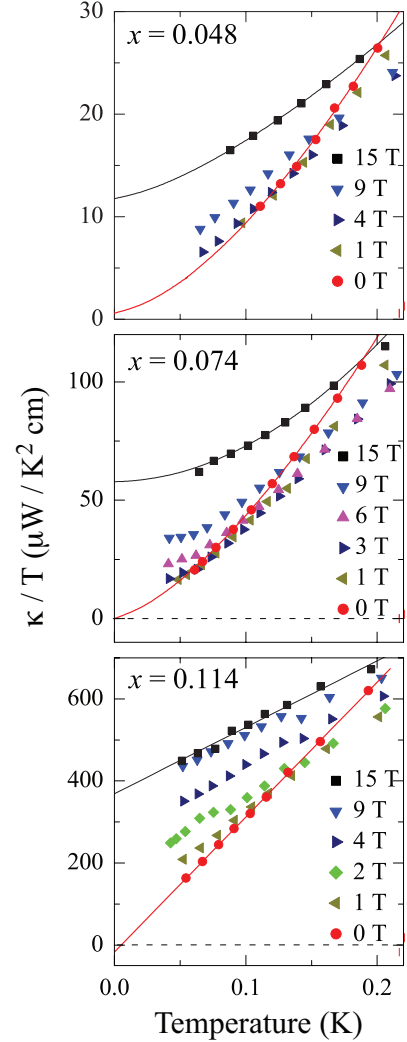}
\caption{(Color online)   Thermal conductivity $\kappa$ divided by temperature $T$ versus $T$ for three crystals of Ba(Fe$_{1-x}$Co$_x)_2$As$_2$ in various applied magnetic fields $H$ as indicated, with the heat flow in the $a$-$b$~plane.\cite{Tanatar2010}  The curves are fits to the data for $H = 0$ (red) and $H = 15$~T (black) by Eq.~(\ref{Eqkappa}).  To convert the vertical scale to  units of W/K$^2$m to compare with Fig.~\ref{KuritaFig2}, divide by 10\,000.  Reprinted with permission from Ref.~\onlinecite{Tanatar2010}.  Copyright (2010) by the American Physical Society.}
\label{TanatarFig2}
\end{figure}

\begin{figure}
\includegraphics [width=3.in]{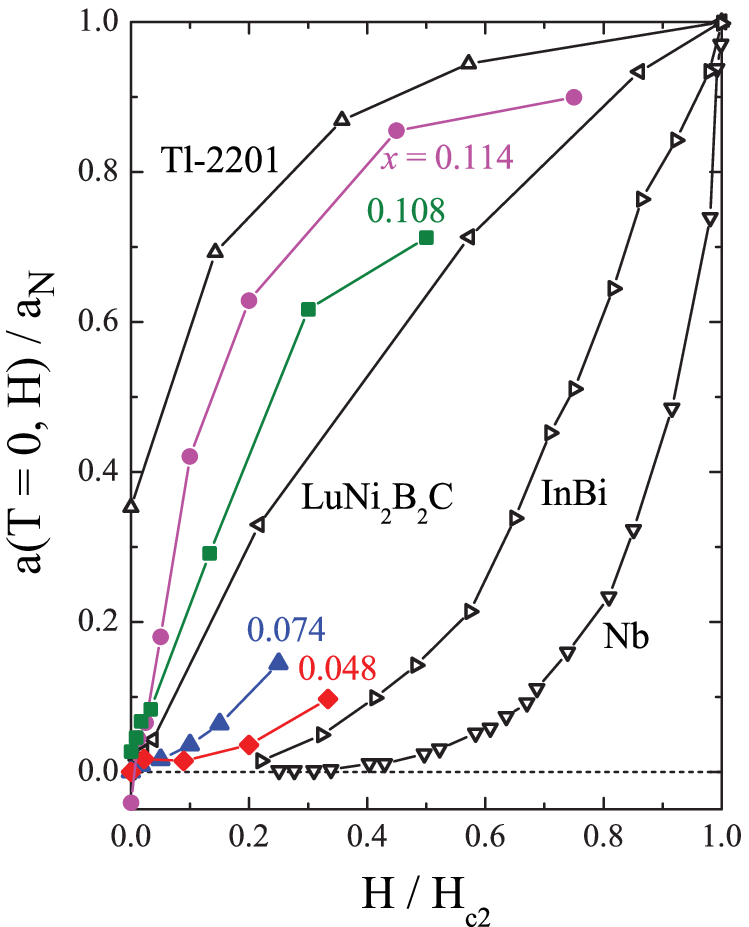}
\caption{(Color online)   Thermal conductivity $\kappa$ divided by temperature $T$ at $T = 0$, $a(T = 0,H)$ [see Eq.~(\ref{Eqkappa})], divided by the normal state value $a_{\rm N}$, versus magnetic field $H$ divided by the superconducting upper critical magnetic field $H_{\rm c2}$ for four crystals of Ba(Fe$_{1-x}$Co$_x)_2$As$_2$ (see Fig.~\ref{TanatarFig2}).\cite{Tanatar2010}  Data for additional materials are included for comparison as described in the text.  The lines are guides to the eye.  Reprinted with slight modifications with permission from Ref.~\onlinecite{Tanatar2010}.  Copyright (2010) by the American Physical Society.}
\label{TanatarFig3}
\end{figure}

Thermal conductivity $\kappa$ measurements have been important to help establish the superconducting gap structure of the Fe-based superconductors.  As a base line, the in-plane thermal conductivity data for an undoped and nonsuperconducting single crystal of BaFe$_2$As$_2$ from measurements down to 40~mK by Kurita and coworkers are shown in Fig.~\ref{KuritaFig2},\cite{Kurita2009a} together with power law fits by
\be
\frac{\kappa}{T} = a + bT^{\alpha-1},
\label{Eqkappa}
\ee
where the first term on the right is from conduction electrons and the second term is from phonons and magnons.  Note that these data were taken in the orthorhombic, SDW phase.  From Table~\ref{NeutDispersionData}, the SDW phase has an anisotropy gap $\sim 110$~K in the spin wave excitation spectrum, so the contribution to $\kappa$ from magnons is negligible at these low temperatures.  One nominally expects the phonon exponent to be $\alpha = 3$ if the phonon scattering mean free path is limited by the sample boundaries, but various mechanisms can lower this exponent.\cite{Kurita2009a}  From Fig.~\ref{KuritaFig2}, for reference, the electronic contribution in the $ab$~plane for $T\to 0$ is 
\be
\lim_{T\to 0}\frac{\kappa}{T} = 252~\mu{\rm W/K^2~cm}.\hspace{0.2in}({\rm BaFe_2As_2})
\label{EqkTBa122}
\ee
The Wiedemann-Franz law states that the thermal and electrical conductivities of the conduction electrons in a metal are proportional to each other, according to $\lim_{T\to 0}[\kappa(T)/T] = L_0/\rho(T= 0)$, which is the parameter $a$ in Eq.~(\ref{Eqkappa}), where the Lorenz number is $L_0= \pi^2k_{\rm B}^2/(3e^2) = 2.45 \times 10^{-8} ~{\rm W}\,\Omega/{\rm K}^2$  and $\rho(T=0)$ is the measured electrical resistivity.\cite{kittel1966, Blatt1968}  Using the measured $\rho(T=0) = 95~\mu\Omega$\,cm in the $a$-$b$ plane, the Wiedemann-Franz law is found to be obeyed for BaFe$_2$As$_2$,\cite{Kurita2009a} as shown by the horizontal dashed line in Fig.~\ref{KuritaFig2}.  The excess thermal conductivity above the horizontal dashed line is the contribution from phonons, the second term in Eq.~(\ref{Eqkappa}).

Due to the energy gap $2\Delta$ for single quasiparticle excitations in superconductors, the contribution of the conduction electrons to the thermal conductivity in  superconductors comes only from quasiparticles that are thermally excited above the gap.  This means that for a fully superconducting sample at very low temperatures, one has $a = 0$ in Eq.~(\ref{Eqkappa}). In general, the coefficient $a=a(T,H)$ is both temperature- and field-dependent in the superconducting state.  The power of low-temperature thermal conductivity measurements, then, is that they are very sensitive to the presence of thermally excited quasiparticles as opposed to the superconducting condensate. For example, an $s$-wave superconductor that is not fully superconducting will show a value of $a(T\to0, H=0)$ in Eq.~(\ref{Eqkappa}) that is proportional to the non-superconducting volume fraction.  Alternatively, for a fully superconducting sample, the magnitude at $T = 0$ and temperature and magnetic field dependences of $a$ can give important information about the presence of nodes in the superconducting order parameter in momentum space.

Thermal conductivity measurements were carried out on four single crystals from the underdoped to the overdoped regime of Ba(Fe$_{1-x}$Co$_x)_2$As$_2$ (see the phase diagram in Fig.~\ref{FigBaKFe2As2_phase_diag}) by Tanatar et al.\ as shown in Fig.~\ref{TanatarFig2}.\cite{Tanatar2010}  Fits to the data by Eq.~(\ref{Eqkappa}) are shown as the solid curves for $H = 0$ and 15~T\@.  For the $H = 0$ fits, the values of $a$ are equal to zero to within the experimental uncertainty of $\pm 5~\mu {\rm W/K}^2$\,cm [compare with Eq.~(\ref{EqkTBa122})], demonstrating the lack of (line) nodes in the superconducting order parameter in the $a$-$b$~plane over the full range of doping from underdoped to overdoped.  The same result was obtained by Dong et al.\ for the even more overdoped composition $x = 0.135$ with $T_{\rm c} = 8.1$~K.\cite{Dong2010}  Machida et al.\ obtained for an optimally doped crystal of Ba(Fe$_{0.93}$Co$_{0.07})_2$As$_2$ the nonzero value $\kappa/T = 12~\mu$W/(K$^2$~cm) for $T\to0$,\cite{Machida2009} but this value is still a factor of 21 smaller than in Eq.~(\ref{EqkTBa122}) for pure ${\rm BaFe_2As_2}$.   These results rule out $d$-wave superconductivity over the whole composition range\cite{Tanatar2010} and are consistent with the $s^\pm$ scenario for the pairing symmetry.  The authors ruled out symmetry-imposed point nodes in the superconducting order parameter unless they are located \emph{on} the $k_z$ axis with respect to at least one of the electron or hole Fermi surfaces.\cite{Tanatar2010}  However, as far as is known, there are no Fermi surfaces in this system with sections that could potentially satisfy this condition (see, e.g., Figs.~\ref{FigFS} and~\ref{FigBa(FeCo)2As2_ARPES} above).  Theoretically, the interaction of a static SDW in the underdoped regime with the superconducting state is not expected to lead to nodes in the superconducting gap if the pairing symmetry is $s^\pm$, but would if the pairing symmetry is the conventional $s^{++}$-type.\cite{Parker2009} From comparison with the experimental results, this treatment thus supports the $s^\pm$ pairing state.

The field dependence of the electronic thermal conductivity in a Type-II superconductor is also a very useful indicator for the symmetry of the superconducting order parameter.  A popular way of presenting such data is to plot the ratio of the extrapolated $a_0(H) \equiv a(H,T\to0)$ to the normal state electronic $a_{\rm N}$ from Eq.~(\ref{Eqkappa}) versus the ratio $H/H_{\rm c2}$.  Such a plot for the three samples in Fig.~\ref{TanatarFig2} is shown in Fig.~\ref{TanatarFig3}, together with corresponding data for the clean and dirty $s$-wave superconductors Nb and InBi, respectively, for a borocarbide superconductor ${\rm LuNi_2B_2C}$, and for a $d$-wave high $T_{\rm c}$ cuprate Tl2201.\cite{Tanatar2010}  The data for the underdoped and optimally doped samples with $x = 0.048$ and~0.074 are similar to those for the $s$-wave superconductors, but the data for the overdoped sample with $x = 0.114$ indicate a substantial reduction in the minimum gap in the system.  The authors estimate that the ratio of the minimum to the maximum gaps for this composition to be about 1/6.\cite{Tanatar2010}  This variation in the gap magnitude may arise from the corrugation in the hole Fermi surfaces as illustrated above in Figs.~\ref{FigFS} and~\ref{FigBa(FeCo)2As2_ARPES}, and the resultant variations in the degree of nesting with the electron Fermi surfaces in the antiferromagnetic spin fluctuation scenario for the pairing mechanism.

Thermal conductivity measurements on crystals of Ba$_{0.72}$K$_{0.28}$Fe$_2$As$_2$ ($T_{\rm c} = 30$~K) at temperatures down to 50~mK ruled out the existence of line and in-plane point gap nodes, thus ruling out an in-plane $d$-wave pairing symmetry.\cite{Luo2009}  However, a significant magnetic field $H$ dependence of the thermal conductivity for $H \ll H_{\rm c2}$ and $T \to 0$ indicated that the minimum gap is very small on part of the Fermi surface,\cite{Luo2009} consistent with the above suggestion of a corrugated dependence of the superconducting gap $|\Delta|$ on $k_z$.  The zero-field measurements followed the relation in Eq.~(\ref{Eqkappa}), where the residual electronic coefficient (expected to be zero for a gap without in-plane nodes) was the almost negligible value $a = 5$--7~$\mu$W/(cm~K$^2)$, which is a factor of 50 smaller than for undoped ${\rm BaFe_2As_2}$ in Fig.~\ref{KuritaFig2}. The phonon scattering exponent was $\alpha = 2.5$--2.65.\cite{Luo2009}

Thermal conductivity measurements down to 1.2~K indicated that single crystal Ba$_{1-x}$K$_{x}$Fe$_2$As$_2$ (with unspecified $x$) is a clean-limit superconductor, where the ratio of the mean free path to the coherence length at 1.2~K is $\ell/\xi \sim 60$.\cite{Checkelsky2008}  

Similar measurements were carried out for single crystals of BaFe$_{1.9}$Ni$_{0.1}$As$_2$ ($T_{\rm c} = 20.3$~K) at temperatures down to 60~mK.\cite{Ding2009}  These authors found the parameter $a = (-3\pm 2)~\mu$W/(cm~K$^2$) and $\alpha = 2.02(1)$, the null result for $a$ again indicating nodeless superconductivity parallel to the $a$-$b$~plane.  The magnetic field dependence of $a$ was consistent with $s$-wave or $s^\pm$-wave superconductivity.\cite{Ding2009}  

Thermal conductivity measurements were carried out in the $ab$-plane on a single crystal of FeSe with onset $T_{\rm c} = 8.8$~K and transition width $\sim 3$~K at temperatures down to 0.12~K and in magnetic fields up to 14.5~T.\cite{Dong2009}  In zero field, the data were fitted from 0.15 to 0.7~K by Eq.~(\ref{Eqkappa}) with $a = 16(2)~\mu$W/(cm~K$^2)$ and $\alpha = 2.47$.  The (residual) value of $a$ is too small to correspond to a superconducting gap with nodes, and suggests that part of the sample was not superconducting.  Together with the field dependence of $a$, the authors concluded that FeSe is a multi-gap nodeless superconductor.\cite{Dong2009}

\begin{figure}
\includegraphics [width=3.3in]{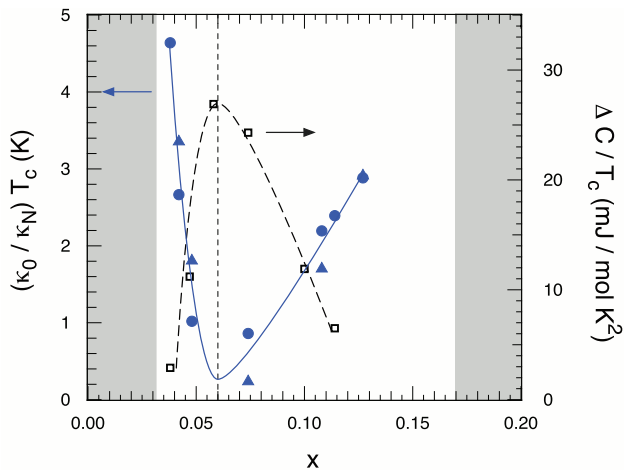}
\caption{(Color online)   Ratio of the thermal conductivity in the superconducting state parallel to the $c$-axis to that in the normal state for $T\to 0$, $\kappa_0/\kappa_{\rm N}$, multiplied by $T_{\rm c}$, versus composition $x$ for Ba(Fe$_{1-x}$Co$_x)_2$As$_2$ crystals  (left-hand scale).\cite{Reid2010}  Also shown is the heat capacity jump $\Delta C$ at $T_{\rm c}$, divided by $T_{\rm c}$, versus $x$ (right-hand scale).\cite{Budko2009}  The white area is the superconducting composition region.  Reprinted with permission from Ref.~\onlinecite{Reid2010}. }
\label{Reid_Kappa0}
\end{figure}

Reid and coworkers reported thermal conductivity measurements both parallel to the $ab$-plane and to the $c$-axis of Ba(Fe$_{1-x}$Co$_x)_2$As$_2$ crystals.\cite{Reid2010}  The $ab$-plane data reproduce the above measurements, showing no evidence for nodes in the superconducting order parameter at zero field.  Along the $c$-axis, the thermal conductivity $\kappa(T\to 0)/T$ was close to zero for the optimum doping concentration $x = 0.06$, but increased rapidly with increasing deviation from this concentration as shown in Fig.~\ref{Reid_Kappa0}.  The authors attributed this increase to the development of nodes in the superconducting order parameter along the $c$-axis as the composition deviates from the optimum one for superconductivity.  Interestingly, the increase in the  $c$-axis thermal conductivity tracks the decrease in the observed heat capacity jump $\Delta C/T_{\rm c}$ at $T_{\rm c}$,\cite{Budko2009} as shown on the right scale in Fig.~\ref{Reid_Kappa0}.  Kogan has interpreted the latter as arising from strong pairbreaking that reduces both $T_{\rm c}$ and $\Delta C/T|_{T_{\rm c}}$ at the same time.\cite{Kogan2009a, Kogan2010}  

An alternative explanation of the two sets of data in Fig.~\ref{Reid_Kappa0} is in terms of a reduction of superconducting volume fraction (and increase in the normal fraction) with decreasing $T_{\rm c}$ upon underdoping or overdoping as discussed more fully below in Secs.~\ref{Sec_SC_Cp} and~\ref{SecInhomo}.  But then one could argue that if the normal fraction of the crystals increased with decreasing $T_{\rm c}$, both the $ab$-plane and $c$-axis thermal conductivities would show $\kappa(T\to 0)/T > 0$, and therefore this explanation cannot be correct because in the $ab$-plane, $\kappa(T\to 0)/T = 0$ for all superconducting values of $x$.  However, one must consider how the normal component might be distributed in the crystals.  Suppose the normal component formed thin plates or rods parallel to the $c$-axis in a superconducting matrix.  Then for the $ab$-plane thermal conductivity, the superconducting regions would encapsulate the normal regions in the direction of thermal transport, in which case the superconducting regions would dominate the thermal transport at low temperatures, giving $\kappa(T\to 0)/T = 0$.  However, for $c$-axis thermal transport, the superconducting and normal regions would be in parallel, so the normal regions would dominate, giving $\kappa(T\to 0)/T > 0$.

\subsubsection{\label{ResonanceMode} Neutron Spin Resonance Mode}

Inelastic neutron scattering measurements have constrained the symmetry of the superconducting order parameter in the doped (Ca,Sr,Ba)Fe$_2$As$_2$ and 11-type compounds.\cite{Osborn2009}  Such measurements observe a magnetic neutron scattering ``spin resonance mode'' in the superconducting state that has the same in-plane wavevector $\left(\frac{1}{2},\frac{1}{2}\right)$~r.l.u.\ as the  magnetic neutron scattering in the normal state above $T_{\rm c}$ arising from nesting between the hole and electron Fermi surface pockets.  For the 122- and 1111-type compounds, this wave vector is the same as the in-plane component of the long-range SDW stripe-$b$ ordering in the undoped parent compounds.  This resonance in the superconucting state is believed to arise from scattering of electrons between different regions of Fermi surface with opposite signs of the superconducting order parameter, thus supporting the $s^\pm$ pairing model.\cite{Korshunov2008, Maier2008}  In the Fe(Se,Te) materials, the wave vector $\left(\frac{1}{2},\frac{1}{2}\right)$~r.l.u.\  of the resonance mode is still the spanning vector between the electron and hole Fermi surface pockets, even though the static ordering wave vector  $\left(\frac{1}{2},0\right)$~r.l.u.\ in Fe$_{1+y}$Te is different.  This commonality between the 122-, 1111- and 11-type compounds strongly suggests that the same superconducting mechanism is at play in all Fe-based superconductors considered in this review.  As discussed previously in this review, Utfeld et al.\ have calculated that the noninteracting $\chi({\bf Q})$ for optimally doped ${\rm Ba(Fe_{0.93}Co_{0.07})_2As_2}$ has a pronounced peak near the nesting wave vector, in spite of significant 3D dispersion of one of the hole Fermi pockets,\cite{Utfeld2010} thus further supporting the $s^\pm$ superconducting pairing model.  Maier and coworkers have suggested that additional neutron scattering experiments be done to search for resonances in other parts of the Brillouin zone to distinguish between different gap structures.\cite{Maier2009}  A summary of literature data for the resonance mode energy in different Fe-based superconductors is given in Table~\ref{ResonanceModeData}.\cite{Inosov2010, Babkevich2010, Christianson2009, Pratt2010, Lumsden2009, Li2010, Argyriou2009, Mook2009, Wang2010, Li2009, Christianson2008, Qiu2009, Bao2010, Shamoto2010}

\begin{figure}
\includegraphics [width=3.5in]{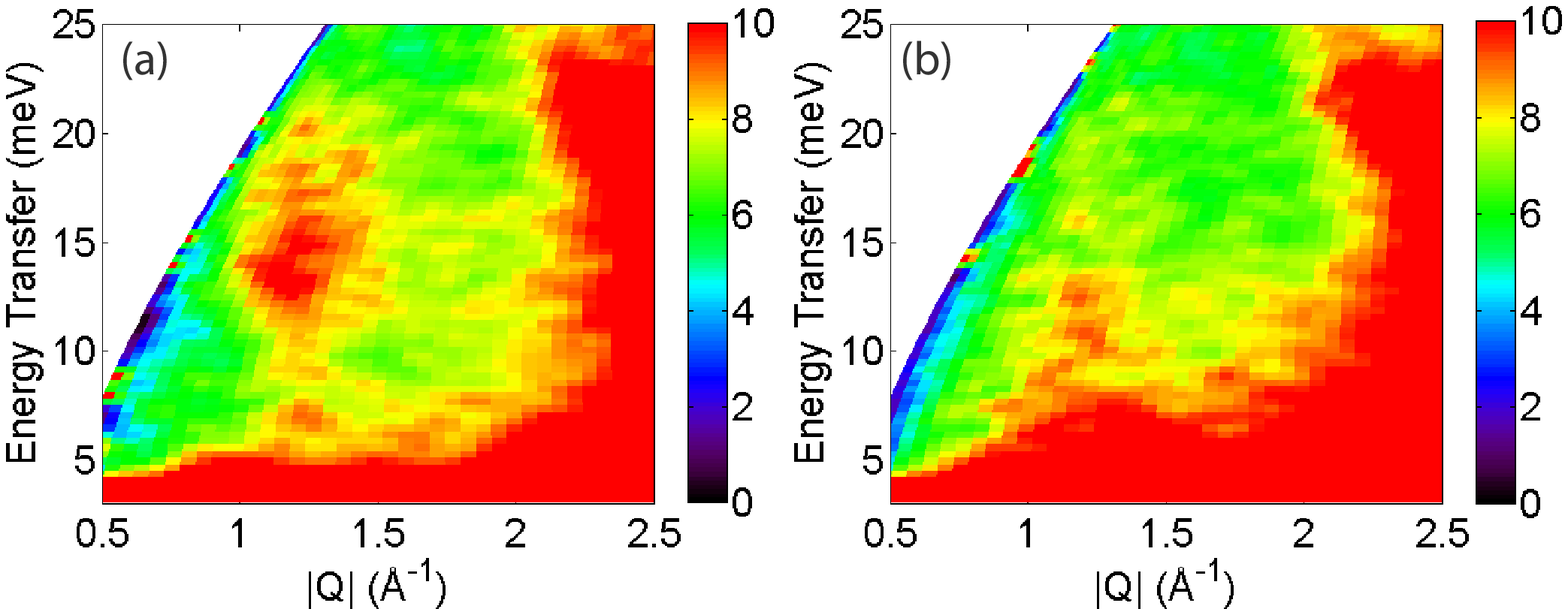}\vspace{0.25in}
\includegraphics [width=2.in]{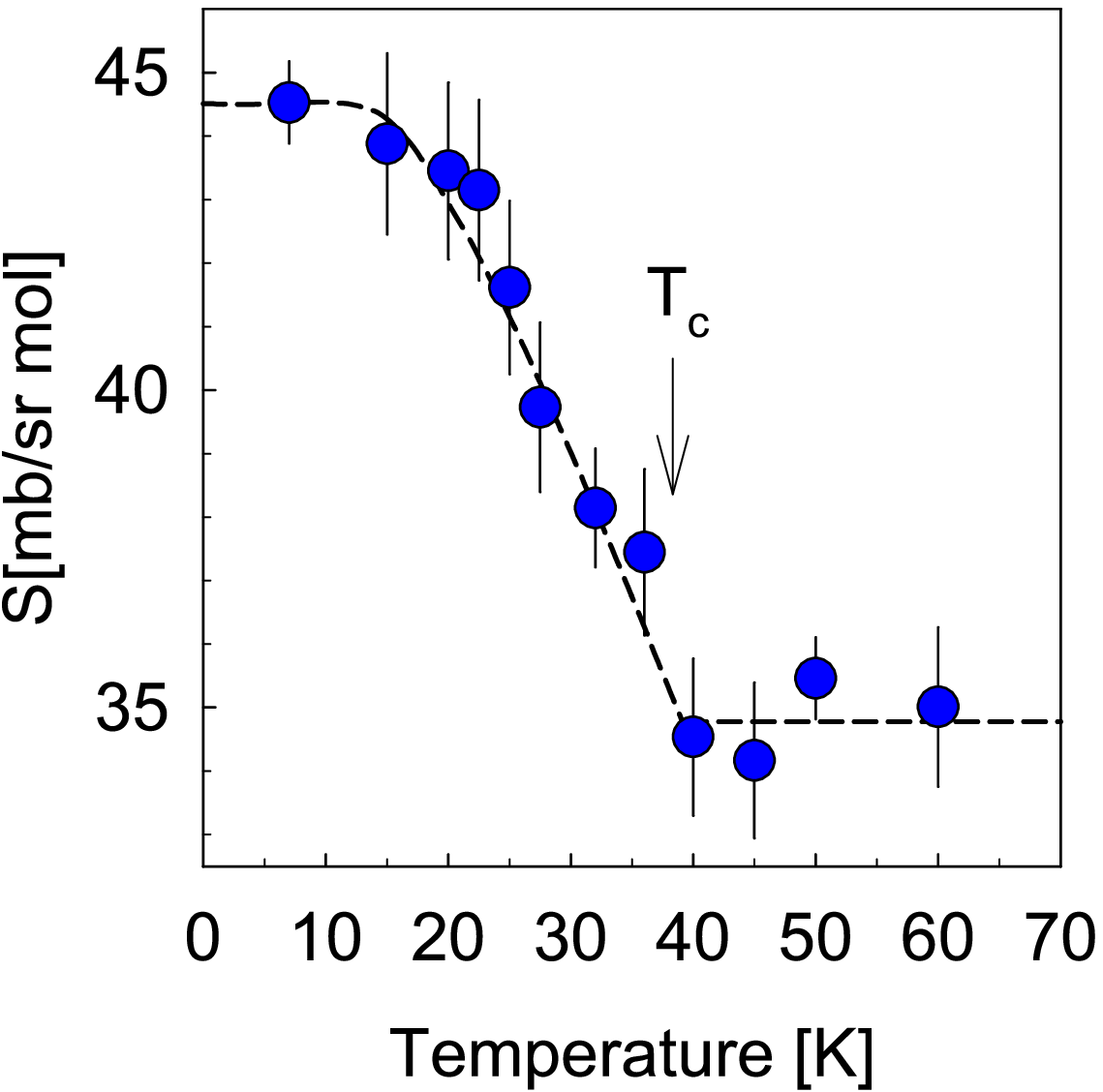}
\caption{(Color online) Top two panels: Contours of neutron scattering energy transfer $S$ versus momentum transfer $|\mathbf{Q}|$ for polycrystalline Ba$_{0.6}$K$_{0.4}$Fe$_2$As$_2$.  The intense scattering (red) at the bottom of parts (a,b) comes from elastic nuclear scattering and the intense scattering near the right edges of (a,b) comes from inelastic nuclear scattering.  The data are taken at temperatures of 7~K (a) and 50~K (b) at incident neutron energies of 60~meV\@.  The color scales on the right sides of (a,b) are in units of mb/(sr~meV~mol).  The increase of scattering centered at energy $\sim 15$~meV and $|\mathbf{Q}| \sim 1.15$\AA$^{-1}$ at 7~K compared to 50~K is due to a ``neutron spin resonance mode'' associated with the superconductivity that develops below $T_{\rm c} = 38$~K\@.  Bottom panel: Neutron scattering intensity of the resonance mode at an incident neutron energy of 60~meV as in (a,b), integrated over the $|\mathbf{Q}|$ range from 1.0 to 1.3 \AA$^{-1}$ and energy range from 12.5 to 17.5 meV, versus temperature. Reproduced by permission from Ref.~\onlinecite{Christianson2008} and from Macmillan Publishers Ltd: Ref.~\onlinecite{Christianson2008}, Copyright (2008).}
\label{FigBaKFe2As2ResMode}
\end{figure}

\begin{figure}
\includegraphics [width=3.3in]{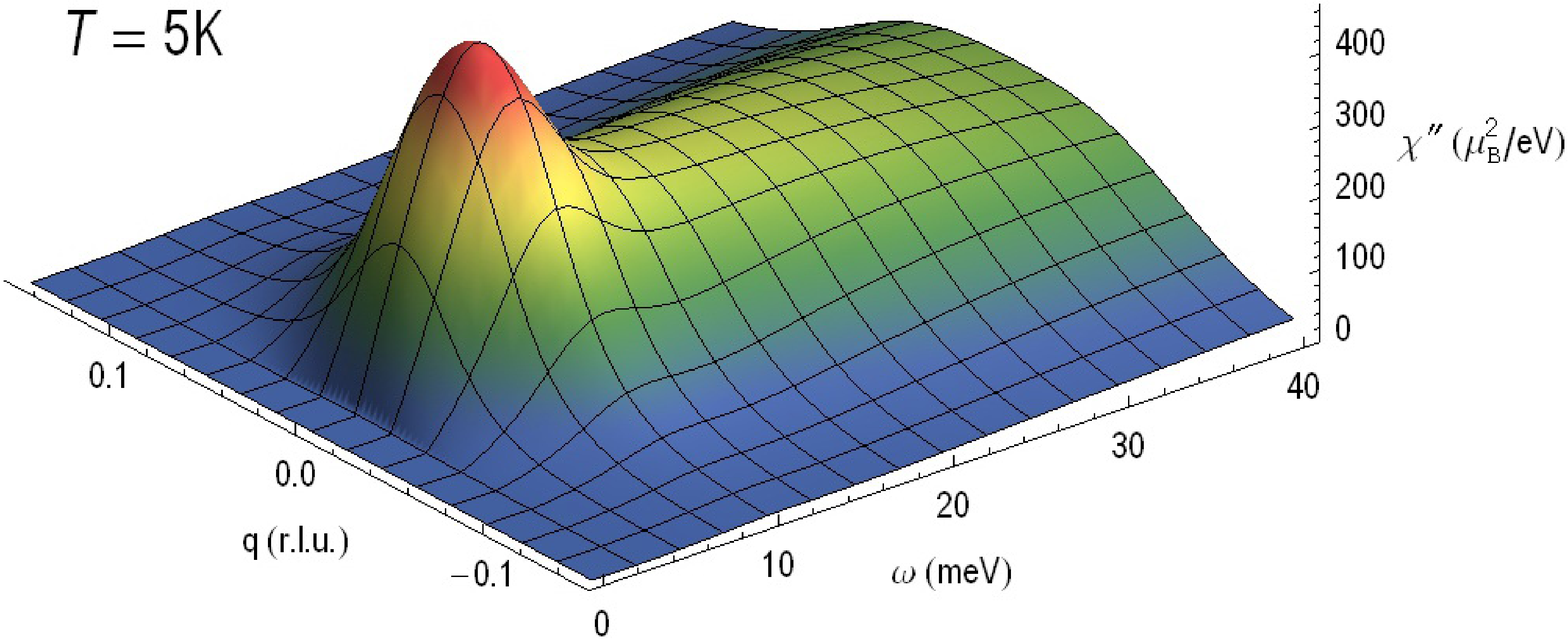}\vspace{0.2in}
\includegraphics [width=3.3in]{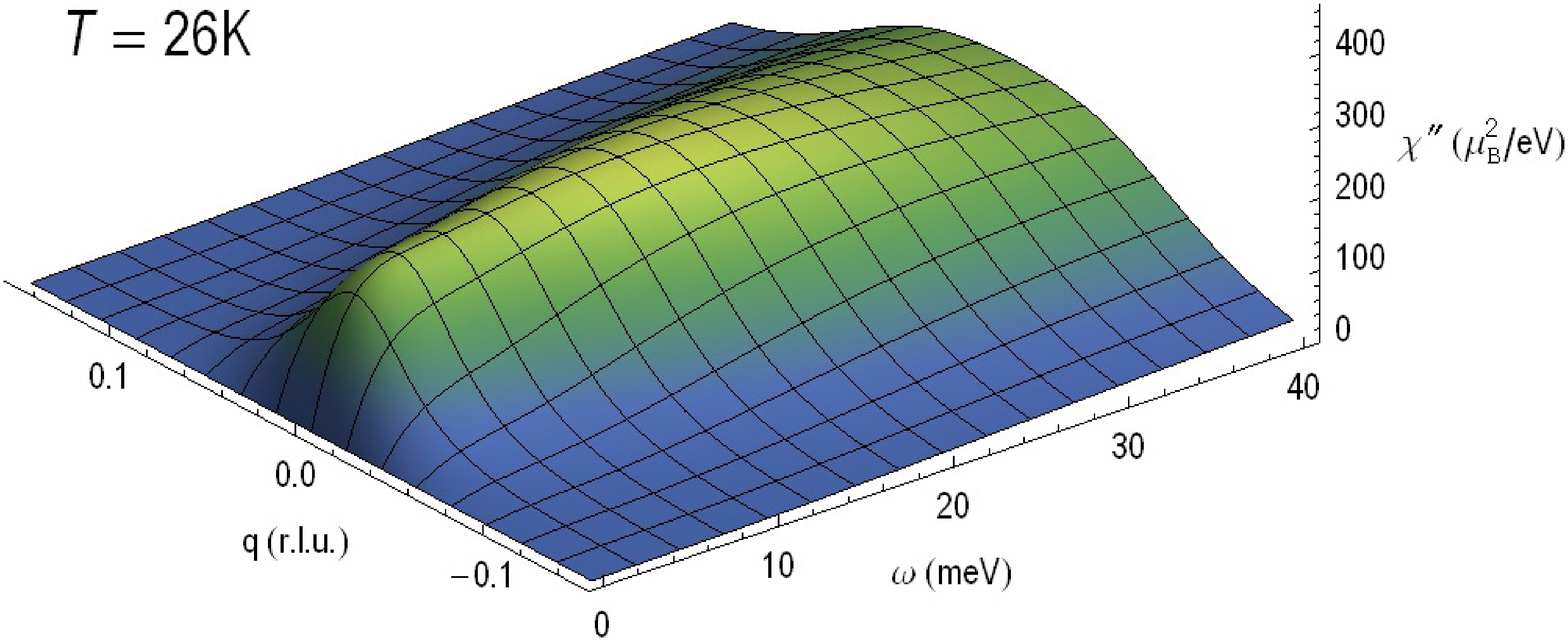}
\caption{(Color online) Imaginary part $\chi^{\prime\prime}$ of the dynamical magnetic susceptibility of a ${\rm Ba(Fe_{0.925}Co_{0.075})_2As_2}$ single crystal at temperatures $T = 5$~K~($< T_{\rm c} = 25$~K, top) and~26~K~($> T_{\rm c}$, bottom)  versus 2D in-plane wave vector $q = |{\bf Q} - {\bf Q}_{\rm AF}|$ and magnetic excitation energy $\hbar \omega$, where ${\bf Q}_{\rm AF} = (\frac{1}{2}, \frac{1}{2}, L)$~r.l.u.\ in tetragonal notation and $\hbar \equiv 1$  in the figure.\cite{Inosov2010}  The units of $\chi^{\prime\prime}$ in the figure are $\mu_{\rm B}^2$~eV$^{-1}$~f.u.$^{-1}$, where f.u.\ means formula unit. Additional 3D isothermal plots for temperatures from 5~to 300~K are available.\cite{Inosov2010}  The  wave vector scans from which the figure was constructed were longitudinal $(\frac{1}{2} + h, \frac{1}{2} + h, L)$~r.l.u.\ scans ($L = 1,3$), i.e.\ ${\bf q} = (h,h,0)$~r.l.u., and the anisotropy of the magnetic excitations in the ${\bf Q}_x$-${\bf Q}_y$ plane was not examined (D. S. Inosov, private communication).  The spin gap and spin resonance mode are both present at 5~K~$\ll T_{\rm c}$ (note that $\omega = 0$ is on the far left edge) and are both absent at 26~K~$> T_{\rm c}$. Reprinted with permission from Ref.~\onlinecite{Inosov2010}. }
\label{FigFeCoResMode}
\end{figure}

%\clearpage
%\squeezetable
\begin{table}
\caption{\label{ResonanceModeData} Neutron spin resonance mode energies $E_{\rm R}$ for Fe-based superconductors and the wave vectors ${\bf Q}_{\rm R}$ at which they were observed in orthorhombic reciprocal lattice unit (r.l.u.) notation.  The designation (1,0,0)~r.l.u.\ is the nesting wave vector between hole and electron Fermi surface pockets, which is $\left(\frac{1}{2},\frac{1}{2},0\right)$~r.l.u.\ in tetragonal notation.  The listed superconducting transition temperature $T_{\rm c}$ values are the onset temperatures for superconductivity and may overestimate the bulk $T_{\rm c}$.  The ratio of $E_{\rm R}$ to $k_{\rm B}T_{\rm c}$ is also shown, where $k_{\rm B}$ is Boltzmann's constant.}
\begin{ruledtabular}
\begin{tabular}{l|cccccc}
Compound & $T_{\rm c}$ & $E_{\rm R}$ & ${\bf Q}_{\rm R}$  & $\frac{E_{\rm R}}{k_{\rm B}T_{\rm c}}$  & Ref. \\
  & (K) & (meV) & (r.l.u.) \\ \hline
${\rm Ba_{0.6}K_{0.4}Fe_2As_2}$\footnotemark[7]  & 38 & 14 & (1,0,1) & 4.3  & \onlinecite{Christianson2008} \\
${\rm Ba(Fe_{0.96}Co_{0.04})_2As_2}$   & 11\footnotemark[4] & 4.5(5)  & (1,0,1) & 4.7 & \onlinecite{Christianson2009} \\
${\rm Ba(Fe_{0.953}Co_{0.047})_2As_2}$   & 17\footnotemark[1] & 5  & (1,0,1) & 3.4 & \onlinecite{Pratt2010} \\
&  & 8  & (1,0,0) & 5.5 & \onlinecite{Pratt2010} \\
${\rm Ba(Fe_{0.926}Co_{0.074})_2As_2}$   & 22.2 & 8.3(1) & (1,0,0) & 4.3 & \onlinecite{Li2010} \\
${\rm Ba(Fe_{0.925}Co_{0.075})_2As_2}$   & 25 & 9.5(5) & (1,0,1) & 4.4 & \onlinecite{Inosov2010} \\
${\rm Ba(Fe_{0.92}Co_{0.08})_2As_2}$   & 22 & 8.6(5) & (1,0,1) & 4.5 & \onlinecite{Lumsden2009} \\
${\rm Ba(Fe_{0.9625}Ni_{0.0375})_2As_2}$\footnotemark[2]   & 12.3\footnotemark[3] & 5 & (1,0,1) & 4.7 & \onlinecite{Wang2010} \\
& & 7 & (1,0,0) & 6.6 & \onlinecite{Wang2010} \\
${\rm Ba(Fe_{0.95}Ni_{0.05})_2As_2}$   & 20 & 8.7(4) & (1,0,1) & 5.0 & \onlinecite{Li2009} \\
& & 7.2(7) & (1,0,0) & 4.2 & \onlinecite{Li2009} \\
${\rm Ba(Fe_{0.95}Ni_{0.075})_2As_2}$\footnotemark[2]   & 15 & 6 & (1,0,1) & 4.6 & \onlinecite{Wang2010} \\
& & 8 & (1,0,0) & 6.2 & \onlinecite{Wang2010} \\
${\rm Fe(Se_{0.4}Te_{0.6})}$   & 14 & 6.51(4) & (1,0,0) & 5.4 & \onlinecite{Qiu2009} \\
${\rm Fe(Se_{0.4}Te_{0.6})}$   & 14.6 & 7.1(1) & (1,0,0) & 5.6 & \onlinecite{Bao2010} \\
${\rm Fe(Se_{0.4}Te_{0.6})}$   & 14 & 6 & $(1,\pm \varepsilon,0)$\footnotemark[6] & 5.0 & \onlinecite{Argyriou2009} \\
${\rm Fe(Se_{0.5}Te_{0.5})}$   & 14\footnotemark[5] & 7 & (1,0,0) & 5.8 & \onlinecite{Mook2009} \\
${\rm Fe_{1.01}(Se_{0.5}Te_{0.5})}$   & 14\footnotemark[5] & 7 & (1,0,0) & 5.8 & \onlinecite{Babkevich2010} \\
${\rm LaFeAsO_{0.92}F_{0.08}}$\footnotemark[7]   & 29 & 13 & (1,0,0) & 5.2 & \onlinecite{Shamoto2010} \\
\end{tabular}
\end{ruledtabular}
\footnotetext[1]{The sample also has a structural transition temperature $T_{\rm S} = 60$~K and an antiferromagnetic ordering temperature $T_{\rm N} = 47$~K\@.}
\footnotetext[2]{Nominal composition.}
\footnotetext[3]{This sample also has an antiferromagnetic ordering temperature $T_{\rm N} = 58$~K\@.}
\footnotetext[4]{This sample also has an antiferromagnetic ordering temperature $T_{\rm N} = 58.0(6)$~K\@.}
\footnotetext[5]{Broad superconducting transition.}
\footnotetext[6]{Slightly incommensurate with $\varepsilon = 0.035(10)$. The transverse scan with a flat top in Fig.~3(a) of Ref.~\onlinecite{Mook2009} might also be interpreted to arise from incommensurability.}
\footnotetext[7]{Polycrystalline sample.}
\end{table}

An early example of neutron spin resonance mode data is shown in the top panel of Fig.~\ref{FigBaKFe2As2ResMode} for \emph{polycrystalline} Ba$_{0.6}$K$_{0.4}$Fe$_2$As$_2$.\cite{Christianson2008}   As shown in the bottom panel of Fig.~\ref{FigBaKFe2As2ResMode}, the intensity of the resonance is zero above $T_{\rm c}$ and grows monotonically with decreasing temperature below $T_{\rm c}$.\cite{Christianson2008}  In Ref.~\onlinecite{Christianson2008}, the authors show that this increase in intensity comes at the expense of intensity at lower energy, which has been confirmed in many subsequent experiments.  This resonance occurs naturally in a $d$-wave superconductor which has both positive and negative superconducting order parameters around a single Fermi surface.  For the FeAs-based superconductors, in view of the above ARPES and other measurements of the nodeless $s$-wave nature within the FeAs plane of the superconductivity on each observed Fermi surface sheet, the observation of the resonance mode in neutron scattering measurements strongly supports the $s^\pm$-wave scenario for superconductivity in which the sign of the superconducting order parameter is the same within each Fermi surface sheet at the $\Gamma$ point or the M~point, but has opposite signs at these two points.\cite{Christianson2008}  
As noted above for the FeAs-based materials, the in-plane wavevector of the resonant scattering below $T_{\rm c}$ is the same as that at which the SDW fluctuations occur above $T_{\rm c}$, and at which the long-range SDW ordering occurs in the undoped parent compounds.  For Ba$_{0.6}$K$_{0.4}$Fe$_2$As$_2$, the horizontal repeat wavelength $\lambda$ between the iron atoms in the vertical spin stripes in Fig.~\ref{Stripe_Mag_Struct} is $\lambda = \sqrt{2} a = 5.53$~\AA, where $a = 3.91$~\AA\ is the tetragonal basal plane lattice parameter (see Fig.~\ref{Struct122}) of this compound.  The wavevector associated with this wavelength is $|\mathbf{Q}| = 2\pi/\lambda = 1.14$~\AA$^{-1}$, in agreement with the observed resonance mode wavevector in the top panel of Fig.~\ref{FigBaKFe2As2ResMode}.  The energy of the resonant mode, $E_{\rm R} \approx 15$~meV, corresponds to $E_{\rm R}/k_{\rm B}T_{\rm c} \approx 4.6$ where $k_{\rm B}$ is Boltzmann's constant, remarkably similar to values of $\approx 5$ obtained for the resonance mode in the layered cuprate high $T_{\rm c}$ superconductors.\cite{Hufner2008, Eschrig2006}  

A more recent inelastic neutron scattering study of a ${\rm Ba(Fe_{0.925}Co_{0.075})_2As_2}$ single crystal is shown in Fig.~\ref{FigFeCoResMode} for temperatures equal to (25~K) and far below (5~K) $T_{\rm c}$.\cite{Inosov2010}  In this study, the characteristics of the resonance mode are very similar to those described above.

\begin{figure}
\includegraphics [width=3.3in]{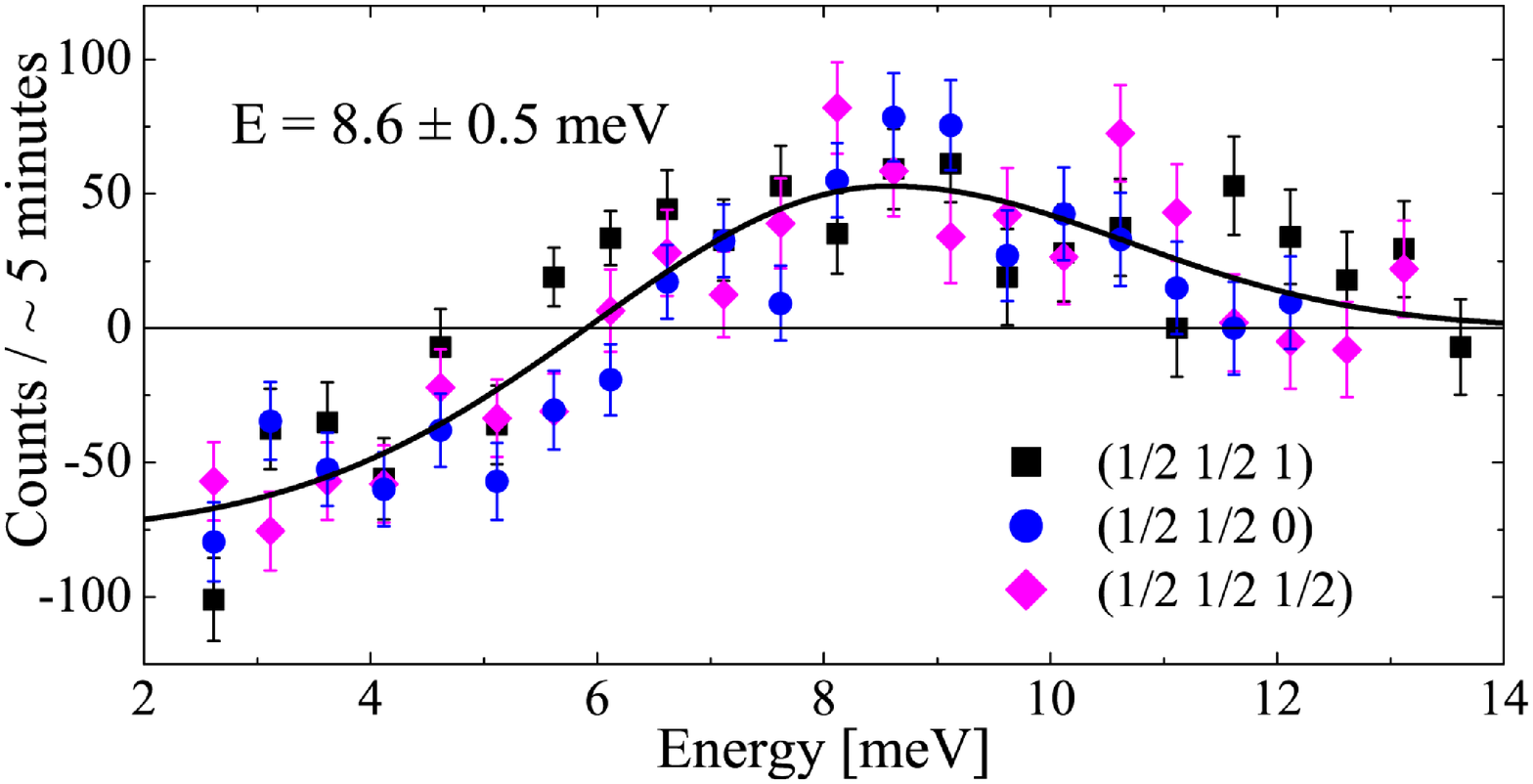}
\caption{(Color online) Difference between 10~K and 30~K scans of neutron scattering intensity versus magnetic excitation energy for the three fixed wavevectors ${\bf Q} = (\frac{1}{2} \frac{1}{2} L)$~r.l.u.\ given in the figure for single crystal BaFe$_{1.84}$Co$_{0.16}$As$_2$ ($T_{\rm c} = 22$~K).  The data show little dispersion of the magnetic excitation energy along the $c$-axis (\emph{i.e.}, with different $L$).  The data also illustrate the redistribution of spectral weight from the spin gap (or spin pseudogap) region at low energies to the resonance energy of about 8.5~meV upon cooling from above $T_{\rm c}$ to below $T_{\rm c}$.  Reprinted with permission from Ref.~\onlinecite{Lumsden2009}.  Copyright (2009) by the American Physical Society.}
\label{Ba(FeCo)2As2_dispersion}
\end{figure}

Additional evidences for the existence of a neutron resonance mode in the superconducting state of the Fe-based superconductors at the nesting wavevector were found from inelastic neutron scattering measurements on single crystals of ${\rm Ba(Fe_{1.935}Co_{0.065})_2As_2}$ ($T_{\rm c} = 23$~K),\cite{Lester2010} ${\rm Ba(Fe_{1.92}Co_{0.08})_2As_2}$ ($T_{\rm c} = 22$~K),\cite{Parshall2009, Lumsden2009}  ${\rm Ba(Fe_{1.95}Ni_{0.05})_2As_2}$ ($T_{\rm c} = 20$~K),\cite{Chi2009, Li2009} FeTe$_{0.5}$Se$_{0.5}$ ($T_{\rm c} \sim 14$~K),\cite{Mook2009} and FeTe$_{0.6}$Se$_{0.4}$ ($T_{\rm c} = 14$~K).\cite{Qiu2009} These compounds have lower $T_{\rm c}$ than Ba$_{0.6}$K$_{0.4}$Fe$_2$As$_2$ (38~K) but showed similar ratios of $E_{\rm R}/k_{\rm B}T_{\rm c}$, confirming a scaling of $E_{\rm R}$ with $T_{\rm c}$, and confirming the opening of a spin gap that compensates the increased scattering at $E_{\rm R}$.  

The transfer of spectral weight to the resonance energy from lower energy within the superconducting gap for ${\rm Ba(Fe_{1.92}Co_{0.08})_2As_2}$ upon cooling below $T_{\rm c}$ is shown in Fig.~\ref{Ba(FeCo)2As2_dispersion}.\cite{Lumsden2009}  The resonance peak thus evidently forms from a redistribution of the spin fluctuation spectral weight that already exists above $T_{\rm c}$ (see also the other references above and/or in Table~\ref{ResonanceModeData}).  

Measurements of the resonance mode for ${\rm Ba(Fe_{1.92}Co_{0.08})_2As_2}$ showed little  dispersion of the resonance peak along $Q_z$ as also shown in Fig.~\ref{Ba(FeCo)2As2_dispersion}.\cite{Lumsden2009}  The in-plane bandwidth was estimated to be 70~meV with a lower limit of 60 meV, whereas the $c$-axis dispersion was very weak with an estimated bandwidth of 0.6 meV and upper limit of 1.5 meV\@.  Thus the ratio of the in-plane to out-of-plane bandwidths was $\sim 120$ with a lower bound of 40.\cite{Lumsden2009}  From comparison with the much more three-dimensional spin wave dispersions found for undoped AF-ordered BaFe$_2$As$_2$, the authors inferred that doping with Co strongly enhances two-dimensionality of the magnetic excitations.  Furthermore, the spin fluctuation scattering above $T_{\rm c}$ in BaFe$_{1.84}$Co$_{0.16}$As$_2$ was also found to be quasi-two-dimensional.\cite{Parshall2009}  Similarly, the magnetic resonance mode in FeTe$_{0.6}$Se$_{0.4}$ was also found to be quasi-two-dimensional.\cite{Qiu2009}  In the latter study, the authors showed theoretically that the resonance is consistent with a bound state associated with $s^\pm$ pairing arising from (imperfect) quasi-two-dimensional Fermi surface nesting.

In contrast, the neutron scattering measurements on underdoped Ba(Fe$_{1-x}$Ni$_x$)$_2$As$_2$ (Refs.~\onlinecite{Chi2009}, \onlinecite{Wang2010}, \onlinecite{Li2009}) and ${\rm Ba(Fe_{0.953}Co_{0.047})_2As_2}$ (Ref.~\onlinecite{Pratt2010}) showed significant $c$-axis dispersions of the resonance mode energy (see Table~\ref{ResonanceModeData}).  The authors of Refs.~\onlinecite{Wang2010}, \onlinecite{Chi2009} and \onlinecite{Li2009} suggested that if the resonance energy reflects the superconducting pairing interaction via spin fluctuations, then one would expect a modulation of the superconducting gap along $k_z$, as is actually suggested from ARPES measurements of the band structure in Sec.~\ref{ARPESSC} above, presumably arising from interlayer spin correlations.\cite{Wang2010, Pratt2010}

The above anisotropy (or lack thereof) in the spin correlations of the resonance mode refer to anisotropy along the $c$-axis.  Li et al.\ showed from inelastic neutron scattering measurements of ${\rm Ba(Fe_{1.926}Co_{0.074})_2As_2}$ single crystals that the in-plane anisotropy of the spin fluctuations in the normal state, reflecting a ``nematic degree of freedom,'' is very similar in the neutron spin resonance mode,\cite{Li2010} which again demonstrates that the resonance mode grows out of, and consists of a redistribution in energy of, the normal state antiferromagnetic spin fluctuations.

In their neutron scattering study of the spin resonance in FeTe$_{0.4}$Se$_{0.6}$, Mook and coworkers suggested that the spin fluctuations giving rise to the spin resonance are distinct from the spin fluctuations giving rise to the superconductivity.\cite{Mook2009}  However, the transverse scan with a flat top in Fig.~3(a) of Ref.~\onlinecite{Mook2009}, on which this suggestion is based, might  also be interpreted to arise from incommensurability (see Table~\ref{ResonanceModeData} and Ref.~\onlinecite{Argyriou2009}).  Lee and coworkers studied the in-plane anisotropy of the neutron spin resonance in FeTe$_{0.5}$Se$_{0.5}$ and found a distinct splitting of the neutron spin resonance in wave vector space along the $\left(\frac{1}{2},\frac{1}{2}\right)$~r.l.u.\ direction.\cite{Lee2009}  They attributed this structure to a coupling of spin and orbital correlations.

An inelastic polarized neutron scattering study of the neutron spin resonance was carried out on single crystals of optimally doped ${\rm Ba(Fe_{0.95}Ni_{0.05})_2As_2}$ with $T_{\rm c} = 20$~K by Lipscombe et al.\cite{Lipscombe2010}  They confirmed that the resonance is purely magnetic in origin and that the resonance is centered within the $a$-$b$ plane at the reciprocal space position $(\frac{1}{2},\frac{1}{2})$~r.l.u. in tetragonal notation.  They also found that there is a preferential orientation of the spin fluctuations to be within the $a$-$b$ plane as opposed to, e.g., being uniformly distributed in direction.  They point out that this is very surprising in the paramagnetic state and suggest that the same anisotropy leading to magnetic anisotropy in the magnetically ordered state in the undoped and underdoped compounds is still present in the optimally doped superconducting composition.  They further point out that this spin-space anisotropy is not expected from current theories of the spin resonance.

Several measurements of the influence of a magnetic field on the neutron spin resonance mode have been reported.\cite{Wen2010, Zhao2010, Bao2010}  The onset temperature of the resonance and its magnitude decrease as the field increases, as does the $T_{\rm c}$ itself, further confirming the close connection between the spin resonance and superconductivity.  

The study by Bao et al.\ was carried out on three co-aligned single crystals of FeTe$_{0.6}$Se$_{0.4}$ with a total mass of 15.3~g and a bulk $T_{\rm c} = 14.6$~K.\cite{Bao2010}  The neutron scattering resonance peak in an applied magnetic field $H = 14$~T splits into three peaks, consisting of an unshifted central lobe plus two side lobes.\cite{Bao2010}  The magnetic field was applied along the $c$-axis, perpendicular the the Fe square lattice planes.  The total integrated magnetic scattering intensity of the resonance peak(s) is independent of magnetic field to within the errors.  The authors fitted their data at both fields by a central peak plus the two field-dependent side lobes at energies
\be
\hbar \omega = \hbar\Omega_0 \pm \sqrt{\Delta^2 + (g\mu_{\rm B}H)^2}
\label{Fitneutres}
\ee
and obtained the values of the $g$-factor $g = 2.54$ and anisotropy energy $\Delta = 1.2(4)$~meV fitting parameters, and the energy level diagram versus field.\cite{Bao2010}  The data and fits are consistent with a transition between a singlet ground state and a triplet excited state of a pair of antiferromagnetically coupled quasiparticles, each with spin $S = 1/2$, but there was no explanation for the unusual value of the $g$-factor.  Interestingly, extrapolating the lowest energy level data to energy $E = 0$ gives a critical field of 47(9)~T, which happens to be the same as the value inferred from high-field resistivity data for the upper critical field $H_{\rm c2}(T = 0)$.\cite{Bao2010}  This suggests an intimate relationship between the spin resonance mode, and in particular the triplet species associated with it, and the superconducting condensate.  The authors conclude, ``The high field neutron data show that magnetic fluctuations play a central role in iron superconductivity and suggest that the formation of a triplet bound state actually drives superconductivity in FeTe$_{0.6}$Se$_{0.4}$.''\cite{Bao2010}
\\

\paragraph*{Summary}  The evidence is strong that the neutron scattering spin resonance mode arises from a redistribution of magnetic scattering spectral weight from energies below the maximum superconducting gap to the resonance peak, where the wavevector of the resonance for all the Fe-based compounds is the same as the nesting wavevector between the electron and hole Fermi surface pockets.  For the FeAs materials, this nesting wave vector is the same as the SDW wave vector in the respective nonsuperconducting parent compounds.  Indeed, the energy dispersion versus wavevector of the in-plane spin excitations around the resonant wavevector is similar to those in the undoped parent compounds around the AF ordering wavevector above $T_{\rm N}$, and many measurements indicate that the resonance mode consists of spin fluctuations that pre-exist in the normal state.  The observation and properties of the neutron spin resonance mode, together with other measurements that indicate a uniform superconducting gap within the $a$-$b$~plane for each electron or hole band, provide strong support for the $s^\pm$ pairing symmetry in which the signs of the superconducting order parameter are reversed between the electron and hole Fermi surface pockets as illustrated in the inset of Fig.~\ref{FigBaFeCoAsMT}(c) in Sec.~\ref{Sec_IntSCMag}.

\subsubsection{\label{SecQPI} Quasiparticle Interference (QPI) Measurements}

\subsubsection*{a. Introduction}

In a scanning tunneling microscopy (STM) image of a metallic surface taken at bias $eV = \hbar\omega$, the differential tunneling conductance $dI/dV$ is proportional to the electronic density of states at energy $\hbar\omega$ in the imaged surface, times the appropriate Fermi-Dirac distribution function.  In the presence of nonmagnetic and/or magnetic impurities in a material, the conduction carriers scatter off (screen) these impurities and set up an electron standing wave pattern resulting from Friedel-like oscillations in the electron density and also in the local density of states at and near the Fermi energy, which shows up as lighter and darker areas in the STM image that are distinct from the positions of the atoms themselves.  These oscillations thus arise from coherent quasiparticle interference (QPI), and the resulting interference pattern is called a QPI pattern.  By Fourier-transforming an STM image that contains a QPI pattern resulting from the presence of impurities in metals, one can obtain a FT-STM image in {\bf q}-space.\cite{Balatsky2006}  

Here we quote from Ref.~\onlinecite{Balatsky2006}.  ``The enhanced signal in the FT-STM image at wave vector {\bf q} and bias $eV = \hbar \omega$ corresponds to a large amplitude for (quasiparticle) scattering off of an impurity.  Qualitatively, this amplitude depends on the number of available initial and final states at a given energy in regions of the Brillouin zone separated by {\bf q}.''  Thus the amplitude $A({\bf q},\omega)$ in the FT-STM image is
\be
A({\bf q},\omega) \propto \int{\cal N}_{\bf k}(\omega){\cal N}_{\bf k+q}(\omega)f(\hbar\omega)[1-f(\hbar\omega)]d{\bf k}, 
\label{EqAqomega}
\ee
where ${\cal N}_{\bf k}(\omega)$ is the momentum- and energy-dependent density of states and $f(\hbar\omega)$ is the Fermi-Dirac distribution function.  The probability that an initial conduction electron state at energy $\hbar\omega$ is occupied is $f(\hbar\omega)$, and the probability that a final state at energy $\hbar\omega$ is unoccupied is $1 - f(\hbar\omega)$.  Then, ``The greater the number of `matching' pairs of initial and final states, the more a quasiparticle (elastically) scatters from one into another, producing a feature in the FT-STM image.''  The integrand in Eq.~(\ref{EqAqomega}) is maximum on the Fermi surface, so the FT-STM technique allows to map the Fermi surface (in wave vector space) at the surfaces of simple metals such as Be(0001) and Cu(111).\cite{Sprunger1997, Peterson1998}  For other metals, the interpretation of the FT-STM images can be more complicated, as discussed in Ref.~\onlinecite{Balatsky2006} and further illustrated in the following section.  The FT-STM technique is complementary to low-energy electron diffraction (LEED) and ARPES measurements.\cite{Sprunger1997}

Wang and coworkers pointed out that FT-STM measurements can distinguish between $s^{++}$ and $s^\pm$ superconducting pairing scenarios.\cite{Wang2009b}  In the following section, the use of this technique for this purpose is described.

\subsubsection*{b. Results: ${\rm Ca(Fe_{0.97}Co_{0.03})_2As_2}$}

Hanaguri reported STM measurements at 0.4~K on a single crystal of tetragonal ${\rm FeTe_{0.60}Se_{0.40}}$ with $T_{\rm c}\sim 14$~K.\cite{Hanaguri2010}  They found peaks in FT-STM images at in-plane wave vectors ${\bf q}_1 = \left(\frac{2\pi}{a},\frac{2\pi}{a}\right)$, ${\bf q}_2 = \left(\frac{\pi}{a},\frac{\pi}{a}\right)$ and ${\bf q}_3 = \left(\frac{2\pi}{a},0\right)$ in tetragonal notation where $a$ is the tetragonal basal plane lattice parameter.  They identified the peak at ${\bf q}_1$ as a Bragg peak arising from electron diffraction from the lattice (${\bf q}_1$ is a reciprocal lattice vector, see Sec.~\ref{SecBZ}) and was not further considered.  They further studied the peak at wave vector ${\bf q}_2$, which is the nesting wave vector between the electron and hole pockets, and ${\bf q}_3$ connecting different electron pockets, and considered these to be QPI peaks.  From the magnetic field dependences of the peaks at ${\bf q}_2$ and ${\bf q}_3$, they concluded that the superconducting pairing must be of the $s^\pm$ type.  

However, Mazin and Singh noticed that ${\bf q}_3$ in the study of Hanaguri et al.\ is also a reciprocal lattice vector of the tetragonal crystal structure and that the peak width according to their calculations was much too narrow for a QPI peak, so they concluded that the peak at ${\bf q}_3$ is another Bragg peak.\cite{Mazin2010}  They also concluded that the peak at ${\bf q}_2$ is also too narrow to be a QPI peak, and suggested that it might arise from surface reconstruction.  In response, Hanaguri et al.\ replied that the claimed QPI peaks ${\bf q}_2$ and ${\bf q}_3$ were too broad to be Bragg peaks, and that ${\bf q}_3$ consisted of overlapping QPI and Bragg peaks.\cite{Hanaguri2010a}

\subsubsection{\label{Sec_IntSCMag} Interaction of Superconductivity with Long-Range Antiferromagnetic Order}

As seen in the phase diagrams in Fig.~\ref{FigBaKFe2As2_phase_diag} and in other phase diagrams discussed in Sec.~\ref{SecIntro}, the tetragonal-to-orthorhombic structural transition and the long-range antiferromagnetic (AF) transition must both be suppressed before the optimum $T_{\rm c}$ is obtained in any of the FeAs-based systems. This suggests that superconductivity SC and \emph{long-range} AF ordering strongly interact, and in fact compete with each other.  Important additional information about this issue was obtained in the systems Ba(Fe$_{1-x}M_x)_2$As$_2$ ($M$ = Co, Ni, Rh) in which the phase diagrams show regions where SC and long-range AF order appear to coexist.  Such coexistence could occur either from macroscopic phase segregation of SC and AF phases or from coexistence of the two phenomena in the same volume element.  Strong evidence for the second possiblility was obtained by measuring the ordered Fe moment in underdoped samples versus temperature upon traversing $T_{\rm c}$.  

In the Ba(Fe$_{1-x}$Co$_x)_2$As$_2$ system, the first direct indications of coupling between the SC and AF order parameters were found by Lester et al., where the temperature $T$ dependence of the AF order parameter, which is the sublattice magnetization $M$ (also called the staggered magnetization), showed subtle slope changes versus $T$ on passing through $T_{\rm c}$ for crystals with $x = 0.045$ and~0.051.\cite{Lester2009}  Then Pratt et al.\cite{Pratt2009} and Christianson et al.\cite{Christianson2009}  published back-to-back papers showing a well-defined \emph{decrease} in $M$ on cooling through $T_{\rm c}$ for $x = 0.047$ and 0.04, respectively.  This latter effect was also subsequently observed in single crystals of ${\rm Ba(Fe_{0.9625}Ni_{0.0375})_2As_2}$ (nominal composition)\cite{Wang2009} and ${\rm Ba(Fe_{0.961}Rh_{0.039})_2As_2}$.\cite{Kreyssig2010}  Christianson et al.\ suggested that the decrease in magnetic spectral weight associated with the loss in Bragg intensity is transferred to the superconducting resonance mode described in the previous section.\cite{Christianson2009} The most likely interpretation of these data is that SC and long-range AF order coexist in the same volume element, because if the SC and AF existed in different regions/grains, it is hard to see how the $M$ in the AF regions would have been affected by SC in other regions, or alternatively how AF regions could be converted to SC regions.  In addition, there is no evidence that the fraction of superconducting material increases with decreasing $T$ below $T_{\rm c}$.  Thus the data again indicate that SC competes with long-range AF order in these compounds.  Furthermore, additional measurements and theory of the phase diagram and especially of the coexistence of SC and AF that are discussed in the following paragraph are consistent with the interpretation in which SC and AF can homogeneously coexist in the same volume element.

\begin{figure*}
\includegraphics [width=5in]{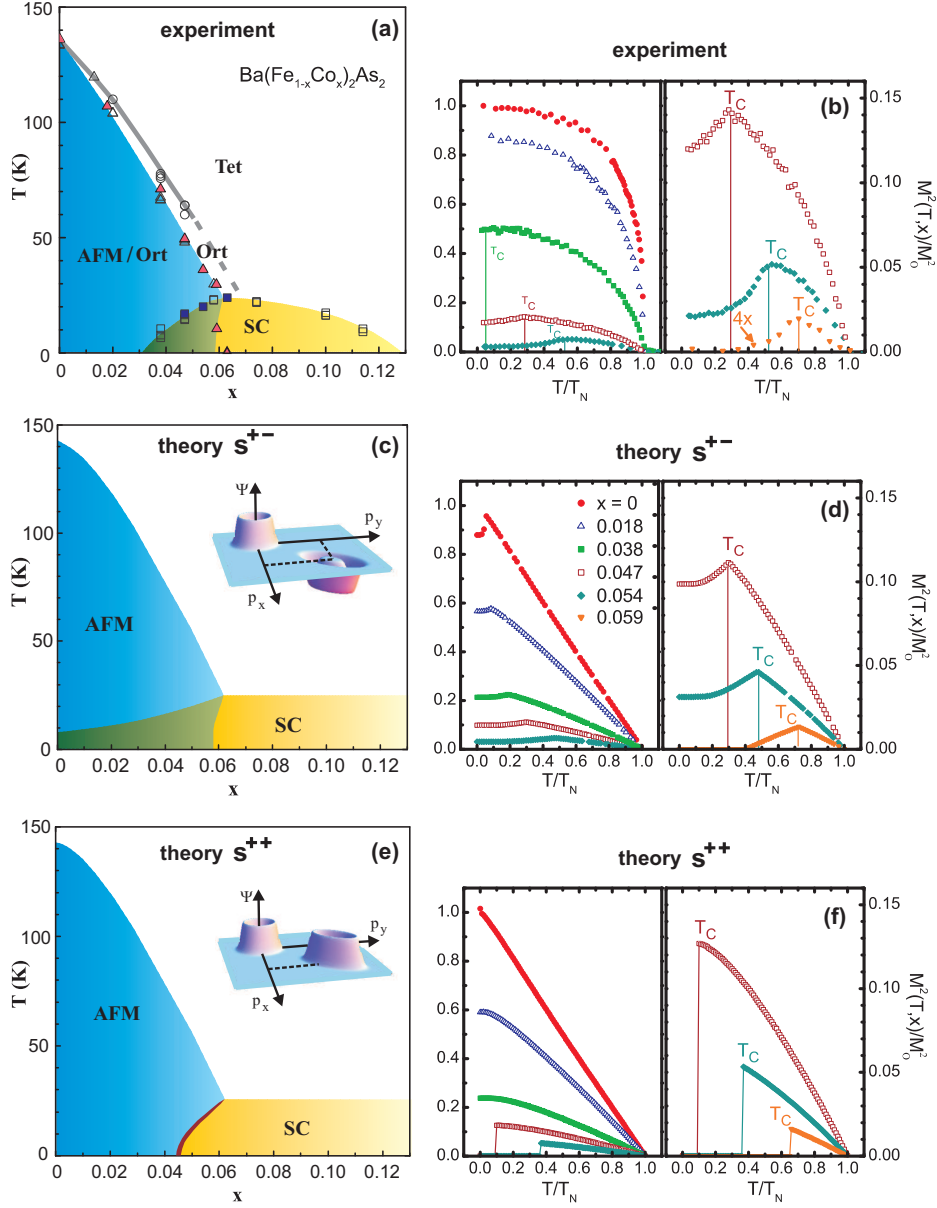}
\caption{(Color online) (a) The $T$-$x$ phase diagram of the Ba(Fe$_{1-x}$Co$_{x}$)$_{2}$As$_{2}$ system determined from neutron diffraction (solid symbols) and bulk thermodynamic and transport (open symbols, Ref.~\onlinecite{Ni2008g}) measurements.  The phase regions are the high temperature tetragonal phase (Tet), the orthorhombic phase (Ort), the combined antiferromagnetic (AFM) and Ort phase, and the superconducting (SC) phase.  See also Fig.~\ref{FigBaKFe2As2_phase_diag} above.  (b) Square of the normalized ordered moment $M/M_0$  versus temperature $T$ measured by neutron diffraction for five different compositions $x$, where $M_0\equiv M(x = 0,T=0)=0.87~\mu_{\rm B}$/(Fe atom). (c), (e): The phase diagrams  obtained theoretically for unconventional $s^{+-}$ and conventional $s^{++}$ pairing states (insets), respectively. In (c), there is a region of coexistence of AFM and SC (shaded green), but in (e) there is not.  In (e), a first order transition occurs in a very narrow regime between the AFM and SC states as shown.  In the insets of (c) and (e), the vertical axis is the superconducting order parameter $\psi = \Delta \exp(i\phi)$ where $\Delta$ is the magnitude and $\phi$ is the phase.  For the $s^{++}$ state, $\phi = 0$ on both hole and electron pockets, whereas for the $s^{+-}$ state $\phi$ is 0 on the electron pockets and $\pi$~rad on the hole pockets (or \emph{vice versa}). The momentum $p_x$ and $p_y$ directions are along the tetragonal $a$- and $b$-axes, respectively.  (d), (f): Theoretical $M^2(T)$ for the $s^{+-}$ and $s^{++}$ states, respectively.  In (b), (d) and (f), the subpanel on the right is an expanded plot of the data in the left subpanel, respectively, for the three samples with the largest Co dopings of $x = 0.047$, 0.054, and 0.059.  Reprinted with permission from Ref.~\onlinecite{Fernandes2010}.  Copyright (2010) by the American Physical Society.}
\label{FigBaFeCoAsMT}
\end{figure*}

A systematic study of the effect of SC on the AF sublattice magnetization $M(T)$ in the Ba(Fe$_{1-x}$Co$_x)_2$As$_2$ system was carried out by Fernandes et al.\ using magnetic neutron diffraction measurements, as shown in Fig.~\ref{FigBaFeCoAsMT}(b).\cite{Fernandes2010, Fernandes2010a}  These remarkable data indicate that $M(T)$ for $x = 0.059$ is \emph{re-entrant} (returns to a zero value) upon cooling to low $T$.  Figure~\ref{FigBaFeCoAsMT}(a) shows the phase diagram, including data from the new neutron diffraction measurements.  Figures~\ref{FigBaFeCoAsMT}(c)--(f) show theoretical calculations of the phase diagram (c, e) and $M(T)$ (d, f) for the conventional $s^{++}$ and unconventional $s^{+-}$ (or $s^\pm$) SC pairing states.  The $s^{++}$ pairing state is one in which the sign of the superconducting order parameter is the same on the electron and hole Fermi surface pockets, whereas in the $s^{+-}$ pairing state the sign of the order parameter is opposite on the electron and the hole pockets.  The calculations for the $s^{+-}$ pairing state are obviously in much better agreement with experiment than for the $s^{++}$ pairing state.  The authors find that AF and conventional phonon-mediated SC cannot coexist.  They conclude that ``$s^{+-}$ pairing and magnetic order (can) compete but coexist microscopically, whereas both phases are mutually exclusive in case of $s^{++}$ pairing. ... This conclusion is robust and independent of specific details of the model.''  The authors further conclude, ``Our findings strongly suggest that superconductivity is unconventional (i.e., $s^{+-}$) in all members of the iron arsenide family.'' 

Vorontsov, Vavilov and Chubukov carried out an extensive theoretical investigation of the possible coexistence between superconductivity and a SDW in a two-band quasi-two-dimensional semimetal.\cite{Vorontsov2010}  In agreement with Refs.~\onlinecite{Fernandes2010} and \onlinecite{Fernandes2010a}, they found that it is much more likely that this coexistence occurs if the orbital pairing symmetry of the superconducting state is $s^\pm$ instead of $s^{++}$.\cite{Vorontsov2010}  Furthermore, they found that coexistence of superconductivity with a \emph{commensurate} SDW is only possible if the hole and electron Fermi surfaces have different Fermi velocities and also different cross-sectional shapes (e.g., circular versus elliptical).  Amazingly, Refs.~\onlinecite{Fernandes2010b} and \onlinecite{Vorontsov2010} both found that a coexisting superconducting and SDW state can develop, on cooling, out of a pre-existing completely gapped (i.e., semiconducting) SDW state.   Parker et al.\ have also theoretically analyzed the coexistence of superconducting and spin density wave order.\cite{Parker2009}  They noted that ``the SDW observed in underdoped pnictide compounds does not have any considerable destructive effect on the $s^\pm$ superconductivity, besides the obvious competition between the two instabilities for the density of states at the Fermi level'' that causes $T_{\rm c}$ to decrease in the presence of the SDW\@.

Chauvi\`ere et al.\ carried out inelastic light (Raman) scattering measurements on crystals in the Ba(Fe$_{1-x}$Co$_x)_2$As$_2$ system, with a focus on the small composition region of coexistence between superconductivity and long-range antiferromagnetic ordering.\cite{Chauviere2010}  They found that the superconducting gap was significantly depressed for samples in the coexistence composition range.  Their results were interpreted as due to  the competition between itinerant antiferromagnetism and superconductivity for the same conduction carriers, consistent with the above theoretical considerations.

\subsubsection{\label{SecPenDepth} Magnetic Penetration Depth}

\subsubsection*{a. Introduction}

For an $s$-wave (or $s^\pm$-wave) superconductor, the superconducting gap exists over the entire Fermi surface(s), and one therefore expects that physical properties involving quasiparticle excitations above the superconducting quasiparticle energy gap $2\Delta$ should decay exponentially for $T \to 0$, \emph{i.e.}, at temperatures sufficiently below the smallest gap.  Such quantities include the electronic specific heat, the London penetration depth, and nuclear spin-lattice relaxation rates $1/T_1$ obtained from NMR measurements.  However, most $1/T_1$ measurements of the Fe-based superconductors below $T_{\rm c}$ indicate a power law dependence on temperature ($\sim T^3$) instead of an exponential decrease (see references cited in Ref.~\onlinecite{DParker2008}).   Similarly, magnetic penetration depth measurements reveal power-law behavior $T^n$ at the lowest temperatures such as for ${\rm BaFe_{1.93}Co_{0.07}ŒAs_2}$ single crystals, where $n = 2.4(1)$ was obtained and attributed to either the existence of gapless regions or point nodes on the Fermi surface in the superconducting state.\cite{Gordon2009}  Electronic heat capacity $C_{\rm e}(T)$ measurements (after subtraction of the phonon contribution from the measured data) have also indicated power law rather than exponential behavior.  For example, Zeng et al.\ reported that near-optimally doped single crystals of both ${\rm Ba(Fe_{0.92}Co_{0.08})_2As_2}$ and ${\rm Ba(Fe_{0.95}Ni_{0.05})_2As_2}$ follow $C_{\rm e}(T) = \gamma_0 T + A T^3$ at $T \ll T_{\rm c}$, where $\gamma_0$ is the ``residual'' Sommerfeld coefficient at $T = 0$ and $A$ is a constant.\cite{Zeng2010}  However, impurity effects and spatial inhomogeneities can also contribute to such behaviors.  

For penetration depth measurements such as rf and microwave measurements, the measured penetration depth $\lambda(0)$ at $T\to 0$ is related to the London penetration depth by\cite{Tinkham1975}
\be
\lambda(0) = \lambda_{\rm L}(0)\sqrt{1 + \frac{\xi_0}{\ell}},
\label{lambdaLambdaL}
\ee
where 
\begin{equation}
\xi_0 = \frac{\hbar v_{\rm F}}{\pi \Delta(0)}
\label{Eqxi0}
\end{equation}
is the BCS coherence length, $\ell$ is the quasiparticle mean free path, $v_{\rm F}$ is the Fermi velocity and $\Delta(0)$ is one-half the zero-temperature energy gap for quasiparticle excitations.  The limit $\xi_0/\ell\to 0$ is called the clean limit and the opposite limit the dirty limit.  An equivalent way to express this is to substitute $\ell = v_{\rm F}\tau$, where $1/\tau$ is the scattering rate of quasiparticles by impurities and $\tau$ is the mean free scattering time.  Thus the criteria for clean and dirty limits can be expressed in equivalent ways as
\bea
\frac{\xi_0}{\ell},\ \frac{\hbar v_{\rm F}}{\pi \Delta(0)\ell},\ \frac{\hbar}{\pi\Delta(0)\tau} &\ll& 1, \hspace{0.2in}{\rm (clean\ limit)}\nonumber\\
\label{EqCleanDirty}\\
\frac{\xi_0}{\ell},\ \frac{\hbar v_{\rm F}}{\pi \Delta(0)\ell},\ \frac{\hbar}{\pi\Delta(0)\tau} &\gg& 1. \hspace{0.2in}{\rm (dirty\ limit)}\nonumber
\eea

In the following we further consider penetration depth measurements and then $1/T_1$ measurements.

\subsubsection*{b. Penetration Depth Measurement Results}

A sensitive measurement of the change in the magnetic penetration depth $\lambda$ versus $T$ can be performed by measuring the change $\Delta f = f - f_0$ in the resonant frequency $f=(2\pi\sqrt{LC})^{-1}$ with respect to the unperturbed resonant frequency $f_0 \approx 14$~MHz of a resonant circuit containing a capacitor $C$ and a coil of inductance $L$ in which the sample is placed using a tunnel diode resonator (TDR) circuit.\cite{Prozorov2009}  Using this circuit, the resolution of the $\Delta f$ measurements is 0.1~Hz out of 14~MHz, i.e., about 7~ppb, which corresponds to a resolution in a change $\Delta \lambda$ in $\lambda$ of about 1~\AA\ for mm-size crystals.  The amplitude of the ac field seen by the sample is $\sim 10$~mOe, which is much less than the lower critical field and the samples are hence in the Meissner state in zero applied dc field.  In terms of the dimensionless differential magnetic volume susceptibility $\chi = dM/dH$ of the sample, for $|\Delta f_{\rm max}|/f_0 \ll 1$ one has
\[
\frac{\Delta f(T)}{f_0} = -G\,4\pi\chi(T),  
\]
where $G$ is a dimensionless calibration constant.  In the application of the technique to penetration depth measurements, one obtains\cite{Prozorov2009}
\be
\frac{\Delta f(T)}{f_0} = G\left\{1-\frac{\sqrt{\mu}\lambda(T)}{R}\tanh\left[\frac{R\sqrt{\mu}}{\lambda(T)}\right] \right\},
\label{EqTDR}
\ee
where $R$ is an ``effective'' linear sample dimension, $\lambda$ is the penetration depth of the corresponding ``nonmagnetic'' material, and $\mu$ is the dimensionless relative magnetic permeability of the sample ($\mu = 1$ in free space) which can be temperature dependent.  Usually $\mu \approx 1$ except for samples containing magnetic ions such as the Nd ions in the compound NdFeAsO$_{0.9}$F$_{0.1}$, for which ignoring the deviation of $\mu(T)$ from unity gives the wrong $T$ dependence of $\lambda$ at the lowest $T$.\cite{Martin2009}   In particular, the activated temperature dependence reported for PrFeAsO$_{1-y}$ by Hashimoto et al.,\cite{Hashimoto2009b} which continues to be cited as evidence for a nodeless superconducting order parameter, is most likely an artifact due to the magnetism of the Pr$^{+3}$ ions.\cite{Martin2009}

As discussed above in Sec.~\ref{SecChiUnits}, complete diamagnetism corresponds to $4\pi\chi = -1$ (in the absence of demagnetization effects which make this value more negative) where $\chi$ is the dimensionless volume susceptibility in Gaussian units.  In Eq.~(\ref{EqTDR}) this situation corresponds to  $\lambda(T \to 0)/R = 0$ and therefore $\Delta f(T\to 0)/f_0 = G$.  Thus, from Eq.~(\ref{EqTDR}) the calibration constant $G$ can be determined by measuring the frequency shift upon pulling the sample out of the coil from a completely superconducting state.  Alternatively, $G$ can be determined by measuring the temperature-dependent skin depth $\delta$ of a sample in the normal state above $T_{\rm c}$ using the TDR technique, which takes the place of $\lambda$ in the above equations, and then  comparing the result with the measured dc resistivity of the same sample.  The $\delta$ and resistivity $\rho$ are related in SI units by\cite{Corson1962}
\be
\delta = ({\rm 504~m})\sqrt{\frac{\rho}{\mu f}},
\label{Eqdelta}
\ee
where the relative permeability $\mu$ is dimensionless as above, $\rho$ is in $\Omega~{\rm m}$ and the frequency $f$ is in Hz. For a typical Fe-based material with a normal state resistivity of 0.1~m$\Omega$~cm, and using an rf frequency of 14~MHz and $\mu = 1$, Eq.~(\ref{Eqdelta}) gives $\delta = 0.13$~mm, which is smaller than a typical linear in-plane dimension of 1--2~mm for crystals used in the measurements.   

Penetration depth measurements are also carried out in the superconducting state using other types of measurements.  Zero-temperature penetration depths $\lambda(T\to0)$ for various Fe-based superconductors determined using the various techniques described in the table caption are summarized in Table~\ref{LambdaTable}.\cite{Fischer2010, Kim2010, Bendele2010, Biswas2010, Luan2010, Kim2010a, Nakajima2010, Li2008c, Gordon2010, Williams2010, Wu2010, Luetkens2008, Malone2009, Weyeneth2009, Drew2008, Inosov2010a}

%\squeezetable
\begin{table*}
\caption{\label{LambdaTable} Magnetic penetration depths at low temperatures for Fe-based superconductors.  The methods used are tunnel diode resonator (TDR), magnetic force microscope (MFM), torque magnetometer (TM), muon spin relaxation ($\mu$SR), infrared spectroscopy (IR), teraHertz (THz) conductivity, and small angle neutron scattering (SANS) measurements.  The second to last column gives the ratio of the zero-temperature superconducting condensate density $\rho_{\rm S}(0)$ to the normal state coherent carrier concentration $n$ calculated using the $\lambda_{ab} (T \to 0)$ data,  the $f_{\rm p}^\prime$ data in Tables~\ref{Opticsdata2} and~\ref{Opticsdata3}, and Eq.~(\ref{EqRhoSomegap2}), denoted ``method 1''.  This method is equally valid for pair-breaking and non-pair-breaking impurities.  The last column gives ``method 2'' of calculating $\rho_{\rm S}(0)/n$ from the largest zero-temperature superconducting gap datum for a given compound $\Delta(0)$ in Table~\ref{SCGapValues} and the quasiparticle scattering rate $1/\tau^\prime$ data in Table~\ref{Opticsdata2} according to Eq.~(\ref{EqRhosn2}).  Method~2 assumes that the quasiparticle scattering centers are not pair-breaking, i.e., that there are no quasiparticle states in the superconducting gap.}
\begin{ruledtabular}
\begin{tabular}{l|cccccc}
 Compound   & $T_{\rm c}$ & $\lambda_{ab} (T \to 0)$ & method & Ref. & $\frac{\rho_{\rm S}(0)}{n}$ & $\frac{\rho_{\rm S}(0)}{n}$ \\
 & (K) & (nm) & & & method 1 & method 2\\ \hline
${\rm Fe_{1.03}Te_{0.63}Se_{0.37}}$\footnotemark[1] & 13 & 560(20) & TDR & \onlinecite{Kim2010} & 0.20\\
${\rm FeTe_{0.5}Se_{0.5}}$\footnotemark[1] & 14.6 & 491(8) & $\mu$SR & \onlinecite{Bendele2010} & 0.26 & 0.60\footnotemark[6]\\
&  & $\lambda_c: $1320(14) & $\mu$SR & \onlinecite{Bendele2010}\\
${\rm FeTe_{0.5}Se_{0.5}}$\footnotemark[2] & 14.6 & 534\footnotemark[3]--703 & $\mu$SR & \onlinecite{Biswas2010} & \\
${\rm RbFe_2As_2}$\footnotemark[2] & 2.5 & 267(5) & $\mu$SR & \onlinecite{Shermadini2010}\\
${\rm Ba_{0.6}K_{0.4}Fe_2As_2}$\footnotemark[1] & 37 & 200(8) & IR & \onlinecite{Li2008c}&&0.41\\
${\rm Ba(Fe_{0.961}Co_{0.039})_2As_2}$\footnotemark[1] & ? & 905(40) & TDR & \onlinecite{Gordon2010}\\
${\rm Ba(Fe_{0.961}Co_{0.039})_2As_2}$\footnotemark[1] & ? & 670(40) & TDR & \onlinecite{Gordon2010}\\
${\rm Ba(Fe_{0.942}Co_{0.058})_2As_2}$\footnotemark[1] & ? & 170(40) & TDR & \onlinecite{Gordon2010}\\
${\rm Ba(Fe_{0.939}Co_{0.061})_2As_2}$\footnotemark[1] & 23.6 & 189.4(11) & $\mu$SR & \onlinecite{Williams2010}\\
${\rm Ba(Fe_{0.926}Co_{0.074})_2As_2}$\footnotemark[1] & 22.1 & 224.2(6) & $\mu$SR & \onlinecite{Williams2010} && 0.16\\
${\rm Ba(Fe_{0.900}Co_{0.100})_2As_2}$\footnotemark[1] & ? & 170(40) & TDR & \onlinecite{Gordon2010}&&0.33\\
${\rm Ba(Fe_{0.899}Co_{0.101})_2As_2}$\footnotemark[1] & 14.1 & 332.2(22) & $\mu$SR & \onlinecite{Williams2010}\\
${\rm Ba(Fe_{0.890}Co_{0.110})_2As_2}$\footnotemark[1] & 10.3 & 453.8(26) & $\mu$SR & \onlinecite{Williams2010}\\
${\rm Ba(Fe_{0.95}Co_{0.05})_2As_2}$\footnotemark[1] & 18.5 & 325(50) & MFM & \onlinecite{Luan2010}\\
${\rm Ba(Fe_{0.94}Co_{0.06})_2As_2}$\footnotemark[1] & 25 & 277(25)\footnotemark[5] & IR & \onlinecite{Nakajima2010}\\
${\rm Ba(Fe_{0.935}Co_{0.065})_2As_2}$\footnotemark[1] & 23.5 & 270 & IR & \onlinecite{Kim2010a}\\
${\rm Ba(Fe_{0.92}Co_{0.08})_2As_2}$\footnotemark[1] & 25 & 350(35)\footnotemark[5] & IR & \onlinecite{Wu2010}\\
${\rm Ba(Fe_{0.92}Co_{0.08})_2As_2}$\footnotemark[1] & 20 & 315(30)\footnotemark[5] & IR & \onlinecite{Nakajima2010}\\
${\rm Ba(Fe_{0.9}Co_{0.1})_2As_2}$\footnotemark[2]  & 23 & 450(20) & THz  &\onlinecite{Fischer2010} & 0.18\footnotemark[4]\\
${\rm Ba(Fe_{0.95}Ni_{0.05})_2As_2}$\footnotemark[1] & 20 & 300(30)\footnotemark[5] & IR & \onlinecite{Wu2010}\\
${\rm Sr(Fe_{0.87}Co_{0.13})_2As_2}$\footnotemark[1] & 16.2 & 325.5(5) & $\mu$SR & \onlinecite{Williams2010}\\
LaFeAsO$_{0.925}$F$_{0.075}$\footnotemark[2] & 22 & 364\footnotemark[3]--447 & $\mu$SR & \onlinecite{Luetkens2008}\\
LaFeAsO$_{0.9}$F$_{0.1}$\footnotemark[2] & 26.0 & 254\footnotemark[3]--333 & $\mu$SR & \onlinecite{Luetkens2008}\\
SmFeAsO$_{1-x}$F$_x$\footnotemark[1] & 44 & 140(20) & TDR & \onlinecite{Malone2009}\\
SmFeAsO$_{0.8}$F$_{0.2}$\footnotemark[1] & 46.0 & 210(30) & TM & \onlinecite{Weyeneth2009}\\
SmFeAsO$_{1-x}$F$_x$\footnotemark[2] & 45 & 190\footnotemark[3] & $\mu$SR & \onlinecite{Drew2008}\\
LiFeAs\footnotemark[1] & 17 & 210(20) & SANS & \onlinecite{Inosov2010a}\\
\end{tabular}
\end{ruledtabular}
\footnotetext[1]{Single crystal sample.}
\footnotetext[2]{Polycrystalline sample.}
\footnotetext[3]{Calculated for the high-anisotropy limit.}
\footnotetext[4]{Calculated from the complex $\sigma(\omega)$.}
\footnotetext[5]{From a fit containing two Drude terms; this value is for the narrow Drude term.}
\footnotetext[6]{From the 1/$\tau^\prime$ obtained in a low-frequency limit of a generalized Drude fit at 18~K.  A much smaller value of $\rho_{\rm S}/n$ is obtained if the 1/$\tau^\prime$ value from a conventional Drude-Lorentz fit at 100~K is used.}
\end{table*}

\begin{figure}%Fig 79
\includegraphics [width=3.3in]{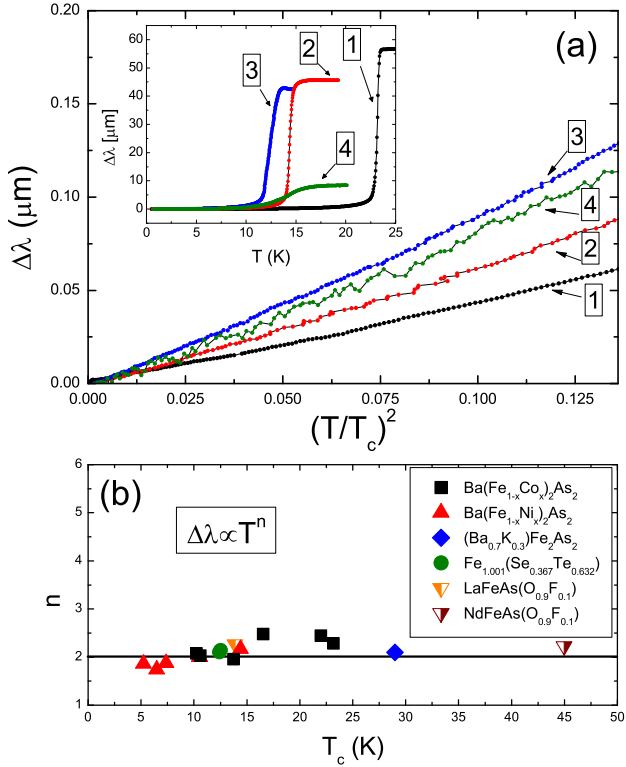}
\caption{(Color online)  (a) Change $\Delta \lambda$  in the London penetration depth $\lambda$ versus the square of the tempererature $T^2$ for a variety of Fe-based superconductors, normalized to the square of the superconducting transition temperature $T_{\rm c}$.  The codes for the plots are: (1) ${\rm Ba(Fe_{0.942}Co_{0.058})_2As_2}$; (2) ${\rm Ba(Fe_{0.941}Co_{0.059})_2As_2}$; (3) ${\rm Fe_{1.001}(Te_{0.632}Se_{0.367})}$; and (4) LaFeAs${\rm O_{0.9}F_{0.1}}$.  Inset: $\Delta\lambda$ versus $T$ up to temperatures above the respective $T_{\rm c}$s.  (b) Fitted exponent $n$ in the power-law dependence $\Delta\lambda = AT^n$ for $T\to 0$ for the various compounds as indicated. A fitted value $n \gtrsim 4$ to the data would be consistent within the errors with an exponential temperature dependence, but that is not observed.  Reprinted with permission from Ref.~\onlinecite{Gordon_Fig1_0912.5346}.  Copyright (2010) by the American Physical Society.}
\label{Gordon_Fig1_0912.5346}
\end{figure}

A survey and comparison of the low-$T$ behaviors of the penetration depths in a variety of Fe-based superconductors measured by the TDR technique is shown in Fig.~\ref{Gordon_Fig1_0912.5346}(a).\cite{Gordon_Fig1_0912.5346}  As exemplified in the figure, all Fe-based crystals examined using TDR by Prozorov's group all had power law behaviors with exponents around 2 at low temperatures as shown in Fig.~\ref{Gordon_Fig1_0912.5346}(b).  In particular, none showed exponential temperature dependences even at the lowest temperatures (R. Prozorov, private communication).  This group developed an Al-coating technique for obtaining absolute values of $\lambda_{ab}(0)$ from the TDR technique, which they used to measure $\lambda_{ab}(0)$ of Ba(Fe$_{1-x}$Co$_x)_2$As$_2$ crystals.\cite{Gordon2010}  They found that $\lambda_{ab}(0)$ is approximately constant ($\approx 260$~nm) for $0.047 \leq x \leq 0.10$, but was much larger (640--920~nm) for $x = 0.039$ and 0.042.

\begin{figure}
\includegraphics [width=3.3in]{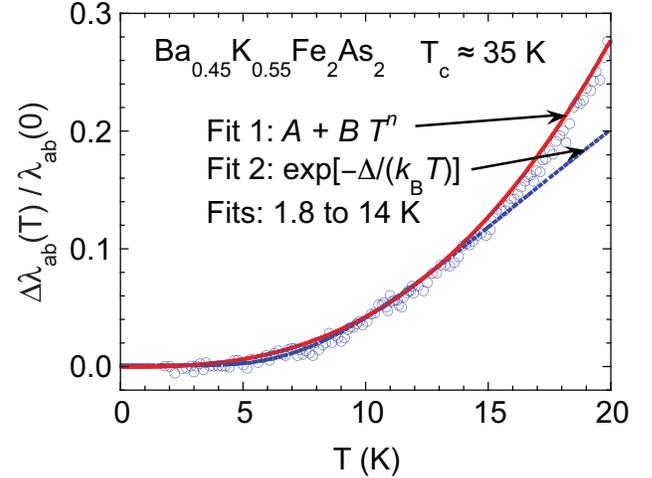}
\caption{(Color online) Change $\Delta\lambda_{ab}(T)$ in the $ab$-plane London penetration depth divided by the zero-temperature value $\lambda_{ab}(0)$ versus temperature $T$ for a single crystal of ${\rm Ba_{0.45}K_{0.55}Fe_2As_2}$ (open blue circles).\cite{Hashimoto2009}  Fits to the data from 1.8 to~14~K by Fits~1 and~2 in Eq.~(\ref{LambdaFitEqs}) with fitting parameters in Eqs.~(\ref{HashFitdata}) are shown as solid red and dotted blue curves, respectively.}
\label{Hashimoto_Fig3_FitPlot}
\end{figure}

\begin{figure}
\includegraphics [width=3.2in,viewport=10 00 560 370,clip]{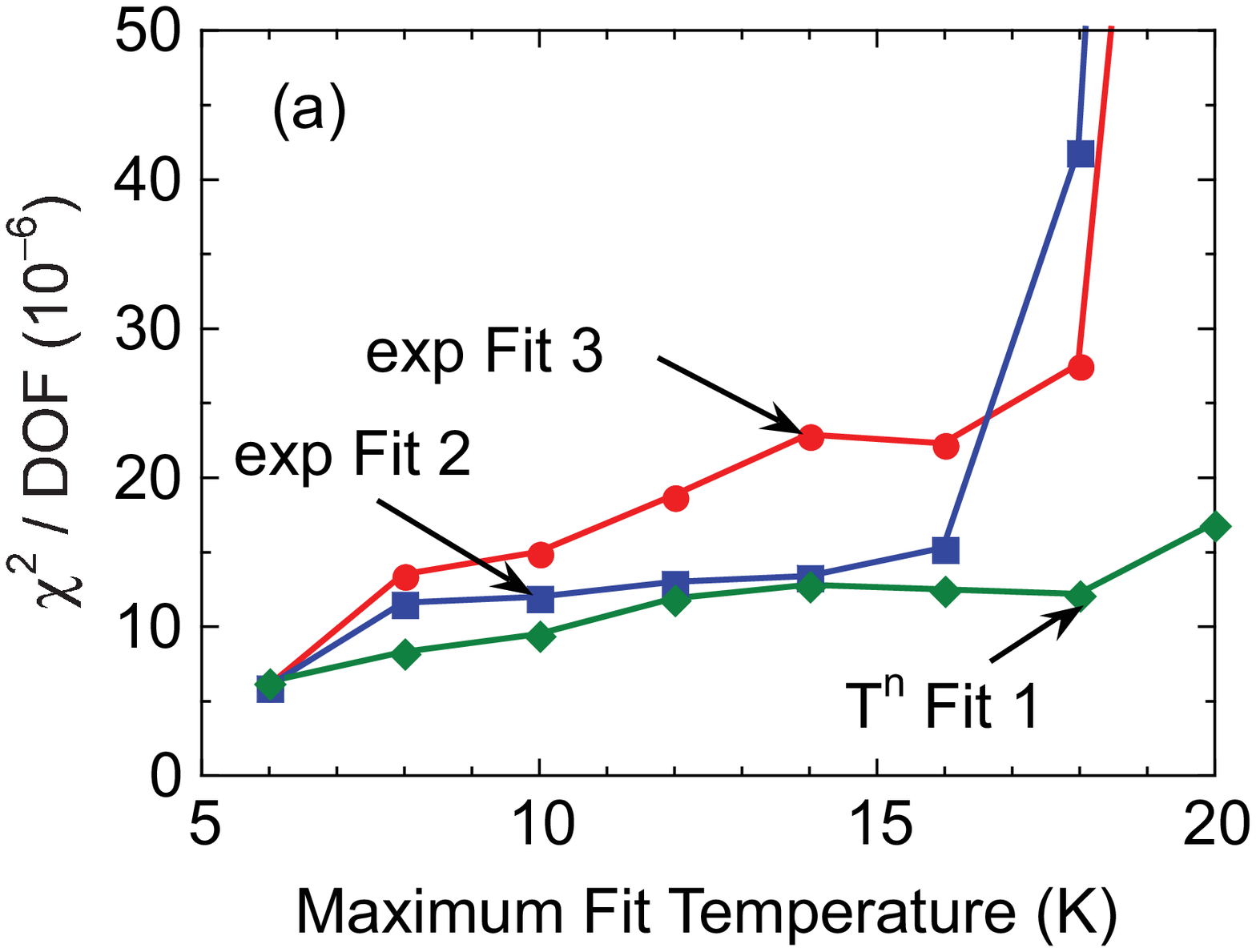}
\includegraphics [width=3.3in]{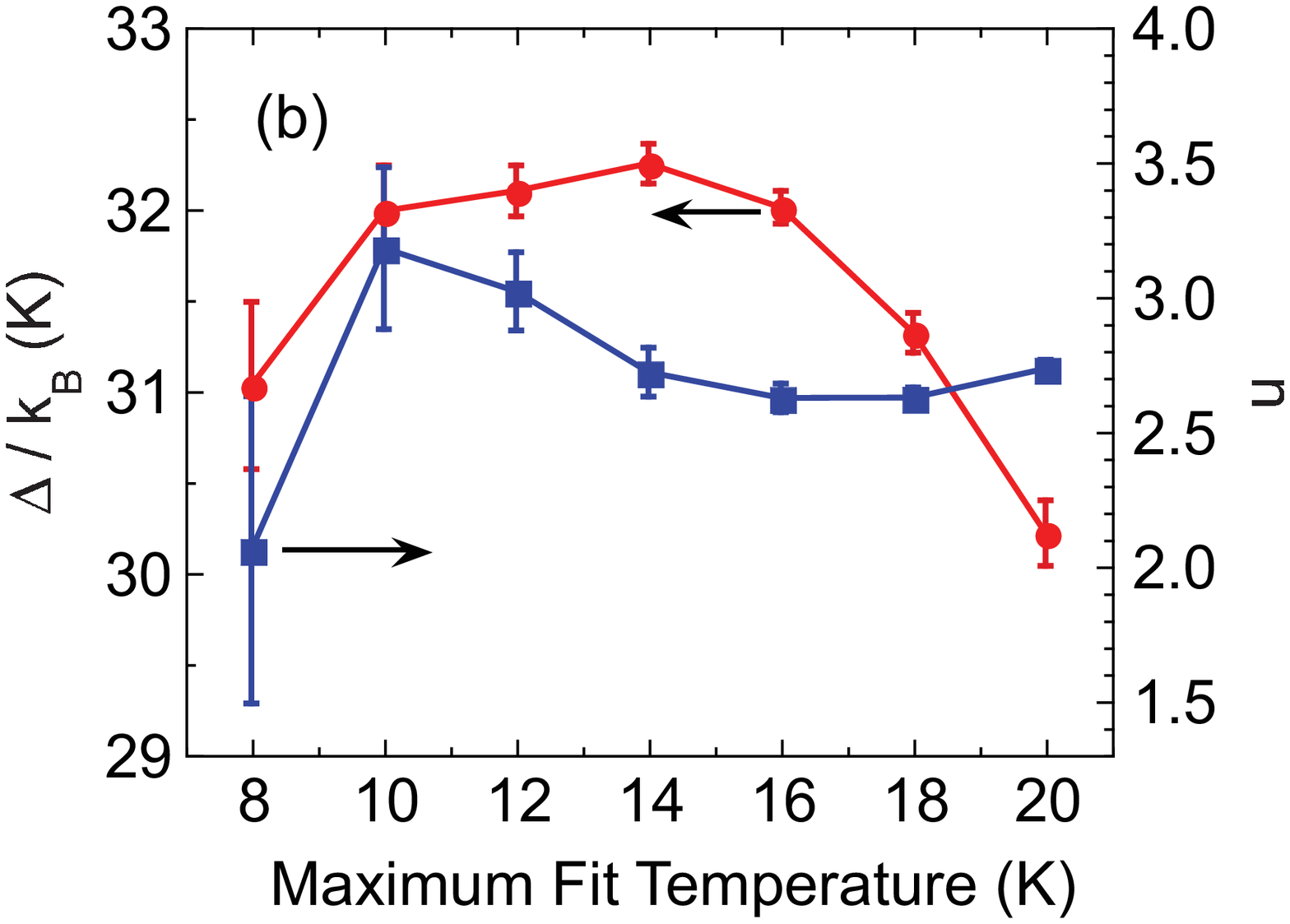}
\caption{(Color online)  (a) Goodness of fit $\chi^2$/DOF for the three fitting expressions in Eqs.~(\ref{LambdaFitEqs}) to the data in Fig.~\ref{Hashimoto_Fig3_FitPlot} over temperature ranges from 1.8~K to the indicated maximum fit temperature, where the latter ranged from 6~K to 20~K\@.  From panel (a), the exp~Fit~2 and $T^n$~Fit~1 fitted the data equally well for maximum fitting temperatures less than 16~K\@.  (b) Fitting parameter $\Delta/k_{\rm B}$ for exp~Fit~2 (left vertical scale) and $n$ from $T^n$~Fit~1 (right vertical scale) versus the maximum fit temperature.}
\label{Hashimoto_Fit_pars}
\end{figure}

Hashimoto and coworkers carried out in-plane penetration depth $\lambda_{ab}$ measurements on Ba$_{0.45}$K$_{0.55}$Fe$_2$As$_2$ crystals ($T_{\rm c} = 25$--33~K) with different impurity concentrations using a 28~GHz cavity perturbation technique and have suggested an exponential dependence of the penetration depth at low temperatures in the cleanest crystal (\#3), but found power law-like dependences in two less pure crystals.\cite{Hashimoto2009}  We digitized the penetration depth data for the cleanest crystal \#3 from the inset of Fig.~3 of Ref.~\onlinecite{Hashimoto2009} and the data are plotted as open circles in Fig.~\ref{Hashimoto_Fig3_FitPlot}.  Then we carried out both exponential and power law fits to the data over various temperature ranges.  The functions fitted to the data were
\bea
T^n~{\rm Fit~1:}\hspace{0.2in} \frac{\Delta\lambda_{ab}(T)}{\Delta\lambda_{ab}(0)} &=& A + BT^n \nonumber\\
{\rm exp~Fit~2:} \hspace{0.2in} \frac{\Delta\lambda_{ab}(T)}{\Delta\lambda_{ab}(0)} &=& \exp[-\Delta_0/(k_{\rm B}T)]\label{LambdaFitEqs}\\
{\rm exp~Fit~3:} \hspace{0.2in} \frac{\Delta\lambda_{ab}(T)}{\Delta\lambda_{ab}(0)} &=& \sqrt{\frac{\pi\Delta_0}{2k_{\rm B}T}}\exp[-\Delta_0/(k_{\rm B}T)],\nonumber
\eea
where $\Delta\lambda \equiv \lambda(T) - \lambda(1.8$~K), $\Delta_0 \equiv \Delta(T = 0)$ and exp~Fit~2 and exp~Fit~3 are single-parameter ($\Delta_0/k_{\rm B}$) fitting functions approximately valid at $T/T_{\rm c}\lesssim 1/2$ for the dirty and clean limits, respectively.\cite{Halbritter1971}  The power law Fit~1 function instead has three adjustable parameters $A$, $B$ and $n$.  

We fitted all three expressions in Eqs.~(\ref{LambdaFitEqs}) to the experimental data in Fig.~\ref{Hashimoto_Fig3_FitPlot} over variable temperature ranges from 1.8~K up to a maximum fit temperature from 6 to 20~K $\approx T_{\rm c}/2$.  The goodness of fit was computed as $\chi^2$/DOF, where $\chi^2/{\rm DOF} = (1/{\rm DOF})\sum_{i = 1}^{N_{\rm fit}}(y_i -y_{i~{\rm fit}})^2$, $N_{\rm fit}$ is the number of data points fitted, and the number of degrees of freedom DOF equals $N_{\rm fit}$ minus the number of fitting parameters.  The $\chi^2$/DOF values for the three expressions in Eqs.~(\ref{LambdaFitEqs}) fitted to the data in Fig.~\ref{Hashimoto_Fig3_FitPlot} over temperature ranges from 1.8~K to the indicated maximum fit temperature are plotted in Fig.~\ref{Hashimoto_Fit_pars}(a).  The $T^n$~Fit~1 and exp~Fit~2 functions fitted the data equally well for maximum fitting temperatures less than 16~K and both are superior to  exp~Fit~3.  For a maximum fitting temperature of 14~K, the important fitting parameters are
\bea
n&=& 2.72(9)\hspace{0.2in}(T^n{\rm ~Fit~1)}\nonumber\\
\label{HashFitdata}\\
\frac{\Delta_0}{k_{\rm B}} &=& 32.3(2)~{\rm K}\hspace{0.2in}{\rm (exp~Fit~2)}.\nonumber
\eea

The second of Eqs.~(\ref{HashFitdata}) gives $2\Delta_0/(k_{\rm B}T_{\rm c}) \approx 1.9$, which is much less than the BCS value of 3.52.  This suggests that if the fit is valid, then there are multiple superconducting gaps in the material with one of them significantly less than predicted by the BCS theory, consistent with ARPES and other data discussed above.  On the other hand, the power law Fit~1 function fits the data equally well, with a power $n$ that is significantly less than expected ($\geq 4$) if the data followed an exponential or two-fluid power law temperature dependence.  However, the experimental data are not sufficiently precise to distinguish between activated and power law temperature dependences.

Most studies of the temperature dependence of $\lambda$ require the presence of two distinct gaps to fit the data, similar to those found by ARPES above.  A summary of the literature values of the two gaps found from different types of measurements, including penetration depth measurements, is given by Evtushinsky et al.\cite{Evtushinsky2009}  A survey of optical and THz measurements of the superconducting gap(s) and penetration depths in the Fe-based materials is given by Dressel et al.\cite{Dressel2010}

\subsubsection*{c. Superconducting Condensate Density}

\begin{figure}
\includegraphics [width=3.3in]{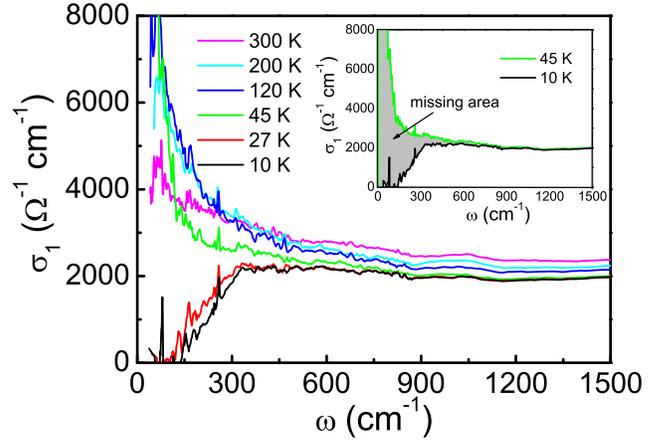}
\caption{(Color online) In-plane optical conductivity $\sigma_1$ versus frequency $f^\prime$ (denoted ``$\omega$'' in the figure) at the indicated temperatures for a single crystal of ${\rm Ba_{0.6}K_{0.4}Fe_2As_2}$ with $T_{\rm c} = 37$~K.\cite{Li2008c}  A superconducting energy gap is seen to develop below $T_{\rm c}$ in the frequency range below 310~cm$^{-1}$.  The inset shows the ``missing area'' $A$ in Eqs.~(\ref{EqMissArea}), (\ref{EqRhos}) and~(\ref{EqLambdaA}).  Reprinted with permission from Ref.~\onlinecite{Li2008c}.  Copyright (2008) by the American Physical Society.}
\label{LiFig3PRL}
\end{figure}

When a material becomes superconducting, at least some fraction of the conduction carriers condense into the superconducting state.  The corresponding part of the optical conductivity spectral weight is transferred into a Dirac delta function at $\omega = 0$ (Ref.~\onlinecite{Tinkham1975}) that is not accessible from optics measurements due to a lower limit on the measured frequency range.  This missing spectral weight (``missing area $A$'') in a plot of $\sigma_1$ versus $\omega$ in cgs units is simply defined by
\be
A = \int_0^\infty\sigma^{\rm n}_1(\omega)d\omega -\int_0^\infty\sigma^{\rm s}_1(\omega)d\omega ,
\label{EqMissArea}
\ee
where $\sigma^{\rm n}_1(\omega)$ and $\sigma^{\rm s}_1(\omega)$ are the real parts of the electronic conductivity in the normal and superconducting states, respectively.  The missing spectral weight generally occurs in the far infrared part of the electromagnetic spectrum where the energy of a photon is less than the superconducting energy gap $\hbar\omega < 2\Delta$.  Measuring the frequency range over which suppression of the optical conductivity occurs allows an estimate of the superconducting energy gap to be determined.   Defining $\rho_{\rm S}$ as the number density of normal state current carriers that have condensed into the superconducting state, from Eqs.~(\ref{eqnomega1}) and~(\ref{EqMissArea}) one gets\cite{Tinkham1975}
\be
\rho_{\rm S} = \frac{2m^*}{\pi e^2}A,
\label{EqRhos}
\ee
where $m^*$ is the effective mass of the individual quasiparticles.  

Thus by measuring the low-frequency optical conductivity spectral weights above and below $T_{\rm c}$ and subtracting them, one can determine the superconducting carrier density.  Because the measured spectral weights are single particle spectral weights, this $\rho_{\rm S}$ is the density of single conduction carriers  in the superconducting condensate, not the density of Cooper pairs which is a factor of two smaller.  Using the expression relating $\rho_{\rm S}$ to the magnetic penetration depth $\lambda$,\cite{Tinkham1975}
\be
\rho_{\rm S} = \frac{m^* c^2}{4\pi e^2\lambda^2},
\label{Eqnslambda}
\ee
one can also derive $\lambda$ from $A$ using Eqs.~(\ref{EqRhos}) and~(\ref{Eqnslambda}), yielding the simple relation\cite{Tinkham1975}
\be
\frac{1}{\lambda} = \frac{\sqrt{8A}}{c},
\label{EqLambdaA}
\ee
in which $m^*$ has dropped out, so Eq.~(\ref{EqLambdaA}) allows an unambiguous measure of $\lambda$.  

We emphasize that the superconducting condensate density $\rho_S$ is not just another name for the inverse square of the penetration depth.  It is a distinct, physically meaningful, and measurable quantity via Eq.~(\ref{EqRhos}).  For example, since $\rho_S$ is a scalar, which is obviously isotropic, the anisotropy in $\lambda$ for ${\rm FeTe_{0.5}Se_{0.5}}$ in Table~\ref{LambdaTable} presumably results from the anisotropy in $m^*$.  Also, from Eqs.~(\ref{lambdaLambdaL}) and~(\ref{Eqnslambda}), a reduction in the mean-free path towards the dirty limit evidently reduces $\rho_S$, assuming that $m^*$ remains constant.  Finally, we remark that the expression~(\ref{EqRhos}) does not depend on the nature of the scattering centers in a material, i.e., whether they are pair-breaking or not.

\begin{figure}
\includegraphics [width=3.3in]{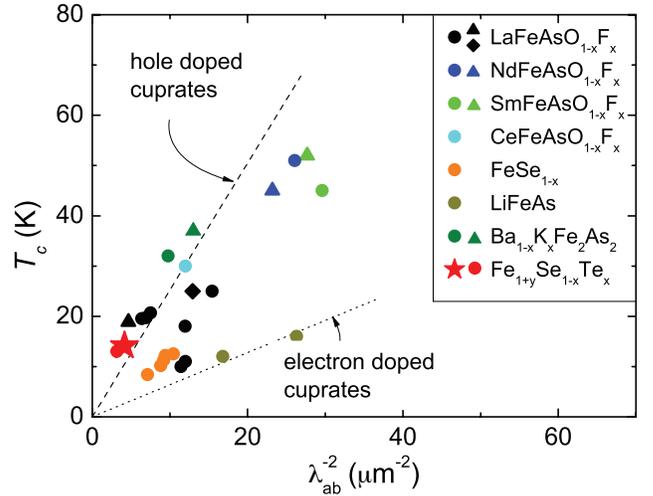}
\caption{(Color online) Uemura plot of the superconducting transition temperature $T_{\rm c}$ versus the inverse square of the zero-temperature in-plane London penetration depth, $\lambda_{ab}^{-2}$, for various Fe-based compounds.  The red star datum for Fe$_{1+y}$Se$_{1-x}$Te$_x$ is from Ref.~\onlinecite{Bendele2010}.  The references for the other data points plotted are cited in Ref.~\onlinecite{Bendele2010}.  Reprinted with permission from Ref.~\onlinecite{Bendele2010}. Copyright (2010) by the American Physical Society.}
\label{Bendele_Fig11}
\end{figure}

An early example of relevant $ab$-plane optical conductivity data is shown in Fig.~\ref{LiFig3PRL} for a single crystal of ${\rm Ba_{0.6}K_{0.4}Fe_2As_2}$ with $T_{\rm c} = 37$~K.\cite{Li2008c}  The data clearly show the opening of a superconducting gap below $T_{\rm c}$. The $\sigma_1$ is zero to within the errors for $f^\prime  < 150~$cm$^{-1}$, suggesting a full $s$-wave-like superconducting gap $2\Delta \approx 150$~cm$^{-1} = 19$~meV = 220~K, which gives $2\Delta/k_{\rm B}T_{\rm c} = 5.9$.\cite{Li2008c}  This $2\Delta$ corresponds to the smaller of the two gaps seen by ARPES, as expected.\cite{Li2008c}  The ``missing area'' $A$ in Eqs.~(\ref{EqMissArea}), (\ref{EqRhos}) and~(\ref{EqLambdaA}) is shown in the inset of Fig.~\ref{LiFig3PRL}.  From the value of $A$, $\lambda = 200(8)$~nm was determined from Eq.~(\ref{EqLambdaA}).\cite{Li2008c}  Several values of $\lambda$ determined this way from optics measurements on other Fe-based superconductors are also listed above in Table~\ref{LambdaTable}.  These values are similar to values determined from other types of measurements on comparable materials.

A plot of the $T_{\rm c}$ versus the in-plane $1/\lambda_{ab}(0)^2 \sim \rho_{\rm S}/m^*$ (``Uemura plot''\cite{Uemura1991}) for Fe-based superconductors by Bendele and coworkers is shown in Fig.~\ref{Bendele_Fig11}.\cite{Bendele2010}  A generally positive correlation is seen between the two quantities, as was previously observed for the cuprate superconductors as indicated by dashed and dotted lines in the figure, which demonstrates that $T_{\rm c}$ increases with $\rho_{\rm S}/m^*$.

An improved correlation between $T_{\rm c}$ and $\rho_{\rm S}$ was suggested for the cuprates by Homes et al., namely $\rho_{\rm S} \propto \sigma_0 T_{\rm c}$,\cite{Homes2004, Homes2005} where $\sigma_0$ is the dc conductivity just above $T_{\rm c}$.  Wu and coworkers\cite{Wu2010} and Li et al.\cite{Li2008c} found that this correlation works well for their samples of Fe-based superconductors.

Using Eq.~(\ref{Eqnslambda}) and the expression for the plasma angular frequency $\omega_{\rm p}$ of the normal state coherent current carriers given above in Eq.~(\ref{EqOmegaP}), one can derive the ratio of the zero-temperature superconducting condensate density $\rho_{\rm S}(0)$ to the normal state coherent carrier density $n$ as
\be
\frac{\rho_{\rm S}(0)}{n} = \left[\frac{c}{\lambda(0)\,\omega_{\rm p}}\right]^2.
\label{EqRhoSomegap}
\ee
Note that the factor $m^*$ has dropped out of the expression, the right-hand side of the expression contains only two reasonably well-defined experimental quantities, and the same expression is valid in both cgs and SI units.  Using Eq.~(\ref{Eqfpp}), one can rewrite the dimensionless ratio~(\ref{EqRhoSomegap}) in commonly used units as
\be
\frac{\rho_{\rm S}(0)}{n} = \left[(2\pi \times 10^{-7})\,\lambda(0)\,f^\prime_{\rm p}\right]^{-2},
\label{EqRhoSomegap2}
\ee
where $\lambda(0)$ is in units of nm and $f^\prime_{\rm p} = \omega_{\rm p}/(2\pi c)$ is the plasma frequency expressed in cgs optics units of cm$^{-1}$ as in Tables~\ref{Opticsdata2} and~\ref{Opticsdata3} above.  For simple conventional superconductors in the clean limit, one would expect this ratio to be close to unity, because all of the conduction carriers are expected to become superconducting.  However, in correlated electron materials and/or superconductors containing  impurities that scatter the current carriers, it is of interest to examine how this ratio varies between compounds because of the possibility that not all current carriers contribute to the superconducting condensate.  

The ratio $\rho_{\rm S}(0)/n$ as determined using Eq.~(\ref{EqRhoSomegap2}) (``method~1'') is given in the next-to-last column of Table~\ref{LambdaTable} for several samples.  These values suggest that a significant fraction of the current carriers do not contribute to the superconducting condensate.  This was also previously suggested by Homes and coworkers who measured the in-plane optical properties versus temperature of a superconducting single crystal of FeTe$_{0.55}$Se$_{0.45}$ with $T_{\rm c} = 14$~K and deduced that only about 18\% of the current carriers became superconducting,\cite{Homes2010} similar to values of $\rho_{\rm S}(0)/n$ for Fe$_{1+y}$Te$_{1-x}$Se$_{x}$ in the next-to-last column of Table~\ref{LambdaTable}.  They inferred that the small ratio resulted because the compound is approaching the dirty limit, where the carrier scattering rate becomes significant compared to the superconducting energy gap $2\Delta$ for quasiparticle excitations.  

\begin{figure}
\includegraphics [width=3.in]{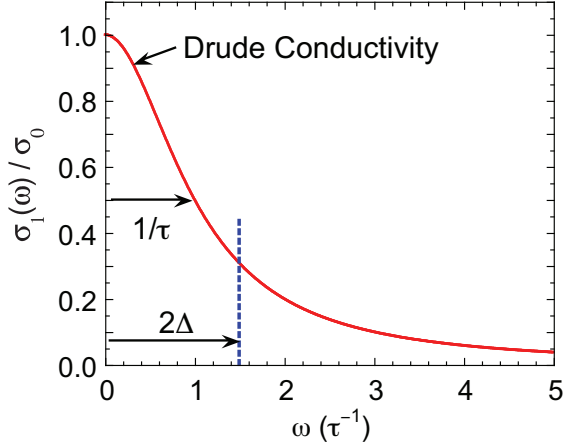}
\caption{(Color online) Real part $\sigma_1$ of the Drude optical conductivity, normalized by the zero frequency value $\sigma_0$, versus angular frequency $\omega$ in units of (1~rad)/$\tau$ (solid red curve). The value of $\omega$ that is equal to (1 rad)/$\tau$ is shown by the upper horizontal arrow.  Note that $\hbar\omega$ is the magnitude of the energy of a quasiparticle as measured from the Fermi energy.  Also shown by the lower horizontal arrow and vertical blue dashed line is an arbitrary value $2\Delta/\hbar$ of the superconducing quasiparticle energy gap in angular frequency units of rad/s.  When a material becomes superconducting, the optical spectral weight for $0 \leq \omega \leq 2\Delta/\hbar$ collapses into a Dirac delta function at $\omega = 0$.  The superconducting carrier density $\rho_{\rm S}$ is thus proportional to this spectral weight, which decreases as $2\Delta/\hbar$ decreases.}
\label{SC_density}
\end{figure}

An explanation follows of how the superconducting condensate density decreases as the quasiparticle scattering rate \emph{from non-pairbreaking impurities} increases in an effective single band model, with assistance from Raphael Fernandes and Christopher Homes (private communications).  Shown in Fig.~\ref{SC_density} is a plot of the Drude optical conductivity in Eq.~(\ref{EqDrudeSigma1}) versus angular frequency, where it is important to note that $\hbar\omega$ is the magnitude of the energy of a quasiparticle as measured from the Fermi energy.  Also shown is an arbitrary value of the quasiparticle energy gap $2\Delta/\hbar$ in angular frequency units of rad/s, and of the angular frequency (1~rad)/$\tau$ corresponding to the fixed mean scattering rate.  When a material becomes superconducting, the quasiparticles within an energy $\Delta$ on either side of the Fermi surface condense, by definition, into the superconducting state at $T = 0$ (here we do not consider pair-breaking impurities that introduce quasiparticle states into the superconducting gap, see e.g.\ Ref.~\onlinecite{Kogan2010}).  In terms of Fig.~\ref{SC_density}, this means that the optical spectral weight up to $\omega = 2\Delta/\hbar$ condenses at $T = 0$ into a Dirac delta function at $\omega = 0$.  By inspection of the figure, if $2\Delta/\hbar \gg 1/\tau$, all of the carriers (with total concentration $n$) condense.  However, as also seen from the figure, as $2\Delta/\hbar$ decreases, so does the superconducting condensate density.  Using Eqs.~(\ref{EqDrudeSigma1}), (\ref{EqSigma3}) and~(\ref{eqnomega1}), one obtains
\bea
\frac{\rho_{\rm S}(T = 0)}{n} &=& \frac{2m^*}{\pi e^2 n}\int_0^{2\Delta(0)/\hbar} \sigma_1(\omega)d\omega \nonumber\\
&=& \frac{2}{\pi}\arctan\left[\frac{2\tau\Delta(0)}{\hbar}\right]\label{EqRhosn}\\
&=& \frac{2}{\pi}\arctan\left[\frac{4\pi\Delta(0)}{h/\tau}\right].\nonumber
\eea
The clean limit corresponds to $2\Delta(0)\tau/\hbar \gg 1$, whereas the dirty limit corresponds to $2\Delta(0)\tau/\hbar \ll 1$.  These criteria are essentially the same as the criteria for the clean and dirty limits given above in Eq.~(\ref{EqCleanDirty}).  For the clean limit, one obtains $\rho_{\rm S}(0) = n$ from Eq.~(\ref{EqRhosn}), as expected.  In the dirty limit $2\Delta/\hbar \ll 1/\tau$, one instead obtains
\be
\frac{\rho_{\rm S}(0)}{n} = \frac{8\Delta(0)}{h/\tau}  \ll 1.\hspace{0.2in}{\rm (dirty\ limit)}
\label{EqDirtyRhos}
\ee

Using Eq.~(\ref{Eqtaup1}), one can express the quasiparticle scattering rate $1/\tau$, which is in units of s$^{-1}$ in Eq.~(\ref{EqRhosn}), in terms of the scattering rate $1/\tau^\prime$ expressed in optics papers in units of cm$^{-1}$.  Then Eq.~(\ref{EqRhosn}) becomes
\be
\frac{\rho_{\rm S}(0)}{n} = \frac{2}{\pi}\arctan\left[16.13\frac{\Delta(0)}{1/\tau^\prime}\right],
\label{EqRhosn2}
\ee
where on the right-hand-side $\Delta(0)$ is in meV and $1/\tau^\prime$ is in cgs optics units of cm$^{-1}$.  Using the largest $\Delta(0)$ datum for a given compound in Table~\ref{SCGapValues} and the $1/\tau^\prime$ data in Table~\ref{Opticsdata2}, several values of $\rho_{\rm S}/n$ were computed using Eq.~(\ref{EqRhosn2}) and are listed in the last column of Table~\ref{LambdaTable} (``method 2'').  

All of the values in the last two columns of Table~\ref{LambdaTable} are significantly smaller than unity.  This indicates that the Fe-based superconductors are approaching the dirty limit.  This quasiparticle scattering mechanism of reducing $\rho_{\rm S}(T = 0)/n$ below unity leaves normal state conduction carriers that do not condense into the superconducting state.  These uncondensed quasiparticles do not exhibit low-energy excitations such as a Sommerfeld electronic heat capacity at low temperatures $T \ll T_{\rm c}$, because according to Fig.~\ref{SC_density} their spectral weight is at thermally inaccessible energies $\hbar\omega > 2\Delta$ for non-pair-breaking impurities.

From their data in Table~\ref{LambdaTable}, Williams et al.\ found that $1/\lambda_{\rm ab}^2(0)\propto T_{\rm c}^2$ for single crystals in the Ba(Fe$_{1-x}$Co$_x)_2$As$_2$ system, instead of varying linearly with the doping concentration $x$, indicating that the superconducting condensate density decreases as the composition $x$ deviates from the optimum value $x_0$.\cite{Williams2010} They suggested that either spatial phase separation into superconducting and normal regions occurs on a fine scale with increasing $x - x_0$, or that there is phase separation in momentum space where some fraction of the current carriers remains normal below $T_{\rm c}$.\cite{Williams2010}  On the other hand, Gordon et al.\ did not observe a decrease in $\rho_{\rm S}(0)$ for $x > x_0$ in their Ba(Fe$_{1-x}$Co$_x)_2$As$_2$ crystals up to $x = 0.10$.\cite{Gordon2010}  They attributed their observed strong decrease in $\rho_{\rm S}(0)$ for $x < x_0$ in the region of coexistence of superconductivity and SDW (see Fig.~\ref{FigBaKFe2As2_phase_diag} and Table~\ref{LambdaTable}) to competition between these two transitions for the same conduction carriers,\cite{Gordon2010} as predicted theoretically.\cite{Fernandes2010b}

\subsubsection{NMR $1/T_1$ Measurements below $T_{\rm c}$}

\begin{figure}
\includegraphics [width=3.3in]{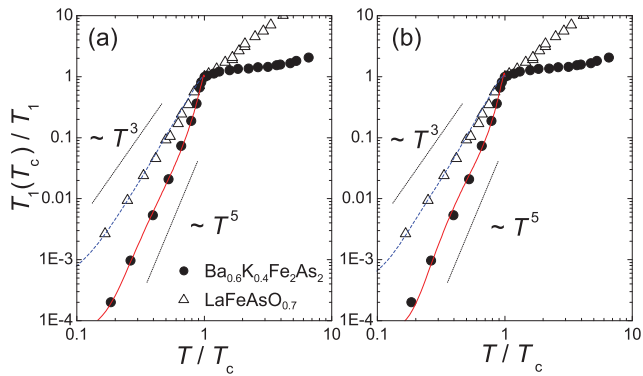}
\caption{(Color online).  $^{57}$Fe nuclear spin-lattice relaxation rate $1/T_1$, divided by its value $1/T_1(T_{\rm c})$ at $T_{\rm c}$, versus temperature $T$ divided by $T_{\rm c}$, for polycrystalline samples of ${\rm Ba_{0.6}K_{0.4}Fe_2As_2}$ and LaFeAsO$_{0.7}$.  Different multiple superconducting gap models in panels (a) and (b), respectively, are used to fit the data for the two compounds as shown by the solid curves.  Reproduced by permission from Ref.~\onlinecite{Yashima2009}.    Copyright (2009) by the Physical Society of Japan.}
\label{Yashima_Fig4}
\end{figure}

Many NMR measurements of the nuclear spin-lattice relaxation rate $1/T_1$ of different nuclei have been carried out below $T_{\rm c}$ for various Fe-based superconductors.  Almost all of them show a power law behavior $1/T_1 \sim T^n$ with $n \sim 3$--5 instead of an exponential dependence expected if there were no nodes in the superconducting order parameter in momentum space.  Shown in Fig.~\ref{Yashima_Fig4} are illustrative  $^{57}$Fe nuclear spin-lattice relaxation rate $1/T_1$ measurements versus $T$ of polycrystalline samples of ${\rm Ba_{0.6}K_{0.4}Fe_2As_2}$ and LaFeAsO$_{0.7}$.\cite{Yashima2009}  The data can be fitted by power laws with $n = 4.4$ and 3, respectively, as shown by the straight lines with slopes corresponding to $n = 5$ and 3 in the figure, suggesting the absence of full gaps in these compounds.  However, the large exponent of 4.4 for ${\rm Ba_{0.6}K_{0.4}Fe_2As_2}$ suggests the presence of a conventional $s$-wave gap.  Indeed, both data sets were fitted by two different models, each with multiple full gaps as shown by the solid curves through the respective data sets in Fig.~\ref{Yashima_Fig4}(a) and~(b).\cite{Yashima2009}

The influence of multiple gap values on $1/T_1$ would be exacerbated by interband scattering between the electron and hole pockets by nonmagnetic impurities which could introduce quasiparticle states within the superconducting gap or even make the superconductor gapless.
\cite{DParker2008, Chubukov2008, Bang2009, Dolgov2009, Vorontsov2009, Onari2009, Matsumoto2009}  
For $s^\pm$ pairing, nonmagnetic impurities are ``pairbreaking'', acting like magnetic impurities in a conventional $s$-wave superconductor.  Thus, these impurities also reduce $T_{\rm c}$ in the $s^\pm$ pairing scenario, in addition to introducting quasiparticle states in the superconducting gap.  Conversely, \emph{intraband} scattering off \emph{magnetic} impurities are pairbreaking whereas \emph{interband} scattering is not with little effect on pairing or $T_{\rm c}$.\cite{JLi2009}

\begin{figure}
\includegraphics [width=3.in]{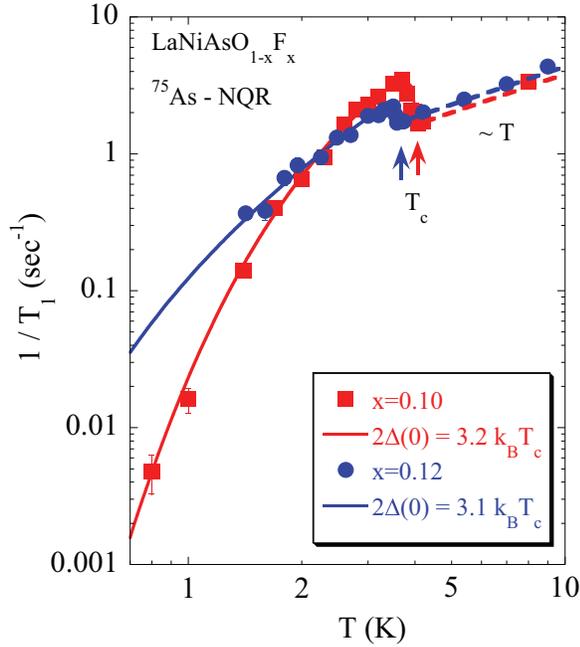}
\caption{(Color online).  $^{75}$As nuclear spin-lattice relaxation rate $1/T_1$ versus temperature $T$ for two polycrystalline samples of LaNiAsO$_{1-x}$F$_x$.  The respective $T_{\rm c}$'s for the two samples are indicated by vertical arrows.  Each sample shows a ``Hebel-Slichter peak'' at $T_{\rm c}$ that is indicative of conventional $s$-wave electron-phonon pairing.  Each solid curve is a fit to the respective data by an isotropic ($s$-wave) superconducting gap with the indicated value.  These data and fits indicate that not all Fe-based superconductors show unconventional behavior of $1/T_1$ below $T_{\rm c}$. Reprinted with permission from Ref.~\onlinecite{Tabuchi2010}.  Copyright (2010) by the American Physical Society.}
\label{TabuchiFig2}
\end{figure}

Not all superconductors crystallizing in the 1111- or 122-type structures show such ambiguities in behavior of $1/T_1$ below $T_{\rm c}$ shown above in Fig.~\ref{Yashima_Fig4}.  Data for two polycrystalline samples of LaNiAsO$_{1-x}$F$_x$ are plotted in Fig.~\ref{TabuchiFig2}.\cite{Tabuchi2010} The data show a peak at $T_{\rm c}$ for each sample, called the Hebel-Slichter peak, which is a signature of a conventional $s$-wave superconducting state.  The data follow exponential temperature dependences below $T_{\rm c}$, indicated as solid curves in the figure,  with superconducting energy gaps $2\Delta$ that are close to the BCS value of 3.53~$k_{\rm B}T_{\rm c}$.

\subsubsection{Summary}

The available evidence indicates that the intrinsic superconductivity in the FeAs-based compounds involves spin-singlet $s$-wave-like ($s^\pm$-wave) pairing with multiple gaps on the different electron and hole Fermi surface sheets in momentum space, but where the minimum superconducting gap may vary strongly with momentum component $k_z$ of the Fermi surface(s).   The non-exponential temperature dependences of the penetration depth and $1/T_1$ at low temperatures, and the lack of a clear superconducting gap  with no in-gap states in the scanning tunneling spectroscopy measurements discussed in Sec.~\ref{SecRSInhomo} below, can be ascribed in this scenario to the influence of these  multiple superconducting gaps in the $k_x$-$k_y$ plane together with a modulation of the superconducting gap in the $k_z$-direction, possibly exacerbated by chemical/spatial inhomogeneity, or impurity scattering effects that introduce quasiparticle states into the minimum superconducting gap.

\subsection{\label{Hc2} Upper Critical Field $H_{\rm c2}$ and Superconducting Coherence Length $\xi$}

%\squeezetable
\begin{table}
\caption{\label{Hc2Table} Upper critical magnetic fields $H_{\rm c2}$ measured at the indicated temperatures $T$ for single crystals of Fe-based superconductors with onset superconducting transition temperature $T_{\rm c}$.  The notation $H_{c2,ab}$ means that the applied magnetic field is in the $ab$-plane, and $H_{c2,c}$ means that the applied field is parallel to the $c$-axis.}
\begin{ruledtabular}
\begin{tabular}{l|ccccc}
 Compound   & $T_{\rm c}$ & $T$ & $H_{c2,c}$ & $H_{c2,ab}$  & Ref. \\
 & (K) & (K) & (T) & (T) &   \\ \hline
${\rm NdFeAsO_{0.7}F_{0.3}}$ & 47 & 35 & 9\footnotemark[1] & 54\footnotemark[1] & \onlinecite{Jaroszynski2008}\\
                             & 47 & 20 & 30\footnotemark[1] & --- & \onlinecite{Jaroszynski2008}\\
${\rm Ba_{0.55}K_{0.45}Fe_{2}As_{2}}$\footnotemark[2] & 32 & 14 & 57 & 68 & \onlinecite{Altarawneh2008}\\
${\rm Ba_{0.6}K_{0.4}Fe_{2}As_{2}}$ & 28.2 & 10 & 56 & 57 &   \onlinecite{Yuan2009}\\
${\rm Ba(Fe_{0.92}Co_{0.08})_2As_{2}}$ & 22.1 & 1 & 50\footnotemark[1] & 53\footnotemark[1] & \onlinecite{Kano2009}\\
${\rm Ba(Fe_{0.9}Co_{0.1})_2As_{2}}$ & 22 & 10 & 41\footnotemark[1] & 25\footnotemark[1] & \onlinecite{Yamamoto2009}\\
${\rm Sr(Fe_{0.9}Co_{0.1})_2As_{2}}$ & 20 & 1 & 46\footnotemark[3] & 46\footnotemark[3] & \onlinecite{Baily2009}\\
${\rm KFe_2As_2}$ & 2.79 & 0.3 & 1.25\footnotemark[1] & 4.40\footnotemark[1] & \onlinecite{Terashima2009a}\\
${\rm Fe_{1.05}(Te_{0.85}Se_{0.15})}$ & 14.1 & 1.5 & 46\footnotemark[3] & 45\footnotemark[3] & \onlinecite{Kida2010}\\
${\rm Fe_{1.11}(Te_{0.6}Se_{0.4})}$ & 14\footnotemark[4] & 1.5 & 43\footnotemark[1] & 42\footnotemark[1] & \onlinecite{Fang2010}\\
${\rm Fe(Te_{0.6}Se_{0.4})}$ & 14\footnotemark[5] & 3 & 46\footnotemark[1] & 46\footnotemark[1] & \onlinecite{Khim2010}\\
${\rm Fe_{1.14}(Te_{0.91}S_{0.09})}$ & 7.8\footnotemark[6] & 0.2 & 27\footnotemark[1] & 26\footnotemark[1] & \onlinecite{Lei2010}\\
${\rm Fe(Te_{0.52}S_{0.48})}$ & 15 & 1.5 & 47\footnotemark[1] & 44\footnotemark[1] & \onlinecite{Braithwaite2010}\\
\end{tabular}
\end{ruledtabular}
\footnotetext[1]{50\% normal resistance criterion for $T_{\rm c}$.}
\footnotetext[2]{Contaminated with Sn from the Sn flux used to grow the crystals.}
\footnotetext[3]{90\% normal resistance criterion for $T_{\rm c}$ ($T_{\rm c}$ onset).}
\footnotetext[4]{The superconducting transition width measured by zero-field-cooled magnetic susceptility in a low (30~Oe)  field along the $c$-axis is comparable to $T_{\rm c}$ itself.}
\footnotetext[5]{The superconducting transition width measured by zero-field-cooled magnetic susceptility in a low (10~Oe) field along the $ab$-plane is about 2~K and the superconducting diamagnetism at 3~K is about 75\% of $-4\pi\chi$.}
\footnotetext[6]{The zero-field resistive transition width (1.3~K) was rather large.}
\end{table}

The Fe-based superconductors are found to be Type-II superconductors.  The upper critical fields $H_{\rm c2}$ of all of them have been measured to some extent, although $H_{\rm c2}(T = 0)$ turns out to be of order 60~T or higher at $T \to 0$, so most laboratory fields are not sufficient to obtain the full $H_{\rm c2}(T)$ behaviors.  Here, except for ${\rm KFe_2As_2}$ which has low $H_{\rm c2}$,  we only consider $H_{\rm c2}$ measurements on single crystals that attain fields $H \gtrsim 50$~T, because these give definitive information on the superconducting anisotropy at low temperatures as well as its temperature dependence.  

Several values of $H_{\rm c2}(T)$ are collected in Table~\ref{Hc2Table}.\cite{Baily2009, Jaroszynski2008, Altarawneh2008, Yuan2009, Kano2009, Yamamoto2009, Terashima2009a, Kida2010, Fang2010, Khim2010, Lei2010, Braithwaite2010}  For ${\rm Ba_{0.6}K_{0.4}Fe_{2}As_{2}}$, the ratio 
\[\gamma \equiv \frac{H_{c2,ab}}{H_{c2,c}}
\]
 decreases monotonically with decreasing temperature from a value of $\approx 2$ near $T_{\rm c}$ to a value of $\approx 1$ at 10~K\@.\cite{Yuan2009}  The data thus indicate that this compound is a nearly isotropic superconductor for $T \to 0$ with respect to $H_{\rm c2}$.  Using the expression\cite{Tinkham1975}
\be
H_{\rm c2} = \frac{\Phi_0}{2\pi \xi^2},
\ee
where $\Phi_0 = hc/(2e) = 2.07 \times 10^{-7}$~G~cm$^2$ is the flux quantum, one obtains superconducting coherence lengths $\xi(10~{\rm K}) \approx 2.4$~nm.  The extrapolated zero-temperature upper critical field is $H_{\rm c2}(T\to 0) \sim 70$~T\@.  For ${\rm Sr(Fe_{0.9}Co_{0.1})_2As_{2}}$ and ${\rm Ba(Fe_{0.92}Co_{0.08})_2As_{2}}$ with $T_{\rm c} = 20$--22~K, the measured $\gamma \approx 1$ at 1~K.\cite{Baily2009, Kano2009}  Thus the 122-type superconductors are isotropic at $T \ll T_{\rm c}$, despite their layered crystal structure.  A similar behavior is seen for Fe$_{1+y}$(Te$_{1+x}$Se$_x$).  However, as noted in the table, the superconducting transition width of ${\rm Fe_{1.11}(Te_{0.6}Se_{0.4})}$ from magnetic measurements at low field can be about the same as $T_{\rm c}$ itself,\cite{Fang2010} so in these cases the reliability of these data is uncertain.  The data of Khim et al.\ for this system show that $\gamma$ decreases from $\approx 3$ near $T_{\rm c} = 14$~K to 0.99 at 3.8~K\@.\cite{Khim2010}

The upper critical fields for ${\rm NdFeAsO_{0.7}F_{0.3}}$ with $T_{\rm c} = 47$~K are so high, and $H_{\rm c2,c}$ shows positive curvature over the observed $T$ range, that no accurate extrapolation of $H_{\rm c2}$ to low temperatures can be made.\cite{Altarawneh2008}  The authors quote $\gamma = 9.2$ at 44~K and 5 at 23~K\@.

\subsection{\label{Sec_SC_Cp} Specific Heat Jump at $T_{\rm c}$}

\begin{figure}
\includegraphics [width=3.3in]{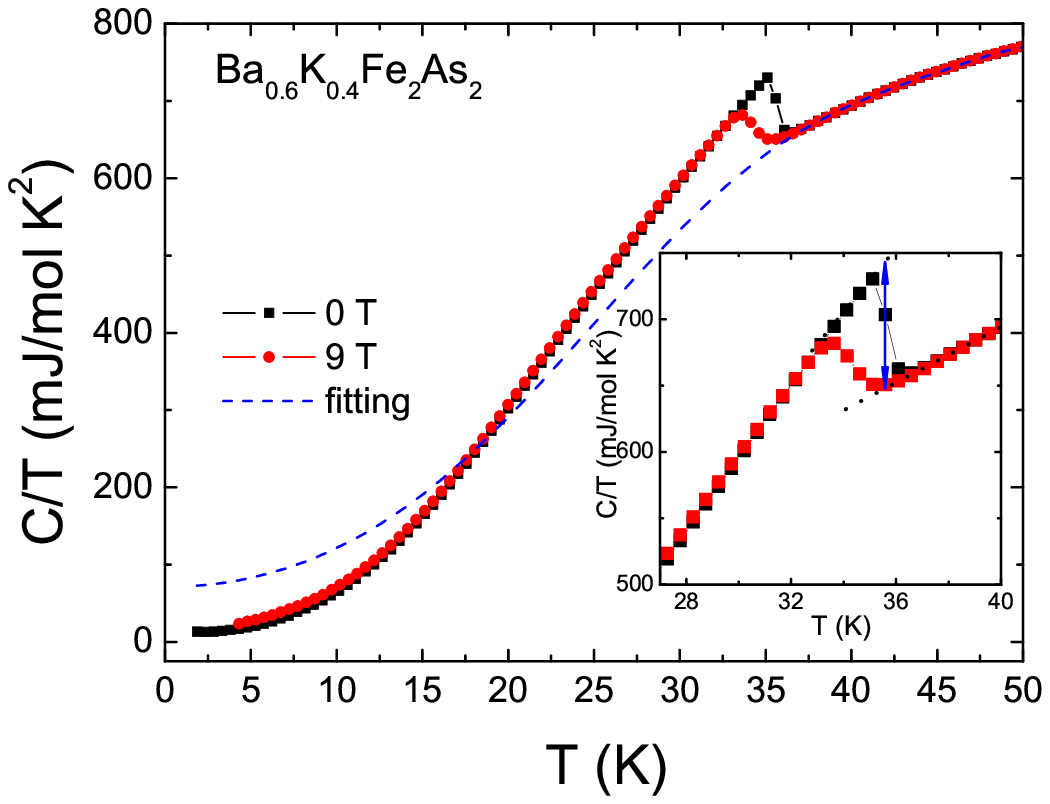}
\includegraphics [width=3.1in]{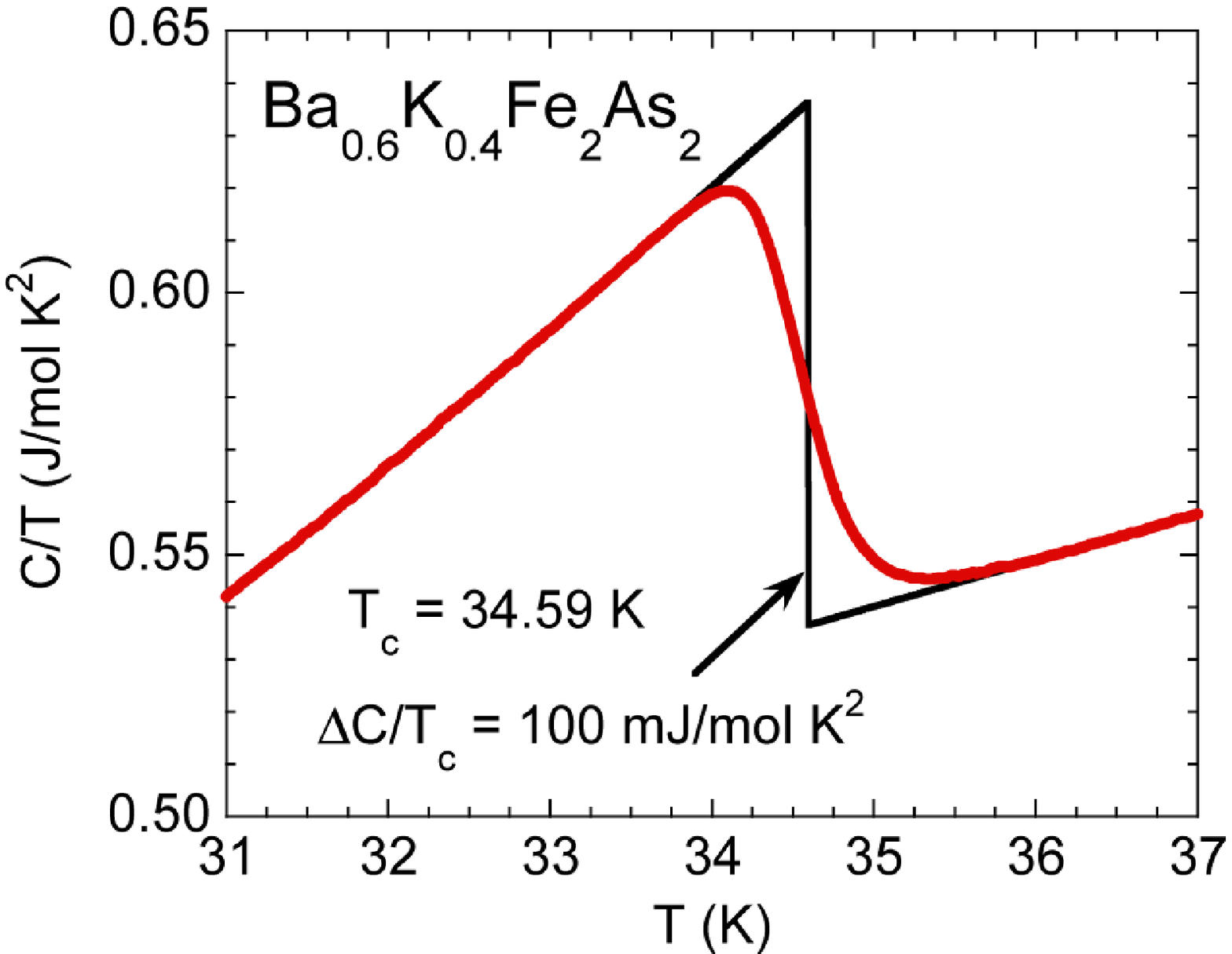}
\caption{(Color online) Top panel: Heat capacity divided by temperature $C/T$ versus temperature $T$ for a single crystal of Ba$_{0.6}$K$_{0.4}$Fe$_{2}$As$_{2}$ and applied magnetic fields of zero and 9~T.\cite{Mu2008} The dashed blue line is an estimate of the normal state heat capacity, where the electronic specific heat coefficient is $\gamma = \lim_{T \to 0}C(T)/T = 63.3$~mJ/mol~K$^2$.\cite{Mu2008}  This normal state heat capacity satisfies the constraint that the entropies of the normal and superconducting states must be the same at $T_{\rm c}$.\cite{Mu2008}  The inset shows an expanded plot of the data near $T_{\rm c} \approx 36$~K\@.  Bottom panel: Expanded plot of the data around $T_{\rm c}$ for a similar crystal.\cite{Welp2009}  The straight black lines show an entropy-conserving construction to determine $T_{\rm c}$ and the idealized heat capacity jump $\Delta C/T_{\rm c}$ at $T_{\rm c}$.\cite{Welp2009}  Reprinted with permission from Refs.~\onlinecite{Mu2008} and~\onlinecite{Welp2009}.  Copyrights (2009) by the American Physical Society.}
\label{FigBaKFe2As2Cp}
\end{figure}

An important thermal property associated with the onset of superconductivity in a material is the ``jump'' $\Delta C$ in the electronic component of the heat capacity $C$ at $T_{\rm c}$.  A nearly discontinuous increase in $C$ at $T_{\rm c}$ only occurs if a sample has a single well-defined $T_{\rm c}$ that is not smeared out due to inhomogeneities in the sample.  For Fe-based superconductors, this ideal behavior is rare.  The nearly ideal type of behavior was observed for single crystals of Ba$_{0.6}$K$_{0.4}$Fe$_{2}$As$_{2}$ as shown in the top panel of Fig.~\ref{FigBaKFe2As2Cp}.\cite{Mu2008}  A similarly sharp transition in $C(T)$ at $T_{\rm c}$ for a polycrystalline sample of the same composition was found in Ref.~\onlinecite{Rotter2009}.  The heat capacity data for a crystal of the same material as studied in Ref.~\onlinecite{Mu2008} is shown in the bottom panel of the figure,\cite{Welp2009} where the idealized vertical heat capacity jump at $T_{\rm c}$ is seen to be $\Delta C/T_{\rm c} = 100$~mJ/mol~K$^2$.  BCS theory predicts that $\Delta C/(\gamma T_{\rm c}) = 1.43$.  The available values of the bare band structure density of states for Ba$_{1-x}$K$_{x}$Fe$_{2}$As$_{2}$ for $x$=0.4--0.5 are 3.9 to 6.2 states/(eV f.u.) for both spin directions,\cite{Wang2009, Shein2008c, djsingh2008} which from Eq.~(\ref{EqGamma0}) give $\gamma_0 = 9.2$ to 14.6 mJ/mol~K$^2$, which in turn give  $\Delta C/(\gamma_0 T_{\rm c}) = 6.8$--10.9, far larger than the BCS value of 1.43.  Such large values of $\Delta C/(\gamma T_{\rm c})$ have not been previously observed for any superconductor, which suggests that the actual $\gamma$ value is much larger than the bare band structure prediction $\gamma_0$.  

Indeed, from an analysis of superconducting state data, Mu \emph{et al.}\ concluded that the normal state Sommerfeld coefficient value is $\gamma = 63.3$~mJ/mol~K$^2$ as shown by the $T = 0$ intercept of the dashed blue line in the top panel of Fig.~\ref{FigBaKFe2As2Cp},\cite{Mu2008} which is 4--5 times larger than the bare band structure prediction, and gives a ratio $\Delta C/\gamma T_{\rm c} = 1.58$ that is close to the BCS value of 1.43.  The dashed blue normal state $C/T$ curve also satisfies the requirement that the superconducting state entopy and the normal state entropy are the same at $T_{\rm c}$.\cite{Mu2008}  Fukazawa et al.\ found that the end member KFe$_{2}$As$_{2}$ has a similarly large normal state $\gamma = 69.1$~mJ/mol~K$^2$,\cite{Fukazawa2009} which is a factor of 6.8 larger than predicted\cite{Terashima2010} from LDA band calculations.  This large discrepancy between the experimental $\gamma$ and the bare band structure $\gamma_0$ points towards the presence of an some mechanism for enhancing the experimental $\gamma$ far above $\gamma_0$.  In view of the antiferromagnetic spin fluctuations observed at and above $T_{\rm c}$ that are converted below $T_{\rm c}$ into the resonance mode observed in neutron scattering, it seems likely that the enhancement of $\gamma_0$ by the necessary factor of 4--7 is due to interaction of the current carriers with antiferromagnetic spin fluctuations, which are also implicated in the superconducting mechanism (see Sec.~\ref{Sec_SC_Mech} below).  From analysis of optical data for a single crystal of Ba(Fe$_{0.92}$Co$_{0.08})_2$As$_2$, Wu et al.\ discovered that this mechanism can indeed explain the enhancement.\cite{Wu2010a}  

Subsequent work has confirmed these experimental findings for the Ba$_{1-x}$K$_{x}$Fe$_{2}$As$_{2}$ system and has been extended to the Eu$_{1-x}$K$_{x}$Fe$_{2}$As$_{2}$ system.  A large value $\Delta C/T_{\rm c} = 70$~mJ/mol~K$^2$ was obtained by Jeevan and Gegenwart for polycrystalline Eu$_{0.5}$K$_{0.5}$Fe$_{2}$As$_{2}$ with $T_{\rm c}=32$~K.\cite{Jeevan2010}  Storey and coworkers obtained $\Delta C/T_{\rm c} = 48$~mJ/mol~K$^2$ for polycrystalline Ba$_{0.7}$K$_{0.3}$Fe$_{2}$As$_{2}$ with $T_{\rm c}=35$~K.\cite{Storey2010}  They also inferred the normal state electronic specific heat coefficient $\gamma = 47$~mJ/mol~K$^2$, yielding $\Delta C/(\gamma T_{\rm c}) = 1.02$ which is significantly smaller than the BCS value of 1.43.  Popovich and coworkers obtained a very large $\Delta C/T_{\rm c} = 120$~mJ/mol~K$^2$ for a 13.6~mg single crystal of Ba$_{0.68}$K$_{0.32}$Fe$_{2}$As$_{2}$ with a very sharp superconducting transition at $T_{\rm c} = 38.5$~K.\cite{Popovich2010}  They also inferred $\gamma = 50$~mJ/mol~K$^2$, yielding $\Delta C/(\gamma T_{\rm c}) = 2.4$, indicating strong-coupling superconductivity.  Furthermore, their electronic specific heat data below $T_{\rm c}$ indicated two-gap superconductivity, consistent with the above ARPES measurements in Sec.~\ref{ARPESSC}.  As pointed out by Popovich et al., and as seen in the above and in many other sets of data, the value of $\Delta C/T_{\rm c}$ is evidently sensitive to the homogeneity and perfection of the sample, which also determines the associated width of the superconducting transition.\cite{Popovich2010}

As mentioned above in Sec.~\ref{SecThermCond} with reference to the right-hand scale of Fig.~\ref{Reid_Kappa0}, the $C(T)$ in the vicinity of $T_{\rm c}$ has been studied for single crystalline samples with different values of $x$ and $T_{\rm c}$ in the systems BaFe$_{2-x}$Co$_x$As$_2$ and BaFe$_{2-x}$Ni$_x$As$_2$ by Bud'ko and coworkers.\cite{Budko2009}  Plotting these values for their crystals together with literature values for FeAs-based superconductors, they found that $\Delta C/T_{\rm c} \sim T_{\rm c}^2$.   Kogan explained this result theoretically in terms of pairbreaking effects,\cite{Kogan2009a} whereas Zaanen proposed that this result arises from ``a normal state that is a quantum critical metal undergoing a pairing instability.''\cite{Zaanen2009}  An alternative interpretation is that this result arises because the superconducting fraction decreases as $T_{\rm c}$ decreases, as discussed next.

\begin{figure}
\includegraphics [width=3.in]{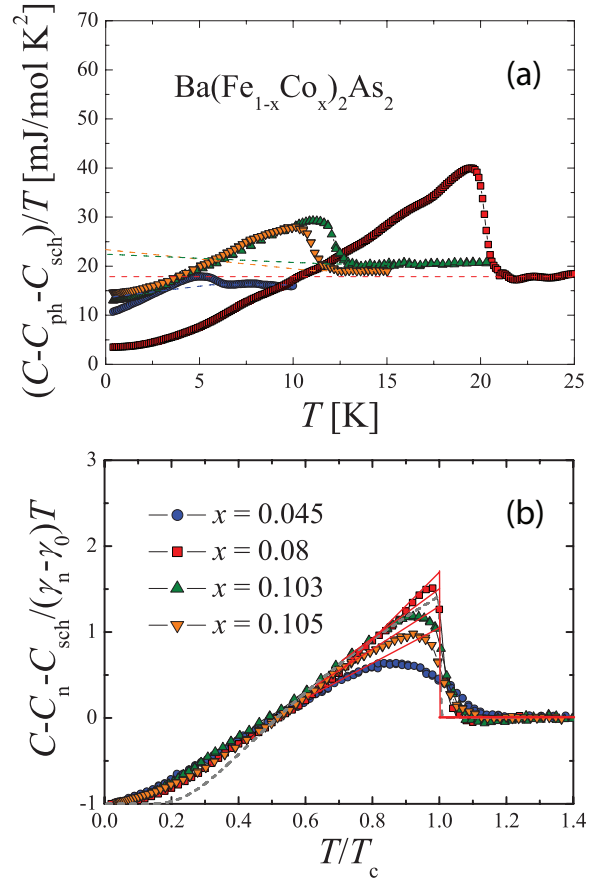}
\caption{(Color online) (a) Heat capacity $C$, corrected for the lattice $C_{\rm ph}$ and Schottky $C_{\rm Sch}$ contributions, divided by temperature $T$, versus $T$ for single crystals of Ba(Fe$_{1-x}$Co$_x)_2$As$_2$ with compositions as indicated.  (b)  $(C - C_{\rm n} - C_{\rm Sch})/T$ divided by $\gamma_{\rm n} - \gamma_0$, versus $T$ where $C_{\rm n}$ is the normal state electronic ($\gamma_{\rm n}T$) plus phonon contribution, $\gamma_{\rm n}$ is the normal state Sommerfeld coefficient and $\gamma_{\rm 0}$ is the measured $T \to 0~\gamma$ value in zero magnetic field.  Reprinted with permission from Ref.~\onlinecite{Gofryk2010}.  Copyright (2010) by the American Physical Society.}
\label{GofrykFigs1_3}
\end{figure}

\begin{figure}
\includegraphics [width=3.in]{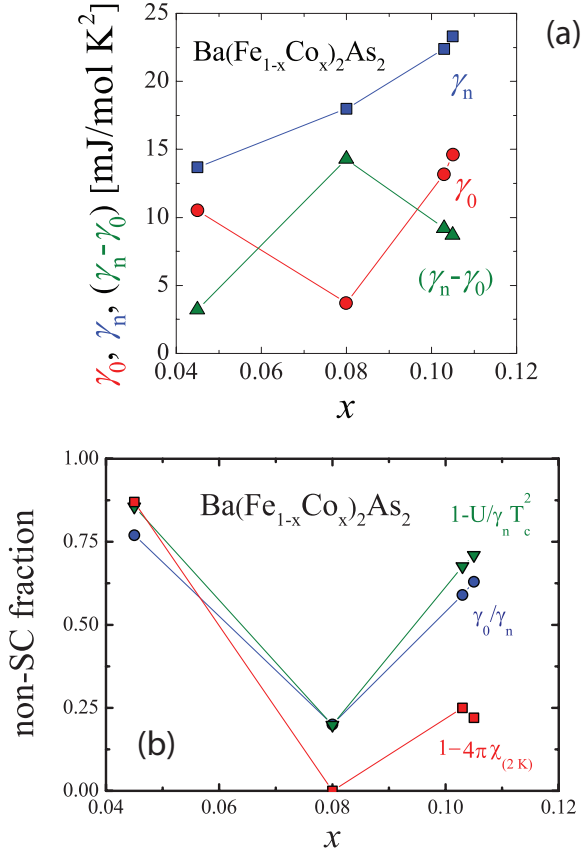}
\caption{(Color online) (a) Normal state Sommerfeld coefficient $\gamma_{\rm n}$, measured $T \to 0$ value $\gamma_{\rm 0}$ and the difference $\gamma_{\rm n} - \gamma_0$ versus composition $x$ of Ba(Fe$_{1-x}$Co$_x)_2$As$_2$ single crystals.  (b) Non-superconducting volume fraction of the crystals versus composition $x$, as deduced from different measurements as indicated.  Reprinted with permission from Ref.~\onlinecite{Gofryk2010}.  Copyright (2010) by the American Physical Society.}
\label{GofrykFigs1_2}
\end{figure}

Gofryk and coworkers have made a detailed study of the low-temperature heat capacity of single crystals of Ba(Fe$_{1-x}$Co$_x)_2$As$_2$ from the underdoped to the overdoped regime.\cite{Gofryk2010}  They corrected their data for the lattice contribution of ${\rm BaFe_2As_2}$ and for a small Schottky-like upturn at the lowest temperatures.  The corrected heat capacity data constitute the electronic contribution and are shown in Fig.~\ref{GofrykFigs1_3}(a).  A pronounced zero-temperature Sommerfeld coefficient $\gamma_0$ appeared in the superconducting state for both underdoped and overdoped compositions, as plotted in Fig.~\ref{GofrykFigs1_2}(a).  Even the optimum doping concentration gives a nonzero $\gamma_0$.  The normal state $\gamma_{\rm n}$ values were determined by entropy balance as indicated by the dashed lines in Fig.~\ref{GofrykFigs1_3}(a).  Then, the data in Fig.~\ref{GofrykFigs1_3}(a) were replotted as $[C - C_{\rm n} - C_{\rm Sch}](T)/T$, normalized by the part $\gamma_{\rm s} \equiv \gamma_{\rm n} - \gamma_0$ of $\gamma$ associated with the superconductivity, as shown in Fig.~\ref{GofrykFigs1_3}(b) where the data are plotted versus $T/T_{\rm c}$.  In these normalized plots, $\Delta C/\gamma_{\rm s}T_{\rm c}$ associated with the superconducting electrons can be read off the plots.  Thus, for the optimally doped composition $x = 0.08$, one sees that $\Delta C/\gamma_{\rm s}T_{\rm c} \approx 1.6$ which is close to the BCS value of 1.43, decreasing to 1.1 for the broad transition for $x = 0.045$.  From the data in Fig.~\ref{GofrykFigs1_3}(b), it appears that the $T_{\rm c}$ is developing a distribution of unknown width as the composition deviates from optimum doping, which may be at least partly responsible for the drop in the value of $\Delta C/\gamma_{\rm s}T_{\rm c}$.  

Gofryk et al.\ interpreted the ratio $\gamma_0/\gamma_{\rm n}$ as an indication of the non-superconducting fraction of the respective crystals, and augmented this result by other measurements as shown in Fig.~\ref{GofrykFigs1_2}(b).\cite{Gofryk2010}  For highly underdoped and overdoped samples, one sees that this analysis indicates that the crystals are less than 50\% superconducting.  The non-superconducting fraction seems to have qualitatively the same composition dependence as the zero-temperature $c$-axis thermal conductivity and the suppression of the heat capacity jump $\Delta C/T_{\rm c}$ at $T_{\rm c}$ in Fig.~\ref{Reid_Kappa0} above, suggesting a possible causal relationship between these three quantities.  From inelastic light scattering (Raman) experiments on single crystals of the Ba(Fe$_{1-x}$Co$_x)_2$As$_2$ system, Chauvi\`ere et al.\ also found that the superfluid density showed a sharp decrease for $x$ values deviating from the optimum doping value $x = 0.065$.\cite{Chauviere2010}  

Mu and coworkers have obtained very similar results for the heat capacity versus temperature and composition of single crystals of BaFe$_{2-x}$Co$_x$As$_2$, including the nonzero value of $\gamma_0$ for the optimum doping concentration, and the minimum in $\gamma_0$ and the maximum of $\Delta C/T_{\rm c}$ at the optimum doping concentration.\cite{Mu2009}

Hardy et al.\ have carried out a detailed heat capacity $C(T)$ study of 13 crystals of Ba(Fe$_{1-x}$Co$_x)_2$As$_2$ over the composition range $0 \leq x \leq 0.203$ from $T =0.4$~K up to above $T_{\rm c}(x)$.\cite{Hardy2010b}  They isolated the electronic contribution $C_{\rm e}(T)$ by subtracting an appropriate lattice contribution.  Their findings for $C_{\rm e}(T)$ were: (i) Crystals with $x = 0,$ 0.035, 0.140, 0.153 and 0.203 showed no evidence for superconductivity; (ii) None of the crystals showed a Schottky-like upturn in $C_{\rm p}(T)$, indicating high purity; (iii) The superconducting transitions were sharp except for the superconducting end-point compositions $x = 0.04$ and~0.12, indicating good crystal homogeneity; (iv) The normal state Sommerfeld coefficient $\gamma_{\rm n}$ increases linearly from $x = 0$ to the optimum $x = 0.0575$, attributed to the suppression of the SDW gap with increasing $x$, and then decreases linearly from $x = 0.0575$ to $x = 0.20$; (v) The superconducting $\Delta C/\gamma_{\rm n}T_{\rm c}$ shows a sharp peak at the optimum doping concentration $x = 0.0575$ and strongly decreases on either side;  (vi) A residual zero-temperature $\gamma_{\rm r}$ is present for all crystals.  The minimum value is found for optimum doping, and strongly increases as $x$ approaches the endpoints of the superconducting composition range; and (vii) The superconducting state data were best fitted by a two-gap model, and the gaps were determined versus composition $x$.  For optimum doping, the gaps were $\Delta = 2.1$ and 5.1~meV.\cite{Hardy2010b}  Results (v) and (vi) are similar to those already noted above.\cite{Gofryk2010, Mu2009}  The result (iv) is consistent with the maximum $T_{\rm c}$ being associated with the maximum normal state density of states at the Fermi energy.  The decrease in $\gamma_{\rm n}$ for larger $x$ is suggested to arise because the hole Fermi pockets are being eliminated due to the electron doping, which is also consistent with the decrease in $T_{\rm c}$ in an interband coupling scenario for the mechanism.\cite{Hardy2010b} 

A critically important question in these studies is whether the superconducting and normal fractions coexist in the same volume element.  If not, what causes the spatial phase separation of superconducting and nonsuperconducting regions, and what are their sizes?  If so, is strong pair-breaking involved?

Regarding the $\Delta C/T_{\rm c}$ versus $T_{\rm c}$ of Ba(Fe$_{1-x}$Co$_x)_2$As$_2$ crystals in the above studies, it would be very interesting to see if annealing the crystals could sharpen up the superconducting transitions, especially in the lower-$T_{\rm c}$ crystals, and to see if the $\gamma_0$ values observed in the heat capacities of the superconducting crystals at low temperatures $T \ll T_{\rm c}$ could be eliminated.

Theoretically, Bang has claimed that in an $s^\pm$ pairing scenario, when the electron and hole pockets have different superconducting gaps and when impurity scattering is present, a nonzero Sommerfeld coefficient in the superconducting state $\gamma_0 = \lim_{T\to 0}C/T$ can occur, whereas due to coherence factors in the superconducting state the electronic thermal conductivity is, nonintuitively, $\lim_{T\to 0}\kappa(T)/T = 0$.\cite{Bang2010}

\subsection{\label{SecInhomo} Real Space and Momentum Space Inhomogeneities}

\subsubsection{Introduction}

Several studies have indicated the presence of spatial inhomogeneities on the nanometer scale as well as momentum space inhomogeneities in the Fe-based superconductors as discussed in the following two sections.  Spatial inhomogeneities, if present, can have potentially large impacts on the macroscopic thermodynamic (heat capacity and magnetic susceptibility), thermal and electronic transport, and superconducting properties of the samples.  The Fe-based superconductivity research community has not paid sufficient attention to these possible effects.  For example, one wonders whether such effects cause underdoped and overdoped crystals of superconducting Ba(Fe$_{1-x}$Co$_{x})_2$As$_2$ to exhibit the large Sommerfeld coefficients $\gamma_0$ observed in their specific heats in $H = 0$ for $T \to 0$ as illustrated above in Figs.~\ref{GofrykFigs1_3}(a) and~\ref{GofrykFigs1_2}(a) and, coincidentally(?), correspondingly small values of the specific heat jump $\Delta C/T_{\rm c}$ at $T_{\rm c}$ as illustrated above in Fig.~\ref{Reid_Kappa0}, broadening of the superconducting transitions as shown in Fig.~\ref{GofrykFigs1_3}(b), and a reduction in the superconducting condensate density.\cite{Williams2010}  Does the broadening of the transitions in Fig.~\ref{GofrykFigs1_3}(b) for underdoped and overdoped crystals result from a distribution of $T_{\rm c}$, and if so, how broad is the $T_{\rm c}$ distribution?  Does it extend to $T_{\rm c} = 0$, thus causing the nonzero $\gamma_0$ values in Figs.~\ref{GofrykFigs1_3}(a) and~\ref{GofrykFigs1_2}(a)?  Are the nonzero $\gamma_0$ values indicative of a large volume fraction of nonsuperconducting material in these crystals as suggested in Fig.~\ref{GofrykFigs1_2}?\cite{Gofryk2010}  Are these nonzero $\gamma_0$ values associated with itinerant conduction carriers, or are they associated with localized electrons/holes?  Are such suggested nonsuperconducing fractions or related effects relevant to explain the nonzero thermal conductivity $\kappa/T(T\to0)$ values along the $c$-axis in Fig.~\ref{Reid_Kappa0} that only occur for underdoped and overdoped crystals and not for the optimally doped composition, or do these $\kappa/T(T\to0)$ results indicate the presence of intrinsic $c$-axis nodes in the superconducting order parameter as concluded in Ref.~\onlinecite{Reid2010}?  Regarding momentum space inhomogeneities, the main issues are related to the effects of impurities on causing electronic states in the superconducting gap, the dependence of the superconducting order parameter in momentum space on its $k_z$-component, and whether there is ``phase separation'' in momentum space between superconducting and normal electrons below $T_{\rm c}$.  These are all critically important questions that remain to be resolved.

\subsubsection{\label{SecRSInhomo} Real-Space Inhomogeneities}

We have referred above to possible nanoscale phase separation between superconducting and normal regions, and/or between regions with differing superconducting condensate densities, to understand the results of heat capacity, penetration depth and superconducting condensate measurements.  Here we discuss several additional evidences for nanoscale spatial inhomogeneities.

As noted above in Sec.~\ref{SecStructOverview}, transmission electron microscopy of the $A$${\rm Fe_2As_2}$ ($A$ = Ca, Sr) compounds in the low-temperature orthorhombic structure showed the presence of twin boundaries that are separated by only 0.1--0.4~$\mu$m.\cite{Ma2009}   In addition, a nanoscale tweed pattern is found in Ca${\rm Fe_2As_2}$.\cite{Ma2009}  

Massee et al.\ examined a clean surface of a crystal of Ba(Fe$_{0.93}$Co$_{0.07})_2$As$_2$ ($T_{\rm c} = 22.0$~K) at 4.2~K using scanning tunneling microscopy (STM) and scanning tunneling spectroscopy (STS).\cite{Massee2008, Yin2009a}  The surface was prepared by cleaving the crystal at room temperature under ultra high vacuum.  No surface reconstruction was observed using low-energy electron diffraction (LEED) just after the STM/STS measurements.  The STM measurements at 4.2~K showed an irregular mottled appearance for both doped Ba(Fe$_{0.93}$Co$_{0.07})_2$As$_2$ and undoped ${\rm BaFe_2As_2}$ crystals, which the authors suggest is a result of ``partial liftoff of the Ba ions upon cleavage.''  The STS data are obtained by measuring the differential conductance $dI/dV$ versus sample-tip voltage $V$.  Two important features of the data were examined as a function of position on the surface: the superconducting gap $2\Delta$ and the zero-bias conductance, i.e., $dI/dV$ at $V=0$.  They found a wide distribution of gap values $2\Delta/k_{\rm B}T_{\rm c} = 5.3$ to~10.7, with the median value~7.4, all of which are larger than the BCS weak-coupling $s$-wave value of 3.53.  The length scale of these variations ($\sim 8$~\AA) was close to the average in-plane Co-Co distance of $\sim 10$~\AA, suggesting that the origin of these gap variations was Co clustering that inevitably occurs statistically.  The zero bias conductance was found to anticorrelate with the gap value, as expected.  

An important finding of Massee et al.\ with respect to the above penetration depth, $1/T_1$ and specific heat results in Secs.~\ref{SecGapsetc} and~\ref{Sec_SC_Cp} was that all of the STS scans showed a large nonzero zero bias conductance.\cite{Massee2008}  That is, none of the STS spectra showed a clear quasiparticle energy gap at the Fermi energy at 4.2~K, which therefore indicated the presence of a significant density of electronic states inside the superconducting gap.  Regions without a superconducting gap at all (normal regions) were not observed for this optimally doped crystal.  Similar spatial inhomogeneity of the superconducting gap and a finite density of states inside the superconducting gap were detected from STS measurements of Ba(Fe$_{0.9}$Co$_{0.1})_2$As$_2$ crystals with $T_{\rm c} = 25.3$~K in zero magnetic field by Yin et al.\cite{Yin2009}  On the other hand, Hanaguri et al.\ observed a clear quasiparticle gap from STS measurements at 0.4~K on a single crystal of ${\rm FeTe_{0.60}Se_{0.40}}$ with $T_{\rm c}\sim 14$~K.\cite{Hanaguri2010}  For comparison, a beautiful textbook example of STS of an $s$-wave superconductor with no states in the gap (Nb, $T_{\rm c} = 9.3$~K) is given in Fig.~1(b) of Ref.~\onlinecite{Pan1998} that shows electron tunneling data at 0.335~K between nonsuperconducting Au and superconducting Nb, together with the theoretical curve.  

Kim and coworkers have studied nonuniformity on the surface of optimally-doped single crystals of approximate composition ${\rm Ba(Fe_{0.925}Co_{0.075})_2As_2}$ using four-probe scanning tunneling spectroscopy and scanning electron microscopy with wavelength-dispersive x-ray compositional analysis.\cite{Kim2009}  They found uniform regions with sharp $T_{\rm c} = 22.1(2)$~K and other regions where the resistively measured transition width was several degrees.  From the microprobe compositional analysis, they found evidence for Co-concentration variations on the surface that are evidently responsible for the variations in $T_{\rm c}$.\cite{Kim2009}

\begin{figure}
\includegraphics [width=3.in]{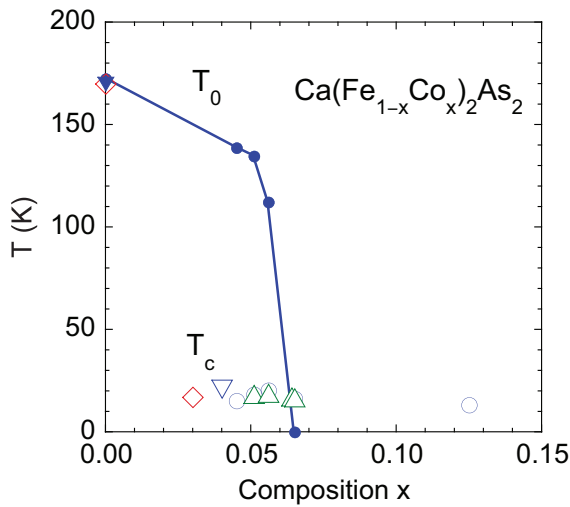}
\caption{(Color online) Preliminary phase diagram for the Ca(Fe$_{1-x}$Co$_x)_2$As$_2$ system constructed from literature data.\cite{Kumar2009, Klingeler2010, Matusiak2010, Pramanik2010}  Here $T_0$ is the crystallographic/magnetic transition temperature (filled symbols) and $T_{\rm c}$ is the superconducting transition temperature (open symbols).}
\label{Ca(FeCo)2As2_phase_diag}
\end{figure}

Chuang et al.\ have detected inhomogeneities on the surface of an underdoped ${\rm Ca(Fe_{0.97}Co_{0.03})_2As_2}$ crystal using spectroscopic imaging-scanning tunneling microscopy (SI-STM) and FT-STM (see Sec.~\ref{SecQPI}) measurements at $T = 4.3$~K.\cite{Chuang2010}  The surface showed electronic structures oriented along the $a$-axis that were about 2.2~nm long.  The authors further found that ``the delocalized electronic states detectable by quasiparticle interference (QPI, see Sec.~\ref{SecQPI}) imaging are dispersive along the $b$~axis only and are consistent with a nematic $\alpha_2$ band with an apparent band folding having (wavelength 2.2~nm) along the $a$-axis.''  The authors concluded that ``As none of these phenomena are expected merely due to crystal symmetry, underdoped ferropnictides may exhibit a more complex electronic nematic state than originally expected.''  This latter statement is puzzling, since the twofold rotational symmetry of the observed effects is the same as the twofold rotational symmetry of the bulk orthorhombic low-temperature structure (see below).  Knolle et al.\ subsequently theoretically explained the QPI peaks as arising from ``quasiparticle scattering between bands involved in the SDW formation.  Because of the ellipticity of the electron pocket and the fact that only one of the electron pockets is involved in the SDW, the resulting QPI has a pronounced one-dimensional structure.''\cite{Knolle2010a}

The nematic features proposed by Chuang et al.\cite{Chuang2010} are a type of real space inhomogeneity/anisotropy in a material.  With regard to the quasiparticle interference (QPI) and scanning tunneling microscopy (STM) images of Chuang et al.,\cite{Chuang2010} Kimber et al.\cite{Kimber2010} objected that the low-temperature crystal structure examined by Chuang et al.\ was an orthorhombic distortion of the high-temperature tetragonal structure.  Thus the in-plane nanoscopic electronic features identified in the SI-STM images had the same twofold rotational asymmetry as the lattice.  Kimber et al.\ carried out band calculations for this phase and concluded ``that this asymmetry is consistent with the underlying long range magnetic order and that LDA electronic structure provides a better description of the QPI images than the 1D band structure conjectured by Chuang et al.''\cite{Kimber2010}

Chuang et al.\ did not characterize the bulk crystallographic or physical properties of their crystal.\cite{Chuang2010}  They claimed from their SI-STM data that their crystal was orthorhombically distorted at 4.3~K\@.  However, they did not report what the tetragonal-to-orthorhombic and SDW transition temperatures  were, or whether or not the crystal was superconducting.  According to the supplementary material, ``Unlike Ba(Fe$_{1-x}$Co$_x)_2$As$_2$ samples, these exhibit significant spatial variations in Co density from location to location but this is the trade off required, at present, to achieve excellent surfaces for SI-STM.''  The influence of this spatial compositional disorder on the SI-STM/QPI measurements was not discussed.  The phase diagram for the Ca(Fe$_{1-x}$Co$_x)_2$As$_2$ system is unknown in detail at present, but we have collected the available literature data\cite{Kumar2009, Klingeler2010, Matusiak2010, Pramanik2010} and constructed a preliminary phase diagram which is presented in Fig.~\ref{Ca(FeCo)2As2_phase_diag}.  According to Fig.~\ref{Ca(FeCo)2As2_phase_diag}, the authors' crystal with composition $x = 0.030(5)$ apparently had $T_0 \sim 150$~K, thus confirming the conclusion of Chuang et al.\ that their crystal had the distorted orthorhombic structure at their measurement temperature of 4.3~K\@.  According to Pramanik et al., the optimum $T_{\rm c}$ occurs for $x = 0.056$--0.065, and the superconducting volume fraction decreases on either side of this maximum,\cite{Pramanik2010} similar to the Ba(Fe$_{1-x}$Co$_x)_2$As$_2$ system discussed above.  In particular, their crystal with $x = 0.051$ exhibited only 30\% diamagnetic shielding.  Therefore it is likely that the crystal with $x = 0.030(5)$ studied by Chuang et al.\cite{Chuang2010} was not a (bulk) superconductor.

\subsubsection{\label{SecPspInhomo} Momentum-Space Inhomogeneities}

Here we first summarize several points previously discussed in this review that are relevant to  momentum-space inhomogeneities in the Fe-based superconductors.  Then the influence of impurities on momentum-space inhomogeneity will be separately discussed.  An intrinsic inhomogeneity that is probably present is the superconducting gap inhomogeneity, or perhaps more properly called ``modulation'', along the $k_z$ axis of momentum space.  If the superconducting mechanism depends on nesting between the electron and hole Fermi surfaces as is widely believed, then the modulation of the Fermi surfaces along the $k_z$ axis as shown theoretically in Figs.~\ref{LaFeAsOBS} and~\ref{FigFS} and experimentally in Fig.~\ref{FigBa(FeCo)2As2_ARPES} above would presumably also cause a modulation in the magnitude of the superconducting energy gap along $k_z$.  Indeed, dispersion of the superconducting energy gap in the $k_z$ direction on at least one of the hole Fermi surface pockets was independently observed for single crystals of ${\rm Ba_{0.6}K_{0.4}Fe_2As_2}$ by two different groups.\cite{Xu2010a, Zhang2010}  This modulation could contribute to the power-law behaviors of the penetration depth and $1/T_1$ in the superconducting state as discussed above.  It would be interesting to study this modulation in other members of the Fe-based superconductor family.

In addition to these intrinsic inhomogeneities, momentum space inhomogeneities in the electronic properties could arise from statistical distributions of impurities.  Most of the Fe-based superconductors are produced by partial substitution of one element for another, which leads to electrostatic inhomogeneities (see below).  Pair-breaking impurities\cite{Kogan2010} would depress $T_{\rm c}$ and introduce states in the superconducting gap at the same time.  Such substitutional impurities disturb both the real space and momentum space homogeneity of the crystals, especially if the impurities reside on the transition metal (Fe or Ni) sublattice.  The $^{59}$Co and $^{75}$As NMR studies versus temperature of a single crystal of ${\rm Ba(Fe_{0.9}Co_{0.1})_2As_2}$ with $T_{\rm c} = 22$~K by Ning et al.\  demonstrated that the substituted Co atoms are nonmagnetic impurities rather than magnetic impurities carrying a local moment.\cite{Ning2008}

The heat capacity data in Fig.~\ref{GofrykFigs1_3}(a) suggest that only part of the conduction carriers become superconducting for non-optimally-doped crystals of Ba(Fe$_{1-x}$Co$_x)_2$As$_2$, and that that fraction has very low energy excitation modes that are sufficiently small to give rise to a nonzero Sommerfeld specific heat coefficient.  A very interesting possibility that deserves further experimental and theoretical attention is whether this differentiation is in momentum space and not in real space.  From optics measurements of FeTe$_{0.55}$Se$_{0.45}$ with $T_{\rm c} = 14$~K, Homes and coworkers concluded that only about one-fourth of the current carriers condensed below $T_{\rm c}$ into the superconducting state and cited this result as evidence that this compound is a dirty superconductor, which is the usual explanation for this suppression.\cite{Homes2010}  That is, Planck's constant times the quasiparticle scattering rate is larger than the quasiparticle gap $2\Delta$.  Williams et al.\cite{Williams2010} and Gordon et al.\cite{Gordon2010} also suggested momentum space segregation of superconducting and normal carriers below $T_{\rm c}$ as a possible explanation of their respective measurements of the magnetic penetration depth versus $x$ in single crystals of Ba(Fe$_{1-x}$Co$_x)_2$As$_2$.

\subsubsection*{\label{Impurities} a. Impurities}

Proposals have been made that the seeming inconsistencies between the nodeless $s^\pm$ pairing scenario and the power law temperature behaviors observed for the penetration depth and many $1/T_1$ measurements below $T_{\rm c}$ can be explained by the presence of pair-breaking impurities that introduce quasiparticle states into the superconducting gap (in momentum space).  Due to its importance in interpreting the superconducting state properties of the Fe-based materials, we discuss the presence and role of impurities here in more detail.

The discussion in Sec.~\ref{BadMetals?} indicates that the Fe-based superconductors are coherent metals at temperature just above $T_{\rm c}$, but impurity scattering at such temperatures is certainly not negligible.  In fact, according to the last column of Table~\ref{Opticsdata2}, (1) the Fe-based superconductors are experimentally found to be in the limit $k_{\rm B}T_{\rm c} \ll h/\tau$, where $1/\tau$ is the quasiparticle transport scattering rate.  Furthermore, the superfluid fraction data in the last two columns of Table~\ref{LambdaTable} indicate that the Fe-based materials are dirty superconductors.  (2) The impurities in most cases are evidently nonmagnetic, i.e., they do not carry local magnetic moments.\cite{Ning2008}  Constraints (1) and (2) are the two main experimental constraints on theory with respect to the impurities.  Theories distinguish between intraband and interband scattering, and these two types of scattering generally have different influences on the superconducting properties.  Several theoretical investigations will now be highlighted that treat impurity scattering.

Chubukov, Efremov, and Eremin formulated a two-orbital semimetallic model leading to small, equally populated electron and hole Fermi surfaces at the center and corners of the Brillouin zone at zero doping.\cite{Chubukov2008}  They noted that no new physics is found in a four-band model except to reveal the correct  antiferromagnetic structure.  They treated antiferromagnetism and superconductivity on an equal footing.  For the undoped case, they found that the SDW with the calculated transition temperature $T_{\rm SDW} > T_{\rm c}$, where $T_{\rm c}$ is for the $s^\pm$ pairing state, whereas in the doped case the order of the transition temperatures is reversed and superconductivity wins, as observed.  They found that when carriers are scattered between hole and electron pockets by nonmagnetic impurities, the result is to break superconducting pairs and the impurity acts like a magnetic impurity in a conventional $s$-wave superconductor, and the $s^\pm$ state can even become gapless.  On the other hand, they found that the nonmagnetic impurities have no effect in the $s^{++}$ pairing state in which the sign of the superconducting order parameter on the electron and hole pockets is the same.  They demonstrated that nonmagnetic impurity scattering in the $s^\pm$ pairing state causes the otherwise expected exponential temperature dependence of $1/T_1$ below $T_{\rm c}$ to change to an approximate  $T^3$ temperature dependence over a wide temperature range below $T_{\rm c}$,\cite{Chubukov2008} as often observed.  Their theory also predicts the observed neutron spin resonance mode for the $s^\pm$ pairing state at energy $<2\Delta$ due to the sign change between the superconducting order parameters on the electron and hole pockets.  Finally, they predicted the temperature dependence of the NMR Knight shift $K$ below $T_{\rm c}$, which was found to be influenced by impurities.  At impurity concentrations that are sufficiently small that the superconducting state is not gapless, both $1/T_1$ and $K$ are expected to follow an exponential temperature dependence at very low temperatures.\cite{Chubukov2008}

Ng formulated a Ginzburg-Landau theory of dirty $s^\pm$ two-band (electron and hole bands) superconductors containing nonmagnetic impurities.\cite{Ng2009}  The author concluded that ``the dirty two-band superconductor behaves as an effective dirty one-band superconductor in the regime $T_{\rm c} \ll 1/\tau_t$ where measurement of superfluid properties cannot distinguish between whether the system is originally a single-band or a two-band superconductor.''  The author stated that because different superconducting gaps were observed in the Fe-based superconductors, this finding implied that ``existing materials are located in the weak interband scattering regime $T_{\rm c} \geq 1/\tau_t$'', which conflicts with the above experimental observations.  Thus the applicability of this theory to the Fe-based superconductors is questionable.  In a related paper, Ng and Avishai predicted that with $s^\pm$ pairing, nonmagnetic impurities always suppress superconductivity and introduce states into the superconducting gap and are thus pair-breaking.\cite{Ng2009b}

Bang, Choi and Won studied the influence of impurity scattering on the $s^\pm$ pairing state using the ${\cal T}$~matrix method.\cite{Bang2009}  They found that strong scattering from impurities can produce a V-shaped density of states versus energy inside the superconducting gap, which can explain the observed power-law temperature dependence of $1/T_1$.  They also found that magnetic and nonmagnetic impurites depress $T_{\rm c}$ by about equal amounts.

The strong pair-breaking theory of Kogan\cite{Kogan2010} can explain various reported anomalous superconducting features such as $\Delta C/T_{\rm c} \sim T_{\rm c}^2$ and penetration depth $\Delta\lambda \sim T^2$.  This theory describes gapless superconductivity of a material with a highly anisotropic superconducting order parameter (such as in $s^\pm$ pairing) in the limit $T_{\rm c}/T_{\rm c0} \ll 1$, where $T_{\rm c0}$ is the $T_{\rm c}$ in the absence of impurities, similar to the theory of Abrikosov and Gor'kov for magnetic impurities in isotropic $s$-wave superconductors in the gapless regime.  This regime was also studied briefly by Chubukov et al.\cite{Chubukov2008}  Both nonmagnetic and magnetic impurities are pairbreaking.  It is further assumed that the average order parameter over all Fermi surfaces in  momentum space is zero, $\langle\Delta({\bf k})\rangle_{\rm FS} = 0$.  Interestingly, the calculated $T_{\rm c0}$ is of order room temperature, but Kogan points out that probably one cannot increase $T_{\rm c}$ to room temperature by removing impurity scattering because the source of the scattering such as antiferromagnetic spin fluctuations may itself be needed for the superconducting mechanism.\cite{Kogan2010}  With respect to the validity of the approach, the main outstanding question is whether the Fe-based superconductors are in the strong pair-breaking limit as required by the theory.

\paragraph*{Summary} Various formulations of the theory for the $s^\pm$ superconducting pairing state are able to explain the anomalous temperature dependences of the penetration depth and $1/T_1$ if the $s^\pm$ pairing state is complemented by impurity scattering.  To quantitatively test the latter theories, estimates of the theoretical scattering parameters in terms of experimental quantities such as resistivities and/or quasiparticle scattering rates are needed to verify that the theoretically required quasiparticle scattering strength is indeed attained.

\subsection{\label{Sec_Anomalous_SC} Unconventional Fe-based Superconductors with Clear Nodes in the Superconducting Order Parameter}

Evidence indicates non-$s$-wave superconductivity with nodes in the superconducting order parameter in momentum space in 1111-type LaFePO and 122-type BaFe$_2$As$_{2-x}$P$_x$ compounds.

In-plane penetration depth measurements using the TDR technique on single crystals of LaFePO with $T_{\rm c} \approx 6$~K indicated a nonexponential temperature dependence at low temperatures,\cite{Fletcher2009} qualitatively similar to those described above for FeAs-based superconductors where the penetration depth varied as a power law $\Delta\lambda \sim T^n$ with $n \approx 2$.  However, for LaFePO crystals with $T_{\rm c} = 5.6$~K, an even smaller exponent $n = 1.2(1)$ was obtained, indicative of nodes in the order parameter.  The exponent is close to the value of unity expected for line nodes in the clean limit.  As noted by the authors, the linear $T$ dependence of $\Delta\lambda$ is difficult to produce from extrinsic sources.  A similar exponent $n = 1.1(1)$ for the penetration depth was subsequently confirmed using a scanning SQUID susceptometer.\cite{Hicks2009}  Furthermore, in-plane thermal conductivity measurements at low temperatures were consistent with nodes in the gap.\cite{Yamashita2009}  The field dependence of these measurements suggested the presence of gapped 3D and 2D Fermi surface(s), and additional 2D Fermi surface(s) that have the nodes.  The authors suggest that ``the nodal $s^\pm$-wave structure can be the best candidate for the gap symmetry of LaFePO.''

Hashimoto and coworkers reported penetration depth and thermal conductivity measurements on BaFe$_2$As$_{0.67}$P$_{0.33}$ crystals with $T_{\rm c} = 30$~K that indicate the presence of line nodes in the superconducting order parameter.\cite{Hashimoto2009a}  Nakai et al.\ reported a \emph{linear} temperature dependence of $^{31}$P $1/T_1$ for a sample of BaFe$_2$As$_{0.67}$P$_{0.33}$ below $T_{\rm c} = 30$~K, indicating a residual density of states at the Fermi energy at temperatures below $T_{\rm c}$, again indicating the presence of line nodes in the gap.\cite{Nakai2010}

\subsection{\label{Sec_SC_Mech}Superconducting Mechanism}

%\squeezetable
\begin{table}
\caption{\label{IsotopeEffects} Isotope effect exponents $\alpha$ in Eq.~(\ref{EqAlpha2}) for various Fe-based compounds for the superconducting transition temperature $T_{\rm c}$ and the antiferromagnetic, or spin density wave, transition temperature $T_{\rm N}$.  The element for which the isotopic mass was changed is indicated.  All samples are polycrystalline.}
\begin{ruledtabular}
\begin{tabular}{l|ccccc}
 Compound   & $T_{\rm c}$ & $T_{\rm N}$ & element & $\alpha$ &  Ref. \\
 & (K) & (K) &    \\ \hline
Hg & 4.15 &&Hg & 0.44(5)\footnotemark[1] & \onlinecite{Reynolds1950} \\
\hline
${\rm SmFeAsO}$ &  & 130 & O & $-0.05(1)$ & \onlinecite{RHLiu2009}\\
${\rm SmFeAsO}$ &  & 130 & Fe & 0.39(2) & \onlinecite{RHLiu2009}\\
${\rm SmFeAsO}$$_{1-y}$ & 54.0 &  & Fe & $-0.024(15)$ & \onlinecite{Shirage2010}\\
${\rm SmFeAsO_{0.85}F_{0.15}}$ & 41 &  & O & $-0.06(1)$ & \onlinecite{RHLiu2009}\\
${\rm SmFeAsO_{0.85}F_{0.15}}$ & 41 &  & Fe & 0.34(3) & \onlinecite{RHLiu2009}\\
\hline
${\rm BaFe_2As_2}$ &  & 137 & Fe & 0.36(2) & \onlinecite{RHLiu2009}\\
${\rm Ba_{0.6}K_{0.4}Fe_2As_2}$ & 37.3 &  & Fe & 0.37(3) & \onlinecite{RHLiu2009}\\
${\rm Ba_{0.6}K_{0.4}Fe_2As_2}$ & 37.8 &  & Fe & $-0.18(3)$ & \onlinecite{Shirage2009}\\
\hline
${\rm FeSe_{0.975}}$ & 8.21 &  & Fe & $0.81(15)$ & \onlinecite{Khasanov2010}\\
\end{tabular}
\end{ruledtabular}
\footnotetext[1]{This work was published in 1950, before the BCS theory was published in 1957.  The quoted $\alpha$ was obtained by us from a fit of the given $T_{\rm c}(M)$ data by Eq.~(\ref{EqAlpha2}).}
\end{table}

Calculations of the electron-phonon coupling in LaNiPO showed that the observed $T_{\rm c} = 3$--4.2~K can be explained as arising from the electron-phonon interaction.\cite{Subedi2008}  On the other hand, similar calculations for LaFeAsO$_{1-x}$F$_x$ compounds showed that this mechanism produces a maximum $T_{\rm c}$ that is much too low.\cite{Boeri2008, Haule2008}  Ultrafast pump-probe optical spectroscopy studies have experimentally confirmed that the electron-phonon coupling constant $\lambda_{\rm ep}$ is quite small in the FeAs-based materials, being in the range of $\approx 0.12$--0.25 for the compounds ${\rm BaFe_2As_2}$, ${\rm SrFe_2As_2}$, SmFeAsO and ${\rm Ba(Fe_{0.92}Co_{0.08})_2As_2}$.\cite{Chia2010, Mertelj2010, Stojchevska2010, Mansart2010}  Several groups have argued that since the strong magnetoelastic coupling in the Fe-type materials (see Sec.~\ref{SecMagnetoElastic}) has not yet been considered within the theory for superconductivity via the electron-phonon mechanism, it is premature to conclude that the electron-phonon mechanism is not relevant to these materials.\cite{Yildirim2009, Kulic2009, Yndurain2009}  

A nonzero effect of changing elemental isotopes on $T_{\rm c}$ or $T_{\rm N}$ (``isotope effect'') is often taken to mean that phonons are involved in the superconducting or magnetic mechanism, respectively.  The isotope effect of a particular element is expressed as 
\be
T_{\rm c~or~N} \sim \frac{1}{M^\alpha}, 
\label{EqAlpha2}
\ee
where $M$ is the mass of the isotope and $\alpha$ is the isotope effect exponent.  The $\alpha$ is the negative of the slope determined from a linear fit to $\ln T_{\rm c~or~N}$ versus $\ln M$.  For pure $sp$-type elemental superconductors such as Hg, one obtains $\alpha \approx 1/2$ (see Table~\ref{IsotopeEffects}),\cite{Reynolds1950} as expected from the conventional electron-phonon mechanism for superconductivity.\cite{Tinkham1975}  The results of isotope effect experiments on the $T_{\rm c}$ and $T_{\rm N}$ of Fe-based superconductors and parent compounds are shown in Table~\ref{IsotopeEffects}.\cite{Khasanov2010, RHLiu2009, Shirage2010, Shirage2009}

Significant Fe isotope effects on $T_{\rm N}$ of BaFe$_2$As$_2$ and SmFeAsO and $T_{\rm c}$ of Ba$_{0.6}$K$_{0.4}$Fe$_2$As$_2$ and SmFeAsO$_{0.85}$F$_{0.15}$ were observed, which indicate that the magnetism and superconductivity are both somehow coupled to phonons, respectively.\cite{RHLiu2009}  Note that the signs of $\alpha$ are opposite to each other between the two studies in Table~\ref{IsotopeEffects} of the Fe isotope effect on $T_{\rm c}$ of Ba$_{0.6}$K$_{0.4}$Fe$_2$As$_2$.  A similar disagreement was found for the Fe isotope effect on $T_{\rm c}$ in the SmFeAsO-type compounds.\cite{RHLiu2009, Shirage2010}  It is difficult to identify the source(s) of these disagreements.  Khasanov et al.\ obtained a large Fe isotope effect on $T_{\rm c}$ of ${\rm FeSe_{0.975}}$ (Table~\ref{IsotopeEffects}), but they cautioned that part of this isotope shift might be due to their observed isotope effect on the low-temperature crystal structures, such as a slight change in the Se-Fe-Se bond angle.\cite{Khasanov2010}  After accounting for this slight structural change, the authors estimated $\alpha \approx 0.4$.\cite{Khasanov2010}

We remark that a nonzero isotope effect on $T_{\rm c}$ does not necessarily mean that the electron-phonon mechanism is responsible for the superconductivity.  The isotope effect could simply reflect the indirect influence of phonons on the actual pairing mechanism.  In the present instance, we know that there is a very strong magnetoelastic interaction in the Fe-based materials (see Sec.~\ref{SecMagnetoElastic}), so a nonzero isotope effect on $T_{\rm c}$ would not be particularly surprising if the actual superconducting mechanism is associated with the exchange of antiferromagnetic bosons.

As noted above in Sec.~\ref{ARPESSC}, the larger superconducting gap on the two hole pockets of both electron- and hole-doped BaFe$_2$As$_2$ occurs on the hole pocket with the better Fermi surface nesting with the electron pockets, suggesting that the superconducting mechanism is associated with Fermi surface nesting between the hole and electron pockets.  

NMR Knight shift measurements consistently indicate that the spin of the Cooper pairs is zero (spin singlets).  The orbital pairing symmetry seems to be consistently $s^\pm$,\cite{mazin2008} on the basis of all experimental data so far.  However, the identification of the pairing symmetry as $s^\pm$ does not itself imply a specific mechanism of superconductivity, although it appears to rule out the electron-phonon interaction.  $^{77}$Se NMR measurements of FeSe under applied pressure $p$ showed that the increase of $T_{\rm c}$ from 9~K to 18~K with increasing $p$ from zero to 2.2 GPa was correlated with an increase in the nuclear spin-lattice relaxation rate $1/T_1T$ at $T_{\rm c}$ which evidently results from enhanced AF fluctuations.\cite{Imai2009}  Thus these measurements ``suggest a link between spin fluctuations and the superconducting mechanism in FeSe.''  

The extensive inelastic neutron scattering studies we have described in detail consistently show that spin fluctuations occur above $T_{\rm c}$ in \emph{all} of the Fe-based superconductors examined, and they \emph{all} peak at the same in-plane wave vector ${\bf Q}_{\rm nesting} = \left(\frac{1}{2},\frac{1}{2}\right)$~r.l.u., which is the nesting wave vector between the electron and hole Fermi surfaces (and for the FeAs-based materials, this is also the in-plane antiferromagnetic ordering wave vector).  This is true even for the system Fe$_{1+y}$Te$_{1-x}$Se$_x$ which exhibits antiferromagnetic ordering at a \emph{different} wave vector ${\bf Q}_{\rm AF} = \left(\frac{1}{2},0\right)$~r.l.u.\ for the non-superconducting compositions with $x\approx 0$, but still shows the magnetic fluctuations above $T_{\rm c}$ at ${\bf Q}_{\rm nesting} = \left(\frac{1}{2},\frac{1}{2}\right)$~r.l.u.\ in the superconducting compositions $x \sim 0.4$--0.6, as predicted by Subedi et al.\cite{Subedi2008a}  Furthermore, the neutron spin resonance that occurs only below $T_{\rm c}$ develops out of the existing antiferromagnetic spin fluctuation background with a peak at wave vector ${\bf Q}_{\rm nesting}$.  The neutron spin resonance is understood to occur between Fermi surfaces, or parts thereof, that have opposite signs of the superconducting order parameter.  In the cuprates, this occurs within a single Fermi surface due to $d$-wave pairing, whereas in the Fe-based compounds it occurs between separate electron and hole Fermi surfaces due to unconventional $s^\pm$ pairing.  

It seems that the occurrence of these universal features in the Fe-based superconductors is much more than just a coincidence, and indeed strongly point towards an antiferromagnetic spin fluctuation model for the pairing mechanism.  Indeed, Wu et al.\ have extracted the electron-boson spectral density $\alpha^2F(\omega)$ versus energy $\hbar\omega$ from optics measurements on optimally doped ${\rm Ba(Fe_{0.92}Co_{0.08})_2As_2}$ crystals and found remarkable agreement with $\chi^{\prime\prime}(\omega)$ obtained from inelastic magnetic neutron scattering measurements on a similar sample.\cite{Wu2010a}  They concluded that this agreement gives ``compelling evidence that in iron-based superconductors spin fluctuations strongly couple to the charge carriers and mediate superconductivity.''\cite{Wu2010a}  Zhang and coworkers have calculated $T_{\rm c}$ on the basis of this mechanism and obtained a value of 70~K,\cite{JZhang2009} which is close to the maximum $T_{\rm c}\approx 56$~K observed experimentally.  Furthermore, they find that their superconducting state is consistent with the fully gapped $s^\pm$ pairing symmetry discussed above.  

As discussed in Sec.~\ref{SecInhomo}, it appears that the nodal states suggested by penetration depth measurements and by some $1/T_1$, $c$-axis thermal conductivity, and heat capacity measurements may be impurity, inhomogeneity, and/or ``accidental'' effects rather than intrinsic features of the Fe-based superconductors.

\section{\label{TheSearch} The Search for New Superconductors: Examples of A Few Possible Stepping Stones to the Future}

As a result of the discovery of high temperature superconductivity in iron-based materials, new materials have been synthesized and studied that were not previously investigated.  Many of the materials already discussed in this review and/or listed in the tables in the Appendix were discovered through this process.  Less well known are studies of materials that did not result in new superconductors but are still important nevertheless.  Here we consider a few such studies.  As repeatedly proven in the past, new materials research is one of the main avenues for quantum leaps to occur in condensed matter and materials physics.  As a community, we should continue to pursue the dream of room-temperature superconductivity  through a continuing program of new-materials research.  As a side benefit, exciting discoveries in other contexts are bound to occur along the way.

\subsection{The Peculiar Case of BaMn$_2$As$_2$ }

\begin{figure}
\includegraphics [width=3.3in]{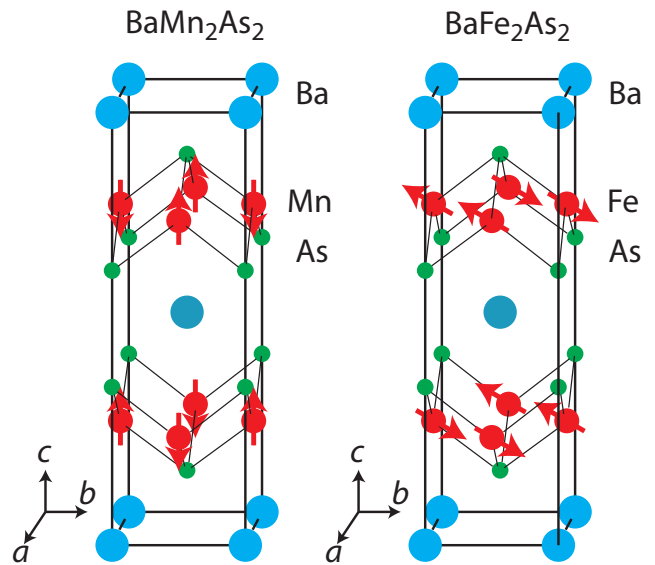}
\caption{(Color online) Comparison of the magnetic structures of BaMn$_2$As$_2$ and BaFe$_2$As$_2$.  The room-temperature tetragonal ThCr$_2$Si$_2$-type crystal structures are shown.  The arrows on the Mn and Fe atoms indicate the directions of the magnetic moments of these atoms in the antiferromagnetically ordered structures.  In BaMn$_2$As$_2$, the magnetic structure is a N\'eel structure with the ordered moments along the $c$-axis, whereas in BaF$_2$As$_2$ the magnetic structure is a stripe structure with the stripes in the tetragonal [1$\bar{1}$0] direction (orthorhombic $b$-axis direction) and with the antiferromagnetic propagation vector in the tetragonal [110] direction (orthorhombic $a$-axis direction).  Due to the limitations of the two-dimensional figure, it appears that the ordered Fe moments in BaF$_2$As$_2$ have a vertical component, but they do not.  The Fe moments are aligned within the basal plane and are oriented along an axis of Fe square lattice (see Fig.~\ref{Stripe_Mag_Struct}).}
\label{Ba(Mn,Fe)2As2_Mag_struct}
\end{figure}

%\clearpage
%\squeezetable
\begin{table}
\caption{\label{FeMnCompare} Comparison of the properties of ${\rm BaMn_2As_2}$ and ${\rm BaFe_2As_2}$.  Here ``AF'' means antiferromagnetic, $T_{\rm N}$ is the N\'eel temperature and $\mu_{\rm B}$ is the Bohr magneton.  The ordered moment is per transition metal (Mn or Fe) atom.  The data for ${\rm BaMn_2As_2}$ are from Refs.~\onlinecite{an2009}, \onlinecite{singh2009} and~\onlinecite{YSingh2009}.}
\begin{ruledtabular}
\begin{tabular}{l|cc}
Property & ${\rm BaMn_2As_2}$ & ${\rm BaFe_2As_2}$ \\
\hline
ground state  & AF insulator & AF metal  \\
ground state crystal structure & tetragonal & orthorhombic \\
$T_{\rm N}$ (K) & 625 & 137 \\
ordered moment ($\mu_{\rm B}$) & 3.9 & 0.9\\
moment direction & $\mu\parallel c$ & $\mu\parallel ab$ \\
magnetic structure & N\'eel & stripe \\
likely magnetic type & local moment & itinerant \\
dominant Fe-Fe interactions  & nearest-neighbor & 2$^{\rm nd}$-neighbor
\end{tabular}
\end{ruledtabular}
\end{table}

\begin{figure}
\includegraphics [width=3.3in]{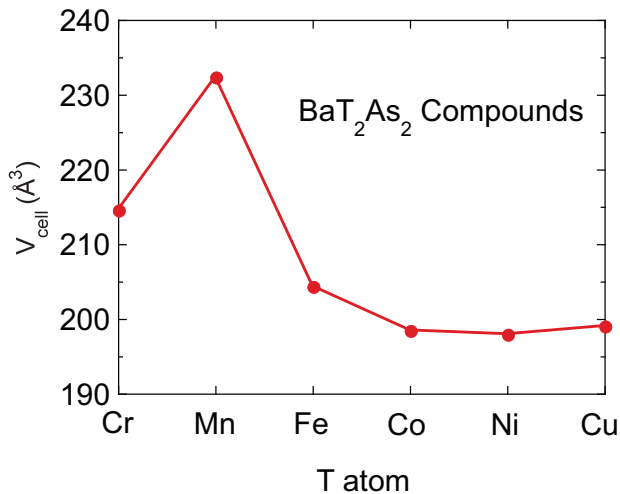}
\caption{(Color online) Tetragonal unit cell volume at room temperature versus 3$d$ transition metal $T$ for 122-type Ba$T_2$As$_2$ compounds.}
\label{122_Cell_volumes}
\end{figure}

The element Mn is a 3$d$ element that is adjacent to Fe in the periodic table, and hence might be expected to have similar chemical properties.  Indeed, the compound BaMn$_2$As$_2$ exists and has the same tetragonal ThCr$_2$Si$_2$-type structure as BaFe$_2$As$_2$ has at room temperature.  However, there the similarity stops.  Although both compounds exhibit antiferromagnetic ordering, the physical properties of BaMn$_2$As$_2$ are highly divergent from those of BaFe$_2$As$_2$ as indicated in Fig.~\ref{Ba(Mn,Fe)2As2_Mag_struct} and Table~\ref{FeMnCompare}.\cite{an2009, singh2009, YSingh2009}  Band theory calculations by An et al.\cite{an2009} correctly predicted the primary magnetic properties before they were measured,\cite{singh2009, YSingh2009} and also the electronic transport properties, including the small band gap semiconducting nature of the compound, the magnetic structure, the high ordered moment and the high N\'eel temperature.  The divergence in the electronic and magnetic properties from those of BaFe$_2$As$_2$ was found to arise from the strong Hund's coupling, the stability of the half-filled $d$-shell of the Mn$^{+2}$ ($d^5$) ion, and strong spin-dependent Mn-As hybridization.\cite{an2009}  

The tetragonal unit cell volume versus 3$d$ transition metal $T$ in the series of 122-type Ba$T_2$As$_2$ compounds is plotted in Fig.~\ref{122_Cell_volumes}.  The Mn compound clearly stands out from the rest.  It is well known that the high-spin state of a Mn, Fe, or Co atom has a larger volume than the lower-spin states.\cite{Cordero2008}  Therefore this figure shows that Mn is special in that it can evidently be considered to be a local moment ion that is in a higher spin state than for any of the other 3$d$ metal $T$ atoms shown.  Furthermore, the $T$ = Cr, Fe, Co, Ni, and Cu compounds all have metallic ground states instead of the insulating ground state for $T$ = Mn.  In particular, the compounds with $T =$~Cr and~Fe, on either side of $T =$~Mn in the periodic table, are predicted and observed to be an itinerant antiferromagnets instead of insulating local moment antiferromagnets.\cite{DJSingh2009}

Thus, with respect to the properties, BaMn$_2$As$_2$ is situated between BaFe$_2$As$_2$ and the high $T_{\rm c}$ cuprates.  On this basis BaMn$_2$As$_2$ was suggested to be a potential superconductor parent compound that might form a bridge between these two classes of high $T_{\rm c}$ superconductors.\cite{singh2009, YSingh2009} 

The magnetism of the Mn spins in the 1111-type compound PrMnSbO is similar to that in BaMn$_2$As$_2$.  In PrMnSbO, the Mn spins exhibit antiferromagnetic ordering at a relatively high temperature $T_{\rm N} = 230$~K with a C-type structure, in which the in-plane Mn spins are antiferromagnetically aligned as in BaMn$_2$As$_2$, but with the ordered moment in the $ab$-plane instead of in the $c$-axis direction, and with ferromagnetic instead of antiferromagnetic alignment between layers.\cite{Kimber2010a}  The compound undergoes a tetragonal to orthorhombic transition below 35~K, and the ordered moments at 4~K are 3.69(3)~$\mu_{\rm B}$/Mn, similar to that of Mn in BaMn$_2$As$_2$, and 2.96(3)~$\mu_{\rm B}$/Pr.\cite{Kimber2010a}

\subsection{The Even More Peculiar Case of Sr$_2$Mn$_3$As$_2$O$_2$-Type Compounds }

\begin{figure}
%arXiv: comment out first line and uncomment 2nd or 3rd line
%\includegraphics [width=3.3in]{Fig91.eps}
%\includegraphics [width=\columnwidth]{Fig91.eps}
\includegraphics* [width=\columnwidth]{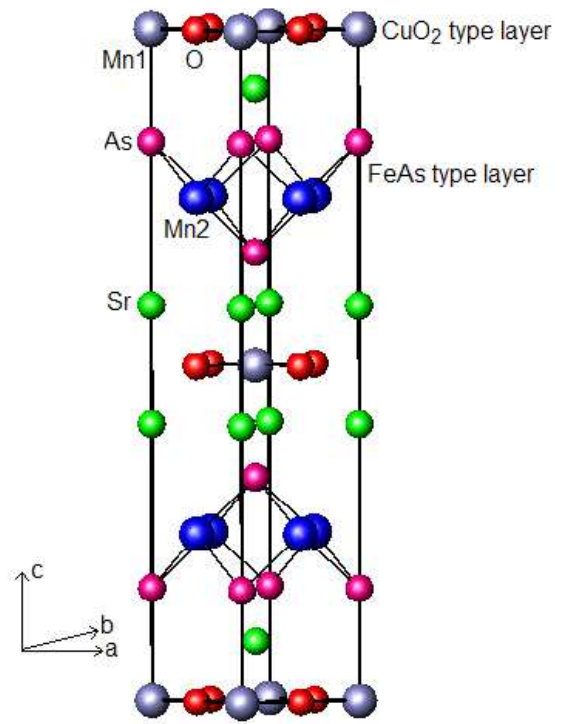}
\caption{(Color online) Body-centered tetragonal crystal structure of ${\rm Sr_2Mn_3As_2O_2}$.\cite{Brechtel1979}  Due to magnetic frustration effects, the Mn(2) moments in ${\rm Sr_2Mn_3As_2O_2}$ do not order long-range, although evidence for two-dimensional short-range ordering is found.  Reprinted with permission from Ref.~\onlinecite{Nath2010}.  Copyright (2010) by the American Physical Society.}
\label{NathfigSr2Mn3As2O2}
\end{figure}

\begin{figure}
\includegraphics [width=3.in]{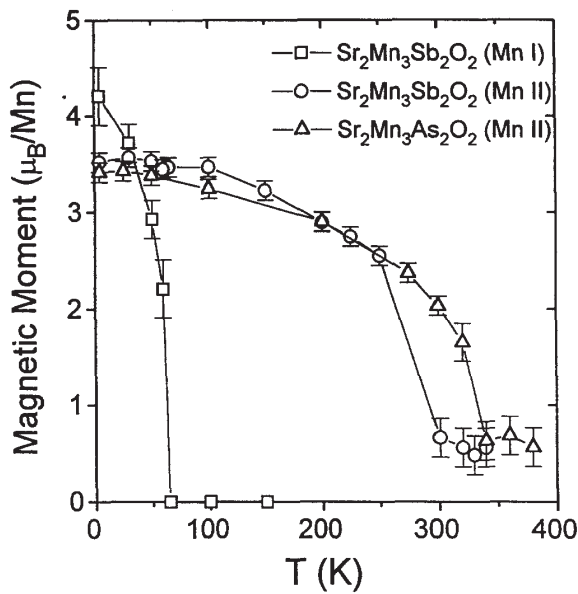}
\caption{Ordered moments versus temperature for Mn(1), in the MnO$_2$ layers, and Mn(2), in the Mn(As,Sb) layers, in ${\rm Sr_2Mn_3As_2O_2}$ and ${\rm Sr_2Mn_3Sb_2O_2}$.  Due to magnetic frustration effects, the Mn(1) moments in ${\rm Sr_2Mn_3As_2O_2}$ do not order long-range, although evidence for two-dimensional short-range ordering is found.  Reproduced with permission from Ref.~\onlinecite{Brock1996}, Copyright (1996), with permission from Elsevier.}
\label{Fig6Brock_Sr2Mn3Pn2O2}
\end{figure}

\begin{figure}
\includegraphics [width=2.5in]{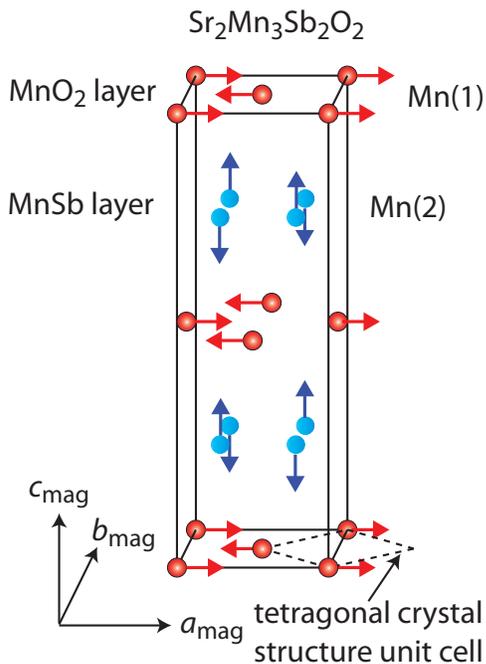}
\caption{(Color online) Magnetic structure of ${\rm Sr_2Mn_3Sb_2O_2}$.  The magnetic unit cell has lattice parameters $a_{\rm mag} = a_{\rm mag} = \sqrt{2}\,a,\ c_{\rm mag} = c$, where $a$ and $c$ are the tetragonal structural unit cell parameters.  The basal plane of the tetragonal unit cell is shown at the bottom as dashed lines.  From the data in this figure and in Fig.~\ref{Fig6Brock_Sr2Mn3Pn2O2}, the two Mn sublattices order independently.  Adapted from Ref.~\onlinecite{Brock1996}, Copyright (1996), with permission from Elsevier.}
\label{Sr2Mn3O2Sb2MagStruct}
\end{figure}

Compounds based on Sr$_2$Mn$_3$As$_2$O$_2$ are body-centered-tetragonal (\emph{I}4/\emph{mmm}) with $a = 4.14$~\AA\ and $c = 18.82$~\AA\ for Sr$_2$Mn$_3$As$_2$O$_2$ and with 2 f.u./unit cell.\cite{Brechtel1979}  As shown in Fig.~\ref{NathfigSr2Mn3As2O2}, the crystal structure consists of Mn(1)O$_2$ layers that are isostructural with the CuO$_2$ layers in the layered cuprate high $T_{\rm c}$ superconductors, that are interlaced with Mn(2)As layers that are isostructural with the FeAs layers in the iron arsenide high $T_{\rm c}$ superconductors.\cite{Nath2010}  These two types of layers are separated by layers of Sr.  This interleaving of the two types of layers along the $c$-axis suggests interesting avenues to search for new high-$T_{\rm c}$ superconductors.

Fascinating similarities and differences were found by Brock and coworkers between the magnetic properties of Sr$_2$Mn$_3$As$_2$O$_2$ and Sr$_2$Mn$_3$Sb$_2$O$_2$.\cite{Brock1996}  The two Mn sublattices in the Sb compound order antiferromagnetically with quite high ordered moments around 4~$\mu_{\rm B}$/Mn~atom as shown in Fig.~\ref{Fig6Brock_Sr2Mn3Pn2O2}, but the two sublattices also have very different ordering temperatures and magnetic structures in Sr$_2$Mn$_3$Sb$_2$O$_2$.  The magnetic structure is shown in Fig.~\ref{Sr2Mn3O2Sb2MagStruct},\cite{Brock1996} where it is seen that the two Mn sublattices seem to order independently of each other.  One sees from Fig.~\ref{Sr2Mn3O2Sb2MagStruct} that the interlayer interactions of the Mn(1) spins are frustrated, both with the adjacent Mn(2) layer and the nearest-neighbor Mn(1) layer.  This frustration is only lifted by interactions with the second-neighbor Mn(1) layer.  For this reason, the long-range Mn(1) ordering temperature is much less than that of Mn(2) in Sr$_2$Mn$_3$Sb$_2$O$_2$, and is completely eliminated in Sr$_2$Mn$_3$As$_2$O$_2$.  However, evidence was found for short-range two-dimensional antiferromagnetic ordering of the Mn(1) spins in the latter compound below about 70~K.\cite{Brock1996, Brock1996a}

\begin{figure}
\includegraphics [width=3.4in]{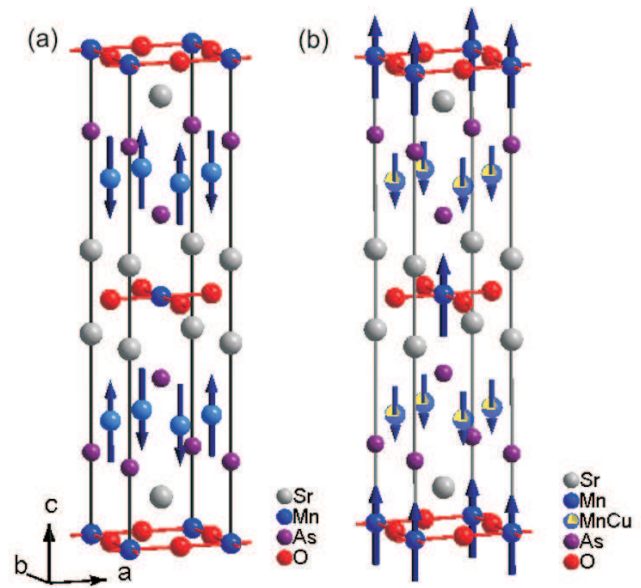}
\caption{(Color online) Magnetic structures of ${\rm Sr_2Mn_3As_2O_2}$ (left) and ${\rm Sr_2Mn_2CuAs_2O_2}$ (right).\cite{Nath2010}  The former compound is a G-type antiferromagnet in which the Mn spins in the MnO$_2$ layers do not exhibit long-range magnetic ordering, whereas the latter is an A-type ferrimagnet in which the Mn spins in both the MnO$_2$-type and Mn$_2$As$_2$ layers participate in long-range magnetic ordering.  Reprinted with permission from Ref.~\onlinecite{Nath2010}.  Copyright (2010) by the American Physical Society.}
\label{Nathfig16}
\end{figure}

Nath and coworkers have succeeded in replacing approximately one out of the three Mn atoms in Sr$_2$Mn$_3$As$_2$O$_2$ by Cu.\cite{Nath2010}  The hope was that the Cu would replace the Mn in the MnO$_2$ layers, giving CuO$_2$ layers as in the layered cuprates, alternating with ${\rm Mn_2As_2}$ layers like the ${\rm Fe_2As_2}$ layers in the Fe-based high $T_{\rm c}$ superconductors. Instead, from refinement of neutron diffraction patterns the Cu was found to replace about half of the Mn in the ${\rm Mn_2As_2}$ layers.  On the other hand, resistivity measurements indicated that this replacement may have resulted in metallic character instead of the insulating character observed for Sr$_2$Mn$_3$As$_2$O$_2$ and Sr$_2$Zn$_2$MnAs$_2$O$_2$.  Single crystal resistivity measurements are needed to confirm this metallic result for Sr$_2$Mn$_2$CuAs$_2$O$_2$.  Remarkably, the partial replacement of Mn by Cu resulted in a completely different magnetic structure, as shown in Fig.~\ref{Nathfig16}.\cite{Nath2010}  The G-type N\'eel (checkerboard) antiferromagnetic ordering in Sr$_2$Mn$_3$As$_2$O$_2$, where the nearest neighbor spins to a given spin are all aligned antiparallel to the given spin, is replaced by A-type ferrimagnetic ordering in which the spin per formula unit per layer in the two types of layers are not equal, and are aligned ferromagnetically within a layer but antiferromagnetically between layers.

Many opportunities to explore a diverse range of other layered oxychalcogenides and oxypnictides have been summarized.\cite{Clarke2008, Ozawa2008, Pottgen2008, Volkova2008}  We have illustrated above the  chemical flexibility in this class of compounds.  Another nice example is the study of pure and Li-doped ${\rm BaTi_2As_2O}$ by Wang et al.\cite{Wang2010a}  Thus there is reason to hope that some members of this class of layered oxypnictide and oxychalcogenide materials might be doped to become superconducting, perhaps even at high temperatures.

\section{\label{SecConclusions} Summary and Outlook}

We have given an overview of the scientific landscape of the Fe-based superconductors.  These superconductors evolve by doping or pressurizing \emph{semimetallic} parent compounds (see Figs.~\ref{semimetal_BS}  and~\ref{FigBaKFe2As2_FS}) that exhibit long-range spin-density-wave (SDW) and concomitant structural transitions at 70--200~K\@.  These parent compounds thus have equal numbers of electron and hole conduction carriers because they are ``valence compounds''.  Optimum superconductivity emerges when these long-range transitions are both completely suppressed by doping or pressure, but where dynamic short-range antiferromagnetic spin correlations are retained, thus suggesting an unconventional electronic/magnetic mechanism for the superconductivity, qualitatively similar to that believed responsible for superconductivity in the cuprate high temperature superconductors.  The magnetic susceptibilities of all of the high-$T_{\rm c}$ Fe-based compounds are strongly enhanced as shown in Fig.~\ref{ChiN(EF)}, which at first sight seems to conflict with the dynamic antiferromagnetic ordering just mentioned which would be expected to \emph{depress} the uniform susceptibility.  This conundrum has not been specifically addressed theoretically.  These observations suggest that to find new superconductors of the Fe-based type, one could profitably look for other semimetals that have enhanced uniform static susceptibilities but which also exhibit (seemingly conflicting) SDW instabilities.   As shown in Fig.~\ref{TcN(EF)}, there is no obvious correlation between $T_{\rm c}$ and the bare density of states at the Fermi energy $N(E_{\rm F})$ from band structure calculations.

In contrast to the cuprates where an effective one-band electronic conduction model is sufficient to describe the low-energy electronic excitations, in the Fe-based compounds it is very clear that at least four bands contribute as evidenced by their distinct Fermi surfaces (see Fig.~\ref{FigNakayama_BaKFeAs_ARPES}).  These Fermi surfaces are corrugated along the $k_z$ axis (Figs.~\ref{LaFeAsOBS}, \ref{FigFS}, \ref{FigBa(FeCo)2As2_ARPES}), indicating an anisotropic three-dimensional band structure, again in contrast to the cuprates which are much more two-dimensional.  These differences are correlated with the different conduction mechanisms in the two types of compounds.  In the cuprates, from structural considerations electron conduction has to occur primarily via hopping from Cu to O to Cu, whereas in the Fe-based compounds direct Fe-Fe hopping dominates.  These features are reflected in the orbital-decomposed densities of states at the Fermi energy $N(E_{\rm F})$ from band calculations.  In the cuprates, there is a large O $2p$ contribution to $N(E_{\rm F})$, whereas for the Fe-based compounds $N(E_{\rm F})$ is dominated by the Fe $3d$ orbitals.

A many-body mechanism is evidently responsible for the mass enhancements of the bands in Fe$_{1+y}$Te$_{1-x}$Se$_x$ as observed from ARPES\cite{Tamai2010} and optical measurements,\cite{Homes2010} for the factor of four to seven enhancement of normal state electronic specific heat of ${\rm Ba_{0.6}K_{0.4}Fe_2As_2}$ (Refs.~\onlinecite{Mu2008} and  \onlinecite{Welp2009}) and ${\rm KFe_2As_2}$,\cite{Terashima2010} and of the magnetic susceptibility and ARPES measurements of the Fe-based superconductors generally.  Many statements in the Fe-based superconductor literature have been made that such many-body mass enhancements (reductions of the LDA band widths) are observed, but there is usually no discussion as to the microscopic origin of these enhancements.  From analysis of optical data for a single crystal of Ba(Fe$_{0.92}$Co$_{0.08})_2$As$_2$, Wu et al.\ showed that these enhancements are consistent with expectation for the interaction of the conduction carriers with antiferromagnetic spin fluctuations.\cite{Wu2010a}

Unlike the cuprates, there appears to be no pseudogap in the spin excitation spectrum above $T_{\rm c}$ for optimally doped Ba(Fe$_{1-x}$Co$_x)_2$As$_2$.\cite{Inosov2010}

There have been some suggestions put forward that the small ordered moments observed at the lowest temperatures in the FeAs-based materials that show antiferromagnetic transitions, which all have the same in-plane component of the antiferromagnetic ordering wave vector ${\bf Q}_{\rm AF} = \left(\frac{1}{2},\frac{1}{2}\right)$~r.l.u.\ in tetragonal notation (Table~\ref{FeOrdering}), arise from frustration and/or fluctuations in a large-local-moment system.  In this regard we note the oft-repeated statement that even for an itinerant electron system, a local moment Heisenberg model can be used to fit the spin wave dispersion relations, at least at low energies.  However, the preponderance of the experimental data  for the \emph{FeAs}-based superconductors, as  extensively discussed in Sec.~\ref{SecItinLocMag}, argue against this possibility and instead indicate that the magnetic ordering transitions are spin density wave (SDW) transitions associated with itinerant electron antiferromagnetism.  

As discussed in Sec.~\ref{Sec122typeneuts}, from inelastic magnetic neutron scattering measurements, direct estimates of lower limits of the instantaneous effective moments in the paramagnetic state of ${\rm CaFe_2As_2}$ ($\mu_{\rm eff} = 0.47~\mu_{\rm B}$/Fe at a maximum integrated energy of 100~meV) and ${\rm Ba(Fe_{1.935}Co_{0.065})_2As_2}$ ($\mu_{\rm eff} = 0.31~\mu_{\rm B}$/Fe at a maximum integrated energy of 80~meV) were obtained by Diallo et al.\cite{Diallo2010} and by us from the data in Fig.~5 of Lester et al.,\cite{Lester2010} respectively.  These values are far smaller than the value $\mu_{\rm eff} = 4.90~\mu_{\rm B}$/Fe expected for a localized spin $S = 2$ with $g$-factor $g = 2$.

In further support of the itinerant magnetism scenario for the FeAs-based compounds we mention the successes in fitting the energy and wave vector dependences of inelastic magnetic neutron scattering data on various FeAs-based systems by itinerant magnetism models as in Figs.~\ref{DialloCaFig3} and~\ref{DialloCaFig8} (Ref.~\onlinecite{Diallo2010}) and Figs.~\ref{FigInosov_3D_60K} and~\ref{FigChipp3Ts}.\cite{Inosov2010}  Additional strong support for the itinerant magnetism model was obtained by us in Sec.~\ref{SecNeutsNMR} of this review by showing that the generalized magnetic susceptibility $\chi^{\prime\prime}({\bf Q},\omega)$ obtained using the nearly antiferromagnetic Fermi liquid model to fit inelastic magnetic neutron scattering data for single crystals of the Ba(Fe$_{1-x}$Co$_x)_2$As$_2$ system\cite{Inosov2010} could be used, \emph{with no adjustable parameters}, to quantitatively fit the temperature dependence and  semiquantitatively fit the magnitude of the antiferromagnetic spin fluctuation component of the $^{75}$As $1/T_1$ NMR data\cite{Ning2010} for crystals of the same system.  This itinerant antiferromagnetism interpretation is further confirmed by a calculation of the wave vector dependent static susceptibility $\chi({\bf Q})$ for LaFeAsO by Mazin et al.\ in Fig.~\ref{Mazin_chi},\cite{mazin2008} which shows a peak at ${\bf Q}_{\rm AF}$.  Furthermore, Schmidt et al.\ have theoretically studied frustrated local-moment models for the magnetism of the Fe-based superconductors and parent compounds in detail, and concluded that ``the anomalously low moment in the pnictides is not explained by quantum fluctuations in effective localized moment models but needs a more microscopic viewpoint including the itinerant multiorbital nature of the magnetic state.''\cite{Schmidt2010}  

On the other hand, as discussed in Secs.~\ref{Sec11BSARPES}, \ref{SecChi} and \ref{Sec11-type}, local-moment physics appears to be valid in the Fe$_{1+y}$(Te$_{1-x}$Se$_x$) system for compositions near the Te-rich end with $x\approx 0$.

This issue of itinerant magnetism versus local moment magnetism bears directly on the more general question of the degree of electron correlations present in the Fe-based superconductors beyond LDA band structure calculations.  Certainly, with the possible exception of Fe$_{1+y}$Te, the Fe-based materials show strong evidence for at least partially coherent normal state electronic conduction.  As discussed in Sec.~\ref{OpticsExp}, the optical conductivities of most of the Fe-based materials at low frequencies have been successfully fitted by a Drude contribution arising from coherent quasiparticles.  Sometimes these fits have also included a phenomenological Drude-like term attributed to incoherent carrier conduction, but the nature and/or origin of these incoherent conduction carriers is unclear. By ``coherent'', one means that the wave vectors of the current carriers are well-defined, and that the mean-free-time and mean-free-path for current carrier scattering are valid concepts to apply.  For ``incoherent'' transport, none of these concepts apply any more.  In any case, because at least four bands contribute to the carrier conduction in the Fe-based superconductors and parent compounds as discussed above, it seems likely that more than a two-band description of the current carriers is needed for a rigorous description of the optical conductivity of these carriers.  A quantitative assessment of the degree of electron correlation in Fig.~\ref{FigOptics_Correlations} based on optical measurements of the conduction carriers\cite{Qazilbash2008} indicates that the Fe-based materials are intermediate between strongly correlated ${\rm La_2CuO_4}$ and weakly correlated silver metal.  This assessment is at least qualitatively corroborated by many other measurements on these materials.

The homogeneity of Fe-based superconductors is currently an open subject as discussed in Secs.~\ref{Sec_SC_Cp} and~\ref{SecInhomo}.  The study of the low-temperature heat capacity $C$ of Ba(Fe$_{1-x}$Co$_x)_2$As$_2$ single crystals revealed values of $\gamma_0 \equiv \lim_{T \to 0}C/T$ in zero magnetic field that grow as the deviation of $x$ from the optimum value $x_0\approx 0.06$ increases [Fig.~\ref{GofrykFigs1_2}(a)].\cite{Gofryk2010, Mu2009}  Even the optimally doped composition $x = x_0$ showed a nonzero $\gamma_0$.  One interpretation of $\gamma_0$ is that it arises from a nonsuperconducting volume fraction that grows as $|x-x_0|$ increases.\cite{Gofryk2010}  This interpretation is consistent with the observed\cite{Budko2009} strong decrease in the heat capacity jump $\Delta C/T_{\rm c}$ at $T_{\rm c}$ with increasing $|x-x_0|$, with the decrease in the superconducting carrier density with increasing $|x-x_0|$,\cite{Chauviere2010, Williams2010} and possibly with the increase in $c$-axis thermal conductivity $\lim_{T\to 0}\kappa/T$ with increasing $|x-x_0|$.\cite{Reid2010}  

A very interesting topic that deserves further experimental and theoretical attention is whether this differentiation between superconducting and normal fractions occurs in momentum space or in real space as discussed in Secs.~\ref{SecPenDepth} and~\ref{SecPspInhomo}.  For example, we are not aware of any studies of the influence, either positive or negative, of annealing Fe-based superconductor crystals on their properties.  
Local surface measurements such as scanning tunneling spectroscopy measurements indicate spatial  inhomogeneity in the superconducting order parameter.  These observations have potential consequences for the interpretation of other data below $T_{\rm c}$ such as the above power law dependences of $1/T_1$ and $\lambda$ and the temperature dependence of the $c$-axis thermal conductivity.  Similarly, for the Fe$_{1+y}$(Te$_{1-x}$Se$_x$) superconductors, the reported superconducting transition widths are often very large, indicating a broad distribution of $T_{\rm c}$ values.  The occurrence of such finite-width $T_{\rm c}$ distributions would impact the analysis of the temperature dependences of various types of data below $T_{\rm c}$ in terms of theory such as for multigap superconductivity.  

Many experiments of various kinds cited in Sec.~\ref{SecSCProps} indicate that the FeAs-based materials are spin-singlet, $s^\pm$-wave\cite{mazin2008} dirty Type-II superconductors with an unconventional spin fluctuation mechanism.  The $s^\pm$-wave pairing means that there are no intrinsic nodes in the superconducting order parameter on any Fermi surface sheet (or ``pocket''), but the superconducting order parameter has opposite signs on the hole versus the electron sheets.  This mechanism depends on nesting between the Fermi surface hole pockets at the center $\Gamma$ point of the Brillouin zone and the electron pockets at the corner M points for the 11-, 111-, 1111-type compounds or the equivalent X points for the 122-type compounds, which follows as an inherent feature of their semimetallic band structure as discussed in Sec.~\ref{SecBanStrucIntro}.  From ARPES measurements (Fig.~\ref{FigBa(FeCo)2As2_ARPES})\cite{Vilmercati2009} and theory, the hole pockets are found to have three-dimensional dispersion that evidently affects the nesting condition with the electron pockets and both it and its consquences need to be further addressed both experimentally and theoretically.  Indeed, dispersion of the superconducting energy gap in the $k_z$ direction on at least one of the hole Fermi surface pockets was independently observed for single crystals of ${\rm Ba_{0.6}K_{0.4}Fe_2As_2}$ by two different groups.\cite{Xu2010a, Zhang2010}  These ARPES and other results such as anisotropic electronic conductivity\cite{Tanatar2009} and near isotropic superconducting upper critical fields for $T\to 0$, especially for the 122-type compounds (Table~\ref{Hc2Table}), indicate that quasi-two-dimensionality as in the layered cuprates is not required for high $T_{\rm c}$ superconductivity in the Fe-based systems.  

The $s^\pm$ pairing scenario would be expected to give rise to an exponential temperature dependence of $1/T_1$ and of the magnetic penetration depth $\lambda$ at low temperatures below $T_{\rm c}$, but power law behaviors are often observed instead: ($\lambda$: Fig.~\ref{Gordon_Fig1_0912.5346})\cite{Gordon_Fig1_0912.5346} and ($1/T_1$: Fig.~\ref{Yashima_Fig4}).\cite{Yashima2009}  This apparent conflict appears to arise because ARPES measurements are sensitive to the maximum gap at each $k_z$ cut of the Fermi surface, where the superconducting gap can and at least sometimes does disperse with $k_z$, whereas the $\lambda$ and $1/T_1$ measurements are sensitive to the minimum gap and also to nonmagnetic impurities that evidently act as pair breakers and introduce quasiparticle states into the superconducting gap as discussed in Sec.~\ref{SecInhomo}.  Thermal conductivity measurements are perhaps the most conclusive regarding the gap in the $ab$-plane and they indicate a superconducting state that is gapless for heat transport in the $ab$-plane (Figs.~\ref{TanatarFig2}, \ref{TanatarFig3}).\cite{Tanatar2010}  On the other hand, the same type of measurements along the $c$-axis versus composition $x$ for single crystals in the Ba(Fe$_{1-x}$Co$_x)_2$As$_2$ show nonzero coefficients $\lim_{T\to 0}\kappa/T$ for $x$ values deviating from the optimum value $x_0$ for superconductivity, suggesting that $k_z$-axis nodes develop in the superconducting order parameter in these compositions.\cite{Reid2010}

Bang has theoretically claimed that in an $s^\pm$ pairing scenario, when the electron and hole pockets have different superconducting gaps and when impurity scattering is present, a nonzero $\gamma_0$ can occur, whereas, surprisingly, due to coherence factors in the superconducting state, the electronic thermal conductivity obeys the seemingly conflicting behavior $\lim_{T\to 0}\kappa(T)/T = 0$.\cite{Bang2010}

The relationships between the theoretical parameters of pair-breaking impurities that are invoked to explain the above $\lambda$ and $1/T_1$ measurements need to be clearly expressed in terms of experimentally measurable quantities such as the quasiparticle scattering rates and/or conductivity values in Tables~\ref{Opticsdata2} and~\ref{Opticsdata3}.  The relationships of traditional measures of impurity effects to the properties of the Fe-based superconductors also need to be clarified.  For example, the superconducting condensate fraction data in the last two columns of Table~\ref{LambdaTable} and associated discussion in Sec.~\ref{SecPenDepth} were interpreted in terms of the relatively benign influence of non-pair-breaking impurities in a dirty superconductor, because the impurities giving rise to the quasiparticle scattering were assumed not to introduce quasiparticle states into the superconducting gap.  But it appears that most impurities in the Fe-based superconductors such as in Ba(Fe$_{1-x}$Co$_x)_2$As$_2$ are nonmagnetic, which theory predicts are pair-breaking in the $s^\pm$ pairing scenario.  These pair-breakers are needed in order to introduce quasiparticle states inside the superconducting gap that can in turn help to explain the above anomalous $1/T_1$ and $\lambda(T)$ measurements. Thus some traditional measures of how ``dirty'' a superconductor is, such as discussed in Sec.~\ref{SecPenDepth}, are probably still qualitatively valid but need to be re-evaluated for their quantitative applicability.  Also, the conditions under which doping produces magnetic versus nonmagnetic impurities remain to be clarified both experimentally and theoretically.  Finally, it seems possible that some fraction of doped carriers could be localized with internal degrees of freedom.  Such carriers might contribute to a zero-temperature Sommerfeld heat capacity coefficient that is observed in, e.g., superconducting samples of the Ba(Fe$_{1-x}$Co$_x)_2$As$_2$ system, but not to the electronic or thermal transport properties.

As discussed in Sec.~\ref{Sec_SC_Mech}, the electron-phonon mechanism appears to be too weak to give rise to the high $T_{\rm c}$ values observed in the Fe-based superconductors.\cite{Boeri2008, Haule2008, Chia2010, Mertelj2010, Stojchevska2010, Mansart2010}  The proximity of the superconducting compositions to those with long-range antiferromagnetic/spin density wave order suggests that an electronic mechanism might be responsible that involves the exchange of antiferromagnetic spin fluctuations.  Indeed, the experimental evidence is strong that the highest-$T_{\rm c}$ Fe-based superconductors are spin singlet, $s^\pm$-wave superconductors with this superconducting mechanism.  Furthermore, a $T_{\rm c} = 70$~K was calculated on the basis of this mechanism with a superconducting state that is consistent with the fully gapped $s^\pm$ pairing symmetry.\cite{JZhang2009}  Wu et al.\ have extracted the electron-boson spectral density $\alpha^2F$ versus energy $\hbar\omega$ from optics measurements on ${\rm Ba(Fe_{0.92}Co_{0.08})_2As_2}$ crystals and found remarkable agreement with $\chi^{\prime\prime}(\omega)$ obtained from inelastic magnetic neutron scattering measurements.\cite{Wu2010a}  They concluded that this agreement gives ``compelling evidence that in iron-based superconductors spin fluctuations strongly couple to the charge carriers and mediate superconductivity.''\cite{Wu2010a}  Convincing evidence has been presented that long-range antiferromagnetic ordering and bulk superconductivity coexist in the same volume element in an underdoped composition region of the system Ba(Fe$_{1-x}$Co$_x)_2$As$_2$, irrespective of the possible spatial inhomogeneity discussed above.\cite{Fernandes2010}  This observed microscopic coexistence has been concluded to rule out conventional $s^{++}$ pairing, but is fully consistent with $s^\pm$ pairing.\cite{Fernandes2010, Fernandes2010a}   

It thus appears that the nodal states suggested by penetration depth, $1/T_1$, and $c$-axis thermal conductivity measurements for most FeAs-based superconductors may be impurity, inhomogeneity, or ``accidental'' effects rather than intrinsic features.

Two special Fe-based superconductors are LaFePO and BaFe$_2$(As$_{1-x}$P$_x$). Convincing experimental evidence of various kinds has been presented by several groups for nodal superconductivity in these compounds,\cite{Fletcher2009, Hicks2009, Hashimoto2009a, Nakai2010} as discussed in Sec.~\ref{Sec_Anomalous_SC}.

Several examples of potential avenues for discovering new high temperature superconductors were given in Sec.~\ref{TheSearch}.  There are unlimited exciting opportunities for new materials research in this field.  There is far too little activity in this area, particularly in view of the fact that the cuprate and the Fe-based high-$T_{\rm c}$ superconductors were both discovered from such new-materials research.

The major goal of this review was to answer the question, ``what is so special about Fe in the Fe-based superconductors?''  There are some clear answers and also some important unanswered aspects to this question.  We hope that this review will be useful to the research community to further develop the field and to others following its development.  The puzzle of high temperature superconductivity in the layered iron pnictides and chalcogenides will certainly provide an interesting and challenging playground for many years to come.

\begin{acknowledgments}
The author is grateful to the following colleagues for providing figures: Jeffrey Lynn (Fig.~\ref{Structures}), Andreas Kreyssig [Ba(Fe$_{1-x}$Co$_x$)$_2$As$_2$ phase diagram in Fig.~\ref{FigBaKFe2As2_phase_diag}], Adam Kaminski (Fig.~\ref{FigBaKFe2As2_FS}), Norman Mannella (Fig.~\ref{FigBa(FeCo)2As2_ARPES}), Danel Orobengoa (Fig.~\ref{BZs}), Christopher Homes (Fig.~\ref{HomesFig3}), Takashi Imai (Fig.~\ref{NingFig4}), Robert McQueeney (Fig.~\ref{McQueeneyfig2}), Yunping Wang (Fig.~\ref{Ba(FeCo)2As2_corner_jcnB}), Andrew Christianson (top two panels of Fig.~\ref{FigBaKFe2As2ResMode}), Rafael Fernandes (Fig.~\ref{FigBaFeCoAsMT}), and Ulrich Welp for sending the data for the bottom panel of Fig.~\ref{FigBaKFe2As2Cp}.  The author also thanks colleagues and collaborators for helpful discussions, especially Dimitri Basov, Elbio Dagotto, Leonardo Degiorgi, Rafael Fernandez, Alan Goldman, Philipp Hansmann, Bruce Harmon, Dmytro Inosov, Adam Kaminski, Vladimir Kogan, David Lynch, Igor Mazin, Robert McQueeney, Danel Orobengoa, Ruslan Prozorov, David Singh, David Vaknin, and Dan Wu.  Special thanks go to Andreas Kreyssig for many discussions clarifying the crystallographic and magnetic structures and the reciprocal lattices of the Fe-based superconductors, and to Thomas Timusk and Christopher Homes for clarifying the units conventions used by the CMP optics community and for clarifying various aspects of optical properties.  Work at the Ames Laboratory was supported by the Department of Energy-Basic Energy Sciences under Contract No.~DE-AC02-07CH11358.

\end{acknowledgments}

%\clearpage
 
\appendix*\section{\label{SecDataTables} Abbreviations and Additional Data Tables}

In this Appendix we first give a list of the most common abbreviations used in the text, and then give tables of additional literature data on the properties of materials described in this review, along with the respective references.\\
\\
{\bf Caption to Table~\ref{LoTStructData1111}}: Low-temperature crystal and magnetic structure data for polycrystalline ZrCuSiAs-type (1111-type) FeAs-based parent compounds $R$FeAs(O,F).  In the high temperature primitive tetragonal $P$4/\emph{nmm} structure, $Z = 2$ formula units per unit cell, one has $c > a$ and the Wyckoff atomic positions are $R$: 2$c$ ($\frac{1}{4}$,$\frac{1}{4}$,$z_R$); Fe: 2$b$ ($\frac{3}{4}$,$\frac{1}{4}$,$\frac{1}{2}$); As: 2$c$ ($\frac{1}{4}$,$\frac{1}{4}$,$z_{\rm As}$); O,F: 2$a$ ($\frac{3}{4}$,$\frac{1}{4}$,0).  In the low temperature 1111-type orthorhombic \emph{Cmma} (the notation was recently changed to \emph{Cmme}) structure, $Z = 4$, the lattice parameters satisfy $c > a > b$ and the Wyckoff atomic positions are $R$: 4$g$ (0,$\frac{1}{4}$,$z_R$); Fe: 4$b$ ($\frac{1}{4}$,0,$\frac{1}{2}$); As: 4$g$ (0,$\frac{1}{4}$,$z_{\rm As}$); O,F: 4$a$ ($\frac{1}{4}$,0,0).  For the orthorhombic crystal structure, some authors define $a$ and $b$ such that $b > a$, which we have reversed here to give $a > b$ so that the same convention is used for all listings.  Also included are the low-temperature ordered magnetic moment per Fe or $R$ atom $\vec{\mu}$ or magnitude $\mu$ and the antiferromagnetic ordering wave vector $Q_{\rm AF}$ with respect to the low-temperature \emph{Fmmm} orthorhombic unit cell (see Figs.~\ref{FigTetrag_Ortho_struct} and~\ref{Stripe_Mag_Struct}) determined from magnetic neutron diffraction measurements.  For the ordered moment direction, the unit vectors $\hat{\bf a}$, $\hat{\bf b}$, and $\hat{\bf c}$ are in the directions of the orthorhombic $a$, $b$ and $c$ axes, respectively.  The $Q_{\rm AF}$ is expressed in orthorhombic reciprocal lattice units (r.l.u.) $(2\pi/a,\,2\pi/b,\,2\pi/c)$.  For $Q_{\rm AF} = (100)$, the nearest-neighbor ordered moments are antiferromagnetically aligned along the orthorhombic $a$ axis, and ferromagnetically aligned along $b$, which is the magnetic stripe axis for the Fe spins, see Fig.~\ref{Stripe_Mag_Struct}.  The last digit ``0'' in the ordering wavevector means that the nearest neighbor Fe spins along $c$ are ferromagnetically aligned.  In tetragonal r.l.u.\ notation, the ordering wave vector is $\left(\frac{1}{2}\frac{1}{2}0\right)$.  For $Q_{\rm AF} = (10\frac{1}{2})$, the last entry ``$\frac{1}{2}$'' in the ordering wavevector means that the nearest neighbor Fe spins along $c$ are antiferromagnetically aligned.  In tetragonal r.l.u.\ notation, the ordering wave vector is $\left(\frac{1}{2}\frac{1}{2}\frac{1}{2}\right)$.

\squeezetable
\begin{table*}
\caption{\label{Symbols} Common symbols used in this review and their units.  The units used in this review are Gaussian (cgs) units unless indicated as SI units.  In Gaussian units, 1~G = 1~Oe and the Tesla (T) is a unit of convenience: 1~T = $10^4$~Oe\@.}
\begin{ruledtabular}
\begin{tabular}{lll}
Symbol   & Physical Quantity Represented & Units \\
\hline
$a,~b,~c$ & lattice parameters & \AA\  \\
$A,~B$ & NMR hyperfine coupling constant & $\mu_{\rm B}$/Oe \\
ARPES & angle-resolved photoemission spectroscopy & ---\\
$B$ & magnetic induction & G \\
bct & body-centered-tetragonal & --- \\
BZ & Brillouin zone & ---\\
$c$ & speed of light in vacuum & cm/s \\
$C$ & Curie constant & cm$^3$~K/mol, cm$^3$~K/atom, cm$^3$~K/f.u. \\
$E_{\rm F}$ & Fermi energy & eV \\
$f$ & frequency & Hz (s$^{-1}$) \\
$f^\prime$ & frequency & cm$^{-1}$ \\
$f_{\rm p}^\prime$ & plasma frequency & cm$^{-1}$ \\
f.u. & formula unit & --- \\
$(h,k,\ell)$ & reduced reciprocal lattice units & ---\\
$(H,K,L)$ & reciprocal lattice units & ---\\
$h$ & Planck's constant & erg~s \\
$\hbar$ & Planck's constant/(2$\pi$) & erg~s \\
$H$ & magnetic field & Oe, G, T \\
$\Im$ & imaginary part & --- \\
$J$ & exchange coupling constant &  eV \\
$k_{\rm B}$ & Boltzmann's constant & erg/K \\
$K$ & NMR shift & --- \\
$K$ & Drude spectral weight & erg \\
$m^*$ & effective or band mass & gram \\
$m_{\rm e}$ & free electron mass & gram \\
$m_{\rm b}$ & band mass & gram \\
$M$ & magnetization & G, G~cm$^3$/mol \\
$n$ & carrier concentration & cm$^{-3}$ \\
$N(E_{\rm F})$ & Density of states at $E_{\rm F}$ & 1/(eV f.u.), both spin directions \\
pt & primitive tetragonal & ---\\
${\bf Q}$ & momentum transfer, momentum & erg s/cm \\ 
$\Re$ & real part & --- \\
$T$ & temperature & K \\
$T_{\rm c}$ & superconducting transition temperature & K \\
$T_{\rm N}$ & N\'eel temperature & K \\
$\Gamma,\gamma$ & relaxation rate & s$^{-1}$ \\
$\gamma$ & Sommerfeld specific heat coefficient & mJ/mol~K$^2$\\
$\gamma_{\rm n}/(2\pi)$ & nuclear gyromagnetic ratio & MHz/T \\
$\gamma_{\rm e}/(2\pi)$ & electron gyromagnetic ratio & MHz/T \\
$\varepsilon_0$ & dielectric permittivity of free space & C$^2$/(J m) = s/($\Omega$~cm) (SI)\\ 
$\theta$ & Weiss temperature & K \\
$\theta_2,~\theta_4$ & bond angle & degrees \\
$\mu$ & magnetic moment & G~cm$^3$ = erg/G \\
$\mu_{\rm B}$ & Bohr magneton & G~cm$^3$ = erg/G \\
$\xi$ & correlation length & \AA, nm \\
$\rho$ & electrical resistivity & $\Omega$~cm (SI) \\
$\sigma = 1/\rho$ & electrical conductivity & ($\Omega$~cm)$^{-1}$ (SI) \\
$\sigma_0$ & dc electrical conductivity & ($\Omega$~cm)$^{-1}$ (SI) \\
$\sigma_1$ & real part of electrical conductivity & ($\Omega$~cm)$^{-1}$ (SI) \\
$\tau$ & relaxation time & s \\
$1/\tau^\prime$ & relaxation rate & cm$^{-1}$ \\
$\chi$ & magnetic susceptibility & cm$^3$, cm$^3$/mol, dimensionless  \\
$\omega = 2\pi f$ & angular frequency & rad/s  \\
$\omega_{\rm p}$ & plasma angular frequency & rad/s  \\
\end{tabular}
\end{ruledtabular}
\end{table*}

%\squeezetable
\begin{table*}
\caption{\label{data4} Crystal data for tetragonal rare-earth and FeAs-based ZrCuSiAs-type $R$FeAs$Y$ (1111-type)  compounds.  Here, $R$ is a rare-earth, $Y$ is O and/or F, $z_{A}$ and $z_{X}$ are the $c$-axis position parameters of atoms $A$ and $X$, respectively, $\theta_2$ and $\theta_4$ are the two-fold and four-fold $X$-$M$-$X$ bond angles within an $M$-centered $MX_4$ tetrahedron, respectively, and $h_{MX}$ is the separation of a plane of $M$ atoms from either of the two adjacent planes of $X$ atoms.  The temperature $T$ at which the diffraction data were obtained is given (``RT'' means room temperature), along with the structural and/or magnetic transition temperature $T_0$ and the superconducting transition temperature $T_{\rm c}$ for the sample if given.  Samples are polycrystalline unless otherwise noted.}
\begin{ruledtabular}
\begin{tabular}{lcccccccccccc}
Compound & $T$ & $T_0$ & $T_{\rm c}$ & $a$ & $c$ & $z_{A}$ & $z_{X}$ & $d_{M-X}$ & $\theta_2$ & $\theta_4$ & $h_{MX}$ &Ref.\\
& (K) & (K) & (K) & (\AA) & (\AA) & &   & (\AA) & (deg) & (deg) & (\AA)\\ \hline
LaFeAsO & RT & 150 & & 4.03533(4) & 8.74090(9) & 0.14154(5) & 0.6512(2) & 2.412 & 113.5 & 107.5 & 1.322 & \onlinecite{kamihara2008}\\
LaFeAsO & 175 & 137, 155 & & 4.03007(9) & 8.7368(2) & 0.1418(3) & 0.6507(4) & 2.407(2) & 113.7(1) & 107.41(7) & 1.317 & \onlinecite{Cruz2008}\\
LaFeAsO & 300 & 160 & & 4.03268(1) & 8.74111(4) & 0.14134(4) & 0.65166(7) & 2.4131 & 113.35 & 107.57 & 1.326 & \onlinecite{nomura2008}\\
LaFeAsO & 300 & 145& & 4.0345(1) & 8.7387(4) & 0.1420(3) & 0.6498(4) & 2.4048 & 114.04 & 107.24 & 1.309 & \onlinecite{Sefat2008h}\\
LaFeAsO & RT & & & 4.0322(2) & 8.7364(4) & 0.1416(4) & 0.6508(5) & 2.4084 & 113.67 & 107.41 & 1.317 & \onlinecite{Qureshi2010}\\
LaFeAsO & RT & 140,152(4) & & 4.0316(1) & 8.7541(1) & 0.1405(2) & 0.6543(3) & 2.4265 & 112.35 & 108.05 & 1.351 & \onlinecite{Li2010c}\\
LaFeAsO$_{0.88}$\footnotemark[1] & RT & & 28 & 4.02291(8) & 8.7121(2) & 0.1453(3) & 0.6527(4) & 2.4116   & 113.04 & 107.72 & 1.330 & \onlinecite{Lee2008}\\
LaFeAsO$_{0.92}$F$_{0.08}$ & 175 & & 26  & 4.0229(1) & 8.7142(2) & 0.1446(3) & 0.6527(4) & 2.412(2)   & 113.0(3) & 107.72(6) & 1.331 & \onlinecite{Cruz2008}\\
LaFeAsO$_{0.89}$F$_{0.11}$ & RT & & 28  & 4.0277(2) & 8.7125(5) & 0.1455(3) & 0.6522(5) & 2.411   & 113.27 & 107.60 & 1.326 & \onlinecite{sefat2008b}\\
LaFeAsO$_{0.87}$F$_{0.13}$ & 1.5 & & 26  & 4.0245(3) & 8.713(1) & 0.1442(8) & 0.6541(8) & 2.419   & 112.6 & 107.9 & 1.343 & \onlinecite{Qiu2008b}\\
LaFeAsO$_{0.86}$F$_{0.14}$ & 300 & & 20  & 4.02460(2) & 8.69525(5) & 0.14725(6) & 0.65319(10) & 2.4132 & 113.00 & 107.74 & 1.332 & \onlinecite{nomura2008}\\
LaFeAsO$_{0.85}$F$_{0.15}$ & RT & &  & 4.02447(3) & 8.6948(1) & 0.1450(4) & 0.6540(4) & 2.417 & 112.72 & 107.87 & 1.339 & \onlinecite{Qureshi2010}\\
LaFe$_{0.89}$Co$_{0.11}$AsO & 300 & & 14.3 & 4.0351(1) & 8.7132(3) & 0.1412(3) & 0.6505(4) & 2.4063   & 113.96 & 107.28 & 1.311 & \onlinecite{Sefat2008h}\\
CeFeAsO & 175 & & & 3.99591(5) & 8.6522(1) & 0.1413(3) & 0.6546(2) & 2.4044   & 112.40 & 108.03 & 1.338 & \onlinecite{Zhao2008}\\
CeFeAsO$_{0.84}$F$_{0.16}$ & 60 & & 40 & 3.98470(3) & 8.6032(1) & 0.1480(4) & 0.6565(3) & 2.4046 & 111.90 & 108.27 & 1.346 & \onlinecite{Zhao2008}\\
PrFeAsO &  RT & & & 3.985(1) & 8.595(3) & 0.1399(2) & 0.6565(3) & 2.404 &  111.95 & 108.24 & 1.345 & \onlinecite{quebe2000}\\ 
PrFeAsO\footnotemark[1] &  175 & & & 3.97716(5) & 8.6057(2) & 0.1397(6) & 0.6559(4) & 2.3988 &  111.99 & 108.23 & 1.342 & \onlinecite{Zhao2008b}\\ 
PrFeAsO$_{0.85}$F$_{0.15}$\footnotemark[1] & 5 & & 52 & 3.9700(1) & 8.5331(4) & 0.1504(1) & 0.6548(5) & 2.3843 & 112.72 & 107.87 & 1.321 & \onlinecite{Zhao2008b}\\ 
PrFeAsO$_{0.85}$\footnotemark[1] & 5 & & 52 & 3.9686(1) & 8.5365(3) & 0.1450(7) & 0.6546(5) & 2.3831 &  112.75 & 107.86 & 1.320 & \onlinecite{Zhao2008b}\\ 
NdFeAsO &  175 & 1.96, 150 & & 3.9611(1) & 8.5724(2) & 0.1393(3) & 0.6580(4) & 2.3994 &  111.27 & 108.58 & 1.354 & \onlinecite{Qiu2008}\\ 
NdFeAsO$_{0.95}$\footnotemark[1] & RT & & & 3.96666(7) & 8.5699(2) & 0.1390(2) & 0.6571(3) & 2.3971   & 111.66 & 108.39 & 1.346 & \onlinecite{Lee2008}\\
NdFeAsO$_{0.92}$\footnotemark[1] & RT & & 35 & 3.95940(6) & 8.5550(2) & 0.1413(2) & 0.6586(3) & 2.4000 & 111.15 & 108.64 & 1.357 & \onlinecite{Lee2008}\\
NdFeAsO$_{0.86}$\footnotemark[1] & RT & & 44 & 3.95365(7) & 8.5581(2) & 0.1429(3) & 0.6587(3) & 2.3984 & 111.02 & 108.70 & 1.358 & \onlinecite{Lee2008}\\
NdFeAsO$_{0.83}$\footnotemark[1] & RT & & 51 & 3.94755(7) & 8.5446(2) & 0.1440(3) & 0.6600(3) & 2.4010 & 110.58 & 108.92 & 1.367 & \onlinecite{Lee2008}\\
NdFeAsO$_{0.83}$\footnotemark[1] & 10 & & 51 & 3.9423(1) & 8.5129(3) & 0.1434(3) & 0.6624(4) & 2.4076 & 109.91 & 109.25 & 1.382 & \onlinecite{Lee2008}\\
NdFeAsO$_{0.8}$F$_{0.2}$ &  1.5 & & 50 & 3.9495(1) & 8.5370(3) & 0.1421(4) & 0.6599(4) & 2.4006 &  110.69 & 108.87 & 1.365 & \onlinecite{Qiu2008}\\
SmFeAsO &  300 & & & 3.9390(1) & 8.4980(1) & 0.1372(1) & 0.6599(2) & 2.3928 &  110.79 & 108.81 & 1.359 & \onlinecite{Martinelli2009}\\
SmFeAsO &  RT & & & 3.9391(2) & 8.4970(4) & 0.1368(2) & 0.6609(4) & 2.3976 &  110.47 & 108.98 & 1.367 & \onlinecite{Martinelli2008}\\ 
SmFeAsO$_{0.93}$F$_{0.07}$ & RT & & 35 & 3.9344(2) & 8.4817(5) & 0.1397(2) & 0.6611(5) & 2.3952 &  110.43 & 108.99 & 1.366 & \onlinecite{Martinelli2008}\\ 
SmFeAsO$_{1-x}$F$_{x}$ &  295 & & 52 & 3.9339(1) & 8.4684(6) & 0.1411(1) & 0.6609(2) & 2.3928 &  110.58 & 108.92 & 1.363 & \onlinecite{Zhigadlo2008}\\ 
SmFeAsO$_{0.8}$F$_{0.2}$ & 20 & & 54 & 3.92699(3) & 8.4413(1) & 0.1420(1) & 0.6608(2) & 2.3871(9) &  110.69 & 108.87 & 1.357 & \onlinecite{Margadonna2009}\\ 
SmFeAsO$_{0.68(3)}$\footnotemark[1] &  RT & & 55 & 3.90236(8) & 8.4315(3) & 0.1437(2) & 0.6628(4) & 2.3854(18) &  109.76(12) & 109.33(6) & 1.373 & \onlinecite{Ju2010}\\ 
SmFeAsO\footnotemark[1]$^,$\footnotemark[2] &  295 & &  & 3.9427(1) & 8.4923(3) & 0.1372(1) & 0.6603(1) & 2.3955(6) &  110.76(4) & 108.83(2) & 1.3613(9) & \onlinecite{Zhigadlo2010}\\ 
Sm$_{0.89}$Th$_{0.11}$FeAsO\footnotemark[1]$^,$\footnotemark[2] &  295 & & 49.5 & 3.9369(1) & 8.4510(6) & 0.1411(1) & 0.6611(1) & 2.3937(9) &  110.64(6) & 108.89(3) & 1.3615(9) & \onlinecite{Zhigadlo2010}\\ 
Sm$_{0.90}$Th$_{0.10(5)}$FeAsO\footnotemark[1] &  295 & & 51.5 & 3.9404(2) & 8.4730(6) & 0.1421(3) & 0.6618(7) & 2.400(3) &  110.3(1) & 109.0(3) & 1.371 & \onlinecite{Zhigadlo2010}\\ 
&  15 & & 51.5 & 3.9357(2) & 8.4327(6) & 0.1424(3) & 0.6595(7) & 2.384(3) & 111.3(1) & 108.6(3) & 1.345 & \onlinecite{Zhigadlo2010}\\ 
GdFeAsO$_{1-y}$\footnotemark[1] & RT & & 54 & 3.90311(9) & 8.4137(3) & 0.1399(1) & 0.6630(1) & 2.3853 &  109.81 & 109.30 & 1.371 & \onlinecite{Lee2008b}\\ 
TbFeAsO$_{0.9}$F$_{0.1}$\footnotemark[1] & RT & & 46 & 3.8634(3) & 8.333(1) & 0.1447(4) & 0.6654(6) & 2.373 &  108.98 & 109.72 & 1.378 & \onlinecite{Bos2008}\\ 
\end{tabular}
\end{ruledtabular}
\footnotetext[1]{high-pressure synthesis}
\footnotetext[2]{single crystal}

\end{table*}

%\squeezetable
\begin{table*}
\caption{\label{data4A} Crystal data for additional tetragonal ZrCuSiAs-type $AMXY$ (1111-type)  compounds.  Here, $A$ is an alkaline earth or rare earth metal, $M$ is a transition element, $X$ is a pnictogen (P, As), $Y$ is O and/or F, $z_{A}$ and $z_{X}$ are the $c$-axis position parameters of atoms $A$ and $X$, respectively, $\theta_2$ and $\theta_4$ are the two-fold and four-fold $X$-$M$-$X$ bond angles within an $M$-centered $MX_4$ tetrahedron, respectively, and $h_{MX}$ is the separation of a plane of $M$ atoms from either of the two adjacent planes of $X$ atoms.  The temperature $T$ at which the diffraction data were obtained is given (``RT'' means room temperature), along with the structural and/or magnetic transition temperature $T_0$ and the superconducting transition temperature $T_{\rm c}$ for the sample if given.}
\begin{ruledtabular}
\begin{tabular}{lcccccccccccc}
Compound & $T$ & $T_0$ & $T_{\rm c}$ & $a$ & $c$ & $z_{A}$ & $z_{X}$ & $d_{M-X}$ & $\theta_2$ & $\theta_4$ & $h_{MX}$ &Ref.\\
& (K) & (K) & (K) & (\AA) & (\AA) & &   & (\AA) & (deg) & (deg) & (\AA)\\ \hline
LaFePO &  RT & & 4\footnotemark[3] & 3.96358(2) & 8.51222(7) & 0.14870(6) & 0.6339(2) & 2.2862 &  120.18(8) & 104.39(4) & 1.140 &\onlinecite{Kamihara2006}\\ 
LaFePO\footnotemark[4] &  RT & & 6 & 3.941(2) & 8.507(5) & 0.14890(2) & 0.63477(10) & 2.280 &  119.62 & 104.65 & 1.146 &\onlinecite{Coldea2008}\\ 
LaFePO &  298 & & $<0.35$ & 3.96306(4) & 8.5087(1) & 0.1487(2) & 0.6348(3) & 2.2895(1) &  119.87 & 104.53 & 1.147 & \onlinecite{McQueen2008}\\ 
LaFePO &  298 & & 7\footnotemark[2] & 3.9610(1) & 8.5158(2) & 0.1496(2) & 0.6362(6) & 2.295(3) &  119.3(2) & 104.8(1) & 1.160 & \onlinecite{Tegel2008c}\\ 
PrFePO\footnotemark[4] & RT & & & 3.9113(6) & 8.345(2) & 0.14830(7) & 0.6396(3) & 2.276(2) & 118.4 & 105.2 & 1.165 & \onlinecite{Zimmer1995}\\ 
SmFePO & RT & 5 & 3 & 3.88069(5) & 8.2054(1) & 0.14502(9) & 0.6423(4) & 2.265 &  117.9 & 105.4 & 1.168 & \onlinecite{Kamihara2008c}\\ 
SmFePO & RT & & & 3.878(1) & 8.205(1) & 0.1482(7) & 0.642(4) & 2.26(2) & 118.0 & 105.4 & 1.165 & \onlinecite{Zimmer1995}\\ 
CaFeAsF &  300 & 120 & & 3.875 & 8.582 & & & 2.392 &  108.2 & 110.1 & &\onlinecite{Nomura2008a}\\ 
CaFe$_{0.88}$Co$_{0.12}$AsF &  300 & & 24 & 3.879 & 8.539 &  &  & 2.386 &  108.7 & 109.9 & & \onlinecite{Nomura2008a}\\ 
SrFeAsF &  297 & 175 & & 3.9930(1) & 8.9546(1) & 0.1598(2) & 0.6527(2) & 2.420(1) &  111.2(1) & 108.7(1) & 1.367 & \onlinecite{Tegel2008b}\\ 
SrFeAsF &  240 & 133,180 & & 3.9996(1) & 8.9618(4) & 0.1583(2) & 0.6515(1) & 2.417(2) & 111.7  & 108.4 & 1.358 & \onlinecite{Xiao2009b}\\ 
SrFeAsF &  300 & & & 3.99938(3) & 8.9727(1) & 0.15903(13) & 0.6528(2) & 2.4246 &  111.13 & 108.65 & 1.371 & \onlinecite{Matsuishi2008}\\ 
SrFe$_{0.875}$Co$_{0.125}$AsF & 300 & & 4.0\footnotemark[1] & 4.00182(2) & 8.94345(8) & 0.15821(12) & 0.6510(2) & 2.4138 &  111.99 & 108.23 & 1.350 & \onlinecite{Matsuishi2008}\\ 
LaCoPO & RT & 43\footnotemark[5] &  & 3.9681(9) & 8.3779(1) & 0.1509(5) & 0.6321(9) &  2.2719  & 121.69 & 103.73 & 1.107 & \onlinecite{Yanagi2008}\\
LaCoPO & RT & & & 3.9678(9) & 8.379(3) & 0.155(1) & 0.617(6) & 2.21(2) & 127.4 & 101.3 & 0.980 & \onlinecite{Zimmer1995}\\
LaNiAsO & 297 & & 2.4 & 4.12309(1) & 8.18848(6) & 0.14697(9) & 0.6368(1) &  2.3463(7)  & 122.95(6) & 103.18(3) & 1.120 & \onlinecite{Watanabe2008}\\
LaNiPO & RT & & 4.3 & 4.0453(1) & 8.1054(3) & 0.15190(9) & 0.6257(5) & 2.265(2) &  126.5(2) & 101.7(1) & 1.019 & \onlinecite{Tegel2008}\\
LaNiPO & 298 & & 4.2 & 4.04669(6) & 8.1089(2) & 0.1530(2) & 0.6244(2) & 2.261 &  127.0 & 101.5 & 1.009 & \onlinecite{McQueen2009}\\
ThCuPO\footnotemark[4] & RT & &  & 3.8995(4) & 8.2939(7) &  &  &  &   &  &  & \onlinecite{Sakai2010}\\
UCuPO\footnotemark[4] & RT & 220 &  & 3.7958(2) & 8.2456(4) &  &  &  &   &  &  & \onlinecite{Sakai2010}\\
CeRuPO & RT & & & 4.026(1) & 8.256(2) & 0.14716(4) & 0.6419(2) & 2.329(1) & 119.6 & 104.7 & 1.172 & \onlinecite{Zimmer1995}\\ 

\end{tabular}
\end{ruledtabular}
\footnotetext[1]{8 vol\% diamagnetic shielding fraction}
\footnotetext[2]{48 vol\% diamagnetic shielding fraction}
\footnotetext[3]{18 vol\% diamagnetic shielding fraction}
\footnotetext[4]{single crystal}
\footnotetext[5]{ferromagnetic Curie temperature}

\end{table*}

%\squeezetable
\begin{table*}
\caption{\label{data5} Crystal data for body-centered-tetragonal ThCr$_2$Si$_2$-type $AM_2$As$_2$ (122-type)  compounds.  Here, $M$ is a 3$d$ transition element possibly mixed with a 4$d$ transition element, $z_{\rm As}$ is the $c$-axis position parameter of As atom, $\theta_2$ and $\theta_4$ are the twofold and fourfold As-$M$-As bond angles within an $M$-centered $M$As$_4$ tetrahedron, respectively, and $h_{M{\rm As}}$ is the separation of a plane of $M$ atoms from either of the two adjacent planes of As atoms.  The temperature $T$ at which the diffraction data were obtained is given (``RT'' means room temperature), along with the magnetic and/or crystallographic transition temperature $T_0$ and the superconducting transition temperature $T_{\rm c}$ for the sample if given.}
\begin{ruledtabular}
\begin{tabular}{lccccccccccc}
Compound & $T$ & $T_0$ & $T_{\rm c}$ & $a$ & $c$ & $z_{\rm As}$ & $d_{M-{\rm As}}$ & $\theta_2$ & $\theta_4$ & $h_{M{\rm As}}$ & Ref.\\
& (K) & (K) & (K) & (\AA) & (\AA) &    & (\AA) & (deg) & (deg) & (\AA) \\ \hline
SrCr$_{2}$As$_2$\footnotemark[1] & RT & & & 3.918(3) & 13.05(1) & 0.367 & 2.484 & 104.1 & 112.2 & 1.527 & \onlinecite{pfisterer1980}\\
BaCr$_{2}$As$_2$\footnotemark[1] & RT & & & 3.963(3) & 13.60(1) & 0.361 & 2.491 & 105.4 & 111.5 & 1.510 & \onlinecite{pfisterer1980}\\
BaCr$_{2}$As$_2$\footnotemark[2] & RT & & $< 2$ & 3.9678(4) & 13.632(3) & 0.361 & 2.429 & 105.3 & 111.6 & 1.513 & \onlinecite{DJSingh2009}\\
BaMn$_{2}$As$_2$\footnotemark[2] & RT & & $< 1.8$ & 4.1686(4) & 13.473(3) & 0.3615(3) & 2.569 & 108.44 & 109.99 & 1.502 & \onlinecite{singh2009}\\
KFe$_2$As$_2$\footnotemark[1] & RT & & 3.6 & 3.842 & 13.838 & 0.3533 & 2.394 & 106.72(14) &  & 1.429 & \onlinecite{Rotter2008c}\\
CaFe$_{2}$As$_2$\footnotemark[2] & 300 & 173 &  & 3.879(3) & 11.740(3) &  &  &  &  &  &\onlinecite{Goldman2008}\\
CaFe$_{2}$As$_2$\footnotemark[2]$^,$\footnotemark[3] & 297 & & 20 & 3.872(9) & 11.730(2) & 0.3665(9) & 2.370(9) & 109.5(2) & 109.4(6) & 1.367 &\onlinecite{wu2008}\\
SrFe$_{2}$As$_2$\footnotemark[1] & 297 & & & 3.9243(1) & 12.3644(1) & 0.3600(1) & 2.388(1) & 110.5(1) & 108.9(1) & 1.360 &\onlinecite{Tegel2008d}\\
SrFe$_{2}$As$_2$\footnotemark[2] & 250 & & & 3.9289(3) & 12.3172(12) & 0.36035(5) & 2.3890(4) & 110.63(3) & 108.90(2) & 1.359 &\onlinecite{Saha2010}\\
${\rm Sr_{0.33}Ca_{0.67}Fe_{2}As_2}$\footnotemark[2] & 250 & & & 3.9066(8) & 11.988(5) & 0.36423(7) & 2.3855(7) & 109.94(4) & 109.24(2) & 1.369 &\onlinecite{Saha2010}\\
SrFe$_{1.8}$Co$_{0.2}$As$_2$\footnotemark[1] & RT & & 19 & 3.9278(2) & 12.3026(2) & 0.3613(1) & 2.3941 & 110.23 & 109.09 & 1.369 & \onlinecite{Leithe-Jasper2008}\\
SrFe$_{1.4}$Ru$_{0.6}$As$_2$\footnotemark[1] & RT & & 19.3 & 3.99178(2) & 12.0635(1) & 0.3599(1) & 2.396 & 112.81 & 107.83 & 1.326 & \onlinecite{Schnelle2009}\\
SrFe$_{1.3}$Ru$_{0.7}$As$_2$\footnotemark[1] & RT & & 19.3 & 4.00507(2) & 12.0087(1) & 0.3598(1) & 2.398 & 113.28 & 107.60 & 1.319 & \onlinecite{Schnelle2009}\\
SrFe$_{1.2}$Ru$_{0.8}$As$_2$\footnotemark[1] & RT & & 17.6 & 4.01090(2) & 11.9835(1) & 0.3598(1) & 2.399 & 113.46 & 107.51 & 1.316 & \onlinecite{Schnelle2009}\\
BaFe$_2$As$_2$\footnotemark[2] & 250 & 136.0 & & 3.948 & 12.947 &  &  &  &  &  & \onlinecite{Wilson2009}\\
BaFe$_2$As$_2$\footnotemark[1] & 297 & 140 & & 3.9625(1) & 13.001(1) & 0.35393(8) & 2.403(1) & 111.1(1) & 108.7(1) & 1.360 & \onlinecite{rotter2008b}\\
BaFe$_2$As$_2$\footnotemark[1] & RT & 140(3)& & 3.963 & 13.016 & 0.3545 & 2.403 & 111.09(15) &  & 1.360 & \onlinecite{Rotter2008c}\\
BaFe$_2$As$_2$\footnotemark[2] & 175 & 140(2) & & 3.9622(4) & 13.001(1) & 0.35393(8) & 2.3980(6) & 111.41(4) &  108.52 & 1.352 & \onlinecite{Rotter2010}\\
BaFe$_2$As$_2$\footnotemark[2] & RT & 137 & & 3.9633(4) & 13.022(2) & 0.35424(6) & 2.402(2) & 111.18(3) &  108.62(2) & 1.358 & \onlinecite{Rullier-Albenque2010}\\
${\rm Ba(Fe_{0.62}Ru_{0.38})_2As_2}$\footnotemark[2] & RT & & 7 & 4.0342(5) & 12.749(2) & 0.35328(8) & 2.409(2) & 113.73(4) &  107.39(2) & 1.317 & \onlinecite{Rullier-Albenque2010}\\
Ba$_{0.89}$K$_{0.11(6)}$Fe$_2$As$_2$\footnotemark[1] & RT & 115(3) & 2.5 & 3.949 & 13.088 & 0.3545 & 2.402 & 110.60(14) &  & 1.367 & \onlinecite{Rotter2008c}\\
Ba$_{0.82}$K$_{0.18(5)}$Fe$_2$As$_2$\footnotemark[1] & RT & 90(5) & 25.4 & 3.937 & 13.155 & 0.3552 & 2.406 & 109.82(16) &  & 1.383 & \onlinecite{Rotter2008c}\\
Ba$_{0.70}$K$_{0.30(6)}$Fe$_2$As$_2$\footnotemark[1] & RT & & 36.4 & 3.919 & 13.263 & 0.3542 & 2.398 & 109.58(14) &  & 1.382 & \onlinecite{Rotter2008c}\\
Ba$_{0.63}$K$_{0.37(6)}$Fe$_2$As$_2$\footnotemark[1] & RT & & 38.6 & 3.915 & 13.294 & 0.3541 & 2.397 & 109.49(16) &  & 1.383 & \onlinecite{Rotter2008c}\\
Ba$_{0.56}$K$_{0.44(6)}$Fe$_2$As$_2$\footnotemark[1] & RT & & 36.8 & 3.907 & 13.335 & 0.3537 & 2.394 & 109.39(14) &  & 1.383 & \onlinecite{Rotter2008c}\\
Ba$_{0.38}$K$_{0.62(6)}$Fe$_2$As$_2$\footnotemark[1] & RT & & 29.6 & 3.887 & 13.506 & 0.3544 & 2.401 & 108.12(14) &  & 1.409 & \onlinecite{Rotter2008c}\\
Ba$_{0.29}$K$_{0.71(6)}$Fe$_2$As$_2$\footnotemark[1] & RT & & 14.6 & 3.880 & 13.569 & 0.3548 & 2.406 & 107.53(14) &  & 1.422 & \onlinecite{Rotter2008c}\\
Ba$_{0.15}$K$_{0.85(6)}$Fe$_2$As$_2$\footnotemark[1] & RT & & 8.9 & 3.852 & 13.735 & 0.3535 & 2.394 & 107.15(14) &  & 1.421 & \onlinecite{Rotter2008c}\\
Ba$_{0.10}$K$_{0.90(6)}$Fe$_2$As$_2$\footnotemark[1] & RT & & 8.9 & 3.848 & 13.793 & 0.3539 & 2.399 & 106.63(14) &  & 1.433 & \onlinecite{Rotter2008c}\\
Ba$_{0.7}$K$_{0.3}$FeAs$_2$\footnotemark[1] & 300 & & 36 & 3.9257(1) & 13.2702(3) & 0.3545(1) & 2.4033 & 109.5 & 109.4 & 1.387 & \onlinecite{Rotter2009}\\
EuFe$_{2}$As$_2$\footnotemark[1] & 297 & & & 3.9062(1) & 12.1247(2) & 0.3625(1) & 2.382(1) & 110.1(1) & 109.1(1) & 1.364 & \onlinecite{Tegel2008d}\\
(Sr$_{3}$Sc$_{2}$O$_{5}$)Fe$_{2}$As$_2$ & 300 & $< 4.2$ & & 4.0781(1) & 26.8386(5)	& & 2.431(1) & 113.8(1) & 107.4(1) & & \onlinecite{Tegel2009}\\
\end{tabular}
\end{ruledtabular}
\footnotetext[1]{polycrystalline sample}
\footnotetext[2]{single crystal sample}
\footnotetext[3]{powder sample with Na partially substituted for Ca}
\end{table*}

%\squeezetable
\begin{table*}
\caption{\label{data5A} Crystal data for body-centered-tetragonal ThCr$_2$Si$_2$-type $AM_2X_2$ (122-type)  compounds.  Here, $M$ is a 3$d$ or 4$d$ transition element, $X$ is a pnictogen (P, As), $z_{X}$ is the $c$-axis position parameter of atom $X$, $\theta_2$ and $\theta_4$ are the twofold and fourfold $X$-$M$-$X$ bond angles within an $M$-centered $MX_4$ tetrahedron, respectively, and $h_{MX}$ is the separation of a plane of $M$ atoms from either of the two adjacent planes of $X$ atoms.  The temperature $T$ at which the diffraction data were obtained is given (``RT'' means room temperature), along with the magnetic and/or crystallographic transition temperature $T_0$ and the superconducting transition temperature $T_{\rm c}$ for the sample if given.}
\begin{ruledtabular}
\begin{tabular}{lccccccccccc}
Compound & $T$ & $T_0$ & $T_{\rm c}$ & $a$ & $c$ & $z_{X}$ & $d_{M-X}$ & $\theta_2$ & $\theta_4$ & $h_{MX}$ & Ref.\\
& (K) & (K) & (K) & (\AA) & (\AA) &    & (\AA) & (deg) & (deg) & (\AA) \\ \hline
CaCo$_{2}$As$_2$\footnotemark[1] & RT & & & 3.989(3) & 10.33(1) & 0.372 & 2.359 & 115.4 & 106.6 & 1.260 & \onlinecite{pfisterer1980}\\
SrCo$_{2}$As$_2$\footnotemark[1] & RT & & & 3.935(7) & 11.83(2) & 0.362 & 2.372 & 112.1 & 108.2 & 1.325 & \onlinecite{pfisterer1980}\\
BaCo$_{2}$As$_2$\footnotemark[1] & RT & & & 3.958(5) & 12.67(2) & 0.361 & 2.428 & 109.2 & 109.6 & 1.406 & \onlinecite{pfisterer1980}\\
BaCo$_{2}$As$_2$\footnotemark[2] & RT & & $<2$ & 3.9537(1) & 12.6524(6) & & & & & & \onlinecite{Sefat2009}\\
CaNi$_{2}$As$_2$\footnotemark[1] & RT & & & 4.053(6) & 9.90(2) & 0.370 & 2.349 & 119.2 & 104.8 & 1.188 & \onlinecite{pfisterer1980}\\
SrNi$_{2}$As$_2$\footnotemark[1] & RT & & & 4.151(6) & 10.23(4) & 0.363 & 2.376 & 121.8 & 103.7 & 1.156 & \onlinecite{pfisterer1980}\\
BaNi$_{2}$As$_2$\footnotemark[1] & RT & & & 4.142(4) & 11.65(4) & 0.355 & 2.405 & 118.9 & 105.0 & 1.223 & \onlinecite{pfisterer1980}\\
BaNi$_{2}$As$_2$\footnotemark[2] & RT & & 0.7 & 4.112(4) & 11.54(2) & 0.3476(3) & 2.344 & 122.6 & 103.3 & 1.126 & \onlinecite{ronning2008}\\
SrNi$_{2}$As$_2$\footnotemark[2] & 124 & & 0.62 & 4.1374(8) & 10.188(4) & 0.3634(1) & 2.369 & 121.64 & 103.75 & 1.155 & \onlinecite{Bauer2008}\\
EuNi$_{2}$As$_2$\footnotemark[2] & 124 & & $< 0.4$ & 4.0964(6) & 10.029(3) & 0.3674(2) & 2.3625 & 120.22 & 104.38 & 1.177 & \onlinecite{Bauer2008}\\
SrCu$_{2}$As$_2$\footnotemark[1] & RT & & & 4.271(1) & 10.18(2) & 0.377 & 2.496 & 117.6 & 105.6 & 1.293 & \onlinecite{pfisterer1980}\\
BaCu$_{2}$As$_2$\footnotemark[1] & RT & & & 4.446(5) & 10.07(1) & 0.374 & 2.550 & 121.4 & 103.9 & 1.249 & \onlinecite{pfisterer1980}\\
SrRu$_{2}$As$_2$\footnotemark[1] & RT & & $< 1.8$ & 4.1713(1) & 11.1845(4) & 0.3612(2) & 2.4283 & 118.38 & 105.21 & 1.244 & \onlinecite{Nath2009}\\
SrRu$_{2}$As$_2$\footnotemark[1] & RT & &  & 4.16911(2) & 11.1706(1) & 0.3591(1) & 2.415 & 119.38 & 104.76 & 1.219 & \onlinecite{Schnelle2009}\\
BaRu$_{2}$As$_2$\footnotemark[1] & RT & & $< 1.8$ & 4.15248(8) & 12.2504(3) & 0.3527(1) & 2.4277 & 117.57 & 105.58 & 1.258 & \onlinecite{Nath2009}\\
BaRh$_{2}$As$_2$\footnotemark[2] & RT & & $< 1.8$ & 4.0564(6) & 12.797(4) & 0.3566(3) & 2.444 & 112.15 & 108.15 & 1.364 & \onlinecite{singh2008}\\\hline
BaMn$_{2}$P$_2$\footnotemark[2] & RT &  & & 4.037(1) & 13.061(1) & 0.3570(3) & 2.455(2) & 110.63(9) & 108.90(5) & 1.398 & \onlinecite{Mewis1980}\\
CaFe$_{2}$P$_2$\footnotemark[2] & RT &  & & 3.855(1) & 9.985(1) & 0.3643(3) & 2.240(2) & 118.74(9) & 105.04(4) & 1.141 & \onlinecite{Mewis1980}\\
SrFe$_{2}$P$_2$\footnotemark[2] & RT &  & & 3.825(1) & 11.612(1) & 0.3521(8) & 2.251(5) & 116.4(3) & 106.1(1) & 1.186 & \onlinecite{Mewis1980}\\
${\rm BaFe_2(As_{0.71}P_{0.29})_2}$\footnotemark[2]$^,$\footnotemark[3] & 175 & & & 3.9178(1) & 12.7610(7) & As: 0.3544(1) & 2.3689(7) & 111.57(5) &  108.44 & 1.332 & \onlinecite{Rotter2010}\\
& & & & & & P: 0.3402(1) & 2.272(4) & 119.1(3) &  104.9 & 1.151 & \onlinecite{Rotter2010}\\
${\rm BaFe_2(As_{0.53}P_{0.47})_2}$\footnotemark[2]$^,$\footnotemark[3] & 175 & & & 3.9065(1) & 12.7355(6) & As: 0.3542(1) & 2.361(1) & 111.6(2) &  108.4 & 1.327 & \onlinecite{Rotter2010}\\
& & & & & & P: 0.3438(4) & 2.289(3) & 117.1(2) &  105.8 & 1.195 & \onlinecite{Rotter2010}\\
${\rm BaFe_2(As_{0.24}P_{0.76})_2}$\footnotemark[2]$^,$\footnotemark[3] & 175 & & & 3.8660(3) & 12.592(1) & As: 0.3571(4) & 2.357(3) & 110.2(2) &  109.1 & 1.349 & \onlinecite{Rotter2010}\\
& & & & & & P: 0.3430(4) & 2.260(2) & 117.6(1) &  105.6 & 1.171 & \onlinecite{Rotter2010}\\
${\rm BaFe_2P_2}$\footnotemark[2] & 175 & & & 3.8435(4) & 12.422(2) & 0.3459(1) & 2.2614(9) & 116.39(7) &  106.12 & 1.191 & \onlinecite{Rotter2010}\\
BaFe$_{2}$P$_2$\footnotemark[2] & RT &  & & 3.840(1) & 12.442(1) & 0.3456(4) & 2.259(3) & 116.4(2) & 106.1(1) & 1.189 & \onlinecite{Mewis1980}\\
EuFe$_{2}$P$_2$\footnotemark[1] & RT & 29.5 (Eu) & --- & 3.8178(1) & 11.2372(3) & 0.3548(2) & 2.243 & 116.7(1) & 106.0 & 1.178 & \onlinecite{Feng2010}\\
CaCo$_{2}$P$_2$\footnotemark[2] & RT &  & & 3.858(1) & 9.593(1) & 0.3721(4) & 2.257(2) & 117.5(1) & 105.62(6) & 1.171 & \onlinecite{Mewis1980}\\
CaNi$_{2}$P$_2$\footnotemark[2] & RT &  & & 3.916(1) & 9.363(1) & 0.3774(5) & 2.293(3) & 117.3(1) & 105.70(7) & 1.193 & \onlinecite{Mewis1980}\\
BaNi$_{2}$P$_2$\footnotemark[2] & RT &  & & 3.947(1) & 11.820(1) & 0.3431(3) & 2.260(2) & 121.71 & 103.72 & 1.100 & \onlinecite{Keimes1997}\\
CaCu$_{1.75}$P$_2$\footnotemark[2] & RT &  & & 4.014(1) & 9.627(1) & 0.3831(4) & 2.381(2) & 114.9(1) & 106.83(5) & 1.281 & \onlinecite{Mewis1980}\\
SrCu$_{1.75}$P$_2$\footnotemark[2] & RT &  & & 4.166(1) & 9.607(1) & 0.3805(4) & 2.431(2) & 117.9(1) & 105.42(5) & 1.254 & \onlinecite{Mewis1980}\\
EuCu$_{1.75}$P$_2$\footnotemark[2] & RT &  & & 4.110(1) & 9.591(1) & 0.3810(5) & 2.409(2) & 117.1(1) & 105.79(6) & 1.256 & \onlinecite{Mewis1980}\\

\end{tabular}
\end{ruledtabular}
\footnotetext[1]{polycrystalline sample}
\footnotetext[2]{single crystal sample}
\footnotetext[3]{The As and P $z$ parameters were refined separately}
\end{table*}

%\squeezetable
\begin{table*}
\caption{\label{data6} Crystal data for primitive tetragonal Fe$X$-type (11-type, anti-PbO structure)  compounds in space group \emph{P}4/\emph{nmm} (No.~129, second setting), $Z = 2$~formula units per unit cell. Atomic positions: Fe in 2(a) (1/4, 3/4, 0); $X = $ Se, Te in 2(c) (1/4, 1/4, $z$) with $z \sim 0.25$.  Here, $z_{X}$ is the $c$-axis position parameter of atom $X$, $\theta_2$ and $\theta_4$ are the twofold and fourfold $X$-$Fe$-$X$ bond angles within an Fe-centered Fe$X_4$ tetrahedron, respectively, and $h_{{\rm Fe}X}$ is the separation of a plane of Fe atoms from either of the two adjacent planes of $X$ atoms.  The temperature $T$ at which the diffraction data were obtained is given (``RT'' means room temperature), along with the magnetic and/or crystallographic transition temperature $T_0$ and the superconducting transition temperature $T_{\rm c}$ for the sample if given.  Not included are crystal data for any excess Fe atoms ($y$ in the formula Fe$_{1+y}$Te$_{1-x}$Se$_{x}$) in the $X$ layers.  The compositions given are those deduced from structure refinements, if available, rather than the nominal compositions.}
\begin{ruledtabular}
\begin{tabular}{lccccccccccc}
Compound & $T$ & $T_0$ & $T_{\rm c}$ & $a$ & $c$ & $z_{X}$ & $d_{{\rm Fe}-X}$ & $\theta_2$ & $\theta_4$ & $h_{{\rm Fe}X}$ & Ref.\\
& (K) & (K) & (K) & (\AA) & (\AA) &    & (\AA) & (deg) & (deg) & (\AA) \\ \hline
Fe$_{1+y}$Te\footnotemark[1] & 300 & 73 &  & 3.8219(1) & 6.2851(1) & 0.2792(4) & 2.594 & 94.88 & 117.23 & 1.755 & \onlinecite{Martinelli2010}\\
Fe$_{1.087}$Te\footnotemark[2] & RT & 67 &  & 3.826(1) & 6.273(3) & 0.28141(8) & 2.603 & 94.60 & 117.38 & 1.765 & \onlinecite{Viennois2010}\\
Fe$_{1+y}$Te$_{0.95}$Se$_{0.05}$\footnotemark[1] & 300 & 50\footnotemark[8] & 11.0 & 3.8184(1) & 6.2617(1) & 0.2763(4) & 2.576 & 95.63 & 116.80 & 1.730 & \onlinecite{Martinelli2010}\\
Fe$_{1+y}$Te$_{0.9}$Se$_{0.1}$\footnotemark[1] & 300 &  & 11.9 & 3.8160(1) & 6.2381(1) & 0.2746(4) & 2.564 & 96.17 & 116.51 & 1.713 & \onlinecite{Martinelli2010}\\
Fe$_{1+y}$Te$_{0.85}$Se$_{0.15}$\footnotemark[1] & 300 &  & 12.7 & 3.8133(1) & 6.2116(1) & 0.2729(4) & 2.551 & 96.72 & 116.20 & 1.695 & \onlinecite{Martinelli2010}\\
Fe$_{1+y}$Te$_{0.80}$Se$_{0.20}$\footnotemark[1] & 300 &  & 13.6 & 3.8114(1) & 6.1843(1) & 0.2714(4) & 2.539 & 97.26 & 115.90 & 1.678 & \onlinecite{Martinelli2010}\\
${\rm Fe_{1.049}Te_{0.79}Se_{0.21}}$\footnotemark[2] & RT & 14 &  & 3.815(2) & 6.187(4) & 0.2789(4) & 2.572 & 95.73 & 116.75 & 1.726 & \onlinecite{Viennois2010}\\
${\rm Fe_{1.035}Te_{0.78}Se_{0.22}}$\footnotemark[2] & RT & --- & 6 & 3.806(3) & 6.187(6) & 0.2779(4) & 2.565 & 95.80 & 116.71 & 1.719 & \onlinecite{Viennois2010}\\
${\rm Fe_{1.053}Te_{0.73}Se_{0.27}}$\footnotemark[2] & RT & 5 &  & 3.807(3) & 6.153(7) & 0.2787(4) & 2.562 & 95.97 & 116.62 & 1.715 & \onlinecite{Viennois2010}\\
${\rm Fe_{1.013}Te_{0.68}Se_{0.32}}$\footnotemark[2] & RT & --- & 11 & 3.803(2) & 6.136(3) & 0.2767(3) & 2.549 & 96.48 & 116.33 & 1.698 & \onlinecite{Viennois2010}\\
FeTe$_{0.56}$Se$_{0.44}$\footnotemark[2] & 173 &  & 13 & 3.7996(2) & 5.9895(6) & Se: 0.2468(7)\footnotemark[3] & 2.407(3) & 104.1(1) & 112.15(6) & 1.478 & \onlinecite{Tegel2010}\\
 &  &  &  &  &  & Te: 0.2868(3)\footnotemark[3] & 2.561(1) & 95.75(4) & 116.74(2) & 1.718 & \\
 &  &  &  &  &  & Se-Fe-Te: & 99.99(9) & 114.32(5) &  & \\
Fe$_{1.03}$Te$_{0.43}$Se$_{0.57}$\footnotemark[1] & 295 & 40 & 13.9 & 3.800742(4) & 5.99263(3) &  0.27388(9) & 2.5110(4) & 98.37(2) & 115.29(1) & 1.641 & \onlinecite{Gresty2009}\\
FeSe$_{0.92}$\footnotemark[1]$^,$\footnotemark[4] & 295 & 70\footnotemark[7] & 8 & 3.77376(2) & 5.52482(5) & 0.2652(1) & 2.389 & 104.34 & 112.10 & 1.465 & \onlinecite{Margadonna2008}\\
Fe$_{0.987}$Se\footnotemark[1]$^,$\footnotemark[5] & 298 &  &  & 3.7747(1) & 5.5229(1) & 0.2669(2) & 2.395 & 104.02 & 112.26 & 1.474 & \onlinecite{McQueen2009b}\\
Fe$_{0.997}$Se\footnotemark[1]$^,$\footnotemark[6] & 298 &  &  & 3.7734(1) & 5.5258(1) & 0.2672(1) & 2.396 & 103.91 & 112.32 & 1.477 & \onlinecite{McQueen2009b}\\
FeSe$_{1.003}$\footnotemark[1]$^,$\footnotemark[4] & 298 & 60--80\footnotemark[7] & 8 & 3.7724(1) & 5.5217(1) & 0.2673(2) & 2.395 & 103.91 & 112.32 & 1.476 & \onlinecite{Phelan2009}\\
Fe$_{1.01}$Se\footnotemark[1]$^,$\footnotemark[5] & 298 & 90\footnotemark[7] & 8.5 &  &  &  &  &  &  &  & \onlinecite{McQueen2009c}\\
Fe$_{1.03}$Se\footnotemark[1]$^,$\footnotemark[5] & 298 & none & none &  &  &  &  &  &  &  & \onlinecite{McQueen2009c}\\
FeSe$_{0.975(3)}$\footnotemark[1]$^,$\footnotemark[9] & 290 & 100 & 8.21 & 3.77381(2) & 5.52330(5) & 0.26732(14) & 2.396 & 103.91 & 112.32 & 1.476 & \onlinecite{Pomjakushina2009}\\
FeSe$_{0.975(4)}$\footnotemark[1]$^,$\footnotemark[9] & 250 & 100 & 8.21 & 3.76988(5) & 5.51637(9) & 0.2674(3) & 2.394 & 103.91 & 112.32 & 1.475 & \onlinecite{Khasanov2010}\\
\end{tabular}
\end{ruledtabular}
\footnotetext[1]{Polycrystalline sample.}
\footnotetext[2]{Single crystal sample.}
\footnotetext[3]{The $z$ parameters of the statistically distributed Se and Te atoms on the 2($c$) sites were found to be different.}
\footnotetext[4]{The structure was refined for Se deficiency, rather than Fe excess.}
\footnotetext[5]{Contained small amounts of Fe and Fe$_3$O$_4$.}
\footnotetext[6]{Contained small amounts of Fe, Fe$_7$Se$_8$ and $\delta$FeSe.}
\footnotetext[7]{Orthorhombic structural transition temperature only.}
\footnotetext[8]{Magnetic and monoclinic structural transition temperature.}
\footnotetext[9]{Contained small amounts of Fe and $\delta$FeSe.}
\end{table*}

%\squeezetable
\begin{table*}
\caption{\label{data7} Crystal data for primitive tetragonal $A$Fe$X$ 111-type compounds in space group \emph{P}4/\emph{nmm} (No.~129, second setting), $Z = 2$~formula units per unit cell. Atomic positions: $A$ in 2(c) (1/4, 1/4, $z_A$); Fe in 2(a) (1/4, 3/4, 0); $X$ in 2(c) (1/4, 1/4, $z_X$).  Here, $A$ is an alkali metal, $X$ is a pnictogen (P, As), $z_{A}$ is the $c$-axis position parameter of atom $A$, $z_{X}$ is the $c$-axis position parameter of atom $X$, $\theta_2$ and $\theta_4$ are the two-fold and four-fold $X$-Fe-$X$ bond angles within an Fe-centered Fe$X_4$ tetrahedron, respectively, and $h_{{\rm Fe-}X}$ is the separation of a plane of Fe atoms from either of the two adjacent planes of $X$ atoms.  The temperature $T$ at which the diffraction data were obtained is given (``RT'' means room temperature), along with the magnetic and/or crystallographic transition temperature $T_0$ and the superconducting transition temperature $T_{\rm c}$ for the sample if given.}
\begin{ruledtabular}
\begin{tabular}{lcccccccccccc}
Compound & $T$ & $T_0$ & $T_{\rm c}$ & $a$ & $c$ & $z_{A}$ & $z_{X}$ & $d_{M-X}$ & $\theta_2$ & $\theta_4$ & $h_{MX}$ &Ref.\\
& (K) & (K) & (K) & (\AA) & (\AA) & &   & (\AA) & (deg) & (deg) & (\AA)\\ \hline
LiFeAs\footnotemark[2] & 215 &  & 18 & 3.7914(7) & 6.364(2) & 0.3459(15) & 0.7635(1) & 2.421 & 103.11 & 112.75 & 1.505 & \onlinecite{Tapp2008R}\\
LiFeAs\footnotemark[1] & 295 & none & 10 & 3.776360(4) & 6.35679(1) & 0.3464(2) & 0.76285(3) & 2.4162(1) & 102.793(6) & 112.910(3) & 1.508 & \onlinecite{Pitcher2008C}\\
NaFeAs\footnotemark[1] & 70 & 37\footnotemark[3],45 & 5 & 3.94481(3) & 6.99680(8) & 0.3535(2) & 0.7976(1) & 2.4282(4) & 108.64(3) & 109.89 & 1.416 & \onlinecite{Li2009b}\\
LiFeP\footnotemark[1] & RT? &  & 6 & 3.69239(2) & 6.03081(2) & 0.3480(1) & 0.7800(5) & 2.273 & 108.59 & 109.91 & 1.327 & \onlinecite{Deng2009}\\
\end{tabular}
\end{ruledtabular}
\footnotetext[1]{Polycrystalline sample.}
\footnotetext[2]{Single crystal sample.}
\footnotetext[3]{Ordered magnetic moment at 5~K = 0.09(4)~$\mu_{\rm B}$/Fe.}
\end{table*}

\squeezetable
\begin{table*}
\caption{\label{LoTStructData1111} The caption for this table is placed at the beginning of the Appendix, due to its length.}
\begin{ruledtabular}
\begin{tabular}{l|ccccccccc}
Compound & $T$ & $a$ & $b$ & $c$ & $z_{\rm As}$ & $z_{R}$ & $\vec{\mu}$ or $\mu$ & $Q_{\rm AF}$\footnotemark[4] & Ref.\\
         & (K) & (\AA) & (\AA) & (\AA) &&&  ($\mu_{\rm B}$) & (r.l.u.)\\
\hline
LaFeAsO  & 175 & 4.03007(9) & $=a$ & 8.7368(2) & 0.6507(4) &  0.1418(3)  &&&\onlinecite{Cruz2008}\\
  & 8\footnotemark[1] & 5.7096 & 5.6819 & 8.7262(5) &  &  & 0.36(5) & $\left(10\frac{1}{2}\right)$  & \onlinecite{Cruz2008}\\
LaFeAsO  & 300 & 4.03268(1) & $=a$ & 8.74111(4) & 0.65166(7) &  0.14134(4)  &&&\onlinecite{nomura2008}\\
  & 120 & 5.71043(3) & 5.68262(3)  & 8.71964(4) & 0.65129(7)  & 0.14171(4) & &  & \onlinecite{nomura2008}\\
LaFeAsO  & 175 & 4.03007(9) & $=a$ & 8.7368(2) & 0.6507(4) &  0.1417(3)  &&&\onlinecite{Huang2008}\\
  & 2 & 5.70988(9) & 5.68195(9) & 8.7265(1) & 0.6506(3) & 0.1430(3) & 0.36(5) & & \onlinecite{Huang2008}\\
LaFeAsO  & 300 & 4.0345(1) & $=a$ & 8.7387(4) & 0.6498(4) &  0.1420(3)  &&&\onlinecite{Sefat2008h}\\
  & 4 & 5.7103(2) & 5.6823(2) & 8.7117(4) & 0.6501(4) & 0.1424(3) &  &  & \onlinecite{Sefat2008h}\\
LaFeAsO  & RT & 4.0322(2) & $=a$ & 8.7364(4) & 0.6508(5) &  0.1416(4)  &&&\onlinecite{Qureshi2010}\\
  & 2 & 5.7063(4) & 5.6788(4) & 8.7094(6) & 0.6505(5) & 0.1420(4) & 0.63(1)\,$\hat{\bf a}$ & $\left(10\frac{1}{2}\right)$ & \onlinecite{Qureshi2010}\\
LaFeAsO\footnotemark[5]  & 9.5 &  &  &  &  &  & 0.78(8)\,$\hat{\bf a}$ & & \onlinecite{Li2010c}\\
CeFeAsO  & 175 & 3.99591(5) & $=a$ & 8.6522(1) & 0.6546(2) &  0.1413(3)  &&&\onlinecite{Zhao2008}\\
  & 1.4 & 5.66263(4) & 5.63273(4) & 8.64446(7) & 0.6553(1) & 0.1402(2) &  &  & \onlinecite{Zhao2008}\\
  & 40 & --- & --- & --- & --- & --- & Fe: 0.8(1)\,$\hat{\bf a}$ & (100) & \onlinecite{Zhao2008}\\
& 1.7 & --- & --- & --- & --- & --- & Fe: 0.94(3)\,$\hat{\bf a}$ & (100) & \onlinecite{Zhao2008}\\
& 1.7 & --- & --- & --- & --- & --- & Ce: 0.83(2)\footnotemark[2] &  & \onlinecite{Zhao2008}\\
PrFeAsO  & 175 & 3.97716(5) & $=a$ & 8.6057(2) & 0.6559(4) &  0.1397(6)  &&&\onlinecite{Zhao2008b}\\
  & 5 & 5.6374(1) & 5.6063(1) & 8.5966(2) & 0.6565(3) & 0.1385(5) & Fe: 0.48(9)\,$\hat{\bf a}$ & (100) & \onlinecite{Zhao2008b}\\
  & 5 &  &  &  &  &  & Pr: 0.84(4)\,$\hat{\bf c}$ & AF & \onlinecite{Zhao2008b}\\
PrFeAsO  & 30 & 5.701 & 5.672 & 8.689 &  &   & Fe: 0.35(5)\,$\hat{\bf a}$ &  &\onlinecite{Kimber2008}\\
  & 1.4 & &  & &  & & Fe: 0.53(20) & & \onlinecite{Kimber2008}\\
  & 1.4 &  &  &  &  &  & Pr: 0.83(9) & $\left(101\right)$\footnotemark[6] & \onlinecite{Kimber2008}\\
NdFeAsO & 30 & --- & --- & --- & --- & --- & Fe: 0.25(7) & $\left(10\frac{1}{2}\right)$? & \onlinecite{Chen2008}\\
NdFeAsO  & 175 & 3.9611(1) & $=a$ & 8.5724(2) & 0.6580(4) &  0.1393(3)  &&&\onlinecite{Qiu2008}\\
  & 0.3 & 5.6159(1) & 5.5870(1) & 8.5570(2) & 0.6584(4) & 0.1389(2) & Fe: 0.9(1)\,$\hat{\bf a}$ & (100) & \onlinecite{Qiu2008}\\
  & 0.3 &  &  &  &  &  & Nd: 1.55(4)\footnotemark[3] & (100) & \onlinecite{Qiu2008}\\
NdFeAsO & 30 & --- & --- & --- & --- & --- & Fe: 0.54(1)\,$\hat{\bf a}$ & $\left(10\frac{1}{2}\right)$ & \onlinecite{Tian2010}\\
 & 10 & --- & --- & --- & --- & --- & Fe: 0.32(2)\,$\hat{\bf a}$ & $(100)$ & \onlinecite{Tian2010}\\
 & 1.5 & --- & --- & --- & --- & --- & Fe: 0.41(5)\,$\hat{\bf a}$ & $(100)$ & \onlinecite{Tian2010}\\
& 1.5 & --- & --- & --- & --- & --- & Nd: 1.30(5)\footnotemark[2] & $(100)$ & \onlinecite{Tian2010}\\
SmFeAsO  & 295 & 3.93880(2) & $=a$ & 8.51111(7) & --- &  ---  &&&\onlinecite{Margadonna2009}\\
  & 20 & 5.55105(5) & 5.57884(5) & 8.47014(9) & 0.6612(2) & 0.13741(8) &  &  & \onlinecite{Margadonna2009}\\
\hline
CaFeAsF & 2 & --- & --- & --- & --- & --- & 0.49(5)\,$\hat{\bf a}$ & $\left(10\frac{1}{2}\right)$ & \onlinecite{Xiao2009c}\\
SrFeAsF  & 297 & 3.9930(1) & $=a$ & 8.9546(1) & 0.6527(2) &  0.1598(2)  &&&\onlinecite{Tegel2008b}\\
  & 10 & 5.6602(1) & 5.6155(1) & 8.9173(2) & 0.6494(2) & 0.1635(2) & & & \onlinecite{Tegel2008b}\\
SrFeAsF  & 240 & 3.9996(1) & $=a$ & 8.9618(4) & 0.6515(1) &  0.1583(2)  &&&\onlinecite{Xiao2009b}\\
  & 2 & 5.6689(1) & 5.6260(1) & 8.9325(2) & 0.6525(1) & 0.1584(1) & 0.58(6)\,$\hat{\bf a}$ & & \onlinecite{Xiao2009b}\\
\end{tabular}
\end{ruledtabular}.

\footnotetext[1]{The low temperature structure was indexed in Ref.~\onlinecite{Cruz2008} on a \emph{P}112/\emph{n} monoclinic (M) unit cell instead of the currently accepted\cite{nomura2008} higher symmetry \emph{Cmma} (or \emph{Cmme}) orthorhombic (O) unit cell.  We have converted the originally reported M lattice parameters to the O ones using the Supplementary Information to Ref.~\onlinecite{Cruz2008} which gives $a_{\rm O} = a_{\rm M}\sqrt{2(1-\cos\gamma_{\rm M}})$, $b_{\rm O} = a_{\rm M}\sqrt{2(1+\cos\gamma_{\rm M}})$ and $c_{\rm O} = c_{\rm M}$ with $a_{\rm M} = 4.0275(2)$\AA, $c_{\rm M} = 8.7262(5)$\AA, and $\gamma_{\rm M} = 90.279(3)^\circ$.  See also Ref.~\onlinecite{Huang2008} by the same group.}
\footnotetext[2]{Noncollinear Ce spin structure (tentative) or noncollinear Nd spin structure, respectively.}
\footnotetext[3]{The observed magnetic ordering transition for both Fe and Nd is at 1.96(3)~K\@.  The Nd magnetic moment points approximately along the orthorhombic [101] axis.  The upper limit for the ordered moment for separate Fe moment ordering at higher temperatures as in LaFeAsO is $0.17~\mu_{\rm B}$/Fe.}
\footnotetext[4]{Whenever a paper listed the propagation vector as (101)~r.l.u.\ in orthorhombic notation, this was replaced by $\left(10\frac{1}{2}\right)$~r.l.u. in orthorhombic notation (see Sec.~\ref{1111-111-122}).  The (101)~r.l.u.\ notation is with respect to the magnetic unit cell, not the crystallographic unit cell as conventionally used.}
\footnotetext[5]{Single crystal.}
\footnotetext[6]{The $c$-axis designation is due to Pr ordering: there are two layers of Pr per unit cell.}
\end{table*}

%\squeezetable
\begin{table*}
\caption{\label{LoTStructData122} Low-temperature crystal and magnetic structure data for ${\rm ThCr_2Si_2}$-type (122-type) $A$${\rm Fe_2As_2}$ parent compounds.  In the high temperature tetragonal $I$4/\emph{mmm} structure, $Z = 2$ formula units per unit cell, the Wyckoff atomic positions are $A$: 2$a$ (0,0,0); Fe: 4$d$ ($\frac{1}{2}$,0,$\frac{1}{4}$); As: 4$e$ (0,0,$z$).  In the low temperature 122-type orthorhombic \emph{Fmmm} structure, $Z = 4$, the lattice parameters satisfy $c > a > b$ and the Wyckoff atomic positions are $A$: 4$a$ (0,0,0); Fe: 8$f$ ($\frac{1}{4}$,$\frac{1}{4}$,$\frac{1}{4}$); As: 8$i$ (0,0,$z_{\rm As}$).  For the orthorhombic crystal structure, some authors define $a$ and $b$ such that $b > a$, which we have reversed here to give $a > b$ so that the same convention is used for all listings.  Other notation is as described in the caption to Table~\ref{LoTStructData1111}.}
\begin{ruledtabular}
\begin{tabular}{l|cccccccc}
Compound & $T$ & $a$ & $b$ & $c$ & $z_{\rm As}$ & $\vec{\mu}$ or $\mu$ & $Q_{\rm AF}$ & Ref.\\
         & (K) & (\AA) & (\AA) & (\AA) &&  ($\mu_{\rm B}$) & (r.l.u.)\\
\hline
${\rm CaFe_2As_2}$\footnotemark[1]  & 300 & 3.879(3) & $=a$ & 11.740(3) & --- &    &&\onlinecite{Goldman2008}\\
  & 10 & 5.506(2) & 5.450(2) & 11.664(6) & 0.3664(5) &   0.80(5)\,$\hat{\bf a}$ & (101)\footnotemark[3] & \onlinecite{Goldman2008}\\
${\rm CaFe_2As_2}$\footnotemark[2]  & 50 & 5.5312(2) & 5.4576(2) & 11.683(1) & 0.3689(5) &    &&\onlinecite{Kreyssig2008}\\
${\rm SrFe_2As_2}$\footnotemark[1]  & 150 & 5.5695(9) & 5.512(1) & 12.298(1) & --- & &&\onlinecite{Zhao2008c}\\
  & 10 & --- & --- & --- & --- & 0.94(4)\,$\hat{\bf a}$ & (101)\footnotemark[3] &\onlinecite{Zhao2008c}\\
${\rm SrFe_2As_2}$\footnotemark[2]  & 297 & 3.9243(1) & $=a$ & 12.3644(1) & 0.3600(1) & &&\onlinecite{Tegel2008d}\\
  & 90 & 5.5783(3) & 5.5175(3) & 12.2965(6) & 0.3612(3) & &&\onlinecite{Tegel2008d}\\
${\rm SrFe_2As_2}$\footnotemark[2]  & 220 & --- & --- & --- & 0.3602(2) & &&\onlinecite{Kaneko2008}\\
  & 1.5 & --- & --- & --- & 0.3604(2) & 1.01(3)\,$\hat{\bf a}$ & (101)\footnotemark[3]&\onlinecite{Kaneko2008}\\
${\rm SrFe_2As_2}$\footnotemark[1]   & 10 & --- & --- & --- &  & 1.04(1) & &\onlinecite{Lee2010}\\
${\rm BaFe_2As_2}$\footnotemark[2]  & 297 & 3.9625(1) & $=a$ & 13.0168(3) & 0.3545(1) & &&\onlinecite{rotter2008b}\\
  & 20 & 5.6146(1) & 5.5742(1) & 12.9453(3) & 0.3538(1) & &&\onlinecite{rotter2008b}\\
${\rm BaFe_2As_2}$\footnotemark[2]  & 175 & 3.95702(4) & $=a$ & 12.9685(2) & 0.35405(8) & &&\onlinecite{Huang2008b}\\
  & 5 & 5.61587(5) & 5.57125(5) & 12.9428(1) & 0.35406(7) &  0.87(3)\,$\hat{\bf a}$ & (101)\footnotemark[3] & \onlinecite{Huang2008b}\\
${\rm BaFe_2As_2}$\footnotemark[1]  & 3 & --- & --- & --- & --- & 0.93(6)\,$\hat{\bf a}$ & (101)\footnotemark[3] &\onlinecite{Wilson2009}\\
${\rm BaFe_2As_2}$\footnotemark[1]  & --- & --- & --- & --- & --- & 0.91(21) &  &\onlinecite{Matan2009}\\
${\rm EuFe_2As_2}$\footnotemark[2]  & 297 & 3.9062(1) & $=a$ & 12.1247(2) & 0.3625(1) & &&\onlinecite{Tegel2008d}\\
  & 10 & 5.5546(2) & 5.4983(2) & 12.0590(4) & 0.3632(1) & &&\onlinecite{Tegel2008d}\\
${\rm EuFe_2As_2}$\footnotemark[1]  & 2.5 & 5.537(2) & 5.505(2) & 12.057(2) & 0.363(5) & Fe: 0.98(8)\,$\hat{\bf a}$ & Fe: (101)\footnotemark[3] &\onlinecite{Xiao2009}\\
  &  2.5 &  &  &  &  & Eu: 6.8(3)\,$\hat{\bf a}$ & Eu: (001)\footnotemark[4] &\onlinecite{Xiao2009}\\
\end{tabular}
\end{ruledtabular}.
\footnotetext[1]{Single crystal sample}
\footnotetext[2]{Polycrystalline sample}
\footnotetext[3]{The nearest-neighbor ordered moments are antiferromagnetically aligned along the orthorhombic $a$ and $c$ axes, and ferromagnetically aligned along $b$, which is the magnetic stripe axis.  See Fig.~\ref{Stripe_Mag_Struct} below.  In tetragonal reciprocal lattice unit notation, the ordering wave vector is $\left(\frac{1}{2}\frac{1}{2}1\right)$.}
\footnotetext[4]{This is so-called A-type collinear antiferromagnetic ordering.  The $R$ spins are aligned ferromagnetically within an $R$ plane parallel to the $a$-$b$ plane and are aligned antiferromagnetically between planes along the $c$ axis.}
\end{table*}

\squeezetable
\begin{table*}
\caption{Physical property data for 1111-type LaFeAsO$_{1-x}$F$_{x}$ compounds.  The data shown are magnetic and/or structural transition temperature(s) $T_0$, superconducting transition onset temperature $T_{\rm c}$, bare band structure density of states for both spin directions $N(E_{F})$, measured magnetic susceptibility $\chi$(300~K), and Sommerfeld coefficient $\gamma$ and Debye temperature $\Theta_{\rm D}$ from heat capacity data.  The samples are polycrystalline unless otherwise noted.}
\label{LaFeAsOFdata}
\begin{ruledtabular}
\begin{tabular}{lcccccc}
 LaFeAsO$_{1-x}$F$_{x}$   & $T_0$ & $T_{\rm c}$ & $N(E_{F})$& $\chi$(300~K) & $\gamma$ & $\Theta_{\rm D}$ \\
$x$ & (K) & (K) & $\left(\frac{\rm states}{\rm eV~f.u.}\right)$ & $\left(10^{-4} \frac{\rm cm^3}{\rm mol}\right)$ & $\left(\frac{\rm mJ}{\rm mol~K^{2}}\right)$ & (K)\\ \hline
 0 & 150 [\onlinecite{kamihara2008}] & & 2.62\footnotemark[2] [\onlinecite{djsingh2008b}]&  4.9 [\onlinecite{kamihara2008}] & 1.6 [\onlinecite{kohama2008}] \\
0 & 155 [\onlinecite{kohama2008}] & & 2.34\footnotemark[3] [\onlinecite{Xu2008}] & 4.1 [\onlinecite{nomura2008}] & 3.7 [\onlinecite{dong2008b}] & 282 [\onlinecite{dong2008b}] \\
0 & 145, 160 [\onlinecite{mcguire2008}] & & 2.43\footnotemark[3] [\onlinecite{Li2008}]& 2.7 [\onlinecite{mcguire2008}] & \\
0 & 138, 156 [\onlinecite{klauss2008}] & & 1.85\footnotemark[2]  [\onlinecite{Ma2008a}] & 3.35\footnotemark[5] [\onlinecite{Klingeler2008}]\\
0 &  & & 2.00\footnotemark[4] [\onlinecite{Nekrasov2008a}] \\
0 &  & & 2.17 [\onlinecite{Nekrasov2008c}] \\
0 &  & & 2.52\footnotemark[2] [\onlinecite{Larson2009}] \\
0 &  & & 3.80\footnotemark[4] [\onlinecite{Wang2009}] \\
0 & 140,154.5\footnotemark[1] [\onlinecite{Yan2009}] & &   & $\chi_{ab}=8.2$\footnotemark[1] [\onlinecite{Yan2009}]  \\
 & & & & $\chi_{c}=3.9$\footnotemark[1] [\onlinecite{Yan2009}]  \\
& & & & $\bar{\chi}=6.8$\footnotemark[1] [\onlinecite{Yan2009}]  \\
0.01 & 134, 151 [\onlinecite{Klingeler2008}] & & & 3.5\footnotemark[5] [\onlinecite{Klingeler2008}]\\
0.02 & 132, 151 [\onlinecite{Klingeler2008}] & & & 3.6\footnotemark[5] [\onlinecite{Klingeler2008}]\\
0.025 & 150 [\onlinecite{kohama2008}] & & & 6.7 [\onlinecite{nomura2008}]& 3.2 [\onlinecite{kohama2008}]  \\
0.04 & 122, 147 [\onlinecite{Klingeler2008}] & & & 3.9\footnotemark[5] [\onlinecite{Klingeler2008}] \\
0.05 & & 25 [\onlinecite{kamihara2008}] & 1.30\footnotemark[2] [\onlinecite{Ma2008a}] & 9.1 [\onlinecite{kamihara2008}] & 3 [\onlinecite{kohama2008}] & 319 [\onlinecite{kohama2008}]\\
0.05 & & 25 [\onlinecite{nomura2008}] & 1.86\footnotemark[2] [\onlinecite{Larson2009}] & 10.0 [\onlinecite{nomura2008}] & \\
0.05 & & 21 [\onlinecite{Klingeler2008}] & &  3.2\footnotemark[5] [\onlinecite{Klingeler2008}] \\
0.06 & & 20 [\onlinecite{Klingeler2008}] & &  3.2\footnotemark[5] [\onlinecite{Klingeler2008}] \\
0.10  & & 27 [\onlinecite{Klingeler2008}] & 1.28\footnotemark[2] [\onlinecite{mazin2008}] & 3.3\footnotemark[5] [\onlinecite{Klingeler2008}] & 0.7 [\onlinecite{Gang2008}] & 316 [\onlinecite{Gang2008}] \\
0.10  & & & 1.56\footnotemark[2] [\onlinecite{Larson2009}] & $\chi^{\rm spin}$ = 1.8 [\onlinecite{Imai2008}] & \\
0.10 &  & & 2.01\footnotemark[4] [\onlinecite{Wang2009}] \\
0.11  & & 26 [\onlinecite{nomura2008}] &  & 6.7 [\onlinecite{nomura2008}] & 4.1 [\onlinecite{sefat2008b}] & 325 [\onlinecite{sefat2008b}] \\
0.11  & &  &  &  &   3 [\onlinecite{kohama2008}] & 308 [\onlinecite{kohama2008}]  \\
0.125  & & 18 [\onlinecite{Klingeler2008}] &  & 3.2\footnotemark[5] [\onlinecite{Klingeler2008}] & \\
0.14  & & 15 [\onlinecite{nomura2008}] &  & 4.1 [\onlinecite{nomura2008}] & 8 [\onlinecite{kohama2008}] & 332 [\onlinecite{kohama2008}]  \\ 
\end{tabular}
\end{ruledtabular}
\footnotetext[1]{single crystals} 
\footnotetext[2]{Experimental lattice parameters and optimized atomic coordinates were used in the calculations}
\footnotetext[3]{Optimized lattice parameters and atomic coordinates were used in the calculations.}
\footnotetext[4]{the experimental crystal structure was used in the calculations}
\footnotetext[5]{the data were corrected for ferromagnetic impurities}
\end{table*}

\squeezetable
\begin{table*}
\caption{Physical property data for various 1111-type compounds.  The data shown are magnetic and/or structural transition temperature(s) $T_0$, superconducting transition onset temperature $T_{\rm c}$, bare band structure density of states for both spin directions $N(E_{F})$, measured magnetic susceptibility $\chi$(300~K), and Sommerfeld coefficient $\gamma$ and Debye temperature $\Theta_{\rm D}$ from heat capacity data.  The samples are polycrystalline unless otherwise noted.}
\label{data}
\begin{ruledtabular}
\begin{tabular}{lcccccc}
 Compound  & $T_0$ & $T_{\rm c}$ & $N(E_{F})$& $\chi$(300~K) & $\gamma$ & $\Theta_{\rm D}$ \\
 & (K) & (K) & $\left(\frac{\rm states}{\rm eV~f.u.}\right)$ & $\left(10^{-4} \frac{\rm cm^3}{\rm mol}\right)$ & $\left(\frac{\rm mJ}{\rm mol~K^{2}}\right)$ & (K)\\ \hline
NdFeAsO & 2.0, 141 [\onlinecite{Qiu2008,Ychen2008}]& & 1.99 [\onlinecite{Nekrasov2008c}] & \\
NdFeAsO$_{0.89}$F$_{0.11}$ &  & 52 [\onlinecite{ren2008a}] \\
\hline
LaFe$_{0.925}$Co$_{0.075}$AsO && 13\footnotemark[1] [\onlinecite{Wang2008b}] \\
LaFe$_{0.89}$Co$_{0.11}$AsO & & 14\footnotemark[2] [\onlinecite{Sefat2008h}] & & [\onlinecite{Sefat2008h}]\footnotemark[3] \\
LaFe$_{0.875}$Co$_{0.125}$AsO &&& 1.63\footnotemark[4] [\onlinecite{Li2008}] \\
LaFe$_{0.75}$Co$_{0.25}$AsO &&& 0.75\footnotemark[4] [\onlinecite{Li2008}] \\
\hline
SrFeAsF & 175 [\onlinecite{Tegel2008b}] & & 2.14 [\onlinecite{Nekrasov2008a}] &  & 1.5 [\onlinecite{Tegel2008b}] & 339 [\onlinecite{Tegel2008b}] \\ 
SrFeAsF & 173\footnotemark[5] [\onlinecite{Han2008}] & & 1.54\footnotemark[4] [\onlinecite{Shein2008}] & 3.3\footnotemark[5] [\onlinecite{Han2008}] \\
SrFeAsF & 122, 175 [\onlinecite{Baker2009}] & & &  3.3 [\onlinecite{Baker2009}] & 3.4\footnotemark[6] [\onlinecite{Baker2009}] \\
SrFeAsF & 180 [\onlinecite{Matsuishi2008}] & & &  4.4 [\onlinecite{Matsuishi2008}] &  \\
SrFe$_{0.875}$Co$_{0.125}$AsF &  & 5\footnotemark[7] [\onlinecite{Matsuishi2008}]  & & 2.4 [\onlinecite{Matsuishi2008}]  \\
CaFeAsF & 120 [\onlinecite{Matsuishi2008a}] & & 1.83 [\onlinecite{Nekrasov2008a}] \\
CaFeAsF &  & & 1.895\footnotemark[4] [\onlinecite{Shein2008}] \\
CaFe$_{0.95}$Ni$_{0.05}$AsF & & 15\footnotemark[8] [\onlinecite{Matsuishi2008a}] \\
CaFe$_{0.9}$Co$_{0.1}$AsF & & 23\footnotemark[8] [\onlinecite{Matsuishi2008a}] \\
\hline
LaFePO\footnotemark[9] &  & 6.6\footnotemark[9] [\onlinecite{Baumbach2008}] & 3.4 [\onlinecite{Kamihara2008b}] & 3\footnotemark[9]$^,$\footnotemark[10] [\onlinecite{Baumbach2008}] & 12.7\footnotemark[9] [\onlinecite{Baumbach2008}] & 268\footnotemark[9] [\onlinecite{Baumbach2008}] \\ 
LaFePO &  & 3.9 [\onlinecite{Kamihara2008}] & 3.2 [\onlinecite{Lebegue2009}] & 4.5 [\onlinecite{Kamihara2008}]  \\ 
LaFePO &  & $<$0.35 [\onlinecite{McQueen2008}] & & 2.3\footnotemark[11] [\onlinecite{McQueen2008}] & 12.5\footnotemark[11] [\onlinecite{McQueen2008}] \\ 
LaNiAsO &  & 2.4 [\onlinecite{Watanabe2008}] & 1.62 [\onlinecite{Xu2008}] \\
LaNiAsO$_{0.9}$F$_{0.1}$ &  & 3.8 [\onlinecite{Li2008b}] &&& 7.3\footnotemark[12] [\onlinecite{Li2008b}] \\
LaNiPO &  & 4.3 [\onlinecite{Tegel2008}] & 2.2 [\onlinecite{Zhang2008}] \\
\end{tabular}
\end{ruledtabular}
\footnotetext[1]{10\% superconducting diamagnetic shielding fraction}
\footnotetext[2]{90\% superconducting diamagnetic shielding fraction}
\footnotetext[3]{A chaotic dependence of $\chi(T)$ on Co content is reported}
\footnotetext[4]{Optimized lattice parameters and atomic coordinates were used in the calculations.}
\footnotetext[5]{Significant contamination by SrF$_2$ is reported}
\footnotetext[6]{Nonconventional fit to the $C(T)$ data}
\footnotetext[7]{8\% superconducting diamagnetic shielding fraction}
\footnotetext[8]{From resistance measurements only}
\footnotetext[9]{Single crystals} 
\footnotetext[10]{Sample contained 1--2\% ferromagnetic Fe$_2$P}
\footnotetext[11]{Sample contained 0.7\% La$_2$O$_3$ and 2\% FeP}
\footnotetext[12]{Obtained in the normal state under magnetic field $H = 10$~T $>H_{\rm c2}$.}
\end{table*}

%\squeezetable
\begin{table*}
\caption{Physical property data for polycrystalline and single crystalline BaFe$_2$As$_2$-based 122 compounds.  The data shown are magnetic and/or structural transition temperature $T_0$; superconducting transition onset temperature $T_{\rm c}$; bare band structure density of states for both spin directions for the nonmagnetic state $N(E_{F})$; measured magnetic susceptibilities $\chi_{ab}$(300~K), $\chi_c$(300~K), the powder-averaged $\bar{\chi}$(300~K); and low-$T$ Sommerfeld coefficient $\gamma$, inferred normal state $\gamma_{\rm n}$ for superconducting samples and low-$T$ Debye temperature $\Theta_{\rm D}$ from heat capacity measurements.}
\label{BaFe122data}
\begin{ruledtabular}
\begin{tabular}{lccccccccc}
Compound & $T_0$ & $T_{\rm c}$ & $N(E_{F})$& $\chi_{ab}$(300~K) & $\chi_c$(300~K) & $\bar{\chi}$(300~K) & $\gamma$ & $\Theta_{\rm D}$\\
& (K) & (K) & $\left(\frac{\rm states}{\rm eV~f.u.}\right)$ & $\left(10^{-4} \frac{\rm cm^3}{\rm mol}\right)$ & $\left(10^{-4} \frac{\rm cm^3}{\rm mol}\right)$ & $\left(10^{-4} \frac{\rm cm^3}{\rm mol}\right)$ & $\left(\frac{\rm mJ}{\rm mol~K^{2}}\right)$ & (K)\\ \hline
BaFe$_{2}$As$_{2}$ & 140\footnotemark[4] [\onlinecite{rotter2008b}] & & $4.59$\footnotemark[1] [\onlinecite{djsingh2008}] &  10.7\footnotemark[5] [\onlinecite{wang2008}] & 7.2\footnotemark[5] [\onlinecite{wang2008}] & 9.5\footnotemark[5] [\onlinecite{wang2008}] & \\
BaFe$_{2}$As$_{2}$  &&&3.06\footnotemark[3] [\onlinecite{djsingh2008}] \\
BaFe$_{2}$As$_{2}$ & & & 4.69\footnotemark[1] [\onlinecite{singh2008}] \\
BaFe$_{2}$As$_{2}$ & 136\footnotemark[5] [\onlinecite{Dong2008a}]  & & 4.76\footnotemark[1] [\onlinecite{Kasinathan2009}] & & & & 6.1\footnotemark[5] [\onlinecite{Dong2008a}] & 186\footnotemark[5] [\onlinecite{Dong2008a}] \\
BaFe$_{2}$As$_{2}$ & 132\footnotemark[5] [\onlinecite{Sefat2009b}]&  & 4.22\footnotemark[6] [\onlinecite{Nekrasov2008}] & & 6.8\footnotemark[5] [\onlinecite{Sefat2009b}] & 6.5\footnotemark[5] [\onlinecite{Sefat2009b}]& 6.7\footnotemark[5] [\onlinecite{Sefat2009b}]&  & \\
BaFe$_{2}$As$_{2}$ & 137\footnotemark[5] [\onlinecite{Wang2008c}]  & & 4.55\footnotemark[1] [\onlinecite{Shein2008c}] & 10.7\footnotemark[5] [\onlinecite{Wang2008c}] & & &  & \\
BaFe$_{2}$As$_{2}$ &  & & 3.93\footnotemark[3] [\onlinecite{Ma2008}] & 8.9\footnotemark[5] [\onlinecite{Wilson2009}] & & &  & \\
BaFe$_{2}$As$_{2}$ &  & & 4.90\footnotemark[1] [\onlinecite{Wang2009}] &  & & &  & \\
BaFe$_{2}$As$_{2}$ &  & & 3.30\footnotemark[3] [\onlinecite{paulraj2009}] &  &  & & &  & \\
BaFe$_{2}$As$_{2}$ & 138\footnotemark[5] [\onlinecite{Sun2009}] & & & & 4.45\footnotemark[5] [\onlinecite{Sun2009}] & & &  & \\
Ba$_{0.7}$K$_{0.3}$Fe$_{2}$As$_{2}$ & & 35\footnotemark[4] [\onlinecite{Storey2010}] & & & &  & 1.8\footnotemark[4] [\onlinecite{Storey2010}] & 257\footnotemark[4] [\onlinecite{Storey2010}] \\
& & & & & &  & $\gamma_{\rm n} = 47$\footnotemark[4] [\onlinecite{Storey2010}] &  \\
Ba$_{0.68}$K$_{0.32}$Fe$_{2}$As$_{2}$ & & 38.5\footnotemark[5] [\onlinecite{Popovich2010}] & & & &  & 1.2\footnotemark[5] [\onlinecite{Popovich2010}] & 270\footnotemark[5] [\onlinecite{Popovich2010}] \\
& & & & & &  & $\gamma_{\rm n} = 50$\footnotemark[5] [\onlinecite{Popovich2010}] &  \\
Ba$_{0.6}$K$_{0.4}$Fe$_{2}$As$_{2}$ & & 38\footnotemark[4] [\onlinecite{rotter2008a}] &  6.2\footnotemark[1] [\onlinecite{Wang2009}]& & &  &  \\
Ba$_{0.6}$K$_{0.4}$Fe$_{2}$As$_{2}$ & & 35.6\footnotemark[5] [\onlinecite{Mu2008}] &  & & &  & 7.7\footnotemark[5] [\onlinecite{Mu2008}] & 274\footnotemark[5] [\onlinecite{Mu2008}] \\
& & & & & &  & $\gamma_{\rm n} = 63$\footnotemark[5] [\onlinecite{Mu2008}] &  \\
Ba$_{0.5}$K$_{0.5}$Fe$_{2}$As$_{2}$ & & 35.6\footnotemark[4] [\onlinecite{Dong2008a}] & 5.53 [\onlinecite{Shein2008c}] & & & & 9.1\footnotemark[4] [\onlinecite{Dong2008a}] & 246\footnotemark[4] [\onlinecite{Dong2008a}] \\
Ba$_{0.5}$K$_{0.5}$Fe$_{2}$As$_{2}$  &&& 3.90\footnotemark[3] [\onlinecite{djsingh2008}] \\
BaFe$_{1.884}$Co$_{0.116}$As$_{2}$ & 30\footnotemark[5] [\onlinecite{Ni2008g}] & 23\footnotemark[5] [\onlinecite{Ni2008g}]&  & 9.8\footnotemark[5] [\onlinecite{Ni2008g}] &  & & \\
BaFe$_{1.852}$Co$_{0.148}$As$_{2}$ & & 22\footnotemark[5] [\onlinecite{Ni2008g}]&  &  7.1\footnotemark[5] [\onlinecite{Ni2008g}] &  & & \\
BaFe$_{1.84}$Co$_{0.16}$As$_{2}$ & & 21\footnotemark[5] [\onlinecite{Sefat2009b}] &  &  10.6\footnotemark[5] [\onlinecite{Sefat2009b}] & 5.9\footnotemark[5] [\onlinecite{Sefat2009b}] & 9.0\footnotemark[5] [\onlinecite{Sefat2009b}] & \\
BaFe$_{1.8}$Co$_{0.2}$As$_{2}$ & & 22\footnotemark[5] [\onlinecite{Sefat2008g}]&  &  7.0\footnotemark[5] [\onlinecite{Sefat2008g}] & 4.9\footnotemark[5] [\onlinecite{Sefat2008g}] & 6.3\footnotemark[5] [\onlinecite{Sefat2008g}]& \\
BaFe$_{1.8}$Co$_{0.2}$As$_{2}$ & & 23\footnotemark[5] [\onlinecite{Wang2008c}]&  &  7.3\footnotemark[5] [\onlinecite{Wang2008c}] & & \\ 
BaFe$_{1.5}$Ru$_{0.5}$As$_{2}$ & & & 5.78 [\onlinecite{paulraj2009}] & &  & & \\ 
BaFe$_{1.25}$Ru$_{0.75}$As$_{2}$ & & 21\footnotemark[2] [\onlinecite{paulraj2009}] & & &  & & \\
\end{tabular}
\footnotetext[1]{uses experimental structural data}
\footnotetext[2]{31 vol\% superconductivity at 4 K}
\footnotetext[3]{the experimental lattice parameters and optimized internal coordinates were used}
\footnotetext[4]{polycrystalline sample}
\footnotetext[5]{single crystal grown using self-flux}
\footnotetext[6]{the crystal structure parameters of 
Ba$_{0.6}$K$_{0.4}$Fe$_{2}$As$_{2}$ were used in the calculations}

\end{ruledtabular}
\end{table*}

%\squeezetable
\begin{table*}
\caption{Physical property data for polycrystalline and single crystalline FeAs-based 122-type and Fe-based 111-type compounds.  The data shown are magnetic and/or structural transition temperature $T_0$; superconducting transition onset temperature $T_{\rm c}$; bare band structure density of states for both spin directions for the nonmagnetic state $N(E_{F})$; measured magnetic susceptibilities $\chi_{ab}$(300~K), $\chi_c$(300~K), the powder-averaged $\bar{\chi}$(300~K); and low-$T$ Sommerfeld coefficient $\gamma$, low-$T$ Debye temperature $\Theta_{\rm D}$ and inferred normal state Sommerfeld coefficient $\gamma_{\rm n}$ from heat capacity measurements.}
\label{Various122}
\begin{ruledtabular}
\begin{tabular}{lccccccccc}
Compound & $T_0$ & $T_{\rm c}$ & $N(E_{F})$& $\chi_{ab}$(300~K) & $\chi_c$(300~K) & $\bar{\chi}$(300~K) & $\gamma$ & $\Theta_{\rm D}$\\
& (K) & (K) & $\left(\frac{\rm states}{\rm eV~f.u.}\right)$ & $\left(10^{-4} \frac{\rm cm^3}{\rm mol}\right)$ & $\left(10^{-4} \frac{\rm cm^3}{\rm mol}\right)$ & $\left(10^{-4} \frac{\rm cm^3}{\rm mol}\right)$ & $\left(\frac{\rm mJ}{\rm mol~K^{2}}\right)$ & (K)\\ \hline
KFe$_{2}$As$_{2}$ & none & 3.3 [\onlinecite{Fukazawa2009}] & 4.28\footnotemark[4] [\onlinecite{Terashima2010}] &  &  & & $\gamma_{\rm n} = $ 69.1(2) [\onlinecite{Fukazawa2009}] &  \\ 
\hline
SrFe$_{2}$As$_{2}$ & 198\footnotemark[1] [\onlinecite{yan2008}]& & 4.91\footnotemark[4]  [\onlinecite{Kasinathan2009}]  & 13.6\footnotemark[1] [\onlinecite{yan2008}]& 7.1\footnotemark[1] [\onlinecite{yan2008}]& 11.4\footnotemark[1] [\onlinecite{yan2008}]& 3.3\footnotemark[1] [\onlinecite{yan2008}] & 248\footnotemark[1] [\onlinecite{yan2008}]\\
SrFe$_{2}$As$_{2}$ & 200\footnotemark[1] [\onlinecite{chen2008d}] &  & 3.27\footnotemark[5] [\onlinecite{Ma2008}] & 8.6 [\onlinecite{Kitagawa2009}] & 7.4 [\onlinecite{Kitagawa2009}]  & 8.2 [\onlinecite{Kitagawa2009}] & 6.5\footnotemark[1] [\onlinecite{chen2008d}] & 245\footnotemark[1] [\onlinecite{chen2008d}]\\ 
SrFe$_{2}$As$_2$ & 203\footnotemark[2] [\onlinecite{Tegel2008d}] &  & \\
SrFe$_{2}$As$_{2}$ & 202\footnotemark[1] [\onlinecite{Sun2009}] & & & & 4.2\footnotemark[1] [\onlinecite{Sun2009}] & & &  & \\
SrFe$_{1.8}$Co$_{0.2}$As$_{2}$ & & 20 [\onlinecite{Leith-Jasper2008}] & \\ 
SrFe$_{1.7}$Co$_{0.3}$As$_{2}$ & & 13 [\onlinecite{Leith-Jasper2008}] & 4.90\footnotemark[4]  [\onlinecite{Leith-Jasper2008},\onlinecite{Kasinathan2009}] \\ 
SrFe$_{1.5}$Ru$_{0.5}$As$_{2}$ & &  & 4.4 [\onlinecite{Schnelle2009}] \\
SrFe$_{1.5}$Ru$_{0.5}$As$_{2}$ & &  & 3.17 [\onlinecite{LZhang2009}] \\
SrFe$_{1.4}$Ru$_{0.6}$As$_{2}$ & & 19.3 [\onlinecite{Schnelle2009}] & 4.2\footnotemark[7] [\onlinecite{Schnelle2009}]& &  & & 6.2 [\onlinecite{Schnelle2009}] & 232 [\onlinecite{Schnelle2009}]\\ 
SrFe$_{1.3}$Ru$_{0.7}$As$_{2}$ & & 19.3 [\onlinecite{Schnelle2009}] & 4.0\footnotemark[7] [\onlinecite{Schnelle2009}]& &  & & 7.3 [\onlinecite{Schnelle2009}] & 229 [\onlinecite{Schnelle2009}]\\ 
SrFe$_{1.2}$Ru$_{0.8}$As$_{2}$ & & 17.6 [\onlinecite{Schnelle2009}] & 3.9\footnotemark[7] [\onlinecite{Schnelle2009}] & &  & & 6.7 [\onlinecite{Schnelle2009}] & 231 [\onlinecite{Schnelle2009}]\\ 
SrFe$_{1.0}$Ru$_{1.0}$As$_{2}$ & & & 3.5 [\onlinecite{Schnelle2009}] & &  & & 6.2 [\onlinecite{Schnelle2009}] & 232 [\onlinecite{Schnelle2009}]\\ 
\hline
CaFe$_{2}$As$_{2}$ & 165\footnotemark[3] [\onlinecite{Wu2008}] & & 6.17\footnotemark[4]  [\onlinecite{Kasinathan2009}]  & 14.1\footnotemark[3] [\onlinecite{Wu2008}] & 8.0\footnotemark[3] [\onlinecite{Wu2008}] & 12.1\footnotemark[3] [\onlinecite{Wu2008}] &  \\ 
CaFe$_{2}$As$_{2}$ & 169\footnotemark[1] [\onlinecite{Curro2009}] & & 3.95\footnotemark[5] [\onlinecite{Ma2008}]&  &  & &  &  \\ 
CaFe$_{2}$As$_{2}$ & 171\footnotemark[1] [\onlinecite{Ronning2008}] & & &  14.7\footnotemark[1] [\onlinecite{Ronning2008}]  & 12.2\footnotemark[1] [\onlinecite{Ronning2008}] & 13.9\footnotemark[1] [\onlinecite{Ronning2008}] & 8.2\footnotemark[1] [\onlinecite{Ronning2008}] & 292\footnotemark[1] [\onlinecite{Ronning2008}]  \\ 
CaFe$_{2}$As$_{2}$ & 170\footnotemark[1] [\onlinecite{Ni2008b}] & & &  &   & &   4.7\footnotemark[1] [\onlinecite{Ni2008b}] & 258\footnotemark[1] [\onlinecite{Ni2008b}]  \\
CaFe$_{2}$As$_{2}$ & 170\footnotemark[1] [\onlinecite{Kumar2009}] & & &  12.9\footnotemark[1] [\onlinecite{Kumar2009}]  & 8.2\footnotemark[1] [\onlinecite{Kumar2009}] & 11.3\footnotemark[1] [\onlinecite{Kumar2009}] &  &  \\
Ca$_{0.5}$Na$_{0.5}$Fe$_{2}$As$_{2}$ & & 18\footnotemark[2] [\onlinecite{Dong2008a}]  &   & & &   &  25.1\footnotemark[2] [\onlinecite{Dong2008a}] & 217\footnotemark[2] [\onlinecite{Dong2008a}] \\ \hline
EuFe$_{2}$As$_2$ & 19\footnotemark[3] [\onlinecite{Jiang2009}], 190\footnotemark[2] [\onlinecite{Tegel2008d}] \\
Eu$_{0.5}$K$_{0.5}$Fe$_{2}$As$_2$ & & 32\footnotemark[2] [\onlinecite{jeevan2008a}] \\
\hline

NaFeAs &  & 9\footnotemark[6]  [\onlinecite{Parker2008}] & 1.97\footnotemark[4] [\onlinecite{Jishi2008}] & & & 4.9\footnotemark[8] [\onlinecite{Parker2008}] \\ 
LiFeAs &  & 18 [\onlinecite{Tapp2008R, Wang2008S}] & 2.26\footnotemark[4] [\onlinecite{Jishi2008}] & & &  \\ 
LiFeAs &  & 16 [\onlinecite{Pitcher2008C}] & 1.93\footnotemark[4] [\onlinecite{Nekrasov2008b}] & & &  \\
LiFeAs &  & & 1.79\footnotemark[5] [\onlinecite{djsingh2008}] & & & &  \\
LiFeAs &  & & 1.91\footnotemark[5] [\onlinecite{Shein2010}] & & & &  \\
LiFeP &  & & 1.68\footnotemark[5] [\onlinecite{Shein2010}] & & & &  \\
\end{tabular}
\footnotetext[1]{crystals grown using Sn flux}
\footnotetext[2]{polycrystalline sample}
\footnotetext[3]{single crystal grown using self-flux}
\footnotetext[4]{uses experimental structural data}
\footnotetext[5]{the experimental lattice parameters and optimized internal coordinates were used}
\footnotetext[6]{5.6 vol\% superconductivity at 2 K}
\footnotetext[7]{interpolated}
\footnotetext[8]{large ($\sim$~2/3) contribution of ferromagnetic impurity to the magnetization}

\end{ruledtabular}
\end{table*}

%\squeezetable
\begin{table*}
\caption{Physical property data for $AT_2$(As,P)$_2$ compounds with $A = $ Ca, Sr, Ba, La and $T = $ Cr, Co, Ni, Ru, Rh, Ir. The data shown are the magnetic and/or structural transition temperature $T_0$, superconducting transition onset temperature $T_{\rm c}$, bare nonmagnetic band structure density of states for both spin directions $N(E_{F})$,  measured magnetic susceptibilities $\chi_{ab}$(300~K), $\chi_c$(300~K), the powder-averaged $\bar{\chi}$(300~K), and Sommerfeld coefficient $\gamma$ and Debye temperature $\Theta_{\rm D}$ from heat capacity data.}
\label{data3}
\begin{ruledtabular}
\begin{tabular}{lcccccccc}
Compound & $T_0$ & $T_{\rm c}$ & $N(E_{F})$& $\chi_{ab}$(300~K) & $\chi_c$(300~K) & $\bar{\chi}$(300~K) & $\gamma$ & $\Theta_{\rm D}$\\
& (K) & (K) & $\left(\frac{\rm states}{\rm eV~f.u.}\right)$ & $\left(10^{-4} \frac{\rm cm^3}{\rm mol}\right)$ & $\left(10^{-4} \frac{\rm cm^3}{\rm mol}\right)$ & $\left(10^{-4} \frac{\rm cm^3}{\rm mol}\right)$ & $\left(\frac{\rm mJ}{\rm mol~K^{2}}\right)$ & (K)\\ \hline
SrRh$_{2}$As$_{2}$ & &  & 3.399\footnotemark[5] [\onlinecite{Shein2009a}] &  &  &  &  &  \\
BaRh$_{2}$As$_{2}$ & & $< 1.8$\footnotemark[1] [\onlinecite{singh2008}] & 3.49 [\onlinecite{singh2008}] & 0.32\footnotemark[1] [\onlinecite{singh2008}] & 0.40\footnotemark[1] [\onlinecite{singh2008}] & 0.35\footnotemark[1] [\onlinecite{singh2008}] & 4.7\footnotemark[1] [\onlinecite{singh2008}] & 171\footnotemark[2] [\onlinecite{singh2008}] \\
BaRh$_{2}$As$_{2}$ & &  & 3.651\footnotemark[5] [\onlinecite{Shein2009a}] &  &  &  &  &  \\
BaRu$_{2}$As$_{2}$ & & $<1.8$\footnotemark[2] [\onlinecite{Nath2009}] & 1.713\footnotemark[5] [\onlinecite{Shein2009a}] & & & $-0.135$\footnotemark[2] [\onlinecite{Nath2009}] & 4.9\footnotemark[2] [\onlinecite{Nath2009}] & 271\footnotemark[2] [\onlinecite{Nath2009}] \\
SrRu$_{2}$As$_{2}$ & & $<1.8$\footnotemark[2] [\onlinecite{Nath2009}]& 1.9 [\onlinecite{Schnelle2009}] & & &  $-0.205$\footnotemark[2] [\onlinecite{Nath2009}] & 4.1\footnotemark[2] [\onlinecite{Nath2009}] & 271\footnotemark[2] [\onlinecite{Nath2009}] \\ 
SrRu$_{2}$As$_{2}$ & &  & 1.708\footnotemark[5] [\onlinecite{Shein2009a}] &  &  &  &  &  \\
BaCo$_{2}$As$_{2}$ & & $<2$ [\onlinecite{Sefat2009}] & 8.5\footnotemark[5] [\onlinecite{Sefat2009}] & & & & 41\footnotemark[3] [\onlinecite{Sefat2009}] & \\
SrNi$_{2}$As$_{2}$ & & 0.62\footnotemark[1] [\onlinecite{Bauer2008}] & 3.17 [\onlinecite{Shein2009}] & & & & 8.7\footnotemark[1] [\onlinecite{Bauer2008}] & 244\footnotemark[1] [\onlinecite{Bauer2008}] \\
BaNi$_{2}$As$_{2}$ & 130--137\footnotemark[1] [\onlinecite{Ronning2008b}] & 0.7\footnotemark[1] [\onlinecite{Ronning2008b}] & 3.97 [\onlinecite{Shein2009}] & 3.02\footnotemark[3] [\onlinecite{Sefat2009b}] & 1.84\footnotemark[3] [\onlinecite{Sefat2009b}] & 2.63\footnotemark[3] [\onlinecite{Sefat2009b}] & 10.8\footnotemark[1] [\onlinecite{Ronning2008b}] & 206\footnotemark[1] [\onlinecite{Ronning2008b}] \\ 
BaNi$_{2}$As$_{2}$ & & 0.68\footnotemark[1] [\onlinecite{Kurita2009}] &  & & & & 12.3\footnotemark[1] [\onlinecite{Kurita2009}] &  \\
BaCr$_{2}$As$_{2}$ & &  & 3.7\footnotemark[5] [\onlinecite{DJSingh2009}] & & & & 19.3\footnotemark[3] [\onlinecite{DJSingh2009}] &  \\
\hline
LaRu$_{2}$P$_{2}$  & & 4.1 [\onlinecite{jeitschko1987}] &  & &  &  $-0.50$ [\onlinecite{jeitschko1987}] & \\ 
BaRh$_{2}$P$_{2}$  &  & 1.0\footnotemark[1] [\onlinecite{Berry2009}] & & &  & & 9.4\footnotemark[1] [\onlinecite{Berry2009}] \\  
CaRh$_{2}$P$_{2}$  & & 1.0\footnotemark[1] [\onlinecite{Berry2009}] & & & &  & 10.9\footnotemark[1] [\onlinecite{Berry2009}] \\  
BaIr$_{2}$P$_{2}$  & & 2.1\footnotemark[1] [\onlinecite{Berry2009}] & & & &  & 10\footnotemark[1] [\onlinecite{Berry2009}] \\  
SrNi$_{2}$P$_{2}$  & 325\footnotemark[4] [\onlinecite{Ronning2009}] & 1.5\footnotemark[4] [\onlinecite{Ronning2009}] & 3.43\footnotemark[6] [\onlinecite{Shein2009}] & & & $1.55$\footnotemark[8]$^,$\footnotemark[4] [\onlinecite{Ronning2009}] & 15\footnotemark[4] [\onlinecite{Ronning2009}] & 348\footnotemark[4] [\onlinecite{Ronning2009}] \\ 
BaNi$_{2}$P$_{2}$  &  & 3 [\onlinecite{Mine2008}] & 3.73\footnotemark[7] [\onlinecite{Terashima2009}] \\ 
BaNi$_{2}$P$_{2}$  &  &  & 3.97\footnotemark[6] [\onlinecite{Shein2009}] \\ 

\end{tabular}
\footnotetext[1]{crystals grown using Pb flux}
\footnotetext[2]{polycrystalline sample}
\footnotetext[3]{single crystal grown using self-flux}
\footnotetext[4]{crystals grown using Sn flux}
\footnotetext[5]{uses experimental lattice parameters, optimized As positions}
\footnotetext[6]{theoretically optimized crystal structure used in calculations}
\footnotetext[7]{uses experimental structural data}
\footnotetext[8]{$T = 350$~K, crystal, isotropic susceptibility}

\end{ruledtabular}
\end{table*}

%\squeezetable
\begin{table*}
\caption{Physical property data for Fe$_{1+y}$Te$_{1-x}$Se$_x$ compounds. The data shown are the magnetic and/or structural transition temperature $T_0$, superconducting transition onset temperature $T_{\rm c}$, bare band structure density of states for both spin directions $N(E_{F})$ (nonmagnetic unless otherwise noted), measured magnetic susceptibilities $\chi_{ab}$(300~K), $\chi_c$(300~K), the powder-averaged $\bar{\chi}$(300~K), and Sommerfeld coefficient $\gamma$ and Debye temperature $\Theta_{\rm D}$ from heat capacity data.}
\label{FeTeSeData}
\begin{ruledtabular}
\begin{tabular}{lcccccccc}
Compound & $T_0$ & $T_{\rm c}$ & $N(E_{F})$& $\chi_{ab}$(300~K) & $\chi_c$(300~K) & $\bar{\chi}$(300~K) & $\gamma$ & $\Theta_{\rm D}$\\
& (K) & (K) & $\left(\frac{\rm states}{\rm eV~f.u.}\right)$ & $\left(10^{-4} \frac{\rm cm^3}{\rm mol}\right)$ & $\left(10^{-4} \frac{\rm cm^3}{\rm mol}\right)$ & $\left(10^{-4} \frac{\rm cm^3}{\rm mol}\right)$ & $\left(\frac{\rm mJ}{\rm mol~K^{2}}\right)$ & (K)\\ \hline
Fe$_{1.05}$Te & 65 &  &  & &  &  & 34 & 141 [\onlinecite{Chen2009a}] \\
FeTe & &  & 1.83\footnotemark[1] [\onlinecite{Subedi2008a}] & &  &  &  &  \\
FeTe & &  & 2.03\footnotemark[1] [\onlinecite{Ma2009a}] & &  &  &  &  \\
FeTe & &  & 1.98\footnotemark[1]$^,$\footnotemark[2] [\onlinecite{Ma2009a}] & &  &  &  &  \\
Fe$_{1.04}$Te\footnotemark[3] & 72 &  &  & &  &  & 33  [\onlinecite{Liu2009a}] \\
Fe$_{1.11}$Te\footnotemark[3] & 65 &  &  & &  &  & 27  [\onlinecite{Liu2009a}] \\
Fe$_{1.12}$Te\footnotemark[3] & 69 &  &  & 27\footnotemark[5] [\onlinecite{Yang2009a}] & 27\footnotemark[5] [\onlinecite{Yang2009a}] & 27\footnotemark[5] [\onlinecite{Yang2009a}]&  &   \\
\hline
Fe$_{1.00}$Te$_{0.95}$Se$_{0.05}$\footnotemark[3] &  & non-bulk (0.8\%\footnotemark[6]) &  &  & 28\footnotemark[5] [\onlinecite{Yang2009a}] & &  &   \\
Fe$_{1.01}$Te$_{0.88}$Se$_{0.12}$\footnotemark[3] &  & non-bulk (7.5\%\footnotemark[6]) &  &  & 20\footnotemark[5] [\onlinecite{Yang2009a}] & &  &   \\
Fe$_{1.07}$Te$_{0.80}$Se$_{0.20}$\footnotemark[3] &  & 9\footnotemark[6] &  &  & 14\footnotemark[5] [\onlinecite{Yang2009a}] & &  &   \\
Fe$_{1.12}$Te$_{0.70}$Se$_{0.30}$\footnotemark[3] &  & non-bulk (8.9\%\footnotemark[6]) &  &  & 14\footnotemark[5] [\onlinecite{Yang2009a}] & &  &   \\
Fe$_{1.04}$Te$_{0.67}$Se$_{0.33}$\footnotemark[3] &  & 9\footnotemark[6], 14.5\footnotemark[7] &  &  & 11\footnotemark[5] [\onlinecite{Yang2009a}] & &  &   \\
Fe$_{1+y}$Te$_{0.55}$Se$_{0.45}$\footnotemark[3] &  &  &  &  & 9.5\footnotemark[5] [\onlinecite{Yang2009a}] & &  &   \\
Fe$_{1+y}$Te$_{0.52}$Se$_{0.48}$\footnotemark[3] &  &  &  &  & 10.5\footnotemark[5] [\onlinecite{Yang2009a}] & &  &   \\
Fe$_{1.04}$Te$_{0.64}$Se$_{0.36}$\footnotemark[3] & none &  &  & &  &  & 39 & 174 [\onlinecite{Sales2009a}] \\
\hline
Fe$_{1.01}$Se-300~$^\circ$C\footnotemark[4] & 90 [\onlinecite{McQueen2009b,McQueen2009c}] & 9 &   & &  &  & 5.4(3) & 204(1) [\onlinecite{McQueen2009b}] \\
Fe$_{1.03}$Se\footnotemark[4] & none [\onlinecite{McQueen2009b, McQueen2009c}] & none [\onlinecite{McQueen2009b, McQueen2009c}] &   & &  &  & 1.3(6) & 200(1) [\onlinecite{McQueen2009b}] \\
FeSe & &  & 0.95\footnotemark[1] [\onlinecite{Subedi2008a}] & &  &  &  &  \\
FeSe & &  & 1.29\footnotemark[1] [\onlinecite{Ma2009a}] & &  &  &  &  \\
FeSe & &  & 0.48\footnotemark[1]$^,$\footnotemark[2] [\onlinecite{Ma2009a}] & &  &  &  &  \\
Fe$_{1+y}$Se\footnotemark[3] &  &  &  &  & 8\footnotemark[5] [\onlinecite{Yang2009a}] & &  &   \\
\end{tabular}
\footnotetext[1]{calculated using experimental lattice parameters, but optimized As positions}
\footnotetext[2]{calculated for the diagonal double stripe antiferromagnetic structure}
\footnotetext[3]{single crystal sample}
\footnotetext[4]{polycrystalline sample}
\footnotetext[5]{corrected for ferromagnetic impurities}
\footnotetext[6]{measured from zero-field-cooled susceptibility (onset)}
\footnotetext[7]{measured from resistivity (onset)}
\end{ruledtabular}
\end{table*}


\clearpage

\begin{thebibliography}{999}

\bibitem{kamihara2008} Y. Kamihara, T. Watanabe, M. Hirano, and
H. Hosono, J. Am. Chem. Soc. \textbf{130}, 3296 (2008).
\bibitem{Johnson1974} V. Johnson and W. Jeitschko, J. Solid State Chem. {\bf 11}, 161 (1974).
\bibitem{Kamihara2006} Y. Kamihara, H. Hiramatsu, M. Hirano, R. Kawamura, H. Yanagi, T. Kamiya, and H. Hosono, J. Am. Chem. Soc. {\bf 128}, 10012 (2006).
\bibitem{Takahashi2008} H. Takahashi, K. Igawa, K. Arii, Y. Kamihara, M. Hirano, and H. Hosono, Nature {\bf 453}, 376 (2008).
\bibitem{ren2008a} Z.-A. Ren, J. Yang, W. Lu, W. Yi, X.-L. Shen,
Z.-C. Li, G.-C. Che, X.-L. Dong, L.-L. Sun, F. Zhou, and Z.-X.
Zhao, Europhys. Lett. \textbf{82}, 57002 (2008).
\bibitem{Kito2008} H. Kito, H. Eisaki, and A. Iyo, J. Phys. Soc. Jpn. {\bf 77}, 063707 (2008).
\bibitem{ren2008} Z.-A. Ren, L. Wei, Y. Jie, Y. Wei, S. X. Li,
Z. Cai, C. G. Can, D. X. Li, S. L. Ling, Z. Fang, and Z. Z. Xian,
Chin. Phys. Lett. \textbf{25}, 2215 (2008).
\bibitem{Wang2008} C. Wang, L. Li, S. Chi, Z. Zhu, Z. Ren, Y. Li, Y. Wang, X. Lin, Y. Luo, S. Jiang, X. Xu, G. Cao, and Z. Xu, Europhys. Lett. {\bf 83}, 67006 (2008).
\bibitem{Takeshita2008} N. Takeshita, A. Iyo, H. Eisaki, H. Kito, and T. Ito, J. Phys. Soc. Jpn. {\bf 77}, 075003 (2008).
\bibitem{Nath2009} R. Nath, Y. Singh, and D. C. Johnston, Phys. Rev. B {\bf 79}, 174513 (2009).
\bibitem{rotter2008a} M. Rotter, M. Tegel, and D. Johrendt,
Phys. Rev. Lett. \textbf{101}, 107006 (2008).
\bibitem{Tapp2008R} J. H. Tapp, Z. Tang, B. Lv, K. Sasmal, B. Lorenz, P. C. W. Chu, and A. M. Guloy, Phys. Rev. B {\bf 78}, 060505(R) (2008).
\bibitem{Wang2008S} X. C. Wang, Q. Q. Liu, Y. X. Lv, W. B. Gao, L. X. Yang, R. C. Yu, F. Y . Li, and C. Q. Jin, Solid State Commun. {\bf 148}, 538 (2008).
\bibitem{Pitcher2008C} M. J. Pitcher, D. R. Parker, P. Adamson, S. J. C. Herkelrath, A. T. Boothroyd, R. M. Ibberson, M. Brunelli, and S. J. Clarke, Chem. Commun. {\bf 2008}, 5918 (2008).
\bibitem{Ogino2009}  H. Ogino, Y. Matsumura, Y. Katsura, K. Ushiyama, S. Horii, K. Kishio, and J.-I. Shimoyama, Supercond. Sci. Technol. {\bf 22}, 075008 (2009).
\bibitem{Hsu2008} F.-C. Hsu, J.-Y. Luo, K.-W. Yeh, T.-K. Chen, T.-W. Huang, P. M. Wu, Y.-C. Lee, Y.-L. Huang, Y.-Y. Chu, D.-C. Yan, and M.-K. Wu, Proc. Nat. Acad. Sci. USA {\bf 105}, 14\,262 (2008).
\bibitem{Yan2009} J.-Q. Yan, S. Nandi, J. L. Zarestky, W. Tian, A. Kreyssig, B. Jensen, A. Kracher, K. W. Dennis, R. J. McQueeney, A. I. Goldman, R. W McCallum, and T. A. Lograsso, Appl. Phys. Lett. {\bf 95}, 222504 (2009).
\bibitem{Lynn2009} J. W. Lynn and P. Dai, Physica C {\bf 469}, 469 (2009).
\bibitem{sasmal2008} K. Sasmal, B. Lv, B. Lorenz, A. M. Guloy,
F. Chen, Y.-Y. Xue, and C. W. Chu, Phys. Rev. Lett. \textbf{101},
107007 (2008).
\bibitem{chen2008d} G. F. Chen, Z. Li, J. Dong, G. Li, W. Z. Hu, X. D. Zhang, X. H. Song, P. Zheng, N. L. Wang, and J. L. Luo, Phys. Rev. B \textbf{78}, 224512 (2008).
\bibitem{Sales2009a} B. C. Sales, A. S. Sefat, M. A. McGuire, R. Y. Jin, D. Mandrus, and Y. Mozharivskyj, Phys. Rev. B {\bf 79}, 094521 (2009).
\bibitem{Tegel2010} M. Tegel, C. L\"ohnert, and D. Johrendt, Solid State Commun. {\bf 150}, 383 (2010).
\bibitem{Louca2010} D. Louca, K. Horigane, A. Llobet, R. Arita, S. Ji, N. Katayama, S. Konbu, K. Nakamura, T.-Y. Koo, P. Tong, and K. Yamada, Phys. Rev. B {\bf 81}, 134524 (2010).
\bibitem{wang2008} X. F. Wang, T. Wu, G. Wu, H. Chen, Y. L. Xie, J. J. Ying, Y. J. Yan, R. H. Liu, and X. H. Chen, Phys. Rev. Lett. {\bf 102}, 117005 (2009).
\bibitem{Blatt1968} F. J. Blatt, \emph{Physics of Electronic Conduction in Solids} (McGraw-Hill, New York, 1968).
\bibitem{Tanatar2009} M. A. Tanatar, N. Ni, G. D. Samolyuk, S. L. Bud'ko, P. C. Canfield, and R. Prozorov, Phys. Rev. B {\bf 79}, 134528 (2009); M. A. Tanatar, N. Ni, A. Thaler, S. L. Bud'ko, P. C. Canfield, and R. Prozorov, arXiv:1006.2087.
\bibitem{Tanatar2009c} M. A. Tanatar, N. Ni, C. Martin, R. T. Gordon, H. Kim, V. G. Kogan, G. D. Samolyuk, S. L. Bud'ko, P. C. Canfield, and R. Prozorov, Phys. Rev. B {\bf 79}, 094507 (2009).
\bibitem{Moon2010a} S. J. Moon, J. H. Shin, D. Parker, W. S. Choi, I. I. Mazin, Y. S. Lee, J. Y. Kim, N. H. Sung, B. K. Cho, S. H. Khim, J. S. Kim, K. H. Kim, and T. W. Noh, Phys. Rev. B {\bf 81}, 205114 (2010).
\bibitem{Song2010} Y. J. Song, J. S. Ghim, B. H. Min, Y. S. Kwon, M. H. Jung, and J.-S. Rhyee, Appl. Phys. Lett. {\bf 96}, 212508 (2010).
\bibitem{Kashiwaya2010} H. Kashiwaya, K. Shirai, T. Matsumoto, H. Shibata, H. Kambara, M. Ishikado, H. Eisaki, A. Iyo, S. Shamoto, I. Kurosawa, and S. Kashiwaya, Appl. Phys. Lett. {\bf 96}, 202504 (2010).
\bibitem{Ma2008a} F. Ma and Z.-Y. Lu, Phys. Rev. B {\bf 78}, 033111 (2008); Z.-Y. Lu, private communication.
\bibitem{Alireza2009} P. L. Alireza, Y. T. C. Ko, J. Gillett, C. M. Petrone, J. M. Cole, G. G. Lonzarich, and S. E. Sebastian, J. Phys.: Condens. Matter {\bf 21}, 012208 (2009).
\bibitem{Colombier2009} E. Colombier, S. L. Bud'ko, N. Ni, and P. C. Canfield, Phys. Rev. B {\bf 79}, 224518 (2009).
\bibitem{Kotegawa2009} H. Kotegawa, T. Kawazoe, H. Sugawara, K. Murata, and H. Tou, J. Phys. Soc. Jpn. {\bf 78}, 083702 (2009).
\bibitem{Matsubayashi2009} K. Matsubayashi, N. Katayama, K. Ohgushi, A. Yamada, K. Munakata, T. Matsumoto, and Y. Uwatoko, J. Phys. Soc. Jpn. {\bf 78}, 073706 (2009).
\bibitem{Ishikawa2009} F. Ishikawa, N. Eguchi, M. Kodama, K. Fujimaki, M. Einaga, A. Ohmura, A. Nakayama, A. Mitsuda, and Y. Yamada, Phys. Rev. B {\bf 79}, 172506 (2009).
\bibitem{Yamazaki2010} T. Yamazaki, N. Takeshita, R. Kobayashi, H. Fukazawa, Y. Kohori, K. Kihou, C.-H. Lee, H. Kito, A. Iyo, and H. Eisaki, Phys. Rev. B {\bf 81}, 224511 (2010).
\bibitem{Margadonna2009a} S. Margadonna, Y. Takabayashi, Y. Ohishi, Y. Mizuguchi, Y. Takano, T. Kagayama, T. Nakagawa, M. Takata, and K. Prassides, Phys. Rev. B {\bf 80}, 064506 (2009).
\bibitem{Medvedev2009} S. Medvedev, T. M. McQueen, I. A. Troyan, T. Palasyuk, M. I. Eremets, R. J. Cava, S. Naghavi, F. Casper, V. Ksenofontov, G. Wortmann, and C. Felser, Nature Mater. {\bf 8}, 630 (2009).
\bibitem{Zhao2008} J. Zhao, Q. Huang, C. de la Cruz, S. Li, J. W. Lynn, Y. Chen, M. A. Green, G. F. Chen, G. Li, Z. Li, J. L. Luo, N. L. Wang, and P. Dai, Nature Mater. {\bf 7}, 953 (2008).
\bibitem{HChen2009} H. Chen, Y. Ren, Y. Qiu, W. Bao, R. H. Liu, G. Wu, T. Wu, Y. L. Xie, X. F. Wang, Q. Huang, and X. H. Chen, Europhys. Lett. {\bf 85}, 17006 (2009).
\bibitem{Ni2008g} N. Ni, M. E. Tillman, J.-Q. Yan, A. Kracher, S. T. Hannahs, S. L. Bud'ko, and P. C. Canfield, Phys. Rev. B {\bf 78}, 214515 (2008).
\bibitem{Chu2009} J.-H. Chu, J. G. Analytis, C. Kucharczyk, and I. R. Fisher, Phys. Rev. B {\bf 79}, 014506 (2009).
\bibitem{Wang2008c} X. F. Wang, T. Wu, G. Wu, R. H. Liu, H. Chen, Y. L. Xie, and X. H. Chen, New J. Phys. {\bf 11}, 045003 (2009).
\bibitem{Lester2009} C. Lester, J.-H. Chu, J. G. Analytis, S. C. Capelli, A. S. Erickson, C. L. Condron, M. F. Toney, I. R. Fisher, and S. M. Hayden, Phys. Rev. B {\bf 79}, 144523 (2009).
\bibitem{Nandi2010} S. Nandi, M. G. Kim, A. Kreyssig, R. M. Fernandes, D. K. Pratt, A. Thaler, N. Ni, S. L. Bud'ko, P. C. Canfield, J. Schmalian, R. J. McQueeney, and A. I. Goldman, Phys. Rev. Lett. {\bf 104}, 057006 (2010).
\bibitem{Chauviere2010} L. Chauvi\`ere, Y. Gallais, M. Cazayous, M. A. M\'easson, A. Sacuto, D. Colson, and A. Forget, arXiv:1007.0720.
\bibitem{Hardy2010b} F. Hardy, P. Burger, T. Wolf, R. A. Fisher, P. Schweiss, P. Adelmann, R. Heid, R. Fromknecht, R. Eder, D. Ernst, H. v. L\"ohneysen, and C. Meingast, arXiv:1007.2218.
\bibitem{Huang2008} Q. Huang, J. Zhao, J. W. Lynn, G. F. Chen, J. L. Luo, N. L. Wang, and P. Dai, Phys. Rev. B {\bf 78}, 054529 (2008).
\bibitem{Luetgens2009} H. Luetkens, H.-H. Klauss, M. Kraken, F. J. Litterst, T. Dellmann, R. Klingeler, C. Hess, R. Khasanov, A. Amoto, C. Baines, M. Kosmala, O. J. Schumann, M. Braden, J. Hamann-Borrero, N. Leps, A. Kondrat, G. Behr, J. Werner, and B. B\"uchner, Nature Mater. {\bf 8}, 305 (2009).
\bibitem{Kamihara2010} Y. Kamihara, T. Nomura, M. Hirano, J. E. Kim, K. Kato, M. Takata, Y. Kobayashi, S. Kitao, S. Higashitaniguchi, Y. Yoda, M. Seto, and H. Hosono, New J. Phys. {\bf 12}, 033005 (2010).
\bibitem{Margadonna2009} S. Margadonna, Y. Takabayashi, M. T. McDonald, M. Brunelli, G. Wu, R. H. Liu, X. H. Chen, and K. Prassides, Phys. Rev. B {\bf 79}, 014503 (2009).
\bibitem{Drew2009} A. J. Drew, Ch. Niedermayer, P. J. Baker, F. L. Pratt, S. J. Blundell, T. Lancaster, R. H. Liu, G. Wu, X. H. Chen, I. Watanabe, V. K. Malik, A. Dubroka, M. R\"ossle, K. W. Kim, C. Baines, and C. Bernhard, Nature Mater. {\bf 8}, 310 (2009).
\bibitem{Ni2009} N. Ni, A. Thaler, A. Kracher, J. Q. Yan, S. L. Bud'ko, and P. C. Canfield, Phys. Rev. B {\bf 80}, 024511 (2009).
\bibitem{Saha2009} S. R. Saha, N. P. Butch, K. Kirshenbaum, and J. Paglione, Phys. Rev. B {\bf 79}, 224519 (2009).
\bibitem{Han2009} F. Han, X. Zhu, P. Cheng, G. Mu, Y. Jia, L. Fang, Y. Wang, H. Luo, B. Zeng, B. Shen, L. Shan, C. Ren, and H.-H. Wen, Phys. Rev. B {\bf 80}, 024506 (2009).
\bibitem{Kasinathan2009} D. Kasinathan, A. Ormeci, K. Koch, U. Burkhardt, W. Schnelle, A. Leithe-Jasper, and H. Rosner, New J. Phys. {\bf 11}, 025023 (2009); D. Kasinathan, private communication.
\bibitem{Sefat2009c} A. S. Sefat, D. J. Singh, L. H. VanBebber, Y. Mozharivskyj, M. A. McGuire, R. Jin, B. C. Sales, V. Keppens, and D. Mandrus, Phys. Rev. B {\bf 79}, 224524 (2009).
\bibitem{Cheng2010} P. Cheng, B. Shen, J. Hu, and H.-H. Wen, Phys. Rev. B {\bf 81}, 174529 (2010).
\bibitem{Jiang2009a} S. Jiang, H. Xing, G. Xuan, C. Wang, Z. Ren, C. Feng, J. Dai, Z. Xu, and G. Cao, J. Phys.: Condens. Matter {\bf 21}, 382203 (2009).
\bibitem{Schnelle2009} W. Schnelle, A. Leith-Jasper, R. Gumeniuk, U. Burkhardt, D. Kasinathan, and H. Rosner, Phys. Rev. B {\bf 79}, 214516 (2009).
\bibitem{paulraj2009} S. Sharma, A. Bharathi, S. Chandra, V. R. Reddy, S. Paulraj, A. T. Satya, V. S. Sastry, A. Gupta, and C. S. Sundar, Phys. Rev. B {\bf 81}, 174512 (2010).
\bibitem{Rullier-Albenque2010} F. Rullier-Albenque, D. Colson, A. Forget, P. Thu\'ery, and S. Poissonnet, Phys. Rev. B {\bf 81}, 224503 (2010). 
\bibitem{Thaler2010} A. Thaler, N. Ni, A. Kracher, J. Q. Yan, S. L. Bud'ko, and P. C. Canfield, Phys. Rev. B {\bf 82}, 014534 (2010).
\bibitem{Luo2010} Y. Luo, Y. Li, S. Jiang, J. Dai, G. Cao, and Z.-A. Xu, Phys. Rev. B {\bf 81}, 134422 (2010).
\bibitem{Cruz2010} C. de la Cruz, W. Z. Hu, S. Li, Q. Huang, J. W. Lynn, M. A. Green, G. F. Chen, N. L. Wang, H. A. Mook, Q. Si, and P. Dai, Phys. Rev. Lett. {\bf 104}, 017204 (2010).
\bibitem{Kohler2009} A. K\"ohler and G. Behr, J. Supercond. Nov. Magn. {\bf 22}, 565 (2009).
\bibitem{Katayama2010} N. Katayama, S. Ji, D. Louca, S.-H. Lee, M. Fujita, T. J. Sato, J. S. Wen, Z. J. Xu, G. D. Gu, G. Xu, Z. W. Lin, M. Enoki, S. Chang, K. Yamada, and J. M. Tranquada, arXiv:1003.4525 (unpublished).
\bibitem{Liu2010} T. J. Liu, J. Hu, B. Qian, D. Fobes, Z. Q. Mao, W. Bao, M. Reehuis, S. A. J. Kimber, K. Prokes, S. Matas, D. N. Argyriou, A. Hiess, a. Rotaru, H. Pham, L. Spinu, Y. Qiu, V. Thampy, A. T. Savici, J. A. Rodriguez, and C. Broholm, Nature Mater. (in press); arXiv:1003.5647.
\bibitem{Bao2009} W. Bao, Y. Qiu, Q. Huang, M. A. Green, P. Zajdel, M. R. Fitzsimmons, M. Zhernenkov, S. Chang, M. Fang, B. Qian, E. K. Vehstedt, J. Yang, H. M. Pham, L. Spinu, and Z. Q. Mao, Phys. Rev. Lett. {\bf 102}, 247001 (2009).
\bibitem{Wen2009} J. Wen, G. Xu, Z. Xu, Z. W. Lin, Q. Li, W. Ratcliff, G. Gu, and J. M. Tranquada, Phys. Rev. B {\bf 80}, 104506 (2009).
\bibitem{Khasanov2009} R. Khasanov, M. Bendele, A. Amato, P. Babkevich, A. T. Boothroyd, A. Cervellino, K. Conder, S. N. Gvasaliya, H. Keller, H.-H. Klauss, H. Luetkens, V. Pomjakushin, E. Pomjakushina, and B. Roessli, Phys. Rev. B {\bf 80}, 140511(R) (2009).
\bibitem{Babkevich2010} P. Babkevich, M. Bendele, A. T. Boothroyd, K. Conder, S. N. Gvasaliya, R. Khasanov, E. Pomjakushina, and B. Roessli, J. Phys.: Condens. Matter {\bf 22}, 142202 (2010).
\bibitem{Xu2010} Z. Xu, J. Wen, G. Xu, Q. Jie, Z. Lin, Q. Li, S. Chi, D. K. Singh, G. Gu, and J. M. Tranquada, arXiv:1005.4856.
\bibitem{Johnston1997} For a review, see D. C. Johnston, in \emph{Handbook of Magnetic Materials}, Vol. 10, edited by K. H. J. Buschow, Ch. 1 (Elsevier, Amsterdam, 1997), pp.\ 1--237.
\bibitem{arXiv} URL: $<$\url{http://arxiv.org/archive/cond-mat}$>$.
\bibitem{Si2008} Q. Si and E. Abrahams, Phys. Rev. Lett. {\bf 101}, 076401 (2008).
\bibitem{Yildirim2009} T. Yildirim, Phys. Rev. Lett. {\bf 101}, 057010 (2008); Physica C {\bf 469}, 425 (2009).
\bibitem{Norman2008} M. Norman, Physics {\bf 1}, 21 (2008).
\bibitem{Sadovskii2008} M. V. Sadovskii, Physics-Uspekhi {\bf 51}, 1201 (2008).
\bibitem{Ivanovskii2008} A. L. Ivanovskii, Physics-Uspekhi {\bf 51}, 1229 (2008).
\bibitem{Izyumov2008} Yu. A. Izyumov and E. Z. Kurmaev, Physics-Uspekhi {\bf 51}, 1261 (2008).
\bibitem{Ishida2009} K. Ishida, Y. Nakai, and H. Hosono, J. Phys. Soc. Jpn. {\bf 78}, 062001 (2009).
\bibitem{Paglione2010}  J. Paglione and R. L. Greene, arXiv:1006.4618.
\bibitem{Mandrus2010} D. Mandrus, A. S. Sefat, M. A. McGuire, and B. C. Sales, Chem. Mater. {\bf 2010}, 715 (2010).
\bibitem{Mizuguchi2010a} Y. Mizuguchi and Y. Takano, arXiv:1003.2696.
\bibitem{PhysicaC} For topical invited reviews on Fe-based superconductors, see the special issue Physica~C {\bf 468} (9--12), (May-June, 2009).
\bibitem{Clarke2008} S. J. Clarke, P. Adamson, S. J. C. Herkelrath, O. J. Rutt, D. R. Parker, M. J. Pitcher, and C. F. Smura, Inorg. Chem. {\bf 47}, 8473 (2008).
\bibitem{Ozawa2008} T. C. Ozawa and S. M. Kauzlarich, Sci. Technol. Adv. Mater. {\bf 9}, 033003 (2008).
\bibitem{Pottgen2008} R. P\"ottgen and D. Johrendt, Z. Naturforsch. {\bf 63b}, 1135 (2008).
\bibitem{Volkova2008} L. M. Volkova, Supercond. Sci. Technol. {\bf 21}, 095019 (2008).
\bibitem{Just1996} G. Just and P. Paufler, J. Alloys Compd. {\bf 232}, 1 (1996).
\bibitem{RomeConf08} Proc. Internat. Conf. FeAs High $T_{\rm c}$ Superconducting Multilayers and Related Phenomena (Superstripes 2008), Rome, Italy, December 9-13, 2008. Guest Editors: Antonio Bianconi, Nicola Poccia, and Alessandro Ricci, J. Supercond. Nov. Magn. {\bf 22} (6), (Aug. 2009).
\bibitem{NewJPhysFocusIssue} Focus issue on Fe-based superconductors, New J. Phys. {\bf 11}, (Feb. 2009).
\bibitem{Lumsden2010} M. D. Lumsden and A. D. Christianson, J. Phys.: Condens. Matter {\bf 22}, 203203 (2010).
\bibitem{Nagata1999} S. Nagata and T. Atake, J. Therm. Anal. Cal. {\bf 57}, 807 (1999).
\bibitem{Canfield2010} P. C. Canfield and S. L. Bud'ko, Annu. Rev. Condens. Matter Phys. {\bf 1}, 27 (2010).
\bibitem{Wang2009a}  Z. Wang, H. Yang, C. Ma, H. Tian, H. Shi, J. Lu, L. Zeng, and J. Li, J. Phys.: Condens. Matter {\bf 21}, 495701 (2009).
\bibitem{Rotter2010}  M. Rotter, C. Hieke, and D. Johrendt, Phys. Rev. B {\bf 82}, 014513 (2010).
\bibitem{Goldman2008} A. I. Goldman, D. N. Argyriou, B. Ouladdiaf, T. Chatterji, A. Kreyssig, S. Nandi, N. Ni, S. L. Bud'ko, P. C. Canfield, and R. J. McQueeney, Phys. Rev. B {\bf 78}, 100506(R) (2008).
\bibitem{Ni2008b} N. Ni, S. Nandi, A. Kreyssig, A. I. Goldman, E. D. Mun, S. L. Bud'ko, and P. C. Canfield, Phys. Rev. B {\bf 78}, 014523 (2008).
\bibitem{Loudon2010} J. C. Loudon, C. J. Bowell, J. Gillett, S. E. Sebastian, and P. A. Midgley, Phys. Rev. B {\bf 81}, 214111 (2010).
\bibitem{Huang2008b} Q. Huang, Y. Qiu, W. Bao, M. A. Green, J. W. Lynn, Y. C. Gasparovic, T. Wu, G. Wu, and X. H. Chen, Phys. Rev. Lett. {\bf 101}, 257003 (2008).
\bibitem{Wilson2009} S. D. Wilson, Z. Yamani, C. R. Rotundu, B. Freelon, E. Bourret-Courchesne, and R. J. Birgeneau, Phys. Rev. B {\bf 79}, 184519 (2009).
\bibitem{Tanatar2009b} M. A. Tanatar, A. Kreyssig, S. Nandi, N. Ni, S. L. Bud'ko, P. C. Canfield, A. I. Goldman, and R. Prozorov, Phys. Rev. B {\bf 79}, 180508(R) (2009).
\bibitem{Chu2009b} J.-H. Chu, J. G. Analytis, D. Press, K. De Greve, T. D. Ladd, Y. Yamamoto, and I. R. Fisher, Phys. Rev. B {\bf 81}, 214502 (2010).
\bibitem{Ma2009} C. Ma, H. X. Yang, H. F. Tian, H. L. Shi, J. B. Lu, Z. W. Wang, L. J. Zeng, G. F. Chen, N. L. Wang, and J. Q. Li, Phys. Rev. B {\bf 79}, 060506(R) (2009).
\bibitem{Lee2008} C.-H. Lee, A. Iyo, H. Eisaki, H. Kito, M. T. Fernandez-Diaz, T. Ito, K. Kihou, H. Matsuhata, M. Braden, and K. Yamada, J. Phys. Soc. Jpn. {\bf 77}, 083704 (2008).
\bibitem{Lee2008b} C.-H. Lee, A. Iyo, H. Eisaki, H. Kito, M. T. Fernandez-Diaz, R. Kumai, K. Miyazawa, K. Kihou, H. Matsuhata, M. Braden, and K. Yamada, J. Phys. Soc. Jpn. {\bf 77}, Suppl. C, 44 (2008).
\bibitem{Kimber2009} S. A. J. Kimber, A. Kreyssig, Y.-Z. Zhang, H. O. Jeschki, R. Valenti, F. Yokaichiya, E. Colombier, J. Yan, T. C. Hansen, T. Chatterji, R. J. McQueeney, P. C. Canfield, A. I. Goldman, and D. N. Argyriou, Nature Mater. {\bf 8}, 471 (2009).
\bibitem{Mukuda2008} H. Mukuda, N. Terasaki, H. Kinouchi, M. Yashima, Y. Kitaoka, S. Suzuki, S. Miyasaka, S. Tajima, K. Miyazawa, P. Shirage, H. Kito, H. Eisaki, and A. Iyo, J. Phys. Soc. Jpn. {\bf 77}, 093704 (2008).
\bibitem{Mukuda2009} H. Mukuda, N. Terasaki, M. Yashima, H. Nishimura, Y. Kitaoka, and A. Iyo, Physica C {\bf 469}, 559 (2009).
\bibitem{Horigane2009} K. Horigane, H. Hiraka, and K. Ohoyama, J. Phys. Soc. Jpn. {\bf 78}, 074718 (2009).
\bibitem{Mizuguchi2010} Y. Mizuguchi, Y. Hara, K. Deguchi, S. Tsuda, T. Yamaguchi, K. Takeda, H. Kotegawa, H. Tou, and Y. Takano, Supercond. Sci. Technol. {\bf 23}, 054013 (2010).
\bibitem{Okabe2010} H. Okabe, N. Takeshita, K. Horigane, T. Muranaka, and J. Akimitsu, Phys. Rev. B {\bf 81}, 205119 (2010).
\bibitem{Kuroki2009} K. Kuroki, H. Usui, S. Onari, R. Arita, and H. Aoki, Phys. Rev. B {\bf 79}, 224511 (2009).
\bibitem{Huang2010} S. X. Huang, C. L. Chien, V. Thampy, and C. Broholm, Phys. Rev. Lett. {\bf 104}, 217002 (2010).
\bibitem{Kreyssig2010} A. Kreyssig, M. G. Kim, S. Nandi, D. K. Pratt, W. Tian, J. L. Zarestky, N. Ni, A. Thaler, S. L. Bud'ko, P. C Canfield, R. J. McQueeney, and A. I. Goldman, Phys. Rev. B {\bf 81}, 134512 (2010).
\bibitem{Gronvold1954} F. Gronvold, H. Haraldsen, and J. Vihovde, Acta Chem. Scand. {\bf 8}, 1927 (1954).
\bibitem{Chen2009a} G. F. Chen, Z. G. Chen, J. Dong, W. Z. Hu, G. Li, X. D. Zhang, P. Zheng, J. L. Luo, and N. L. Wang, Phys. Rev. B {\bf 79}, 140509(R) (2009).
\bibitem{Liu2009a} T. J. Liu, X. Ke, B. Qian, J. Hu, D. Fobes, E. K. Vehstedt, H. Pham, J. H. Yang, M. H. Fang, L. Spinu, P. Schiffer, Y. Liu, and Z. Q. Mao, Phys. Rev. B {\bf 80}, 174509 (2009).
\bibitem{Johnston1972} D. C. Johnston and A. R. Moodenbaugh, Phys. Lett. {\bf 41A}, 447 (1972).
\bibitem{Moodenbaugh1978} A. R. Moodenbaugh, D. C. Johnston, R. Viswanathan, R. N. Shelton, L. E. DeLong, and W. A. Fertig, J. Low Temp. Phys. {\bf 33}, 175 (1978).
\bibitem{McQueen2009b} T. M. McQueen, Q. Huang, V. Ksenofontov, C. Felser, Q. Xu, H. Zandbergen, Y. S. Hor, J. Allred, A. J. Williams, D. Qu, J. Checkelsky, N. P. Ong, and R. J. Cava, Phys. Rev. B {\bf 79}, 014522 (2009).
\bibitem{Pomjakushina2009} E. Pomjakushina, K. Conder, V. Pomjakushin, M. Bendele, and R. Khasanov, Phys. Rev. B {\bf 80}, 024517 (2009).
\bibitem{Yang2009a} J. Yang, M. Matsui, M. Kawa, H. Ohta, C. Michioka, C. Dong, H. Wang, H. Yuan, M. Fang, and K. Yoshimura, arXiv:0911.4758 (unpublished); J. Phys. Soc. Jpn. {\bf 79}, 074704 (2010).
\bibitem{Viennois2010} R. Viennois, E. Giannini, D. van der Marel, and R. \v{C}ern\'y, J. Solid State Chem. {\bf 183}, 769 (2010).
\bibitem{Okamoto1991} H. Okamoto, J. Phase Equilib. {\bf 12}, 383 (1991).
\bibitem{Margadonna2008} S. Margadonna, Y. Takabayashi, M. T. McDonald, K. Kasperkiewicz, Y. Mizuguchi, Y. Takano, A. N. Fitch, E. Suard, and K. Prassides, Chem. Commun. {\bf 2008}, 5607 (2008).
\bibitem{Fang2008} M. H. Fang, H. M. Pham, B. Qian, T. J. Liu, E. K. Vehstedt, Y. Liu, L. Spinu, and Z. Q. Mao, Phys. Rev. B {\bf 78}, 224503 (2008).
\bibitem{Chiba1955} S. Chiba, J. Phys. Soc. Jpn. {\bf 10}, 837 (1955).
\bibitem{Li2010b} Z. Li, J. Ju, J. Tang, K. Sato, M. Watahiki, and K. Tanigaki, J. Phys. Chem. Solids {\bf 71}, 495 (2010).
\bibitem{McQueen2009c} T. M. McQueen, A. J. Williams, P. W. Stephens, J. Tao, Y. Zhu, V. Ksenofontov, F. Casper, C. Felser, and R. J. Cava, Phys. Rev. Lett. {\bf 103}, 057002 (2009).
\bibitem{Phelan2009} D. Phelan, J. N. Millican, E. L. Thomas, J. B. Le\~ao, Y. Qiu, and R. Paul, Phys. Rev. B {\bf 79}, 014519 (2009).
\bibitem{Khasanov2010} R. Khasanov, M. Bendele, K. Conder, H. Keller, E. Pomjakushina, and V. Pomjakushin, New J. Phys. {\bf 12}, 073024 (2010).
\bibitem{Ratcliff2010} W. Ratcliff II, P. A. Kienzle, J. W. Lynn, S. Li, P. Dai, G. F. Chen, and N. L. Wang, Phys. Rev. B {\bf 81}, 140502(R) (2010).
\bibitem{Lee2010} Y. Lee, D. Vaknin, H. Li, W. Tian, J. L. Zarestky, N. Ni, S. L. Bud'ko, P. C. Canfield, R. J. McQueeney, and B. N. Harmon, Phys. Rev. B {\bf 81}, 060406(R) (2010).
\bibitem{Sawatzky2008} G. A. Sawatzky, I. S. Elfimov, J. van den Brink, and J. Zaanen, Europhys. Lett. {\bf 86}, 17006 (2009); M. Berciu, I. Elfimov, and G. A. Sawatzky, Phys. Rev. B {\bf 79}, 214507 (2009).
\bibitem{Liu2008} C. Liu, G. D. Samolyuk, Y. Lee, N. Ni, T. Kondo, A. F. Santander-Syro, S. L. Bud'ko, J. L. McChesney, E. Rotenberg, T. Valla, A. V. Fedorov, P. C. Canfield, B. N. Harmon, and A. Kaminski, Phys. Rev. Lett. {\bf 101}, 177005 (2008).
\bibitem{mazin2008} I. I. Mazin, D. J. Singh, M. D. Johannes, and M. H. Du, Phys. Rev. Lett. \textbf{101}, 057003 (2008).
\bibitem{dong2008b} J. Dong, H. J. Zhang, G. Xu, Z. Li, G. Li, W. Z. Hu, D. Wu, G. F. Chen, X. Dai, J. L. Luo, Z. Fang, and N. L. Wang, Europhys. Lett. \textbf{83}, 27006 (2008).
\bibitem{kittel1966} C. Kittel, {\it Introduction to Solid State Physics}, 4th ed. (Wiley, New York, 1966).
\bibitem{BZFigs} M. I. Aroyo and H. Wondratschek, Z. Kristallogr. {\bf 210}, 243 (1995); Danel Orobengoa (private communication).  Postscript figures with symmetry line and point designations in the first Brillouin zones for all fourteen three-dimensional Bravais lattices are available for download from the Bilbao Crystallographic Server at $<$\url{http://www.cryst.ehu.es/cryst/get_kvec.html}$>$.  These online figures of the Brillouin zones were endorsed by the recognized primary crystallographic authority:  \emph{International Tables for Crystallography}, Vol.~B, Ed. U. Shmueli, 2$^{\rm nd}$~Edition (Kluwer, Dordrecht, 2001).
\bibitem{Cracknell1979} A. P. Cracknell, B. L. Davies, S. C. Miller, and W. F. Love, \emph{Kronecker Product Tables}, Vol. 1 (IFI/Plenum, New York, 1979).
\bibitem{Jorgensen1997} J. D. Jorgensen, H.-B. Sch\"uttler, D. G. Hinks, D. W. Capone, II, K. Zhang, M. B. Brodsky, and D. J. Scalapino, Phys. Rev. Lett. {\bf 58}, 1024 (1987).
\bibitem{Graser2010} S. Graser, A. F. Kemper, T. A. Maier, H.-P. Cheng, P. J. Hirschfeld, and D. J. Scalapino, Phys. Rev. B {\bf 81}, 214503 (2010).
\bibitem{Xia2009} Y. Xia, D. Qian, L. Wray, D. Hsieh, G. F. Chen, J. L. Luo, N. L. Wang, and M. Z. Hasan, Phys. Rev. Lett. {\bf 103}, 037002 (2009).
\bibitem{Lee2009} S.-H. Lee, G. Xu, W. Ku, J. S. Wen, C. C. Lee, N. Katayama, Z. J. Xu, S. Ji, Z. W. Lin, G. D. Gu, H.-B. Yang, P. D. Johnson, Z.-H. Pan, T. Valla, M. Fujita, T. J. Sato, S. Chang, K. Yamada, and J. M. Tranquada, Phys. Rev. B {\bf 81}, 220502(R) (2010).
\bibitem{Lumsden2010a} M. D. Lumsden, A. D. Christianson, E. A. Goremychkin, S. E. Nagler, H. A. Mook, M. B. Stone, D. L. Abernathy, T. Guidi, G. J. MacDougall, C. de la Cruz, A. S. Sefat, M. A. McGuire, B. C. Sales, and D. Mandrus, Nature Phys. {\bf 6}, 182 (2010).
\bibitem{djsingh2008b} D. J. Singh and M.-H. Du, Phys. Rev. Lett. \textbf{100}, 237003 (2008).
\bibitem{Zhang2009} L. Zhang and D. J. Singh, Phys. Rev. B {\bf 79}, 174530 (2009).
\bibitem{Mazin2009} I. I. Mazin and J. Schmalian, Physica C {\bf 469}, 614 (2009).
\bibitem{Vilmercati2009} P. Vilmercati, A. Fedorov, I. Vobornik, U. Manju, G. Panaccione, A. Goldoni A. S. Sefat, M. A. McGuire, B. C. Sales, R. Jin, D. Mandrus, D. J. Singh, and N. Mannella, Phys. Rev. B {\bf 79}, 220503(R) (2009); {\bf 81}, 029901(E) 2010.
\bibitem{Malaeb2009} W. Malaeb, T. Yoshida, A. Furimori, M. Kubota, K. Ono, K. Kihou, P. M. Shirage, H. Kito, A. Iyo, H. Eisaki, Y. Nakajima, T. Tamegai, and R. Arita, J. Phys. Soc. Jpn. {\bf 78}, 123706 (2009).
\bibitem{Thirupathaiah2010} S. Thirupathaiah, S. de Jong, R. Ovsyannikov, H. A. D\"urr, A. Varykhalov, R. Follath, Y. Huang, R. Huisman, M. S. Golden, Y.-Z. Zhang, H. O. Jeschke, R. Valent\'i, A. Erb, A. Gloskovskii, and J. Fink, Phys. Rev. B {\bf 81}, 104512 (2010).
\bibitem{Sato2009} T. Sato, K. Nakayama, Y. Sekiba, P. Richard, Y.-M. Xu, S. Souma, T. Takahashi, G. F. Chen, J. L. Luo, N. L. Wang, and H. Ding, Phys. Rev. Lett. {\bf 103}, 047002 (2009).
\bibitem{Terashima2010} T. Terashima, M. Kimata, N. Kurita, H. Satsukawa, A. Harada, K. Hazama, M. Imai, A. Sato, K. Kihou, C.-H. Lee, H. Kito, H. Eisaki, A. Iyo, T. Saito, H. Fukazawa, Y. Kohori, H. Harima, and S. Uji, J. Phys. Soc. Jpn. {\bf 79}, 053702 (2010).
\bibitem{Utfeld2010} C. Utfeld, J. Laverock, T. D. Haynes, S. B. Dugdale, J. A. Duffy, M. W. Butchers, J. W. Taylor, S. R. Giblin, J. G. Analytis, J.-H. Chu, I. R. Fisher, M. Itou, and Y. Sakurai, Phys. Rev. B {\bf 81}, 064509 (2010).
\bibitem{Shein2010} I. R. Shein and A. L. Ivanovskii, Solid State Commun. {\bf 150}, 152 (2010).
\bibitem{Subedi2008a} A. Subedi, L. Zhang, D. J. Singh, and M. H. Du, Phys. Rev. B {\bf 78}, 134514 (2008).
\bibitem{Tamai2010} A. Tamai, A. Y. Ganin, E. Rozbicki, J. Bacsa, W. Meevasana, P. D. C. King, M. Caffio, R. Schaub, S. Margadonna, K. Prassides, M. J. Rosseinsky, and F. Baumberger, Phys. Rev. Lett. {\bf 104}, 097002 (2010).
\bibitem{Homes2010} C. C. Homes, A. Akrap, J. S. Wen, Z. J. Xu, Z. W. Lin, Q. Li, and G. D. Gu, Phys. Rev. B {\bf 81}, 180508(R) (2010); C. C. Homes, A. Akrap, J. Wen, Z. Xu, Z. W. Lin, Q. Li, and G. Gu, arXiv:1007.1447.
\bibitem{an2009} J. An, A. S. Sefat, D. J. Singh, and 
M.-H. Du, Phys. Rev. B {\bf 79}, 075120 (2009).
\bibitem{singh2009} Y. Singh, A. Ellern, and D. C. Johnston, Phys. Rev. B {\bf 79}, 094519 (2009).
\bibitem{YSingh2009} Y. Singh, M. A. Green, Q. Huang, A. Kreyssig, R. J. McQueeney, D. C. Johnston, and A. I. Goldman Phys. Rev. B {\bf 80}, 100403(R) (2009).
\bibitem{Harrison1970} W. A. Harrison, \emph{Solid State Theory} (McGraw-Hill, New York, 1970).
\bibitem{Dressel2002} M. Dressel and G. Gr\"uner, \emph{Electrodynamics of Solids} (Cambridge University Press, Cambridge, 2002).
\bibitem{JYang2009} J. Yang, D. H\"uvonen, U. Nagel, T. R\~o\~om, N. Ni, P. C. Canfield, S. L. Bud'ko, J. P. Carbotte, and T. Timusk, Phys. Rev. Lett. {\bf 102}, 187003 (2009).
\bibitem{Wu2009} D. Wu, N. Bara\v{s}i\'c, N. Drichko, S. Kaiser, A. Faridian, M. Dressel, S. Jiang, Z. Ren, L. J. Li, G. H. Cao, Z. A. Xu, H. S. Jeevan, and P. Gegenwart, Phys. Rev. B {\bf 79}, 155103 (2009).
\bibitem{Qazilbash2008} M. M. Qazilbash, J. J. Hamlin, R. E. Baumbach, L. Zhang, D. J. Singh, M. B. Maple, and D. N. Basov, Nature Phys. {\bf 5}, 647 (2009).
\bibitem{Stojilovic2010} N. Stojilovic, A. Koncz, L. W. Kohlman, R. Hu, C. Petrovic, and S. V. Dordevic, Phys. Rev. B {\bf 81}, 174518 (2010).
\bibitem{Barasic2010} N. Bara\v{s}i\'c, D. Wu, M. Dressel, L. J. Li, G. H. Cao, and Z. A. Xu, arXiv:1004.1658.
\bibitem{Puchkov1996} A. V. Puchkov, D. N. Basov, and T. Timusk, J. Phys.: Condens. Matter {\bf 8}, 10049 (1996).
\bibitem{Hu2009} W. Z. Hu, Q. M. Zhang, and N. L. Wang, Physica C {\bf 469}, 545 (2009).
\bibitem{Ma2008} F.-J. Ma, Z.-Y. Lu, and T. Xiang, Front. Phys. China {\bf 5}, 150 (2010).
\bibitem{Hu2008} W. Z. Hu, J. Dong, G. Li, Z. Li, P. Zheng, G. F. Chen, J. L. Luo, and N. L. Wang, Phys. Rev. Lett. {\bf 101}, 257005 (2008).
\bibitem{Chen2009} Z. G. Chen, R. H. Yuan, T. Dong, and N. L. Wang, Phys. Rev. B {\bf 81}, 100502(R) (2010).
\bibitem{Drechsler2009} S.-L. Drechsler, H. Rosner, M. Grobosch, G. Behr, F. Roth, G. Fuchs, K. Koepernik, R. Schuster, J. Malek, S. Elgazzar, M. Rotter, D. Johrendt, H.-H. Klauss, B. B\"uchner, and M. Knupfer, arXiv:0904.0827 (unpublished).
\bibitem{Akrap2009} A. Akrap, J. J. Tu, L. J. Li, G. H. Cao, Z. A. Xu, and C. C. Homes, Phys. Rev. B {\bf 80}, 180502(R) (2009).
\bibitem{Fischer2010} T. Fischer, A. V. Pronin, J. Wosnitza, K. Iida, F. Kurth, S. Haindl, L. Schultz, B. Holzapfel, and E. Schachinger, arXiv:1005.0692.
\bibitem{Wu2010} D. Wu, N. Bari\v{s}i\'c, P. Kallina, A. Faridian, B. Gorshunov, N. Drichko, L. J. Li, X. Lin, G. H. Cao, Z. A. Xu, N. L. Wang, and M. Dressel, Phys. Rev. B {\bf 81}, 100512(R) (2010); Dan Wu (private communication); D. Wu, N. Bari\v{s}i\'c, N. Drichko, P. Kallina, A. Faridian, B. Gorshunov, M. Dressel, L. J. Li, X. Lin, G. H. Cao, and Z. A. Xu, Physica C (in press); arXiv:0912.0849.
\bibitem{Nakamura2009} D. Nakamura, Y. Imai, A. Maeda, T. Katase, H. Hiramatsu, and H. Hosono, arXiv:0912.4351v3.
\bibitem{Ni2008h} N. Ni, S. L. Bud'ko, A. Kreyssig, S. Nandi, G. E. Rustan, A. I. Goldman, S. Gupta, J. D. Corbett, A. Kracher, and P. C. Canfield, Phys. Rev. B {\bf 78}, 014507 (2008).
\bibitem{Baumbach2008} J. J. Hamlin, R. E. Baumbach. D. A. Zocco, T. A. Sayles, and M. B. Maple, J. Phys.: Condens. Matter {\bf 20}, 365220 (2008); R. E. Baumbach, J. J. Hamlin, L. Shu, D. A. Zocco, N. M. Crisosto, and M. B. Maple, New J. Phys. {\bf 11}, 025018 (2009).
\bibitem{Nakajima2010} M. Nakajima, S. Ishida, K. Kihou, Y. Tomioka, T. Ito, Y. Yoshida, C. H. Lee, H. Kito, A. Iyo, H. Eisaki, K. M. Kojima, and S. Uchida, Phys. Rev. B {\bf 81}, 104528 (2010).
\bibitem{Lucarelli2010} A. Lucarelli, A. Dusza, F. Pfuner, P. Lerch, J. G. Analytis, J.-H. Chu, I. R. Fisher, and L. Degiorgi, arXiv:1004.3022.
\bibitem{Chen2009b} Z. G. Chen, G. Xu, W. Z. Hu, X. D. Zhang, P. Zheng, G. F. Chen, J. L. Luo, Z. Fang, and N. L. Wang, Phys. Rev. B {\bf 80}, 094506 (2009).
\bibitem{Moon2010} C.-Y. Moon and H. J. Choi, Phys. Rev. Lett. {\bf 104}, 057003 (2010).
\bibitem{Dai2009} J. Dai, Q. Si, J.-X. Zhu, and E. Abrahams, Proc. Nat. Acad. Sci. USA {\bf 106}, 4118 (2009).
\bibitem{Si2009} Q. Si, E. Abrahams, J. Dai, and J.-X. Zhu, New J. Phys. {\bf 11}, 045001 (2009).
\bibitem{Chen2010} Z. G. Chen, T. Dong, R. H. Ruan, B. F. Hu, B. Cheng, W. Z. Hu, P. Zheng, Z. Fang, X. Dai, and N. L. Wang, arXiv:1001.1689 (unpublished).
\bibitem{Pallecchi2009} I. Pallecchi, G. Lamura, M. Tropeano, M. Putti, R. Viennois, E. Giannini, and D. Van der Marel, Phys. Rev. B. {\bf 80}, 214511 (2009).
\bibitem{Crako2009} L. Craco, M. S. Laad, and S. Leoni, arXiv:0910.3828 (unpublished).
\bibitem{Craco2009a} L. Craco and M. S. Laad, arXiv:1001.3273 (unpublished).
\bibitem{DJSingh2009a} For a review, see D. J. Singh, Physica C {\bf 469}, 418 (2009).
\bibitem{Belashchenko2008} K. D. Belashchenko and V. P. Antropov, Phys. Rev. B {\bf 78}, 212505 (2008).
\bibitem{Zbiri2009} M. Zbiri, H. Schober, M. R. Johnson, S. Rols, R. Mittal, Y. Su, M. Rotter, and D. Johrendt, Phys. Rev. B {\bf 79}, 064511 (2009).
\bibitem{Hahn2009} S. E. Hahn, Y. Lee, N. Ni, P. C. Canfield, A. I. Goldman, R. J. McQueeney, B. N. Harmon, A. Alatas, B. M. Leu, E. E. Alp, D. Y. Chung, I. S. Todorov, and M. G. Kanatzidis, Phys. Rev. B {\bf 79}, 220511(R) (2009).
\bibitem{Kreyssig2008} A. Kreyssig, M. A. Green, Y. Lee, G. D. Samolyuk, P. Zajdel, J. W. Lynn, S. L. Bud'ko, M. S. Torikachvili, N. Ni, S. Nandi, J. B. Le\~ao, S. J. Poulton, D. N. Argyriou, B. N. Harmon, R. J. McQueeney, P. C. Canfield, and A. I. Goldman, Phys. Rev. B {\bf 78}, 184517 (2008).
\bibitem{Goldman2009} A. I. Goldman, A. Kreyssig, K. Proke\u{s}, D. K. Pratt, D. N. Argyriou, J. W. Lynn, S. Nandi, S. A. J. Kimber, Y. Chen, Y. B. Lee, G. Samolyuk, J. B. Le\~ao, S. J. Poulton, S. L. Bud'ko, N. Ni, P. C. Canfield, B. N. Harmon, and R. J. McQueeney, Phys. Rev. B {\bf 79}, 024513 (2009).
\bibitem{Garbarino2008} G. Garbarino, P. Toulemonde, M. \'Alvarez-Murga, A. Sow, M. Mezouar, and M. N\'u\~nez-Regueiro, Phys. Rev. B {\bf 78}, 100507(R) (2008).
\bibitem{Martinelli2009b} A. Martinelli, M. Ferretti, A. Palenzona, and M. Merlini, Physica C {\bf 469}, 782 (2009).
\bibitem{Liu2009} C. Liu, T. Kondo, N. Ni, A. D. Palczewski, A. Bostwick, G. D. Samolyuk, R. Khasanov, M. Shi, E. Rotenberg, S. L. Bud'ko, P. C. Canfield, and A. Kaminski, Phys. Rev. Lett. {\bf 102}, 167004 (2009).
\bibitem{Shimojima2009} T. Shimojima, K. Ishizaka, Y. Ishida, N. Katayama, K. Ohgushi, T. Kiss, M. Okawa, T. Togashi, X.-Y. Wang, C.-T. Chen, S. Watanabe, R. Kadota, T. Oguchi, A. Chainani, and S. Shin, Phys. Rev. Lett. {\bf 104}, 057002 (2010).
\bibitem{Sebastian2008} S. E. Sebastian, J. Gillett, N. Harrison, P. H. C. Lau, D. J. Singh, C. H. Mielki, and G. G. Lonzarich, J. Phys.: Condens. Matter {\bf 20}, 422203 (2008).
\bibitem{Analytis2009} J. G. Analytis, R. D. McDonald, J.-H. Chu, S. C. Riggs, A. F. Bangura, C. Kucharczyk, M. Johannes, and I. R. Fisher, Phys. Rev. B {\bf 80}, 064507 (2009).
\bibitem{Klingeler2008} R. Klingeler, N. Leps, I. Hellmann, A. Popa, U. Stockert, C. Hess, V. Kataev, H.-J. Grafe, F. Hammerath, G. Lang, S. Wurmehl, G. Behr, L. Harnagea, S. Singh, and B. B\"uchner, Phys. Rev. B {\bf 81}, 024506 (2010); A. Kondrat, J. E. Hamann-Borrero, N. Leps, M. Kosmala, O. Schumann, A. K\"ohler, J. Werner, G. Behr, M. Braden, R. Klingeler, B. B\"uchner, and C. Hess, Eur. Phys. J. B {\bf 70}, 461 (2009); R. Klingeler, private communication.
\bibitem{Klingeler2010} R. Klingeler, N. Leps, I. Hellmann, A. Popa, U. Stockert, C. Hess, V. Kataev, H.-J. Grafe, F. Hammerath, G. Lang, S. Wurmehl, G. Behr, L. Harnagea, S. Singh, and B. B\"uchner, Phys. Rev. B {\bf 81}, 024506 (2010).
\bibitem{GMZhang2008} G. M. Zhang, Y. H. Su, Z. Y. Lu, Z. Y. Weng, D. H. Lee, and T. Xiang, Europhys. Lett. {\bf 86}, 37006 (2009).
\bibitem{Laad2009} M. S. Laad and L. Craco, arXiv:0903.3732 (unpublished).
\bibitem{Korshunov2009} M. M. Korshunov, I. Eremin, D. V. Efremov, D. L. Maslov, and A. V. Chubukov, Phys. Rev. Lett. {\bf 102}, 236403 (2009).
\bibitem{Rullier-Albenque2009} F. Rullier-Albenque, D. Colson, A. Forget, and H. Alloul, Phys. Rev. Lett. {\bf 103}, 057001 (2009).
\bibitem{Yi2009} M. Yi, D. H. Lu, J. G. Analytis, J.-H. Chu, S.-K. Mo, R.-H. He, R. G. Moore, X. J. Zhou, G. F. Chen, J. L. Luo, N. L. Wang, Z. Hussain, D. J. Singh, I. R. Fisher, and Z.-X. Shen, Phys. Rev. B {\bf 80}, 024515 (2009).
\bibitem{Johnston2010} D. C. Johnston, Y. Lee, and B. N. Harmon, (unpublished).
\bibitem{Sales2009} B. C. Sales, M. A. McGuire, A. S. Sefat, and D. Mandrus, Physica C {\bf 470}, 304 (2010).
\bibitem{Skornyakov2010a} S. L. Skornyakov, A. A. Katanin, and V. I. Anisimov, arXiv:1006.1509.
\bibitem{Zhang2009a} L. Zhang, D. J. Singh, and M. H. Du, Phys. Rev. B {\bf 79}, 012506 (2009).
\bibitem{singh2008} Y. Singh, Y. Lee, S. Nandi, A. Kreyssig, A. Ellern, S. Das, R. Nath, B. N. Harmon, A. I. Goldman, and D. C. Johnston, Phys. Rev. B \textbf{78}, 104512 (2008).  The quoted value of $\chi_{\rm core}$ in this paper is not correct.  It should read $\chi_{\rm core} = -0.303 \times 10^{-4}$~cm$^3$/mol.
\bibitem{djsingh2008} D. J. Singh,
Phys. Rev. B \textbf{78}, 094511 (2008); private communication.
\bibitem{Kohama2008} Y. Kohama, Y. Kamihara, M. Hirano, H. Kawaji, T. Atake, and H. Hosono, Phys. Rev. B {\bf 78}, 020512(R) (2008).
\bibitem{Haule2009} K. Haule and G. Kotliar, New J. Phys. {\bf 11}, 025021 (2009).
\bibitem{Parshall2009} D. Parshall, K. A. Lokshin, J. Niedziela, A. D. Christianson, M. D. Lumsden, H. A. Mook, S. E. Nagler, M. A. McGuire, M. B. Stone, D. L. Abernathy, A. S. Sefat, B. C. Sales, D. G. Mandrus, and T. Egami, Phys. Rev. B {\bf 80}, 012502 (2009).
\bibitem{JZhang2009} J. Zhang, R. Sknepnek, R. M. Fernandes, and J. Schmalian, Phys. Rev. B {\bf 79}, 220502(R) (2009).
\bibitem{Inosov2010} D. S. Inosov, J. T. Park, P. Bourges, D. L. Sun, Y. Sidis, A. Schneidewind, K. Hradil, D. Haug, C. T. Lin, B. Keimer, and V. Hinkov, Nature Phys. {\bf 6}, 178 (2010).  In this paper, the units of $\chi^{\prime\prime}$ are $\mu_{\rm B}^2$~eV$^{-1}$~f.u.$^{-1}$, where f.u.\ means formula unit of ${\rm BaFe_{1.85}Co_{0.15}As_2}$.  The  wave vector scans are longitudinal $(\frac{1}{2} + h, \frac{1}{2} + h, L)$ scans ($L = 1,3$) through the magnetic zone centers at tetragonal $(\frac{1}{2}, \frac{1}{2}, L)$~r.l.u., i.e.\ ${\bf q} = (h,h)$~r.l.u., and the anisotropy of the magnetic excitations in the ${\bf Q}_x$-${\bf Q}_y$ plane was not examined (D. S. Inosov, private communication).  The $\chi^{\prime\prime}(q,\omega)$ frames at different temperatures are available in the source files of arXiv:0907.3632.
\bibitem{Diallo2010} S. O. Diallo, D. K. Pratt, R. M. Fernandes, W. Tian, J. L. Zarestky, M. Lumsden, T. G. Perring, C. L. Broholm, N. Ni, S. L. Bud'ko, P. C. Canfield, H.-F. Li, D. Vaknin, A. Kreyssig, A. I. Goldman, and R. J. McQueeney, Phys. Rev. B {\bf 81}, 214407 (2010).
\bibitem{Slichter1963} C. P. Slichter, \emph{Principles of Magnetic Resonance} (Harper \& Row, New York, 1963).
\bibitem{Abragam1961} A. Abragam, \emph{Principles of Nuclear Magnetism} (Clarendon Press, Oxford, 1978).
\bibitem{Imai2008} T. Imai, K. Ahilan, F. Ning, M. A. McGuire, A. S. Sefat, R. Jin, B. C. Sales, and D. Mandrus, J. Phys. Soc. Jpn. {\bf 77}, Suppl. C, 47 (2008). 
\bibitem{Ning2009} F. Ning, K. Ahilan, T. Imai, A. S. Sefat, R. Jin, M. A. McGuire, B. C. Sales, and D. Mandrus, J. Phys. Soc. Jpn. {\bf 78}, 013711 (2009).
\bibitem{Kitagawa2008} K. Kitagawa, N. Katayama, K. Ohgushi, M. Yoshida, and M. Takigawa, J. Phys. Soc. Jpn. {\bf 77}, 114709 (2008).
\bibitem{Imai2009} T. Imai, K. Ahilan, F. L. Ning, T. M. McQueen, and R. J. Cava, Phys. Rev. Lett. {\bf 102}, 177005 (2009).
\bibitem{Michioka2009} C. Michioka, H. Ohta, M. Matsui, J. Yang, K. Yoshimura, and M. Fang, arXiv:0911.3729 (unpublished).
\bibitem{Nath2009a} R. Nath, Y. Furukawa, F. Borsa, E. E. Kaul, M. Baenitz, C. Geibel, and D. C. Johnston, Phys. Rev. B {\bf 80}, 214430 (2009), and references cited therein.
\bibitem{Moreo1990} A. Moreo, E. Dagotto, T. Jolicoeur, and J. Riera, Phys. Rev. B {\bf 42}, 6283 (1990).
\bibitem{Hotta2003} T. Hotta, M. Moraghebi, A. Feiguin, A. Moreo, S. Yunoki, and E. Dagotto, Phys. Rev. Lett. {\bf 90}, 247203 (2003).
\bibitem{Dong2008} S. Dong, R. Yu, S. Yunoki, J.-M. Liu, and E. Dagotto, Phys. Rev. B {\bf 78}, 155121 (2008).
\bibitem{Cruz2008} C. de la Cruz, Q. Huang, J. W. Lynn, J. Li, W. Ratcliff II, J. L. Zarestky, H. A. Mook, G. F. Chen, J. L. Luo, N. L. Wang, and P. Dai, Nature {\bf 453}, 899 (2008).
\bibitem{Qureshi2010} N. Qureshi, Y. Drees, J. Werner, S. Wurmehl, C. Hess, R. Klingeler, B. B\"uchner, M. T. Fern\a'andez-D\'iaz, and M. Braden, arXiv:1002.4326 (unpublished).
\bibitem{mcguire2008} M. A. McGuire, A. D. Christianson, A. S. Sefat, B. C. Sales, M. D. Lumsden, R. Jin, E. A. Payzant, D. Mandrus, Y. Luan, V. Keppens, V. Varadarajan, J. W. Brill, R. P. Hermann, M. T. Sougrati, F. Grandjean, and G. J. Long, Phys. Rev. B \textbf{78}, 094517 (2008).
\bibitem{Li2010c} H.-F. Li, W. Tian, J.-Q. Yan, J. L. Zarestky, R. W. McCallum, T. A. Lograsso, and D. Vaknin, arXiv:1007.2197.
\bibitem{Zhao2008b} J. Zhao, Q. Huang, C. de la Cruz, J. W. Lynn, M. D. Lumsden, Z. A. Ren, J. Yang, X. Shen, X. Dong, Z. Zhao, and P. Dai, Phys. Rev. B {\bf 78}, 132504 (2008).
\bibitem{Kimber2008} S. A. J. Kimber, D. N. Argyriou, F. Yokaichiya, K. Habicht, S. Gerischer, T. Hansen, T. Chatterji, R. Klingeler, C. Hess, G. Behr, A. Kondrat, and B. B\"uchner, Phys. Rev. B {\bf 78}, 140503(R) (2008).
\bibitem{Chen2008} Y. Chen, J. W. Lynn, J. Li, G. Li, G. F. Chen, J. L. Luo, N. L. Wang, P. Dai, C. de la Cruz, and H. A. Mook, Phys. Rev. B {\bf 78}, 064515 (2008).
\bibitem{Qiu2008} Y. Qiu, W. Bao, Q. Huang, T. Yildirim, J. M. Simmons, M. A. Green, J. W. Lynn, Y. C. Gasparovic, J. Li, T. Wu, G. Wu, and X. H. Chen, Phys. Rev. Lett. {\bf 101}, 257002 (2008).
\bibitem{Tian2010} W. Tian, W. Ratcliff II, M. G. Kim, J.-Q. Yan, P. A. Kienzle, Q. Huang, B. Jensen, K. W. Dennis, R. W. McCallum, T. A. Lograsso, R. J. McQueeney, A. I. Goldman, J. W. Lynn, and A. Kreyssig, arXiv:1006.1135.
\bibitem{Xiao2009c} Y. Xiao, Y. Su, R. Mittal, T. Chatterji, T. Hansen, C. M. N. Kumar, S. Matsuishi, H. Hosono, and Th. Brueckel, Phys. Rev. B {\bf 79}, 060504(R) (2009).
\bibitem{Xiao2009b} Y. Xiao, Y. Su, R. Mittal, T. Chatterji, T. Hansen, S. Price, C. M. N. Kumar, J. Persson, S. Matsuishi, Y. Inoue, H. Hosono, and Th. Brueckel, Phys. Rev. B {\bf 81}, 094523 (2010).
\bibitem{Zhao2008c} J. Zhao, W. Ratcliff II, J. W. Lynn, G. F. Chen, J. L. Luo, N. L. Wang, J. Hu, and P. Dai, Phys. Rev. B {\bf 78}, 140504(R) (2008).
\bibitem{Kaneko2008} K. Kaneko, A. Hoser, N. Caroca-Canales, A. Jesche, C. Krellner, O. Stockert, and C. Geibel, Phys. Rev. B {\bf 78}, 212502 (2008).
\bibitem{Matan2009} K. Matan, R. Morinaga, K. Iida, and T. J. Sato, Phys. Rev. B {\bf 79}, 054526 (2009).
\bibitem{Xiao2009} Y. Xiao, Y. Su, M. Meven, R. Mittal, C. M. N. Kumar, T. Chatterji, S. Price, J. Persson, N. Kumar, S. K. Dhar, A. Thamizhavel, and Th. Brueckel, Phys. Rev. B {\bf 80}, 174424 (2009).
\bibitem{Li2009b} S. Li, C. de la Cruz, Q. Huang, G. F. Chen, T.-L. Xia, J. L. Luo, N. L. Wang, and P. Dai, Phys. Rev. B {\bf 80}, 020504(R) (2009).
\bibitem{Martinelli2010} A. Martinelli, A. Palenzona, M. Tropeano, C. Ferdeghini, M. Putti, M. R. Cimberle, T. D. Nguyen, M. Affronte, and C. Ritter, Phys. Rev. B {\bf 81}, 094115 (2010).
\bibitem{Li2009c} S. Li, C. de la Cruz, Q. Huang, Y. Chen, J. W. Lynn, J. Hu, Y.-L. Huang, F.-C. Hsu, K.-W. Yeh, M.-K. Wu, and P. Dai, Phys. Rev. B {\bf 79}, 054503 (2009).
\bibitem{Iikubo2009} S. Ikkubo, M. Fujita, S. Niitaka, and H. Takagi, J. Phys. Soc. Jpn. {\bf 78}, 103704 (2009).
\bibitem{Roth1958}  W. L. Roth, Phys. Rev. {\bf 110}, 1333 (1958).
\bibitem{Rodriguez2009} J. P. Rodriguez and E. H. Rezayi, Phys. Rev. Lett. {\bf 103}, 097204 (2009).
\bibitem{Manousakis2010} E. Manousakis, J. Ren, S. Meng, and E. Kaxiras, Solid State Commun. {\bf 150}, 62 (2010).
\bibitem{Mazin2009a} I. I. Mazin and M. D. Johannes, Nature Phys. {\bf 5}, 141 (2009).
\bibitem{Hansmann2010} P. Hansmann, R. Arita, A. Toschi, S. Sakai, G. Sangiovanni, and K. Held, Phys. Rev. Lett. {\bf 104}, 197002 (2010).
\bibitem{Diallo2009} S. O. Diallo, V. P. Antropov, T. G. Perring, C. Broholm, J. J. Pulikkotil, N. Ni, S. L. Bud'ko, P. C. Canfield, A. Kreyssig, A. I. Goldman, and R. J. McQueeney, Phys. Rev. Lett. {\bf 102}, 187206 (2009).  Equations~(1) and~(2) refer to the scattering wave vector ${\bf Q}$ instead of relative wave vector ${\bf q}$.  In Fig.~4, the wave vectors plotted are the reduced wave vectors ${\bf q} = (h,k,\ell)$~r.l.u. The expression for $E[{\bf q} = (010)]$ was derived assuming that $D$ is negligible and $J_{1c} \ll 2\delta J_1$.  The stability conditions for the stripe-$b$ state are given in Eqs.~(\ref{EqStripeStab}) of this review.
\bibitem{Schmidt2010} B. Schmidt, M. Siahatgar, and P. Thalmeier, Phys. Rev. B {\bf 81}, 165101 (2010).
\bibitem{Kitagawa2009} K. Kitagawa, N. Katayama, K. Ohgushi, and M. Takigawa, J. Phys. Soc. Jpn. {\bf 78}, 063706 (2009).
\bibitem{Monni2010} M. Monni, F. Bernardini, G. Profeta, A. Sanna, S. Sharma, J. K. Dewhurst, C. Bersier, A. Continenza, E. K. U. Gross, and S. Massidda, Phys. Rev. B {\bf 81}, 104503 (2010).
\bibitem{Ma2008b} F. Ma, Z.-Y. Lu, and T. Xiang, Phys. Rev. B {\bf 78}, 224517 (2008).
\bibitem{Yaresko2009} A. N. Yaresko, G.-Q. Liu, V. N. Antonov, and O. K. Andersen, Phys. Rev. B {\bf 79}, 144421 (2009).
\bibitem{Daghofer2008} M. Daghofer, A. Moreo, J. A. Riera, E. Arrigoni, D. J. Scalapino, and E. Dagotto, Phys. Rev. Lett. {\bf 101}, 237004 (2008); A. Moreo, M. Daghofer, J. A. Riera, and E. Dagotto, Phys. Rev. B {\bf 79}, 134502 (2009).  See also the cited references.
\bibitem{Herrero-Martin2009} J. Herrero-Mart\'in, V. Scagnoli, C. Mazzoli, Y. Su, R. Mittal, Y. Xiao, T. Brueckel, N. Kumar, S. K. Dhar, A. Thamizhavel, and L. Paolasini, Phys. Rev. B. {\bf 80}, 134411 (2009).
\bibitem{Free2010} D. G. Free and J. S. O. Evans, Phys. Rev. B {\bf 81}, 214433 (2010).
\bibitem{Zhu2010} J.-X. Zhu, R. Yu, H. Wang, L. L. Zhao, M. D. Jones, J. Dai, E. Abrahams, E. Morosan, M. Fang, and Q. Si, Phys. Rev. Lett. {\bf 104}, 216405 (2010).
\bibitem{Ma2009a} F. Ma, W. Ji, J. Hu, Z.-Y. Lu, and T. Xiang, Phys. Rev. Lett. {\bf 102}, 177003 (2009).
\bibitem{Han2010} M. J. Han and S. Y. Savrasov, Phys. Rev. Lett. {\bf 103}, 067001 (2009); Reply to Comment {\bf 104}, 099702 (2010).
\bibitem{Fang2009a} C. Fang, B. A. Bernevig, and J. Hu, Europhys. Lett. {\bf 86}, 67005 (2009).
\bibitem{Haule2008} K. Haule, J. H. Shim, and G. Kotliar, Phys. Rev. Lett. {\bf 100}, 226402 (2008).
\bibitem{Craco2008} L. Craco, M. S. Laad, S. Leoni, and H. Rosner, Phys. Rev. B {\bf 78}, 134511 (2008); M. S. Laad, L. Craco, S. Leoni, and H. Rosner, Phys. Rev. B {\bf 79}, 024515 (2009).
\bibitem{Johannes2009} M. D. Johannes and I. I. Mazin, Phys. Rev. B {\bf 79}, 220510(R) (2009).
\bibitem{Wang2009} G. Wang, Y. Qian, G. Xu, X. Dai, and Z. Fang, Phys. Rev. Lett. {\bf 104}, 047002 (2010); Zong Fang, private communication.
\bibitem{Tesanovic2009} Z. Tesanovic, Physics {\bf 2}, 60 (2009).
\bibitem{Anisimov2009} V. I. Anisimov, E. Z. Kurmaev, A. Moewes, and I. A. Izyumov, Physica C {\bf 469}, 442 (2009).
\bibitem{Skornyakov2009} S. L. Skornyakov, A. V. Efremov, N. A. Skorikov, M. A. Korotin, Yu. A. Izyumov, V. I. Anisimov, A. V. Kozhevnikov, and D. Vollhardt, Phys. Rev. B {\bf 80}, 092501 (2009).
\bibitem{Skornyakov2010} S. L. Skornyakov, N. A. Skorikov, A. V. Lukoyanov, A. O. Shorikov, and V. I. Anisimov, Phys. Rev. B {\bf 81}, 174522 (2010).
\bibitem{Aichhorn2009} M. Aichhorn, L. Pourovskii, V. Vildosola, M. Ferrero, O. Parcollet, T. Miyake, A. Georges, and S. Biermann, Phys. Rev. B {\bf 80}, 085101 (2009).
\bibitem{Emery1995} V. J. Emery and S. A. Kivelson, Phys. Rev. Lett. {\bf 74}, 3253 (1995).
\bibitem{Allen2002} P. B. Allen, Physica B {\bf 318}, 24 (2002).
\bibitem{Fang2009} L. Fang, H. Luo, P. Cheng, Z. Wang, Y. Jia, G. Mu, B. Shen, I. I. Mazin, L. Shan, C. Ren, and H.-H. Wen, Phys. Rev. B {\bf 80}, 140508(R) (2009).
\bibitem{Kasahara2009} S. Kasahara, T. Shibauchi, K. Hashimoto, K. Ikada, S. Tonegawa, R. Okazaki, H. Shishido, H. Ikeda, H. Takeya, K. Hirata, T. Terashima, and Y. Matsuda, Phys. Rev. B {\bf 81}, 184519 (2010).
\bibitem{Coldea2008} A. I. Coldea, J. D. Fletcher, A. Carrington, J. G. Analytis, A. F. Bangura, J.-H. Chu, A. S. Erickson, I. R. Fisher, N. E. Hussey, and R. D. McDonald, Phys. Rev. Lett. {\bf 101}, 216402 (2008).
\bibitem{Coldea2009} A. I. Coldea, C. M. J. Andrew, J. G. Analytis, R. D. McDonald, A. F. Bangura, J.-H. Chu, I. R. Fisher, and A. Carrington, Phys. Rev. Lett. {\bf 103}, 026404 (2009).
\bibitem{Analytis2009a} J. G. Analytis, C. J. Andrew, A. I. Coldea, A. McCollam, J.-H. Chu, R. D. McDonald, I. R. Fisher, and A. Carrington, Phys. Rev. Lett. {\bf 103}, 076401 (2009).
\bibitem{Ychen2008} Y. Chen, J. W. Lynn, J. Li, G. Li, G. F. Chen, J. L. Luo, N. L. Wang, P. Dai, C. dela Cruz, and H. A. Mook, Phys. Rev. B {\bf 78}, 064515 (2008).
\bibitem{Kondo2009} T. Kondo, R. M. Fernandes, R. Khasanov, C. Liu, A. D. Palczewski, N. Ni, M. Shi, A. Bostwick, E. Rotenberg, J. Schmalian, S. L. Bud'ko, P. C. Canfield, and A. Kaminski, Phys. Rev. B {\bf 81}, 060507(R) (2010) .
\bibitem{Fernandes2010b} R. M. Fernandes and J. Schmalian, Phys. Rev. B {\bf 82}, 014520 (2010).
\bibitem{Yin2008} Z. P. Yin, S. Leb\`egue, M. J. Han, B. P. Neal, S. Y. Savrasov, and W. E. Pickett, Phys. Rev. Lett. {\bf 101}, 047001 (2008).
\bibitem{Knolle2010} J. Knolle, I. Eremin, A. V. Chubukov, and R. Moessner, Phys. Rev. B {\bf 81}, 140506(R) (2010).
\bibitem{Kaneshita2010} E. Kaneshita and T. Tohyama, arXiv:1002.2701 (unpublished).
\bibitem{DJSingh2009b} D. J. Singh, M.-H. Du, L. Zhang, A. Subedi, and J. An, Physica C {\bf 469}, 886 (2009).
\bibitem{Yanagi2008} H. Yanagi, R. Kawamura, T. Kamiya, Y. Kamihara, M. Hirano, T. Nakamura, H. Osawa, and H. Hosono, Phys. Rev. B {\bf 77}, 224431 (2008).
\bibitem{Millis2005} A. J. Millis, A. Zimmers, R. P. S. M. Lobo, N. Bontemps, and C. C. Homes, Phys. Rev. B {\bf 72}, 224517 (2005).
\bibitem{Yang2009} W. L. Yang, A. P. Sorini, C-C. Chen, B. Moritz, W.-S. Lee, F. Vernay, P. Olalde-Velasco, J. D. Denlinger, B. Delley, J.-H. Chu, J. G. Analytis, I. R. Fisher, Z. A. Ren, J. Yang, W. Lu, Z. X. Zhao, J. van den Brink, Z. Hussain, Z.-X. Shen, and T. P. Devereaux, Phys. Rev. B {\bf 80}, 014508 (2009).
\bibitem{Hsieh2008} D. Hsieh, Y. Xia, L. Wray, D. Qian, K. K. Gomes, A. Yazdani, G. F. Chen, J. L. Luo, N. L. Wang, and M. Z. Hasan, arXiv:0812.2289 (unpublished).
\bibitem{Lovesey1984} S. W. Lovesey, \emph{Theory of Neutron Scattering from Condensed Matter}, Vol.~2 (Clarendon Press, Oxford, 1984).
\bibitem{Collins1989} M. F. Collins, \emph{Magnetic Critical Scattering} (Oxford University Press, Oxford, 1989).
\bibitem{Ewings2008} R. A. Ewings, T. G. Perring, R. I. Bewley, T. Guidi, M. J. Pitcher, D. R. Parker, S. J. Clarke, and A. T. Boothroyd, Phys Rev. B {\bf 78}, 220501(R) (2008).
\bibitem{Lester2010} C. Lester, J.-H. Chu, J. G. Analytis, T. G. Perring, I. R. Fisher, and S. M. Hayden, Phys. Rev. B {\bf 81}, 064505 (2010).  The derived values of $J_1$ and $J_2$ listed on page~4 are evidently values of $SJ_1$ and $SJ_2$, respectively.
\bibitem{White1970} R. M. White, \emph{Quantum Theory of Magnetism} (McGraw-Hill, New York, 1970).
\bibitem{McQueeney2008} R. J. McQueeney, S. O. Diallo, V. P. Antropov, G. D. Samolyuk, C. Broholm, N. Ni, S. Nandi, M. Yethiraj, J. L. Zarestky, J. J. Pulikkotil, A. Kreyssig, M. D. Lumsden, B. N. Harmon, P. C. Canfield, and A. I. Goldman, Phys. Rev. Lett. {\bf 101}, 227205 (2008).
\bibitem{Baily2009} S. A. Baily, Y. Kohama, H. Hiramatsu, B. Maiorov, F. F. Balakirev, M. Hirano, and H. Hosono, Phys. Rev. Lett. {\bf 102}, 117004 (2009).
\bibitem{JZhao2009} J. Zhao, D. T. Adroja, D.-X. Yao, R. Bewley, S. Li, X. F. Wang, G. Wu, X. H. Chen, J. Hu, and P. Dai, Nature Phys. {\bf 5}, 555 (2009).
\bibitem{Christianson2009} A. D. Christianson, M. D. Lumsden, S. E. Nagler, G. J. MacDougall, M. A. McGuire, A. S. Sefat, R. Jin, B. C. Sales, and D. Mandrus, Phys. Rev. Lett. {\bf 103}, 087002 (2009).
\bibitem{Applegate2010} R. Applegate, J. Oitmaa, and R. R. P. Singh, Phys. Rev. B {\bf 81}, 024505 (2010).
\bibitem{Nakayama1987} K. Nakayama and T. Moriya, J. Phys. Soc. Jpn. {\bf 56}, 2918 (1987); A. Ishigaki and T. Moriya, \emph{ibid}.\ {\bf 65}, 3402 (1996).
\bibitem{Osborn2009} R. Osborn, S. Rosenkranz, E. A. Goremychkin, and A. D. Christianson, Physica C {\bf 469}, 498 (2009).
\bibitem{JZhao2008} J. Zhao, D.-X. Yao, S. Li, T. Hong, Y. Chen, S. Chang, W. Ratcliff II, J. W. Lynn, H. A. Mook, G. F. Chen, J. L. Luo, N. L. Wang, E. W. Carlson, J. Hu, and P. Dai, Phys. Rev. Lett. {\bf 101}, 167203 (2008).  The quoted value $v_a = 0.28$~eV~\AA\ is a factor of two too small.  The authors corrected this error in Ref.~\onlinecite{Lynn2009}.
\bibitem{Pratt2010} D. K. Pratt, A. Kreyssig, S. Nandi, N. Ni, A. Thaler, M. D. Lumsden, W. Tian, J. L. Zarestky, S. L. Bud'ko, P. C. Canfield, A. I. Goldman, and R. J. McQueeney, Phys. Rev. B {\bf 81}, 140510(R) (2010).
\bibitem{Fawcett1988} E. Fawcett, Rev. Mod. Phys. {\bf 60}, 209 (1988).
\bibitem{Han2008a} M. J. Han, Q. Yin, W. E. Pickett, and S. Y. Savrasov, Phys. Rev. Lett. {\bf 102}, 107003 (2009).
\bibitem{Harriger2009} L. W. Harriger, A. Schneidewind, S. Li, J. Zhao, Z. Li, W. Lu, X. Dong, F. Zhou, Z. Zhao, J. Hu, and P. Dai, Phys. Rev. Lett. {\bf 103}, 087005 (2009).
\bibitem{Wakimoto2009} S. Wakimoto, K. Kodama, M. Ishikado, M. Matsuda, R. Kajimoto, M. Arai, K. Kakurai, F. Esaka, A. Iyo, H. Kito, H. Eisaki, and S. Shamoto, J. Phys. Soc. Jpn. {\bf 79}, 074715 (2010).
\bibitem{Ishikado2009} M. Ishikado, R. Kajimoto, S. Shamoto, M. Arai, A. Iyo, K. Miyazawa, P. M. Shirage, H. Kito, H. Eisaki, S. Kim, H. Hosono, T. Guidi, R. Bewley, and S. M. Bennington, J. Phys. Soc. Jpn. {\bf 78}, 043705 (2009).
\bibitem{Lumsden2009} M. D. Lumsden, A. D. Christianson, D. Parshall, M. B. Stone, S. E. Nagler, G. J. MacDougall, H. A. Mook, K. Lokshin, T. Egami, D. L. Abernathy, E. A. Goremychkin, R. Osborn, M. A. McGuire, A. S. Sefat, R. Jin, B. C. Sales, and D. Mandrus, Phys. Rev. Lett. {\bf 102}, 107005 (2009).
\bibitem{Li2010} H.-F. Li, C. Broholm, D. Vaknin, D. L. Abernathy, D. K. Pratt, W. Tian, R. M. Fernandes, Y. Qiu, N. Ni, M. B. Stone, S. O. Diallo, J. L. Zarestky, S. L. Bud'ko, P. C. Canfield, and R. J. McQueeney, arXiv:1003.1687 (unpublished).
\bibitem{Argyriou2009} D. N. Argyriou, A. Hiess, A. Akbari, I. Eremin, M. M. Korshunov, J. Hu, B. Qian, Z. Mao, Y. Qiu, C. Broholm, and W. Bao, Phys. Rev. B {\bf 81}, 220503(R) (2010).
\bibitem{Mook2009} H. A. Mook, M. D. Lumsden, A. D. Christianson, B. C. Sales, R. Jin, M. A. McGuire, A. Sefat, D. Mandrus, S. E. Nagler, T. Egami, and C. de la Cruz, arXiv:0904.2178 (unpublished).
\bibitem{Mook2009a} H. A. Mook, M. D. Lumsden, A. D. Christianson, S. E. Nagler, B. C. Sales, R. Jin, M. A. McGuire, A. S. Sefat, D. Mandrus, T. Egami, and C. dela Cruz, Phys. Rev. Lett. {\bf 104}, 187002 (2010).
\bibitem{Li2010a} S. Li, C. Zhang, M. Wang, H.-Q. Luo, E. Faulhaber, A. Schneidewind, J. Hu, T. Xiang, and P. Dai, arXiv:1001.1505 (unpublished).
\bibitem{Vignolle2007} B. Vignolle, S. M. Hayden, D. F. McMorrow, H. M. Ronnow, B. Lake, C. D. Frost, and T. G. Perring, Nature Phys. {\bf 3}, 163 (2007).
\bibitem{HbPC1972} \emph{Handbook of Chemistry and Physics}, 53$^{\rm rd}$ edition (CRC Press, Cleveland, 1972), p.~E-57.
\bibitem{moriya1963} T. Moriya, J. Phys. Soc. Jpn. \textbf{18}, 516 (1963).
\bibitem{foot3} In Eq.~(\ref{t1form}), the hyperfine form factor corresponds 
to $\frac{A_{\rm hf}}{g\mu_{\rm B}}$ in Ref.~\onlinecite{moriya1963}.
\bibitem{mahajan1998} A. V. Mahajan, R. Sala, E. Lee, F. Borsa, S. Kondo, 
and D. C. Johnston, Phys. Rev. B \textbf{57}, 8890 (1998).
\bibitem{Ning2010} F. L. Ning, K. Ahilan, T. Imai, A. S. Sefat, M. A. McGuire, B. C. Sales, D. Mandrus, P. Cheng, B. Shen, and H.-H. Wen, Phys. Rev. Lett. {\bf 104}, 037001 (2010); Fanlong Ning, private communication.  $^{75}$As $1/(T_1T)$ versus $T$ data for $H\parallel c$ are available in arXiv:0907.3875v1. 
\bibitem{Kimber2010} S. A. J. Kimber, D. N. Argyriou, and I. I. Mazin, arXiv:1005.1761 (unpublished).
\bibitem{Chuang2010} T.-M. Chuang, M. P. Allan, J. Lee, Y. Xie, N. Ni, S. L. Bud'ko, G. S. Boebinger, P. C. Canfield, and J. C. Davis, Science {\bf 327}, 181 (2010).
\bibitem{Chu2010} J.-H. Chu, J. G. Analytis, K. De Greve, P. L. McMahon, Z. Islam, Y. Yamamoto, and I. R. Fisher, arXiv:1002.3364v2.
\bibitem{Prozorov2009b} R. Prozorov, M. A. Tanatar, N. Ni, A. Kreyssig, S. Nandi, S. L. Bud'ko, A. I. Goldman, and P. C. Canfield, Phys. Rev. B {\bf 80}, 174517 (2009).
\bibitem{Tanatar2010b} M. A. Tanatar, E. C. Blomberg, A. Kreyssig, M. G. Kim, N. Ni, A. Thaler, S. L. Bud'ko, P. C. Canfield, A. I. Goldman, I. I. Mazin, and R. Prozorov, Phys. Rev. B {\bf 81}, 184508 (2010).
\bibitem{Dusza2010} A. Dusza, A. Lucarelli, F. Pfuner, J.-H. Chu, I. R. Fisher, and L. Degiorgi, arXiv:1007.2543.
\bibitem{Fernandes2010d} R. M. Fernandes, L. H. VanBebber, S. Bhattacharya, P. Chandra, V. Keppens, D. Mandrus, M. A. McGuire, B. C. Sales, A. S. Sefat, and J. Schmalian, arXiv:0911.3084.
\bibitem{BCS57} J. Bardeen, L. N. Cooper, and J. R. Schrieffer, Phys. Rev. {\bf 106}, 162 (1957); {\bf 108}, 1175 (1957).
\bibitem{Cooper56} L. N. Cooper, Phys. Rev. {\bf 104}, 1189 (1956).
\bibitem{Van_Harlingen1995} D. J. Van Harlingen, Rev. Mod. Phys. {\bf 67}, 515 (1995); C. C. Tsuei and J. R. Kirtley, Rev. Mod. Phys. {\bf 72}, 969 (2000).
\bibitem{Maeno2001} Y. Maeno, T. M. Rice, and M. Sigrist, Physics Today {\bf 54}, 42 (2001).
\bibitem{Terasaki2009} N. Terasaki, H. Mukuda, M. Yashima, Y. Kitaoka, K. Miyazawa, P. M. Shirage, H. Kito, H. Eisaki, and A. Iyo, J. Phys. Soc. Jpn. {\bf 78}, 013701 (2009).
\bibitem{Ning2008} F. Ning, K. Ahilan, T. Imai, A. S. Sefat, R. Jin, M. A. McGuire, B. C. Sales, and D. Mandrus, J. Phys. Soc. Jpn. {\bf 77}, 103705 (2008).
\bibitem{Matano2008} K. Matano, Z. A. Ren, X. L. Dong, L. L. Sun, Z. X. Zhao, and G.-Q. Zheng, Europhys. Lett. {\bf 83}, 57001 (2008).
\bibitem{Kawabata2008} A. Kawabata, S. C. Lee, T. Moyoshi, Y. Kobayashi, and M. Sato, J. Phys. Soc. Jpn. {\bf 77}, 103704 (2008); {\bf 77}, Suppl. C 147 (2008).
\bibitem{Yashima2009} M. Yashima, H. Nishimura, H. Mukuda, Y. Kitaoka, K. Miyazawa, P. M. Shirage, K. Kihou, H. Kito, H. Eisaki, and A. Iyo, J. Phys. Soc. Jpn. {\bf 78}, 103702 (2009).
\bibitem{Zhou2008} Y.-R. Zhou, Y.-R. Li, J.-W. Zuo, R.-Y. Liu, S.-K. Su, G. F. Chen, J. L. Lu, N. L. Wang, and Y.-P. Wang, arXiv:0812.3295v1 (unpublished).  In Fig.~3 of this paper, the units of critical current should be $\mu$A instead of mA (Yunping Wang, private communication).
\bibitem{CTChen2009} C.-T. Chen, C. C. Tsuei, M. B. Ketchen, Z.-A. Ren, and Z. X. Zhao, Nature Phys. {\bf 6}, 260 (2010).
\bibitem{Nakayama2009} K. Nakayama, T. Sato, P. Richard, Y.-M. Xu, Y. Sekiba, S. Souma, G. F. Chen, J. L. Luo, N. L. Wang, H. Ding, and T. Takahashi, Europhys. Lett. {\bf 85}, 67002 (2009).
\bibitem{Yakoya2001} T. Yokoya, T. Kiss, A. Chainani, S. Shin, M. Nohara, and H. Takagi, Science {\bf 294}, 2518 (2001).
\bibitem{Tsuda2003} S. Tsuda, T. Yokoya, Y. Takano, H. Kito, A. Matsushita, F. Yin, J. Itoh, H. Harima, and S. Shin, Phys. Rev. Lett. {\bf 91}, 127001 (2003).
\bibitem{Kogan2009} V. G. Kogan, C. Martin, and R. Prozorov, Phys. Rev. B {\bf 80}, 014507 (2009).
\bibitem{Evtushinsky2009} D. V. Evtushinsky, D. S. Inosov, V. B. Zabolotnyy, M. S. Viazovska, R. Khasanov, A. Amato, H.-H. Klauss, H. Luetkens, Ch. Niedermayer, G. L. Sun, V. Hinkov, C. T. Lin, A. Varykhalov, A. Koitzsch, M. Knupfer, B. B\"uchner, A. A. Kordyuk, and S. V. Borisenko, New J. Phys. {\bf 11}, 055069 (2009).
\bibitem{Malone2009} L. Malone, J. D. Fletcher, A. Serafin, A. Carrington, N. D. Zhigadlo, Z. Bukowski, S. Katrych, and J. Karpinski, Phys. Rev. B {\bf 79}, 140501(R) (2009).
\bibitem{Kondo2008} T. Kondo, A. F. Santander-Syro, O. Copie, C. Liu, M. E. Tillman, E. D. Mun, J. Schmalian, S. L. Bud'ko, M. A. Tanatar, P. C. Canfield, and A. Kaminski, Phys. Rev. Lett. {\bf 101}, 147003 (2008).
\bibitem{Popovich2010} P. Popovich, A. V. Boris, O. V. Dolgov, A. A. Golubov, D. L. Sun, C. T. Lin, R. K. Kremer, and B. Keimer, Phys. Rev. Lett. {\bf 105}, 027003 (2010).
\bibitem{Ding2008} H. Ding, P. Richard, K. Nakayama, K. Sugawara, T. Arakane, Y. Sekiba, A. Takayama, S. Souma, T. Sato, T. Takahashi, Z. Wang, X. Dai, Z. Fang, G. F. Chen, J. L. Luo, and N. L. Wang, Europhys. Lett. {\bf 83}, 47001 (2008).
\bibitem{Xu2010a}  Y.-M. Xu, Y.-B. Huang, X.-Y. Cui, R. Elia, R. Milan, M. Shi, G.-F. Chen, P. Zheng, N.-L. Wang, P.-C. Dai, J.-P. Hu, Z. Wang, and H. Ding, arXiv:1006.3958.
\bibitem{Zhang2010} Y. Zhang, L. X. Yang, F. Chen, B. Zhou, X. F. Wang, X. H. Chen, M. Arita, K. Shimada, H. Namatame, M. Taniguchi, J. P. Hu, B. P. Xie, and D. L. Feng, arXiv:1006.3936.
\bibitem{Luan2010} L. Luan, O. M. Auslaender, T. M. Lippman, C. W. Hicks, B. Kalisky, J.-H. Chu, J. G. Analytis, I. R. Fisher, J. R. Kirtley, and K. A. Moler, Phys. Rev. B {\bf 81}, 100501(R) (2010).
\bibitem{Kim2010a} K. W. Kim, M. R\"ossle, A. Dubroka, V. K. Malik, T. Wolf, and C. Bernhard, Phys. Rev. B {\bf 81}, 214508 (2010).
\bibitem{Gofryk2010a} K. Gofryk, A. S. Sefat, E. D. Bauer, M. A. McGuire, B. C. Sales, D. Mandrus, J. D. Thompson, and F. Ronning, New J. Phys. {\bf 12}, 023006 (2010).
\bibitem{Williams2009} T. J. Williams, A. A. Aczel, E. Baggio-Saitovitch, S. L. Bud'ko, P. C. Canfield, J. P. Carlo, T. Goko, J. Munevar, N. Ni, Y. J. Uemura, W. Yu, and G. M. Luke, Phys. Rev. B {\bf 80}, 094501 (2009).
\bibitem{Hardy2010} F. Hardy, T. Wolf, R. A. Fisher, R. Eder, P. Schweiss, P. Adelmann, H. v. L\"ohneysen, and C. Meingast, Phys. Rev. B {\bf 81}, 060501(R) (2010).
\bibitem{Perucchi2010} A. Perucchi, L. Baldassarre, S. Lupi, J. Jiang, J. D. Weiss, E. E. Hellstrom, S. Lee, C. W. Bark, C. B. Eom, M. Putti, I. Pallecchi, C. Marini, and P. Dore, Eur. Phys. J. B (accepted); arXiv:1003.0565v2.
\bibitem{Inosov2010b} D. S. Inosov, J. S. White, D. V. Evtushinsky, I. V. Morozov, A. Cameron, U. Stockert, V. B. Zabolotnyy, T. K. Kim, A. A. Kordyuk, S. V. Borisenko, E. M. Forgan, R. Klingeler, J. T. Park, S. Wurmehl, A. N. Vasiliev, G. Behr, C. D. Dewhurst, and V. Hinkov, Phys. Rev. Lett. {\bf 104}, 187001 (2010).
\bibitem{Kato2009} T. Kato, Y. Mizuguchi, H. Nakamura, T. Machida, H. Sakata, and Y. Takano, Phys. Rev. B {\bf 80}, 180507(R) (2009).
\bibitem{Nakayama2009a} K. Nakayama, T. Sato, P. Richard, T. Kawahara, Y. Sekiba, T. Qian, G. F. Chen, J. L. Luo, N. L. Wang, H. Ding, and T. Takahashi, arXiv:0907.0763.
\bibitem{Bendele2010} M. Bendele, S. Weyeneth, R. Puzniak, A. Maisuradze, E. Pomjakushina, K. Conder, V. Pomjakushin, H. Luetkens, S. Katrych, A. Wisniewski, R. Khasanov, and H. Keller, Phys. Rev. B {\bf 81}, 224520 (2010).
\bibitem{Terashima2009b} K Terashima, Y. Sekiba, J. H. Bowen, K. Nakayama, T. Kawahara, T. Sato, P. Richard, Y.-M. Xu, L. J. Li, G. H. Cao, Z.-A. Xu, H. Ding, and T. Takahashi, Proc. Nat. Acad. Sci. USA {\bf 106}, 7330 (2009).
\bibitem{Chi2009} S. Chi, A. Schneidewind, J. Zhao, L. W. Harriger, L. Li, Y. Luo, G. Cao, Z. Xu, M. Loewenhaupt, J. Hu, and P. Dai, Phys. Rev. Lett. {\bf 102}, 107006 (2009).
\bibitem{Kurita2009a} N. Kurita, F. Ronning, C. F. Miclea, E. D. Bauer, J. D. Thompson, A. S. Sefat, M. A. McGuire, B. C. Sales, D. Mandrus, and R. Movshovich, Phys. Rev. B {\bf 79}, 214439 (2009).
\bibitem{Tanatar2010} M. A. Tanatar, J.-Ph. Reid, H. Shakeripour, X. G. Luo, N. Doiron-Leyraud, N. Ni, S. L. Bud'ko, P. C. Canfield, R. Prozorov, and L. Taillefer, Phys. Rev. Lett. {\bf 104}, 067002 (2010).
\bibitem{Dong2010} J. K. Dong, S. Y. Zhou, T. Y. Guan, X. Qiu, C. Zhang, P. Cheng, L. Fang, H. H. Wen, and S. Y. Li, Phys. Rev. B {\bf 81}, 094520 (2010).
\bibitem{Machida2009} Y. Machida, K. Tomokuni, T. Isono, K. Izawa, Y. Nakajima, and T. Tamegai, J. Phys. Soc. Jpn. {\bf 78}, 073705 (2009).
\bibitem{Parker2009} D. Parker, M. G. Vavilov, A. V. Chubukov, and I. I. Mazin, Phys. Rev. B {\bf 80}, 100508(R) (2009).
\bibitem{Luo2009} X. G. Luo, M. A Tanatar, J.-Ph. Reid, H. Shakeripour, N. Doiron-Leyraud, N. Ni, S. L. Bud'ko, P. C. Canfield, H. Luo, Z. Wang, H.-H. Wen, R. Prozorov, and L. Taillefer, Phys. Rev. B {\bf 80}, 140503(R) (2009).
\bibitem{Checkelsky2008} J. G. Checkelsky, L. Li, G. F. Chen, J. L. Luo, N. L. Wang, and N. P. Ong, arXiv:0811.4668 (unpublished).
\bibitem{Ding2009} L. Ding, J. K. Dong, S. Y. Zhou, T. Y. Guan, X. Qiu, C. Zhang, L. J. Li, X. Lin, G. H. Cao, Z. A. Xu, and S. Y. Li, New J. Phys. {\bf 11}, 093018 (2009).
\bibitem{Dong2009} J. K. Dong, T. Y. Guan, S. Y. Zhou, X. Qiu, L. Ding, C. Zhang, U. Patel, Z. L. Xiao, and S. Y. Li, Phys. Rev. B {\bf 80}, 024518 (2009).
\bibitem{Reid2010} J.-Ph. Reid, M. A. Tanatar, X. G. Luo, H. Shakeripour, N. Doiron-Leyraud, N. Ni, S. L. Bud'ko, P. C. Canfield, R. Prozorov, and L. Taillefer, arXiv:1004.3804 (unpublished).
\bibitem{Budko2009} S. L. Bud'ko, N. Ni, and P. C. Canfield, Phys. Rev. B {\bf 79}, 220516(R) (2009).
\bibitem{Kogan2009a} V. G. Kogan, Phys. Rev. B {\bf 80}, 214532 (2009).
\bibitem{Kogan2010} V. G. Kogan, Phys. Rev. B {\bf 81}, 184528 (2010).
\bibitem{Korshunov2008} M. M. Korshunov and I. Eremin, Phys. Rev. B {\bf 78}, 140509(R) (2008).
\bibitem{Maier2008} T. A. Maier and D. J. Scalapino, Phys. Rev. B {\bf 78}, 020514(R) (2008).
\bibitem{Maier2009} T. A. Maier, S. Graser, D. J. Scalapino, and P. Hirschfeld, Phys. Rev. B {\bf 79}, 134520 (2009).
\bibitem{Wang2010} M. Wang, H. Luo, J. Zhao, C. Zhang, M. Wang, K. Marty, S. Chi, J. W. Lynn, A. Schneidewind, S. Li, and P. Dai, Phys. Rev. B {\bf 81}, 174524 (2010).
\bibitem{Li2009} S. Li, Y. Chen, S. Chang, J. W. Lynn, L. Li, Y. Luo, G. Cao, Z. Xu, and P. Dai, Phys. Rev. B {\bf 79}, 174527 (2009).
\bibitem{Christianson2008} A. D. Christianson, E. A. Goremychkin, R. Osborn, S. Rosenkranz, M. D. Lumsden, C. D. Malliakas, I. S. Todorov, H. Claus, D. Y. Chung, M. G. Kanatzidis, R. I. Bewley, and T. Guidi, Nature {\bf 456}, 930 (2008).
\bibitem{Qiu2009} Y. Qiu, W. Bao, Y. Zhao, C. Broholm, V. Stanev, Z. Tesanovic, Y. C. Gasparovic, S. Chang, J. Hu, B. Qian, M. Fang, and Z. Mao, Phys. Rev. Lett. {\bf 103}, 067008 (2009).
\bibitem{Bao2010} W. Bao, A. T. Savici, G. E. Granroth, C. Broholm, K. Habicht, Y. Qiu, J. Hu, T. Liu, and Z. Q. Mao, arXiv:1002.1617 (unpublished).
\bibitem{Shamoto2010} S.-I. Shamoto, M. Ishikado, A. D. Christianson, M. D. Lumsden, S. Wakimoto, K. Kodama, A. Iyo, and M. Arai, arXiv:1006.4640.
\bibitem{Hufner2008} S. H\"ufner, M. A. Hossain, A. Damascelli, and G. A. Sawatzky, Rep. Prog. Phys. {\bf 71}, 062501 (2008).
\bibitem{Eschrig2006} M. Eschrig, Adv. Phys. {\bf 55}, 47 (2006).
\bibitem{Lipscombe2010} O. J. Lipscombe, L. W. Harriger, P. G. Freeman, M. Enderle, C. Zhang, M. Wang, T. Egami, J. Hu, T. Xiang, M. R. Norman, and P. Dai, arXiv:1003.1926 (unpublished).
\bibitem{Wen2010} J. Wen, G. Xu, Z. Xu, Z. W. Lin, Q. Li, Y. Chen, S. Chi, G. Gu, and J. M. Tranquada, Phys. Rev. B {\bf 81}, 100513(R) (2010).
\bibitem{Zhao2010} J. Zhao, L.-P. Regnault, C. Zhang, M. Wang, Z. Li, F. Zhou, Z. Zhao, C. Fang, J. Hu, and P. Dai, Phys. Rev. B {\bf 81}, 180505(R) (2010).
\bibitem{Balatsky2006}  For a review, see A. V. Balatsky, I. Vekhter, and J.-X. Zhu, Rev. Mod. Phys. {\bf 78}, 374 (2006).
\bibitem{Sprunger1997} P. T. Sprunger, L. Petersen, E. W. Plummer, E. Laegsgaard, and F. Besenbacher, Science {\bf 275}, 1764 (1997).
\bibitem{Peterson1998} L. Petersen, P. T. Sprunger, Ph. Hofmann, E. Laegsgaard, B. G. Briner, M. Doering, H.-P. Rust, A. M. Bradshaw, F. Besenbacher, and E. W. Plummer, Phys. Rev. B {\bf 57}, R6858 (1998).
\bibitem{Wang2009b} F. Wang, H. Zhai, and D.-H. Lee, Europhys. Lett. {\bf 85}, 37005 (2009); Phys. Rev. B {\bf 81}, 184512 (2010).
\bibitem{Hanaguri2010} T. Hanaguri, S. Niitaka, K. Kuroki, and H. Takagi, Science {\bf 328}, 474 (2010).
\bibitem{Mazin2010} I. I. Mazin and D. J. Singh, arXiv:1007.0047v2.
\bibitem{Hanaguri2010a} T. Hanaguri, S. Niitaka, K. Kuroki, and H. Takagi, arXiv:1007.0307.
\bibitem{Pratt2009} D. K. Pratt, W. Tian, A. Kreyssig, J. L. Zarestky, S. Nandi, N. Ni, S. L. Bud'ko, P. C. Canfield, A. I Goldman, and R. J. McQueeney, Phys. Rev. Lett. {\bf 103}, 087001 (2009).
\bibitem{Fernandes2010} R. M. Fernandes, D. K. Pratt, W. Tian, J. Zarestky, A. Kreyssig, S. Nandi, M. G. Kim, A. Thaler, N. Ni, P. C. Canfield, R. J. McQueeney, J. Schmalian, and A. I. Goldman, Phys. Rev. B {\bf 81}, 140501(R) (2010).
\bibitem{Fernandes2010a} R. M. Fernandes ad J. Schmalian, Phys. Rev. B {\bf 82}, 014521 (2010).
\bibitem{Vorontsov2010} A. B. Vorontsov, M. G. Vavilov, and A. V. Chubukov, Phys. Rev. B {\bf 81}, 174538 (2010).
\bibitem{DParker2008} D. Parker, O. V. Dolgov, M. M. Korshunov, A. A. Golubov, and I. I. Mazin, Phys. Rev. B {\bf 78}, 134524 (2008).
\bibitem{Gordon2009} R. T. Gordon, N. Ni, C. Martin, M. A. Tanatar, M. D. Vannette, H. Kim, G. D. Samolyuk, J. Schmalian, S. Nandi, A. Kreyssig, A. I. Goldman, J. Q. Yan, S. L. Bud'ko, P. C. Canfield, and R. Prozorov, Phys. Rev. Lett. {\bf 102}, 127004 (2009).
\bibitem{Zeng2010} B. Zeng, G. Mu, B. Shen, P. Cheng, H. Luo, H. Yang, L. Shan, C. Ren, and H. H. Wen, arXiv:1006.2785.
\bibitem{Tinkham1975} M. Tinkham, \emph{Introduction to Superconductivity} (McGraw-Hill, New York, 1975), p.~61.
\bibitem{Prozorov2009} R. Prozorov, M. A. Tanatar, R. T. Gordon, C. Martin, H. Kim, V G. Kogan, N. Ni, M. E. Tillman, S. L. Bud'ko, and P. C. Canfield, Physica C {\bf 469}, 582 (2009).
\bibitem{Martin2009} C. Martin, M. E. Tillman, H. Kim, M. A. Tanatar, S. K. Kim, A. Kreyssig, R. T. Gordon, M. D. Vannette, S. Nandi, V. G. Kogan, S. L. Bud'ko, P. C. Canfield, A. I. Goldman, and R. Prozorov, Phys. Rev. Lett. {\bf 102}, 247002 (2009).
\bibitem{Hashimoto2009b} K. Hashimoto, T. Shibauchi, T. Kato, K. Ikada, R. Okazaki, H. Shishido, M. Ishikado, H. Kito, A. Iyo, H. Eisaki, S. Shamoto, and Y. Matsuda, Phys. Rev. Lett. {\bf 102}, 017002 (2009).
\bibitem{Corson1962} D. Corson and P. Lorrain, \emph{Introduction to Electromagnetic Fields and Waves} (W. H. Freeman, San Francisco, 1962).
\bibitem{Kim2010} H. Kim, C. Martin, R. T. Gordon, M. A. Tanatar, J. Hu, B. Qian, Z. Q. Mao, R. Hu, C. Petrovic, N. Salovich, R. Giannetta, and R. Prozorov, Phys. Rev. B {\bf 81}, 180503(R) (2010).
\bibitem{Biswas2010} P. K. Biswas, G. Balakrishnan, D. M. Paul, C. V. Tomy, M. R. Lees, and A. D. Hillier, Phys. Rev. B {\bf 81}, 092510 (2010).
\bibitem{Shermadini2010} Z. Shermadini, J. Kanter, C. Baines, M. Bendele, Z. Bukowski, R. Khasanov, H.-H. Klauss, H. Luetkens, H. Maeter, G. Pascua, B. Batlogg, and A. Amato, arXiv:1005.3989v3 (unpublished).
\bibitem{Li2008c} G. Li, W. Z. Hu, J. Dong, Z. Li, P. Zheng, G. F. Chen, J. L. Luo, and N. L. Wang, Phys. Rev. Lett. {\bf 101}, 107004 (2008).
\bibitem{Gordon2010} R. T. Gordon, H. Kim, N. Salovich, R. W. Giannetta, R. M. Fernandes, V. G. Kogan, T. Prozorov, S. L. Bud'ko, P. C. Canfield, M. A. Tanatar, and R. Prozorov, arXiv:1006.2068.
\bibitem{Williams2010} T. J. Williams, A. A. Aczel, E. Barrio-Saitovitch, S. L. Bud'ko, P. C. Canfield, J. P. Carlo, T. Goko, H. Kageyama, A. Kitada, J. Munevar, N. Ni, S. R. Saha, K. Kirschenbaum, J. Paglione, D. R. Sanchez-Candela, Y. J. Uemura, and G. M. Luke, arXiv:1005.2136 (unpublished).
\bibitem{Luetkens2008} H. Luetkens, H.-H. Klauss, R. Khasanov, A. Amato, R. Klingeler, I. Hellmann, N. Leps, A. Kondrat, C. Hess, A. K\"ohler, G. Behr, J. Werner, and B. B\"uchner, Phys. Rev. Lett. {\bf 101}, 097009 (2008).
\bibitem{Weyeneth2009} S. Weyeneth, R. Puzniak, U. Mosele, N. D. Zhigadlo, S. Katrych, Z. Bukowski, J. Karpinski, S. Kohout, J. Roos, and H. Keller, J. Supercond. Nov. Magn. {\bf 22}, 325 (2009).
\bibitem{Drew2008} A. J. Drew, F. L. Pratt, T. Lancaster, S. J. Blundell, P. J. Baker, R. H. Liu, G Wu, X. H. Chen, I. Watanabe, V. K. Malik, A. Dubroka, K. W. Kim, M. R\"ossle, and C. Bernhard, Phys. Rev. Lett. {\bf 101}, 097010 (2008).
\bibitem{Inosov2010a} D. S. Inosov, J. S. White, D. V. Evtushinsky, I. V. Morozov, A. Cameron, U. Stockert, V. B. Zabolotnyy, T. K. Kim, A. A. Kordyuk, S. V. Borisenko, E. M. Forgan, R. Klingeler, J. T. Park, S. Wurmehl, A. N. Vasiliev, G. Behr, C. D. Dewhurst, and V. Hinkov, Phys. Rev. Lett. {\bf 104}, 187001 (2010).
\bibitem{Gordon_Fig1_0912.5346} R. T. Gordon, H. Kim, M. A. Tanatar, R. Prozorov, and V. G. Kogan, Phys. Rev. B {\bf 81}, 180501(R) (2010).
\bibitem{Hashimoto2009} K. Hashimoto, T. Shibauchi, S. Kasahara, K. Ikada, S. Tonegawa, T. Kato, R. Okazaki, C. J. van der Beek, M. Konczykowski, H. Takeya, K. Hirata, T. Terashima, and Y. Matsuda, Phys. Rev. Lett. {\bf 102}, 207001 (2009); T. Shibauchi, K. Hashimoto, R. Okazaki, and Y. Matsuda, Physica C {\bf 469}, 590 (2009).
\bibitem{Halbritter1971} J. Halbritter, Z. Phys. {\bf 243}, 201 (1971).
\bibitem{Dressel2010} M. Dressel, D. Wu, N. Bari\v{s}i\'c, and B. Gorshunov, arXiv:1004.2962 (unpublished).
\bibitem{Uemura1991} Y. J. Uemura, L. P. Le, G. M. Luke, B. J. Sternlieb, W. D. Wu, J. H. Brewer, T. M. Riseman, C. L. Seaman, M. B. Maple, M. Ishikawa, D. G. Hinks, J. D. Jorgensen, G. Saito, and H. Yamochi, Phys. Rev. Lett. {\bf 66}, 2665 (1991); Erratum {\bf 68}, 2712 (1992).
\bibitem{Homes2004} C. C. Homes, S. V. Dordevic, M. Strongin, D. A. Bonn, R. Liang, W. N. Hardy, S. Komiya, Y. Ando, G. Yu, N. Kaneko, X. Zhao, M. Greven, D. N. Basov, and T. Timusk, Nature {\bf 430}, 539 (2004).
\bibitem{Homes2005} C. C. Homes, S. V. Dordevic, T. Valla, and M. Strongin, Phys. Rev. B {\bf 72}, 134517 (2005).
\bibitem{Chubukov2008} A. V. Chubukov, D. V. Efremov, and I. Eremin, Phys. Rev. B {\bf 78}, 134512 (2008).  See also cited references.
\bibitem{Bang2009} Y. Bang, H.-Y. Choi, and H. Won, Phys. Rev. B {\bf 79}, 054529 (2009); Y. Bang, Europhys. Lett. {\bf 86}, 47001 (2009).
\bibitem{Dolgov2009} O. V. Dolgov, A. A. Golubov, and D. Parker, New J. Phys. {\bf 11}, 075012 (2009).
\bibitem{Vorontsov2009} A. B. Vorontsov, M. G. Vavilov, and A. V. Chubukov, Phys. Rev. B {\bf 79}, 140507(R) (2009).
\bibitem{Onari2009} S. Onari and H. Kontani, Phys. Rev. Lett. {\bf 103}, 177001 (2009).
\bibitem{Matsumoto2009} M. Matsumoto, M. Koga, and H. Kusunose, J. Phys. Soc. Jpn. {\bf 78}, 084718 (2009).
\bibitem{JLi2009} J. Li and Y. Wang, Europhys. Lett. {\bf 88}, 17009 (2009).
\bibitem{Tabuchi2010} T. Tabuchi, Z. Li, T. Oka, G. F. Chen, S. Kawasaki, J. L. Luo, N. L. Wang, and G.-Q. Zheng, Phys. Rev. B {\bf 81}, 140509(R) (2010).
\bibitem{Jaroszynski2008} J. Jaroszynski, F. Hunte, L. Balicas, Y.-J. Jo, I. Rai\v{c}evi\'c, A. Gurevich, D. C. Larbalestier, F. F. Balakirev, L. Fang, P. Cheng, Y. Jia, Nnd H. H. Wen, Phys. Rev. B {\bf 78} 174523 (2008).
\bibitem{Altarawneh2008} M. M. Altarawneh, K. Collar, C. H. Mielke, N. Ni, S. L. Bud'ko, and P. C. Canfield, Phys Rev. B {\bf 78}, 220505(R) (2008).
\bibitem{Yuan2009} H. Q. Yuan, J. Singleton, F. F. Balakirev, S. A. Baily, G. F. Chen, J. L. Luo, and N. L. Wang, Nature {\bf 457}, 565 (2009).
\bibitem{Kano2009} M. Kano, Y. Kohama, D. Graf, F. Balakirev, A. S. Sefat, M. A. McGuire, B. C. Sales, D. Mandrus, and S. W. Tozer, J. Phys. Soc. Jpn. {\bf 78}, 084719 (2009).
\bibitem{Yamamoto2009} A. Yamamoto, J. Jaroszynski, C. Tarantini, L. Balicas, J. Jiang, A. Gurevich, D. C. Larbalestier, R. Jin, A. S. Sefat, M. A. McGuire, B. C. Sales, D. K. Christen, and D. Mandrus, Appl. Phys. Lett. {\bf 94}, 062511 (2009).
\bibitem{Terashima2009a} T. Terashima, M. Kimata, H. Satsukawa, A. Harada, K. Hazama, S. Uji, H. Harima, G.-F. Chen, J.-L. Luo, and N.-L. Wang, J. Phys. Soc. Jpn. {\bf 78}, 063702 (2009).
\bibitem{Kida2010} T. Kida, M. Kotani, Y. Mizuguchi, Y. Takano, and M. Hagiwara, J. Phys. Soc. Jpn. {\bf 79}, 074706 (2010).
\bibitem{Fang2010} M. Fang, J. Yang, F. F. Balakirev, Y. Kohama, J. Singleton, B. Qian, Z. Q. Mao, H. Wang, and H. Q. Yuan, Phys. Rev. B {\bf 81}, 020509(R) (2010).
\bibitem{Khim2010} S. Khim, J. W. Kim, E. S. Choi, Y. Bang, M. Nohara, H. Takagi, and K. H. Kim, Phys. Rev. B {\bf 81}, 184511 (2010).
\bibitem{Lei2010} H. Lei, R. Hu, E. S. Choi, J. B. Warren, and C. Petrovic, Phys. Rev. B {\bf 81}, 184522 (2010).
\bibitem{Braithwaite2010} D. Braithwaite, G. Lapertot, W. Knafo, and I. Sheikin, J. Phys. Soc. Jpn. {\bf 79}, 053703 (2010).
\bibitem{Mu2008} G. Mu, H. Luo, Z. Wang, L. Shan, C. Ren, and H.-H. Wen, Phys. Rev. B {\bf 79}, 174501 (2009).
\bibitem{Rotter2009} M. Rotter, M. Tegel, I. Schellenberg, F. M. Schappacher, R. P\"ottgen, J. Deisenhofer, A. G\"unther, F. Schrettle, A. Loidl, and D. Johrendt, New J. Phys. {\bf 11}, 025014 (2009).
\bibitem{Welp2009} U. Welp, R. Xie, A. E. Koshelev, W. K. Kwok, H. Q. Luo, Z. S. Wang, G. Mu, and H. H. Wen, Phys. Rev. B {\bf 79}, 094505 (2009).
\bibitem{Shein2008c} I. R. Shein and A. L. Ivanovskii, JETP Lett. {\bf 88}, 107 (2008).
\bibitem{Fukazawa2009} H. Fukazawa, Y. Yamada, K. Kondo, T. Saito, Y. Kohori, K. Kuga, Y. Matsumoto, S. Nakatsuji, H. Kito, P. M. Shirage, K. Kihou, N. Takeshita, C.-H. Lee, A. Iyo, and H. Eisaki, J. Phys. Soc. Jpn. {\bf 78}, 083712 (2009).
\bibitem{Wu2010a} D. Wu, N. Bari\v{s}i\'c, M. Dressel, G. H. Cao, Z.-A. Xu, E. Schachinger, and J. P. Carbotte, arXiv:1006.5468.
\bibitem{Boeri2008} L. Boeri, O. V. Dolgov, and A. A. Golubov, Phys. Rev. Lett. {\bf 101}, 026403 (2008); Physica C {\bf 469}, 628 (2009).
\bibitem{Jeevan2010} H. S. Jeevan and P. Gegenwart, Proc. ICM2009, J. Phys.: Conf. Series {\bf 200}, 012060 (2010).
\bibitem{Storey2010} J. G. Storey, J. W. Loram, J. R. Cooper, Z. Bukowski, and J. Karpinski, arXiv:1001.0474 (unpublished).
\bibitem{Zaanen2009} J. Zaanen, Phys. Rev. B {\bf 80}, 212502 (2009).
\bibitem{Gofryk2010} K. Gofryk, A. S. Sefat, M. A. McGuire, B. C. Sales, D. Mandrus, J. D. Thompson, E. D. Bauer, and F. Ronning, Phys. Rev. B {\bf 81}, 184518 (2010).
\bibitem{Mu2009} G. Mu, B. Zeng, P. Cheng, Z.-S. Wang, L. Fang, B. Shen, L. Shan, C. Ren, and H.-H. Wen, Chin. Phys. Lett. {\bf 27}, 037402 (2010).
\bibitem{Bang2010} Y. Bang, Phys. Rev. Lett. {\bf 104}, 217001 (2010).
\bibitem{Massee2008} F. Massee, Y. Huang, R. Huisman, S. de Jong, J. B. Goedkoop, and M. S. Golden, and M. S. Golden, Phys. Rev. B {\bf 79}, 220517(R) (2009).
\bibitem{Yin2009a}  For a review of STM/STS measurements on Fe-based superconductors, see Y. Yin, M. Zech, T. L. Williams, and J. E. Hoffman, Physica C {\bf 469}, 535 (2009).
\bibitem{Yin2009} Y. Yin, M. Zech, T. L. Williams, X. F. Wang, G. Wu, X. H. Chen, and J. E. Hoffman, Phys. Rev. Lett. {\bf 102}, 097002 (2009).
\bibitem{Pan1998} S. H. Pan, E. W. Hudson, and J. C. Davis, Appl. Phys. Lett. {\bf 73}, 2\,992 (1998).
\bibitem{Kim2009} T.-H. Kim, R. Jin, L. R. Walker, J. Y. Howe, M. H. Pan, J. F. Wendelken, J. R. Thompson, A. S. Sefat, M. A. McGuire, B. C. Sales, D. Mandrus, and A. P. Li, Phys. Rev. B {\bf 80}, 214518 (2009).
\bibitem{Knolle2010a} J. Knolle, I. Eremin, A. Akbari, and R. Moessner, Phys. Rev. Lett. {\bf 104}, 257001 (2010).
\bibitem{Kumar2009} N. Kumar, R. Nagalakshmi, R. Kulkarni, P. L. Paulose, A. K. Nigam, S. K. Dhar, and A. Thamizhavel, Phys. Rev. B {\bf 79}, 012504 (2009).
\bibitem{Matusiak2010} M. Matusiak, Z. Bukowski, and J. Karpinski, Phys. Rev. B {\bf 81}, 020510(R) (2010).
\bibitem{Pramanik2010} A. K. Pramanik, L. Harnagea, S. Singh, S. Aswartham, G. Behr, S. Wurmehl, C. Hess, R. Klingeler, and B. B\"uchner, Phys. Rev. B {\bf 82}, 014503 (2010).
\bibitem{Ng2009} T.-K. Ng, Phys. Rev. Lett. {\bf 103}, 236402 (2009).
\bibitem{Ng2009b} T. K. Ng and Y. Avishai, Phys. Rev. B {\bf 80}, 104504 (2009).
\bibitem{Fletcher2009} J. D. Fletcher, A. Serafin, L. Malone, J. G. Analytis, J.-H. Chu, A. S. Erickson, I. R. Fisher, and A. Carrington, Phys. Rev. Lett. {\bf 102}, 147001 (2009); C. W. Hicks, T. M. Lippman, M. E. Huber, J. G. Analytis, J.-H. Chu, A. S. Erickson, I. R. Fisher, and K. A. Moler, Phys. Rev. Lett. {\bf 103}, 127003 (2009).
\bibitem{Hicks2009} C. W. Hicks, T. M. Lippman, M. E. Huber, J. G. Analytis, J.-H. Chu, A. S. Erickson, I. R. Fisher, and K. A. Moler, Phys. Rev. Lett. {\bf 103}, 127003 (2009).
\bibitem{Yamashita2009} M. Yamashita, N. Nakata, Y. Senshu, S. Tonegawa, K. Ikada, K. Hashimoto, H. Sugawara, T. Shibauchi, and Y. Matsuda, Phys. Rev. B {\bf 80}, 220509(R) (2009).
\bibitem{Hashimoto2009a} K. Hashimoto, M. Yamashita, S. Kasahara, Y. Senshu, N. Nakata, S. Tonegawa, K. Ikada, A. Serafin, A. Carrington, T. Terashima, H. Ikeda, T. Shibauchi, and Y. Matsuda, Phys. Rev. B {\bf 81}, 220501(R) (2010).
\bibitem{Nakai2010} Y. Nakai, T. Iye, S. Kitagawa, K. Ishida, S. Kasahara, T. Shibauchi, Y. Matsuda, and T. Terashima, Phys. Rev. B {\bf 81}, 020503(R) (2010).
\bibitem{Subedi2008} A. Subedi, D. J. Singh, and M.-H. Du, Phys. Rev. B {\bf 78}, 060506(R) (2008).
\bibitem{Chia2010}  E. E. M. Chia, D. Talbayev, J.-X. Zhu, H. Q. Yuan, T. Park, J. D. Thompson, C. Panagopoulos, G. F. Chen, J. L. Luo, N. L. Wang, and A. J. Taylor, Phys. Rev. Lett. {\bf 104}, 027003 (2010).
\bibitem{Mertelj2010}  T. Mertelj, P. Kusar, V. V. Kabanov, L. Stojchevska, N. D. Zhigadlo, S. Katrych, Z. Bukowski, J. Karpinski, S. Weyeneth, and D. Mihailovic, Phys. Rev. B {\bf 81}, 224504 (2010).
\bibitem{Stojchevska2010}  L. Stojchevska, P. Kusar, T. Mertelj, V. V. Kabanov, X. Lin, G. H. Cao, Z. A. Xu, and D. Mihailovic, Phys. Rev. B {\bf 82}, 012505 (2010).
\bibitem{Mansart2010}  B. Mansart, D. Boschetto, A. Savoia, F. Rullier-Albenque, F. Bouquet, E. Papalazarou, A. Forget, D. Colson, A. Rousse, and M. Marsi, Phys. Rev. B {\bf 82}, 024513 (2010).
\bibitem{Kulic2009} M. L. Kuli\'c and A. A. Haghighirad, Europhys. Lett. {\bf 87}, 17007 (2009).
\bibitem{Yndurain2009} F. Yndurain and J. M. Soler, Phys. Rev. B {\bf 79}, 134506 (2009).
\bibitem{Reynolds1950} E. Maxwell, Phys. Rev. {\bf 78}, 477 (1950); C. A. Reynolds, B. Serin, W. H. Wright, and L. B. Nesbitt, Phys. Rev. {\bf 78}, 487 (1950).
\bibitem{RHLiu2009} R. H. Liu, T Wu, G. Wu, H. Chen, X. F. Wang, Y. L. Xie, J. J. Ying, Y. J. Yan, Q. J. Li, B. C. Shi, W. S. Chu, Z. Y. Wu, and X. H. Chen, Nature {\bf 459}, 64 (2009).
\bibitem{Shirage2010} P. M. Shirage, K. Miyazawa, K. Kihou, H. Kito, Y. Yoshida, Y. Tanaka, H. Eisaki, and A. Iyo, Phys. Rev. Lett. {\bf 105}, 037004 (2010).
\bibitem{Shirage2009} P. M. Shirage, K. Kihou, K. Miyazawa, C.-H. Lee, H. Kito, H. Eisaki, T. Yanagisawa, Y. Tanaka, and A. Iyo, Phys. Rev. Lett. {\bf 103}, 257003 (2009).
\bibitem{Cordero2008} B. Cordero, V. G\'omez, A. E. Platero-Prats, M. Rev\'es, J. Echeverr\'ia, E. Cremades, F. Barrag\'an, and S. Alvarez, Dalton Trans. {\bf 2008}, 2832 (2008).
\bibitem{DJSingh2009} D. J. Singh, A. S. Sefat, M. A. McGuire, B. C. Sales, D. Mandrus, L. H. VanBebber, and V. Keppens, Phys. Rev. B {\bf 79}, 094429 (2009).
\bibitem{Kimber2010a} S. A. J. Kimber, A. H. Hill, Y.-Z. Zhang, H. O. Jeschke, R. Valent\'i, C. Ritter, I. Schellenberg, W. Hermes, R. P\"ottgen, and D. N. Argyriou, arXiv:1006.0341.
\bibitem{Brechtel1979} E. Brechtel, G. Cordier, and H. Sch\"afer, Z. Naturforsch. {\bf 34b}, 777 (1979).
\bibitem{Nath2010} R. Nath, V. O. Garlea, A. I. Goldman, and D. C. Johnston, Phys. Rev. B {\bf 81}, 224513 (2010).
\bibitem{Brock1996} S. L. Brock, N. P. Raju, J. E. Greedan, and S. M. Kauzlarich, J. Alloys Compd. {\bf 237}, 9 (1996).
\bibitem{Brock1996a} S. L. Brock and S. M. Kauzlarich, J. Alloys Compd. {\bf 241}, 82 (1996).
\bibitem{Wang2010a} X. F. Wang, Y. J. Yan, J. J. Ying, Q. J. Li, M. Zhang, N. Xu, and X. H. Chen, J. Phys.: Condens. Matter {\bf 22}, 075702 (2010).

%**************************************
% Appendix References

\bibitem{nomura2008} T. Nomura, S. W. Kim, Y. Kamihara, M. Hirano, P. V. Sushko, K. Kato, M. Takata, A. L. Shluger, and H. Hosono, Supercond. Sci. Technol. \textbf{21}, 125028 (2008).
\bibitem{Sefat2008h} A. S. Sefat, A. Huq, M. A. McGuire, R. Jin, B. C. Sales, D. Mandrus, L. M. D. Cranswick, P. W. Stephens, and K. H. Stone, Phys. Rev. B {\bf 78}, 104505 (2008).
\bibitem{sefat2008b} A. S. Sefat, M. A. McGuire, B. C. Sales, R. Jin, J. Y. Howe, and D. Mandrus, Phys. Rev. B {\bf 77}, 174503 (2008).
\bibitem{Qiu2008b} Y. Qiu, M. Kofu, W. Bao, S.-H. Lee, Q. Huang, T. Yildirim, J. R. D. Copley, J. W. Lynn, T. Wu, G. Wu, and X. H. Chen, Phys. Rev. B {\bf 78}, 052508 (2008).
\bibitem{quebe2000} P. Quebe, L. Terbuchte, and W. Jeitschko,
J. Alloys Compd. \textbf{302}, 70 (2000).
\bibitem{Martinelli2009} A. Martinelli, A. Palenzona, C. Ferdeghini, M. Putti, and H. Emerich, J. Alloys Compd. {\bf 477}, L21 (2009).
\bibitem{Martinelli2008} A. Martinelli, M. Ferretti, P. Manfrinetti, A. Palenzona, M. Tropeano, M. R. Cimberle, C. Ferdeghini, R. Valle, C. Bernini, M. Putti, and A. S. Siri, Supercond. Sci. Technol. {\bf 21}, 095017 (2008).
\bibitem{Zhigadlo2008} N. D. Zhigadlo, S. Katrych, Z. Bukowski, S. Weyeneth, R. Puzniak, and J. Karpinski, J. Phys.: Condens. Matter {\bf 20}, 342202 (2008).
\bibitem{Ju2010} J. Ju, K. Huynh, J. Tang, Z. Li, M. Watahiki, K. Sato, H. Terasaki, E. Ohtani, H. Takizawa, and K. Tanigaki, J. Phys. Chem. Solids {\bf 71}, 491 (2010).
\bibitem{Zhigadlo2010} N. D. Zhigadlo, S. Katrych, S. Weyeneth, R. Puzniak, P. Moll, Z. Bukowski, J. Karpinski, H. Keller, and B. Batlogg, arXiv:1006.0126 (unpublished).
\bibitem{Bos2008} J.-W. G. Bos, G. B. S. Penny, J. A. Rodgers, D. A. Sokolov, A. D. Huxley, and J. P. Attfield, Chem. Commun. {\bf 2008}, 3634 (2008).
\bibitem{McQueen2008} T. M. McQueen, M. Regulacio, A. J. Williams, Q. Huang, J. W. Lynn, Y. S. Hor, D. V. West, M. A. Green, and R. J. Cava, Phys. Rev. B {\bf 78}, 024521 (2008).
\bibitem{Tegel2008c} M. Tegel, I. Schellenberg, R. P\"ottgen, and D. Johrendt, Z. Naturforsch. {\bf 63b}, 1057 (2008).
\bibitem{Zimmer1995} B. I. Zimmer, W. Jeitschko, J. H. Albering, R. Glaum, and M. Reehuis, J. Alloys Compds. {\bf 229}, 238 (1995).
\bibitem{Kamihara2008c} Y. Kamihara, H. Hiramatsu, M. Hirano, Y. Kobayashi, S. Kitao, S. Higashitaniguchi, Y. Yoda, M. Seto, and H. Hosono, Phys. Rev. B {\bf 78}, 184512 (2008).
\bibitem{Nomura2008a} T. Nomura, Y. Inoue, S. Matsuishi, M. Hirano, J. E. Kim, K. Kato, M. Takata, and H. Hosono, Supercond. Sci. Technol. {\bf 22}, 055016 (2009).
\bibitem{Tegel2008b} M. Tegel, S. Johansson, V. Weiss, I. Schellenberg, W. Hermes, R. P\"ottgen, and D. Johrendt, Europhys. Lett. {\bf 84}, 67007 (2008).
\bibitem{Matsuishi2008} S. Matsuishi, Y. Inoue, T. Nomura, M. Hirano, and H. Hosono, J. Phys. Soc. Jpn. {\bf 77}, 113709 (2008).
\bibitem{Watanabe2008} T. Watanabe, H. Yanagi, Y. Kamihara, T. Kamiya, M. Hirano, and H. Hosono, J. Solid State Chem. {\bf 181}, 2117 (2008).
\bibitem{Tegel2008} M. Tegel, D. Bichler, and D. Johrendt, Solid State Sci. {\bf 10}, 193 (2008).
\bibitem{McQueen2009} T. M. McQueen, T. Klimczuk, A. J. Williams, Q. Huang, and R. J. Cava, Phys. Rev. B {\bf 79}, 172502 (2009).
\bibitem{Sakai2010} H. Sakai, N. Tateiwa, T. D. Matsuda, T. Sugai, E. Yamamoto, and Y. Haga, J. Phys. Soc. Jpn. {\bf 79}, 074721 (2010).
\bibitem{pfisterer1980} M. Pfisterer and G. Nagorsen,
Z. Naturforsch. \textbf{35b}, 703 (1980); \textbf{38b}, 811 (1983).
\bibitem{Rotter2008c} M. Rotter, M. Pangerl, M. Tegel, and D. Johrendt, Angew. Chem. Int. Ed. {\bf 47}, 7949 (2008).
\bibitem{wu2008} G. Wu, H. Chen, Y. L. Xie, Y. J. Yan,
R. H. Liu, X. F. Wang, J. J. Ying, and X. H. Chen,
J. Phys.: Condens. Matter \textbf{20}, 422201 (2008).
\bibitem{Tegel2008d} M. Tegel, M. Rotter, V. Wei\ss, F. M. Schappacher, R. P\"ottgen, and D. Johrendt, J. Phys.: Condens. Matter {\bf 20}, 452201 (2008).
\bibitem{Saha2010} S. R. Saha, K. Kirshenbaum, N. P. Butch, J. Paglione, and P. Y. Zavalij, arXiv:1006.2147.
\bibitem{Leithe-Jasper2008} A. Leithe-Jasper, W. Schnelle, C. Geibel, and H. Rosner, Phys. Rev. Lett. {\bf 101}, 207004 (2008).
\bibitem{rotter2008b} M. Rotter, M. Tegel, D. Johrendt,
I. Schellenberg, W. Hermes, and R. P\"{o}ttgen,
Phys. Rev. B \textbf{78}, 020503(R) (2008).
\bibitem{Tegel2009} M. Tegel, I. Schellenberg, F. Hummel, R. P\"ottgen, and D. Johrendt, Z. Naturforsch. {\bf 64b}, 815 (2009).
\bibitem{Sefat2009} A. S. Sefat, D. J. Singh, R. Jin, M. A. McGuire, B. C. Sales, and D. Mandrus, Phys. Rev. B {\bf 79}, 024512 (2009).
\bibitem{ronning2008} F. Ronning, N. Kurita, E. D. Bauer,
B. L. Scott, T. Park, T. Klimczuk, R. Movshovich, and
J. D. Thompson, J. Phys.: Condens. Matter \textbf{20},
342203 (2008).
\bibitem{Bauer2008} E. D. Bauer, F. Ronning, B. L. Scott, and J. D. Thompson, Phys. Rev. B {\bf 78}, 172504 (2008).
\bibitem{Mewis1980} A. Mewis, Z. Naturforsch. {\bf 35b}, 141 (1980).
\bibitem{Feng2010} C. Feng, Z. Ren, S. Xu, Z. Xu, G. Cao, I. Nowik, I. Felner, K. Matsubayashi, and Y. Uwatoko, arXiv:1005.0516 (unpublished).
\bibitem{Keimes1997} V. Keimes, D. Johrendt, A. Mewis, C. Huhnt, and W. Schlabitz, Z. Anorg. Allg. Chem. {\bf 623}, 1699 (1997).
\bibitem{Gresty2009} N. C. Gresty, Y. Takabayashi, A. Y. Ganin, M. T. McDonald, J. B. Claridge, D. Giap, Y. Mizuguchi, Y. Takano, T. Kagayama, Y. Ohishi, M. Takata, M. J. Rosseinsky, S. Margadonna, and K. Prassides, J. Am. Chem. Soc. {\bf 131}, 16944 (2009).
\bibitem{Deng2009} Z. Deng, X. C. Wang, Q. Q. Liu, S. J. Zhang, Y. X. Lv, J. L. Zhu, R. C. Yu, and C. Q. Jin, Europhys. Lett. {\bf 87}, 37004 (2009).
\bibitem{kohama2008} Y. Kohama, Y. Kamihara, M. Hirano, H. Kawaji, T. Atake, and H. Hosono, Phys. Rev. B \textbf{78}, 020512 (2008).
\bibitem{Xu2008} G. Xu, W. Ming, Y. Yao, X. Dai, S.-C. Zhang, and Z. Fang, Europhys. Lett. {\bf 82}, 67002 (2008).
\bibitem{Li2008} H. Li, J. Li, S. Zhang, W. Chu, D. Chen, and Z. Wu, arXiv:0807.3153v2 (unpublished).
\bibitem{klauss2008}  H.-H. Klauss, H. Luetkens, R. Klingeler,
C. Hess, F. J. Litterst, M. Kraken, M. M. Korshunov, I. Eremin,
S.-L. Drechsler, R. Khasanov, A. Amato, J. Hamann-Borrero,
N. Leps, A. Kondrat, G. Behr, J. Werner, and B. B\"uchner,
Phys. Rev. Lett. \textbf{101}, 077005 (2008).
\bibitem{Nekrasov2008a}I. A. Nekrasov, Z. V. Pchelkina, and M. V. Sadovskii, JETP Lett. {\bf 88}, 679 (2008).
\bibitem{Nekrasov2008c}I. A. Nekrasov, Z. V. Pchelkina, and M. V. Sadovskii, JETP Lett. {\bf 87}, 560 (2008).
\bibitem{Larson2009} P. Larson and S. Satpathy, Phys. Rev. B {\bf 79}, 054502 (2009).
\bibitem{Gang2008} G. Mu, X.-Y. Zhu, L. Fang, L. Shan, C. Ren, and H.-H. Wen, Chin. Phys. Lett. {\bf 25}, 2221 (2008).
\bibitem{Wang2008b} C. Wang, Y. K. Li, Z. W. Zhu, S. Jiang, X. Lin, Y. K. Luo, S. Chi, L. J. Li, Z. Ren, M. He, H. Chen, Y. T. Wang, Q. Tao, G. H. Cao, and Z. A. Xu, Phys. Rev. B {\bf 79}, 054521 (2009).
\bibitem{Han2008} F. Han, X. Zhu, G. Mu, P. Cheng, and H.-H. Wen, Phys. Rev. B {\bf 78}, 180503(R) (2008).
\bibitem{Shein2008} I. R. Shein and A. L. Ivanovskii, JETP Lett. {\bf 88}, 683 (2008).
\bibitem{Baker2009} P. J. Baker, I. Franke, T. Lancaster, S. J. Blundell, L. Kerslake, and S. J. Clarke, Phys. Rev. B {\bf 79}, 060402(R) (2009).
\bibitem{Matsuishi2008a} S. Matsuishi, Y. Inoue, T. Nomura, Y. Kamihara, M. Hirano, and H. Hosono, New J. Phys. {\bf 11}, 025012 (2009).
\bibitem{Kamihara2008b} Y. Kamihara, M. Hirano, H. Yanagi, T. Kamiya, Y. Saitoh, E. Ikenaga, K. Kobayashi, and H. Hosono, Phys. Rev. B {\bf 77}, 214515 (2008).
\bibitem{Kamihara2008} Y. Kamihara, H. Hiramatsu, M. Hirano, H. Yanagi, T. Kamiya, and H. Hosono, J. Phys. Chem. Solids {\bf 69}, 2916 (2008).
\bibitem{Lebegue2009} S. Leb\`egue, Z. P. Yin, and W. E. Pickett, New J. Phys. {\bf 11}, 025004 (2009).
\bibitem{Li2008b} Z. Li, G. Chen, J. Dong, G. Li, W. Hu, D. Wu, S. Su, P. Zheng, T. Xiang, N. Wang, and J. Luo, Phys. Rev. B {\bf 78}, 060504(R) (2008).
\bibitem{Zhang2008} W.-B. Zhang, X.-B. Xiao, W.-Y. Yu, N. Wang, and B.-Y. Tang, Phys. Rev. B {\bf 77}, 214513 (2008); W.-B. Zhang, private communication.
\bibitem{Dong2008a} J. K. Dong, L. Ding, H. Wang, X. F. Wang, T. Wu, G. Wu, X. H. Chen, and S. Y. Li, New J. Phys. {\bf 10}, 123031 (2008).
\bibitem{Wu2008} G. Wu, H. Chen, T. Wu, Y. L. Xie, Y. J. Yan, R. H. Liu, X. F. Wang, J. J. Ying, and X. H. Chen, J. Phys.: Condens. Matter {\bf 20}, 422201 (2008).
\bibitem{Ronning2008} F. Ronning, T. Klimczuk, E. D. Bauer, H. Volz, and J. D. Thompson, J. Phys.: Condens. Matter {\bf 20}, 322201 (2008).
\bibitem{Sefat2009b} A. S. Sefat, D. J. Singh, R. Jin, M. A. McGuire, B. C. Sales, F. Ronning, and D. Mandrus, Physica C {\bf 469}, 350 (2009).
\bibitem{Nekrasov2008} I. A. Nekrasov, Z. V. Pchelkina, and M. V. Sadovskii, JETP Lett. {\bf 88}, 144 (2008).
\bibitem{Sun2009} G. L. Sun, D. L. Sun, M. Konuma, P. Popovich, A. Boris, J. B. Peng, K.-Y. Choi, P. Lemmens, and C. T. Lin, arXiv:0901.2728v3 (unpublished).
\bibitem{Sefat2008g} A. S. Sefat, R. Jin, M. A. McGuire, B. C. Sales, D. J. Singh, and D. Mandrus, Phys. Rev. Lett. {\bf 101}, 117004 (2008).
\bibitem{yan2008} J.-Q. Yan, A. Kreyssig, S. Nandi, N. Ni,
S. L. Budko, A. Kracher, R. J. McQueeney, R. W. McCallum,
T. A. Lograsso, A. I. Goldman, and P. C. Canfield,
Phys. Rev. B \textbf{78}, 024516 (2008).
\bibitem{Leith-Jasper2008} A. Leithe-Jasper, W. Schnelle, C. Geibel, and H. Rosner, Phys. Rev. Lett. {\bf 101}, 207004 (2008).
\bibitem{LZhang2009} L. Zhang and D. J. Singh, Phys. Rev. B {\bf 79}, 174530 (2009).
\bibitem{Curro2009} N. J. Curro, A. P. Dioguardi, N. ApRoberts-Warren, A. C. Shockley, and P. Klavins, New J. Phys. {\bf 11}, 075004 (2009).
\bibitem{Jiang2009} S. Jiang, Y. Luo, Z. Ren, Z. Zhu, C. Wang, X. Xu, Q. Tao, G. Cao, and Z. Xu, New J. Phys. {\bf 11}, 025007 (2009).
\bibitem{jeevan2008a} H. S. Jeevan, Z. Hossain, D. Kasinathan,
H. Rosner, C. Geibel, and P. Gegenwart, Phys. Rev. B
\textbf{78}, 092406 (2008).
\bibitem{Parker2008} D. R. Parker, M. J. Pitcher, P. J. Baker, I. Franke, T. Lancaster, S. J. Blundell, and S. J. Clarke, Chem. Commun. {\bf 2009}, 2189 (2009).
\bibitem{Jishi2008} R. A. Jishi and H. M. Alyahyaei, Adv. Cond. Mat. Phys. {\bf 2010}, 804343 (2010).
\bibitem{Nekrasov2008b} I. A. Nekrasov, Z. V. Pchelkina, and M. V. Sadovskii, JETP Lett. {\bf 88}, 543 (2008).
\bibitem{Shein2009a} I. R. Shein and A. L. Ivanovskii, Solid State Commun. {\bf 149}, 1860 (2009).
\bibitem{Shein2009} I. R. Shein and A. L. Ivanovskii, Phys. Rev. B {\bf 79}, 054510 (2009).
\bibitem{Ronning2008b} F. Ronning, N. Kurita, E. D. Bauer, B. L. Scott, T. Park, T. Klimczuk, R. Movshovich, and J. D. Thompson, J. Phys.: Condens. Matter {\bf 20}, 342203 (2008).
\bibitem{Kurita2009} N. Kurita, F. Ronning, Y. Tokiwa, E. D. Bauer, A. Subedi, D. J. Singh, J. D. Thompson, and R. Movshovich, Phys. Rev. Lett. {\bf 102}, 147004 (2009).
\bibitem{jeitschko1987} W. Jeitschko, R. Glaum, and
L. Boonk, J. Solid State Chem. \textbf{69}, 93 (1987).
\bibitem{Berry2009} N. Berry, C. Capan, G. Seyfarth, A. D. Bianchi, J. Ziller, and Z. Fisk, Phys. Rev. B {\bf 79}, 180502(R) (2009).
\bibitem{Ronning2009} F. Ronning, E. D. Bauer, T. Park, S.-H. Baek, H. Sakai, and J. D. Thompson, Phys. Rev. B {\bf 79}, 134507 (2009).
\bibitem{Mine2008} T. Mine, H. Yanagi, T. Kamiya, Y. Kamihara, M. Hirano, and H. Hosono, Solid State Commun. {\bf 147}, 111 (2008).
\bibitem{Terashima2009} T. Terashima, M. Kimata, H. Satsukawa, A. Harada, K. Hazama, M. Imai, S. Uji, H. Kito, A. Iyo, H. Eisaki, and H. Harima, J. Phys. Soc. Jpn. {\bf 78}, 033706 (2009).

\end{thebibliography}
\end{document}